\documentclass[10pt,twoside,openright]{report}
 \RequirePackage[paperwidth=17cm,paperheight=24cm,inner=15mm,outer=20mm,top=25mm,bottom=30mm]{geometry}

\usepackage[utf8]{inputenc}
\usepackage{amsmath}
\usepackage{amssymb}
\usepackage{slashed}

\usepackage[font=small,labelfont=bf]{caption} 

\usepackage{perpage}

\usepackage{cite}
\usepackage{nameref}

\usepackage[
colorlinks=true,urlcolor=blue,anchorcolor=blue,citecolor=blue,filecolor=blue,linkcolor=blue,menucolor=blue,pagecolor=blue,linktocpage=true,pdfproducer=medialab,pdfa=true]{hyperref}

\usepackage{cite} 

\usepackage{fancyhdr}

\fancypagestyle{plain}{%
	\fancyhf{}%
	\fancyhead[LE,RO]{\textbf{\thepage}}   
	
}
\addtocontents{toc}{\protect\thispagestyle{plain}}

\usepackage{graphicx}
\usepackage{cite}
\usepackage{relsize}
\usepackage{subcaption}

\usepackage{float} 

\usepackage[dvipsnames]{xcolor}

\usepackage{tcolorbox}
\makeatletter
\newcommand{\mybox}[1]{%
  \setbox0=\hbox{#1}%
  \setlength{\@tempdima}{\dimexpr\wd0+13pt}%
  \begin{tcolorbox}[colframe=red,boxrule=0.5pt,arc=4pt,
      left=6pt,right=6pt,top=6pt,bottom=6pt,boxsep=0pt,width=\@tempdima]
    #1
  \end{tcolorbox}
}
\makeatother

\usepackage{tikz}
\usetikzlibrary{positioning,arrows}
\usetikzlibrary{decorations.pathmorphing}
\usetikzlibrary{decorations.markings}
\usepackage{pgfplots}
\usepackage{xparse}
\usetikzlibrary{positioning,arrows,patterns}
\usetikzlibrary{decorations.markings}
\usetikzlibrary{calc}

\usepackage{xspace}
\usepackage{bm}

\usepackage{ifthen} 
\newboolean{uprightparticles}
\setboolean{uprightparticles}{false} 

\usepackage{longtable,array}
\usepackage{tabularx}

\usepackage{xspace} 
\usepackage{upgreek}

\def\lhcb {\mbox{LHCb}\xspace}

\def\lhc    {\mbox{LHC}\xspace}

\def\MagUp {\mbox{\em Mag\kern -0.05em Up}\xspace}

\def\hltone {HLT1\xspace}
\def\hlttwo {HLT2\xspace}

\ifthenelse{\boolean{uprightparticles}}%
{

 \def\Pmu         {\ensuremath{\upmu}\xspace}

 \def\Ppi         {\ensuremath{\uppi}\xspace}

 \def\Ptau        {\ensuremath{\uptau}\xspace}

 \def\Ppsi        {\ensuremath{\uppsi}\xspace}

 \def\PDelta      {\ensuremath{\Delta}\xspace}                 
 \def\PXi      {\ensuremath{\Xi}\xspace}                 
 \def\PLambda      {\ensuremath{\Lambda}\xspace}                 
 \def\PSigma      {\ensuremath{\Sigma}\xspace}                 
 \def\POmega      {\ensuremath{\Omega}\xspace}                 
 \def\PUpsilon      {\ensuremath{\Upsilon}\xspace}                 
 
 \def\PB      {\ensuremath{\mathrm{B}}\xspace}                 
                  
 \def\PD      {\ensuremath{\mathrm{D}}\xspace}

 \def\PH      {\ensuremath{\mathrm{H}}\xspace}                 
                  
 \def\PJ      {\ensuremath{\mathrm{J}}\xspace}                 
 \def\PK      {\ensuremath{\mathrm{K}}\xspace}

 \def\PW      {\ensuremath{\mathrm{W}}\xspace}

 \def\Pb      {\ensuremath{\mathrm{b}}\xspace}                 
 \def\Pc      {\ensuremath{\mathrm{c}}\xspace}                 
                  
 \def\Pe      {\ensuremath{\mathrm{e}}\xspace}

 \def\Pi      {\ensuremath{\mathrm{i}}\xspace}

 \def\Pp      {\ensuremath{\mathrm{p}}\xspace}                 
 \def\Pq      {\ensuremath{\mathrm{q}}\xspace}                 
                  
 \def\Ps      {\ensuremath{\mathrm{s}}\xspace}

}
{

 \def\Pmu         {\ensuremath{\mu}\xspace}

 \def\Ppi         {\ensuremath{\pi}\xspace}

 \def\Ptau        {\ensuremath{\tau}\xspace}

 \def\Ppsi        {\ensuremath{\psi}\xspace}                 
                  
 \mathchardef\PDelta="7101
 \mathchardef\PXi="7104
 \mathchardef\PLambda="7103
 \mathchardef\PSigma="7106
 \mathchardef\POmega="710A
 \mathchardef\PUpsilon="7107
                  
 \def\PB      {\ensuremath{B}\xspace}                 
                  
 \def\PD      {\ensuremath{D}\xspace}

 \def\PH      {\ensuremath{H}\xspace}                 
                  
 \def\PJ      {\ensuremath{J}\xspace}                 
 \def\PK      {\ensuremath{K}\xspace}

 \def\PW      {\ensuremath{W}\xspace}

 \def\Pb      {\ensuremath{b}\xspace}                 
 \def\Pc      {\ensuremath{c}\xspace}                 
                  
 \def\Pe      {\ensuremath{e}\xspace}

 \def\Pi      {\ensuremath{i}\xspace}

 \def\Pp      {\ensuremath{p}\xspace}                 
 \def\Pq      {\ensuremath{q}\xspace}                 
                  
 \def\Ps      {\ensuremath{s}\xspace}

}

\makeatletter
\ifcase \@ptsize \relax
  \newcommand{\miniscule}{\@setfontsize\miniscule{4}{5}}
\or
  \newcommand{\miniscule}{\@setfontsize\miniscule{5}{6}}
\or
  \newcommand{\miniscule}{\@setfontsize\miniscule{5}{6}}
\fi
\makeatother

\DeclareRobustCommand{\optbar}[1]{\shortstack{{\miniscule (\rule[.5ex]{1.25em}{.18mm})}
  \\ [-.7ex] $#1$}}

\def\ep         {{\ensuremath{\Pe^+}}\xspace}
\def\epm        {{\ensuremath{\Pe^\pm}}\xspace}

\def\mup        {{\ensuremath{\Pmu^+}}\xspace}

\def\taup       {{\ensuremath{\Ptau^+}}\xspace}

\def\W      {{\ensuremath{\PW}}\xspace}
\def\Wp     {{\ensuremath{\PW^+}}\xspace}

\def\quark     {{\ensuremath{\Pq}}\xspace}
\def\quarkbar  {{\ensuremath{\overline \quark}}\xspace}
\def\qqbar     {{\ensuremath{\quark\quarkbar}}\xspace}

\def\squark    {{\ensuremath{\Ps}}\xspace}

\def\cquark    {{\ensuremath{\Pc}}\xspace}

\def\bquark    {{\ensuremath{\Pb}}\xspace}

\def\pion   {{\ensuremath{\Ppi}}\xspace}
\def\piz    {{\ensuremath{\pion^0}}\xspace}

\def\pip    {{\ensuremath{\pion^+}}\xspace}
\def\pim    {{\ensuremath{\pion^-}}\xspace}
\def\pipm   {{\ensuremath{\pion^\pm}}\xspace}

\def\kaon    {{\ensuremath{\PK}}\xspace}
  \def\Kbar    {{\kern 0.2em\overline{\kern -0.2em \PK}{}}\xspace}

\def\KorKbar    {\kern 0.18em\optbar{\kern -0.18em K}{}\xspace}
\def\Kz      {{\ensuremath{\kaon^0}}\xspace}

\def\Kp      {{\ensuremath{\kaon^+}}\xspace}
\def\Km      {{\ensuremath{\kaon^-}}\xspace}
\def\Kpm     {{\ensuremath{\kaon^\pm}}\xspace}

\def\KS      {{\ensuremath{\kaon^0_{\mathrm{ \scriptscriptstyle S}}}}\xspace}

\def\Kstarzb {{\ensuremath{\Kbar{}^{*0}}}\xspace}

  \def\Dbar    {{\kern 0.2em\overline{\kern -0.2em \PD}{}}\xspace}
\def\D       {{\ensuremath{\PD}}\xspace}

\def\DorDbar    {\kern 0.18em\optbar{\kern -0.18em D}{}\xspace}
\def\Dz      {{\ensuremath{\D^0}}\xspace}

\def\Dp      {{\ensuremath{\D^+}}\xspace}

\def\Dsp     {{\ensuremath{\D^+_\squark}}\xspace}

\def\B       {{\ensuremath{\PB}}\xspace}
\def\Bbar    {{\ensuremath{\kern 0.18em\overline{\kern -0.18em \PB}{}}}\xspace}

\def\BorBbar    {\kern 0.18em\optbar{\kern -0.18em B}{}\xspace}
\def\Bz      {{\ensuremath{\B^0}}\xspace}
\def\Bzb     {{\ensuremath{\Bbar{}^0}}\xspace}
\def\Bu      {{\ensuremath{\B^+}}\xspace}

\def\Bp      {{\ensuremath{\Bu}}\xspace}

\def\Bd      {{\ensuremath{\B^0}}\xspace}

\def\jpsi     {{\ensuremath{{\PJ\mskip -3mu/\mskip -2mu\Ppsi\mskip 2mu}}}\xspace}

  \def\Y#1S{\ensuremath{\PUpsilon{(#1S)}}\xspace}

\def\proton      {{\ensuremath{\Pp}}\xspace}

\def\Xires       {{\ensuremath{\PXi}}\xspace}

\def\Lz          {{\ensuremath{\PLambda}}\xspace}
\def\Lbar        {{\ensuremath{\kern 0.1em\overline{\kern -0.1em\PLambda}}}\xspace}
\def\LorLbar    {\kern 0.18em\optbar{\kern -0.18em \PLambda}{}\xspace}

\def\Lb      {{\ensuremath{\Lz^0_\bquark}}\xspace}

\def\Lc      {{\ensuremath{\Lz^+_\cquark}}\xspace}

\def\Xib     {{\ensuremath{\Xires_\bquark}}\xspace}

\def\Xibm    {{\ensuremath{\Xires^-_\bquark}}\xspace}

\def\Xic     {{\ensuremath{\Xires_\cquark}}\xspace}
\def\Xicz    {{\ensuremath{\Xires^0_\cquark}}\xspace}
\def\Xicp    {{\ensuremath{\Xires^+_\cquark}}\xspace}

\def\Xicbarz {{\ensuremath{\Xiresbar{}_\cquark^0}}\xspace}

\newcommand{\decay}[2]{\ensuremath{#1\!\to #2}\xspace}         

\def\to                 {\ensuremath{\rightarrow}\xspace}

\def\order   {{\ensuremath{\mathcal{O}}}\xspace}

\def\CP                {{\ensuremath{C\!P}}\xspace}
\def\CPT               {{\ensuremath{C\!PT}}\xspace}

\def\AT#1     {\ensuremath{A_{\mathrm{T}}^{#1}}\xspace}           

\def\C#1      {\ensuremath{\mathcal{C}_{#1}}\xspace}                       
\def\Cp#1     {\ensuremath{\mathcal{C}_{#1}^{'}}\xspace}                    
\def\Ceff#1   {\ensuremath{\mathcal{C}_{#1}^{\mathrm{(eff)}}}\xspace}        
\def\Cpeff#1  {\ensuremath{\mathcal{C}_{#1}^{'\mathrm{(eff)}}}\xspace}       
\def\Ope#1    {\ensuremath{\mathcal{O}_{#1}}\xspace}                       
\def\Opep#1   {\ensuremath{\mathcal{O}_{#1}^{'}}\xspace}                    

\newcommand{\tev}{\ensuremath{\mathrm{\,Te\kern -0.1em V}}\xspace}
\newcommand{\gev}{\ensuremath{\mathrm{\,Ge\kern -0.1em V}}\xspace}
\newcommand{\mev}{\ensuremath{\mathrm{\,Me\kern -0.1em V}}\xspace}
\newcommand{\kev}{\ensuremath{\mathrm{\,ke\kern -0.1em V}}\xspace}
\newcommand{\ev}{\ensuremath{\mathrm{\,e\kern -0.1em V}}\xspace}
\newcommand{\gevc}{\ensuremath{{\mathrm{\,Ge\kern -0.1em V\!/}c}}\xspace}
\newcommand{\mevc}{\ensuremath{{\mathrm{\,Me\kern -0.1em V\!/}c}}\xspace}
\newcommand{\gevcc}{\ensuremath{{\mathrm{\,Ge\kern -0.1em V\!/}c^2}}\xspace}
\newcommand{\gevgevcccc}{\ensuremath{{\mathrm{\,Ge\kern -0.1em V^2\!/}c^4}}\xspace}
\newcommand{\mevcc}{\ensuremath{{\mathrm{\,Me\kern -0.1em V\!/}c^2}}\xspace}

\def\m    {\ensuremath{\mathrm{ \,m}}\xspace}

\def\cm   {\ensuremath{\mathrm{ \,cm}}\xspace}

\def\mm   {\ensuremath{\mathrm{ \,mm}}\xspace}

\def\mum  {\ensuremath{{\,\upmu\mathrm{m}}}\xspace}

\def\invfb   {\ensuremath{\mbox{\,fb}^{-1}}\xspace}

\def\sec  {\ensuremath{\mathrm{{\,s}}}\xspace}

\def\ns   {\ensuremath{{\mathrm{ \,ns}}}\xspace}
\def\ps   {\ensuremath{{\mathrm{ \,ps}}}\xspace}

\def\mhz  {\ensuremath{{\mathrm{ \,MHz}}}\xspace}

\def\degk {\ensuremath {\mathrm{ K}}\xspace}

\def\order{{\ensuremath{\mathcal{O}}}\xspace}

\def\gsim{{~\raise.15em\hbox{$>$}\kern-.85em
          \lower.35em\hbox{$\sim$}~}\xspace}
\def\lsim{{~\raise.15em\hbox{$<$}\kern-.85em
          \lower.35em\hbox{$\sim$}~}\xspace}

\def\pt         {\mbox{$p_{\mathrm{ T}}$}\xspace}

\def\mrad{\ensuremath{\mathrm{ \,mrad}}\xspace}
\def\rad{\ensuremath{\mathrm{ \,rad}}\xspace}

\def\evtgen     {\mbox{\textsc{EvtGen}}\xspace}

\def\pythia     {\mbox{\textsc{Pythia}}\xspace}

\def\roofit     {\mbox{\textsc{RooFit}}\xspace}
\def\root       {\mbox{\textsc{Root}}\xspace}

\def\tell1  {TELL1\xspace}
\def\ukl1   {UKL1\xspace}

\newcommand{\eg}{\mbox{\itshape e.g.}\xspace}
\newcommand{\ie}{\mbox{\itshape i.e.}\xspace}

\newcommand{\br}{\ensuremath{\mathcal{B}} }

\def\pr          {{\ensuremath{p}}\xspace}
\def\PSigmap      {\ensuremath{\PSigma^+}\xspace}                 
\def\Lb      {{\ensuremath{\Lz^0_\bquark}}\xspace}

\def\Lc      {{\ensuremath{\Lz^+_\cquark}}\xspace}

\def\Xicz    {{\ensuremath{\PXi^0_\cquark}}\xspace}
\def\Xicp    {{\ensuremath{\PXi^+_\cquark}}\xspace}

\def\Xibm    {{\ensuremath{\PXi^-_\bquark}}\xspace}

\def\Xiz    {\ensuremath{\PXi^0}\xspace}
\def\Xim    {\ensuremath{\PXi^-}\xspace}

\def\C               {{\ensuremath{C}}\xspace}
\def\P               {{\ensuremath{P}}\xspace}
\def\T               {{\ensuremath{T}}\xspace}

\def\W               {{\ensuremath{\rm W}}\xspace}

\def\murad  {\ensuremath{{\,\upmu\mathrm{rad}}}\xspace}

\newcommand{\lz}{\ensuremath{{\mathit{\Lambda}}}\xspace}
\newcommand{\Lprp}{\Lambda \to p \pi^-}

\newcommand{\lengthtwo}{\ensuremath{L_{C2}}\xspace}
\newcommand{\Lone}{\ensuremath{L_1}\xspace}
\newcommand{\Ltwo}{\ensuremath{L_2}\xspace}
\newcommand{\lindone}{\ensuremath{\theta_{L1}}\xspace}
\newcommand{\lindtwo}{\ensuremath{\theta_{L2}}\xspace}
\newcommand{\ecm}{\ensuremath{e\,\text{cm}}\xspace}

\newcommand{\LctoLpi}{\ensuremath{\Lc\to\Lz(p\pim)\pip}\xspace}

\newcommand{\pot}{\ensuremath{\mathrm{\,PoT}}\xspace}
\def\Sb              {{\ensuremath{\rm S_2}}\xspace}

\newcommand{\CHa}{}
\newcommand{\CHb}{}
\newcommand{\CHc}{}

\def\at   {{\ensuremath{ a }}\xspace}

\def\dt   {{\ensuremath{ d }}\xspace}

\def\deltat   {{\ensuremath{ \delta }}\xspace}

\def\Ptau        {\ensuremath{\tau}\xspace}                 
\def\Pds         {\ensuremath{\D_\squark}\xspace}

\def\thetay   {{\ensuremath{ \theta_{\y} }}\xspace}
\def\thetayDsTau   {{\ensuremath{ \theta_{\y,\Pds\Ptau} }}\xspace}

\def\xaxis   {{\ensuremath{x}}\xspace}
\def\yaxis   {{\ensuremath{y}}\xspace}
\def\zaxis   {{\ensuremath{z}}\xspace}

\def\y   {{\ensuremath{\sc y}}\xspace}
\def\z   {{\ensuremath{\sc z}}\xspace}

\def\Lc   {{\ensuremath{ L }}\xspace}
\def\thc   {{\ensuremath{ \theta_C }}\xspace}
\def\Ltarc   {{\ensuremath{ L_{\mathrm{tar}} }}\xspace}
\def\phad   {{\ensuremath{ p_{\mathrm 3\pi} }}\xspace}

\def\W               {{\ensuremath{\rm W}}\xspace}
\def\Si               {{\ensuremath{\rm Si}}\xspace}
\def\Ge              {{\ensuremath{\rm Ge}}\xspace}

\def\Sb              {{\ensuremath{\rm \sf S2}}\xspace}

\def\PSigmap      {\ensuremath{\PSigma^+}\xspace}                 
\def\Lb      {{\ensuremath{\Lz^0_\bquark}}\xspace}

\def\Lc      {{\ensuremath{\Lz^+_\cquark}}\xspace}
\def\Xicz    {{\ensuremath{\PXi^0_\cquark}}\xspace}
\def\Xicbarz    {{\ensuremath{\overline{\PXi}^0_\cquark}}\xspace}
\def\Xicp    {{\ensuremath{\PXi^+_\cquark}}\xspace}

\def\Xibm    {{\ensuremath{\PXi^-_\bquark}}\xspace}

\def\Xibp    {{\ensuremath{\overline{\PXi}^+_\bquark}}\xspace}  
\def\Omegabp    {{\ensuremath{\overline{\POmega}^+_\bquark}}\xspace}   
\def\Xiz    {\ensuremath{\PXi^0}\xspace}
\def\Xim    {\ensuremath{\PXi^-}\xspace}
\def\Xip    {\ensuremath{\overline{\PXi}^+}\xspace}  
  
\newcommand{\muvec}{\ensuremath{\bm{\mu}}\xspace}
\newcommand{\Svec}{\ensuremath{\bm{S}}\xspace}
\newcommand{\Evec}{\ensuremath{\bm{E}}\xspace}
\newcommand{\Bvec}{\ensuremath{\bm{B}}\xspace}
\newcommand{\s}{\ensuremath{\text{s}}\xspace}

\def\BdToJpsiKS{$\Bz\rightarrow\jpsi\KS$\xspace}
\def\LbToJpsiLz{$\Lb\rightarrow\jpsi\Lz$\xspace}

\newcommand{\spol}{\ensuremath{\bm{s}}\xspace}

\newcommand{\thC}{\ensuremath{\theta_C}\xspace}
\newcommand{\lenC}{\ensuremath{L_C}\xspace}
\newcommand{\tilt}{\ensuremath{\theta_{y,C}}\xspace}

\newcommand{\thy}{\ensuremath{\theta_y}\xspace}
\newcommand{\LcpKpi}{\ensuremath{\Lc\to\pr\Km\pip}\xspace}

\newcommand{\lag}{\ensuremath{\mathcal{L}}\xspace}

\newcommand{\abbr}[2]{#1 - #2}

\def\threepi{{\ensuremath{\Lc\to\Lz \pip \pip \pim}}\xspace}
\def\kpipi{{\ensuremath{\Xicp\to\Lz \Km \pip \pim}}\xspace}
\def\Lzppi{{\ensuremath{\Lz \to p \pim}}\xspace}

\def\pol{{\ensuremath{\bm{P_{\Lambda}}}\xspace}}

\def\Sigmacp{{\ensuremath{\mathit\Sigma_c^+}}\xspace}

\def\Sigmapm{{\ensuremath{\mathit\Sigma^\pm}}\xspace}

\def\lDD{{\ensuremath{\Lz_{\rm DD}}}\xspace}
\def\lLL{{\ensuremath{\Lz_{\rm LL}}}\xspace}

\newcommand{\p}[2]{{\bf p}_{#1}^{(#2)}} 

\def\mLc{{\ensuremath{m_\Lc}}\xspace}

\usepackage{marvosym}
\usepackage{wasysym}

\newcommand{\itcut}{\item[\color{black}\large \Rightscissors]}
\newcommand{\itcont}{\item[\color{blue}\normalsize \XBox]}
\newcommand{\itfun}{\item[\color{red}$\blacktriangle$]}

\newcommand{\firstitem}{\color{black}$\blacksquare$~}

\newcommand{\firstcont}{{\Large \color{blue} \XBox~}}
\newcommand{\firstfun}{{\color{red}$\blacktriangle$~}}

\newcommand{\elzero}{L0\xspace}

\newcommand{\als}{\ensuremath{\alpha_s}\xspace}
\newcommand{\ale}{\ensuremath{\alpha_e}\xspace}

\newcommand{\Rqed}{{\hat R}^{\text{qed}}}
\newcommand{\Uqcd}{{\hat U}^{\text{qcd}}}

\newcommand{\Uzero}{{\hat U}^{0,\text{qcd}}}
\newcommand{\Uone}{{\hat U}^{1,\text{qcd}}}

\newcommand{\cOale}{\ensuremath{{\cal O}(\ale)}\xspace}

\newcommand{\dq}{\ensuremath{d_q}\xspace}
\newcommand{\dtildeq}{\ensuremath{\widetilde{d}_q}\xspace}
\newcommand{\dc}{\ensuremath{d_c}\xspace}
\newcommand{\dtildec}{\ensuremath{\widetilde{d}_c}\xspace}
\newcommand{\db}{\ensuremath{d_b}\xspace}
\newcommand{\dtildeb}{\ensuremath{\widetilde{d}_b}\xspace}

\newcommand{\Tr}{\mathrm{Tr}}

\newcommand{\imEtaU}{{\rm Im}(\eta_U)}

\newcommand{\imEtaQ}{{\rm Im}(\eta_Q)}
\newcommand{\reEtaU}{{\rm Re}(\eta_U)}

\newcommand{\reEtaQ}{{\rm Re}(\eta_Q)}

\def\be{\begin{equation}}
\def\ee{\end{equation}}

\newcommand{\mpdiag}[2]{
	\hspace*{-0.5cm}\begin{minipage}[c]{0.36\textwidth}
		#1
	\end{minipage}
	\begin{minipage}[c]{0.61\textwidth}
		#2
	\end{minipage}
}

\definecolor{dGray}{gray}{0.4}

\newcommand{\rev}[1]{{\color{red} {#1}}}
\newcommand{\revmaybe}[1]{{\color{dGray} {#1}}}

\newcommand{\copypasted}[2]{{\color{dGray} \textbf{copy-pasted from #1:} \\ {#2}}}

\newcommand{\red}[1]{\textcolor{red}{#1}}

\newcommand{\revafter}[1]{} 

\makeatletter
\def\input@path{{analysis/}}
\makeatother

\graphicspath{{analysis/}}

\usepackage{comment}
\includecomment{mynote}
\excludecomment{mynote}

\includecomment{LARGETABLESFIGURES}
\excludecomment{LARGETABLESFIGURES}

\author{Joan Ruiz Vidal }

\title{ 
Experiments and phenomenology of\\ electric dipole moments \\
}

\begin{document}


\pagenumbering{roman}

\setcounter{page}{1}

\newpage
\begin{titlepage}
\thispagestyle{empty}
\begin{center}
  {\huge \bf  Experiments and phenomenology of}\\ {\huge \bf electric dipole moments}
    \vfill
  

   \begin{center}
   \includegraphics[scale=0.5]{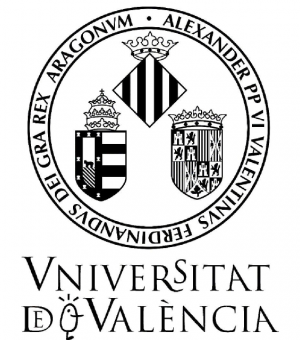}
   \end{center}

  \vfill

  {\Large Tesi doctoral\\[0mm]Programa de Doctorat en F\'isica\\[1mm] Juny 2022}
  
  \vspace*{0.9cm}

  {\large\bf Joan Ruiz Vidal}

  {\large IFIC, Universitat de Val\`encia - CSIC\\
          Departament de F\'isica Atòmica, Molecular i Nuclear}
          
  \vspace*{0.9cm}
          
  {\large Director de tesi: \\ Fernando Martínez Vidal}

\end{center}

		\end{titlepage} 

\newpage
\thispagestyle{empty} 

\tableofcontents


\def\titol {Abstract}
\chapter*{\titol} \addcontentsline{toc}{chapter}{\titol}

The Standard Model (SM) is the best description of fundamental particles and their interactions we have to date. From this theory, all phenomena in the macroscopic world (except for gravity) can be explained, and it has successfully predicted all outcomes of particle experiments on Earth.
However, cosmological observations of the early Universe yield a large imbalance between its content of matter and antimatter, which is several orders of magnitude above the SM prediction. 
To explain these observations, new interactions that do not respect the charge-parity symmetry must exist beyond the SM. Such interactions would induce electric dipole moments (EDMs) in known particles. 
In Part I of the thesis, we propose to extend the active experimental program of EDM searches to charm and bottom baryons, $\tau$ leptons, and \Lz hyperons; also allowing the measurement of their corresponding magnetic dipole moments (MDMs).
The EDM and MDM of short-lived particles can be accessed with a bent-crystal experiment to be installed in the Large Hadron Collider (LHC), while longer-lived \Lz particles can be measured at the LHCb experiment with no additional instrumentation. In Part II, an analysis of LHCb data to measure the \Lz polarization in \threepi decays, essential ingredient for the proposed measurement, is presented. In the last part of the thesis, we derive new indirect bounds on heavy quark EDMs with already available data and explore the phenomenological implications of these and other EDM limits on New Physics models, with special emphasis on extensions of the SM with colour-octet scalars.

\newpage
\thispagestyle{plain} 

\def\titol {List of Publications}
\chapter*{\titol} \addcontentsline{toc}{chapter}{\titol} 

This thesis is partially based on the following research articles:

	\begin{enumerate} \small
		
		\item[\cite{Botella:2016ksl}] F.J. Botella, L. M. Garc\'ia Mart\'in, D. Marangotto, F. Mart\'inez Vidal, A. Merli, N. Neri, A. Oyanguren, J.R.V., {\bf On the search for the electric dipole moment of strange and charm baryons at LHC}, \href{https://inspirehep.net/literature/1505173}{Eur.Phys.J. C77 (2017) no.3, 181.}
		
		\item[\cite{internalnote}] L. Henry, D. Marangotto, F. Martinez Vidal, A. Merli, N. Neri, P. Robbe, J.R.V., {\bf Proposal to search for baryon EDMs with bent crystals at LHCb}, LHCb-INT-2017-011 (restricted access).
		
		\item[\cite{Bagli:2017foe}] E. Bagli, L. Bandiera, G. Cavoto, V. Guidi, L. Henry, D. Marangotto, F. Mart\'inez Vidal, A.Mazzolari, A. Merli, N. Neri, J.R.V., {\bf Electromagnetic dipole moments of charged baryons with bent crystals at the LHC}, \href{https://inspirehep.net/literature/1620469}{Eur.Phys.J. C77 (2017) no.12, 828.}
		
		\item[\cite{Fu:2019utm}] J. Fu, M.A. Giorgi, L. Henry , D. Marangotto, F. Martínez Vidal, A. Merli, N. Neri, J.R.V., {\bf Novel Method for the Direct Measurement of the  $\tau$ Lepton Dipole Moments}, \href{https://inspirehep.net/literature/1713541}{Phys.Rev.Lett. 123 (2019) no.1, 011801.}

		\item[\cite{Aiola:2020yam}] F. Mart\'inez Vidal, N. Neri, J.R.V., et al., {\bf Progress towards the first measurement of charm baryon dipole moments}, \href{https://inspirehep.net/literature/1824423}{Phys.Rev.D 103 (2021) 7, 072003.}

		\item[\cite{Biryukov:2021cml}] V.M. Biryukov, J.R.V., {\bf Improved experimental layout for dipole moment measurements at the LHC}, \href{https://inspirehep.net/literature/1938263}{ Eur.Phys.J.C 82 (2022) 2, 149.}

		\item[\cite{DEMONSTRATOR}] LHCb collaboration, {\bf Long-lived particle reconstruction downstream of the \lhcb magnet},  \href{https://cds.cern.ch/record/2841793}{CERN-LHCb-DP-2022-001.}

		\item[\cite{Gisbert:2019ftm}] H. Gisbert, J.R.V., {\bf Improved bounds on heavy quark electric dipole moments}, \href{https://inspirehep.net/literature/1733692}{Phys.Rev.D 101 (2020) 11, 115010.}
				
		\item[\cite{Gisbert:2021htg}] H. Gisbert, V. Miralles, J.R.V., {\bf Electric dipole moments from colour-octet scalars}. \href{https://inspirehep.net/literature/1970917}{JHEP 04 (2022) 077.}
%
%
%
		
	\end{enumerate}

The content partially reproduces text and figures from these references. 
%
%
%
%
The PhD theses of my collaborators in Part I~\cite{Merli:2019hyz,Marangotto:2020tzf} may reproduce part of this content as well and there may be some overlap with them. If I was not directly involved in the production of a figure, the text of the caption will reflect the original reference starting with "From Ref.~[$x$].", following the criterion of Ref.~\cite{Marangotto:2020tzf}.
In Part III, there is some overlap with the PhD theses of H.~Gisbert~\cite{GisbertMullor:2019vwg} and V.~Miralles~\cite{ThesisVictor}.

\newpage
\thispagestyle{plain} 

\def\titol {Acknowledgements}
\chapter*{\titol} \addcontentsline{toc}{chapter}{\titol} 


First, I would like to thank my supervisor Fernando Martínez for introducing me to the world of particle physics. You gave me a most exciting topic of master thesis with the \Lz EDM proposal. From there, we started together with the bent-crystal experiment and, in all the unending discussions and brainstorming, you always made me feel as an equal. In those intense first years I learned from you much more than physics and, although we also suffered the tremendous speed, I am forever grateful for this accelerated course on \textit{how to propose an experiment at CERN}.

All these and other ideas have been (and continue to be) developed together with other members of our group in Valencia and also at the LHCb group of Milano. I am grateful to all of them for enjoyable collaborations and especially to Nicola Neri for his continuous feedback along these years. Specific mentions are also deserved to Luis Miguel García, who was always ready to help me with the LHCb framework; and Miguel Rebollo, whose work in the last months has allowed the completion of the LHCb preliminary analysis presented in the thesis. I am also grateful to Valery Biryukov for all the open discussions on crystal lenses and for making our fast-paced collaboration so easy and enjoyable for me.

I would also like to acknowledge the common effort of all members of the LHCb collaboration. None the work presented in Part II of this thesis and partially in Part I would have been possible without the contributions of hundreds of people to this phenomenal experiment. Particularly, I would like to thank those members that put effort in documenting the internal codes in the twiki pages and specially to those that contributed to the great \href{https://lhcb.github.io/starterkit-lessons/first-analysis-steps/README.html}{Starter Kit lessons}. I am also grateful to all the colleagues that provided feedback on my analysis at the Charm Working Group.

For many of us, there is no clear boundary between our personal lives and research work. Starting the theoretical studies in my PhD was only possible thanks to my friend Hector Gisbert who 
did not hesitate to start a collaboration with \textit{an experimentalist} on quite a new topic for both of us. I have had so much fun and learned a lot working with you. I hope that we never stop collaborating, wherever we are! The best possible addition to this little team was Victor Miralles who, with his model, has paved the way to produce a potentially large amount of unciteable papers. Thank you for all the laughs during our long summer of tensor decompositions and absurd numerical checks. I cannot imagine a better combination of colleague and friend than you!

I would also like to thank Toni Pich for listening to my (often naive) ideas on these topics and giving us the final support we needed to move forward with the projects of Part III. 

Of course, none of this would have been possible without the continuous support of my family and friends. Gràcies a Adrià 
per tantes vesprades i nits de converses interminables, i per donar-me eixos \textit{boosts} de motivació cada vegada que ens vegem. En la nostra col·laboració disfuncional he aprés que és més important disfrutar el viatge que la destinació. A Andreu 
pel teu suport continu des de la carrera i totes les vesprades de distraccions tan necessàries. Gracias por tan buenos momentos a todas las personas en las comidas de \textit{Escuadrón IFIC} y las cenas de \textit{Garch'n'Go}.

A Judith, gracias por compartir conmigo cada día. Por escucharme siempre \linebreak
y {acompañarme en los momentos} más duros estos años. Los obstáculos del doctorado y la pandemia se hicieron mucho más fáciles contigo. A Marina, Carlos, Rosalia i Silvia, gràcies pel vostre suport i carinyo incondicional.
Gràcies a ma mare Asun i a mon pare Josep per tot l'esforç i el carinyo. Per espentar-nos a Marina i a mi a fer el que més ens agradava, inclús als moments més complicats. A ma mare, gràcies per fer d'amiga i confident, i animarme sempre que ho he necessitat. 

\thispagestyle{plain}

\newpage

\pagenumbering{arabic}
\setcounter{page}{1}

\part{Search for dipole moments of unstable particles at the LHC} \label{part:edm}

\chapter{Introduction: electric and magnetic dipole moments} \label{ch:introexp}

\section{Introduction } \label{sec:introintro}


%

In physics we are used to making questions about the natural world such as \textit{what is matter made of?}, \textit{how does this phenomenon work?}, \textit{what are space and time?}, ... Step by step, we make \textit{theories} or \textit{models} that can describe natural phenomena and, successively, we find more and more general theories that simultaneously describe some of these phenomena, plus others that initially could not be explained. In this sense, a theory is more \textit{fundamental} if it contains the underlying explanation to more phenomena. Today, at the end of the chain of \textit{fundamentality} we find the Standard Model (SM) of particle physics, and Einstein's theory of General Relativity.

The SM, born in the 1970s, has been extremely successful in explaining and predicting the results of past and present particle physics experiments. Ultimately, this agreement between theory predictions and experimental evidence is the only thing that counts when judging the success of a theory. 
However, another aspect of fundamental theories stands out when studying the SM. Weinberg called it \textit{inevitability}~\cite{Weinberg:1992nd}; it is the fact that all the elements of the theory follow from very few initial assumptions or principles that cannot be modified. 

Despite its success, there are a few experimental observations that the SM is not able to accommodate, and others of a (more disputable) theoretical nature. To explain these, new theories \textit{beyond} the SM (BSM), also referred to as New Physics (NP), are called for. These theories contain new fundamental particles interacting with SM particles whose effects could be measured at particle physics experiments. 
There is, however, no guarantee for finding these new particles within the range of masses accessible by current or near-future particle experiments\footnote{Some NP extensions predict light particles (dark matter candidates, axion-like particles, ...) that are \textit{feebly interacting} with SM particles, or only interact with a few of them (\textit{portals}). In both of these cases, there is no guarantee that their couplings produce observable effects in finite-precision experiments either.  }. 
However, many of the most interesting theories, that also address some of the theoretical problems of the SM, 
predict the range of masses of these new particles to be of the order of few {\tev}s, which are energies, in principle, within reach of current technology.

Besides looking for signals of these particles when they are produced \textit{on shell}, in direct searches, we may also narrow down the list of candidate theories with precision measurements which are sensitive to the effects of \textit{off shell} particles running in loops. To systematically study these effects in a model-independent way it is useful to construct Effective Field Theories (EFT), valid at energies below the mass of the new particles, which can greatly simplify the calculations.

In this thesis, low-energy observables will be treated from different points of view. These include an experiment proposal to measure electric (EDM) and magnetic (MDM) dipole moments of unstable particles using bent crystals (in the rest of Part I); an experimental analysis of LHCb data, focused on production and polarization properties of a multihadronic charm baryon decay (in Part II); and two phenomenological works on EDM observables (in Part III) that use both a model-independent approach (Chapter \ref{ch:improvedbounds}) and a specific BSM theory (Chapter \ref{ch:edmsmw}).
Although some of these projects stand in quite different research frameworks, they arose as natural continuations of one another. The common thread connecting all of them will be made apparent in the introduction to each study. 

A first relevant connection can be drawn between Part I (Chapters~\ref{ch:introexp}-\ref{ch:lambdasLHCb}) and Chapter~\ref{ch:improvedbounds}.
With the bent-crystal experiment proposal, introduced in Chapter~\ref{ch:crystals}, it is possible to directly access, among other observables, the EDM of charm baryons for the first time. The motivations for this experiment will be outlined in the remainder of Chapter \ref{ch:introexp} together with the current context of EDM 
 experiments. A more quantitative discussion on the BSM predictions of heavy quark EDMs will be presented in Chapter \ref{ch:improvedbounds}, where new indirect limits on these quantities are derived using available data on the neutron EDM. Thus, with the combination of Part I and Chapter~\ref{ch:improvedbounds} a complete picture of heavy-baryon EDM observables will emerge: from the details of an experimental configuration to the predictions of specific BSM theories and the restrictive power of these observables in comparison to others in NP phenomenology. 

In between, in Part II (whose initial motivation arose from the studies presented Chapter~\ref{ch:lambdasLHCb}) we will show how fundamental polarization observables of charm baryon decays can be determined from experimental data, which are key ingredients for the EDM and MDM measurements.

\section{A bit of history} \label{sec:history}

\begin{mynote}
include:
- first paragraph: catchy story / the history of astonishing improvemtn in sensitivity
- end paragraph: this section historical perspective. The current motivation in the following subsections...
- First time observables considered
- Probably in QM, in the 30's?
- context of the time
- Back-and-forth between theoretical and experimentalists, which were not separated.
- g = 1 / 2 in QM, QFT, and first experiments ...
- baryons: role endorsing/confirming the quark model, predecessor of QCD... nowadays helping to test and build reliable models describing QCD at low energies 
- the improvements are overwhelming; analogy with crystals, just starting...

\end{mynote}

Since the 1950s, searches for the electric dipole moment of fundamental particles have improved their sensitivity by about eight orders of magnitude~\cite{Pendlebury:2000an}. In the first experiments~\cite{Smith:1957ht}, the interest on the neutron EDM was motivated solely by parity violation, which was observed in 1957~\cite{Wu:1957my} in $\beta$ decays. While the deep implications of testing this observable were not well understood at the time, this early program of experiments gave an initial momentum to the field of EDM searches. Nowadays, these low-energy experiments compete with the Large Hadron Collider (LHC) on the exploration of \CP violation beyond the SM and provide exceptional null tests of the SM.

Moreover, if EDM searches saw a trajectory of 70 years leading to their remarkable importance in today's particle physics phenomenology, the historical relevance of MDMs cannot be overstated. This observable has played a central role in the development of quantum mechanics and its relativistic version with the Dirac equation. It also gave unequivocal confirmation of quantum electrodynamics (QED) and consequently of quantum field theory itself. Even nowadays, the muon MDM represents one of the few smoking guns of NP in particle-accelerator experiments. Furthermore, the MDM of baryons played a key role establishing the quark model, 
which would lead to the development of quantum chromodynamics (QCD).

To start, the magnetic moment provided direct experimental evidence of space quantization, and allowed a direct visualization of the spin angular momentum. In 1922, with the growing experience in the use of molecular beams, Stern and Gerlach built an experiment to study the interaction of the magnetic moment of silver atoms with an inhomogeneous magnetic field~\cite{SternGerlach1,SternGerlach2,SternGerlach3}. Contradicting the classical-mechanics expectation of an uniform spread of the beam, in the experiment the beam splits in discrete lines, showing the space quantization of angular momentum. According to Schrödinger's wave theory, the beam would split in $2l+1$ lines, where $l\in\mathbb{Z}$ is the atom's angular momentum, thought to be generated by the \textit{orbiting} electron in the last shell. However, only two lines (an even number!) were visible in the experiment. The correct interpretation of this result came years later, after the proposal of the spin by Goudsmit and Uhlenbeck to explain atomic spectral lines \cite{GoudsmitUhlenbeck}. In reality, an orbital angular momentum could not generate this effect since the electron is in the $5s$ orbital, with $l=0$. However, an intrinsic angular momentum of the electron, with quantum number $s=\frac{1}{2}$, produced the $2s+1=2$ observed lines.

The magnetic moment \muvec of the electron is related to the spin angular momentum $\Svec$ as 
\begin{equation}
\muvec = g \frac{e }{2 m} \Svec ~,
\end{equation}
where $g$ is the so-called $g$-factor. 
From the analogy with a classical rotating body that is uniformly charged, one expects $g=1$, however, it was measured to be $g=2$ within uncertainty, puzzling the physics community at the time. After much theoretical development, it was shown by Dirac in 1928~\cite{Dirac:1928hu} that the exact value of $g=2$ for the electron was a consequence of the relativistically-invariant formulation of quantum mechanics, settling the debate for several years. 
However, the exciting history of the magnetic moment in twentieth-century physics does not stop there. With the transformation of Dirac's theory into a quantum theory of fields,
$g$ was no longer predicted to be exactly 2. In 1948, Schwinger published the very first calculation of a radiative correction in QED~\cite{Schwinger:1948iu}, precisely on the \textit{anomalous} magnetic moment of the electron, yielding $(g-2)/2\approx0.001$. His prediction was confirmed experimentally by Kush and Foley~\cite{Kusch:1948mvb} the same year. From there, theory and experiment kept increasing the accuracy of their predictions and measurement, making the electron magnetic moment one of the most accurate tests of the SM, with its impressive agreement at the 11th significant digit. A similar trajectory has been followed by the muon magnetic moment. With its also impressive accuracy of eight significant figures, this observable is today hinting at the existence of physics beyond the SM, showing a long-standing tension with its prediction exceeding the $3\sigma$ level.

For hadrons, the MDMs have been measured for the lowest-lying baryon octet with $J^P = \frac{1}{2}^+$. Historically, the agreement between the measured MDM and predictions of the quark model for the baryon octet was crucial to assess the constituent quark models of hadrons~\cite{Fritzsch:2015jfa}.

In the following sections, we will briefly see the motivations to measure these observables within today's context of particle physics. We will also make the case for the EDM and MDM of heavy baryons and $\tau$ lepton, which can be measured with a bent-crystal experiment, explored in the next chapters.

\begin{mynote}

\rev{pot ser incloure també alguna paraula sobre el MDM dels barions, que serveix per confirmar el model quark....}


\revmaybe{quark model and mdm of baryons: \\
	Observation shows that p and n are not point-like $g_{p,n} \neq 2$ $\to$ evidence for quarks.
	
	Veure slide 31 de T. Potter.}

\copypasted{TFM}{Precise definition (not to put here)\\
	The magnetic and electric dipole moment operators, $\bm{\hat \mu}$ and $\bm{\hat \delta}$, are proportional to the spin, $\bm{\hat S}$. The proportionality constants are the gyromagnetic and gyroelectric ratios, respectively. They can be parametrized as
	\begin{equation*}
	\bm{\hat \mu} = g \mu_N \frac{\bm{\hat S}}{\hbar} ~~, ~~~~  \bm{\hat \delta} = d \mu_N \frac{\bm{\hat S}}{\hbar} ~~,
	\end{equation*}
	where the constants $g$ and $d$ are the dimensionless magnetic and electric dipole moments. They contain all the information about these quantities and we shall refer to them simply as MDM, for $g$, and EDM, for $d$. The constant $\mu_N$ is the nuclear magneton, $\mu_N = e \hbar / 2 m_p $ in SI units ($\mu_N = e \hbar / 2 m_p c$ in Gaussian units).
	
	Taking $z$ as the quantization axis, there are $2s+1$ eigenvalues for the third component of the spin, with $s \hbar$ being its maximum value. Consequently, the maximum third components of $\bm \mu$ and $\bm \delta$ are
	
	%
	%
	%
	\begin{equation*}
	\mu \equiv \mu_{z,\text{max}} = g \mu_N s ~~, ~~~~ \delta \equiv \delta_{z,\text{max}} = d \mu_N s ~~.
	\end{equation*}

	In the case of spin-1/2 particles (e.g. the $\lz$ baryon), and using the basis of eigenvalues of $S_z$, $| \chi_{+,-} \rangle = \left|  \frac{1}{2}, \pm \frac{1}{2} \right\rangle $, the spin operator is represented by the Pauli matrices as $\bm{\hat S} \to \frac{\hbar}{2} \bm \sigma $.
	
	The expectation values of the spin operators form the polarization vector $\langle \bm{\hat S} \rangle = \frac{\hbar}{2} \textbf{P}$.
	
	}

~\\
~\\
~\\

\end{mynote}

\section{Electric dipole moments} \label{sec:edm}

We shall start by briefly reviewing the role of \CP violation in the first instant after the Big Bang and the observables in particle-physics experiments that can test it. Then, we will focus on EDM experiments exploring the motivations to start a new program of measurements for very-short-lived particles.

\subsection{Matter-antimatter imbalance}



There is a large imbalance between the number of particles and antiparticles in the Universe. Even if an asymmetry was initially present at the Big Bang ($t=0$) it would have been wiped out by the end of the inflationary epoch ($t\sim10^{-32}\,\s$)~\cite{Guth:1980zm}. Also, from the success of Big Bang Nucleosynthesis~\cite{Iocco:2008va} and our general understanding of low-energy phenomena we know that this asymmetry should have been generated dynamically at the electroweak phase transition or before ($t\lesssim10^{-11}\,\s$). Whatever the explicit mechanism for baryogenesis is, it must satisfy three conditions formulated by Sakharov in 1967~\cite{Sakharov:1967dj}:

	\begin{itemize}
		\item baryon number violation;
		\item \C and \CP violation;
		\item out-of-equilibrium dynamics.
	\end{itemize}
	These three conditions are met by the SM but the amount of \CP violation induced by the phase of the Cabibbo-Kobayashi-Maskawa (CKM) matrix cannot account for the observed baryon asymmetry, here normalized to the photon density~\cite{Bennett:2012zja}:
	\begin{equation}
	\left.\frac{n_\B - n_{\Bbar} }{n_{\gamma}} \right|_{\rm obs} = (6.079 \pm 0.090)\times 10^{-10}\; , \quad 
	\left.\frac{n_\B - n_{\Bbar} }{n_{\gamma}}\right|_{\rm SM} \sim 10^{-18}~.
	\end{equation}
	Thus, new sources of  \CP violation beyond the SM must exist. To narrow down \textit{which} of these sources is underlying, observables sensitive to \CP violation need to be measured. 
	To date, \CP violation has only been observed in the mixing and/or decay of neutral mesons ($\Kz$, $B^0_{(s)}$ and $\Dz$) and in the decay of charged B mesons.
	All of the experimental results are in agreement with the SM-CKM prediction within experimental and theoretical uncertainties. Reducing these is necessary to continue searching for deviations from the SM.
	
	Another approach to \CP violation is possible through EDM searches. The situation is quite different in that case since the SM prediction is tiny \revafter{(as we will see in more detail in Chapter \ref{ch:introtheo})}and it represents a negligible background. Thus, any signal of a non-zero EDM in current and planned experiments would be a sign of BSM physics. In addition, a direct signal of an EDM would be an undisputed sign of \T violation, which has only been observed in \Kz decays and entangled $\Bz\Bzb$ systems~\cite{Angelopoulos:1998dv,Lees:2012uka,Bernabeu:2014mva}.

\subsection{\P, \T and \CP violation}

\begin{figure*}[htb]
	\centering
	{ \includegraphics[width=0.45\linewidth]{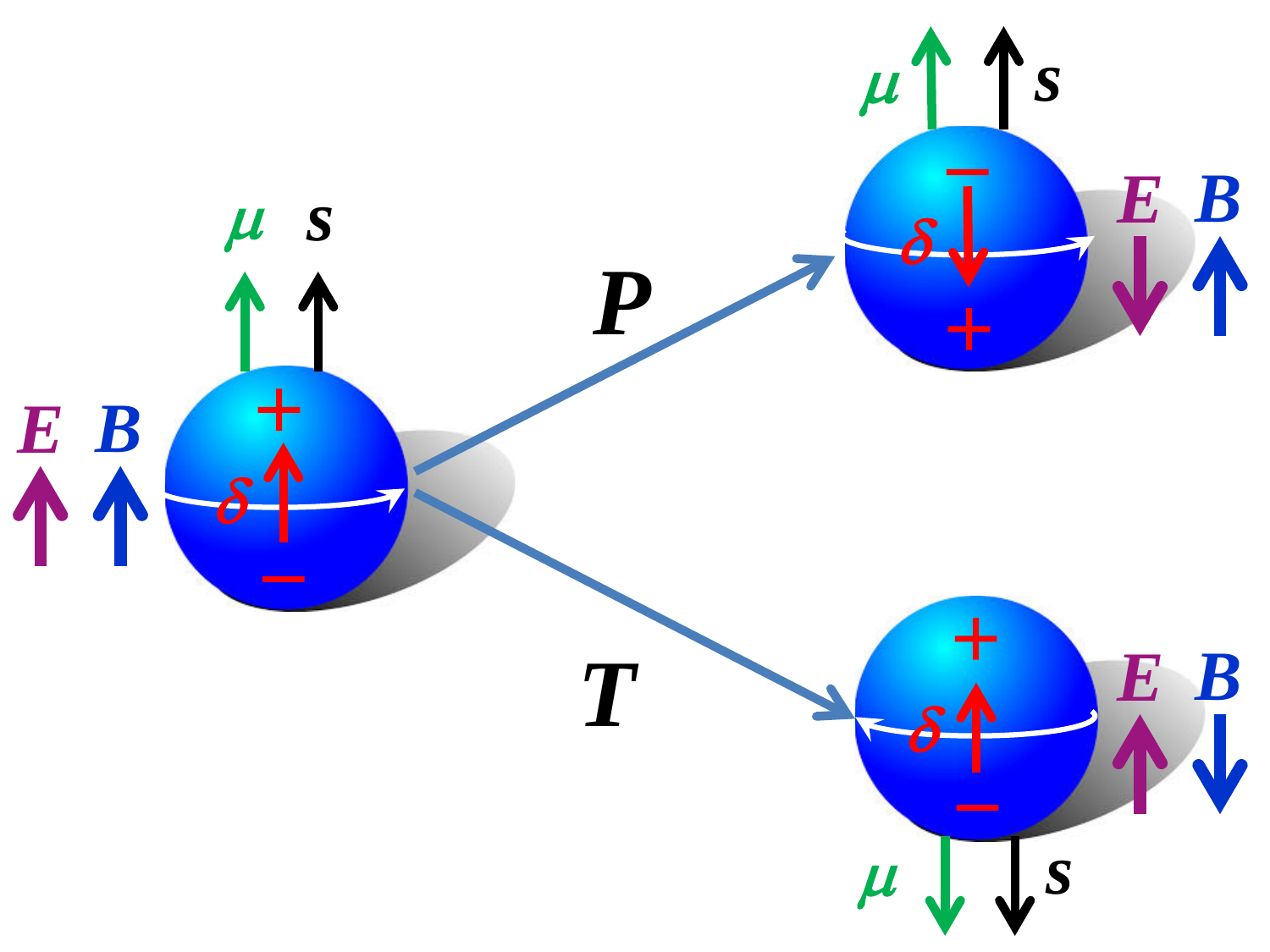} }
	\caption{ A particle with spin ${\bf s}$ is represented as a sphere
		with a spinning charge distribution. Its images through \P and \T are also shown, together with the corresponding
		particle magnetic $\bm \mu$ and electric $\bm{\delta}$ moments, and the external magnetic ${\bm B}$ and electric ${\bm E}$ fields. The represented EDM direction $\delta$ follows the distribution of charge, as in its classical defintion. In quantum mechanics it transforms as the spin.}
	\label{fig:EDM_Symmetry}
\end{figure*}

In classical mechanics, the EDM of a charge distribution $\rho(\bm{r})$ is defined as $\bm{\delta} \equiv \int \bm{r} \rho(\bm{r}) d^3 r$ and 
quantifies the separation of positive and negative electric charges in the system. In subatomic systems, this vector-like quantity is 
either parallel or antiparallel to the spin-polarization vector\footnote{The spin-polarization vector
	is defined as 
	$\bm s \equiv \langle \bm{\hat S} \rangle / (\hbar/2)$, 
	where $\bm{\hat S}$ is the spin operator.
}, $\bm{\delta} \propto \bm{s}$, which is the only intrinsic direction defined by the system. 
The former transforms under parity and time reversal as a polar vector, 
while the later does as an axial vector. The existence of an EDM in fundamental particles therefore requires the violation 
of both \P and \T symmetries (and, relying on the \CPT theorem, \CP). 
This can be seen directly by applying these transformations to the interaction term of the classical Hamiltonian. Considering also the magnetic dipole moment $\bm{\mu}$, which is already an axial vector in its classical description, we have 
%
%
\begin{align} \label{eq:hamiltonianclassic}
& \mathcal{H} = -\bm \mu \cdot {\bf B} -\bm \delta \cdot {\bf E} \ \stackrel{P}\longrightarrow \   
\mathcal{H} = -\bm \mu \cdot {\bf B} +\bm \delta \cdot {\bf E}~,  \\
& \mathcal{H} = -\bm \mu \cdot {\bf B} -\bm \delta \cdot {\bf E} \ \stackrel{T}\longrightarrow \   
\mathcal{H} = -\bm \mu \cdot {\bf B} +\bm \delta \cdot {\bf E}~. \nonumber 
\end{align}
The term proportional to $\bm{\delta}$ changes sign under \P and \T.
Hence, the existence of a  $\bm{\delta}\neq 0$ requires a breaking of the \T and \P symmetries~\cite{Pospelov:2005pr}.
Figure~\ref{fig:EDM_Symmetry} illustrates the effect of these two symmetries on a system with a magnetic and electric dipole moment.

In quantum field theory the EDM of any spin-$\frac{1}{2}$ particle is defined as the coupling 
constant $\delta$ of the operator\footnote{Along the thesis, the EDM of baryons is noted with the symbol $\delta$. Its corresponding adimensional quantity, defined later, will be noted $d$. The (dimensional) Wilson coefficient for the quark EDM will be $d_q$.}
\begin{equation} \label{eq:EDMoperator}
-\frac{i}{2} ~ \delta ~ \bar \psi  \sigma^{\mu\nu} \gamma_5 \psi  F_{\mu \nu} ~,
\end{equation}
where $\psi$ is the Dirac spinor, $F^{\mu \nu}=\partial_\mu A_\nu - \partial_\nu A_\mu$ the electromagnetic strength tensor, and $\gamma_5$ and $\sigma^{\mu\nu}$ are the commonly defined products of Dirac matrices.
Thus, in terms of QFT vertices, the EDM of a particle is just a \CP-violating coupling of a fermionic line with an external photon. In general, all photon couplings can be parameterized by four linearly independent electromagnetic form factors~\cite{Nowakowski:2004cv}, $F_i(q^2),~i=1,2,3,4$, where $q^2$ is the squared four-momentum of the photon. For instance, for a spin-$1/2$ baryon $B$, its transition amplitude with the general electromagnetic current $J_{em}^{\mu}$ can be written as (notation taken from \cite{Guo:2012vf}) 

{ \footnotesize{
	\begin{align}\label{eq:formfactors}
	&\raisebox{-0.8cm}{\includegraphics[width=0.3\textwidth]{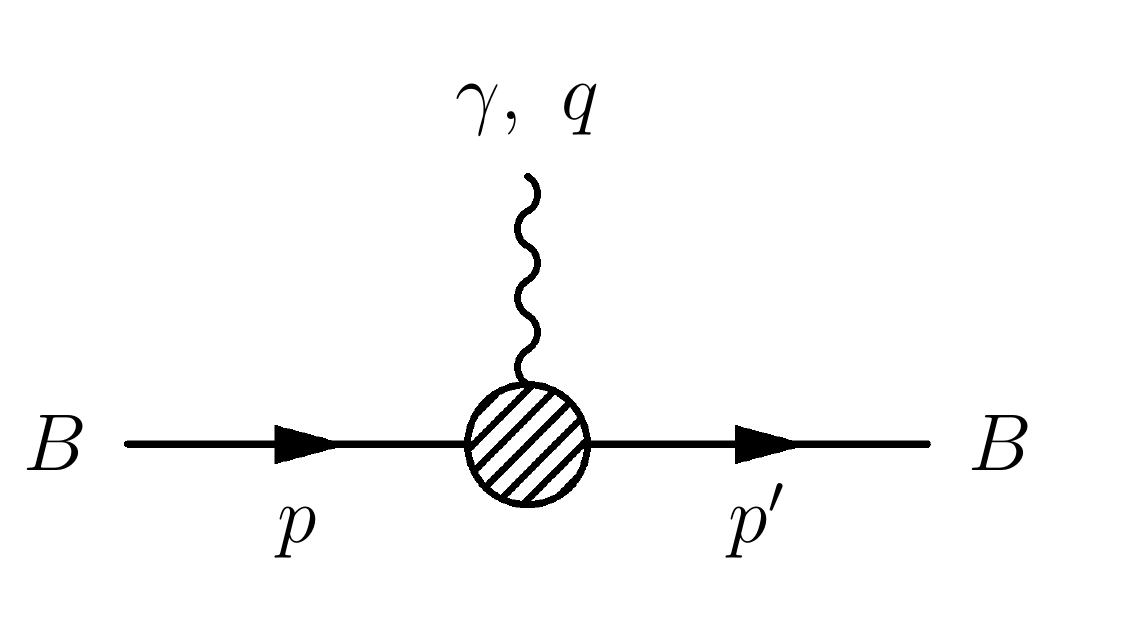}} = \langle B(p') | J_{\textrm em}^\nu | B(p) \rangle ~ \\
	&  = ~ \bar u(p') \left\lbrace
	\gamma^{\nu}F_1\left(q^2\right) - \frac{i\, F_2\left(q^2\right)}{2m_B}
	\sigma^{\mu\nu}q_\mu  - \frac{F_3\left(q^2\right)}{2m_B}
	\sigma^{\mu\nu}q_\mu\gamma_5 +   i \left( \gamma^{\nu}q^2\gamma_5 - 2m_B
	q^{\nu}\gamma_5\right)F_A\left(q^2\right)  \right\rbrace u(p)~,   \nonumber
	\end{align}}}%
where $m_B$ is the baryon mass.
With the photon momentum on shell, $q^2=0$, we can identify the different form factors as the electric charge, $Q=F_1(0)$; the magnetic moment, $\mu = \frac{1}{2m_B} [ F_1(0) + F_2(0)]$; the anapole moment, $F_A(0)$; and the electric dipole moment, $\delta = \frac{1}{2m} F_3(0)$, which is the only term in Eq.~(\ref{eq:formfactors}) that violates \CP symmetry.
The translation between the Lorentz invariant operator in Eq. \eqref{eq:EDMoperator} and the classical EDM interaction term in Eq.~\eqref{eq:hamiltonianclassic} can be found by substituting spinors and Dirac matrices by their Weyl representation, and taking the center-of-mass frame of the particle.

It is important to note that a non-zero EDM only implies \P- and \T-violation in systems of definite parity~\cite{Wirzba:2014mka}. For example, polar molecules like $\text{H}_2 \text{O}$ have degenerated ground states with different parity. They typically have large dipole moments $\order(10^{-8}\,\ecm)$ as compared to \textit{e.g.} the upper limit on the electron EDM  $\order( 10^{-29}\,\ecm)$. 

\begin{figure}[h]
	\centering
	\includegraphics[width=0.99\linewidth]{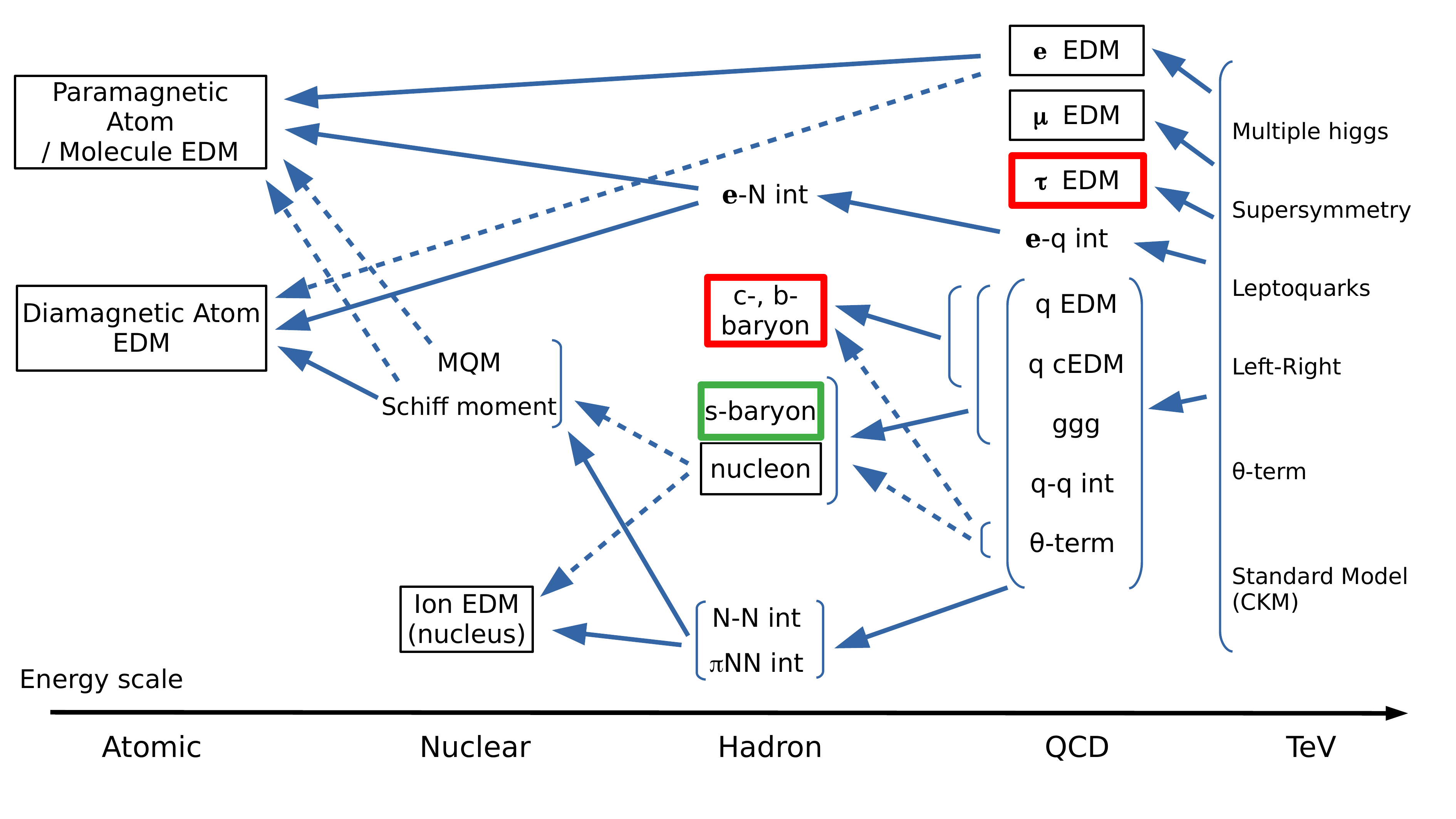}
	\caption{Types of EDM observables (boxed) and their contributions from higher-energy scales. \CP violation sources in fundamental theories (TeV) may induce EDMs in the atomic regime through a series of enhanced (solid arrows) or suppressed (dashed arrows) contributions. Heavy-baryon and $\tau$ EDMs (in red) are accessible with the bent-crystal experiment and strange baryons (in green) can be accessed with the current LHCb detector.  This figure has been adapted from Ref.~\cite{Yamanaka:2014mda}, including the proposed observables. 
	}
	\label{fig:contrib}
\end{figure}

\subsection{Current and future experiments}

An extensive experimental program is taking place worldwide to search for non-zero EDMs of different systems. A schematic summary of the systems being explored is presented in Figure~\ref{fig:contrib} with the contributions from higher-energy scales. Typically, the EDM observables putting the most stringent constraints in models of \CP violation are the neutron and electron EDM, although the relative importance of the different observables strongly depends on the NP model. 
Bounds on the electron EDM 
are obtained by exploiting the strong electric field ($\sim 84\,\text{GV/cm}$) that the unpaired electron feels in the polar molecule $\text{ThO}$~\cite{Baron:2013eja}, achieving limits of $d_e \lesssim10^{-29}\,\ecm$. 
To date, direct measurements of baryon EDMs only exist for neutrons, which with $d_n\lesssim10^{-26}\,\ecm$ \cite{Afach:2015sja} represent the experimental limit closest to the SM prediction ($d_n^{\rm SM}\sim10^{-31}\,\ecm$~\cite{Mannel:2012qk}), and $\Lz$ hyperons, with $d_\Lz\lesssim 10^{-16}\,\ecm$~\cite{Pondrom:1981gu}, which could be significantly improved at the LHCb Run III with the analysis method presented in Chapter~\ref{ch:lambdasLHCb}~\cite{Botella:2016ksl}.
The proton EDM, in turn, has only been bounded indirectly using measurements on atoms and molecules. To first approximation, the nucleus EDM information is not accessible in these systems because of the Schiff screening. This, however, is violated due to finite-size (relativistic) effects in diamagnetic atoms (paramagnetic systems)~\cite{Pospelov:2005pr} and also due to the magnetic quadrupole moment of the nucleus (MQM in Figure~\ref{fig:contrib}) in the case of paramagnetic molecules~\cite{Flambaum:2014jta}. 
%
A direct measurement of the proton EDM might be carried out at CERN with a future storage-ring facility~\cite{Anastassopoulos:2015ura,CPEDM:2019nwp} that could also explore the EDM of light nuclei such as D or $^3$He. This experiment would reach an astonishing sensitivity on the proton EDM of $10^{-29}e\text{cm}$. Finally, the EDM of the muon is also under experimental scrutiny~\cite{Muong-2:2008ebm,Chislett:2016jau} and new techniques~\cite{Abe:2019thb,Adelmann:2021udj} are being considered to bring its uncertainty down to a truly interesting level for BSM physics.

Overall, with the ongoing effort on the field of EDM searches there is a very real possibility to find a signal in the coming years. However, only a positive signal on different systems can disentangle which is the underlying \CP-violating mechanism~\cite{Dekens:2014jka}\footnote{The free parameters of the fundamental NP model can tipically accomodate any EDM value. It is the relation between several EDM (and other) observables what is precisely predicted by BSM theories. }. In this thesis we will present the possibility to extend this experimental program to very-short-lived baryons (notably charm and bottom) and $\tau$ lepton, highlighted in Figure~\ref{fig:contrib}, and significantly improve current limits on strange baryons.


\subsection{Heavy baryons}

To date, heavy-baryon EDMs have not been explored experimentally and the theoretical literature on this topic is scarce, only starting to develop now, triggered by the proposed experimental program. With the extensive literature on the neutron EDM, however, it is easy to identify the potential sources of heavy-baryon EDMs and present here a schematic discussion.

In general, any process involving an external photon and a flavour-conserving \CP-violating interaction in the baryon contributes to the baryon EDM. 
%
%
%
%
%
%
%
%
This process can be investigated from different energy scales and the  descriptions must be related to each other or \textit{matched}. Specifically, the Lagrangian describing these interactions can be constructed below the hadronic scale ($\lesssim 0.7 \gev$), with baryon and meson fields; above the chiral breaking scale ($\gtrsim 1.2 \gev$), with quarks and gluons; and at the NP scale ($\gtrsim200 \gev$), with the complete particle content of the fundamental theory. In Chapter~\ref{ch:edmsmw} we will see an explicit example of working with a fundamental Lagrangian to study EDM phenomenology.

Below the electroweak scale, where heavy degrees of freedom (in the SM and beyond) have been integrated out, the effective operators need not respect the complete gauge symmetry of the SM, as we will see in Chapter~\ref{ch:introtheo}. The sources of baryon EDM at this scale are comprised in the \CP-odd flavour-diagonal effective Lagrangian (notation adapted from Ref.~\cite{Dekens:2014jka}), 

%
\begin{align} \label{eq:lagrangianEDMs}
\mathcal{L}^{\not\text{P}\not\text{T}}_{\text{eff}}  = 
&-\frac{i}{2} \sum_{q={u,d,s,c,b}}\left. d_q \, \bar q \sigma^{\mu\nu} \gamma_5 q\, F_{\mu\nu}  \right.  & 
\includegraphics[width=0.15\textwidth]{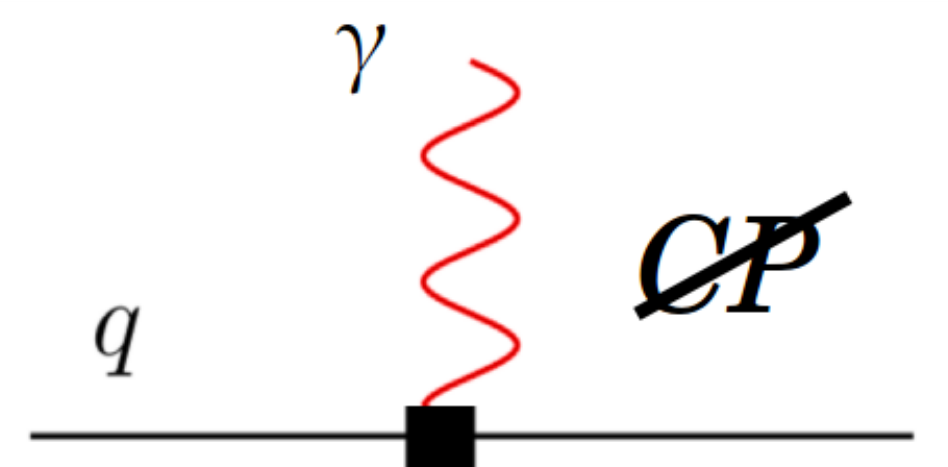} & 
~~~ \text{qEDM} ~~~~ &\nonumber \\
& -\frac{i}{2} \, \sum_{q={u,d,s,c,b}} \left. \tilde d_q\, \bar q \sigma^{\mu\nu} \gamma_5 T_a q\, G^a_{\mu\nu} \right. & 
\includegraphics[width=0.15\textwidth]{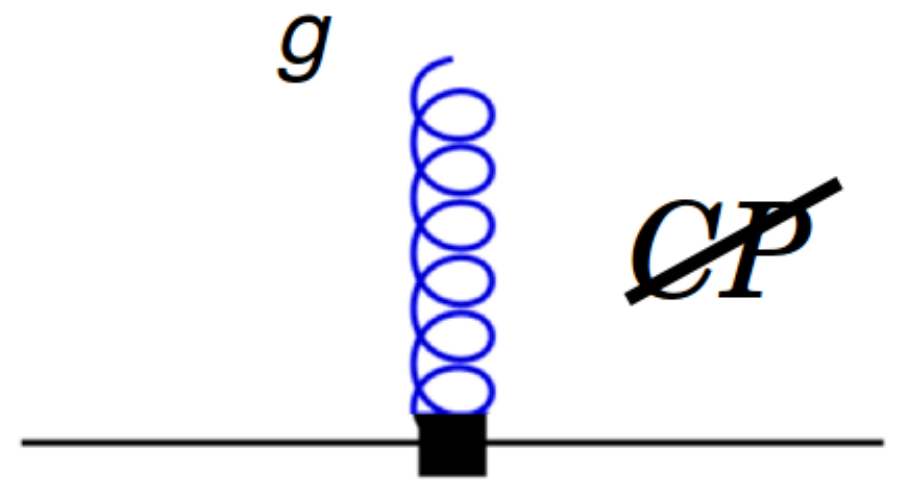} & 
~~~ \text{qCEDM} ~~~~ &\nonumber  \\
& + \sum_{i,j,k,l = {u,d,s,c,b}}C_{ijkl}\, \bar q_i \Gamma  q_j \, \bar q_k \Gamma^\prime  q_l\,   & 
\includegraphics[width=0.10\textwidth]{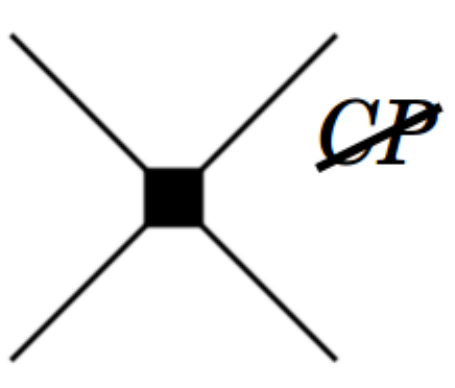} & 
~~~ \text{4q int} ~~~~ & \\
& + \frac{w}{6} f_{abc}\varepsilon^{\mu\nu\alpha\beta}G^a_{\alpha\beta}G^b_{\mu\rho}G_{\nu}^{c \, \rho}   & 
\includegraphics[width=0.12\textwidth]{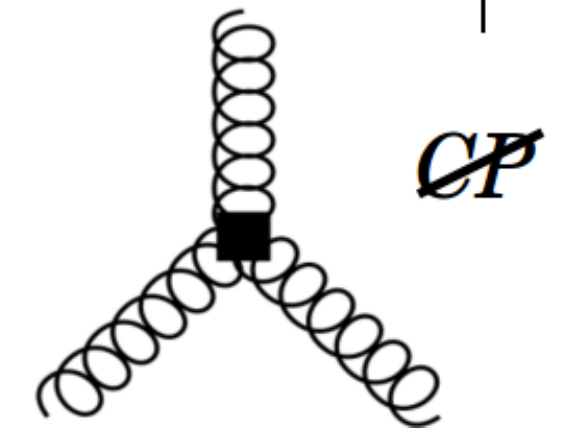} & 
~~~ \text{ggg (Weinberg op.)} ~~~~ &\nonumber \\
&- \bar{\theta} \frac{g^2}{64\pi^2}\epsilon^{\mu\nu\alpha\beta} G^a_{\mu \nu}G^a_{\alpha \beta} \quad & 
\includegraphics[width=0.12\textwidth]{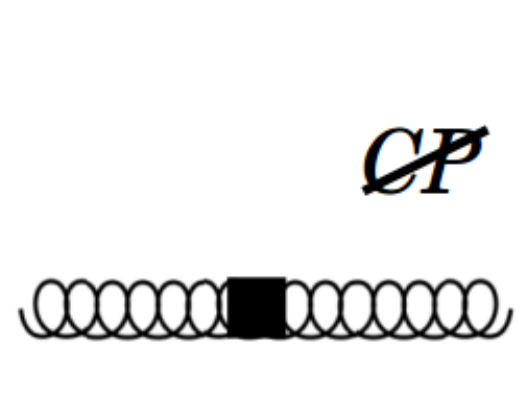} & 
~~~ \theta\text{-QCD term} . ~~~~ & \nonumber 
\end{align} 
These operators are also included in Figure~\ref{fig:contrib} at the ``QCD" scale. We can recognize the quark EDM (qEDM) and, by analogy, the chromo-EDM (qCEDM) with the gluon field strength tensor $G^a_{\mu\nu}$\footnote{Note that due to the non-Abelian character of QCD, the expansion of $G^a_{\mu\nu}$ yields terms with two gluon fields as $\tilde d_q\,g_s\,f_{abc}\, \bar q \sigma^{\mu\nu} \gamma_5 T_a q\,G_\mu^{b} G_\nu^{c}$, not depicted in the diagram.}. Next, we have many four-quark (4q) contact operators where the different Dirac structures are represented by $\Gamma$ and $\Gamma'$ and the quark flavors by the indices $i,j,k,l$. Finally we have the Weinberg operator with three gluons (ggg) and the $\theta$-QCD term. Parts of this Lagrangian will be further studied in Part III (see Eq.~\eqref{eq:lagrangian}). 

This effective Lagrangian plays an intermediate role between the fundamental theories and the EDM of baryons. The various extensions 
of the SM present very different contributions to these operators, which in turn do not contribute at equal parts to the EDM of baryons.
Evaluating the contribution of these operators to the hadronic EDM requires non-perturbative techniques to determine the effects of the strong interaction at low energy. Different approaches exist such as chiral theories, QCD sum rules or lattice QCD. The reliability of these techniques can be tested experimentally through low-energy observables, among which the magnetic moment of charm baryons could play 
an important role~\cite{Sharma:2010vv}, as discussed in Section~\ref{sec:mdm}. We should emphasize that these calculations linking the heavy-baryon EDM to the quark EDM and other effective operators are of utmost importance for the interpretation of the experiment itself and any phenomenological study of NP would need these expressions to link high-energy scale predictions to the low-energy observable. Only recently, the first works on heavy-baryon EDMs have been published within covariant chiral perturbation theory~\cite{Unal:2020ezc,Unal:2021lhb} although the evaluation of the Lagrangian coefficients (low energy constants) still needs lattice QCD input that is not available.

Among the different operators in Eq.~\eqref{eq:lagrangianEDMs}, those involving gluons or light quarks are, in principle, strongly constrained by the neutron or mercury EDM. We say \textit{in principle} because these operators could contribute to the experimental observable with different signs, adding up to tiny hadronic EDMs, compatible with observations. However, for these operators to have a relevant effect in heavy-baryon EDMs at the expected sensitivity level, cancellations of several orders of magnitude would be required in the neutron EDM.
Then, we shall consider only heavy-quark operators as the primary source of heavy-baryon EDMs. Using the counting of naive dimensional analysis~\cite{Georgi:1992dw}, we can obtain a rough estimation of the contribution of charm q(C)EDM to the charm baryon \Lc,
\begin{equation}
\delta_{\Lc} \approx ~~ \pm ~ d_c ~\pm ~\frac{e}{4\pi} \tilde{d_c}~,
\end{equation}
which holds also for bottom and strange baryons. Thus, the contribution of the charm quark EDM is approximately of order one, although non-perturbative effects could enhance this contribution \eg as we are seeing recently in the estimations of $\Delta A_{\CP}$~\cite{LHCb:2019hro,Schacht:2021jaz,Bediaga:2022sxw}.
The prediction on the charm quark EDM itself depends on the NP model, while the SM contribution is negligible, at the level of $10^{-31}\,\ecm$~\cite{Pospelov:2005pr}. It is possible to set upper limits on this quantity based on already available data. We will see all the attempts to do this for charm and bottom EDMs in Chapter~\ref{ch:improvedbounds}, together with some explicit predictions from NP models.

Another source of charm baryon EDM related to the charm valence component arises from four-quark operators, which may be accessed with the decay of charmonium states or more indirect processes. We did not find in the literature a systematic study with the bounds on all flavour-diagonal four-quark operators with heavy quarks.

Regarding the \Lz EDM, calculations exist within different frameworks. We can find expressions as a function of the $\theta$-QCD term \cite{Pich:1991fq, Borasoy:2000pq, Guo:2012vf}, the qCEDM and 4q interactions~\cite{Faessler:2006at}, and the qEDM~\cite{Anselm:1978vu}. The later uses the non-relativistic quark model to compute the EDM of the \Lz in terms of the EDM of the constituent quarks which yields 
null contributions from the $u$ and $d$ quarks, obtaining $\delta_{\Lz}^{\textrm{qEDM}} = d_{s}$.

\section{Magnetic dipole moments} \label{sec:mdm}

In this section we summarize the phenomenological motivations for heavy-baryon and $\tau$-lepton MDM measurements.

\subsection{Hadronic MDMs: probes of low-energy strong interactions}

%

The magnetic moment of baryons has served to test different techniques of low-energy strong interactions. Particularly, the magnetic moments of the lowest-lying baryon octet is recurrently used to assess the predictions of hadronic models going beyond the lowest order in chiral perturbation theory (see \eg Ref.~\cite{MartinCamalich:2010nab}).

\begin{figure}
		\centering
		\includegraphics[height=0.45\columnwidth]{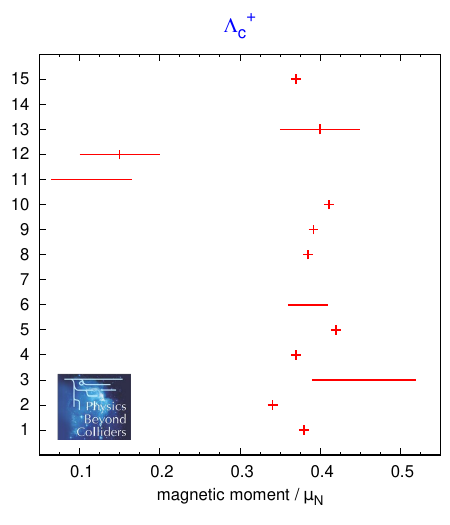}
		\includegraphics[height=0.45\columnwidth]{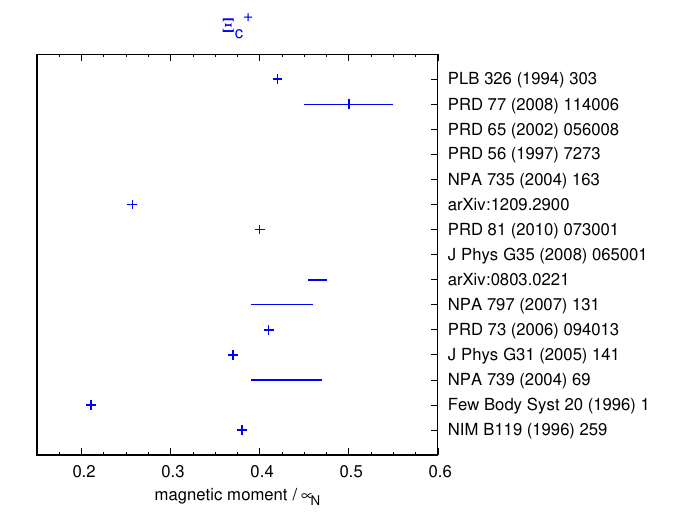}
		\caption{From Ref.~\cite{QCDWorkingGroup:2019dyv}. MDM of the (left) \Lc and (right) \Xicp as predicted from different hadronic theories, with the primary references quoted on vertical axis.}
		\label{fig:mdmcharm}
	\end{figure}

The prediction of heavy-baryon MDMs needs of further assumptions since their effective description cannot be matched directly to the fundamental theory, QCD. However, in this case other methods like the heavy quark expansion can be useful to address the calculation. In the literature (see Refs.~\cite{Sharma:2010vv,Dhir:2013nka} and references therein), many models of low-energy strong interaction have been used to study these observables, whose predictions lie in a broad range of values, shown in Figure~\ref{fig:mdmcharm}. The measurement of baryon magnetic moments with just a $10\%$ accuracy, reachable in the first stage of the bent-crystal experiment, will provide anchor points for these models and ultimately improve our understanding of the internal structure of hadrons.

It has also been argued that a measurement of the charm baryon magnetic moment can provide access to the magnetic moment of the charm quark, and thereby test NP predictions motivated by the anomaly on the muon $g-2$~\cite{Fomin:2019wuw}. However, the relation $\mu_\Lc = \mu_c$ holds only for simplified models and, overall, the interference with other contributions and the systematic uncertainty of these low-energy calculations precludes its interpretation in terms of NP.

Finally, the possibility to measure magnetic moments of baryons and antibaryons and to compare their $g-2$ values provides a new tool for testing \CPT invariance, a cornerstone of the SM and many BSM theories. Similar tests have been performed only recently for the proton by the BASE~\cite{Nagahama:2017eqh} and ATRAP~\cite{DiSciacca:2013hya} collaborations, 
and previously for the electron~\cite{VanDyck:1987ay} and the muon~\cite{Bennett:2004pv}.

\subsection{Leptonic MDMs: probes of New Physics }

The MDM of leptons can be predicted in the SM with much higher accuracy due to the perturbative nature of electroweak interactions. In fact, the leading systematic uncertainties of the SM prediction~\cite{Aoyama:2020ynm} come again from low-energy QCD methods. These enter as suppressed contributions from mesons in the loops\footnote{Find a review \textit{e.g.} in Ref.~\cite{Miller:2007kk}.}. A very high precision on the theoretical predictions is needed to match the astonishing experimental accuracy on the muon\cite{Muong-2:2021ojo} and electron~\cite{Mohr:2015ccw} anomalous magnetic moment, $a_\ell =  (g_\ell - 2)/2$,
\begin{align}
&a_\mu = 0.00116592061 \pm 0.00000000041~,\\
a_e &= 0.00115965218091 \pm 0.00000000000026~,
\end{align}
respectively.
Recently, the muon $g-2$ is attracting a lot of attention since the combination of FNAL~\cite{Muong-2:2021ojo} and BNL~\cite{Muong-2:2006rrc} measurements has a tension with the SM prediction of 4.2$\sigma$.
In contrast, the experimental precision on the $\tau$ $g-2$ is rather poor. Its short lifetime ($\sim 10^{-13}\sec$) prevents the use of the spin-precession technique adopted in the muon $g-2$ experiment and 
the best limit is coming from the comparison of the $e^+ e^- \to e^+ e^- \tau^+ \tau^-$ cross section measured at LEP2~\cite{DELPHI:2003nah} with QED calculations at fourth order in $\alpha$,
\begin{equation}
 -0.052 < a_\tau < 0.013~.
\end{equation}
This precision is several orders of magnitude worse than that of the SM prediction~\cite{Eidelman:2007sb}. 

In the case of the $\tau$ EDM, indirect limits have been obtained from the angular distribution~\cite{Belle:2021ybo,Belle:2002nla} and total cross section~\cite{Blinov:2008mu,DELPHI:2003nah} of the 
$e^+ e^- \to \tau^+ \tau^-$ reaction, and from the electron EDM through light-by-light scattering diagrams~\cite{Grozin:2009jq}. The most restrictive ones are at the level of 
\begin{equation}
	|d_\tau| \lesssim 10^{-17}\ecm~.
\end{equation}
These limits are comparable to some of the most optimistic NP predictions, at the level of $10^{-20} - 10^{-17} \ecm$~\cite{Dekens:2018bci,Gutierrez-Rodriguez:2013eaa,Ibrahim:2010va,Gutierrez-Rodriguez:2009weo,Iltan:2005iy}, whereas the SM represents a negligible background with its first contribution appearing at four loops~\cite{Pospelov:2013sca}, at the level of $10^{-41}\,\ecm$.

This situation may change upon the construction of a dedicated experiment at the LHC which can directly measure the spin precession of $\tau$ leptons using bent-crystal techniques, presented in the next chapters.


\chapter{The experiment with bent crystals} \label{ch:crystals}

In recent years, a novel experimental method has been proposed to measure the EDM and MDM of charm and bottom baryons~\cite{Botella:2016ksl,Fomin:2017ltw} and $\tau$ lepton~\cite{Fomin:2018ybj,Fu:2019utm} at the LHC. The experimental setup relies on the spin precession of positively charged particles produced in a fixed target and subjected to the electric field between the atomic planes of a bent crystal.

Since the publication of our first paper \cite{Botella:2016ksl}, many additional studies have been performed by our group and others: the proposal has been extended to other particles~\cite{Bagli:2017foe,Fomin:2018ybj,Fu:2019utm}, alternative layouts have been explored to increase the efficiency~\cite{Biryukov:2021phs,Biryukov:2021cml}, methods to estimate~\cite{Fomin:2017ltw,Fomin:2019wuw} and reconstruct~\cite{Aiola:2020yam} the polarization have been refined, and many detector effects have been considered~\cite{internalnote}. Furthermore, the possible sites to install the experiment at the LHC have been studied thanks to realistic simulations of the LHC beam optics~\cite{Mirarchi:2019vqi}. 

To contain the relevant information in a structured way we will organize this chapter as follows. In Sections~\ref{sec:concepts} to \ref{sec:spinprecess} we will discuss the basic physics ideas at the core of the experiment: crystal channeling, particle polarization, and spin precession. 
%
%
This experiment could be realised at two different locations of the LHC ring (IR3 and IR8/LHCb), and three different crystal/target configurations have been proposed (for heavy baryons, $\tau$ leptons, and with focusing crystals). We will compare these layouts and make their differences apparent with illustrations in Section~\ref{sec:configurations}. The sensitivity to the dipole moments in each of the configurations is affected by the same key factors. We will introduce these in Section~\ref{sec:keyfactors} primarily focusing on the charm baryon case and pointing out the differences with the other layouts when needed. With all these notions at hand, the detailed optimization of the layout and sensitivity will be discussed for each case separately, later in Chapter~\ref{ch:sensitivity}.

This chapter will hopefully be complementary to the more specialized discussions in Chapter~\ref{ch:sensitivity} or any of our publications.

\begin{figure}[h]
	\centering
	\includegraphics[width=0.99\linewidth]{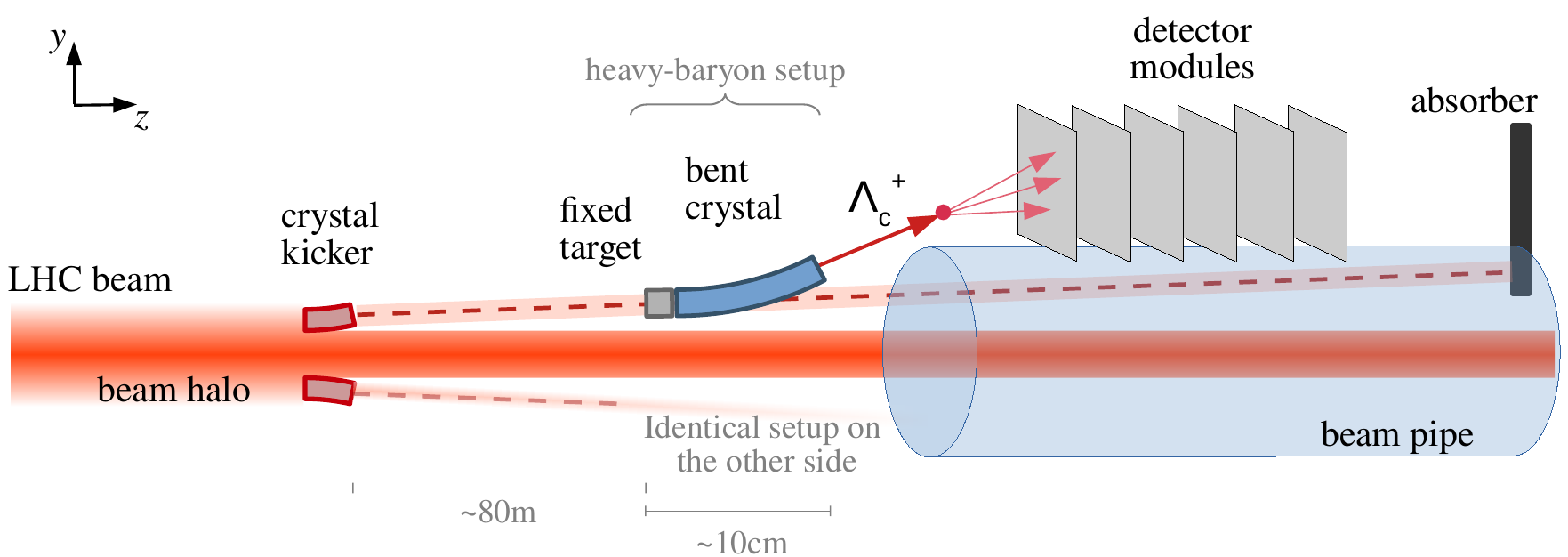}
	\caption{Side view of the proposed experiment layout. The setup may be installed either in front of the LHCb detector or in the LHC interaction region 3, with a dedicated experiment. The proton-target interaction products are produced mostly in the forward direction, staying contained within the beam pipe. In turn, channeled particles are steered and put in the detector acceptance. In reality, the trajectory of the deflected beam halo is not straight, as it has been precisely determined from beam-optics simulations. }
	\label{fig:layoutgeneral}
\end{figure}

\section{Experimental concept} \label{sec:concepts}

The magnetic and electric dipole moments of a spin-1/2 particle is given, in Gaussian units, by
$\bm{\mu} = g \mu_B {\bm s}/2$ and \mbox{$\bm{ \delta} = d \mu_B {\bm s}/2$}, respectively,
where $\bm{s}$ is the spin-polarization vector and $\mu_B=e \hbar / (2 m c)$ is the particle magneton, $m$ being its mass.
The interaction of the magnetic and electric dipole moments with external electromagnetic fields causes the change of the particle spin direction as
\begin{equation}
\label{eq:precesionClassical}
\frac{d \bm s}{d \tau} = \bm \mu \times \bm B^* + \bm \delta \times \bm E^* ~,
\end{equation}
which is obtained from the classical Hamiltonian $H=-\bm \delta \cdot \bm E^* -\bm \mu \cdot \bm B^*$, where $\bm E^*$ and $\bm B^*$ are the external fields in the rest frame of the system. This equation coincides with the full (non-classical) spin equation of motion in some limit as introduced later in Section~\ref{sec:spinprecess}.

Any experimental setup to measure this effect relies on three main elements: (1) a source of polarized particles; (2) an electromagnetic field to induce spin precession; and (3) an analyser of the final polarization vector.
This concept has never been realised with very-short-lived particles ($\sim10^{-13}\,\s$) as they present striking complications with respect to other (meta)stable systems. Now, with cutting-edge bent-crystal technology and the powerful LHC beam, we have a unique opportunity to measure these elusive observables. All three elements can be realised in the following way:

\begin{enumerate}
	\item \textbf{Source of polarized particles}\\
	Part of the LHC protons in the external region of the beam (\textit{beam halo}) can be deflected from the main beam trajectory with a crystal \textit{kicker} and directed to a fixed target, as shown in Figure~\ref{fig:layoutgeneral}. Highly-energetic protons of $7\,\tev$ interact with the target nucleons at a centre-of-mass energy of $115\,\gev$, enough to produce copious amounts of charm and bottom hadrons, that are produced in the forward beam direction with energies of $\sim 1\tev$. These particles are naturally polarized, although the total magnitude of the polarization strongly depends on the initial particle direction, which must be reconstructed. A source of polarized $\tau$ leptons can be found in the weak decay of charm hadrons.
	
	\item \textbf{Electromagnetic field to induce spin precession} \\
	The short lifetime of these ultra-relativistic particles is largely extended in the laboratory frame making them travel a few centimetres after the target before decaying. In this space, a bent crystal is placed with high angular accuracy. Positively-charged particles that enter the crystal are repelled by the positively-charged atomic planes, formed by nuclei, being confined or \textit{trapped} between the crystal atomic planes. These particles are \textit{channeled} along a curved path, subjected to a net electromagnetic field inducing sizeable rotation of the spin-polarization vector. 
	
	\item \textbf{Analyser of the final polarization vector}\\
	After the crystal exit, the surviving particles have been deflected enough for their decay products to exit the beam pipe and be reconstructed in a detector. With a precise reconstruction of the particle directions and kinematics, the polarization can be analysed in a statistical way
	%
	%
	provided that the \P-violating decay asymmetry is significant enough to induce preferential directions of the decay products, as seen in the rest frame of the mother particle.
\end{enumerate}

The possibility to measure spin precession in particles channeled in a bent crystal was first proposed in the 1980's \cite{Lyuboshits:1979qw,Kim:1982ry} and realised a decade later at Fermilab by the E761 collaboration, which measured the MDM of the strange \PSigmap baryon~\cite{Chen:1992wx}.
Following this experiment, the possibility to measure also the MDM of charm baryons was explored in Refs.~\cite{Baublis:1994ku,Samsonov:1996ah}, although its feasibility was limited by the "small" available momentum of the beams at the time, in the range of hundreds of \gevc. 
The \lhc offers a three-fold advantage in this respect, thanks to the large Lorentz factor $\gamma$ of the produced particles. First, they live long enough to go through a crystal of a few centimetres, which is needed to steer the particles outside the beam pipe. Second, the electric and magnetic field in the particle rest frame, $\bm{E^*}\approx\gamma \bm{E}$ and $\bm{B^*}\approx-\gamma \bm\beta\times\bm{E} /c$, is strong enough to induce spin precession. Finally, as shown in Ref.~\cite{Baryshevsky:2016cul}, the amount of produced and channeled particles scales approximately as $\gamma^{3/2}$. Besides the large proton energies, the \lhc also offers advantages regarding the luminosity. In fact, only using a tiny portion of the beam protons, the instantaneous luminosity of a fixed-target experiment at the \lhc quickly reaches its maximum, determined by the occupancy and readout capabilities of the particle detectors.

As opposed to the neutron EDM experiments~\cite{nEDM:2020crw} where the external \Evec and \Bvec fields are prepared in (anti)parallel directions, in our case, the magnetic field $\Bvec^*$ is perpendicular to $\Evec^*$, as it emerges from the external electric field \textit{seen} in the rest frame of the particle, similarly to the muon storage-ring experiments~\cite{Muong-2:2021ojo}. Thus, the signature of a non-zero EDM can be identified as a change in the perpendicular direction to the main precession, driven by the magnetic moment around $\Bvec^*$, as we will see in more detail in Section~\ref{sec:spinprecess}.

\section{Crystal channeling} \label{sec:channeling}

In a crystal, the strong electric field experienced by a charged particle in the proximity of the ordered structure of atoms exerts a strong confinement force onto the particle itself. The particle trajectory can be bound to stay parallel to a crystalline plane or to an atomic string. This phenomenon is called {\it channeling} and can occur if the angle between the particle trajectory and a crystal plane ({\it planar } channeling)  or  a crystal axis ({\it axial} channeling)  is lower than a critical angle, referred to as Lindhard angle. When the crystal is mechanically bent, its planes or atomic strings are bent too. The incoming particle direction is then deflected by an angle equal to that of the crystal bending.

To become familiar with the physics of crystal channeling we shall derive the expression of the Lindhard angle explicitly. Later, we will summarize all the conditions for channeling, providing the relevant analytic expressions. These will be used to obtain the channeling efficiency and optimize the crystal parameters with simulations.
For the sake of clarity, we will treat only planar channeling with positively charged particles, on which the experiment is based. The possibility to use axial channeling in spin-precession experiments is discussed in Ref.~\cite{Bagli:2017foe}.

\begin{figure}[tb]
	\centering
	\includegraphics[width=1.0\linewidth]{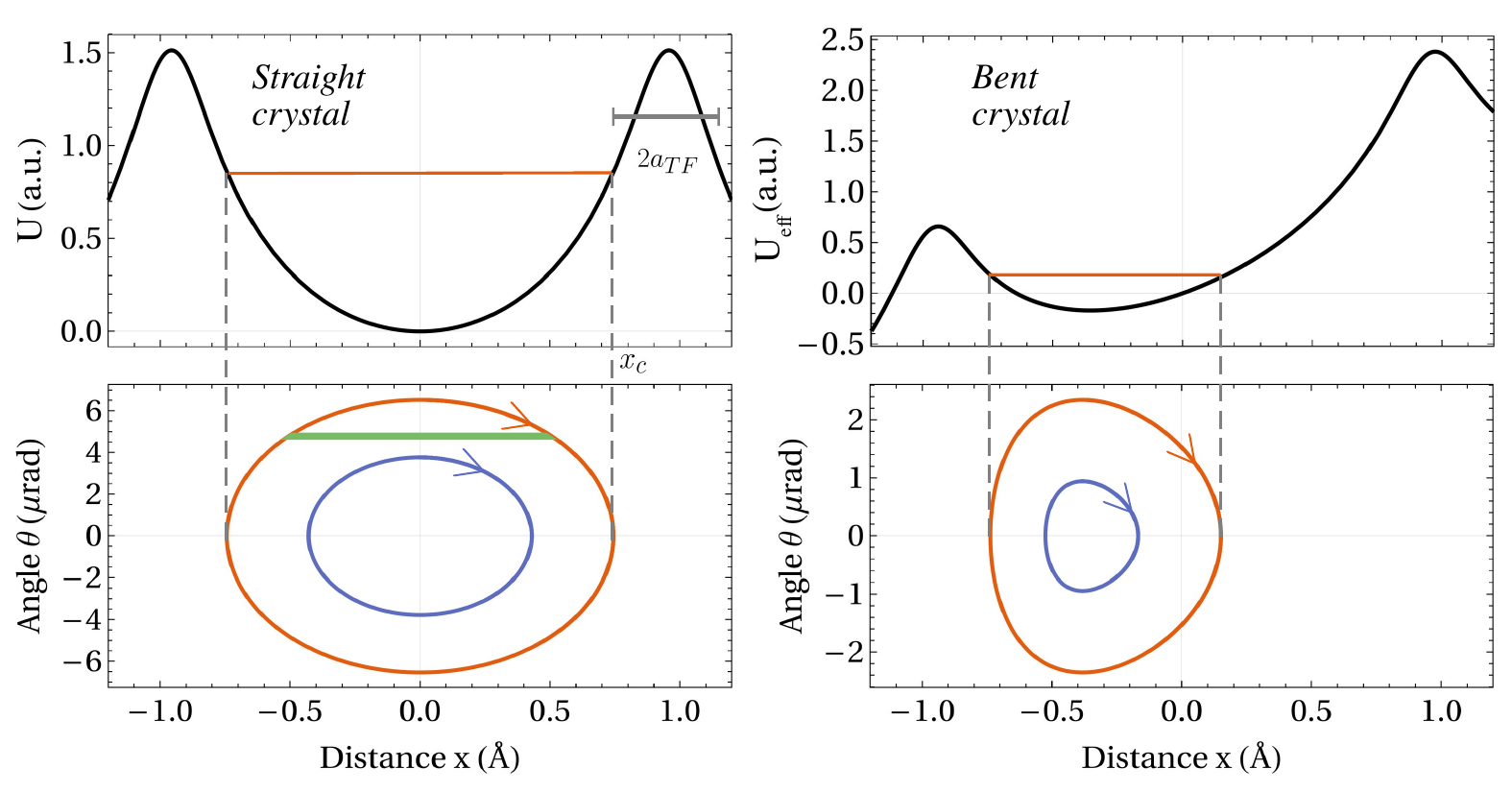}
	\caption{Phase diagram depicting the (top left) electric potential well between atomic planes, in the analytic Moliére approximation~\cite{Biryukov1997}, and (bottom left) two orbits associated with different energy levels. The area of the orange orbit, with maximum possible energy, represents the available phase space for trapping. If the incident angle is fixed at $\theta\approx 5\murad$, the available phase space is represented by the green horizontal band.
		(Right panels) including the centrifugal force in a bent crystal, the effective potential becomes asymmetrical and the trapping probability is reduced. Particles in higher-energy orbits than the orange one would collide with the crystal-plane nuclei, with thermal fluctuations of amplitude $2 a_{TF}$.	}
	\label{fig:phasediagram}
\end{figure}

The crystal atomic planes, separated by a distance $d_p$, generate an (electric) potential well $U(x)$ that can trap a particle if its transverse energy $E_T$ (corresponding to the motion in the normal direction to the crystal plane) is low enough. The origin of the $x$ coordinate is in the middle point between planes. The maximum of the potential, at the plane itself, is ${U_0 = U(d_p / 2)}$. However, due to the thermal motion of the atomic nuclei, with half-amplitude $a_{TF}$, the particles that get too close to the atomic plane will scatter with the nuclei, leaving the channeling mode. Then we shall consider the maximum of the potential to be at a distance $a_{TF}$ away from the crystal plane, \textit{i.e.} $U(x_c)$ where $x_c=d_p / 2 - a_{TF}$, as shown in Figure~\ref{fig:phasediagram} (top left). By operating at low temperature the thermal fluctuations are reduced, increasing the channeling efficiency. The condition for trapping a particle entering the crystal lattice at position $x$ is
\begin{equation} \label{eq:conditionenergy}
E_T + U(x) \leq U(x_c).
\end{equation}
From the classical\footnote{The derivation with relativistic mechanics is fully shown in Ref.~\cite{Biryukov1997}, yielding the same result.


} kinetic energy $(1/2)m v^2$ we can write the transverse energy of the particle as $E_T = (1/2) p v \theta^2 $, where $p$ and $v$ are the particle momentum and velocity and $\theta$ the angle with respect to the atomic planes\footnote{This angle will be noted $\theta_y$ once the axes are introduced, in Figure~\ref{fig:axes}.}. The limiting angle of capture is obtained from Eq.~\eqref{eq:conditionenergy} by setting $U(x)=0$ and reads 
\begin{equation} \label{eq:lindhard}
\theta_L = \sqrt{\frac{2U(x_c)}{pv}}.
\end{equation}
Thus, for being trapped, the angle and position of the particles at the crystal entrance must meet the conditions
\begin{equation}
-\theta_L  \leq \theta \leq \theta_L~~\text{and}~ -x_c\leq x\leq x_c.
\end{equation}
However, these two variables are related and the trapping condition is not rectangular. The relation between $\theta$ and $x$ can be seen through the phase diagram of the oscillations in this harmonic potential, in Figure~\ref{fig:phasediagram} (bottom left). These are in fact the oscillations of a particle on its way through the channel. The area of the outermost ellipse represents the available phase space for trapping. Then, we can calculate the trapping efficiency simply as this area divided by the total phase space. For the two rectangular conditions, we have the efficiency factors $x_c/(d_p/2)$ and $\theta_L/\varTheta$, where $\varTheta$ is the beam divergence. The portion of the remaining rectangle occupied by the ellipse is equivalent to the area of a circle $\pi r^2$ divided by the surrounding square $(2 r)^2$, giving an additional factor of $\pi / 4$. Altogether, the trapping efficiency for a beam of divergence $\varTheta$ is
\begin{equation}
A_\varTheta = \frac{2 x_c}{d_p} \frac{\theta_L}{\varTheta} \frac{\pi}{4}.
\end{equation}
reproducing Eq.~(1.29) of Ref.~\cite{Biryukov1997}. However, if the beam divergence is much smaller than the Lindhard angle (\ie the beam is essentially parallel), $\varTheta\ll\theta_L$, the available phase space is just an horizontal band within the ellipse (green in Figure~\ref{fig:phasediagram}), and the trapping efficiency can be obtained from the equation of the ellipse, ${(x/x_c)^2 + (\theta / \theta_L)^2 =1}$. Therefore, the trapping efficiency for a parallel beam entering a (straight) crystal at an angle $|\theta|\leq\theta_L$ is
\begin{equation} \label{eq:Astr}
A_{\rm straight} = \frac{2 x_c}{d_p} \sqrt{1-\frac{\theta^2}{\theta_L^2}}
\end{equation}
The trapping probability for a single particle, which may reach the crystal entry at any distance $x$ from the atomic planes, is the same as for a parallel beam, and we will use Eq.~\eqref{eq:Astr} in the per-event simulations. 
When the crystal is bent, the centre of the particle path is no longer at $x=0$ due to the constant centrifugal force $p v / R$, where $R$ is the (constant) crystal bending radius. The effective potential $U_{\rm eff}(x)$ seen by the particle becomes asymmetrical and the well depth is reduced, as shown in Figure~\ref{fig:phasediagram} (top right). Consequently, the trapping efficiency and the maximum angle are also reduced. Moreover, a new condition on the maximum longitudinal momentum arises since the electric field at the channel border $U^\prime(x_c)$ should compensate for the centrifugal force. This condition is usually presented as a per-event critical radius $R_c\leq R$, where
\begin{equation} \label{eq:criticalR}
R_c = \frac{p v}{U^\prime(x_c)}~.
\end{equation}
The
The trapping efficiency of a parallel beam of particles in a bent crystal is modified as (find the derivation in Ref.~\cite{Biryukov1997}, chapter 2)
\begin{equation} \label{eq:trappingprobablity}
A_{\rm bent} = A_{\rm straight} \left( 1-\frac{R_c}{R}\right)~
\end{equation}
and, for a divergent beam, it is
\begin{equation} \label{eq:trappingprobablitydiv}
A_{\rm bent} = A_\varTheta \left( 1-\frac{R_c}{R}\right)^2~.
\end{equation}

Even if the particle is trapped in the potential well, scattering processes may remove it from channeling mode. This process of \textit{dechanneling} follows an exponential distribution as $\exp(-L/L_D)$, where $L$ is the length travelled by the particle and $L_D$ the mean dechanneling length. We shall give the expression for the dechanneling probability at the crystal exit, \textit{i.e.} at $L=R\theta_C$, where $\theta_C$ is the crystal bending angle. In the harmonic potential approximation, and for long crystals of a length comparable to the dechanneling length, the probability of avoiding dechanneling is~\cite{Biryukov1997}
\begin{equation}
\label{eq:dechanneling}
{\it w} =\exp{\left(-\frac{\theta_C}{\theta_D \frac{R_c}{R} (1-\frac{R_c}{R})^2}\right)},
\end{equation}
where $\theta_D = \frac{256}{9\pi}\frac{N Z a_{\rm TF} d_p^2}{\ln{\left(2m_e c^2 \gamma/I\right)}-1}$, $N = N_A \frac{\rho}{A}$ is the 
number of atoms per unit volume, $N_A$ the Avogadro number, $A$ the atomic mass (g/mol), $\rho$ the density, $I$ the ionization 
potential, $Z$ the atomic number, $m_e$ the electron mass, $d_p$ the interplanar distance and $a_{\rm TF}$ the half-amplitude of nuclei thermal vibrations in the crystal lattice. We should note that the channeling efficiency is greatly affected by the crystal material and the family of crystal planes, \textit{i.e.} the crystal orientation. We will recurrently compare silicon and germanium crystals exploiting the \{110\} plane family. The potential depth, interplanar distances, and other material-specific parameters are found in Table~\ref{tab:material parameters}.

\subsubsection{Channeling conditions} 
In our analytically-based simulations, we consider a particle to be fully channeled if it meets the following conditions. First, we impose these hard cuts:
\begin{enumerate}
	\item the particle decays after the crystal exit;
	\item the incident angle is below the Lindhard angle, $|\theta|\leq\theta_{L}$, in Eq.~\eqref{eq:lindhard};\footnote{This is one of the efficiency bottlenecks of our experiment, as the Lindhard angle of \Lc particles produced in the target is $\theta_L\approx 7\,\murad$, about three orders of magnitude smaller than their initial angular divergence $\varTheta$. }
	\item the critical radius is below the bending radius $R_c \leq R$, in Eq.~\eqref{eq:criticalR}.
\end{enumerate}
Second, each event has an associated probability to enter/stay in channeling mode. In practice, two uniformly-distributed random numbers $r_i\in[0,1]$ are produced for each event and they are compared with
\begin{enumerate}
	\item[4.] the trapping probability for a parallel beam (single-particle direction) $ r_1 \leq A_{\rm bent}$, in Eq.~\eqref{eq:trappingprobablity};
	\item[5.] the probability to avoid dechanneling $r_2\leq {\it w}$, in Eq.~\eqref{eq:dechanneling}.
\end{enumerate}

More sophisticated crystal channeling simulations going beyond the analytical formulas~\cite{Sytov:2019gad} were used in our most recent study~\cite{Aiola:2020yam}.

\begin{table}[t]
	\centering
	\small
	\caption{From Ref.~\cite{Biryukov1997}. Material parameters used in the crystal optimization. See text for their definitions. \label{tab:material parameters}
	}
	\resizebox{1.\columnwidth}{!}{
		\begin{tabular}{cccccccccc}
			\hline \hline
			Material & $d_p$ [\AA] & $a_{\rm TF}$ [\AA] & $u_{\rm T}$[\AA] & $U(x_c) [\ev]$ & $U'(x_c) [\gev/\cm]$ & $I [\ev]$ & $\rho [\text{g}/\cm^3]$ & $A[\text{g}/\text{mol}]$ & $Z$ \\
			\hline
			Si 110   & 1.92        & 0.194            & 0.075          & 16             & 5.7                  & 173      & 2.329 & 28.0855 & 14 \\
			Ge 110   & 2.00        & 0.148            & 0.085          & 27             & 10                   & 350      & 5.323 & 72.630  & 32 \\
			\hline \hline
		\end{tabular}
	}
\end{table}

\section{Initial polarization}
\label{sec:initialpolarization}

%
%
%
%
%
%
%

The polarization \spol is defined as the expectation value of the three spatial components of the spin operator $\bm{\hat S}$, normalized to 1. In the case of spin-$1/2$ particles, $\bm{s}=\langle \bm{\hat S} \rangle / (\hbar/2)$. The polarization of a sample of particles in a mixture of states is characterized by the spin density matrix $\rho$, from which the polarization can be calculated as
\begin{equation}
 s_i=\text{Tr} ~\rho \sigma_i 
~~~, ~~~\text{where  } \rho =~ W_+ | \chi_+ \rangle \langle \chi_+ |  ~+~   W_- | \chi_- \rangle \langle \chi_- |~~,
\end{equation}
and $W_\pm$ is the fraction of particles in the state  $| \chi_\pm \rangle$.
By direct matrix manipulations, the density matrix can be parametrized as 
\begin{equation}
\rho = \frac{1}{2} 
\left( \begin{array}{cc}
1+s_z & s_x - i s_y \\
s_x+is_y & 1-s_z
\end{array} \right),
\end{equation}
where $\bm s = ( s_x, s_y, s_z)$.

The polarization is always defined in the rest frame of the particle although this can be accessed in different ways, especially in a complex decay chain. Ultimately, these differences reduce to a rotation of the coordinate system that nevertheless can greatly enhance or dilute the overall polarization. We will see an explicit example of polarization measurements in the decay chain \threepi in Part II. 

To measure the spin precession in the bent crystal experiment it is imperative to have initial polarization on the particles that are channeled. In the angular analysis of their decay products, it is possible to extract the initial polarization plus the dipole moments without previous knowledge of its magnitude. However, the sensitivity to EDM and MDM strongly depends on the magnitude and direction of the initial polarization and, overall, we need to ascertain that there will be polarization in the targeted systems to be measured.

\subsubsection{Heavy baryons}

We shall focus on the $\Lc$ case, as it is the most abundant spin-$1/2$ particle among the charm and bottom hadrons produced in the target. Later we will make the appropriate considerations for other systems.

\begin{figure}
	\centering
	\includegraphics[width=0.40\linewidth]{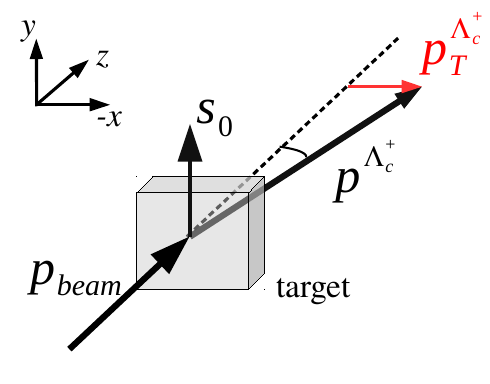}~~~
	\includegraphics[width=0.47\columnwidth]{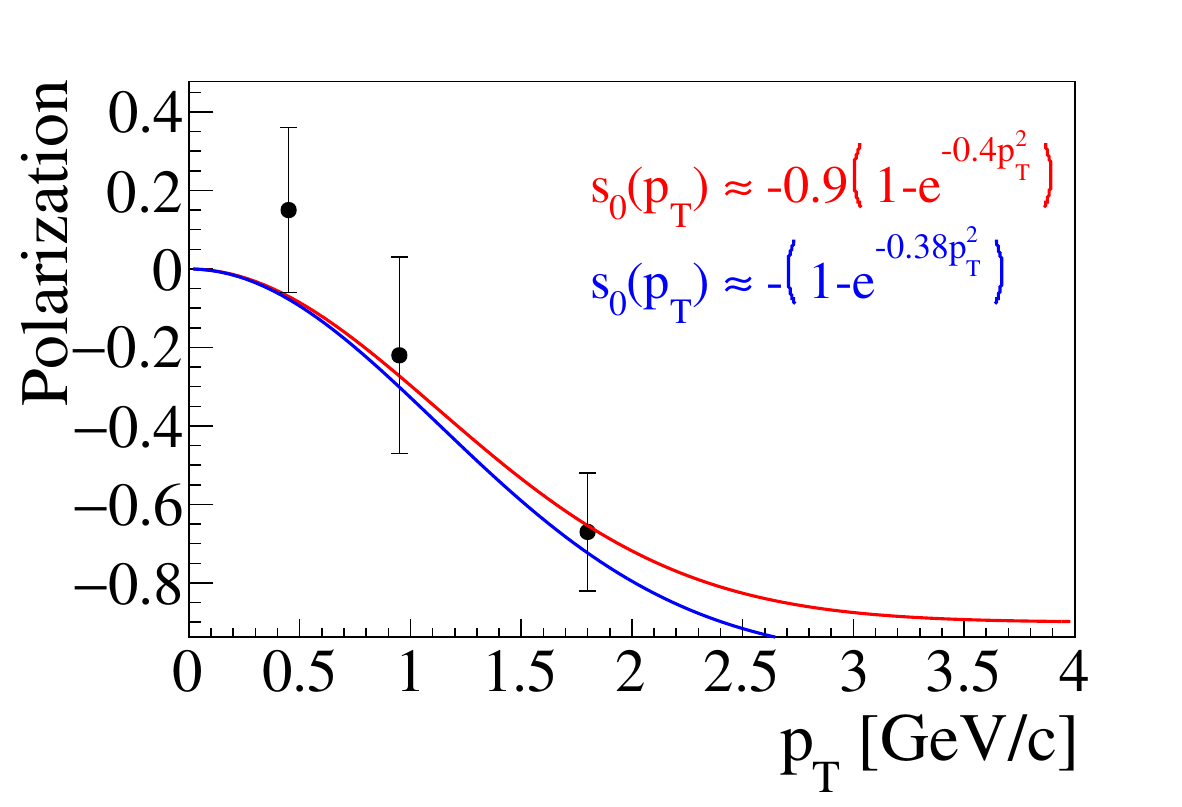}
	\caption{ (Left) the initial polarization $\spol_0$ of a \Lc baryon produced in proton-target interactions is perpendicular to its production plane, defined by the beam and \Lc directions. (Right) dependence of the polarization with the \Lc transverse momentum, depicted in red on the left panel. The data points were measured in Ref.~\cite{Aitala:1999uq}.}
	%
	%
	\label{fig:polarizationtarget}
\end{figure}

Due to parity conservation in strong interactions, the particles produced in proton-target interactions have polarization perpendicular to their production plane. Given the direction of the incoming proton $\bm{\hat p}_{\rm beam}$ and that of the produced particle $\bm{\hat p}_\Lc$, the only vector that is invariant under parity transformation is their cross product (see Figure~\ref{fig:polarizationtarget} (left)), 
\begin{equation}
\spol_0 \propto \bm{\hat p}_{\rm beam} \times \bm{ \hat p}_\Lc~.
\end{equation}
For this reason, particles produced in opposite directions with respect to the incoming protons will have opposite polarization. It is crucial then to reconstruct this direction to avoid a total dilution of the polarization. In practice, the angular resolution of the considered setups is more than enough to separate events with different initial particle directions, as we will see in Section~\ref{sec:keyfactors}.  More precisely, the initial polarization directly depends on the transverse momentum of the particle $p_T$ with respect to the proton beam, correlated to the production angle.

The magnitude of the \Lc polarization is unknown for 7 \tev protons on a fixed target. 
However, a measurement with 40-70 \mevc neutrons on a carbon target gives $s_0=0.5\pm0.2$~\cite{Szwed:1981rr},
and a measurement from interaction of 230 \mevc~\pim on copper target 
yields $s_0=-0.65^{+0.22}_{-0.18}$~\cite{Jezabek:1992ke}.
Moreover, with 500~\gevc $\pim$ on a combination of platinum and diamond targets,
the polarization of the \Lc was measured as a function of the \pt~\cite{Aitala:1999uq}, resulting on the data points shown in Figure~\ref{fig:polarizationtarget} (right).
The measured average polarization is about $-10\%$, reaching $-0.67 \pm 0.15$ for $\pt^2 = [1.24,5.20]\,\gev^2/c^2$. Using a phenomenological dependence based on \Lz hyperons~\cite{ACCMOR:1994fcu} to describe these experimental results~\cite{Aitala:1999uq}, the initial polarization is estimated as~\cite{Aiola:2020yam} 
\begin{equation}
\label{eq:pol_pt}
s_0(\pt)\approx A\left(1-e^{-B p^2_T}\right),
\end{equation}
with $A \approx -0.9$ and $B \approx 0.4$~$(\gevc)^{-2}$, as shown in Figure~\ref{fig:polarizationtarget} (right).
The \Lc and \Xicp baryon polarization versus \pt can be measured precisely in
fixed-target collisions at \lhcb using the SMOG system~\cite{LHCb:2014vhh,Bursche:2018orf} to further improve the polarization model.

Besides the transverse momentum, the polarization also varies with the Feynman-x kinematic variable $x_F=\frac{p_L}{\sqrt{{s}}}$, where $p_L$ is the longitudinal momentum of the produced baryon in the center-of-mass frame of the collision and $\sqrt{s}$ is the energy of the collision. It is observed from \Lz production data \cite{HERA-B:2006rds,Ramberg:1994tk,Fanti:1998px,ATLAS:2014ona} that $\spol_0$ vanishes for $x_F\approx 0$. In the proposed bent crystal setup, $x_F$ varies from 0.1 to 0.5, whereas SMOG data covers the range between $-1$ and 0. Accounting for these dependences, a sizeable \Lc polarization of around 20\% is expected in the proposed experiment.

%
%

\subsubsection{$\bm\tau$ lepton}

\begin{figure}
	\centering
	\raisebox{0.5cm}{\includegraphics[width=0.45\linewidth]{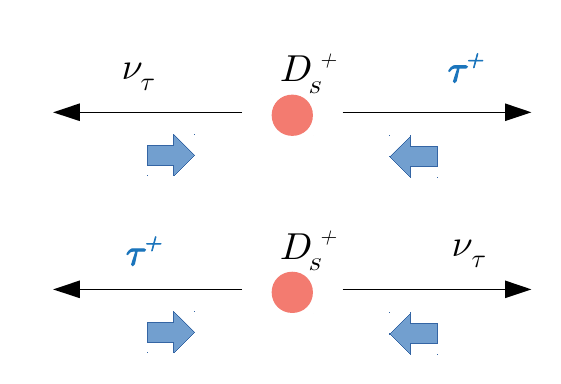}}
	\includegraphics[width=0.45\linewidth]{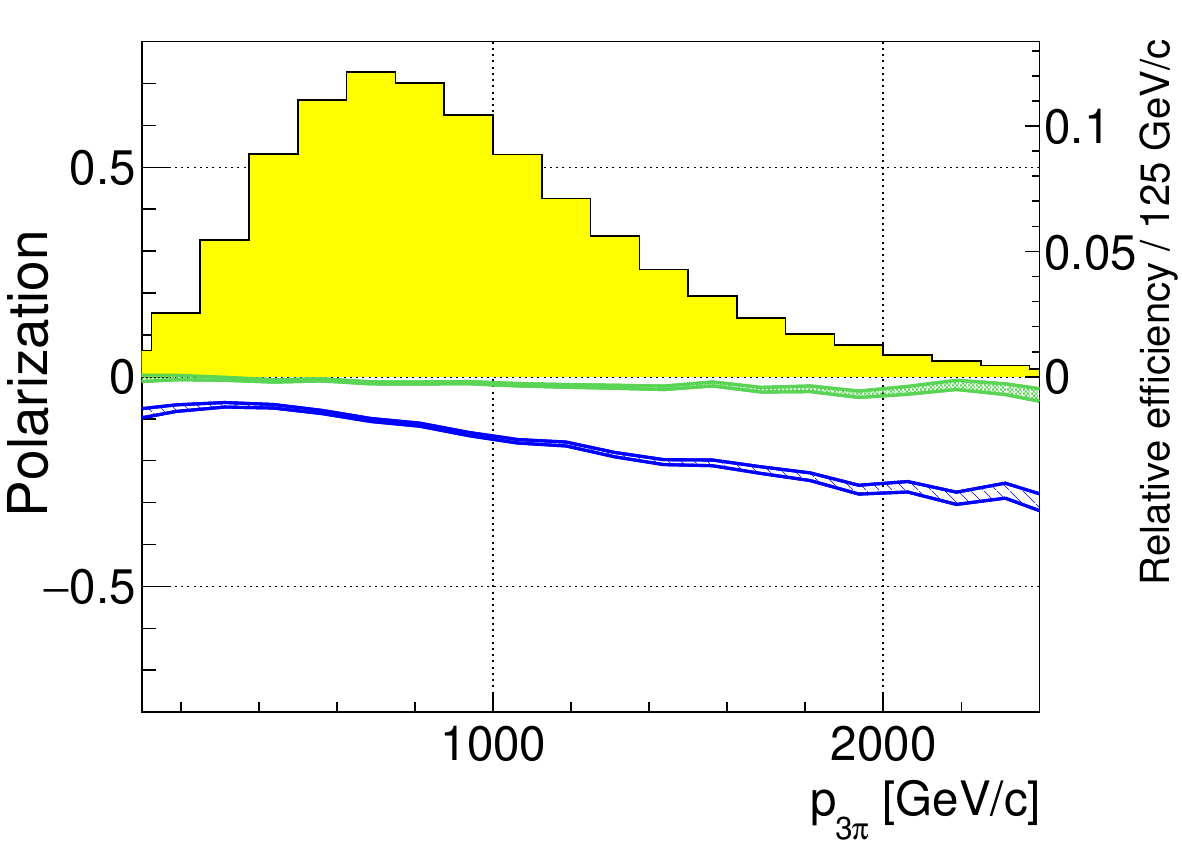}
	\caption{ From Ref.~\cite{Fu:2019utm} (right). (Left) schematic representation of the spin direction (blue arrows) for \taup leptons and neutrinos. In the laboratory frame (\textit{e.g.} consider the \Dsp boosted towards the right), the momentum transfer in the $\Dsp\to\taup\nu_\tau$ decay gives slightly more longitudinal momentum to the forward \taup leptons (in the top panel). For this reason, the longitudinal \taup polarization (right panel, blue band) is negative and increases with the momentum of the $3\pi$ system in the $\taup\to \pip \pip \pim \bar{\nu}_\tau$ decay. The momentum distribution from \textsc{Pythia} simulations is shown (right panel, yellow histogram).
	}
	\label{fig:taupolarization}
\end{figure}

The main source of $\tau$ leptons in hadronic machines like the \lhc is, by far, the decay of charm mesons and specifically the $\Dsp \to \taup \nu_\tau$ decay. Since the $\Dsp$ meson has spin 0, the \taup and $\nu_\tau$ spins are antialigned and, since the neutrino is left-handed, both spins are pointing \textit{inwards}, as schematically shown in Figure~\ref{fig:taupolarization} (left).  Averaging over all \taup directions, the net polarization is zero. Thus, a method to select the $\tau$ direction in the \Dsp rest frame is needed. However, without the neutrino kinematic information, we cannot access the \Dsp frame.

One possibility was proposed in Ref.~\cite{Fomin:2018ybj}. By fixing the relative direction (in the laboratory frame) between the \Dsp and \taup it is possible to obtain statistical information on the polarization. In particular, by selecting \taup particles produced always to the same side of the \Dsp direction, an initial transverse polarization can be achieved. However, the method proposed in Ref.~\cite{Fomin:2018ybj} to literally \textit{fix} the \Dsp direction needs an additional bent crystal to channel the \Dsp particles and thereby know their direction at the first-crystal exit. This induces huge efficiency losses. Instead of using an additional crystal, the \Dsp direction could be correlated to the direction of other particles produced in the same proton-target interaction. This method has been partially explored in Ref.~\cite{NegreSimoMasterThesis}.

A different possibility was investigated in our article~\cite{Fu:2019utm}. Instead of selecting \taup particles going \textit{sideways} from the \Dsp direction, we can focus on discerning \textit{forward} and \textit{backward} \taup directions. For an event where \taup and \Dsp are aligned in the laboratory frame, it is impossible to differentiate if the \taup is forward or backward in the \Dsp rest frame. However, statistically, forward \taup particles have slightly more momentum. Through a simple kinematic cut on the visible part of the \taup decay, \eg $p_{3\pi} \geq 800 \gev$ using $\taup\to 3\pipm \bar{\nu}_\tau$ decays, an initial longitudinal polarization of $\approx-18\%$ can be obtained for channeled $\taup$ leptons~\cite{Fu:2019utm} (see Figure~\ref{fig:taupolarization} (right)).

\section{Spin precession} \label{sec:spinprecess}


In an homogeneous magnetic field \Bvec, the classical motion of the spin-polarization vector \spol is described by the Larmor precession, commonly studied in general-physics courses,
\begin{equation}
\frac{d \spol}{d t} = \muvec \times \Bvec~,
\end{equation}
where \muvec is the magnetic moment. The covariant version of this equation was obtained by Frenkel~\cite{FRENKEL} using an antisymmetric tensor as the relativistic generalization of the spin. Bargmann, Michel and Telegdi~\cite{Bargmann:1959gz} significantly simplified it by using a spin four-vector and included the effect of the electric dipole moment. For charged particles, the Lorentz force curves the particle trajectory and a Wigner rotation at every instant is needed to track the rest frame of the particle. This induces the Thomas precession~\cite{Thomas:1926dy,Thomas:1927yu}. Altogether, we have the Thomas-Bargmann-Michel-Telegdi (TBMT) equation. In its covariant form (see explicit derivation in Ref.~\cite{Fukuyama:2013ioa}),

{
\begin{align}
\label{eq:TBMTCov}
\frac{da^\mu}{d\tau} 
= ~&\frac{g \mu_B}{\hbar} \left[ F^{\mu\nu}a_\nu + \frac{1}{c^2} \left( a_\alpha F^{\alpha\beta} u_\beta \right) u^\mu \right]
- \frac{1}{c^2} \left( a_\alpha \dot{u}^\alpha \right) u^\mu \\
&- \frac{d \mu_B}{\hbar} \left[  F^{*\mu\nu}a_\nu + \frac{1}{c^2} \left( a_\alpha F^{*\alpha\beta} u_\beta \right) u^\mu  \right] ,
\end{align}}

\noindent where $F^{\mu\nu}$ is the electromagnetic tensor, $a^\mu = (a^0,\bm a)$ is the spin
4-pseudovector, $p^\mu = m u ^\mu = \left(E/c,\bm p \right)$ is
the momentum 4-vector, and $\tau$ the proper time.
For homogeneous fields,
the  velocity derivative is given by the Lorentz force,
\begin{equation}
\dot{u}^\mu \equiv \frac{du^\mu}{d\tau} = \frac{q}{mc} F^{\mu\nu} u_\nu .
\end{equation}
In the rest frame of the particle, $a^\mu=(0, \bm s)$, $p^\mu=(mc,\bm 0)$, where $\bm s$ is the non-relativistic spin-polarization 
vector.
Therefore, in any frame $a^\mu p_\mu=0$ and $a_\mu a^\mu=-{\bm s}^2$.

In a the laboratory frame, where the particle has velocity $\bm \beta = \bm p / m \gamma $,
$a^\mu$ is given by~\cite{Jackson:1998nia,Leader2011}
\begin{equation}
\bm a = \bm s + \frac{\gamma^2}{\gamma+1} (\bm \beta \cdot \bm s) \bm \beta~,~~
a^0 = \bm \beta \cdot \bm a = \gamma(\bm \beta \cdot \bm s) ,
\label{eq:SpinLab}
\end{equation}
where the components of the momentum 4-vector are $p^0 = \gamma m c^2$ and $\bm p =\gamma m \bm \beta c$. 
Substituting in the covariant Eq.(\ref{eq:TBMTCov}), the
spin precession equation is~\cite{Jackson:1998nia,Leader2011,Fukuyama:2013ioa,Silenko:2014uca},
\begin{equation} 
\label{eq:TBMTgeneral}
\frac{d \bm s}{ d t} =  \bm s \times \bm \Omega ~, ~~~ \bm \Omega = \bm \Omega_{\rm MDM} + \bm \Omega_{\rm EDM} + \bm \Omega_{\rm TH} ,
\end{equation}
where $t$ is the time in the laboratory frame, and
the precession angular velocity vector $\bm \Omega$ has been split into three contributions,
\begin{equation} 
\label{eq:OMEGAgeneral}
\bm \Omega_{\rm MDM} 
= \frac{g \mu_B}{\hbar} \left( \bm B - \frac{\gamma}{\gamma+1}(\bm \beta \cdot \bm B)\bm \beta - \bm \beta \times \bm E\right) , 
\end{equation}
\begin{equation}
\nonumber
\bm \Omega_{\rm EDM} 
= \frac{d \mu_B}{\hbar} \left( \bm E - \frac{\gamma}{\gamma+1}(\bm \beta \cdot \bm E)\bm \beta + \bm \beta \times \bm B\right) , 
\end{equation}
\begin{equation}
\nonumber
\bm \Omega_{\rm TH}
= \frac{\gamma^2}{\gamma+1} \bm \beta  \times \frac{d \bm \beta}{d t}
= \frac{q}{mc} \left[ \left( \frac{1}{\gamma} - 1 \right) \bm B + \frac{\gamma}{\gamma + 1} (\bm \beta \cdot \bm B)\bm \beta  -  \left(  \frac{1}{\gamma+1} -1 \right) \bm \beta \times \bm E  \right],
\end{equation}
corresponding to the MDM, EDM and Thomas precession.
The electric and magnetic fields, $\bm E$ and $\bm B$, respectively, are expressed in the laboratory frame.

For a neutral particle ($q=0$) the Thomas precession term does not contribute and we obtain the classical 
equation, $d \bm s/ d \tau = \bm \mu \times \bm B^* + \bm \delta \times \bm E^*$,
where $\bm E^*$ and $\bm B^*$ are the external fields in the rest frame of the particle~\cite{Jackson:1998nia}. 
Equations~(\ref{eq:TBMTgeneral}) and~(\ref{eq:OMEGAgeneral}) can be generalized to account for field gradient effects as described in Refs.~\cite{Good:1962zza,Metodiev:2015gda}. These effects are always negligible in our cases as described in Appendix A of Ref.~\cite{Botella:2016ksl}.

\begin{figure}
	\centering
	\includegraphics[width=0.55\linewidth]{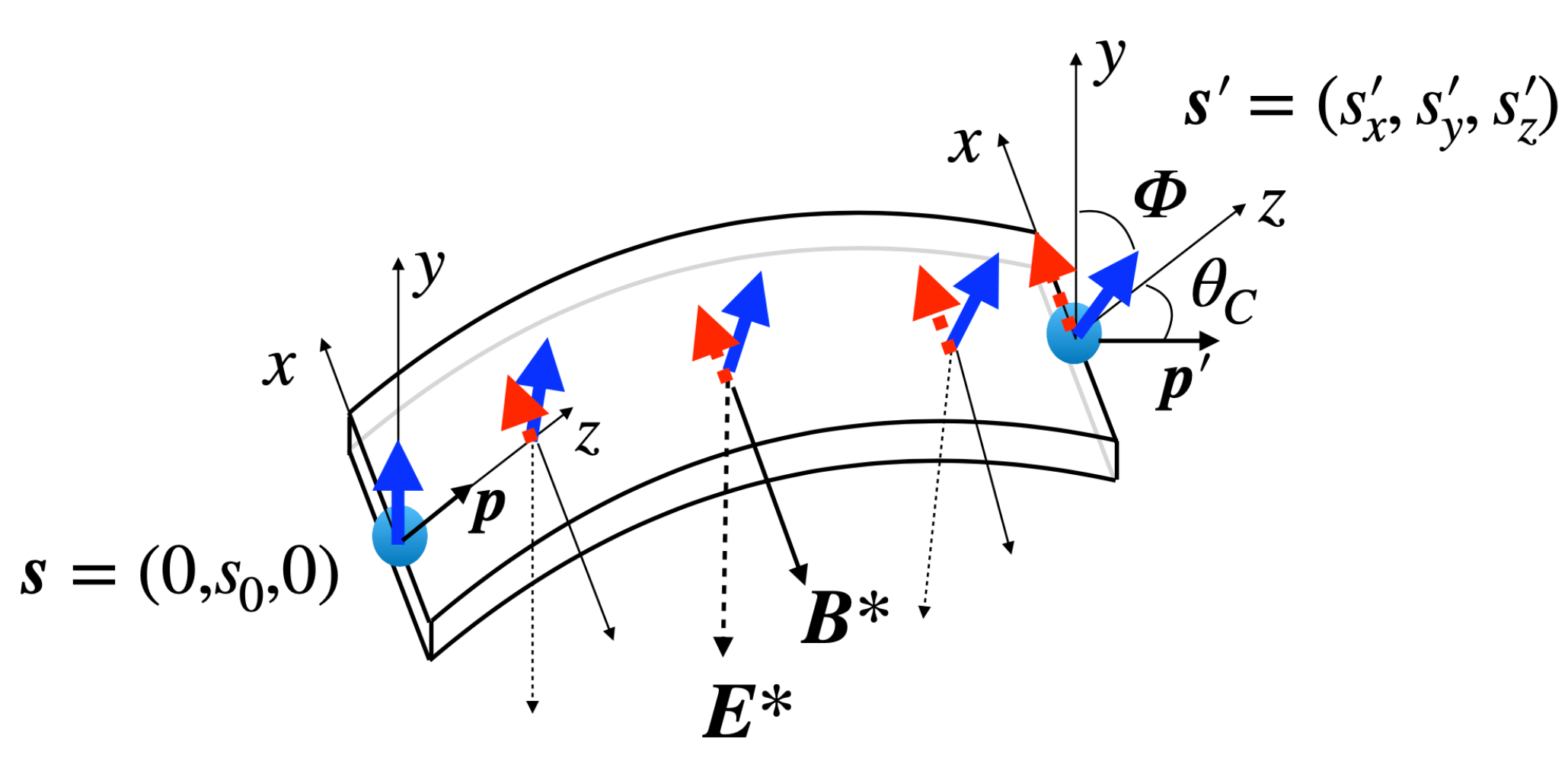}
	\caption{From Ref.~\cite{Aiola:2020yam}.
		Rotation of the (blue) polarization vector around the magnetic field $\Bvec^*$ and build-up of the (red) $s_x$ component due to the EDM precession around $\Evec^*$. The reference frame $(xyz)$ is defined by the crystal axes at the crystal entry. The axes are transported parallel to themselves, along the (curved) particle trajectory.
	}
	\label{fig:axes}
\end{figure}

An illustrative solution to the TBMT equation can be found for the bent crystal experiment with heavy baryons (pure transversal polarization) with some approximations~\cite{Botella:2016ksl},
\begin{equation}
\bm s ~=~
\left\lbrace
\begin{array}{l}
s_{x} \approx   s_0 \dfrac{d}{g-2}  (\cos{\Phi}-1)  \\
s_{y} \approx   s_{0} \cos\Phi \\
s_{z} \approx   s_{0} \sin\Phi
\end{array}
\right.
\text{,~where~} \Phi \approx \frac{g-2}{2}\gamma \theta_C.
\label{eq:precessionsimplified}
\end{equation}
The main precession occurs in the crystal $yz$ plane (channeling plane), while a component along the crystal $x$ axis arises in the presence of an EDM, $d\neq 0$. The total precession angle $\Phi$ is proportional to both the anomalous magnetic moment $(g-2)$ and the integrated magnetic field along the crystal channel. This, in turn, is determined by the crystal bending angle $\theta_C$ and the Lorentz boost factor $\gamma$. The coordinates $(x,y,z)$ are defined in the rest frame of the particle moving along the curved trajectory, as explained in the caption of Figure~\ref{fig:axes}.

\section{Opportunities at the LHC} \label{sec:configurations}


We present in the following a summary of the considered layouts and a comparison of the possible sites for the installation of the bent crystal experiment.

\subsubsection{Three layouts}

\begin{figure}
	\centering
	\includegraphics[width=0.9\linewidth]{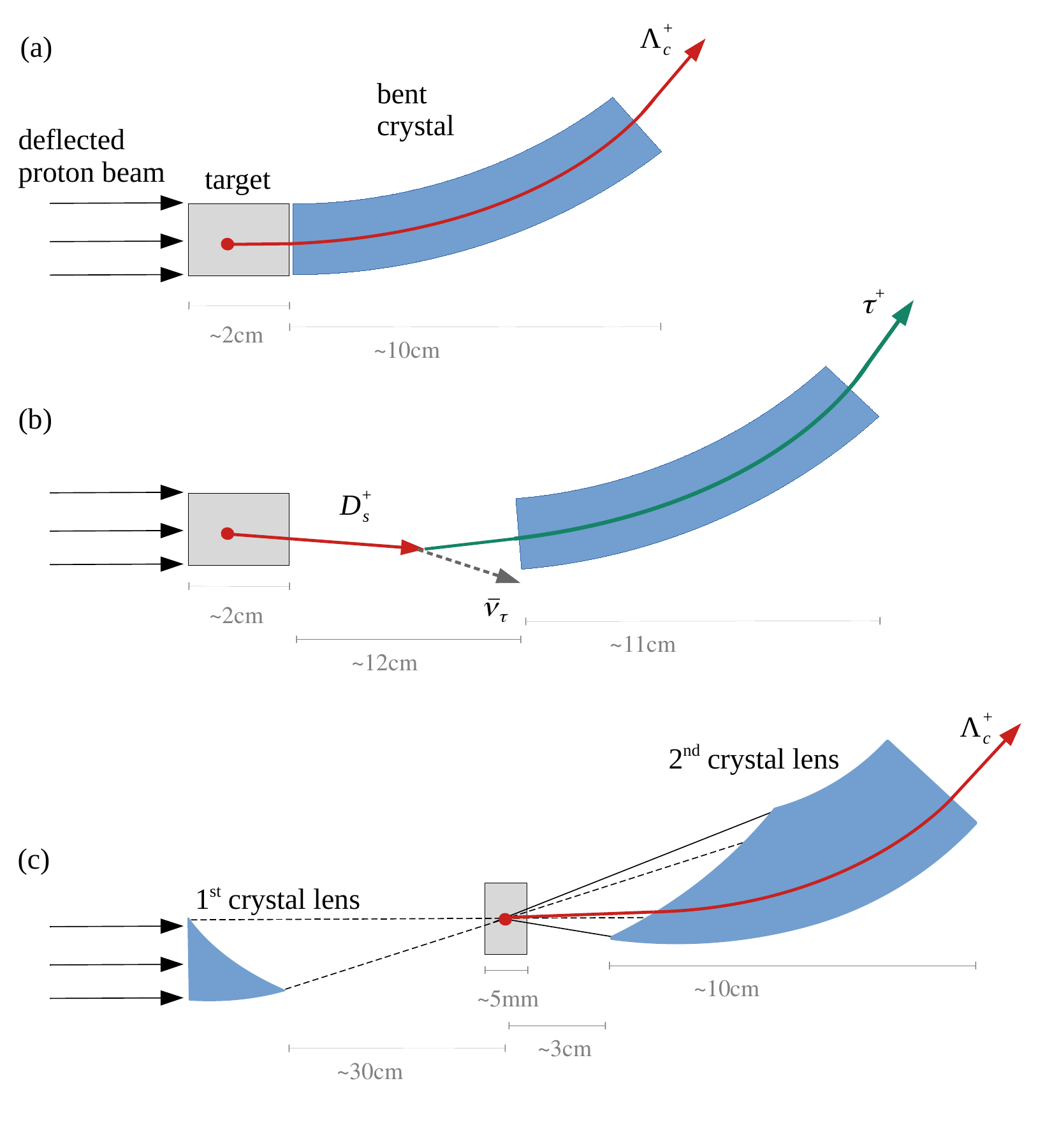}
	\caption{Target/crystal configurations optimized to measure (a) heavy baryons, with attached target and crystal; (b) \taup leptons, with free room for the \Dsp decay; and (c) an alternative (more technically challenging) layout based on crystal lenses. The displayed lengths represent the optimal values as obtained in Chapter~\ref{ch:sensitivity}. 
		In Figure~\ref{fig:layoutgeneral}, only configuration (a) is displayed, while the rest of the layout would be identical for cases (b) and (c).}
	\label{fig:threelayouts}
\end{figure}

Depending on the targeted particle to be measured and the type of crystal geometry, the following target/crystal configurations have been studied, represented in Figure~\ref{fig:threelayouts}. The rest of the general layout (crystal kicker, absorber, vertical position within the beampipe) is conceptually identical to that presented in Figure~\ref{fig:layoutgeneral}. 

\begin{enumerate}
	\item[a.] Heavy baryons
	
	The main physics case of the proposal, charm baryon EDM and MDM, would be explored with this configuration. It provides a unique possibility to directly measure these observables and it is also the simplest setup to be realised. For this reason, sometimes it will be referred to as nominal or baseline layout. To maximize the number of charmed baryons that survive until the end of the crystal, target 
	and crystal 
	are attached. This setup will be explored in detail in Section~\ref{sec:optimbaryons}
	
	\item[b.] $\tau$ lepton
	
	Due to the small size of $(g-2)_{\tau}$ and the relatively low \taup production through the $\Dsp\to\taup\nu_\tau$ decay, to access the first digit of the SM $(g-2)_{\tau}$ prediction we would need extended periods of data taking. The \taup measurement would then be most compatible with a dedicated experiment at the LHC. The \Dsp meson, whose lifetime is actually larger than that of the $\taup$, has to decay before reaching the crystal. Thus, a separation ($\approx 12 \cm$) between target and crystal is required. The optimization of this distance and other setup parameters will be presented in Section~\ref{sec:optimtau} along with the sensitivity studies.
	
	\item[c.] Focusing crystals
	
	The channeling probability in the nominal layout, $\order(10^{-4})$, is mostly affected by the low trapping efficiency. This is determined by the small Lindhard angle for $\sim 1 \tev$ \Lc particles ($\theta_L\approx 7 \murad$) as compared to their initial divergence (within $\varTheta_{\Lc}\approx 1.5\,\mrad$). The Lindhard angle cannot be increased but the overall geometry of the setup can be changed to trap particles in a wider range of directions. With a crystal lens (right side of the target in Figure~\ref{fig:threelayouts} (c)), the atomic planes are not parallel at the crystal entrance, but all point towards the target, where the particles are produced. However, the exact \Lc production point within the target is extremely relevant to determining whether the particle will be trapped, and it is highly correlated with the outgoing angle. To reach a genuine gain in trapping efficiency, of about a factor 20, a first crystal lens is needed to focus the protons onto the centre of the target, coinciding with the focal point of the second lens. The higher complexity of this layout in terms of crystal manufacturing and precision alignment makes the double-lens scheme more feasible for a second stage of the bent crystal experiment.
	
	The general layout with focusing crystals was presented by Biryukov in Refs.~\cite{Biryukov:2021gsd,Biryukov:2021phs}. The geometrical details and realistic estimations of the trapping efficiency were obtained in Ref.~\cite{Biryukov:2021cml}, which is mostly reproduced in Section~\ref{sec:optimfocusing} of this thesis. A detailed study of the gain for \taup leptons does not exist yet.

\end{enumerate}

\subsubsection{Two sites}

The bent crystal experiment may be realised in two possible sites ({interaction regions}, IR) at the LHC ring, each of them with its advantages and disadvantages.

\begin{enumerate}
	\item IR8 / LHCb: 
	the target and bent crystal could be installed in front of the LHCb detector, which is fully instrumented in the forward region like a fixed-target experiment. The device would be around $1.2\,\m$ before the nominal $pp$ collision point. This option would profit from the excellent tracking resolution of the Vertex Locator (VELO), and the high performance of the PID system of LHCb. Furthermore, the LHCb collaboration already has experience in fixed-target events with the SMOG system, which releases inert gasses in the beam pipe to record $p$-gas events. However, with a solid target, the number of interactions per proton is much larger and precise control of the proton flux is critical to ensure radiation safety. Nevertheless, precisely because the LHCb is a running experiment, all the software tools to evaluate the detector occupancy are already in place. Studies on this front were presented in Ref.~\cite{internalnote}, demonstrating the compatibility of the bent-crystal device with the LHCb experiment.

	\item IR3:
	a straight section of the beam at IR3 would also be compatible with the installation of a fixed-target experiment at the LHC. Besides the target plus crystal device, to be inserted in the beam pipe, this option would require a completely new compact detector for the reconstruction of the decay products. The basic layout would consist of a few ($6-8$) tracking stations and a dipole magnet to reconstruct the particle momentum. No dedicated PID systems and/or calorimeters will be required in the first stage of the experiment provided that heavy baryons can be fully reconstructed with similar invariant mass and angular resolution as in LHCb~\cite{Alves:2008zz}. The advantages of this option would be the possibility to have continuous data-taking from the beginning of the experiment, a significantly lower bending angle that would largely increase the event yields, a higher rate of protons on target, easier access to the setup, and less machine/detector safety risks. Moreover, there is significantly more flexibility in the machine parameters and optics to fine-tune the proton extraction method and intensity~\cite{Mirarchi:2019vqi}.

	A proof-of-principle experimental test (combining the accelerator, target/crystal device and detector) to produce the first physics results is under preparation and could be installed during a technical stop of the LHC Run III~\cite{proofofprincipletalk}.
	

\end{enumerate}

\section{The experiment's key factors} \label{sec:keyfactors}

Moving from the conceptual layout to the actual setup optimization we encounter many free setup parameters, detector effects, and analysis strategies that play a major role on the final sensitivity. In the following, these experiment's key factors are introduced, pointing out the main ideas behind them. This may serve as a sort of glossary of concepts for the detailed numerical discussions of Chapter~\ref{ch:sensitivity}.

We will focus on the \Lc baryon case and, if there are relevant differences with the other layouts they will be pointed out.

\paragraph{Target material.} The target would be made of tungsten (W), a standard material in fixed target experiments due to its high density and short nuclear interaction length, while at the same time being relatively easy to manufacture.

\paragraph{Target thickness.}

\begin{figure}[h]
	\begin{center}\
		\includegraphics[width=0.45\columnwidth]{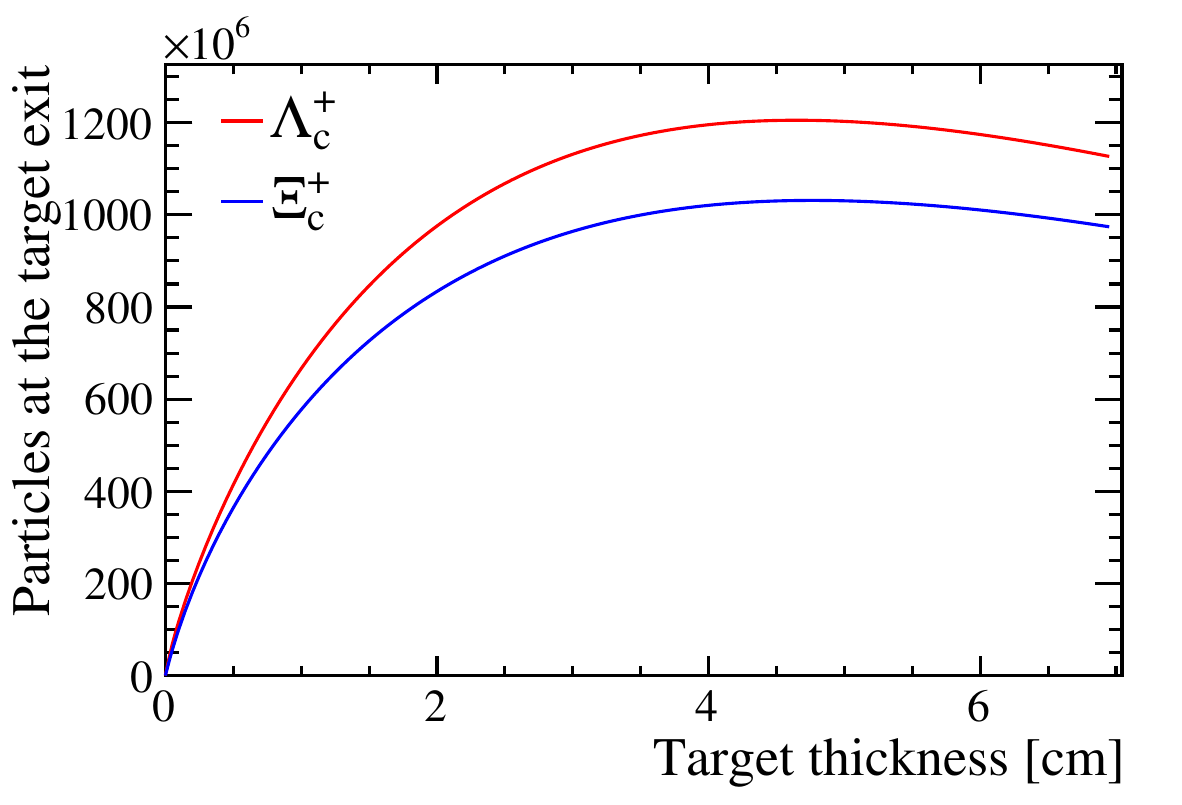}
		\caption{\label{fig:targetNLc} From Ref.~\cite{Aiola:2020yam}. Number of produced \Lc and \Xicp baryons exiting
			the \W target as a function of the target thickness. The case of 
			$1.37\times 10^{13}$ \pot, corresponding to two years running in \lhcb~\cite{Mirarchi:2019vqi}, is shown.
		}
	\end{center}
\end{figure}

Increasing the target thickness $T$ along the beam direction increases the number of \Lc baryons produced at the target. However, these \Lc may be reabsorbed or even decay before the target exit. Furthermore, the proton flux is attenuated for long targets. Accounting for these effects, the number of \Lc baryons at the target exit is~\cite{Aiola:2020yam}
\begin{equation}
\label{eq:Nc_target}
N_{\Lc}(T,\beta\gamma) = \frac{N_\pot}{\lambda_{\W,\Lc}}\beta\gamma c\tau \left( e^{-T/\lambda_\W}  - e^{-T/\lambda'} \right),
\end{equation}
where ${N_\pot}$ is the number of protons on target, $\lambda_{\W,\Lc}\approx 81.35\m$ is the mean free path for \Lc production and ${\beta \gamma c \tau \approx 3.0\cm}$ the mean free path for \Lc decay (at 1\,\tev).
In the exponentials, ${\lambda_\W \approx 8.87\cm}$ is the tungsten nuclear interaction length at $\sqrt{s} \approx 115\gev$ and ${1/\lambda'=1/\lambda_\W^{(\Lc)}+1/(\beta \gamma c \tau)}$ combines the mean free path of absorption $(\lambda_\W^{(\Lc)} \approx \lambda_\W )$ and decay ${(\beta \gamma c \tau})$ of the \Lc.
The optimal length, as shown in Figure~\ref{fig:targetNLc}, is around 2-6\,\cm. In this figure, $N_\Lc(T,\beta\gamma)$ (Eq.~\eqref{eq:Nc_target}), is convoluted with the \Lc baryon spectrum as obtained from \pythia simulations \cite{Aiola:2020yam}. To minimize the background and detector occupancy from secondary interactions while still maximizing the number of potential signal events, we will take $T=2\,\cm$.
In the case of focusing crystals, most of the channeled \Lc baryons are produced at the centre of the target favouring shorter target lengths of $T\approx5\mm$. This optimization will be presented in Section~\ref{sec:optimfocusing}.

\paragraph{Crystal parameters.}


The crystal bending angle \thC and length \lenC are the only crystal parameters that can be tuned to increase the channeling efficiency for a given crystal material, atomic lattice orientation, and operation temperature. It is also common to characterize the curvature through the crystal bending radius,
\begin{equation}
R = \lenC / \thC ~.
\end{equation}
However, increasing the channeling efficiency does not necessarily lead to increasing the sensitivity. The most useful information is in the \Lc particles that have experienced a large precession angle, $\Phi \propto \gamma \thC$. However, increasing the bending \thC, the trapping probability for high-energy particles is reduced, thus reducing the average $\gamma$. This effect may be relaxed by increasing \lenC proportionally and keeping a constant $R$. However, for larger crystal lengths, the total dechanneling probability increases as well as the decay probability. As we can see, the optimal point of $(\thC,\lenC)$ is a trade-off between many effects. For this reason, we will use Monte Carlo simulations that can account for all of them at once. In these simulations, it is crucial to have a realistic distribution of initial particle momentum, that we will obtain from \pythia. 


Crystal lenses have additional degrees of freedom due to their complex geometry. In particular, the shape of the focusing surface and thickness in the transverse direction also play a role in the trapping efficiency. These will be optimized in Section~\ref{sec:optimfocusing}.

\paragraph{Crystal manufacturing. }

%
%
Evidently, an absolute requirement to do the experiment is the possibility to manufacture the crystals. The INFN-Ferrara has already produced several crystal prototypes of silicon and germanium with the required characteristics for the experiment. These prototypes have been tested on beam using $180$-\gev hadrons at the H8 external beam line of the Super Proton Synchrotron (SPS) at CERN. The results of the first test show a good agreement between data and simulations, as reported in Ref.~\cite{Aiola:2020yam}.

\paragraph{Crystal tilt.}

We define the crystal tilt $\tilt$ as the angle between the proton beam and the crystal atomic planes, projected in the $y_L z_L$ plane. This makes that the crystal frame axes $(xyz)$ are rotated with respect to the laboratory frame axes $(x_L y_L z_L)$ around the $x=x_L$ axis, with the beam direction defined by $z_L$. The crystal tilt was initially introduced~\cite{internalnote} to avoid channeling of beam protons in the crystal, $\tilt \geq \theta_{L} \approx 2\,\murad$, but in fact this was not a real issue since 7-\tev protons would not get channeled for the range of bending radius being considered. However, the initial tilt turns out to be critical to maximising the sensitivity to the EDM (irrelevant for MDM). By adjusting this tilt, we can modify the direction of the initial polarization as seen in the crystal reference frame. Specifically, without tilt the initial polarization is all in the $y$ direction, parallel to $\Evec^*$ (Figure~\ref{fig:axes}), preventing the rotation of the EDM around $\Evec^*$; but with tilt, we can induce an initial $s_x$ component of the polarization which allows the EDM precession to start building up since the beginning of the channeling. More precisely, $s_x$ and $s_y$ depend on the tilt as
\begin{equation}
\label{eq:pol_init}
{\boldsymbol s} = (s_x,s_y,0) \approx \frac{s_0(\pt)}{\pt} \left( -p_{y_L}, p_{x_L}, 0 \right),
\end{equation}
where $p_{x_L}$ and $p_{y_L}=p\sin\theta_{y,C}$ are the transverse momentum components along the laboratory $x_L$ and $y_L$axis, respectively, and $p$ 
is the total momentum of the charm baryon.

However, a large crystal tilt also has a drawback. The initial \Lc direction with respect to the proton beam is highly correlated with the \Lc momentum. For $\tilt\geq 200\,\murad$ the average momentum of channeled particles starts to decrease, although the gain on the EDM measurement compensates for this effect, with an optimal tilt around $\tilt\approx500\,\murad$. Nevertheless, there is a very broad region of $\tilt \in [300,700]\,\murad$ with only a 20\%-variation in EDM sensitivity.

\paragraph{Crystal angular positioning.}

Applications of crystal channeling for beam deflection require high angular precision on the crystal position. This is particularly important to deflect LHC protons, as their Lindhard angle is only of $\approx 2\,\murad$. Thus, high precision goniometers are required for the crystal kicker (in Figure~\ref{fig:layoutgeneral}). These requirements are relaxed for the angular positioning of the crystal for spin-precession. In fact, a crystal misalignment of up to $200\,\murad$ has no visible effect on the channeling efficiency.
Thus, the required angular precision in the nominal layout is determined by the desired precision on the crystal tilt which, as mentioned in the previous paragraph, can lie in a region of hundreds of $\murad$ with similar EDM sensitivity.


With focusing crystals the angular precision is much more important, as the focal points of the first and second crystal must be made to coincide (see Figure~\ref{fig:threelayouts} (c)). However, we can achieve the same result by adjusting the vertical position of the crystals itself, for which nanometer-precision positioning systems are in principle available with current technology. This is discussed in detail in Section~\ref{sec:optimfocusing}.

\paragraph{Crystal position.}

Accurate simulations of the LHC beam optics project the deflected proton beam to be about 4 (6) \mm above the main beam when it arrives to the fixed-target setup at IR8 (IR3)~\cite{Mirarchi:2019vqi}. Thus, target and crystal must be positioned with high accuracy to avoid interference with the LHC beam core. 

With the detailed detector layout of the upgraded LHCb we can fine-tune the crystal position at IR8 already. The support structure of the SciFi, a tracking station after the LHCb dipole magnet, has a 4-\mm wide column along the vertical axis~\cite{LHCb-TDR-015}, reducing the detector efficiency for highly-energetic particles, which experience little bending. To avoid these dead regions, the crystal is rotated about its longitudinal axis by about 25 degrees, sending the channeled particles into the detector region with maximum acceptance~\cite{internalnote}.

\paragraph{Proton flux.}


From the total LHC beam flux\footnote{The LHC beam stores a total of $3.2\times10^{14}$ protons corresponding to a current of 0.58 A. These numbers will be increased by a factor of 1.9 at the HL-LHC~\cite{Apollinari:2017lan}.} $\sim 10^{18} \,p/s$ as few as $\sim10^6 \,p/s$ are deflected from the beam halo and directed to the target~\cite{Mirarchi:2019vqi}. Still, since these arrive at a solid target (plus a solid crystal), their interaction probability is much greater than that of the $pp$ head-on collisions. The quantity of interaction products leaving signals in the detector (detector occupancy and readout bandwidth) ultimately determines the feasible proton flux.

Regarding the feasibility of beam deflection with the crystal kicker, the UA9 collaboration has realised successful tests with an essentially identical setup at SPS~\cite{Scandale:2021zbn}	and LHC~\cite{Scandale:2016krl}.

\paragraph{Background rejection strategy.}

\begin{figure}[t]
	\centering
	\includegraphics[width=0.45\linewidth]{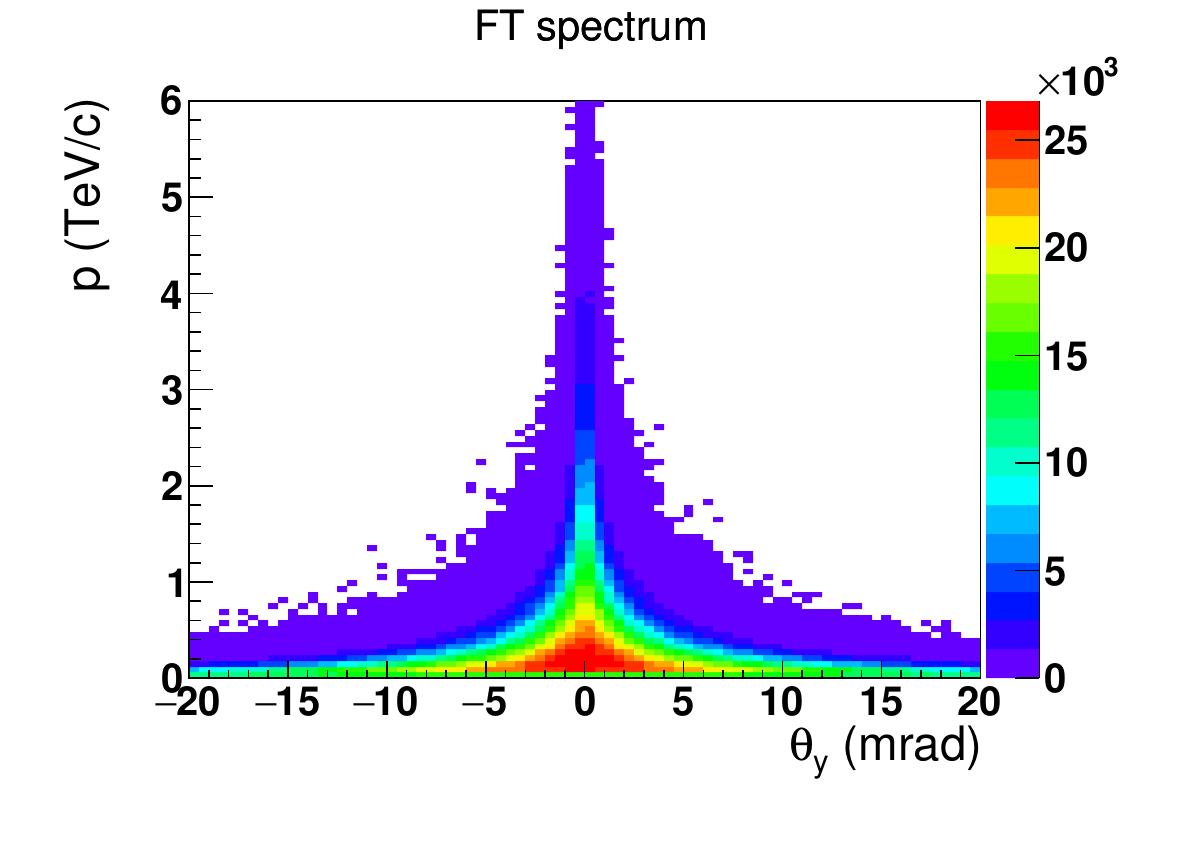}
	\includegraphics[width=0.45\linewidth]{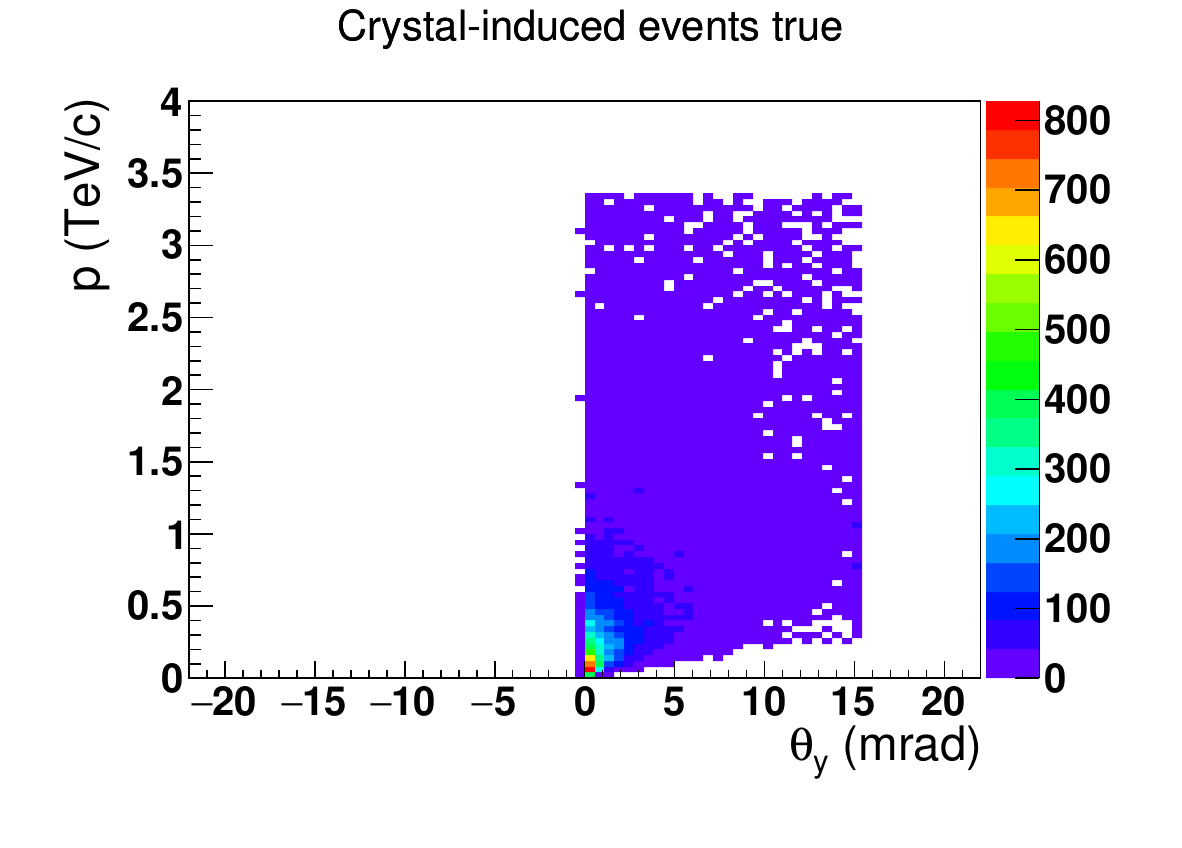}
	\caption{ 
		Two-dimensional distribution of momentum and vertical angle of \Lc particles for (left) non-channeled particles, \textit{i.e.} initial spectrum, and (right) (partially-)channeled \Lc particles. 	
		\revafter{to-do: add signal box in both plots} 
	}
	\label{fig:bkgrejection}
\end{figure}

Among all \Lc particles produced in the fixed target only a tiny fraction 
(order $10^{-5}$) are channeled through the whole crystal. It is essential to find a signature of these events that distinguishes them from other \Lc backgrounds. In Figure~\ref{fig:bkgrejection} (left) we see the distribution in momentum and \thy\footnote{This ($\thy$) is the angle of the \Lc with respect to the atomic planes, projected in the $x_L z_L$ plane. In the introduction to crystal channeling (Section~\ref{sec:channeling}) it was noted $\theta$.} angle of all produced \Lc. A crystal of $\thC=15\,\mrad$ traps particles at $\thy=0$ (unless there is a tilt) and steers them to $\thy=\thC$ preserving the initial momentum. As we can see, in the initial spectrum there are no particles of high momentum ($p_\Lc \geq800\gev$) at such large angles ($\thy=15\,\mrad$). This will be our signature.

Among the \Lc backgrounds we will have particles produced in either the target or the crystal. The majority are not channeled and follow the initial spectrum of Figure~\ref{fig:bkgrejection} (left). Partially channeled particles that do not reach the end of the crystal, because of dechanneling or decay, will have a \thy angle lower than $\thC$, as shown in Figure~\ref{fig:bkgrejection} (right). By applying the cuts $p_\Lc\geq800\,\gev$ and $\thy = \thC \pm 5\sigma_{\thy}$, where the angular resolution is $\sigma_{\thy} \approx 25\, \murad$ at LHCb, we can retain 81\% of fully-channeled \Lc particles with a background contamination of only 4\% of the candidates. The remaining backgrounds are in fact signal-like events: these particles travel through almost the whole crystal, experiencing similar precession angles.

Another type of partially channeled backgrounds comprises \Lc particles that reach the crystal exit but are produced in the middle of the crystal, aligned with the local direction of the atomic planes. These are harder to separate since they have the same final angle. However, after the momentum cut, only high-energy particles survive, which are produced mostly forward. These, to be trapped, had to be produced at the beginning of the crystal channel, again experiencing almost the same deflection, and spin precession, as the signal events. From simulations, we obtain that the induced bias on the precession angle $\Phi/\gamma$ is 4.1\% and could be reduced by adding vertex information on the \Lc production point.

\begin{figure}[t]
	\centering
	\includegraphics[width=0.45\linewidth]{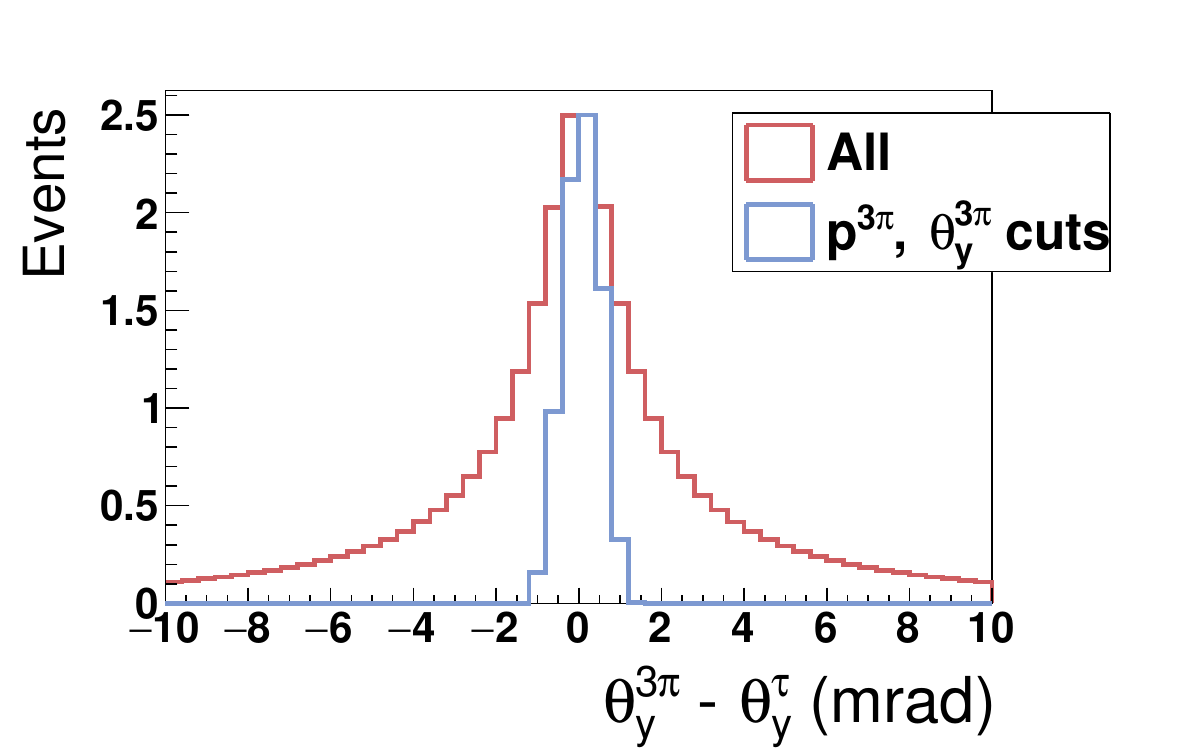}
	\caption{Difference between the vertical angle \thy of the combined $3\pi$ system and the true angle of the \taup lepton. After the background-rejection cuts, the signal candidates have a significantly narrower distribution.}
	\label{fig:resolThetay}
\end{figure}

Regarding the measurement with \taup leptons, the angular cut on $\thy$ is not as precise. Even though the detectors may have a great angular resolution on single tracks ($\approx25\,\murad$), the invisible part of the $\tau\to\pip\pip\pim\nu_\tau$ decay induces an intrinsic resolution ($\approx3\,\mrad$) between the 3$\pi$ system and the true $\tau$ lepton, as shown in Figure~\ref{fig:resolThetay}. Fortunately, the momentum cut selects \taup particles with collimated decay products, which resolution on $\thy$ is much improved ($\approx500\murad$). This ensures that the selected candidates travel almost the entire length of the crystal channel.

Due to the lower production of \taup leptons, many more charm mesons, with similar mass to the $\tau$ lepton, are produced and channeled. These physical backgrounds have a very similar signature to the signal events, especially when they are partially reconstructed and do not create peaking backgrounds. Methods to deal with these backgrounds in a dedicated experiment are under study, and would probably require high-granularity calorimeters to reconstruct neutral particles, PID systems for charged particles, and a further separation of target and crystal to favour long decays (as $\Dsp$ plus $\taup$).

\paragraph{Reconstruction efficiency and resolution.}

The products of proton-target interaction in the very-forward region, where the crystal may intercept them, may have momenta of several TeV. Their decay products are extremely collimated and their trajectories are only slightly bent in the tracking magnetic field for momentum reconstruction. Fortunately, we can evaluate precisely the resolution and detector efficiencies at LHCb with full Monte Carlo simulations of the detector response.
These studies were presented in Ref.~\cite{internalnote} and yield a reconstruction efficiency for $\Lc\to p \Km \pip$ fixed-target events of about 35\% (before offline selections) and \Lc invariant-mass resolution of 20\,\mev.

\paragraph{Final polarization reconstruction.}

To reconstruct the spin-polarization vector, the decay products must have a preferential direction with respect to the initial spin.  The differential distribution in the mother-particle rest frame,
\begin{equation}
\label{eq:AngDist}
\frac{dN}{d\Omega'} \propto 1 + \alpha \bm s \cdot \hat{\bm k} ~,
\end{equation}
valid for any $\frac{1}{2} \to \frac{1}{2} \,J$ ($J$ integer) and $\frac{1}{2} \to J\, 0 $ ($J$ half-integer) transition,
peaks when the daughter-particle direction $ \hat{\bm k}$ is aligned with the spin-polarization vector \spol. This is parametrized through the \P-violating decay-asymmetry parameter $\alpha$ of the decay.
For two-body decays, this parameter is a single value and many measurements are listed in the PDG~\cite{PDG}. However, in decays with three or more decay products, parametrizing the decay asymmetry becomes more involved, as it depends on the intermediate strong resonances of the decay, \textit{i.e.} on the point of the Dalitz phase space.

In the decay $\Lc\to p \Km \pip $, it is in principle possible to select regions of phase space corresponding to the quasi-two-body decays
$\decay{\Lc}{\Delta^{++} \left(\proton \pip\right) \Km$}, 
$\Kstarzb (\Km \pip) \proton $
or
$\Lambda(1520)\left(\proton \Km\right) \pip$, and compute an effective $\alpha_{\rm eff}$ based on the helicity amplitudes of previous Dalitz analyses~\cite{Aitala:1999uq}. These $\alpha_{\rm eff}$ are given in Appendix B of Ref.~\cite{Botella:2016ksl}. However, this method reduces the available statistics and introduces new systematic errors from the large uncertainties of the previous amplitude model~\cite{Aitala:1999uq}. To improve them, an amplitude analysis of the $\Lc\to p \Km \pip $ decay has been performed with large LHCb samples and it is partially public in Ref.~\cite{Marangotto:2020tzf}. Furthermore, with the precise knowledge of the full decay dynamics, the analysis of channeled events can be done with a conditional probability density function (PDF) that accounts for the point of phase space in each event. This method, presented recently in Ref.~\cite{Aiola:2020yam}, utilizes the full sample of $\Lc \to p \Km \pip $ decays.  Moreover, the addition of other decay channels is considered in the same reference, maximizing the available statistics for the EDM and MDM measurement.

In the case of \taup leptons, extracting information on the final polarization only from the 3$\pi$ system is even more challenging. 
A technique based on multivariate classifiers was explored in Ref.~\cite{Fu:2019utm}. Using the invariant masses, momenta and directions of the $3\pi$ system it is possible to have sensitivity to the \taup polarization without previous knowledge of the decay dynamics, neither on the \taup production point or momentum. This sensitivity is parametrized through the \textit{event information} $S$, which, for two-body decays, is related to the decay-asymmetry parameter as $S^2 = \alpha^2 /3$ . 
The achieved event information is, depending on the polarization component with respect to the crystal axes,  $S_x\approx S_y \approx 0.42$ and $S_z \approx 0.23$, which can be compared to the theoretical maximum, extracted with the full kinematic information of the decay, $S=0.58$~\cite{Davier:1992nw}.

\chapter{Optimization and sensitivity} \label{ch:sensitivity}

The uncertainty on the magnetic moment can be estimated through the simplified expression (derived from Eq.~\eqref{eq:precessionsimplified})~\cite{Botella:2016ksl}
\begin{eqnarray}
\sigma_{a} \approx \frac{1}{\alpha s_{0,\zaxis} \gamma \theta_C } \frac{1}{\sqrt{N}} . \ \ \
\label{eq:sensitivitygeneral}
\end{eqnarray}
It directly depends on the initial polarization $\spol_0$, decay-asymmetry parameter $\alpha$, Lorentz factor~$\gamma$, crystal bending angle \thC and the number of channeled and reconstructed particles $N$. The yield itself also depends on the crystal length \lenC, target-crystal separation \Ltarc, impinging proton flux, reconstruction efficiency, and background-rejection cuts. Similarly, $\gamma$ is affected by the crystal parameters (directly by the critical radius $R=\lenC/\thC$ or indirectly through the minimum flight distance $c\tau\beta\gamma\gsim\lenC$), and selection cuts. In turn, the kinematics are also correlated with the initial polarization $\spol_0$.

Thus, in this uncertainty many factors come into play which are related to each other. To obtain reliable estimates of the projected sensitivity we use Monte Carlo simulations and extract the dipole moments by fitting simulated events to a probability density function (PDF), reproducing to a large extent the analysis technique of the actual experiment. 

More specifically, we will use \pythia~\cite{Sjostrand:2006za} and \evtgen~\cite{Lange:2001uf} to simulate the event kinematics in proton-target interactions. 
On these events, we will impose the channeling conditions and, for each of the remaining event candidates, we will randomize the direction of the decay products following the PDF, which contains the spin-precession equations, angular distribution of the decay products and initial polarization. The final sample is fitted to the same PDF and the dipole moments are extracted together with their statistical uncertainties.
\footnote{The procedure of generating and fitting events with the same PDF is usually referred to as Monte Carlo pseudoexperiments or, in jargon, \textit{toy} Monte Carlo. For this part of the studies we will always use the \roofit toolkit~\cite{Verkerke:2003ir} of the ROOT data analysis framework~\cite{Brun:1997pa}.}

Among all the dependences of the PDF, we can distinguish between \textit{fixed parameters} (\textit{e.g.} crystal parameters and decay-asymmetry $\alpha$), \textit{variables} (helicity angles of the decay products), \textit{observables} (the dipole moments), and \textit{conditional parameters} (particle momentum).
%
%
In the generation phase, the PDF uses the particle true momentum and a fixed value for the dipole moments to generate the decay angles. To account for the detector resolutions, the particle momentum and decay angles are smeared before the fitting phase, in which the dipole moments are extracted with an associated statistical uncertainty.

The process is repeated for different values of the setup parameters starting from the channeling conditions (without redoing the \pythia simulations). Then, based on the final uncertainty, we select the optimal configuration. Ideally, we would do a multidimensional optimization scanning over a fine grid of values for the setup parameters. However, note that the channeling efficiency is $\order(10^{-5})$ and we need to process large samples at each point to obtain statistically meaningful results. Thus, different strategies will be adopted to deal with the limited computational resources.


In Sections~\ref{sec:optimbaryons} and \ref{sec:optimtau} we will follow this procedure to optimize the heavy-baryon and $\tau$ setups, respectively. In Section~\ref{sec:optimfocusing} we will estimate the sensitivity of the layout with focusing crystals by comparing its enhanced channeling efficiency to the heavy-baryon layout.




\section{Heavy baryons} \label{sec:optimbaryons}

\begin{flushright}
	This section is partially based on Refs.~\cite{Aiola:2020yam,Bagli:2017foe,internalnote,Botella:2016ksl}\\
\end{flushright}

Since our first article~\cite{Botella:2016ksl}, the method to estimate the sensitivity of the experiment has undergone several improvements getting ever more complex and closer to the actual experiment. The simulation of channeling has been refined going beyond the analytical conditions and the fitting PDF includes now more dependences on the initial polarization and decay dynamics. Moreover, different sites and crystal materials have been considered, and the studies have been extended to a variety of charged baryons. In Section~\ref{sec:optimizationlhcb}
we will show the optimization of crystal parameters $(\lenC,\thC)$ for the case of LHCb (dedicated experiment), referred to as \texttt{S1} (\texttt{S2}). We will consider germanium crystals, which have larger channeling efficiencies than silicon, and \Lc particles. In Section~\ref{sec:sensitivityheavy} we will point out the differences to other cases and will summarize the sensitivity studies.

\subsection{Crystal optimization for \Lc baryons} \label{sec:optimizationlhcb}


As discussed in Chapter~\ref{ch:crystals}, the bent crystal accomplishes two goals: it induces spin precession on the particle and steers it outside the beam pipe and into the detector acceptance. Thus, the bending angle is related to both the channeling efficiency and the geometrical efficiency of the detector.
It is crucial then to have a model for the detector acceptance. In the \texttt{S2} scenario of a dedicated experiment, a threshold on the bending angle is imposed to consider the particles in acceptance. For the LHCb (\texttt{S1}), we use the detector geometry to create our acceptance model.

\paragraph{LHCb geometrical model.}

\lhcb is a single-arm forward spectrometer~\cite{Alves:2008zz,LHCb-DP-2014-002} 
dedicated to the study of particles containing \bquark or \cquark quarks at the \lhc. The tracking system of the upgraded detector~\cite{LHCb-TDR-013,LHCb-TDR-015} (operative since 2022) consists of three main devices: a vertex locator (VELO), surrounding the \pr\pr interaction region; the UT detector, composed by two large-area detectors, upstream of the dipole magnet; and the SciFi detector, compose of three stations (T1-T3) downstream 
of the magnet. A more detailed description of the \lhcb is provided later in Section~\ref{sec:LHCb}.

In this study, the position and shape of the tracking stations are parametrized with the information in the Technical Design Reports (TDR) of the upgraded LHCb detector~\cite{LHCb-TDR-013,LHCb-TDR-015}. A sample of \LcpKpi decays produced with \pythia proton-target interactions is used. The initial position and angle of the \Lc is varied, simulating different target positions and crystal bending angles. 

\begin{figure}
	\centering
	\includegraphics[width=0.54\textwidth]{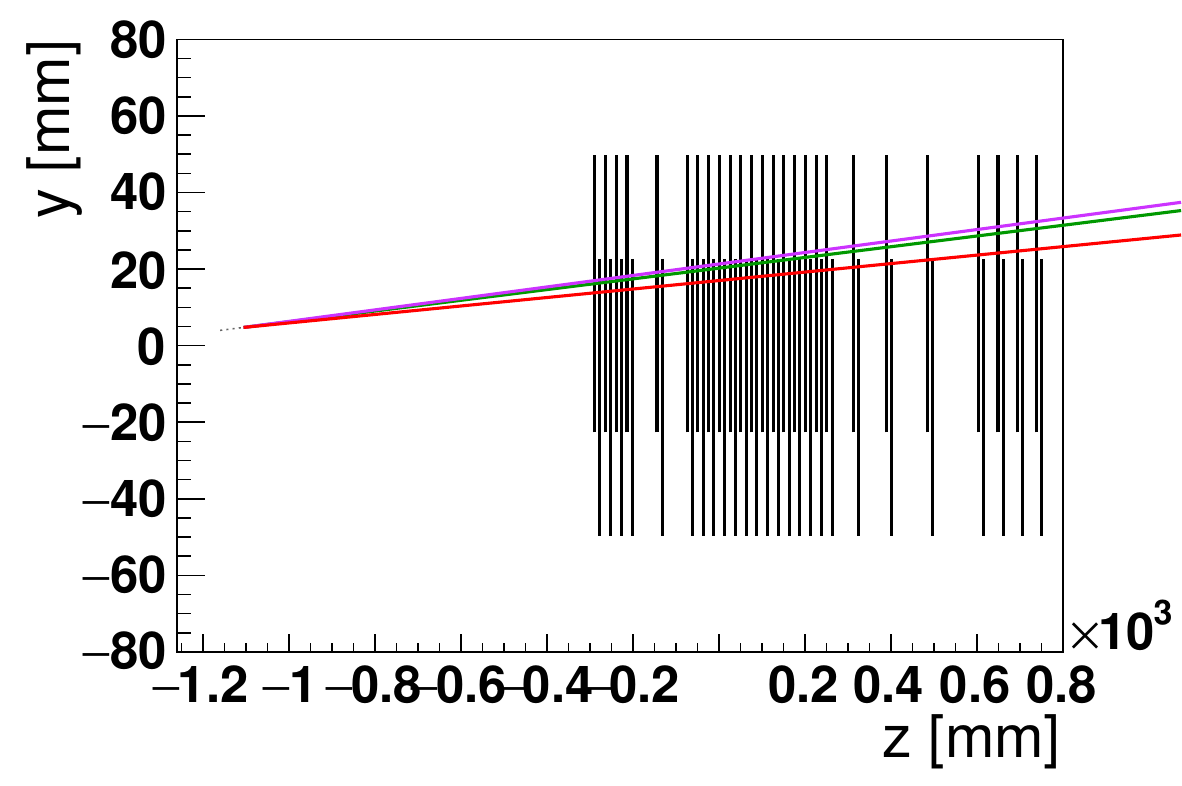}
	\includegraphics[width=0.38\textwidth]{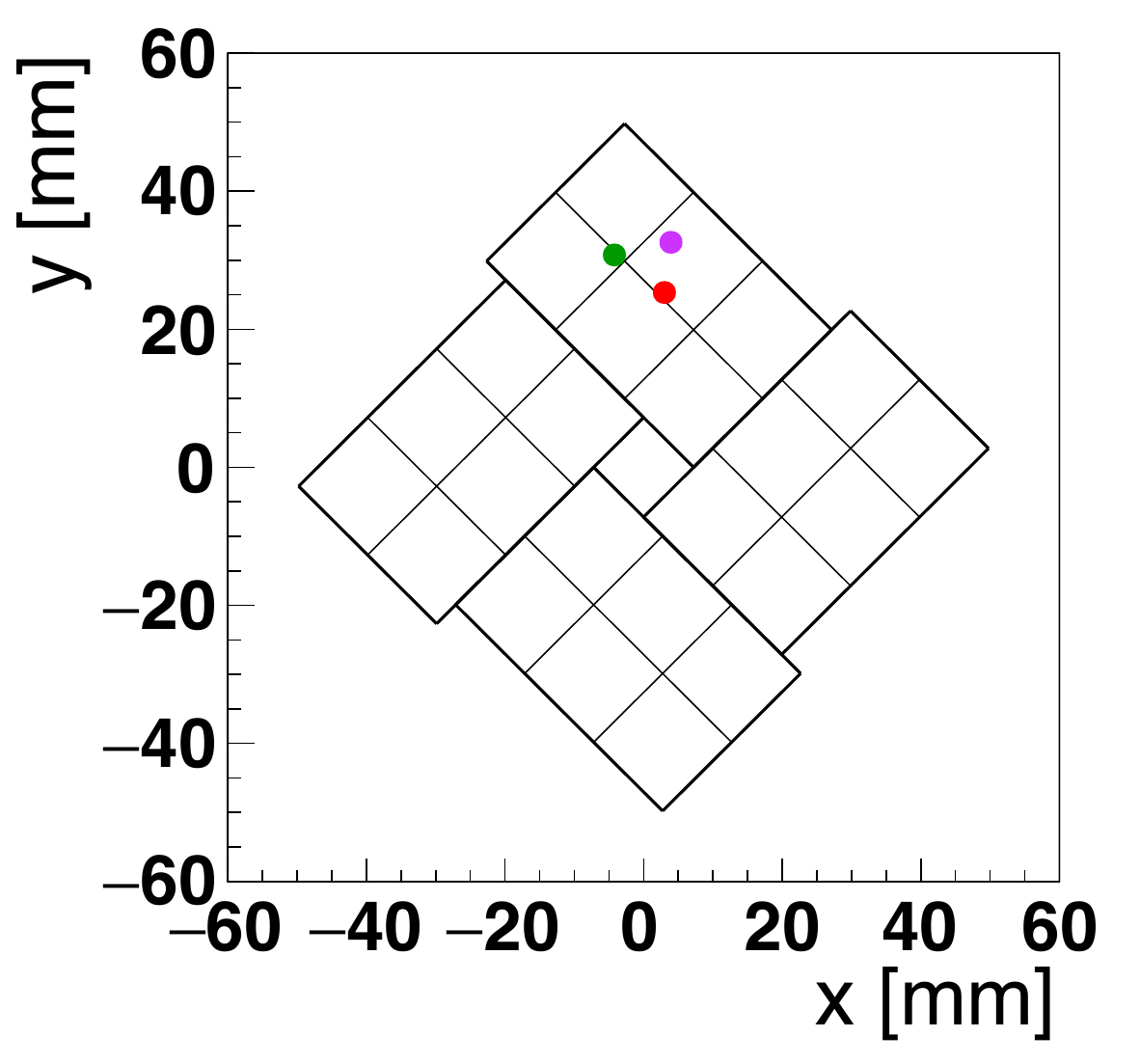} \\
	\includegraphics[width=0.54\textwidth]{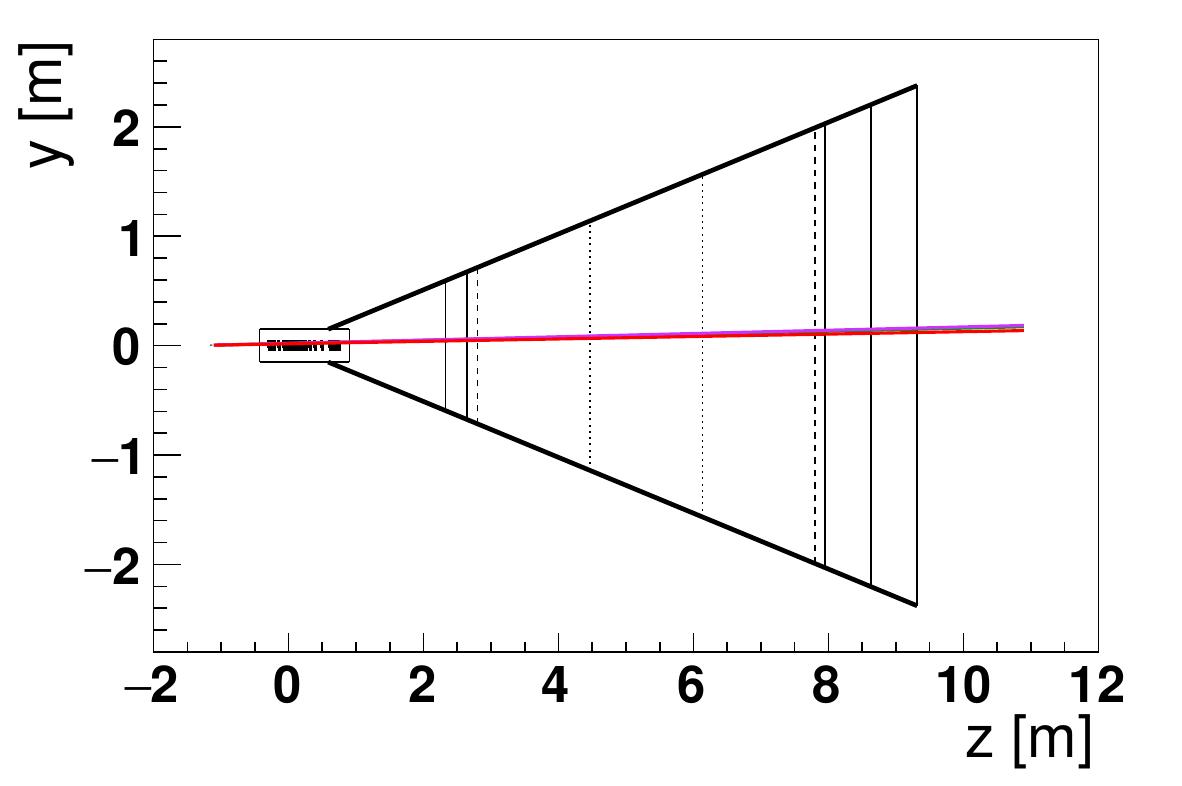}
	\includegraphics[width=0.38\textwidth]{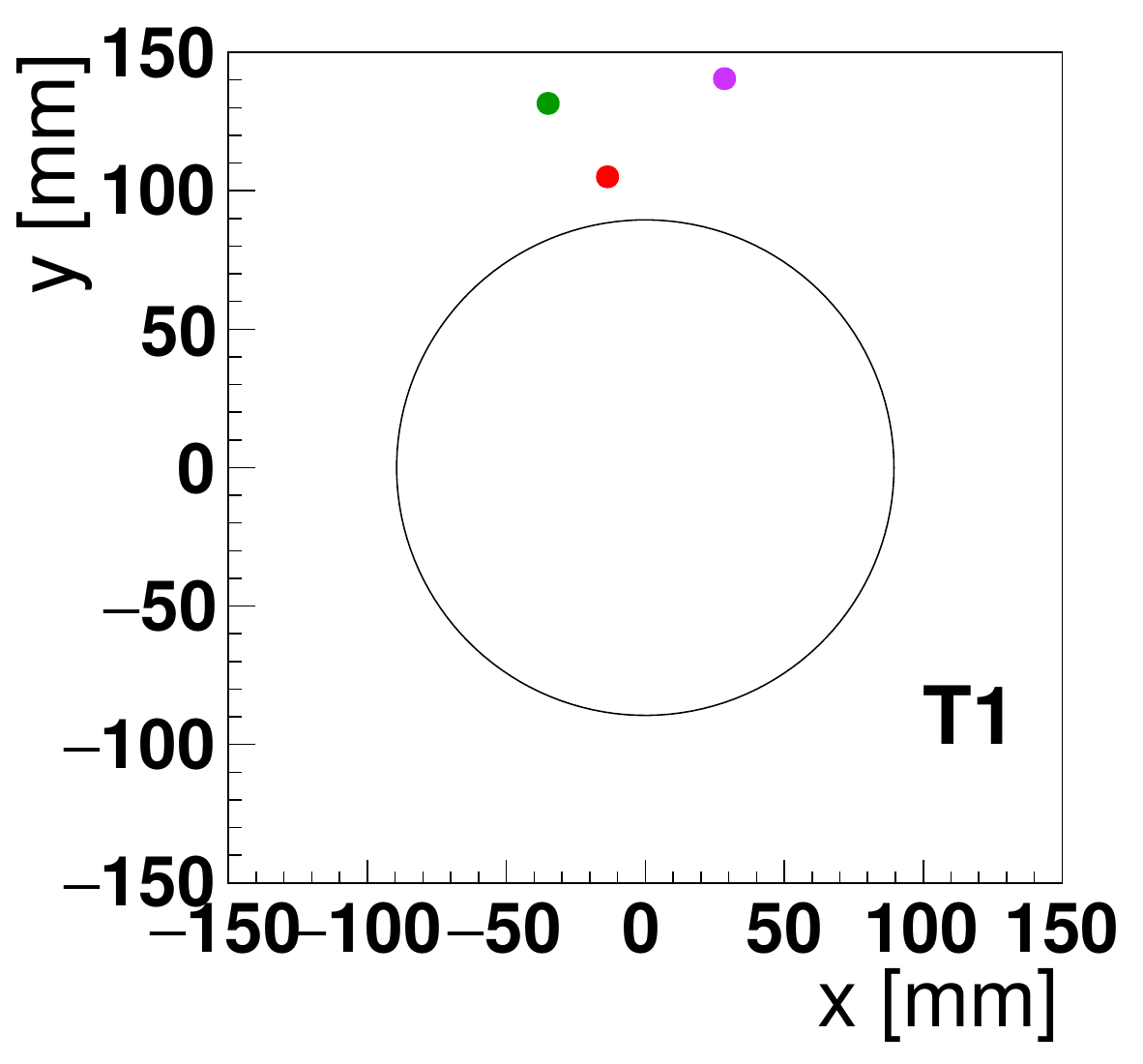}
	\caption{\label{fig:geometricalModel} 
		Geometrical model of the LHCb tracking system and projection of the particle trajectories in a $\Lc\to\pr\Km\pip$ decay, where the \Lc is produced at $(0,0.4,-116)\cm$ in the laboratory (detector) coordinate system and exits the crystal at an angle $\theta_C=14\mrad$. 
		The coloured lines represent the trajectory of the proton (green), pion (violet) and kaon (red); and the points (with the same colours) represent the position of the respective tracks at the tracking modules. \textbf{Top left:} side view of the VELO, which contains 26 (A- and B-side) modules. 
		\textbf{Top right:} Front view of the last VELO module, at $z=73.9\,\cm$, composed of two L-shaped sides. Note that the modules are rotated by 45 degrees with respect to the original design in the TDR~\cite{LHCb-TDR-013}. \textbf{Bottom left:} side view of the whole \lhcb detector, where the solid vertical lines represent tracking stations and the dashed lines, the limits of the dipole magnet. \textbf{Bottom right:} front view of the T1 station, at $z=7.948\,\m$ with a conservative beam clearance of $30\,\mm$ in addition to the beam pipe radius. The bending of the tracks due to the magnetic field is considered, with the pion track being most affected, as it carries the lowest momentum.
	}
\end{figure}

A particle is considered in acceptance if it crosses at least three VELO modules and the three T stations in the SciFi. Both the outer edges of the detectors and the inner boundaries with the beam pipe are considered. To validate the code, a graphical interface was developed to visually compare the particle trajectories with the implemented detector geometry. Four panels of this interface are shown in Figure~\ref{fig:geometricalModel}.

\begin{figure}
	\centering
	\includegraphics[width=0.45\linewidth]{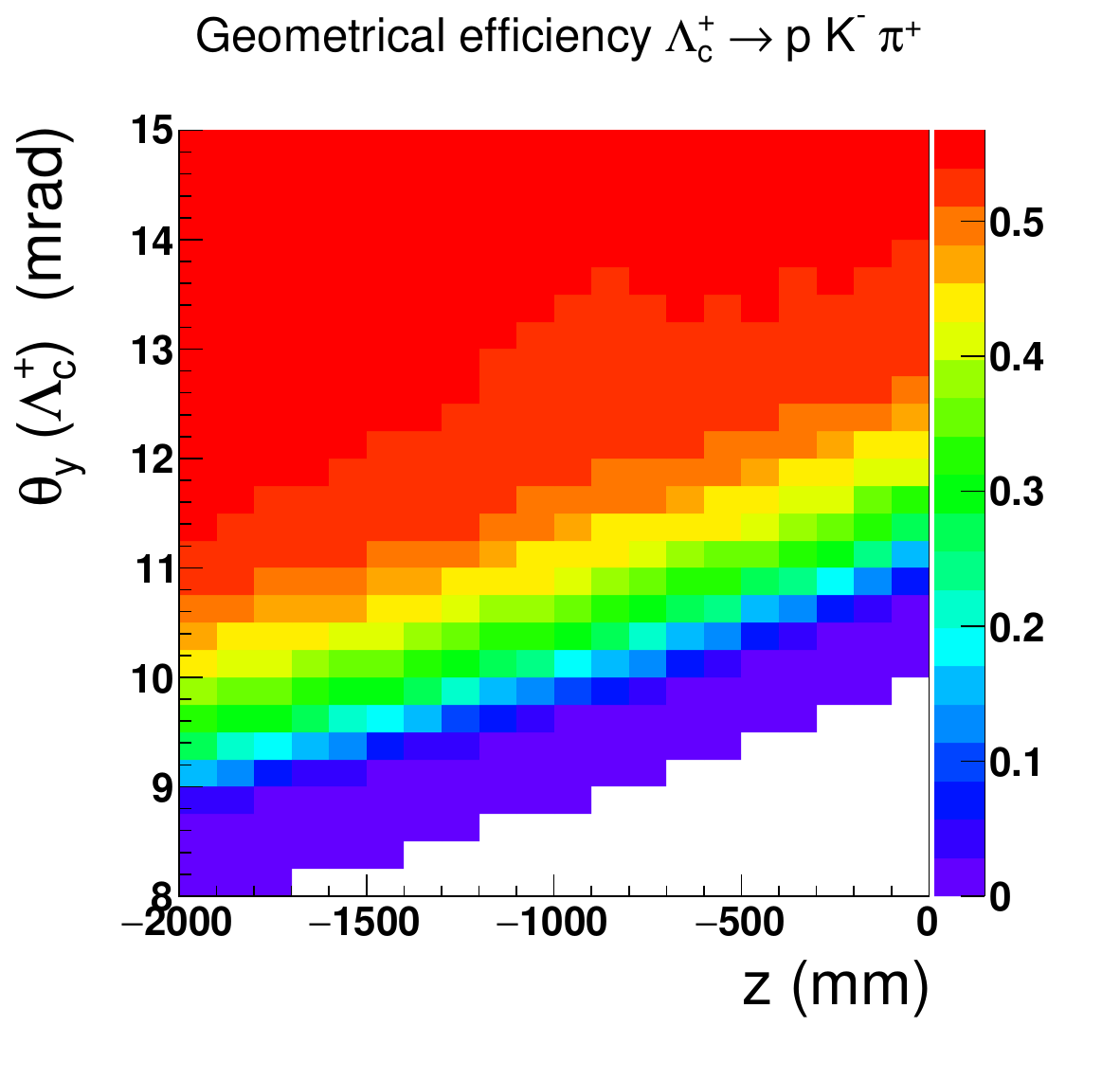}
	\includegraphics[width=0.54\linewidth]{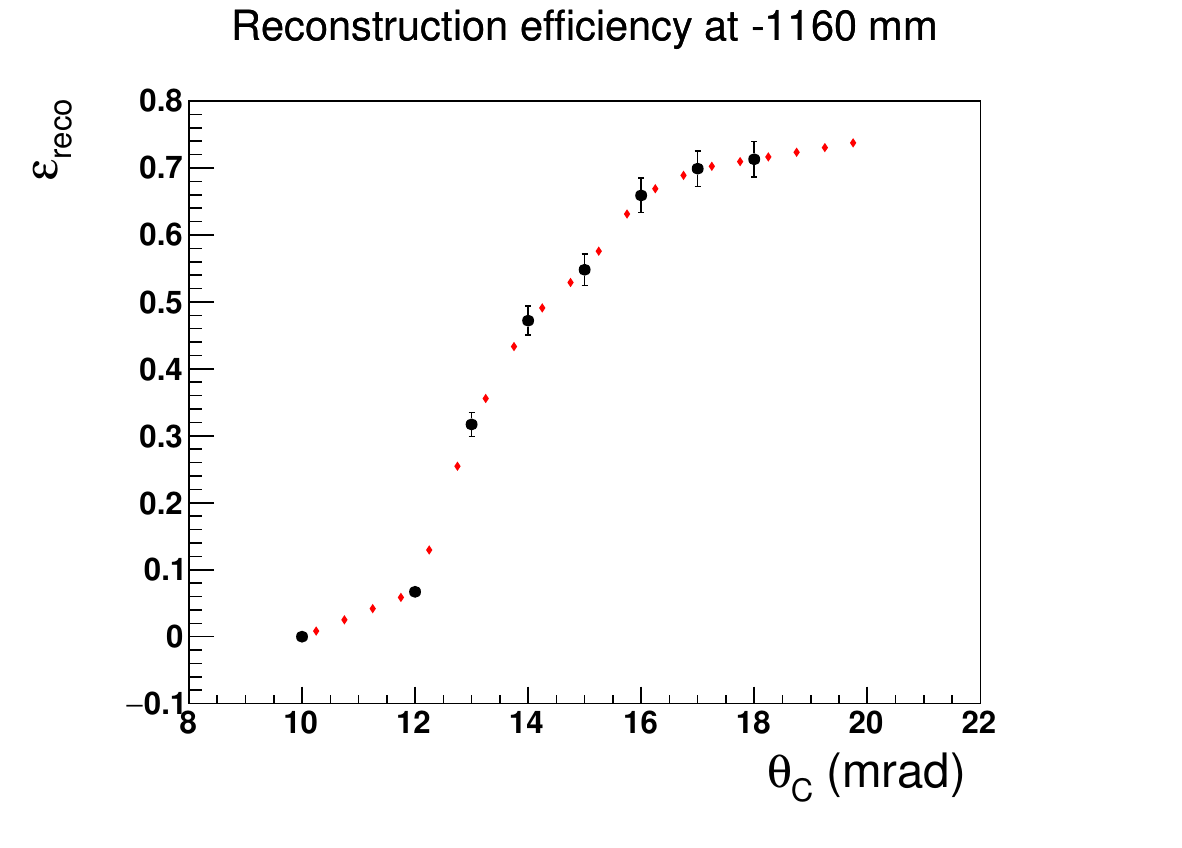}
	\caption{(Left) geometrical efficiency for channeled \Lc baryons as a function of the
		target $z$ position and crystal bending angle, and (right) 1-dimensional projection at $z=-116\,\cm$.
		The acceptace reaches a plateau at $\varepsilon_{\rm geo} \approx 70\%$ since many particles cross the SciFi modules through a 4 \mm wide vertical column. Rotating the crystal by $\approx 25^o$ around the beam direction avoids this dead region, and the plateau reaches $\varepsilon_{\rm geo} = 100\%$, with the same bending angle.
	}
	\label{fig:geoeffztheta}
\end{figure}

This method has also been validated by comparing the obtained geometrical efficiency with the results of full simulations within the LHCb framework~\cite{internalnote}. For this, \LcpKpi events were simulated using \textit{ParticleGun}, that allows to fix the \Lc momentum (tested for 1 and 2 TeV) and direction (scanned for $\theta_y=[12,18]\,\mrad$). To improve the agreement of the geometrical model with the full simulation, an additional 3-\cm beam clearance region around the beam pipe was considered in the T stations. The detector acceptance is shown in Figure~\ref{fig:geoeffztheta} as a function of the bending angle and the $z$ position of the target\footnote{The origin of coordinates is taken at the nominal \pr\pr interaction point.}. To minimize the impact on the normal LHCb operation, the fixed-target setup would be positioned outside the VELO vessel container~\cite{LHCb-TDR-013} $\approx 1.16$\m upstream of the nominal \pr\pr collision point.

%

%

\begin{figure}\centering
	\includegraphics[width=0.45\linewidth]{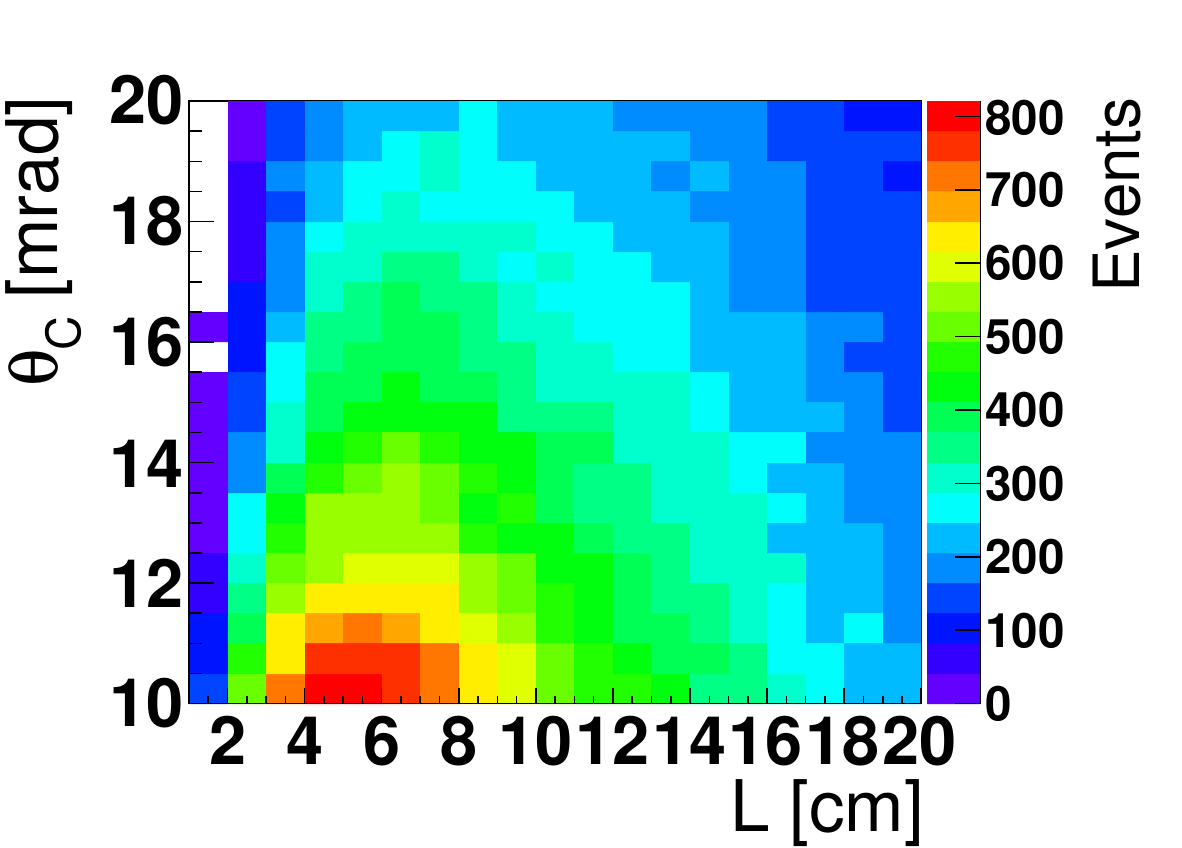}
	\includegraphics[width=0.45\linewidth]{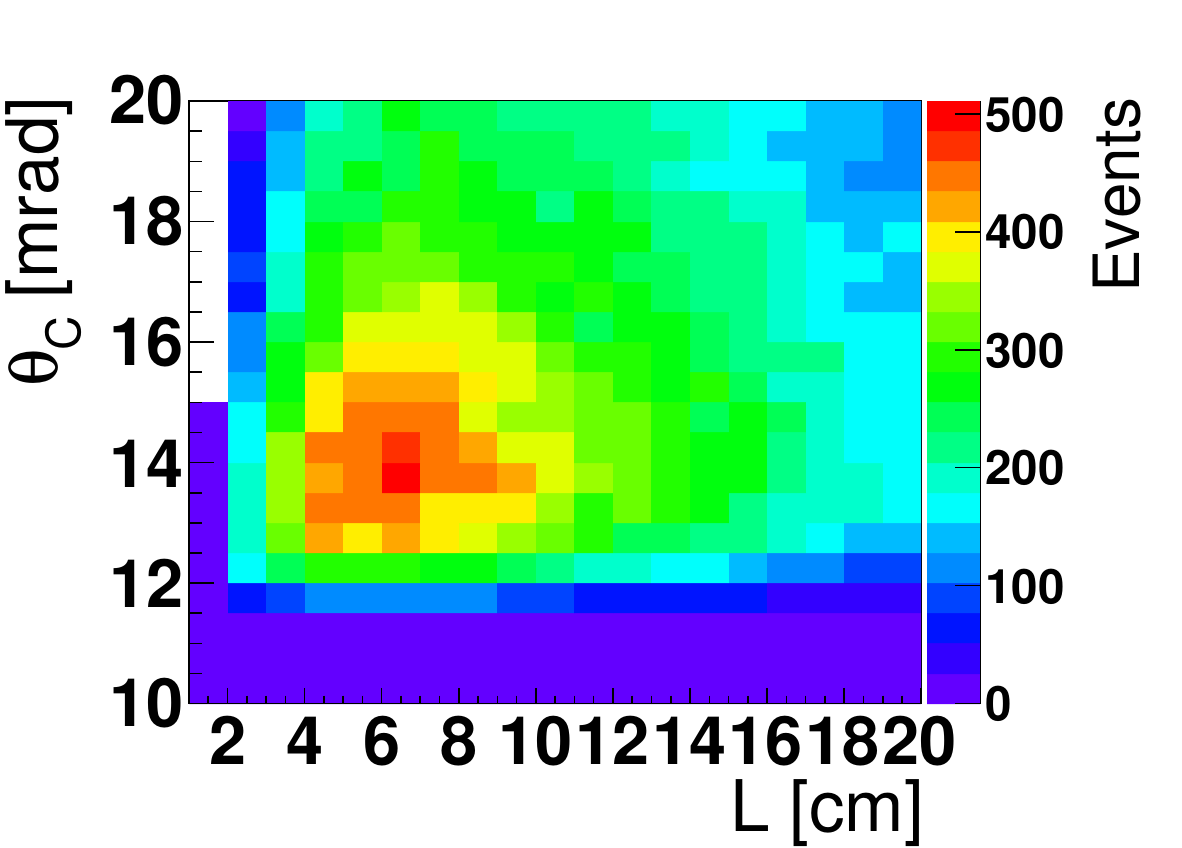}
	\includegraphics[width=0.45\linewidth]{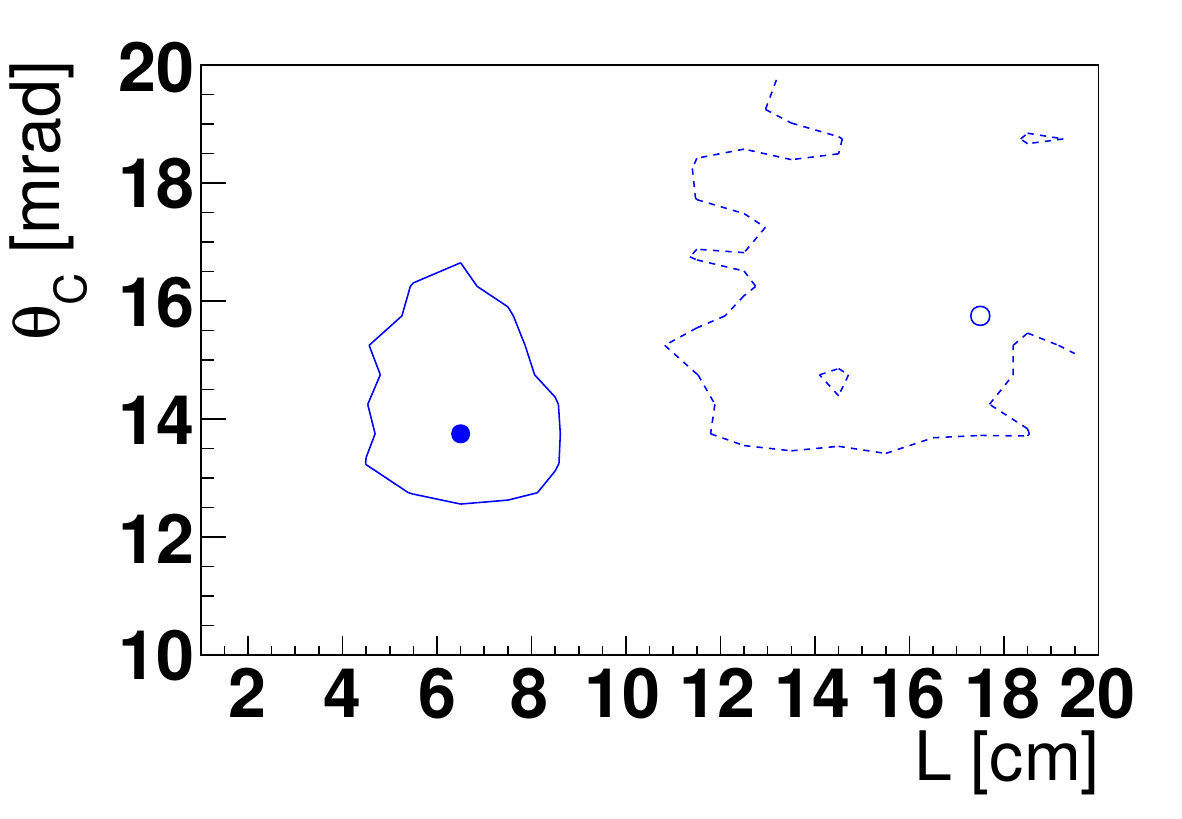}
	\includegraphics[width=0.45\linewidth]{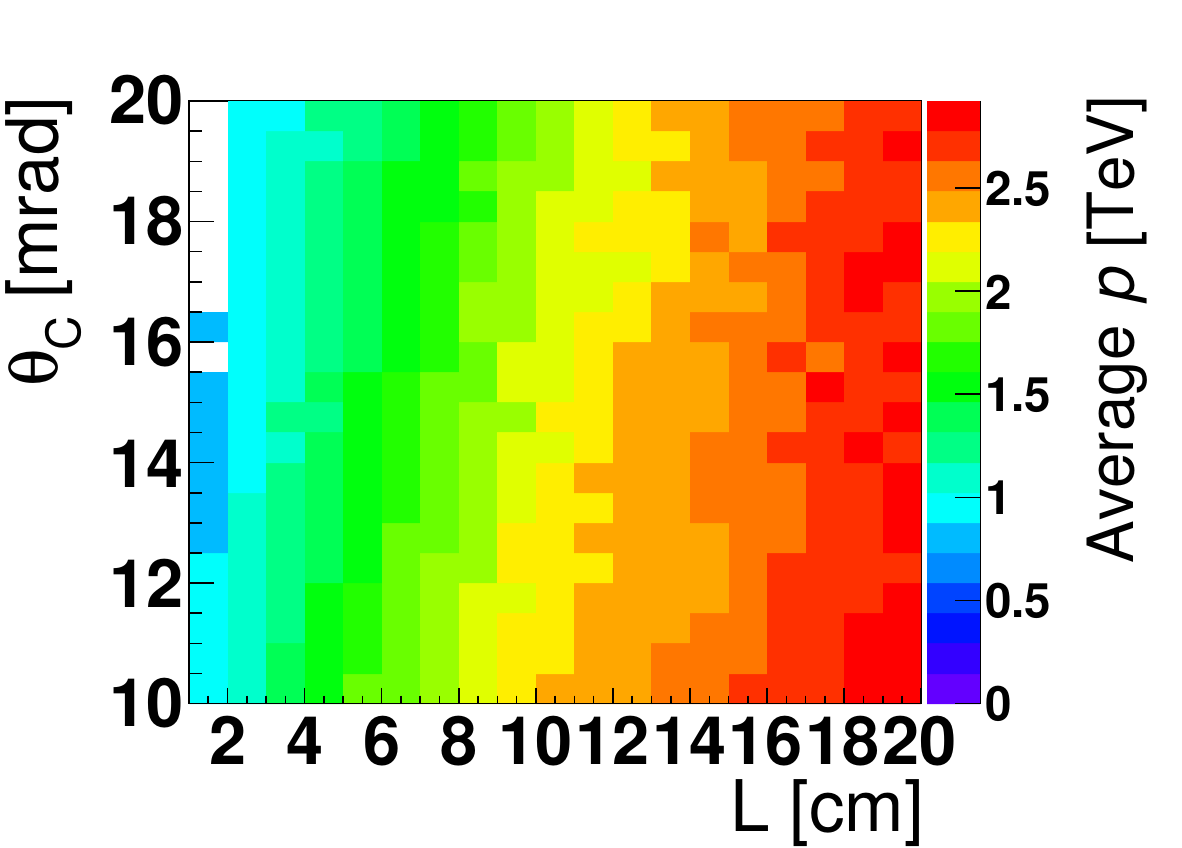}
	\caption{Optimization for a germanium crystal at \lhcb (\texttt{S1}). \textbf{Top left:} channeling efficiency as a function of the crystal bending angle and length, including the background-rejection cut $p_\Lc>800\,\gev$.  \textbf{Top right:} same plot overlaying the geometrical efficiency of LHCb with a target position at $z=-1.16\,{\rm m} $. \textbf{Bottom left:} crystal parameters with maximum sensitivity for EDM (blue dot) and MDM (blue circle) and regions with increased uncertainty of 20\% with respect to the minimum. \textbf{Bottom right:} average momentum, in TeV, of the channeled particles in each combination of $(\lenC,\thC)$.}
	\label{fig:optimizationLHCb}
\end{figure}

\begin{figure}[H] \centering
	\includegraphics[width=0.45\linewidth]{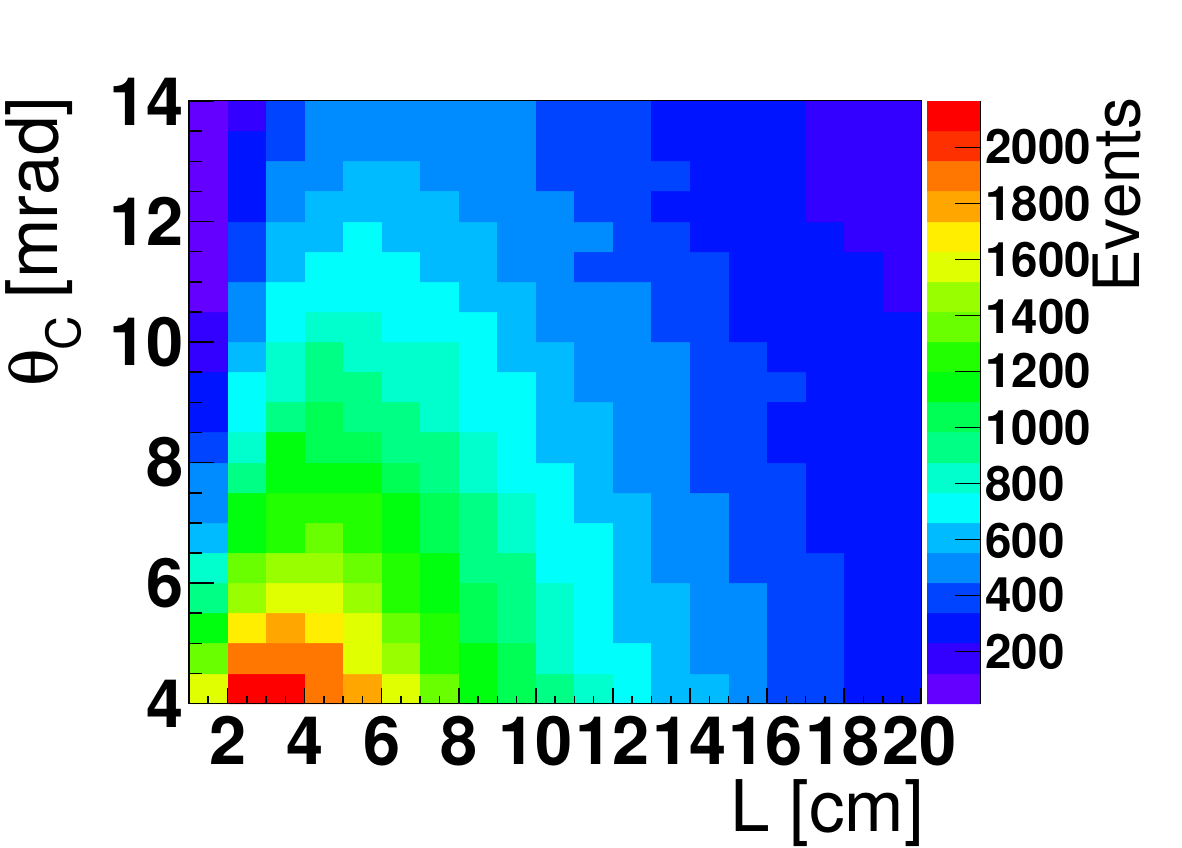}
	\includegraphics[width=0.45\linewidth]{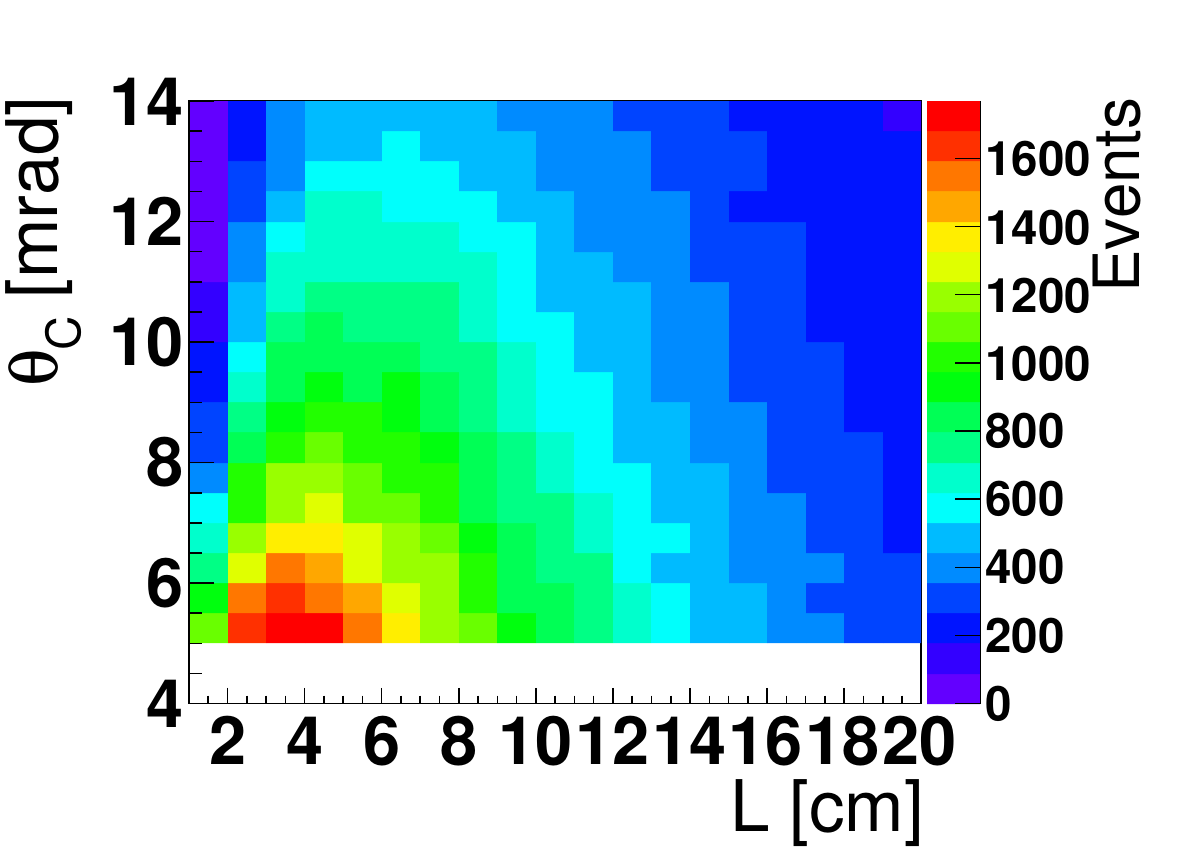}
	\includegraphics[width=0.45\linewidth]{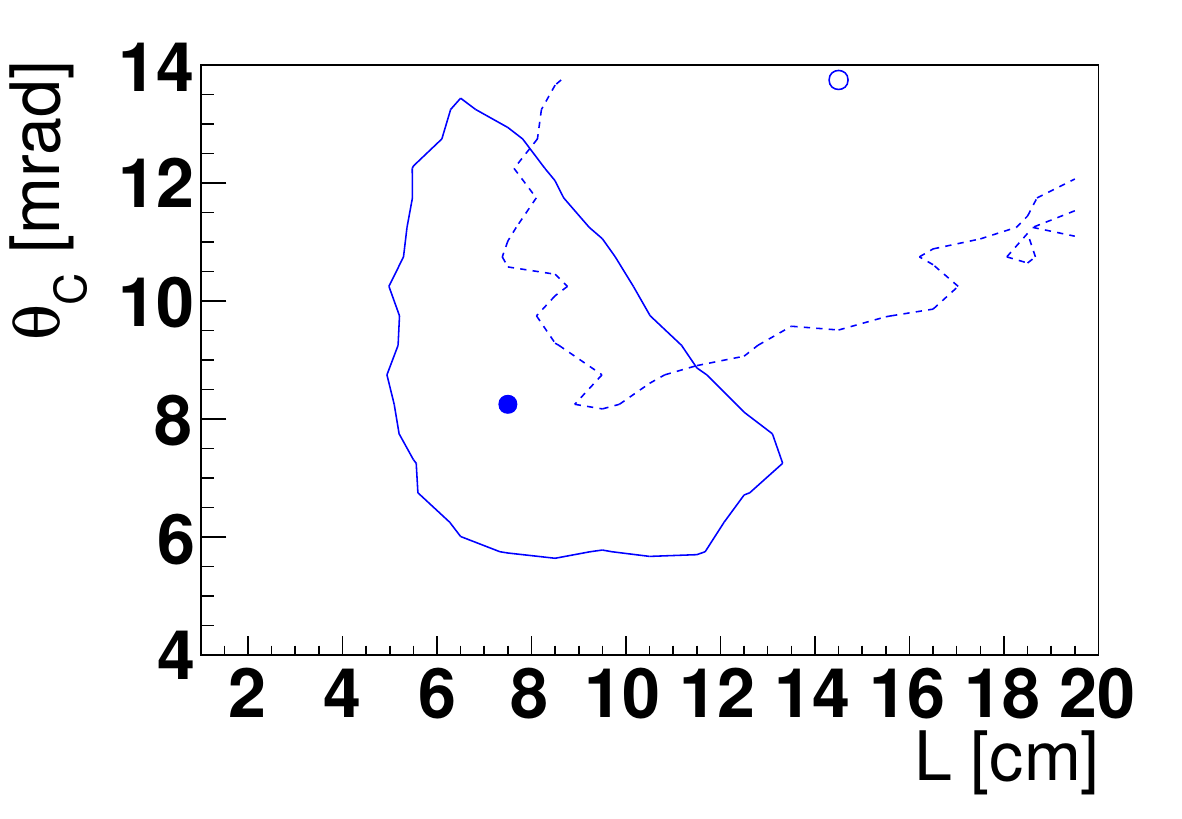}
	\includegraphics[width=0.45\linewidth]{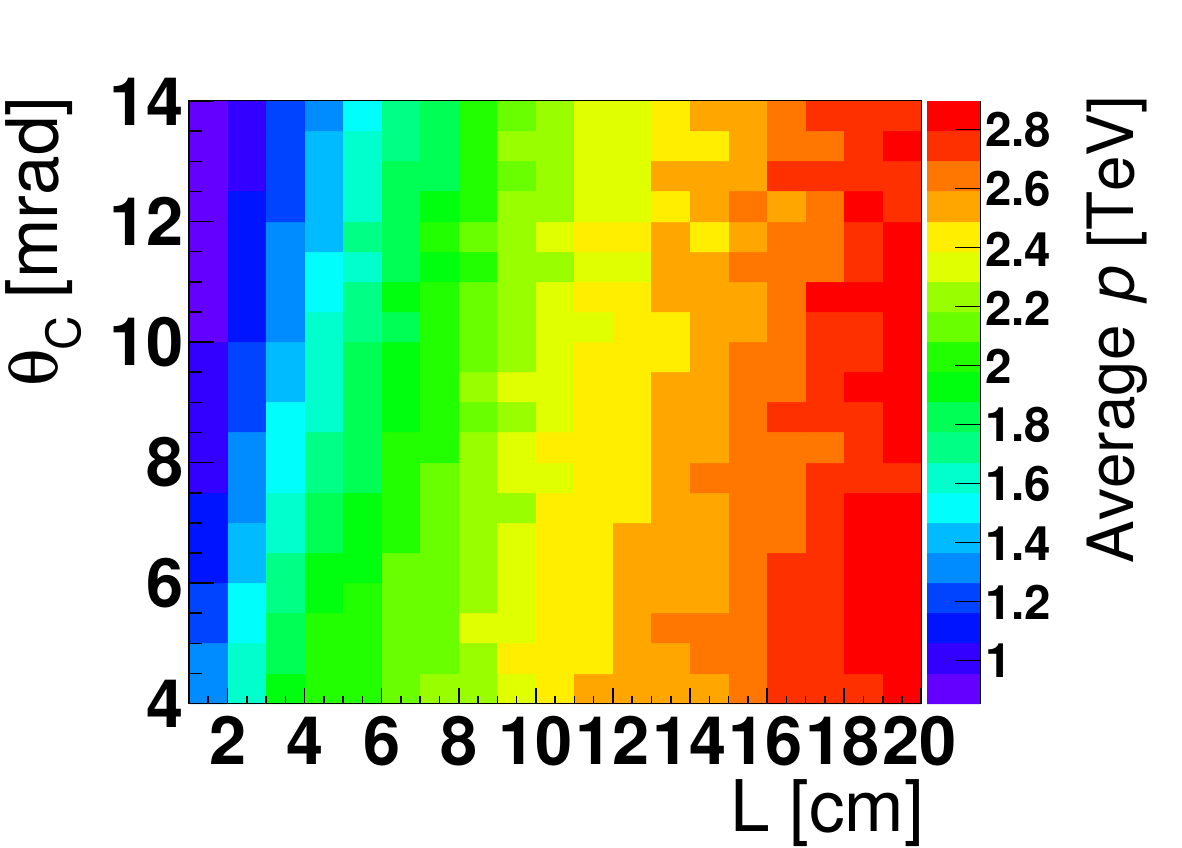}
	\caption{Same as Figure~\ref{fig:optimizationLHCb} but for a dedicated experiment with 100\% acceptance at $\thC\geq5\,\mrad$. Note the range of the $y$ axes with respect to  Figure~\ref{fig:optimizationLHCb}. }
	\label{fig:optimizationDedicated}
\end{figure}

\paragraph{Efficiency and sensitivity maps.} Next, the channeling efficiency is determined as a function of $(\lenC,\thC)$. By imposing the channeling conditions (Section~\ref{sec:channeling}) on the same sample of \LcpKpi decays, we obtain the efficiency map in Figure~\ref{fig:optimizationLHCb} (top left). Overlaying the LHCb geometrical efficiency as a function of \thC, we find the efficiency map in Figure~\ref{fig:optimizationLHCb} (top right), and similarly for a dedicated experiment in Figure~\ref{fig:optimizationDedicated}. At LHCb, the maximum efficiency is found for $\lenC\approx6\,\cm$ and $\thC\approx14\,\mrad$. Longer crystals are disfavoured due to the enhanced probability of decay inside the crystal. In turn, shorter crystals reduce the bending radius, increasing the dechanneling probability. 


Pseudoexperiments are conducted with the events that meet the channeling conditions and are within detector acceptance. The statistical uncertainty for EDM and MDM is extracted at each $(\lenC,\thC)$ point, and plotted in Figure~\ref{fig:optimizationLHCb} and~\ref{fig:optimizationDedicated} (bottom left). To interpret these results, it is important to account for the momentum variation of channeled particles across the plane, in the same figures (bottom right). This pattern of average momentum as a function of $(\lenC,\thC)$ can be explained in the following way: longer crystals require a larger particle lifetime, which favours events with larger Lorentz boost. On the other hand, larger bending angles reduce the bending radius $R=\lenC / \thC$, which disfavours the channeling of high-momentum \Lc baryons

With these notions at hand, we can interpret the contour maps with the regions of minimal uncertainty in $(\lenC,\thC)$: 


\begin{itemize}
	\item The MDM uncertainty is driven by the amount of precession $\Phi\approx ((g-2)/2) \gamma \thC$. A wider $\gamma$ spectrum induces a wider distribution of precession angles, providing \textit{lever arm} for the fit to $g-2$. For this reason, the MDM prefers larger crystal radii (larger \lenC at fixed \thC) to increase the density of events with large boost $\gamma$, even if it is at the cost of reducing the total sample size.
	
	\item The EDM signal, in turn, is proportional to $\cos\Phi-1$ (Eq.~\eqref{eq:precessionsimplified}), and the sensitivity is maximal for events with $\Phi\approx\pi + 2\pi n $. Provided that the distribution of precession angles includes this value, the sensitivity to EDM does not benefit from a wider distribution of $\gamma$ and it is only driven by the total amount of events. For this reason, the minimum EDM uncertainty essentially coincides with the maximum detector efficiency (Figure~\ref{fig:optimizationLHCb}). In a dedicated experiment (Figure~\ref{fig:optimizationDedicated}), the sample can be much larger at low bending angles, but implying also precession angles $\Phi<\pi$. For this reason, the minimum EDM uncertainty extends towards larger $\thC$ and $\gamma$.

\end{itemize}

This discussion on the optimization of the crystal parameters was partially included in Ref.~\cite{Bagli:2017foe}. With these results, the first crystal prototypes were manufactured and tested on beam. The results of these tests were presented in Ref.~\cite{Aiola:2020yam}, together with an updated study of the experiment's sensitivity, whose final results are reproduced in the next subsection. In Ref.~\cite{Aiola:2020yam}, the optimal target thickness and crystal tilt (\textit{crystal orientation angle} therein) were determined to be $T=2\,\cm$ and $\tilt\approx0.3\,\mrad$, respectively. The crystal tilt is beneficial for the EDM since the tilt induces an initial polarization along the $x$ axis, and the (MDM-induced) spin precession does not need a full $\Phi=\pi$ to reach the maximum EDM sensitivity. This advantage disappears with larger magnetic moment ($(g-2)/2 \approx -0.76$ with respect to $\approx-0.03$), as the spin precession is much faster.

\subsection{Absolute sensitivity} \label{sec:sensitivityheavy}

%
%

\subsubsection{Charm baryons}

\begin{figure}
	\centering
	\includegraphics[width=0.55\linewidth]{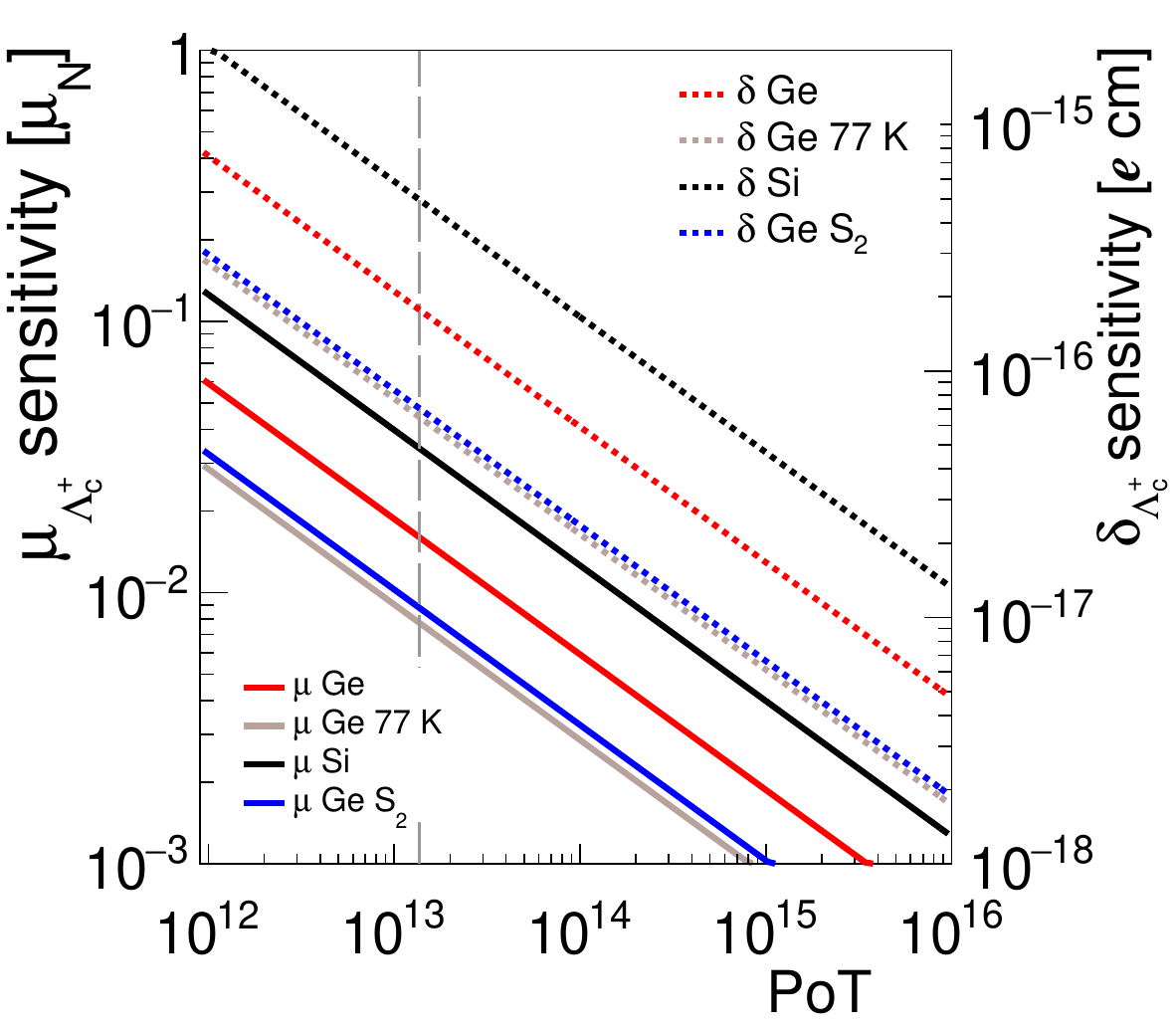}
	\caption{ From Ref.~\cite{Aiola:2020yam}. Uncertainties on the MDM and EDM of \Lc baryons
		as a function of \pot at \lhcb and at a dedicated experiment (\Sb) with increased forward acceptance.
		All three- and four-body \Lc decays from Table II of Ref.~\cite{Aiola:2020yam} are considered,
		with anomalous magnetic moment $a = (g-2)/2$ assumed to be $\approx -0.03$.
		The vertical long-dashed lines refer to $1.37\times10^{13}$~\pot, corresponding to two years of data taking. Similar projections are obtained for \Xicp baryons.
	}
	\label{fig:physreach}
\end{figure}

The final sensitivity for the \Lc dipole moments is shown in Figure~\ref{fig:physreach} as a function of the protons on target  (PoT). Four scenarios were considered~\cite{Aiola:2020yam}:
\begin{itemize}
	\item \texttt{S1} with germanium crystals 
	\item \texttt{S1} with germanium crystals cooled at 77 \degk
	\item \texttt{S1} with silicon crystals
	\item \texttt{S2} with germanium crystals
\end{itemize}
The considered setup parameters are:
\begin{itemize}
	\item \texttt{S1}: $(\lenC,\thC,\tilt,T) = (10\,\cm,16\,\mrad,0.3\,\mrad,2\,\cm)$ 
	\item \texttt{S2}: $(\lenC,\thC,\tilt,T) = (7\,\cm,7\,\mrad,0.3\,\mrad,2\,\cm)$
\end{itemize}

In summary, with $1.37\times 10^{13}$ \pot integrated in two years of data taking at a proton flux $F=10^6\,p/s$~\cite{Mirarchi:2019vqi}, the achievable sensitivity on the \Lc MDM (EDM) is $2\times 10^{-2}\ {\mu_{ N}}$ ($ 3\times 10^{-16}~e\cm$), and similar for the $\Xic^+$ baryon, where $\mu_N$ represents the nuclear magneton. With a fully new dedicated experiment, or using germanium crystal cooled at 77 \degk~\cite{Bezshyyko:2017var} at LHCb, the sensitivity would improve by a factor of two\footnote{The sensitivity improvement with a dedicated experiment was estimated to be of a factor $\sim 10$ in Ref.~\cite{Bagli:2017foe} due to an overly optimistic estimation of the proton flux.}.

With this uncertainty, the magnetic moment of the \Lc and \Xicp baryons could be measured for the first time with 2\% accuracy, much below the systematic uncertainties of the theoretical predictions (that quote any uncertainty), in Figure~\ref{fig:mdmcharm}. The indirect bounds on the charm quark EDM, derived in Chapter~\ref{ch:improvedbounds}, are more restrictive than the projected uncertainty of the direct measurement by a factor $\sim10^{5}$~\cite{Gisbert:2019ftm}. These are based on the limits on the neutron EDM and, in its extraction, several possible cancellation effects are neglected. Besides the charm-quark (C)EDM, many four-quark operators also contribute to the charm baryon EDM, which may have weaker constraints. Moreover, the connection between quark operators and charm baryon EDMs has not been studied yet (a first step evaluating the $\theta$-term contribution was published in Ref.~\cite{Unal:2020ezc}). In conclusion, while finding an EDM signal in this first-stage measurement would be extremely surprising, the experiment will provide valuable direct information on this observable, free of theory assumptions, and, in any case, provide positive information on the MDM of charm baryons, uncharted so far.

\subsubsection{Bottom baryons}

The sensitivity for bottom baryons $\Xibp$  is roughly a factor $\sim10^2$ worse as shown in Figure 13 of Ref.~\cite{Bagli:2017foe}. Firstly, the production of $\Xib^-$ baryons is suppressed with respect to neutral $\Lb$ baryons, and the production of the positively-charged anti-baryon \Xibp could be even more suppressed. Furthermore, the only suitable two-body decay modes are $\Xibp\to \Xip \jpsi$ and $\Xibp \to \Xicbarz \pip$, of which the second one has not even been observed to date~\cite{Bagli:2017foe}. Preliminary estimates show that exclusive semileptonic decays of b-baryons can be exploited, which would increase by several orders of magnitude the yields of channeled and reconstructed baryons while allowing an experimental determination of their polarization~\cite{Korner:1991ph,Konig:1993wz,Bonvicini:1994mr,Korner:1995my,Diaconu:1995mp,Korner:1998nc}.



\pagebreak

\section{$\mathbf{\tau}$ lepton} \label{sec:optimtau}

\begin{flushright}
	This section is partially based on Ref.~\cite{Fu:2019utm}\\
\end{flushright}

The measurement of the \taup dipole moments with bent crystals presents additional challenges due to the suppressed production of \taup particles and its reconstruction, along with the small anomalous magnetic moment of leptons, which reduces the precession angle $\Phi\approx10^{-2}\,\rad$.

The different methods to address these challenges were presented in the previous chapter, and are discussed in more detail in our original reference~\cite{Fu:2019utm}. In this section we will introduce the basic pieces for the sensitivity estimation, and discuss the optimization of the setup for initial longitudinal polarization. At the end of the section, the main results will be summarized.

\begin{figure}
	\centering
	\includegraphics[width=0.55\textwidth]{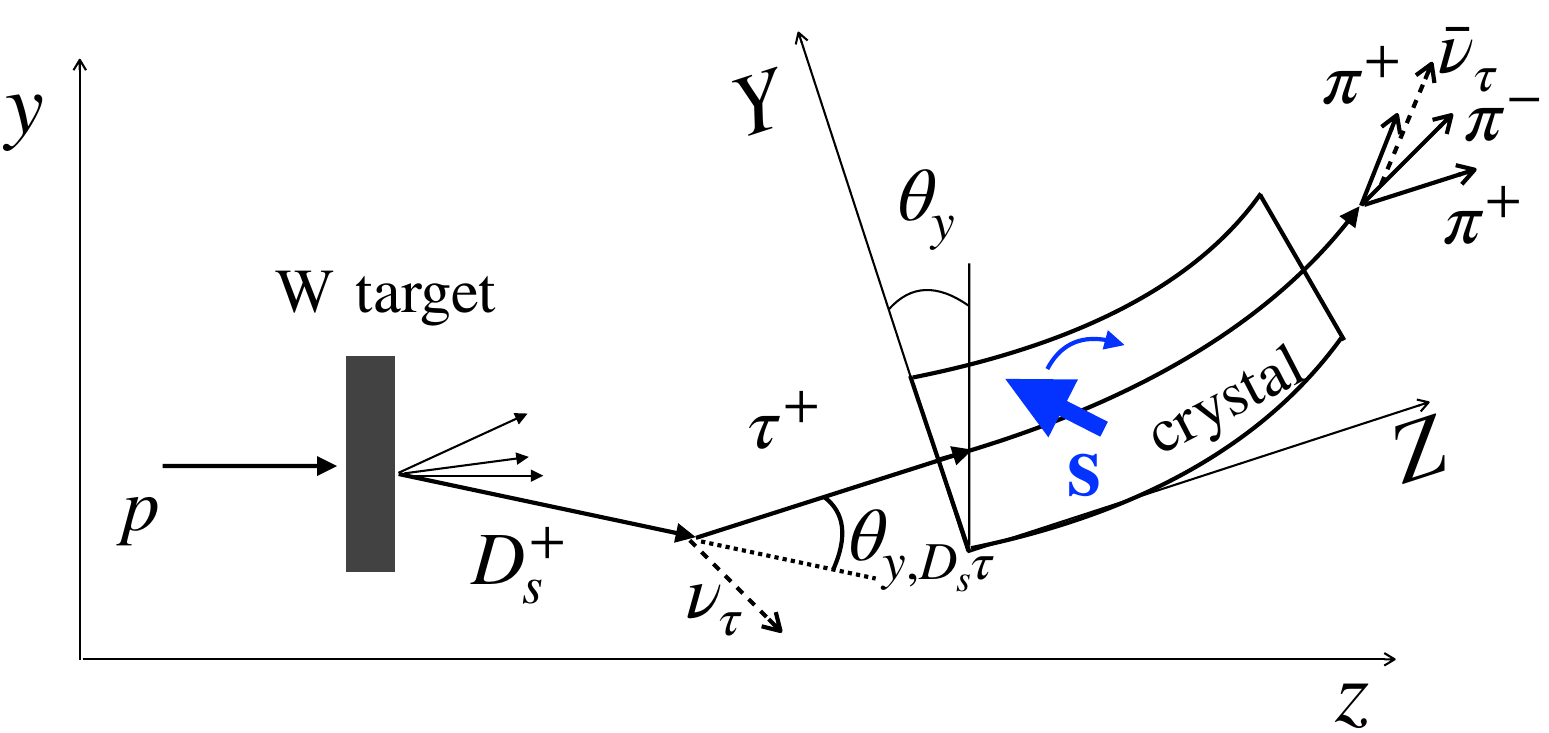} 
	\caption{From Ref.~\cite{Fu:2019utm}.
		Sketch of the fixed-target setup along with the \taup production and decay processes (not to scale). The crystal frame (\xaxis\yaxis\zaxis) is tilted
		in the laboratory frame ($xyz$), previously denoted ($x_Ly_Lz_L$), by \thetay (denoted as $\tilt$ in this thesis) to avoid channeling of non-interacting protons.} 
	\label{fig:PolCrystal}
\end{figure}

\subsection{Polarization}


As discussed in Section~\ref{sec:initialpolarization}, longitudinal polarization can be achieved by selecting the highest-energy candidates. More specifically, in a reference frame defined by the crystal axes at the crystal entry face and comoving with the channeled particle,
the \taup initial polarization ${\bm s}_0$ is given by the unit vector along the \Dsp momentum
in the \taup rest frame~\cite{Halzen:1984mc,Berestetskii:1982qgu},
\begin{equation}
{\bm s}_0 = \frac{1}{\omega} \left( m_{\Ptau}{\bm q} - q_0 {\bm p} + \frac{{\bm q} \cdot {\bm p}}{p_0+m_{\Ptau}} {\bm p} \right),
\label{eq:s}
\end{equation}
where ${\bm p}$ (${\bm q}$) is the momentum of the \taup (\Dsp) and $p_0$ ($q_0$) its energy in the laboratory frame,
$\omega=(m^2_{\Pds}-m^2_{\Ptau})/2$, and $m_{\Pds}$ is the \Dsp mass.
The projection along the (longitudinal) $z$ direction is
\begin{eqnarray}
s_{0,\zaxis} & \approx & \frac{1}{\omega} \left(\left|{\bm q}\right| p_0 -q_0\left|{\bm p}\right| \right),
\label{eq:s_proj_z}
\end{eqnarray}
producing a polarization of $s_{0,\zaxis}\approx -18\%$ when requiring $\phad>800\gevc$. This level of polarization, however, is only obtained when imposing the channeling conditions, that select $\taup$ particles in the very-forward direction. The setup parameters are correlated to the \taup momentum distribution, which in turn is related to the polarization. 
Here we used $[\thC, \lenC, \Ltarc] = [16\mrad,8(11)\cm,12\cm]$ for Ge (Si) crystals, which will be determined later as the optimal parameters, yielding an average Lorentz factor $\gamma\approx 800$.

The projection of $\spol_0$ along the $y$ axis gives the transversal polarization,
\begin{equation}
s_{0,\yaxis}  \approx \frac{m_{\Ptau} \left|{\bm q}\right|}{\omega} \thetayDsTau, 
\label{eq:s_proj_y}
\end{equation}
where
\thetayDsTau is the angle between the \Dsp and the \taup\ {\CHa momenta} in the \y\z plane (see Figure~\ref{fig:PolCrystal}).
A $s_{0,\yaxis}\approx\mp40\%$ polarization can be achieved with a \thetay-tagging that discriminates between positive and negative \thetayDsTau angles. More information on the use of  $s_{0,\yaxis}$ can be found in our original Ref.~\cite{Fu:2019utm} and another study with a different setup~\cite{Fomin:2018ybj}. However, given the difficulties to reach an efficient $\theta_y$-tagging, we will only treat the case for initial longitudinal polarization, maximizing the sensitivity to the EDM, without impact on the MDM uncertainty.


The MDM (EDM) signature is given by the spin rotation in the \yaxis\zaxis bending plane (appearance of a spin component along the \xaxis axis). In these events, with $\Dsp\to\taup(\to3\pi^{\pm}\bar{\nu}_\tau)~\nu_\tau$ decays, the incomplete kinematic information due to missing energy, and the absence of \taup production vertex, make the polarization reconstruction very challenging. As explained in our original work~\cite{Fu:2019utm}, and briefly at the end of Section~\ref{sec:keyfactors}, we developed a novel analysis technique based on multivariate classifiers to determine the rotation of the spin-polarization vector. The average \textit{event information} $S$ parametrizes the ability to reconstruct the polarization and can be different for the different polarization components. With our method, $S_X\approx S_Y \approx 0.42$, while the theoretical maximum (with complete event kinematics) is $S=0.58$~\cite{Davier:1992nw}. Find more details in Ref.~\cite{Fu:2019utm} and references therein.

\subsection{Parameter optimization}


\begin{figure}
	\centering
	\includegraphics[width=0.32\textwidth]{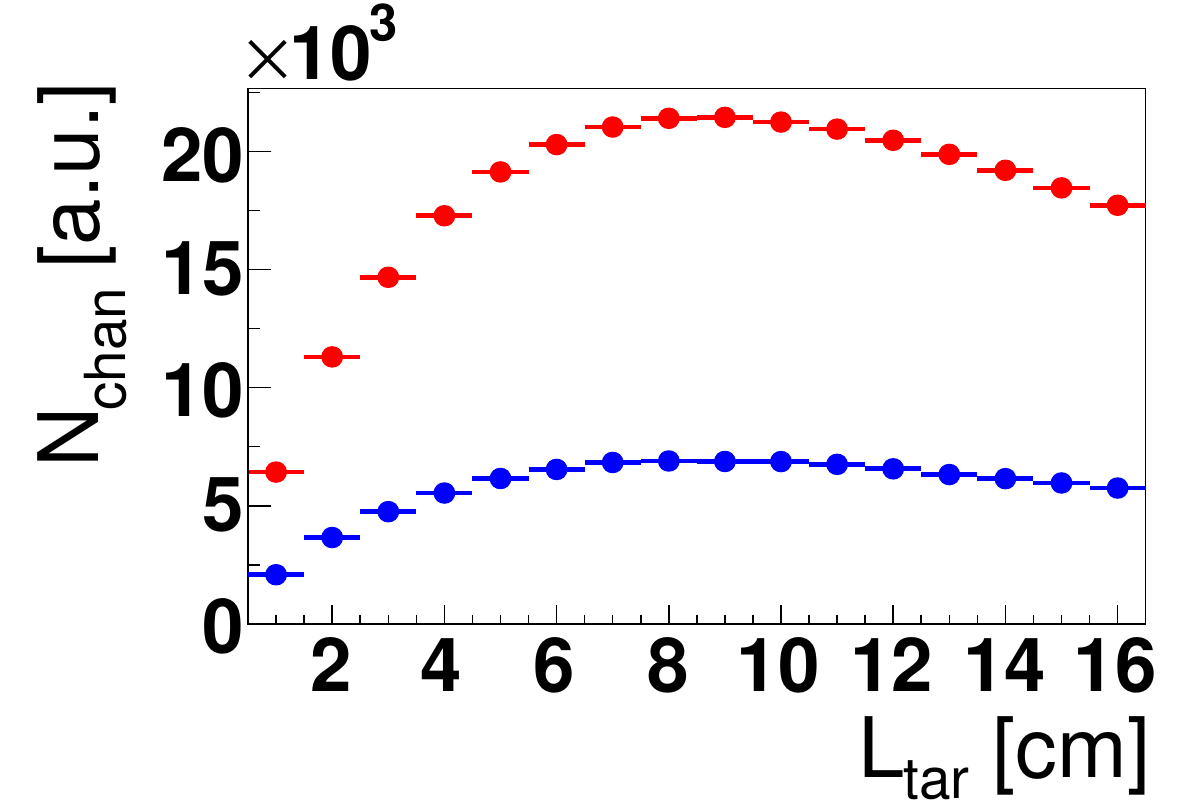}
	\includegraphics[width=0.32\textwidth]{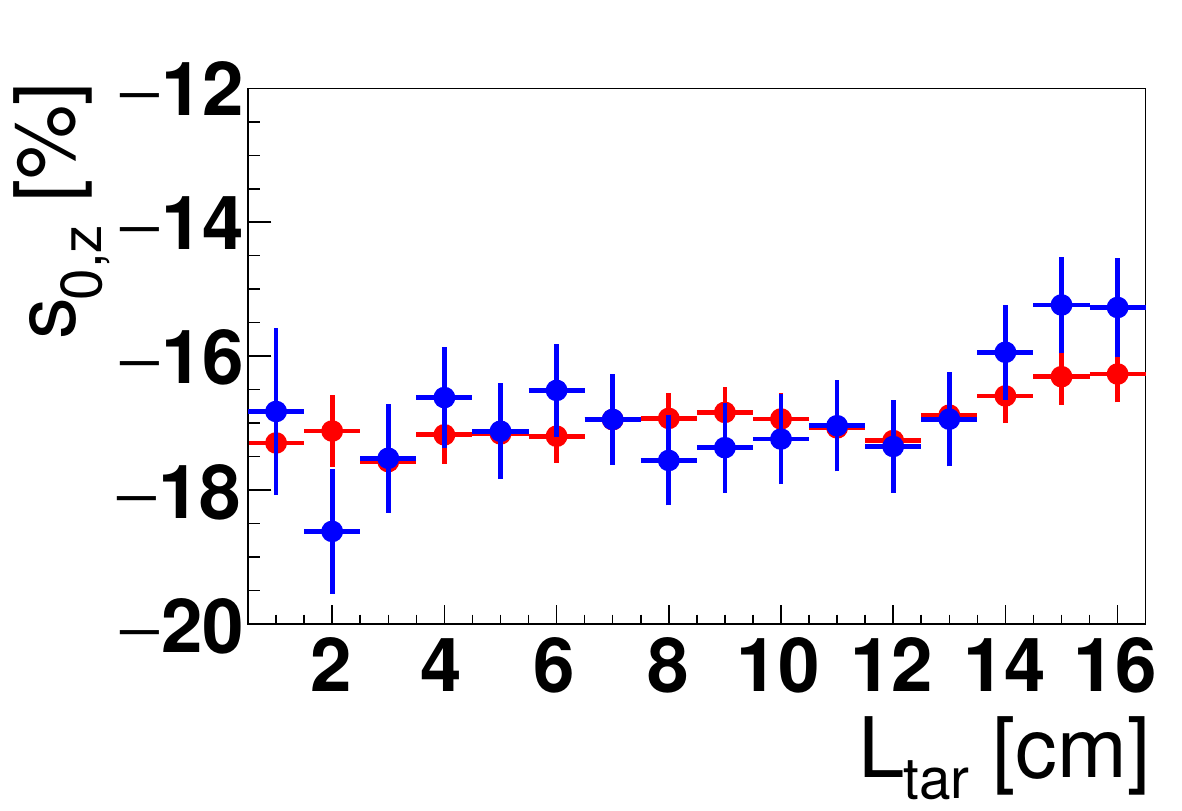}
	\includegraphics[width=0.32\textwidth]{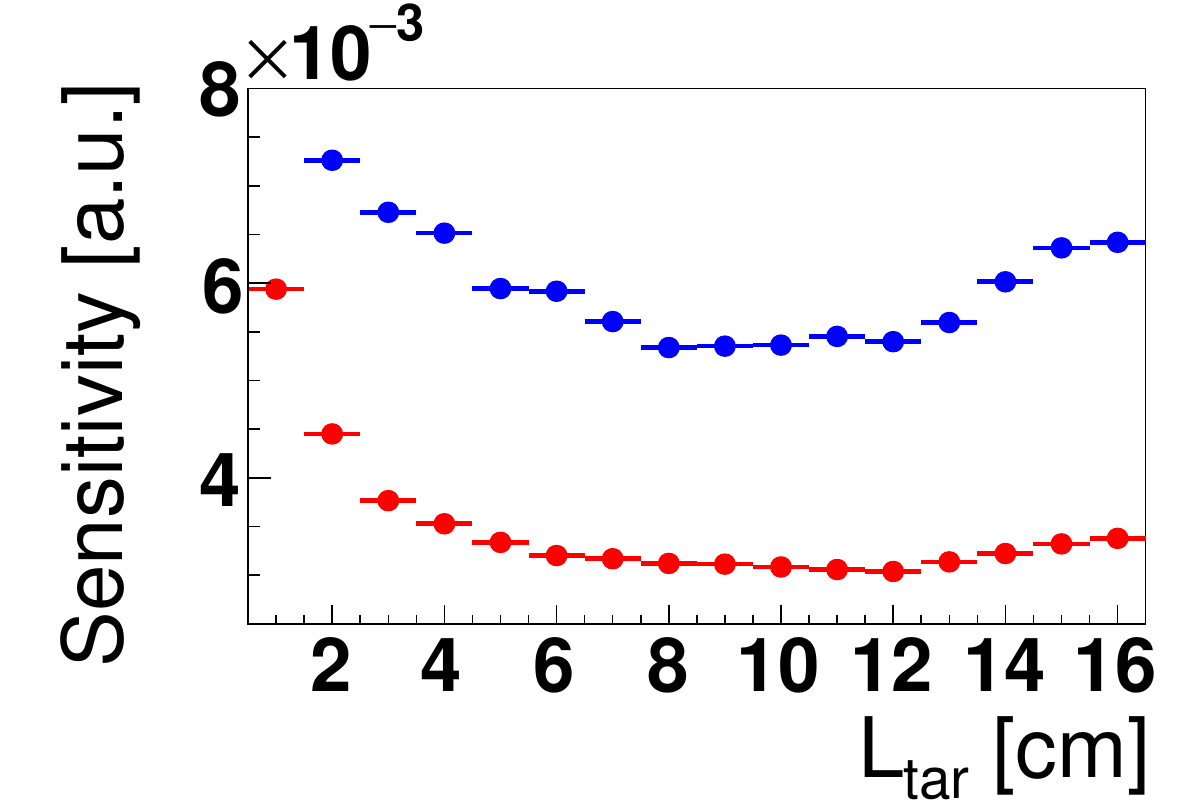}	\caption{\label{fig:optimization_Ltar} 
		From left to right, number of channeled events, longitudinal polarization, and relative uncertainty on the dipole moments as a function of \Ltarc, for \Ge (red) and \Si (blue) crystals.
	}
\end{figure}

There are four setup parameters to be optimized: the bending angle \thC and length \lenC of the crystal, the target-crystal separation $\Ltarc$ (to allow the flight and decay of the \Dsp meson) and the crystal tilt $\tilt$. The crystal tilt \tilt is correlated to the angle between the \Dsp and \taup directions \thetayDsTau (see Figure~\ref{fig:PolCrystal}), and can enhance the transverse polarization. However, as indicated before, we will focus on the case of longitudinal polarization, fixing $\tilt=0.1\,\mrad$\footnote{A small tilt would avoid the channeling of beam protons into the detector with the associated radiation damage. This is actually already prevented by the strong crystal bending of the optimal parameters. However, regions of the initially considered parameter space do allow the channeling of $7\,\tev$ protons and it was fixed to a non-zero value for consistency, with no relevant effect on the sensitivity. }. The statistical uncertainty on the anomalous magnetic moment $a$ and the EDM $d$ can be analytically estimated (for small $\Phi$) as
\begin{eqnarray}
\sigma_{a} \approx \frac{1}{{\CHa S_\yaxis} s_{0,\zaxis} \gamma \theta_C } \frac{1}{\sqrt{N_\taup^{\rm rec}}} , \ \ \
\sigma_{d} \approx \frac{2}{{\CHa S_\xaxis} s_{0,\zaxis} \gamma \theta_C } \frac{1}{\sqrt{N_\taup^{\rm rec}}} .
\label{eq:sensitivitytauanalytical}
\end{eqnarray}
where $N_\taup^{\rm rec}$ is the number of channeled and reconstructed \taup leptons. 
Based on the reconstruction efficiency of single charged pion tracks, a detector reconstruction efficiency of 40\% is assumed. 
This simplified expression gives a good description of the relative sensitivity for different choices of setup parameters. Thus, we will use it for the optimization of the setup, whereas the absolute sensitivity will be determined through pseudoexperiments. Nevertheless, to determine the mean value of $N_\taup^{\rm rec}$, $s_{0,\zaxis}$, and $ \gamma$ we use Monte Carlo simulations as before.

\begin{figure}
	\centering
	\includegraphics[width=0.45\textwidth]{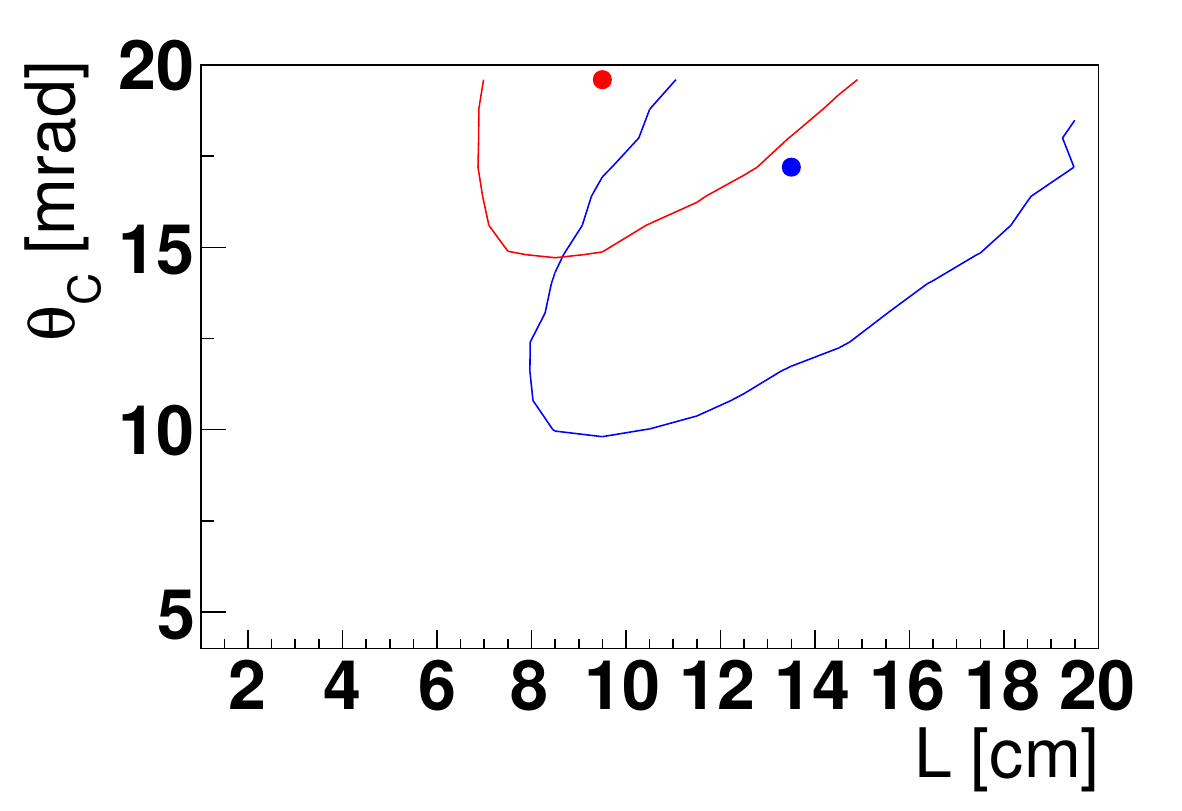}
	\caption{\label{fig:optimization_supp} 
		Regions of minimal uncertainty for both \at and \dt as a function of the crystal parameters \Lc and \thc (left) for \Ge (red) and \Si (blue),
		for initial  $s_{0,\zaxis}$ polarization. The contour lines represent regions whose uncertainties on \at and \dt are increased by 10\% with respect to the minimum (points). Note that the EDM and MDM uncertainty, in Eq.~\eqref{eq:sensitivitytauanalytical}, follow the same dependence with the setup parameters.
	}
\end{figure}

We can see the dependence of $N_\taup^{\rm rec}$ ($\propto N_{\rm chan}$) and $s_{0,\zaxis}$ with $\Ltarc$ in Figure~\ref{fig:optimization_Ltar}. The relative sensitivity, shown also in Figure~\ref{fig:optimization_Ltar}, is essentially flat at $\Ltarc=7-12\,\cm$. To help the rejection of physical channeled backgrounds (notably of $\Dp$ mesons), we exploit the large lifetime of the sequential $\Dsp$ and $\taup$ decays and select the maximum $\Ltarc=12\,\cm$. Fixing this value, we proceed to optimize the crystal parameters. Analogously to the studies conducted for heavy baryons, in Section~\ref{sec:optimbaryons}, the sensitivity is scanned across the $(\lenC,\thC)$ plane. The regions of minimum uncertainty are displayed in Figure~\ref{fig:optimization_supp}. In practice, we repeated the sequence of $\Ltarc$ and $(\lenC,\thC)$ optimization, fixing the other parameter(s) to the last found value. However, since the regions of minimum uncertainty are so broad, the final results are essentially identical to the first iteration. The optimal setup parameters, used to extract the absolute uncertainty, are
\begin{equation}
\thC=16\mrad, ~\lenC=8(11)\cm, ~\Ltarc=12\cm,~\tilt=0.1\,\mrad,
\end{equation}
for Ge (Si) crystals.

\subsection{Absolute sensitivity}

\begin{figure}[htb]
	\centering
	\includegraphics[width=0.46\textwidth]{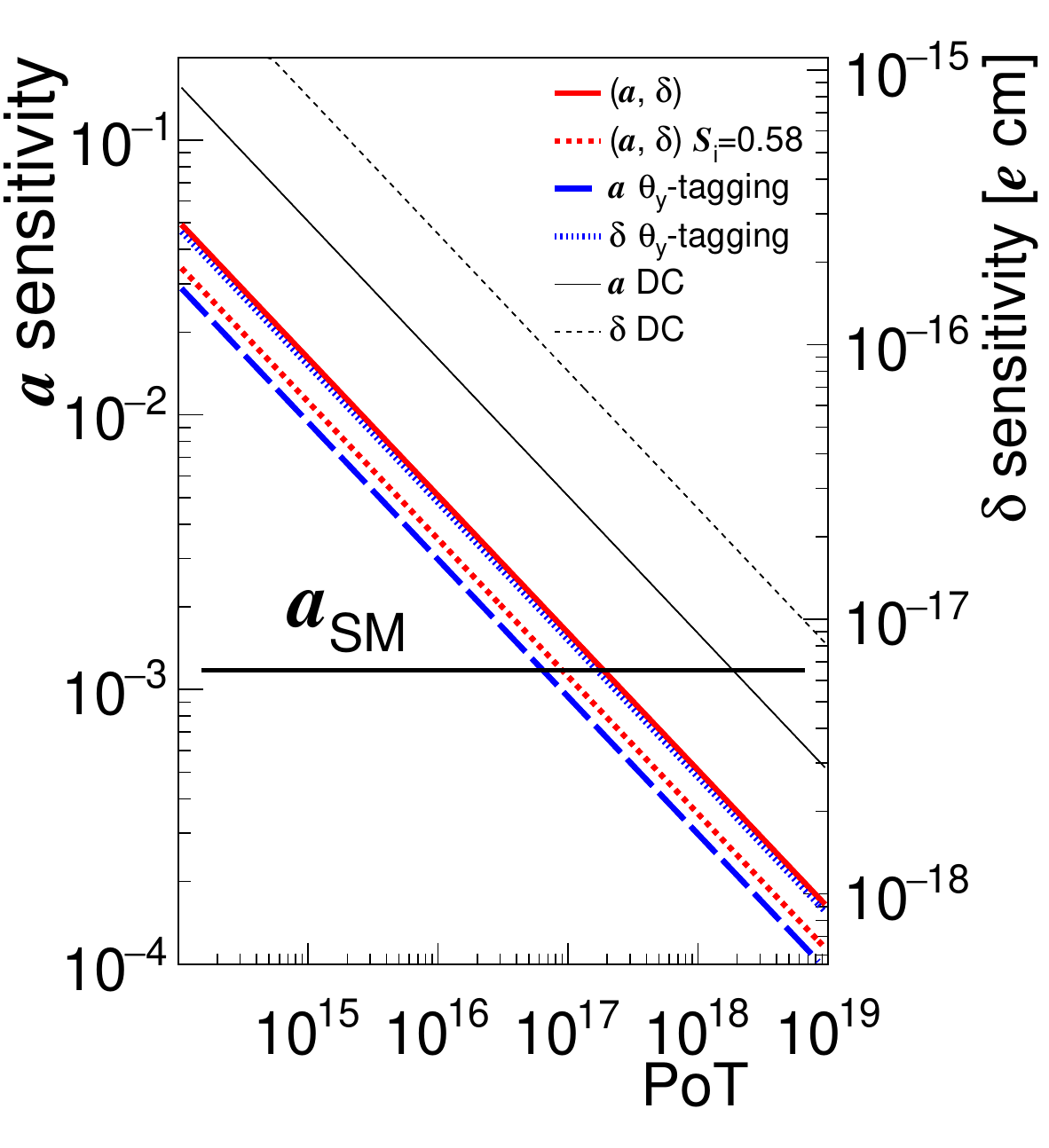}
	\caption{\label{fig:sensitivitytau} From Ref.~\cite{Fu:2019utm}.
		Estimated sensitivities for \at and \deltat as a function of the number of protons on
		{\CHc a 2.5 cm thick W}
		target (\pot) for a \Ge crystal with optimal parameters {\CHb (thick solid red line)},
		compared to other configurations (see text).
		{\CHb These are labeled as (\at, \deltat) when the corresponding lines overlap.}
		The SM model prediction for $a$~\cite{Eidelman:2007sb} is also indicated.
	}
\end{figure}

{\CHb The absolute sensitivity to the dipole moments} is determined with a large number of pseudoexperiments generated and fitted using a probability density function based on the spin precession
equation of motion, and our novel method for the polarization reconstruction.
Figure~\ref{fig:sensitivitytau} illustrates the estimated sensitivities
as a function of the number of impinging protons for a \Ge crystal with optimal parameters {\CHb (thick solid red line)}.
Sensitivities for other configurations with maximum average event information {\CHa $S_i=0.58$} {\CHb (thick dotted red line)},
\thetay-tagging based on a discrimination between positive and negative \thetayDsTau\ {\CHb with} ideal
{\CHb tagging} efficiency of 100\% {\CHb (thick dashed and hatched blue lines)},
and the double crystal (DC) option proposed in Ref.~\cite{Fomin:2018ybj} {\CHb (thin solid and dotted black lines)}, are also shown for comparison.
The corresponding sensitivities for \Si are about a factor two worse. 
%
%
%

%

\subsection{Conclusions}

%
%
%
%
%

In summary, a novel method for the direct measurement of the \Ptau MDM and EDM
has been
presented with interesting perspectives for a stringent test of the SM and search for new physics.
The fixed-target setup and the analysis technique
have been discussed along
with sensitivity projections for possible future scenarios. 
The SM prediction for the \Ptau MDM could be verified experimentally
with
{\CHb a sample of}
around
{\CHc $10^{17}$\pot,}
whereas at the same time a search for the \Ptau EDM
at the level of $10^{-17}~e\cm$ or below
could be performed.
{\CHc This would require about 10\% of the protons storaged during a decade of \lhc operation~\cite{Apollinari:2017lan}.}
This method could be
tested
using the fixed-target setup
proposed for the study of heavy baryons, optimized in Section~\ref{sec:optimbaryons}.

\newpage

\section{Focusing crystals} \label{sec:optimfocusing}

\begin{flushright}
	This section is based on Ref.~\cite{Biryukov:2021cml}\\
\end{flushright}

The use of crystal lenses for spin-precession experiments has only been considered recently, in Refs.~\cite{Biryukov:2021gsd,Biryukov:2021phs,Biryukov:2021cml}, and represents a new application of bent-crystal channeling. With respect to plain crystals, focusing-crystal configurations can substantially improve the trapping efficiency. However, they pose new technical challenges for the experiment. 

\begin{figure}[t]
	\centering
	\includegraphics[width=0.45\linewidth]{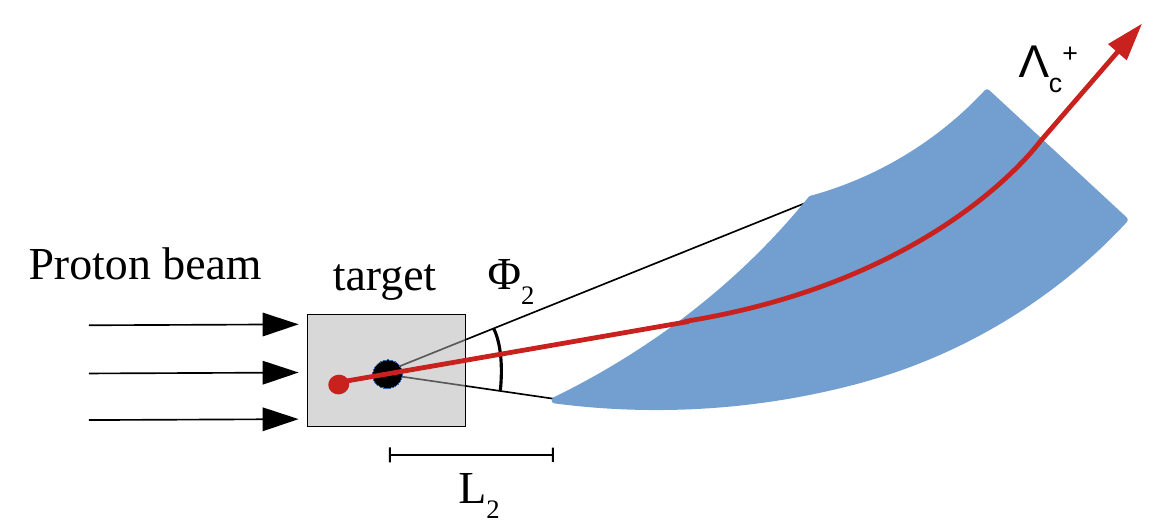}
	\caption{Single-lens scheme. The proton beamlet arrives directly to the target. The produced \Lc particles (red dot) may be channeled if their direction is parallel to the atomic planes at the crystal entry point. }
	\label{fig:generaldiagramsinglelens}
\end{figure}

\begin{figure*}[t]
	\centering
	\includegraphics[width=0.8\textwidth]{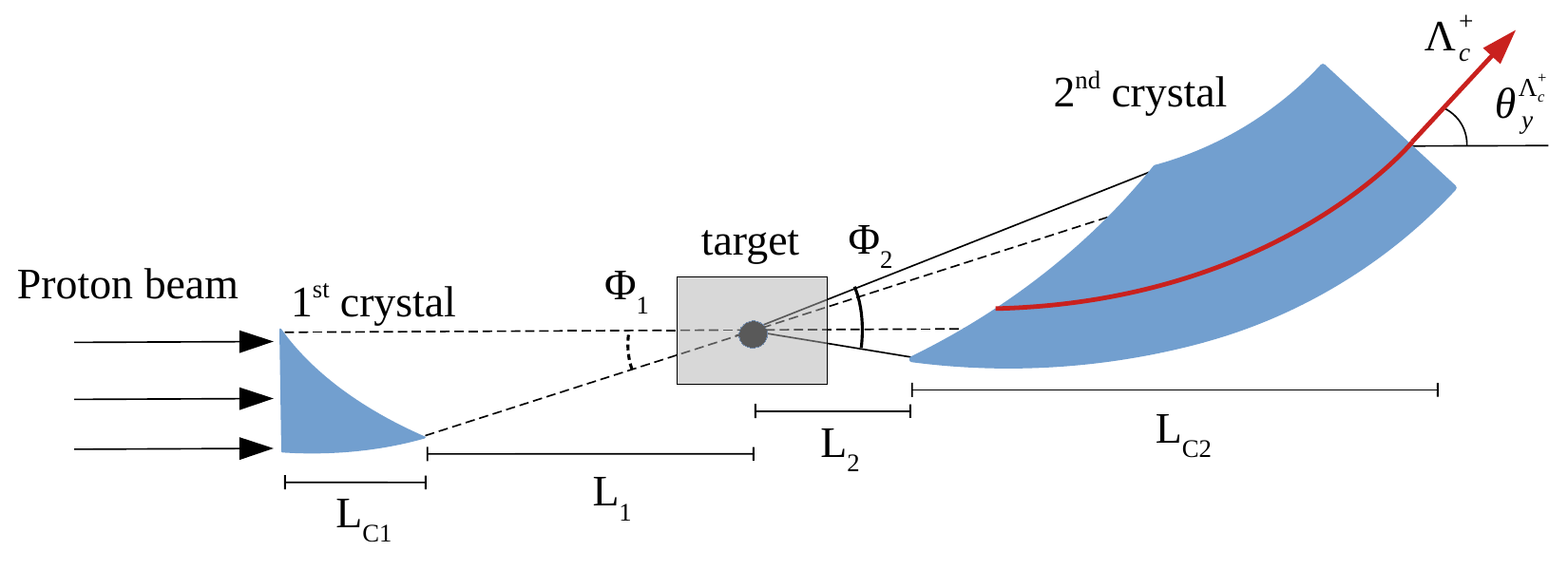}
	\caption{Double-lens scheme. The protons are focused onto the target by means of a crystal lens. 
		%
	}
	\label{fig:generalDiagram}
\end{figure*}

When comparing the channeling efficiency, in this section we will recurrently refer to three schemes:
\begin{itemize}
	\item Plain-crystal scheme: nominal layout for heavy baryons, in Figure~\ref{fig:threelayouts}(a).
	\item Single-lens scheme: replacing the plain-faced crystal by a focusing crystal, in Figure~\ref{fig:generaldiagramsinglelens}.
	\item Double-lens scheme: adding a first lens right before the target to focus the proton beamlet\footnote{The secondary proton beam, deflected by the crystal kicker will often be referred to as \textit{beamlet}.}, in Figure~\ref{fig:generalDiagram} and \ref{fig:threelayouts}(c).
\end{itemize}

We will start this section by describing the experimental layout and giving a first approximation to the trapping efficiency through the notion of \textit{focal window}, in Section~\ref{sec:scheme}. The compatibility of this setup with the LHC beam is addressed in Section~\ref{sec:parameters}, where realistic values for the setup parameters are obtained. 
The trapping efficiency is determined in Section~\ref{sec:efficiency}, with the help of the new trapping condition, presented in Appendix~\ref{app:lensestrapping}. The results for the single- and double-lens scheme are summarized in Section~\ref{sec:conclusion}.

\subsection{Experimental layout} \label{sec:scheme}

As described in Chapter~\ref{ch:crystals}, the interesting \Lc candidates, with $E\gtrsim700\text{GeV}$, are produced approximately in a cone of $\pm 1.5 \mrad$, while only those that are produced aligned to crystal planes, within an angle of $\pm 7\murad$ (Lindhard angle), are trapped between the crystal atomic planes\footnote{We will take the Lindhard angle $\lindtwo = 7 \murad$ in both schemes, corresponding to 700 GeV particles in a silicon crystal oriented in the (110) direction. Since the efficiency is always computed relative to the plain-crystal scheme, this assumption cancels out to a large extent.}.
To increase this efficiency one could consider replacing the plain crystal by a crystal lens, whose atomic planes point towards the target~\cite{Biryukov:2021phs,Biryukov:2021gsd}. Naively thinking, this would allow capturing all \Lc baryons produced at the target. However, in reality, the target volume is much larger than the focal volume, defined as the region in which the produced \Lc particles can be trapped in the crystal.  
Thus, in qualitative terms, while the plain crystal traps particles produced across the whole target volume but restricted to a small range of directions, the crystal lens traps the particles from all directions, but restricted to a small volume. To perform a quantitative comparison of the two schemes, we need to derive the trapping condition for the crystal lens anew. The complete derivation is in Appendix~\ref{app:lensestrapping}. However, we shall introduce the basic concept already, which will serve to justify the need for a double-lens scheme and estimate the crystal parameters.

\subsubsection{Simplified trapping condition }

The requirement for trapping, as always, is that the particle direction must be within the Lindhard angle ($\theta_{L2}$ in Figure \ref{fig:crystalParameters}) when it reaches the crystal entry face. However, in crystal lenses, the direction of the atomic planes varies across the crystal entry face. We know that the tangent of all atomic planes points toward the focal point, where the projection of the Lindhard angle defines a \textit{focal window}. Thus, the trapping condition can be redefined in a more practical way: the (extended) \Lc trajectory must cross the focal window for it to be trapped in the crystal. The height of the focal window can be determined with the help of Figure~\ref{fig:crystalParameters} and reads
\begin{equation}
w_{F2} = 2\lindtwo (\Ltwo + l_2) \approx 0.4-0.7\,\mum, 
\label{eq:window}
\end{equation} 
where \Ltwo is the target-crystal distance and $l_2$ depends on the entrance point, as defined in Figure \ref{fig:crystalParameters}. 

The probability of the \Lc to cross the focal window is necessarily correlated to that of the impinging proton. However, with the beam size $\sigma_{\rm beam} \approx 50 \mum$ only few of the protons cross the focal window\footnote{Even if this probability is small, so is the trapping efficiency in the nominal plain-crystal scheme, and the comparison is not straightforward. Detailed evaluations of this efficiency, described in Section \ref{sec:efficiency}, show an improvement of around 40\% with respect to the plain-crystal scheme.}. 
Independently of the absolute trapping efficiency, it is easy to see that reducing the beam size at the focal window could massively improve it. For this reason, we will explore the double-lens scheme too.

\subsection{Setup parameters and LHC constraints} \label{sec:parameters}

\begin{figure}[t]
	\centering
	\includegraphics[width=0.6\columnwidth]{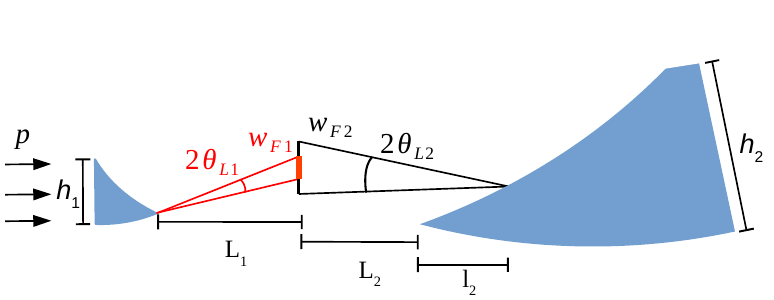}
	\caption{The extended trajectory of the particles must cross the focal window $w_{F2}$ (vertical black line) for these to be trapped in the second crystal. To maximize this probability, the focal window $w_{F1}$ (thick red line), analogously defined for the first crystal, must be embedded within $w_{F2}$.}
	\label{fig:crystalParameters}
\end{figure}

The focused protons converge at the focal point. However, their directions can differ from the tangent to the atomic plane by up to a Lindhard angle. Thus, the focused-proton directions also define a focal window. The shorter the focal distance $\Lone$ is, the smaller the focal window $w_{F1} = 2\lindone \Lone$ will be, resulting in more protons crossing the focal window $w_{F2}$. For a first estimation of the setup distances we shall impose $w_{F1} \lesssim w_{F2}$ (in Figure~\ref{fig:crystalParameters}), leading to $\Lone \lesssim (\lindtwo / \lindone) \Ltwo$. Taking a short target-crystal distance $\Ltwo \approx 3 \cm$ to avoid efficiency losses due to the exponential decay of the \Lc baryon, and $(\lindtwo / \lindone)\approx 2.7$ corresponding to 7 TeV (1 TeV) protons (\Lc), we obtain $\Lone\lesssim 8 \cm$.

After the focal point, the proton beamlet starts to diverge and we should make sure that it stays contained within the beam pipe. Those protons will be stopped by an absorber, positioned around 60~m downstream of the setup~\cite{Barschel:2020drr}, at which point the beamlet width should not exceed $\sim 1\,\cm$. This imposes a hard condition on the focusing angle $\Phi_1 \leq (1\cm/60\,{\rm m}) \approx 170 \murad$.
This, however, excludes the possibility to have $\Lone\approx8\,\cm$. With such a small focusing angle, the protons require longer distances to converge. Specifically, now $\Lone = \sigma_{\rm beam} /\Phi_1 \approx 30 \cm$. The impact on the efficiency is only mild, with 18\% loss, as discussed in Figure 4 of the original Ref.~\cite{Biryukov:2021cml}.

The second lens should be able to capture \Lc particles produced at any angle $\theta$ with respect to the impinging proton. Thus, the focusing angle $\Phi_2$ should cover all possible proton directions (within $\Phi_1$) plus some extra room for the \Lc aperture angle with respect to the proton ($\varTheta_{\Lc} \approx 1.5 \mrad $), \textit{i.e.} $\Phi_2 = \Phi_1 + 2 \varTheta_{\Lc}\approx 3.2 \mrad$ (see Figure~\ref{fig:generalDiagram}), although $\Phi_2$ can be increased arbitrarily without affecting the trapping efficiency. 
The length and bending angle of the crystal lens is not relevant for the trapping efficiency, and the optimal values can be adopted from other studies.

\subsection{Trapping efficiency} \label{sec:efficiency}

\subsubsection{Double-lens scheme}

The concept of focal window was useful to compare the different schemes and determine the setup parameters with first estimations. However, to evaluate the trapping efficiency reliably, also in the case of misaligned crystals, it is convenient to use a trapping condition based not on the focal window but on its corresponding solid of revolution. 
For the sake of clarity and readability, the explicit geometrical details are included only in Appendix~\ref{app:lensestrapping}. 

The trapping condition for misaligned crystals reads (see Appendix \ref{app:lensestrapping})
\begin{equation}\label{eq:conditionDisplacedMainText}
|\theta - \arctan\frac{d}{r} |\leq 
\arctan \frac{\theta_{L2}(\Ltwo+l_2)}{\sqrt{d^2 + r^2}} ~,
\end{equation}
where $r$ is the (signed) distance between the proton-target interaction point and the first-lens focal point; $\theta$ is the \Lc production angle with respect to the impinging proton; $d$ is the vertical distance between the first- and second-lens focal points; and $\lindtwo$ ($\lengthtwo$) was already defined as the Lindhard angle (length) of the second lens. The distance $l_2$ was defined in Figure~\ref{fig:crystalParameters}.

\begin{figure}[t]
	\centering
	\includegraphics[height=0.37\linewidth]{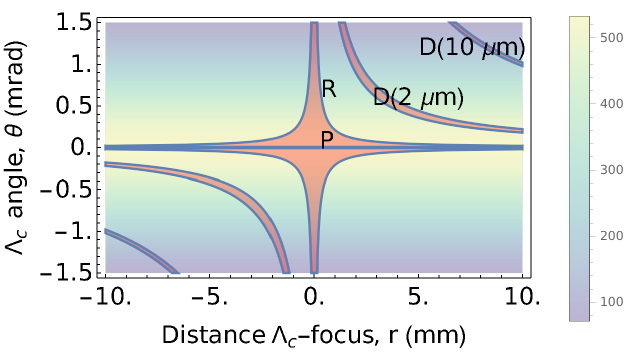}
	\caption{Regions of the $(r,\theta)$ phase space where the trapping condition is met for the plain-crystal scheme (P), the double-lens scheme (R), and the double-lens scheme with a vertical displacement between the two focal points of $d$ ($D(d)$). The background colour gradient shows the probability distribution of the \Lc aperture angle, $\theta$, in Eq~\eqref{eq:ntrappedevents}. Even though the available phase space for the plain-crystal scheme (P) is smaller than the one for vertically displaced focal points (D), it is centred at the peak of the $\theta$ distribution, partially compensating this difference.}
	\label{fig:phasespace}
\end{figure}

The final efficiency can be evaluated as the portion of phase space $(r,\theta)$ fulfilling the trapping condition. However, this is not equally populated, as $\theta$ follows a Gaussian distribution with $\mu = 0$ and $\sigma=1.5 \mrad$. The number of trapped events is proportional to the integral of $G(\theta;~\mu, \sigma)$ over the phase space region $X$,
\begin{equation} \label{eq:ntrappedevents}
N_X = \int_X G(\theta;~0, ~1.5~\mrad) ~dr ~d\theta~.
\end{equation}
For misaligned crystals by a distance $d$, this region is  (see Figure~\ref{fig:phasespace})
\begin{equation}
D(d) = \lbrace(r,\theta)~ : ~ |r| \leq 1 \cm, ~\text{Eq.~\eqref{eq:conditionDisplacedMainText}} \rbrace~.
\end{equation}
For perfectly aligned crystals ($d=0$), we have
\begin{equation}
R = \lbrace(r,\theta)~ : ~ |r| \leq 1 \cm, ~|\theta| \leq \lindtwo (\Ltwo+l_2)/|r| \rbrace~.
\end{equation}
Analogously, in the plain-crystal scheme,
\begin{equation}
P = \lbrace(r,\theta)~ : ~ |r| \leq 1 \cm, ~|\theta| \leq \lindtwo  \rbrace~.
\end{equation}
All these different regions of the \Lc-production phase space $(r,\theta)$ that result in the trapping of the \Lc baryon by the bent crystal are illustrated in Figure~\ref{fig:phasespace}. From this figure, it is already apparent the large potential gain of the double-lens scheme (R) with respect to the (nominal) plain-crystal scheme (P), presented in Section~\ref{sec:optimbaryons}.

Before carrying out the integration and obtaining the final efficiency, there is one more effect needed to accurately estimate the trapping efficiency of the double-lens scheme. In Eq.~\eqref{eq:conditionDisplacedMainText}, we considered all the proton trajectories to intersect exactly at the focal point, but due to the margin of the Lindhard angle at the crystal exit, these directions are diluted within the focal window of the first crystal, $w_{F1}$. 
To account for this smearing without Monte Carlo simulations, we evaluate the trapping efficiency for small vertical displacements $\delta$ of the proton trajectories within the focal window $w_{F1}$ and take their average through the integral
\begin{equation}\label{eq:Nwindow}
N(d) = \frac{1}{w_{F1}}  \int_{-w_{F1}/2}^{w_{F1}/2} N_{D(d + \delta)}  ~\text d \delta~~,
\end{equation}
where $N_{D(d)}$ is defined as in Eq. \eqref{eq:ntrappedevents}, and $d$ is the vertical displacement between the central points of the two crystal foci. Then, using our analytical method, the number of events in acceptance is the result of integrating sequentially over the phase space that meets the trapping condition, in Eq. \eqref{eq:ntrappedevents}, and over the small displacements within the focal window of the first crystal, in Eq. \eqref{eq:Nwindow}. 

\begin{figure}[t]
	\centering
	\includegraphics[height=0.37\linewidth]{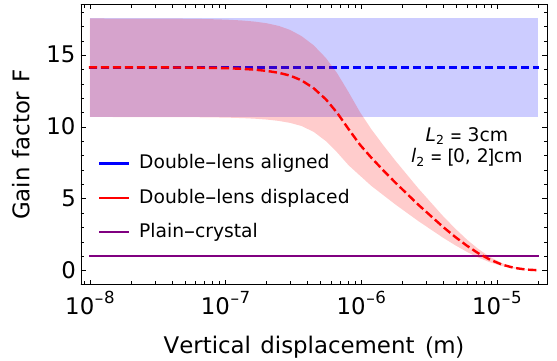}
	\caption{Trapping efficiency with respect to the plain-crystal scheme, as a function of the vertical displacement of the two focal points, with a 2-cm target. The shaded bands represent the variation of the trapping efficiency across the crystal entry face, due to the small change on the distance to the focal point, noted as $l_2$ in Figure \ref{fig:crystalParameters}.
	}
	\label{fig:trappingefficiency}
\end{figure}

Finally, the trapping efficiency gain of the double-lens scheme with respect to the plain-crystal scheme,
\begin{equation}
F = \frac{N(d)}{N_P}~,
\end{equation}
is shown in Figure \ref{fig:trappingefficiency} as a function of the vertical displacement $d$. In the evaluation we used a target thickness of 2\,\cm (\ie $|r|<1\,\cm$). The maximum gain reaches about a factor $F=15$, but decreases rapidly for small vertical displacements of the two focal points, falling below the plain-crystal scheme for $d \gtrsim 8 \mum$. In principle, this accuracy is achievable with current technology and the proposed setup would not represent an efficiency loss in any case, as discussed below in Section~\ref{sec:technology}.

To consider the possible inefficiencies of the first lens, the gain of the double-lens scheme, in Figure~\ref{fig:trappingefficiency}, should be re-scaled by the channeling efficiency of protons in the first lens.
The proton bending efficiency measured at SPS was 77-83\% for bendings of $0.05-0.2\mrad$ for beams of low divergence \cite{sps-rep}. 
In our case, the beam divergence is equal to the Lindhard angle, as induced by the crystal kicker (in Figure~\ref{fig:layoutgeneral}), which spreads the proton directions at the crystal exit. In the introduction to crystal channeling, in Section~\ref{sec:channeling}, we saw that the difference in trapping efficiency between parallel and divergent beams is related to the area of the elliptical orbit in the phase diagram. In this case, for an initial divergence equal to the Lindhard angle the difference is of a factor $\pi/4$.

\subsubsection{Target thickness}

\begin{figure}
	\centering
	\includegraphics[height=0.40\linewidth]{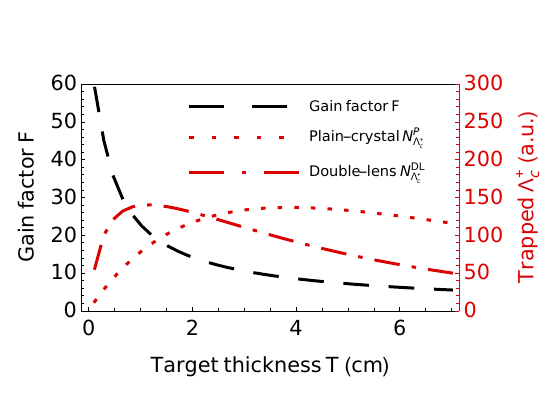}
	\caption{Showing, as a function of the target thickness, the number of \Lc exiting the target (dotted red), as in Figure~\ref{fig:targetNLc}; the trapping-efficiency gain F (dashed black); and the number of produced-survived-and-trapped \Lc in the double-lens scheme (dot-dashed red). The optimal target thickness for the double-lens (plain-crystal) scheme is at 0.5\cm (2.0 cm), corresponding to a gain factor $F \approx 35$.}
	\label{fig:opttarget}
\end{figure}

In the plain crystal scheme, the number of channeled particles is proportional to the target thickness. Conversely, in the plain-crystal scheme, most of the trapped particles are produced around the centre of the target, at $|r|\approx0$, and any reduction in the target thickness would impact the channeling efficiency much less. For instance, if the target thickness was reduced from 2 to 0.5\cm along the beam direction, the efficiency improvement of the proposed layout would increase from a factor 15 to 35. A detailed optimization of the target thickness $T$ is presented in Figure~\ref{fig:opttarget}. The number of \Lc baryons after the target was evaluated as a function of $T$ in Figure~\ref{fig:targetNLc} accounting for the proton flux attenuation, and the absorption and decay of \Lc within the target. Reproducing this dependence for the plain-crystal scheme $N_{\Lc}^P (T)$ (dotted line), and multiplying it by the gain factor of the double-lens scheme $F(T)$ (dashed line), we obtain the variation with $T$ of the number of produced-survived-and-trapped \Lc baryons in the double-lens scheme as $N_{\Lc}^{DL} (T) \propto F(T)~N_{\Lc}^P (T)$ (dash-dotted line). 
The maximum efficiency is found at $T=1\cm$. However, to reduce the background from proton-target interactions, we take the shortest target within a $10\%$-difference in efficiency from the optimal point, yielding $T\approx0.5\cm$.

Before moving on to the single-lens scheme, we shall summarize all the numbers of the double-lens scheme. With a target thickness $T\approx0.5\cm$, there is a gain of a factor $F\approx 35$ in trapping efficiency with respect to the plain-crystal scheme. Accounting for the first-lens efficiency ($\sim 80\%$) and the decay of the \Lc in the additional target-crystal separation of 2\cm in $\Ltwo$ ($\sim 70\%$), the final gain on the number of channeled events would be of a factor $\sim 20$.

\subsubsection{Single-lens scheme}

Evaluating the efficiency of the single-lens scheme is straightforward with the tools developed for the double-lens scheme. To further optimize the setup, we allowed in this case an extra degree of freedom: the target-crystal distance $d_{TC}$. Above, it was always fixed to the focal length (\ie the focal point was assumed at the centre of the target). In the single-lens scheme, the target should be positioned closer to the crystal to avoid losses due to the exponential decay of the \Lc particles. At the optimal point, we find a 40\% increase in trapping efficiency with respect to the plain-crystal scheme.

If the \Lc particles were stable, the gain with the single-lens scheme would reach around a factor $7$ for $L_2\approx d_{TC}\approx20\,\cm$.  Thus, the single-lens scheme is better suited for dipole moment measurements of particles with longer decay times such as $\Xi_c^+$, $\Xibp$ or $\Omegabp$.
Nevertheless, even if the improvement with this scheme is modest, of 40\%, it needs no additional instrumentation with respect to the nominal layout besides the replacement of the plain crystal by a crystal lens, and it has no additional interference with the LHC beam.

\subsection{Technical challenges} \label{sec:technology}

The potential realisation of the presented experimental configuration is subjected to the available crystal technology and presents additional challenges with respect to the plain-crystal scheme.
The specifications for the needed crystal lenses, goniometers, and particle detectors are discussed in the section \textit{Technical challenges} of the original article~\cite{Biryukov:2021cml}.

\subsection{Conclusions} \label{sec:conclusion}

The experimental proposal to measure the electric and magnetic dipole moment of short-lived particles at the LHC suffers from very low efficiencies. In this section, new layouts based on the use of crystal lenses have been explored, finding an improved sensitivity with the same number of protons on target.

The potential statistical gain with the single-lens scheme can reach up to 40\% for $\Lc$ baryons, without any technical limitation besides the construction of the crystal lens itself, although this result is strongly dependent on the transverse size of the deflected beam. Instead, by introducing a short lens before the target to reduce this beam size, the statistical gain can reach up to a factor 20 in the double-lens scheme.
However, this solution distorts the shape of the proton beamlet and requires high accuracy on the positioning of the two crystals.

For bottom baryons, $\overline{\Xi}_b^+$ and $\overline{\Omega}_b^+$ a similar or larger statistical gain can be achieved due to their longer lifetime. 
The improvement for \taup leptons is expected to be of similar size. However, the trapping condition derived in Appendix~\ref{app:lensestrapping} does not apply to the case with an intermediate $\Dsp$ meson in the production of $\taup$ leptons, and the gain should be evaluated with specific Monte Carlo simulations.

\chapter{The experiment with long-lived particles at LHCb} \label{ch:lambdasLHCb}

For longer-lived strange baryons, the LHCb experiment offers a different opportunity to measure their electric and magnetic dipole moments, this time without additional instrumentation. By reconstructing events in which longer-lived particles such as \Lz hyperons go through the magnetic field, it is possible to compare their polarization before and after the magnet, extracting the dipole moments. The biggest challenge to realising this idea~\cite{Botella:2016ksl} is the reconstruction of these events with only the part of the LHCb tracking system located downstream of the magnet. However, having access to this type of event topologies can expand the physics program for the LHCb experiment through the direct measurements of electromagnetic dipole moments. Furthermore, reconstructing these events greatly enhances the lifetime coverage for long-living particle (LLP) searches, predicted by a plethora of NP models.

This chapter is organized as follows. After introducing the LHCb experiment in Section \ref{sec:LHCb}, we will discuss the experimental concept and projected sensitivity of the \Lz EDM and MDM measurements, in Section \ref{sec:lambdaedm}. In Section \ref{sec:ttracks} the challenges for the reconstruction of the interesting events (with T tracks) are described. To finish, in Section \ref{sec:LLP}, we will start studying the possibilities for LLP searches at LHCb, comparing the geometrical acceptance and lifetime coverage of the different subdetectors.

\section{The LHCb experiment} \label{sec:LHCb}

The LHC is the most powerful particle accelerator ever built. Located at the European Organisation for Nuclear Research (CERN) in Switzerland, it is a circular collider of approximately 27 km in circumference constructed in the tunnel of the former Large Electron Positron (LEP) collider, between 50 and 150 m underground. Two separated beams of particles travel in opposite directions intersecting and colliding at four different points along the ring. Namely, in the collision points of the four major experiments: ALICE, ATLAS, CMS, and LHCb.


During most of its working time, the LHC produces proton-proton ($pp$) collisions. However, also lead-lead, xenon-xenon, and proton-lead collisions have been recorded. These heavy-ion runs are primarily motivated for the study of the quark-gluon plasma, in which the ALICE experiment is specialized. Due to the delivered luminosity, exceeding all plans, these runs are providing plenty of physics results, a remarkable example being the first observation of photon-photon scattering \cite{ATLAS:2017fur}. Moreover, proton-gas collisions in fixed-target mode have been recorded by LHCb using the SMOG \cite{LHCb:2014vhh} system, providing unique results like antiproton and charm production in proton-Helium interactions \cite{LHCb-PAPER-2018-031,LHCb-PAPER-2018-023} of interest for cosmic ray physics.

The maximum energy in $pp$ collisions will be reached in Run III, with 14 TeV centre-of-mass energy, a fundamental figure for ATLAS and CMS, which are general-purpose detectors aiming at the direct detection of NP particles produced on-shell. In each (proton) bunch crossing, these experiments record dozens of $pp$ collisions, reaching instantaneous-luminosity values at the level of $10^{34} ~\text{cm}^{-2} \text{s}^{-1}$.

The LHCb experiment~\cite{JINSTLHCB,LHCb-DP-2014-002} specializes in flavour physics and its main focus is the decays of charm and bottom hadrons. These are sensitive to NP effects induced by heavy virtual particles running in the loops.
Given the complexity of the observables being analysed, which often include chains of successive particle decays, together with the small statistical rate of the most interesting events, especially high precision in the reconstruction of events is required, which includes a high-resolution tracking system and excellent particle identification. To achieve this precision with current hardware possibilities it is necessary to reduce the detector occupancy by lowering the instantaneous luminosity with respect to the other experiments at LHC. For that reason, in the LHCb collision point the beams intersect laterally, producing an instantaneous luminosity of $4 \times 10^{32} \text{ cm}^{-s}\text{s}^{-1}$ and a mean number of visible $pp$ collisions per bunch crossing of $\mu=1.8$. Even though the amount of protons in the beam gets smaller during a fill, this collision rate is maintained constant by adjusting the transversal beam overlap, a procedure referred to as luminosity levelling.

\begin{figure}
	\centering
	\includegraphics[width=0.9\linewidth]{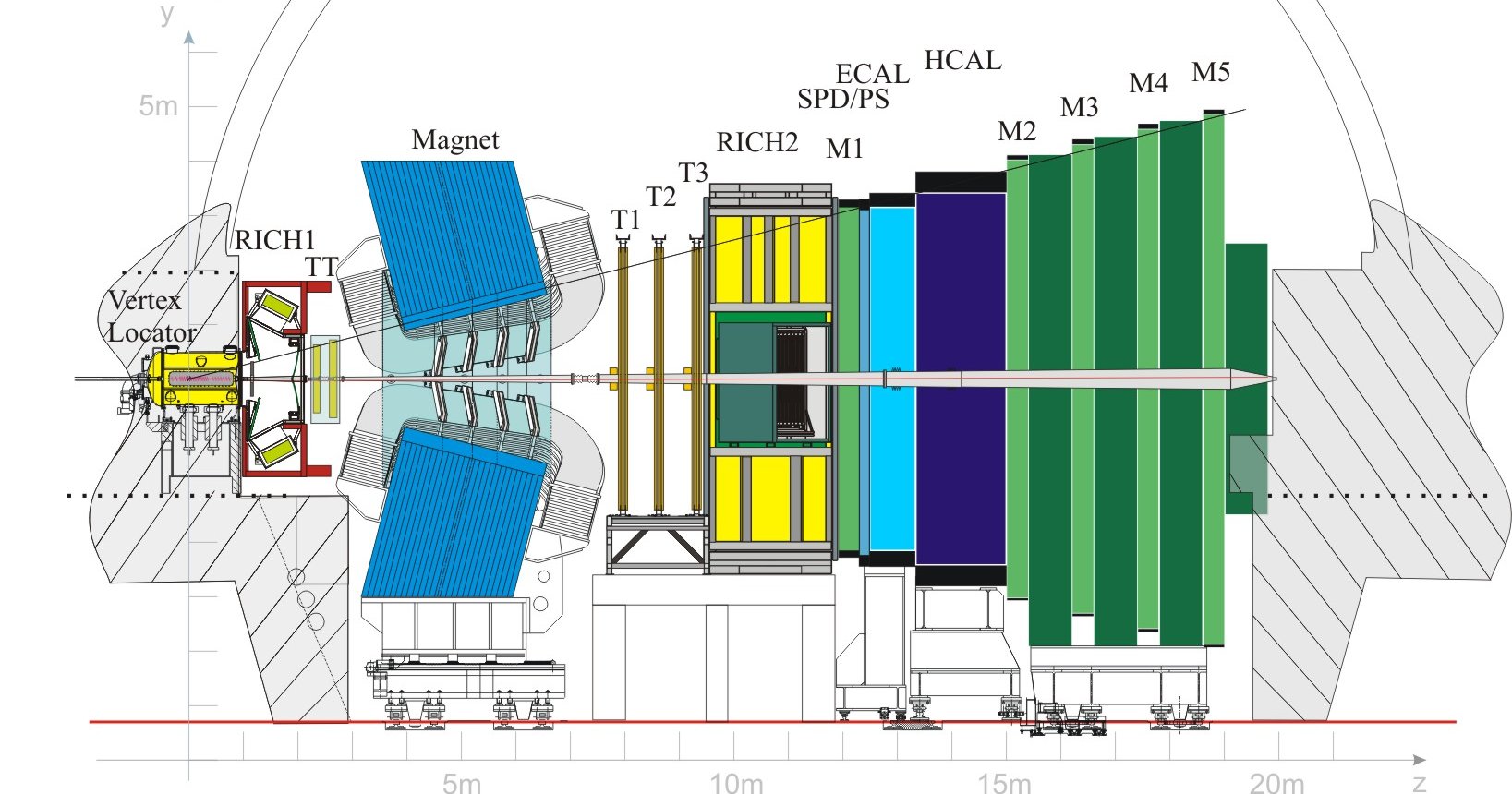}
	\caption{From Ref.~\cite{JINSTLHCB}. LHCb layout during Run I and Run II~\cite{JINSTLHCB}. The $z$ direction is defined along the beam and $y$ is the vertical direction.}
	\label{fig:LHCbDetector}
\end{figure}

The shape of the LHCb detector is also particular compared to the other experiments. It consists of a single-arm detector covering an angular acceptance of 250 and 300 \mrad (14.3 and 17.2 degrees)
in the vertical and horizontal planes, respectively. A strong magnetic field of up to 1.05\,T  bends the trajectory of the charged particles in the horizontal plane, thus the need for a slightly larger angular coverage in the horizontal plane. The reason for this small angular acceptance is that $b\bar b$ and $c\bar c$ pairs are mostly produced at small angles with respect to the beam direction. The dominant production mechanism of heavy-quark pairs is gluon-fusion, in which one of the gluons typically has much larger momentum fraction than the other, dominating the final boost of the quark pair, which stays close to the beam direction.
The layout of the detector is shown in Figure \ref{fig:LHCbDetector}. In the following, we will briefly describe the subsystems of the detector as they were during the Run I (2009-2013) and Run II (2015-2018). For Run III (2022-2025), a major upgrade of the LHCb detector has been done~\cite{LHCbupgrade}. The main differences with respect to the previous detector will be pointed out at the end of the section.

\begin{figure}
	\centering
	\includegraphics[width=0.7\linewidth]{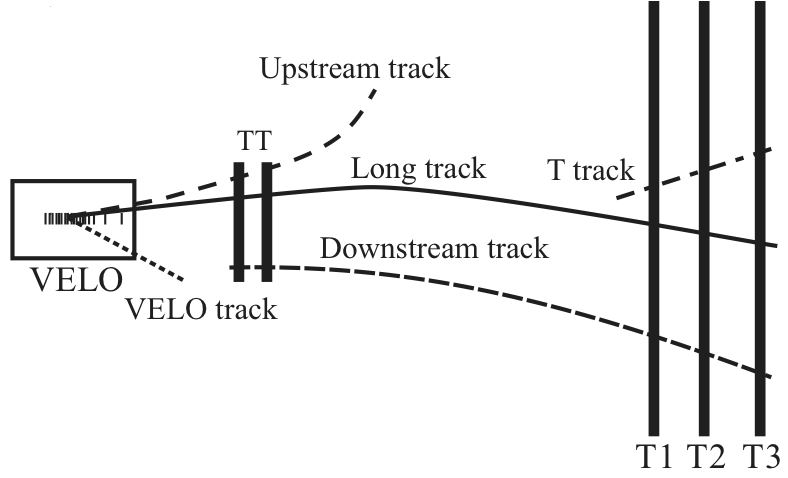}
	\caption{From Ref.~\cite{JINSTLHCB}. Track types and schematic tracking system of the LHCb. }
	\label{fig:tracktype}
\end{figure}

\subsubsection{Tracking system}

The tracking system serves to reconstruct the (curved) trajectory of the charged particles. It consists of three tracking detectors (or \textit{trackers}), VELO, TT, and T-stations; and the dipole magnet. The tracks are classified into different types depending on the combination of trackers involved in their reconstruction. These are represented in Figure~\ref{fig:tracktype} together with a schematic view of the tracking system:


\begin{itemize}
	\item \textbf{VELO:}
	The silicon vertex locator (VELO) has the purpose of identifying with high precision the $pp$ interaction point (primary vertex, PV) and decay points of the heavy hadrons (secondary vertex, SV). Due to the small lifetime of charm and bottom hadrons, \order(1 \ps), a precise reconstruction of PV and SV is challenging even with the large boost available at the LHC. For this reason, the 21 modules of the VELO are located only $8.2\,\mm$ away from the beam. To avoid radiation damage during the beginning of a fill, when the beam is not stable yet, the VELO is \textit{opened}, bringing its modules about 3 \cm away from the beam~\cite{JINSTLHCB}.

	\item \textbf{TT:} 
	The two modules of the TT stations, composed of two layers of microstrip silicon sensors each, locate a total of 4 points of the particle trajectory, allowing to reconstruct its direction before it enters the magnetic field region. This detector is particularly important for the reconstruction of longer-lived particles such as \Lz or \KS hadrons, which often decay after the VELO and their (charged) decay products must be reconstructed with downstream tracks. In fact, in Part II, we will analyse real LHCb data with this type of events.

	\item \textbf{Magnet:} A warm dipole magnet creates a strong magnetic field in the vertical direction $B_y$ that curves the path of the charged particles, which allows the reconstruction of their momenta.
	This was, at least, the purpose in its design. However, we can also profit from the intense magnetic field of the LHCb magnet by measuring its interaction with the electromagnetic dipole moments of long-lived particles such as \Lz hyperons. The three components of the magnetic field ($B_x, B_y, B_z$) are shown in Figure~\ref{fig:bfield} with respect to the $z$ coordinate, fixing $x=y=0$ in this projection. The complete \Bvec map will be used in Section~\ref{sec:lambdaedm} for the sensitivity study of \Lz EDM and MDM.
	
	
	\begin{figure}[t]
		\centering
		\includegraphics[width=1.\textwidth]{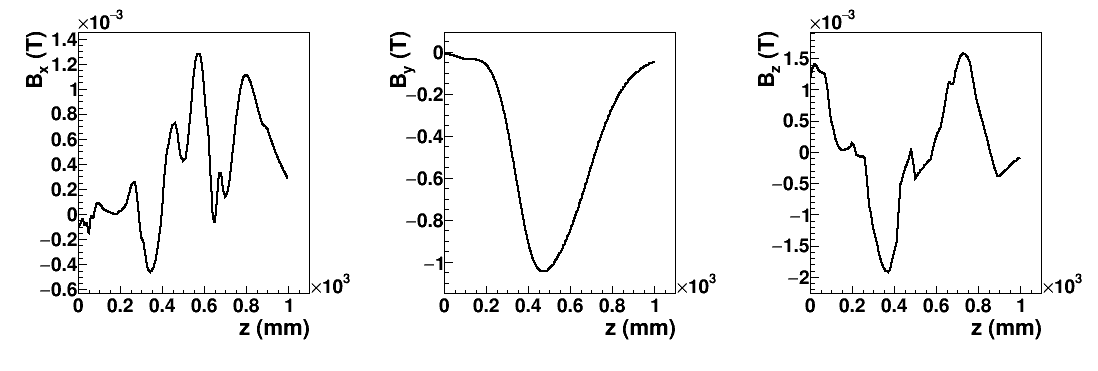}
		\caption{Mapping of the magnetic field in the LHCb detector. The three components $B_x,~B_y~\text{and}~B_z$ are represented with respect to the $z$ coordinate, being evaluated at $x=y=0$. \revafter{Better resolution plots in AFS cern user} }
		\label{fig:bfield}
	\end{figure}

	\item \textbf{T-stations:} 
	After the magnet, the charged particles go through the three modules of the T-stations. These are composed of an external part (outer tracker) and a submodule around the beam pipe (inner tracker), with different granularity due to the different occupancy of each region. 
	The main purpose of this last tracking subsystem is to reconstruct the direction of a particle after it has been curved, to match it with the upstream signals and allow the reconstruction of the momentum. They play a fundamental role in the proposal presented in the next section since the $\Lprp$ decays of interest take place inside the magnetic field region and their decay products have to be reconstructed only using T tracks. Direct information on the $p$ and $\pi^-$ directions and momenta can only be extracted thanks to the residual magnetic field in between the T-stations. 
	This and other challenges in the reconstruction of $\Lzppi$ decays with T tracks are described in Section~\ref{sec:ttracks}.
	
	
\end{itemize}


\subsubsection{Calorimeters}

Except for muons, all particles are absorbed in the calorimeters that measure their total energy. Photons and electrons are captured by the electromagnetic calorimeter (ECAL) after they cross the SPD/PS stations, which help in distinguishing converted neutral pions ($\piz \to \gamma \gamma $) from single photons, and these from electrons. Hadrons are absorbed by the hadronic calorimeter (HCAL). The calorimeters also provide direction and position information for neutral particles.

\subsubsection{Particle identification}

One of the unique features of the LHCb detector when compared to the other experiments at the LHC is its excellent system of particle identification (PID). LHCb contains two Ring Imaging Cherenkov detectors (RICH1 and RICH2) that measure the Cherenkov radiation emitted by charged particles moving faster than the speed of light in the medium (radiators) where they propagate~\cite{LHCbRICHGroup:2012mgd}. By measuring the angle of the emitted photons with respect to the particle trajectory it is possible to access its velocity and, with the measured momentum of the particle, identify its mass. In practice, single Cerenkov photons cannot be assigned to specific tracks with absolute certainty and a global reconstruction of the event is performed. The output is processed by a Neural Network (NN) to obtain the probability of a track to be each of the five (meta)stable charged-particle species: $\pipm, ~\Kpm, ~\pr^\pm, ~\epm, ~\text{and}~ \mu^{\pm}$. This system provides charged particle identification over a wide momentum range, from 2 to 100\,\gevc.


\subsubsection{Trigger system}

The trigger system of LHCb decides whether or not an event must be recorded for its subsequent analysis. It is organized in three different levels, with ever more restrictive requirements. In Section~\ref{sec:trigger} we will determine the most efficient selections in each of these levels to reconstruct \threepi decays.

\begin{itemize}
	\item \textbf{L0:} The Level-0 (L0) trigger is implemented on hardware. It \textit{fires} when high-\pt signals, associated to large centre-of-mass energy in the partonic interaction (\textit{hard collisions}), are detected either in the muon chamber or in the calorimeters. It acts at full collision rate of 40 \mhz and fires at an average rate below 1 \mhz.
	
	\item \textbf{HLT1:} Below 1 \mhz the full detector can be read out and a software-based reconstruction is performed in the High-Level trigger (HLT). In a first stage, HLT1, a partial event reconstruction is performed and the events are saved based on loose selections.
	
	\item \textbf{HLT2:} The full event is reconstructed in the HLT2 stage, and it is compared to a large number of different selections (\textit{trigger lines}), which are often defined for exclusive decay channels.

\end{itemize}

The quality of the HLT2 and offline reconstruction is essentially the same and one could ask if it is possible to do the analyses directly on HLT2-reconstructed objects, without waiting for the centralized offline reconstruction (in the stripping campaigns), and avoiding saving the raw event data (which occupies $\sim 10$ times more). This is the idea behind the \textbf{Turbo stream}, which was implemented in Run II and has been extensively used in analyses.


\subsubsection{LHCb upgrade}

Since the end of Run II, in 2018, a practically new LHCb detector has been installed to replace the one that has been described above~\cite{LHCbupgrade}. In Part II we will exploit real data collected with the previous detector in Run II. However, in the proposed measurements with bent crystals in Chapters \ref{ch:crystals} and \ref{ch:sensitivity} we considered the geometry of the upgraded detector, and we shall briefly describe the main differences.
To start, the tracking system has been completely changed~\cite{LHCb-TDR-013, LHCb-TDR-015}. The new VELO is based on hybrid pixel sensors with $5.5\,\mu m$ pitch. It will also get much closer to the beam, from $8.2$ to $5.1\,\mm$. The T-stations, essential for the reconstruction of T tracks, treated in the rest of the chapter, are replaced by the Scintillating Fiber (SciFi) tracker. The diameter of the sensors will go from $4.9 \mm$ (in the previous drift-time based tubes of the outer tracker) to $0.25\,\mm$. A major improvement has also been made to the trigger system~\cite{LHCb-TDR-016,LHCb-TDR-017}. The whole detector will be read out at the same speed as the bunch crossing frequency (30 MHz), allowing to reconstruct the complete events at full collision rate. The new software-based trigger will have much more flexibility on the requirements to save the events. The electronics of the calorimeters, RICH and muon chambers have also been replaced to allow the new readout~\cite{LHCb-TDR-014}.
These improvements were also needed to operate at much larger instantaneous luminosities, as the number of $pp$ collisions per bunch crossing will increase from {$\mu=1.8$ to $7.6$}.

\section{Dipole moments of the strange $\Lambda$ hyperon}  \label{sec:lambdaedm}

\begin{flushright}
	This section is partially based on Ref.~\cite{Botella:2016ksl}\\
\end{flushright}

As discussed in the introduction, early measurements of the lowest-lying baryon octet magnetic moments played a key role in supporting the quark model initially. Still nowadays, these observables are recurrently used as benchmarks to compare the predictions of low-energy hadronic theories (see \textit{e.g.} Ref.~\cite{Yang:2020rpi}).
In the case of the strange \Lz hyperon~\footnote{In principle, we could also consider the dipole moment measurements of the charged $\mathit\Xi^\pm$, $\mathit\Sigma^{\pm}$ and $\mathit{\Omega}^\pm$ hyperons at LHCb. However, the bending of their own trajectories in the magnetic field region and especially the presence of neutral particles in the main decay channels add substantial difficulties to their reconstruction. Thus, here we only treat the case of neutral \Lz hyperons.}, direct spin-precession experiments were conducted at Fermilab in the 1970s, the most precise experiment being described in Ref.~\cite{MAGN78,PONDROM}. Using a proton beam colliding with a Be target, \Lz hyperons were produced with an average polarization of $8.5\%$. The spin precession was induced by a total integrated magnetic field $\int B dl\approx 10 - 15 \,\rm{Tm}$, reaching precession angles higher than $\Phi =\pi / 2$. Approximately $3\cdot 10^6$ $\Lprp$ decays were registered in this experiment. Comparing the angular distribution of the decay with and without magnetic field, the electric~\cite{PONDROM} and magnetic~\cite{MAGN78} dipole moments were extracted, yielding
\begin{equation}
\delta_{\lz} \leq 1.5 \cdot 10^{-16}~ \ecm ~(\text{at 95\%~CL})~\text{ and }~  \mu_\lz = (-0.6138 \pm 0.0047 )\,~ \mu_N.
\end{equation}

This level of accuracy can be challenged with the existing layout of the LHCb experiment using the nominal $pp$ collision. Conceptually, at the LHCb we would be comparing $\Lprp$ decays taking place before and after the magnet, where the \Lz spin-polarization vector precesses.
With respect to the dedicated experiments at Fermilab, the LHCb presents improvements in terms of statistics as well as initial polarization. Moreover, with LHC energies, both $\Lz$ and $\bar{\Lz}$ are produced in similar amounts, allowing to test \CPT invariance through the comparison of their corresponding magnetic moments~\cite{DiSciacca:2013hya,VanDyck:1987ay,Bennett:2004pv}.

We will start this section by discussing the possible sources of polarized $\Lz$ hyperons in $pp$ collisions. Setting the basic requirements on the \Lz-production channels, the number of events per \invfb and the detector efficiencies are estimated. Subsequently, the spin-precession equations are specified for this case, and projections on the final sensitivity are provided.

\subsubsection{Initial polarization}

The \Lz hyperon is the lightest baryon with strangeness and it is copiously produced in high-energy hadronic machines like the LHC. There are two main production mechanisms:
\begin{itemize}
	\item \textbf{Strong production (\textit{prompt})}\\
	The partons from $pp$ collision hadronize directly into \Lz particles (or some $\Lz^*$ state that swiftly decays into \Lz). Selecting high-$p_T$ kinematic regions, some polarization could be achieved. However, the small $x_F$ of reconstructible \Lz baryons at the LHC anticipates a (negligibly) small polarization (see Section~\ref{sec:initialpolarization}), confirmed by measurements~\cite{ATLAS:2014ona}. Moreover, the  $\Lprp$ decay is the only element of the event and the discrimination of real-$\Lz$ backgrounds, coming from weak decays or material interactions, is extremely challenging, especially with the reduced resolution and trigger efficiency of long-lived \Lz particles.
	
	\item \textbf{Production from weak decays (\textit{secondary})}\\
	The \Lz particles are produced in weak decays of heavier hadrons $H \to \Lambda X$ with longitudinal polarization. The mother particle $H$ lives long enough to generate a displaced vertex which can be reconstructed with the (charged) tracks in $X$, improving the trigger efficiency. Moreover, by fitting simultaneously the complete-event kinematics with the \textit{Decay Tree Fitter} algorithm~\cite{Hulsbergen:2005pu} the angular and invariant-mass resolutions can be improved. 
\end{itemize}

While the total amount of $\Lz$ particles is higher from strong production, the benefits of weak decays in terms of resolution and trigger overcompensate this difference. Moreover, the \Lz polarization in weak decays can be much larger than in strong production. For instance, in the process $\LctoLpi$ ($J^P = \frac{1}{2} ^+ \to \frac{1}{2} ^+ 0^-$), the \Lz polarization can be identified with the decay-asymmetry parameter~\cite{LINK05} as $\bm s_0 = (0,0,\alpha_{\Lambda_c})$, where $\alpha_{\Lambda_c} = 0.91 \pm 0.15$ \cite{PDG}. Thus, we will only consider the second case.

However, decay-asymmetry parameters and $\lz$ polarization have been measured only in few processes. Moreover, the most abundant $\lz$ production channels include three- and four-body decay modes with complex dynamics making any estimation of the initial \Lz polarization very difficult. Thus, there is not enough information available to propose specific channels based on polarization criteria and $\bm s_0$ will have to be measured directly for each channel as a previous step\footnote{
	In Part II of this thesis, we will analyse the \lz polarization from one of the most abundant channels using real LHCb data.}. Thus, in the following section, we will identify the potential channels based only on the production rate.

\subsubsection{Channels and efficiencies}

To identify the most abundant \Lz production channels from heavier baryons, we consider decays containing
only charged particles in the final state which branching ratios have measured values in the PDG (neither upper limits nor marked as \textit{seen}). 
The mother particle of the \Lz may be produced in the $pp$ collision or come from yet another weak decay of an even heavier baryon $H'$, with the complete decay chain $H'\to H X', H\to \Lz X, \Lz\to\pr\pim$. With these criteria, the direct \Lz mother particles can be \Lb, \Lc, \Xicz, \Xicp, \Xim, \Xiz, and $\mathit{\Omega}^-$. The grandmother particles can be those same baryons, plus \Xibm and  $\mathit{\Omega}_b^-$. The exhaustive list of channels is reported in Appendix~\ref{app:channelsLambda}, while the dominant ones are included in Table~\ref{tab:LambdaChannels}. In this reduced table, we have also required at least one charged track originating from the \Lz production vertex. 

The number of \Lz particles produced can be estimated as 
\begin{equation}
N_\Lz=2 \mathcal{L} \sigma_\qqbar f(\quark \to \PH)\br(\PH \to \Lz X')\br(\Lz\to\pr\pim)\br(X'\to\mathrm{charged}) , 
\end{equation}
where $\mathcal{L}$ is the total integrated luminosity, $\sigma_\qqbar$ ($\quark=\cquark,\bquark$) are the heavy-quark 
production cross sections from \pr\pr collisions at $\sqrt{s}=13$\tev~\cite{Aaij:2015bpa,FONLLWEB,Aaij:2010gn,Aaij:2015rla},
and $f$ is the fragmentation fraction into the heavy 
baryon \PH~\cite{Lisovyi:2015uqa,Gladilin:2014tba,Amhis:2014hma,Galanti:2015pqa}.
All branching fractions \br are taken from Ref.~\cite{Olive:2016xmw} and, if
they are given relative to other decay modes, we assume that the sum of branching ratios of all listed decays adds up to one.
In Table~\ref{tab:LambdaChannels} the dominant production channels and the estimated
yields are summarised.
Overall, there are about $1.5\times 10^{11}$ \Lz baryons per \invfb produced directly from heavy baryon decays 
(referred hereafter as short-lived, or SL events), 
and $3.8\times 10^{11}$ from charm baryons decaying through an intermediate \Xim particle (long-lived, or LL events).
The yield of \Lz baryons experimentally available can then be evaluated as
$N_\Lz^{\rm reco} = \epsilon_{\rm geo} \epsilon_{\rm trigger} \epsilon_{\rm reco} N_\Lz$,
where $\epsilon_{\rm geo}$, $\epsilon_{\rm trigger}$ and $\epsilon_{\rm reco}$ are
the geometric, trigger and reconstruction efficiencies of the detector system.

\begin{table}[htb]
	\centering
	\caption{Dominant \Lz production mechanisms from heavy baryon decays and estimated yields
		produced per \invfb at $\sqrt{s}=13$\tev,
		shown separately for SL and LL topologies.
		The \Lz baryons from \Xim decays, produced promptly in 
		the \pr\pr collisions, are given in terms of the unmeasured production cross section. 
	}
	\label{tab:LambdaChannels}
	\renewcommand{\arraystretch}{1.1}
	\resizebox{1\textwidth}{!}{\begin{tabular}{lc lc}
			\hline \hline
			SL events &  $N_{\Lz}/\invfb~(\times 10^{10})$  & LL events, $\Xim\to\Lz\pim$ &  $N_{\Lz}/\invfb~(\times 10^{10})$  \\ 
			\hline
			$\Xicz\to\Lz\Km\pip$       & 7.7 & $\Xicz\to\Xim\pip\pip\pim$ & 23.6 \\
			$\Lc\to\Lz\pip\pip\pim$    & 3.3 & $\Xicz\to\Xim\pip$         & 7.1 \\
			$\Xicp\to\Lz\Km\pip\pip$   & 2.0 & $\Xicp\to\Xim\pip\pip$     & 6.1 \\
			$\Lc\to\Lz\pip$            & 1.3 & $\Lc\to\Xim\Kp\pip$        & 0.6 \\
			$\Xicz\to\Lz\Kp\Km$ (no $\phi$)  & 0.2 & $\Xicz\to\Xim\Kp$              & 0.2 \\
			$\Xicz\to\Lz\phi(\Kp\Km)$  & 0.1 & Prompt $\Xim$              & $0.13\times\sigma_{\pr\pr\to\Xim}~[\mu \rm b]$ \\
			\hline \hline
		\end{tabular}
	}
\end{table}

\begin{figure}[htb]
	\centering
	{ \includegraphics[width=0.48\linewidth]{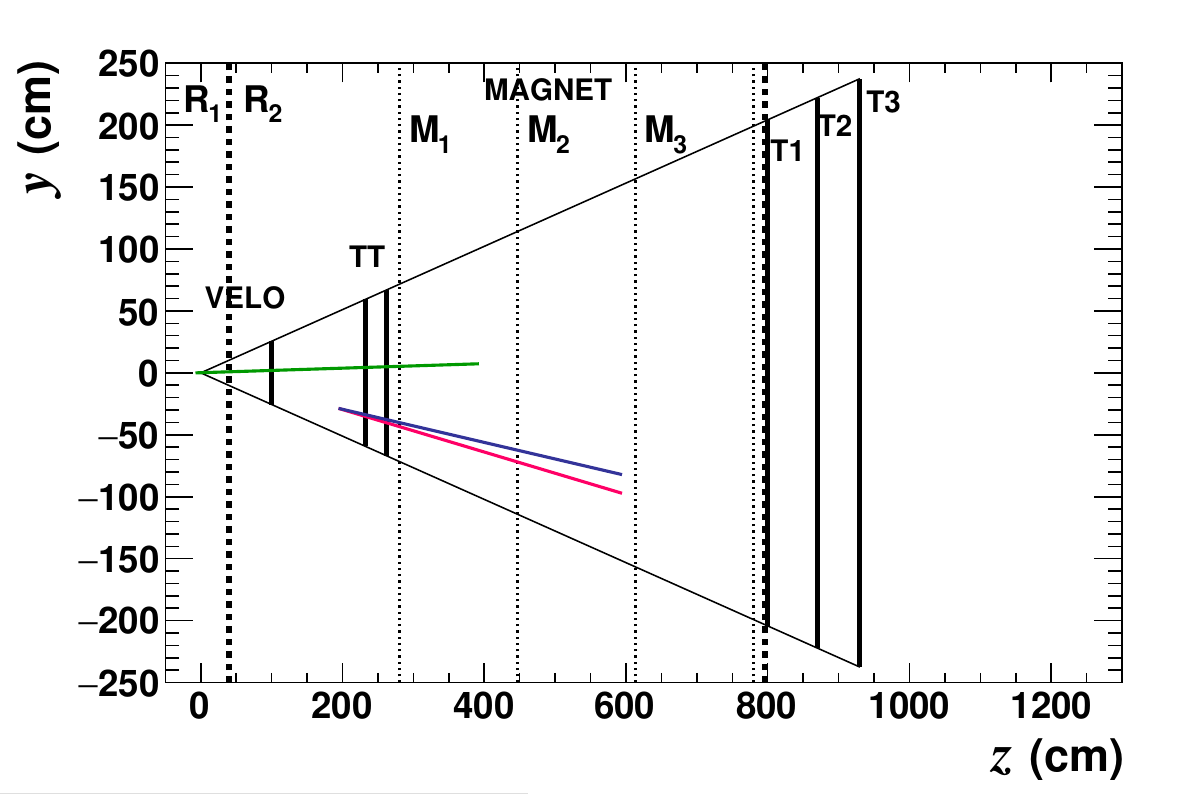} }
	{ \includegraphics[width=0.48\linewidth]{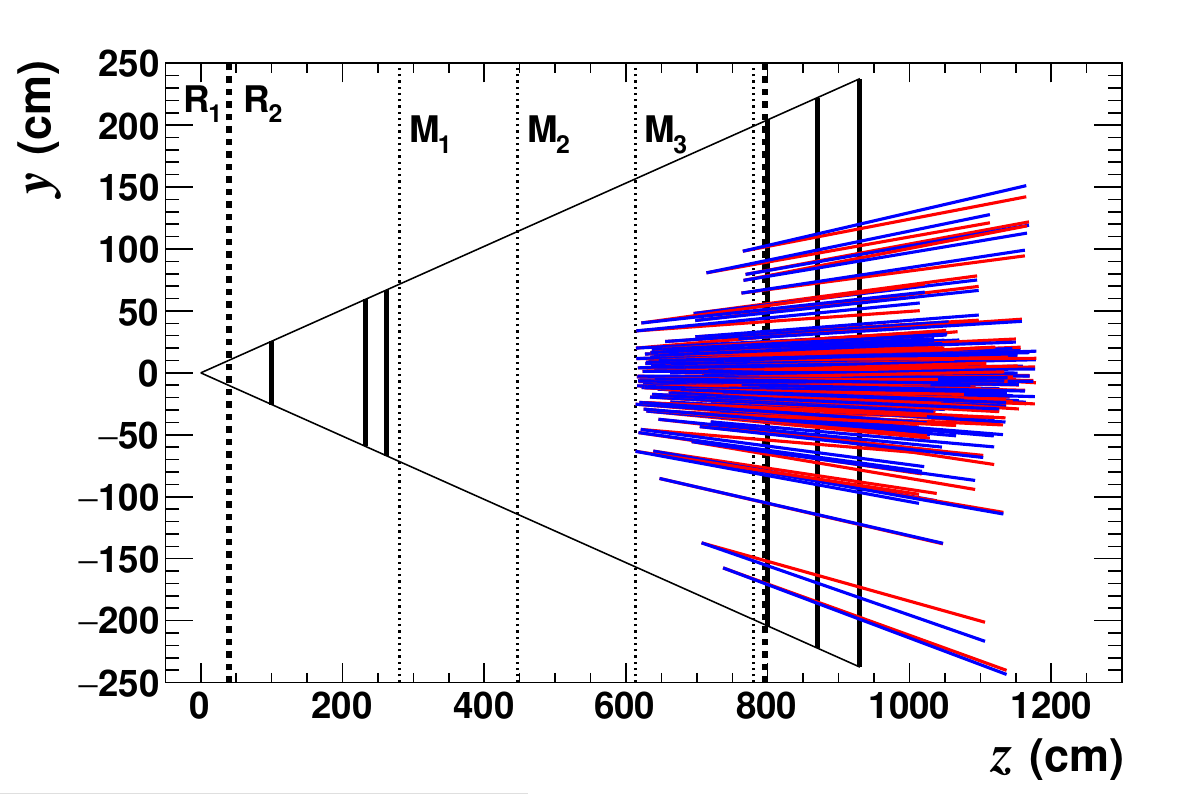} }
	\caption{(Left) Sketch of the simplified geometry of the LHCb tracking system in the laboratory $yz$ plane.
		The oblique lines represent the angular acceptance. 
		The tracking layers and the limits
		of the R$_1$ and R$_2$ regions are shown as solid and dotted thick lines, respectively.
		The magnet is divided in three regions by thin dotted lines. 
		A simulated $\Lc \to\Lz (p \pim)\pip$ decay with corresponding $\pip$ (green), $\pim$ (blue) and $p$ (red) tracks is overlaid. 
		(Right) decay products from \Lz baryons decaying in the
		last region of the magnet, M$_3$. 
	}
	\label{fig:detectorDiagram}
\end{figure}

The geometric efficiency for SL topology has been estimated using a 
Monte Carlo simulation of $\pr\pr$ collisions at $\sqrt{s}=13$\tev and the decay of heavy hadrons, 
using \pythia~\cite{Sjostrand:2006za} and \evtgen~\cite{Lange:2001uf} standalone\footnote{Only generator-level samples were used in these (public) results~\cite{Botella:2016ksl}, without simulations of the reconstruction within the LHCb framework.} toolkits, together with the simplified geometrical model of the LHCb detector introduced in Figure~\ref{fig:geometricalModel}. However, in this (previous) version of the model, the height and width of the tracking stations (VELO, TT and T-stations) were computed according to the detector angular acceptance of 250 (300 \mrad) in the vertical (horizontal) direction, and the beam pipe was assumed to be a cone of 10 \mrad aperture starting in the collision point at $(0,0,0)$. This layout is illustrated in Figure~\ref{fig:detectorDiagram}.

Table~\ref{tab:LambdaGeoEfficiencies} summarizes the geometric efficiencies for \Lz baryons decaying in different regions of the
detector volume, for three different SL topologies. Region R$_1$ is defined such that the $z$ position of the \Lz decay vertex 
is in the range [0-40]\cm from the collision point
and the decay products are within the detector acceptance.
Events in the R$_2$ region have a \Lz decay $z$ position in the range [40-800]\cm.
Charged particles produced together with the \Lz baryon are required to be within the
VELO and T1-T3 stations, or the VELO and TT acceptances, to ensure a precise reconstruction of
the \Lz origin vertex.
Events in the R$_1$ region provide the measurement of the initial \Lz polarization vector;
events in the R$_2$ region allow determining the polarization as a function of the \Lz decay
length in the magnetic field region. Among the latter, \Lz baryons decaying towards the end
of the magnet (M$_3$ region in Table~\ref{tab:LambdaGeoEfficiencies}) provide most of
the sensitivity to the EDM and MDM.
These events are sketched in Figure~\ref{fig:detectorDiagram} (right).
The total geometric efficiency for R$_1$ and R$_2$ regions is about 16\%, with small differences among SL topologies, 
and about $2.4\times 10^{10}$ \Lz baryons per \invfb can be
reconstructed.

\begin{table}[htb]
	\centering
	\caption{Geometric efficiencies (in \%) for \Lz baryons decaying in different regions
		of the \lhcb detector, for the three most abundant channels, simulated at $\sqrt{s}=13$\tev.}
	\label{tab:LambdaGeoEfficiencies}
	\renewcommand{\arraystretch}{1.1}
	\resizebox{1\textwidth}{!}{\begin{tabular}{l ccccc}
			\hline \hline
			Region  &  R$_1$  & R$_2$ &  M$_1$ & M$_2$ & M$_3$ \\ 
			\Lz decay vertex $z$ position (cm)   &  [0-40] & [40-800] & [280-450] & [450-610] &[610-780]  \\  
			\hline
			$\Lc\to\Lz\pip\pip\pim$    & 4.7 & 10.5 & 1.3 & 0.7 & 0.3  \\
			$\Xicz\to\Lz\Km\pip$       & 5.2 & 12.2 & 1.7 & 1.0 & 0.6  \\
			$\Xicp\to\Lz\Km\pip\pip$   & 5.3 & 11.9 & 1.6 & 0.9 & 0.4  \\
			\hline  \hline
		\end{tabular}
	}
\end{table}

\subsubsection{Spin precession}

The precession of the spin-polarization vector is governed by the same equation of motion presented in Eq.~\eqref{eq:TBMTgeneral}. This time it becomes substantially simpler as \Evec=0 and $q=0$,

\begin{equation}\label{TBMTLHCb}
\frac{d \bm s}{ d t} = \bm s \times \bm \Omega ~, 
\end{equation}
\begin{equation}
\bm \Omega
= \frac{\mu_N}{\hbar} \left[ g \left( \bm B -\frac{\gamma - 1}{\gamma}(\bm u \cdot \bm B)\bm u   \right)  + d \beta \bm u \times \bm B  \right] .
\end{equation}

We can solve this differential equation by considering the average magnetic field along the \Lz flight path, and thus ignoring B-field gradient effects (shown to be negligible in the Appendix A.1.1 of the original Ref.~\cite{Botella:2016ksl}), 
\begin{equation}\label{sAnalytical}
\bm s (t)
= (\bm s_0  \cdot \bm \omega) \bm \omega 
+ \left[ \bm s_0 - (\bm s_0 \cdot \bm \omega) \bm \omega \right] \cos(\Omega t) 
+ (\bm s_0 \times \bm \omega) \sin (\Omega t) ~~, 
\end{equation}
where
\begin{equation*}
~\Omega = |\bm \Omega | ~~, ~~ \bm \omega = \bm \Omega / \Omega  ~~.
\end{equation*}

For the particular case of \Lz and \PH baryons flying along the $z$ axis, 
initial longitudinal polarization $\mathbf s_0=(0,0,s_0)$\footnote{The definition of the \Lz rest-frame coordinates and their relation to the laboratory-frame magnetic field is provided in the original Ref.~\cite{Botella:2016ksl}. },
and $\mathbf B = (0,B_y,0)$, 
the solution is
\begin{equation}
\label{eq:sSimpleCase}
\mathbf s ~=~
\left\lbrace
\begin{array}{l}
s_x = - s_{0} \sin\Phi  \\
s_y = - s_{0} \dfrac{d \beta }{g} \sin\Phi \\
s_z =   s_{0} \cos\Phi \\
\end{array}
\right. 
\text{,~~~where~} {\Phi = \Omega t =\frac{D_y\mu_B}{\beta \hbar c} \sqrt{d^2 \beta^2 + g^2}
	~~ \approx ~~ \frac{g D_y \mu_B}{\beta \hbar c} }~,
\end{equation}
with $D_y = \int_0^l B_y dl'$ the integrated magnetic field
along the \Lz flight path.

As it can be seen, the polarization vector precesses in the $xz$ plane, normal to the magnetic field, with angular velocity $\Omega$ proportional to the magnetic moment of the particle, $g$. It is interesting to note that, for charged particles, the angular velocity is proportional to $(g-2)/2$ (not just $g$), as shown \textit{e.g.} in Eq.~\eqref{eq:precessionsimplified}.
The presence of a non-zero EDM $d$ would introduce a non-zero, periodical change on the $s_y$ component. At LHCb $D_y \approx \pm 4~\mathrm{T m}$~\cite{LHCb-DP-2014-002} and
the maximum precession angle for particles traversing the entire magnetic field region yields $\Phi_{\rm max} \approx \pm \pi/4$, 
and allows to achieve about 70\% of the maximum $s_y$ component.

\subsubsection{Sensitivity}

\begin{figure}[htb]
	\centering
	{ \includegraphics[width=0.48\linewidth]{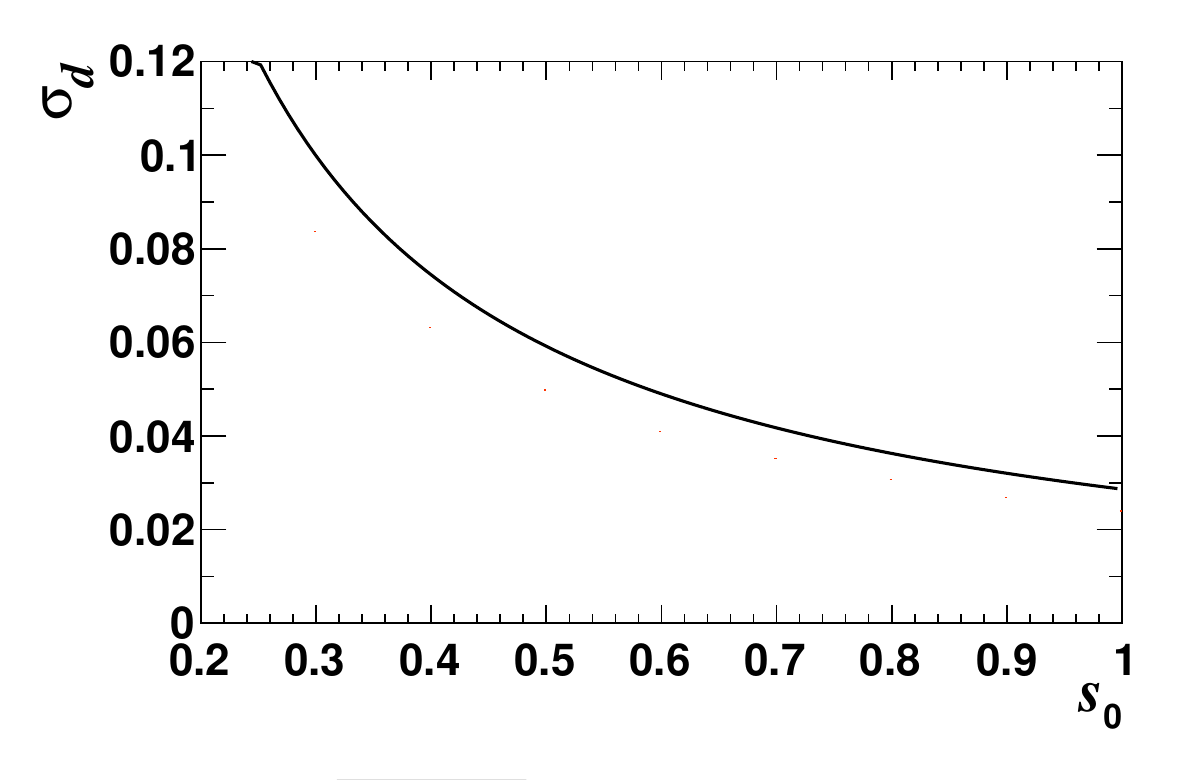} }
	{ \includegraphics[width=0.48\linewidth]{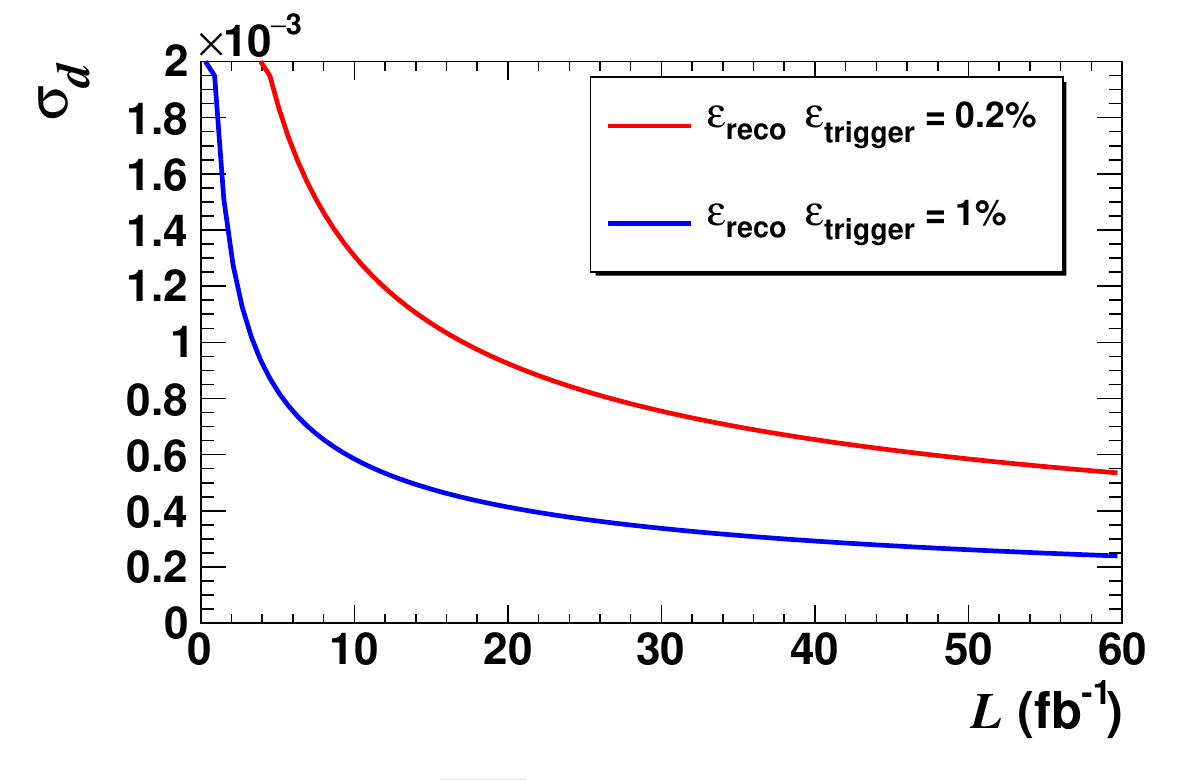} }
	\caption{(Left) Dependence of the $d$ uncertainty with the initial polarization for $N_\Lz^{\rm reco}=10^6$ events, and (right) as a function 
		of the integrated luminosity
		assuming reconstruction efficiency of  0.2\% and 1\%.}
	\label{fig:Lambda_sensitivity}
\end{figure}

To assess the EDM sensitivity, pseudoexperiments have been generated using the \lhcb magnetic field mapping~\cite{LHCb-DP-2014-002,Hicheur:2007jfk} (in Figure~\ref{fig:bfield}) to obtain the integrated magnetic field along the \Lz path. These events were previously selected to be in the detector acceptance.
The decay angular distribution and spin dynamics have been simulated 
using Eqs.~\eqref{eq:AngDist} and \eqref{sAnalytical} as a function of the \Lz flight length\footnote{The precession angle in the bent-crystal experiment was only dependent on the $\gamma$ boost factor since all signal particles go through the full crystal channel. For \Lz particles precessing in the LHCb magnetic field, the precession depends on the $\gamma$ factor as well as on the integrated magnetic field between production and decay points. }.
%
For this study initial polarization vector $\mathbf s_0 = (0,0,s_0)$, 
with $s_0$ varying between 20\% and 100\%, and factors $g=-1.458$~\cite{Olive:2016xmw} and $d=0$,
were used. Each generated sample was adjusted using an unbinned maximum likelihood fitting method with $d$, 
$g$ and $\mathbf s_0$ (or $\alpha\mathbf s_0$) as free parameters. The $d$-factor uncertainty 
scales with the number of events $N_\Lz^{\rm reco}$ and the initial longitudinal polarization $s_0$ as
$\sigma_d \propto 1/(s_0 \sqrt{N_\Lz^{\rm reco}} )$. 
The sensitivity almost saturates at large values of $s_0$, as shown in Figure~\ref{fig:Lambda_sensitivity} (left),
and it partially relaxes 
the requirements on the initial polarization.
Similarly, Figure~\ref{fig:Lambda_sensitivity} (right) shows the expected sensitivity on the EDM as a function 
of the integrated luminosity, summing together SL and LL events, assuming global trigger
and reconstruction efficiency $\epsilon_{\rm trigger} \epsilon_{\rm reco}$ 
of 1\% (estimated for the upgrade detector~\cite{LHCb-TDR-016,LHCb-TDR-015}) 
and 0.2\% (previous detector~\cite{LHCb-DP-2014-002}), where the efficiency estimates are based on a educated guess.
An equivalent sensitivity is obtained for the gyromagnetic factor.
\revafter{\textbf{CHANGE UNITS:}}Therefore, with 8~\invfb a sensitivity $\sigma_d \approx 1.5\times 10^{-3}$ could be achieved (previous detector), 
to be compared to the present limit, $1.7\times 10^{-2}$~\cite{Pondrom:1981gu}. 
With 50~\invfb (upgraded detector) the sensitivity on the gyroelectric factor can reach $\approx 3\times 10^{-4}$.

The reconstruction of long-lived \Lz baryons decaying inside and after the magnet
represents a challenge for the \lhcb experiment,
introducing significant backgrounds and a limited resolution on the measurement of the 
\Lz momentum and decay point. Including resolution effects in the pseudo-experiments by smearing the helicity angles and \Lz decay point, and fitting the events to the signal PDF convoluted with gaussian-resolution functions, we observe that the EDM and MDM uncertainties do not degrade significantly even for poor resolutions. More detailed studies including the resolution effects, updated efficiency estimates and more sophisticated fitting procedures that optimize the sensitivity are in progress.

After the publication of these exploratory studies~\cite{Botella:2016ksl} the reconstruction of long-lived \Lz particles decaying in the magnetic field region has been thoroughly developed and studied using the full LHCb simulation and data framework. Some of the encountered challenges are briefly presented in the next section.

\section{Reconstruction with T tracks} \label{sec:ttracks}

\begin{flushright}
	Part of this chapter is based on Ref.~\cite{DEMONSTRATOR}
\end{flushright}

To demonstrate the feasibility of reconstructing long-lived particles with T tracks several studies have been developed mainly in the LHCb groups of Milano and Valencia. The first results with real data~\cite{DEMONSTRATOR} have focused on the reconstruction of \Lz and \KS in the exclusive decays \LbToJpsiLz and \BdToJpsiKS\footnote{In this section, we will focus the discussion on \LbToJpsiLz decays, especially treating the vertex reconstruction of $\Lprp$.}, respectively, profiting from the dimuon trigger that records all events with $\jpsi\to \mup \mu^-$ decays. While reconstructing the T tracks themselves is possible with the default reconstruction sequence of LHCb, outlined in Refs.~\cite{Alves:2008zz,LHCb-DP-2014-002}, the extrapolation of their trajectory across the intense and non-uniform magnetic field and, notably, the construction of vertices only with T track pairs required non-trivial adaptations of the reconstruction algorithms.


The track transport for Long and Downstream tracks uses a cubic approximation of the full equation of motion for a charged particle in a magnetic field. For short distances and small magnetic fields, the results are practically indistinguishable from the full equation, albeit computationally much faster. However, T tracks are detected in the T1-T3 stations (approximately at 8 m from the $pp$ collision point) and need to be extrapolated in the region where the intensity of the magnetic field is at its peak (3-8 m). The polynomial approximation yields very low reconstruction efficiencies and strong biases on the vertex position. Thus, the first step was to change the particle transporter to use the full equation of motion. Fortunately, a tool based on the Runge-Kutta method for differential equations~\cite{Hairer1993} was already available within the LHCb reconstruction software~\cite{LHCb-2007-140}.
Next, due to the small magnetic field in between the T1-T3 stations, we find a relatively low momentum resolution on the proton and pion, of around 20-30\% as compared to $\lesssim1\%$ with Long tracks. By refitting the whole decay chain simultaneously, including the Long tracks of the $\jpsi\to\mup\mu^-$ decay, and constraining the \Lz and \jpsi masses to their PDG values and the origin of the \Lb/\Bd at the associated PV using the \textit{Decay Tree Fitter} algorithm~\cite{Hulsbergen:2005pu}\footnote{The performance of Decay Tree Fitter is studied in some detail for \threepi decays later in Section~\ref{sec:DTF}.}, a significant improvement of around a factor $2.5$ is found on the momentum resolution.

\begin{figure}
	\centering
	\includegraphics[width=0.45\linewidth]{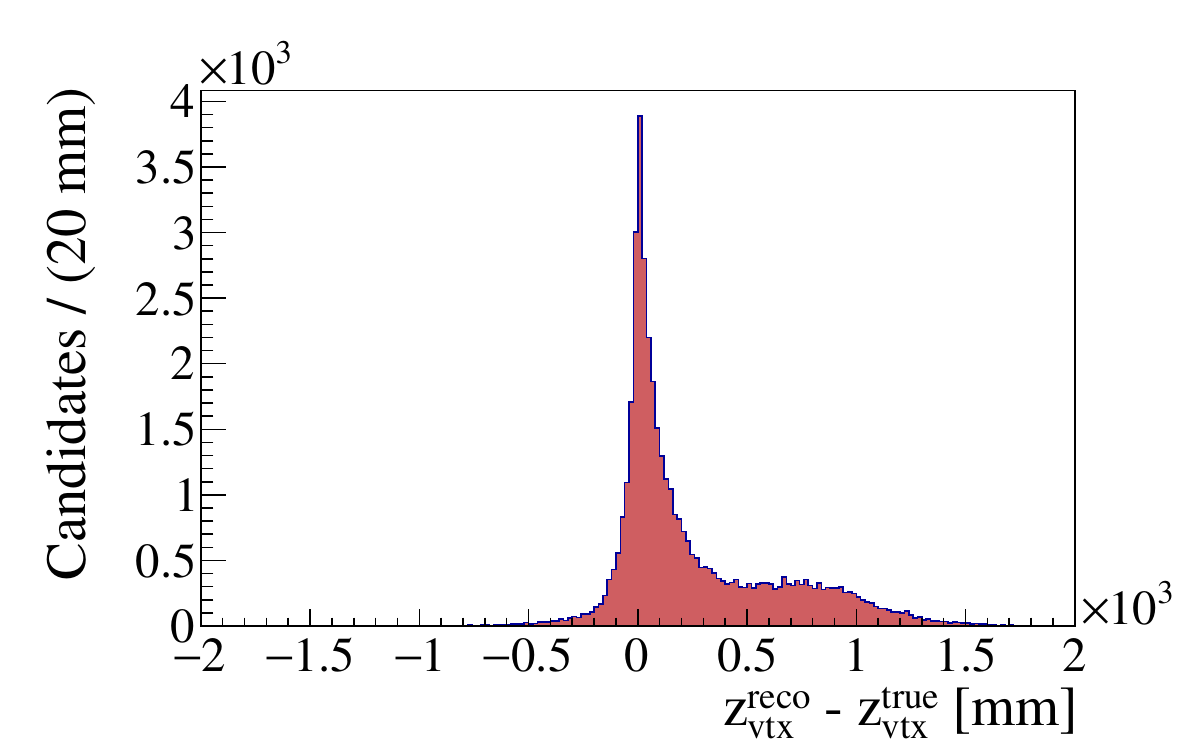}
	\includegraphics[width=0.45\linewidth]{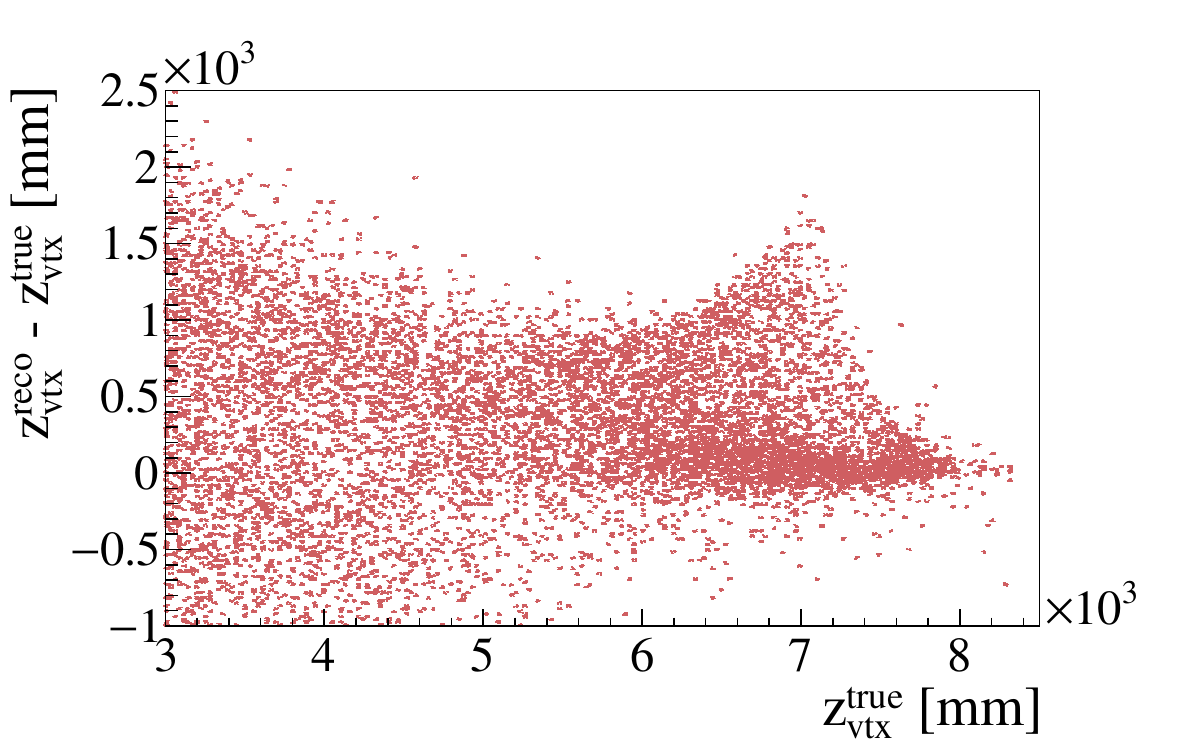}
	\caption{(Left) residuals of the reconstructed vertex $z$ position and (right) dependence of these residuals with respect to the true $z$ vertex. The biased structure approximately follows the shape of the magnetic field. \revafter{Test printing and prescale data set in scatter plot if needed}}
	\label{fig:zbias}
\end{figure}

However, after these substantial improvements in the reconstruction, some of the reconstructed candidates, in the MC samples, show a bias on the reconstructed vertex position, which is displaced by about $0.5-1\,\m$ in the forward (downstream) direction with respect to the true vertex. This bias shows a dependence on the vertex position itself, having the minimum displacement at the centre of the magnet, and much larger residuals where the magnetic field decreases, as shown in Figure~\ref{fig:zbias}.

\begin{figure}
	\centering
	\includegraphics[width=0.45\linewidth]{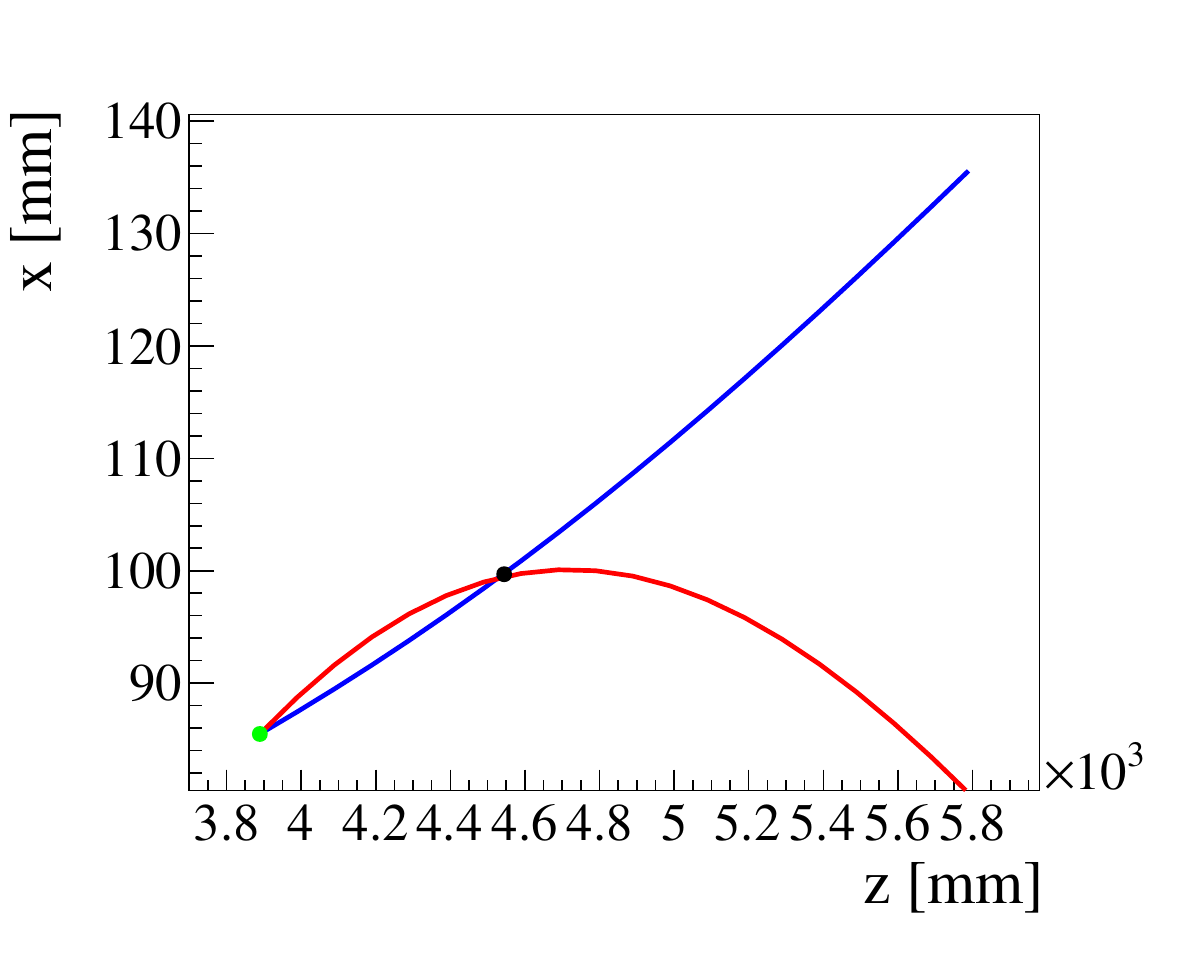}
	\includegraphics[width=0.45\linewidth]{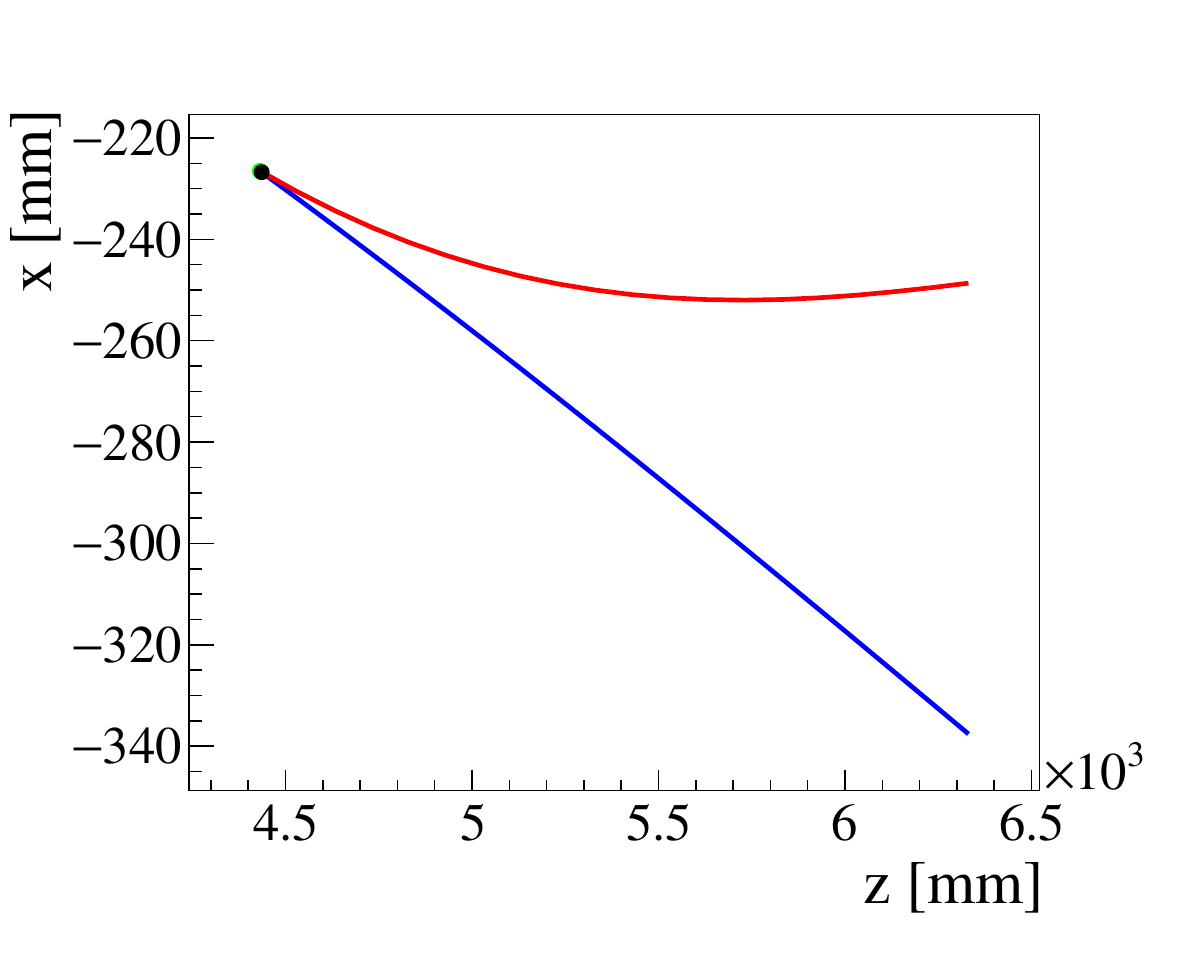}
	\caption{Event displays with (blue) proton and (red) pion bent trajectories together with the (green dot) true vertex and (black dot) reconstructed vertex for typical events from the simulation. The reconstructed vertex is wrongly assigned to the crossing point between particle trajectories in most of the events with closing-track topologies (left), while it is found correctly in opening-track topologies (right). 
	}
	\label{fig:eventdisplays}
\end{figure}

\begin{figure}
	\centering
	\includegraphics[width=0.45\linewidth]{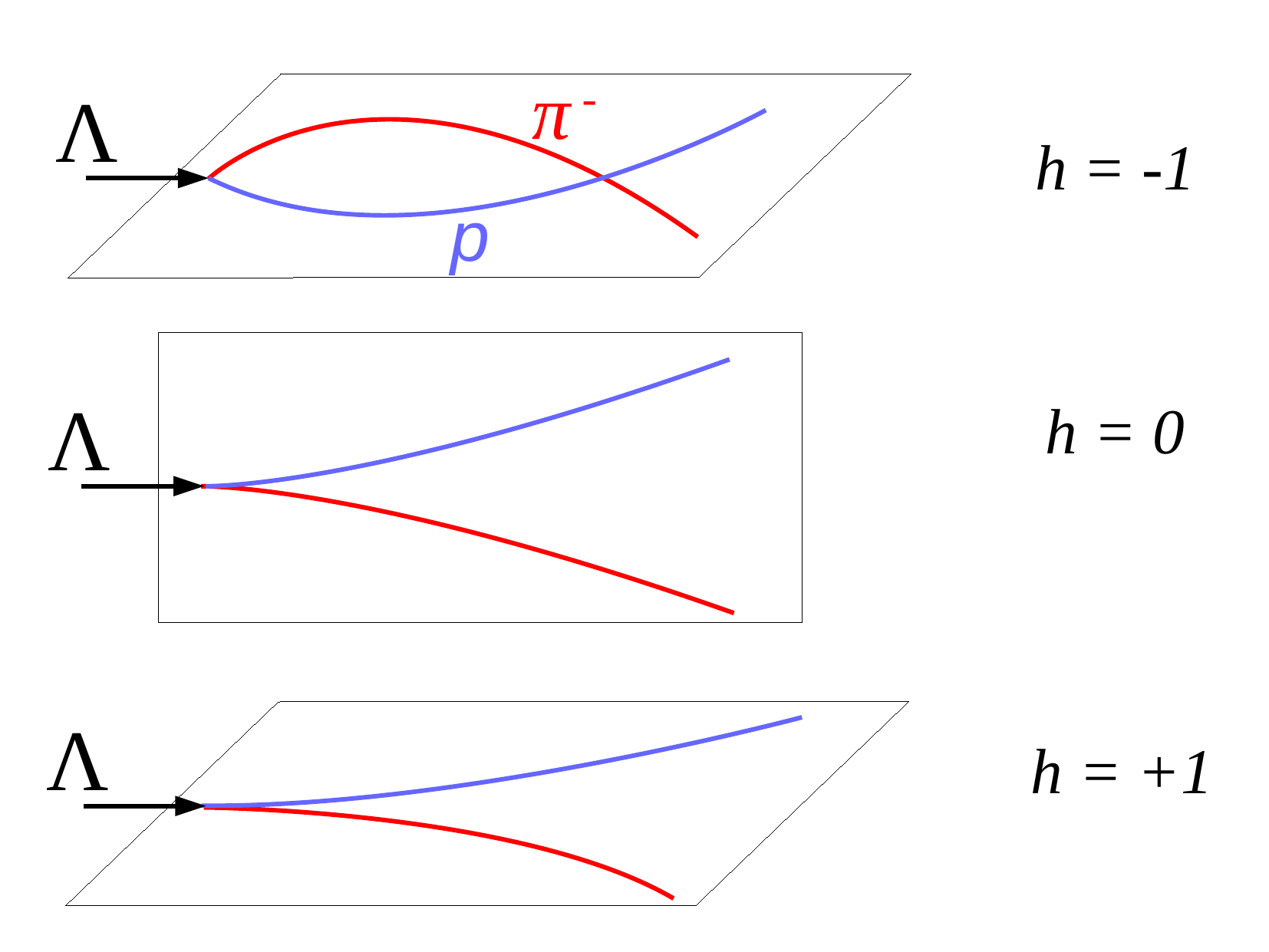}
	\raisebox{0.6cm}{\includegraphics[width=0.45\linewidth]{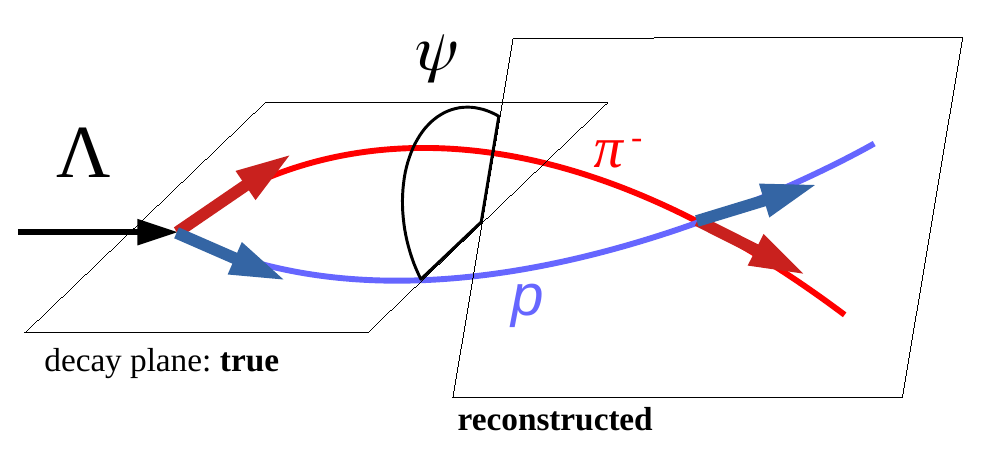}}
	\caption{(Left) schematic representation of events with limiting values of $h$ and (right) definition of $\psi$ as the angle between the true and reconstructed decay planes.
	}
	\label{fig:horizontality}
\end{figure}

Different hypotheses were considered and tested on these events. We tried to systematically study the correlations between the residuals of the $z$ vertex position and other event variables, without any clear interpretation of the results. Eventually, a new hypothesis was considered simply based on the shape of the affected events: when the $p$ and \pim trajectories are curved towards each other, the vertex may be assigned to the crossing point of the tracks, which is always downstream from the true vertex. By plotting the trajectories of \pr and \pim together with the true and reconstructed vertex positions, in Figure~\ref{fig:eventdisplays}, the results unequivocally support this view. In these event displays, the trajectories are calculated using the inhomogeneous magnetic field map in Figure~\ref{fig:bfield} and taking the initial position and momentum of the tracks at the true vertex (green dot). 

To further analyse these events we need to separate them from the rest of the sample. However, using the event display of each event is impractical and we need new variables that encapsulate the relevant information on the event topology. 
%
First, we would like to determine if the two tracks tend to approach each other (closing tracks) or move further away (opening tracks), as in Figure~\ref{fig:eventdisplays} (left) and (right), respectively. However, this absolute distinction can only be done for events embedded in the horizontal (bending) plane. To account for the continuum of possible track geometries also with $p_y\neq 0$, we define the \textit{horizontality}, illustrated in Figure~\ref{fig:horizontality}, which takes the extreme values $h=-1$ ($h=+1$) for completely closing- (opening-)track geometries, both in the horizontal plane, and $h=0$ for initial momenta in the vertical plane. The variable $h$ is defined as the $y$ component of the normalized vector  $\bm{p}_p \times \bm{p}_\pim$, normal to the decay plane, times the proton/antiproton charge and the dipole magnet polarity. 

\begin{figure}
	\centering
	\includegraphics[width=0.45\linewidth]{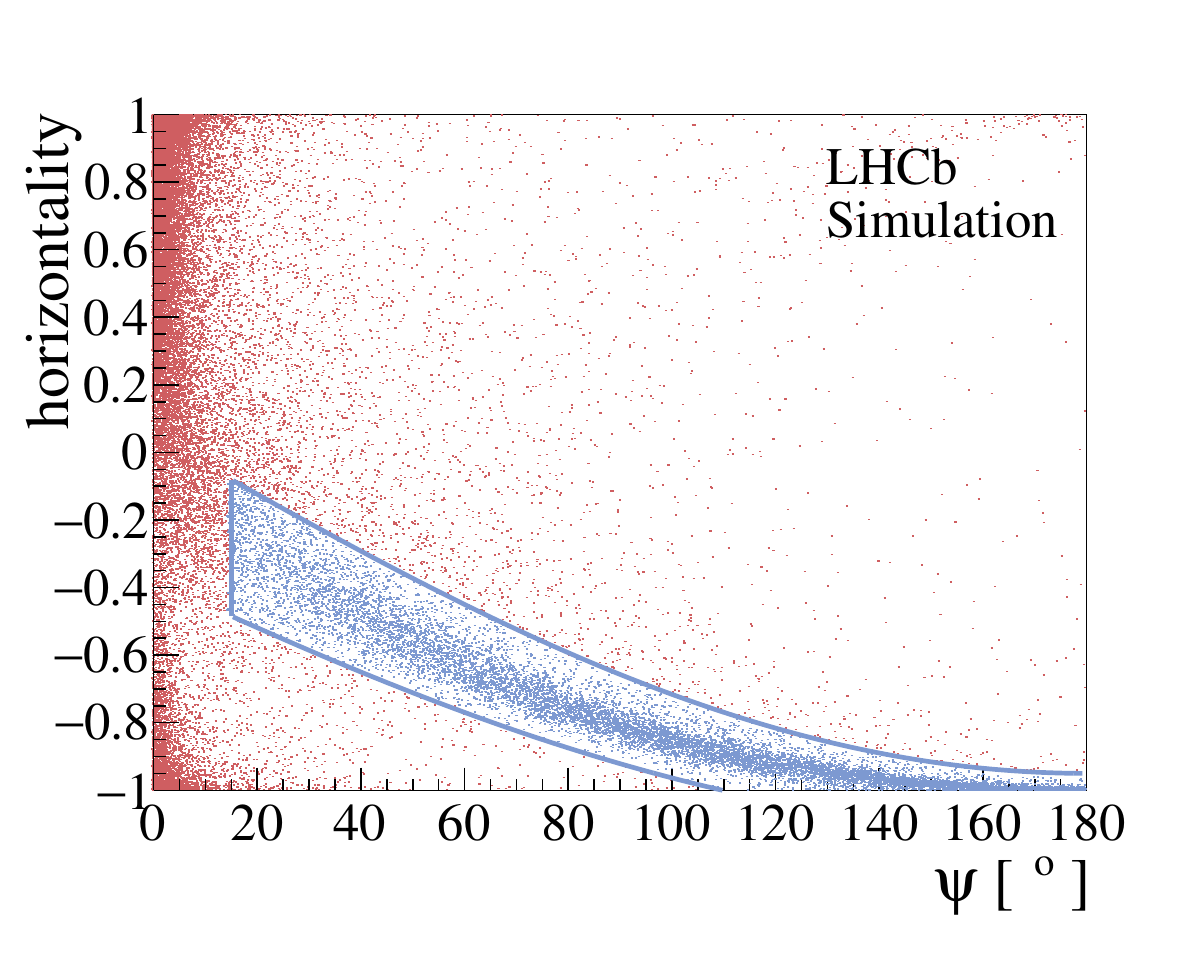}
	\includegraphics[width=0.45\linewidth]{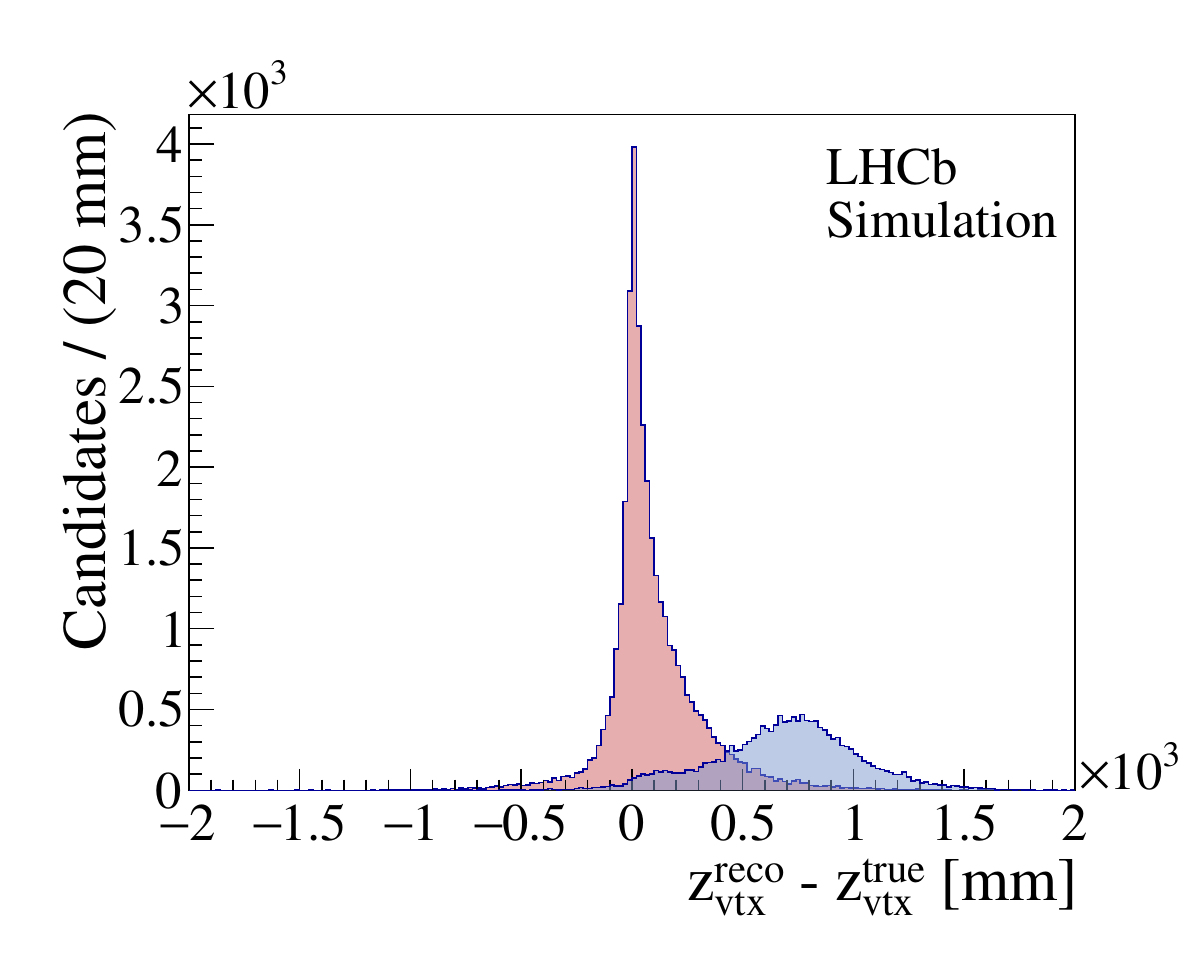}
	\caption{(Left) distribution of true horizontality and $\psi$ for signal \mbox{\LbToJpsiLz} simulated events and (right) residual of the reconstructed vertex $z$ position.
		A strong correlation between $h$ and $\psi$ arises for events with closing-track topologies, $h<0$, for which the \Lz decay vertex is wrongly assigned to the crossing point. By selecting events between the blue lines, amounting to about 30\%, the dominant source of the bias on the vertex $z$ position is clearly identified.
	}
	\label{fig:bananaplot}
\end{figure}

\begin{figure}
	\centering
	\includegraphics[width=0.31\linewidth]{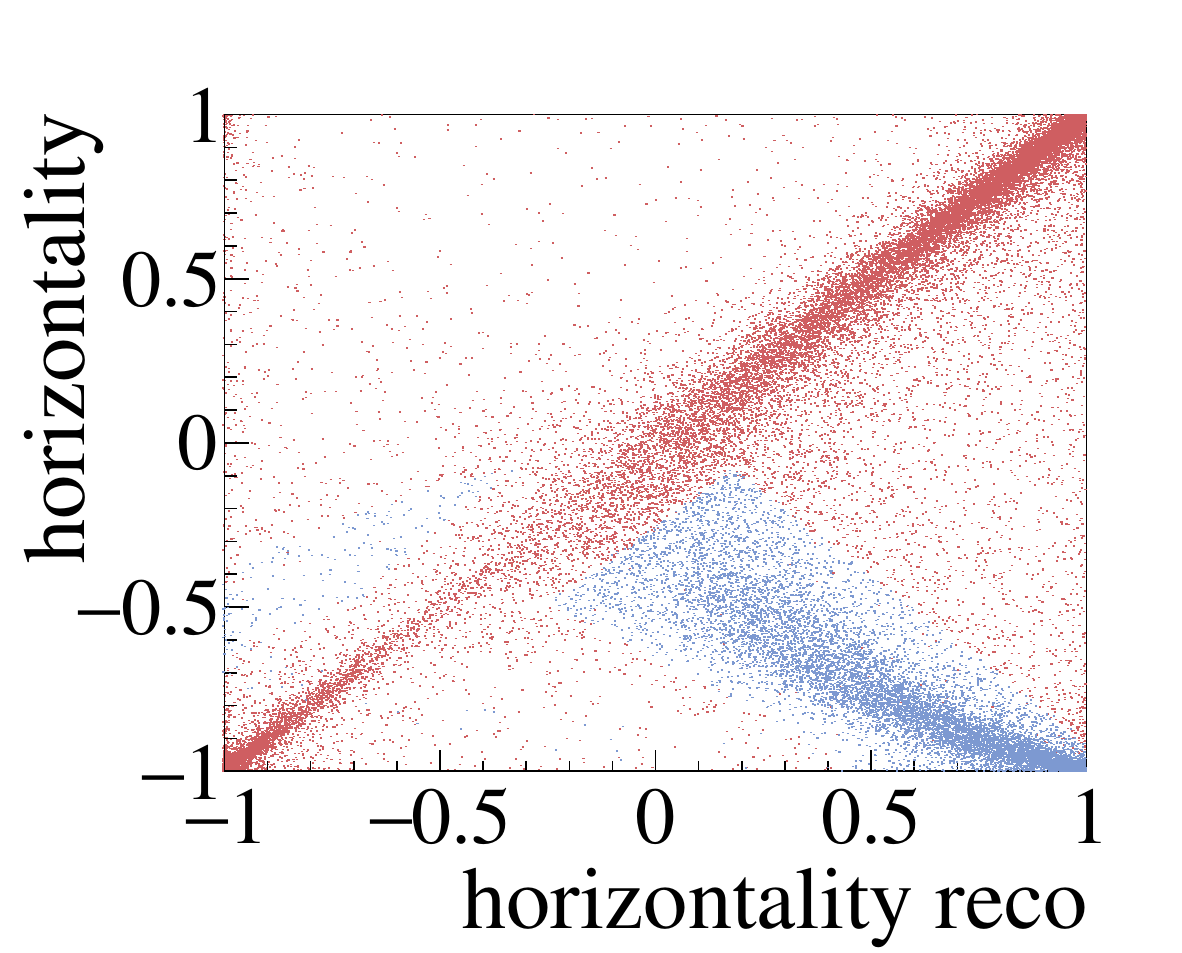}
	\includegraphics[width=0.31\linewidth]{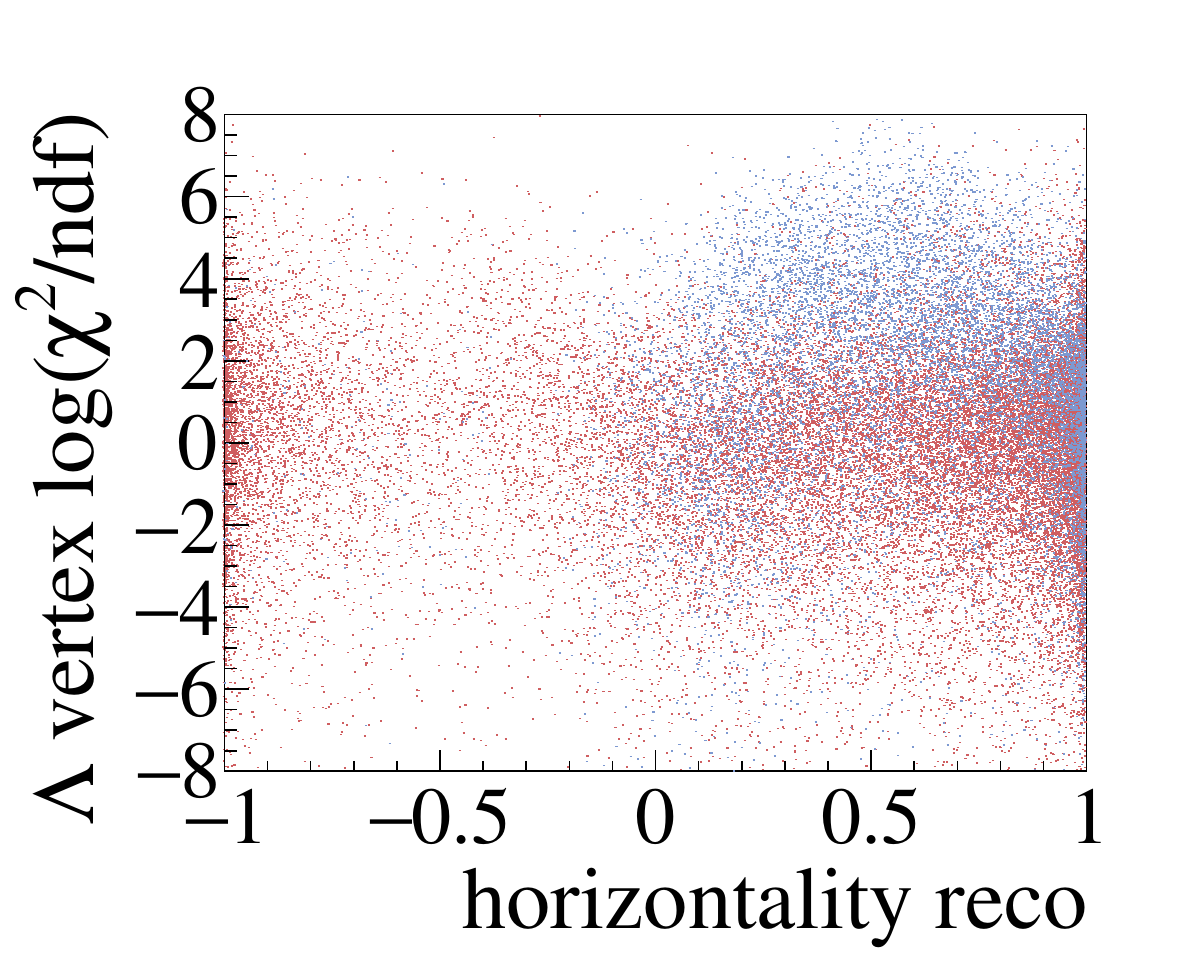}
	\includegraphics[width=0.31\linewidth]{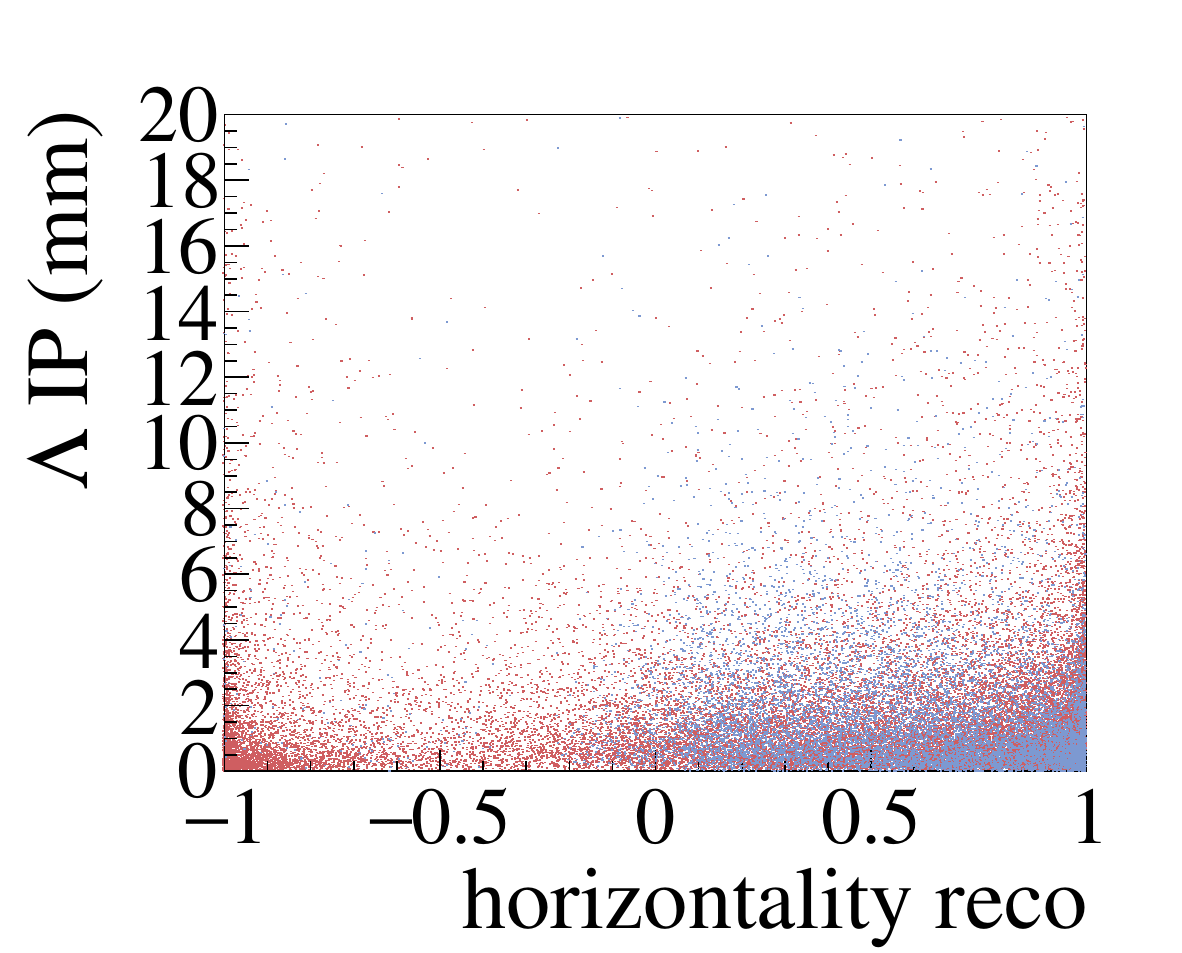}
	\caption{From left to right, correlation of the reconstructed horizontality ($x$ axes) with the true horizontality, the $\chi^2/{\rm ndf}$ of the \Lz vertex, and the \Lz impact parameter.
	}
	\label{fig:variablesDiscrVertex}
\end{figure}

Although having a closing-track geometry ($h<0$) is necessary to identify the problematic events, this condition is not sufficient. To know if the vertex was assigned to the crossing point, we can look at the relative direction of the \pr and \pim in the true and reconstructed vertex. For $h=-1$, the decay-plane vector $\bm{p}_p \times \bm{p}_\pim$ is reversed between the two vertices. Generalizing for the continuum of geometries, we can define the variable $\psi$ as the angle between the true and reconstructed decay planes, as shown in Figure~\ref{fig:horizontality}. Now, plotting the two-dimensional distribution of the new variables $h$ and $\psi$, in Figure~\ref{fig:bananaplot} (left), the biased events (about 30\%) stick out with a clear correlation in the $h<0$ region. 
This plot allows to select the biased events all at once to analyse their distributions in other variables. Particularly, the $z$ position residuals can be plotted with this selection, in Figure~\ref{fig:bananaplot} (right), confirming the closing-track geometry as the cause of the bias. 
The interpretation of the $h$ vs. $\psi$ plot tells us that the decay-plane orientation of true and reconstructed vertices are related. This relation also flips the sign of the reconstructed horizontality, in Figure~\ref{fig:variablesDiscrVertex} (left).
In these events with misreconstructed decay planes, the resolution on the helicity angles is degraded, improving by a factor $2.5$ when they are excluded~\cite{DEMONSTRATOR}.

Possible solutions are currently under study. In the best-case scenario, the vertexing algorithms would be improved to find the real vertex, possibly by forcing the reconstruction of two vertices in each event and differentiating them offline. In the worst-case scenario, the reconstruction cannot be improved or corrected offline and we have to simply remove the problematic events. This is in principle already possible by selecting $h_{\rm reco}<0$ but with a poor signal efficiency of 20\%. 
In either case, we need to identify new variables that can discriminate between real and closing-track vertices without using true (MC) information. In Figure~\ref{fig:variablesDiscrVertex} (middle) and (right) the distributions of some example variables against $h$ is shown. We did not find any variable showing significant discriminant power by itself. Possible lines of investigation include combining them in a multivariate classifier and keep searching for other variables related to the decay topology.

It is relevant to note that, if the T track momentum resolution was substantially improved the default vertexing algorithm would always find the true vertex and the crossing-track effect would be completely avoided. Moreover, for decays with a larger $Q$-value than $\Lzppi$, the increased aperture angle of the decay products also evades the crossing-track vertex even with the current momentum resolution. This feature is observed with $K_S^0 \to \pip \pim$ decays, from the process \BdToJpsiKS~\cite{DEMONSTRATOR}.
Custom vertexing algorithms exploiting the kinematical constraints of the decay chain and the momentum resolution provided by the RICH2 detector are currently being explored.

\pagebreak

\section{Opportunity for long-living particle searches} \label{sec:LLP}

The reconstruction of decay vertices with T tracks opens the possibility to search for undiscovered long-living particles (LLPs) that decay several meters away from the $pp$ interaction point at LHCb. On the theory side, LLPs are predicted in many NP scenarios\footnote{Comprehensive reviews of NP models predicting LLPs can be found in Refs.~\cite{Borsato:2021aum} and \cite{Aielli:2019ivi}, the latter shortly summarized in Ref.~\cite{Aielli:2022awh}.} and the phenomenological constraints on their couplings and masses are often dominated by direct LLP searches, meaning that they could be observed in forthcoming experiments or analyses of already collected data.
On the experimental side, there is a growing interest in the community to extend the LLP searches to broader regions of lifetime, masses and branching ratios. Experiments like \textsc{Mathusla}~\cite{MATHUSLA:2020uve} or \textsc{Codex-b}~\cite{Aielli:2022awh,Gligorov:2017nwh} have been proposed at the LHC (other proposals in Ref.~\cite{Alimena:2019zri}) to detect LLPs decaying dozens of meters away from their production point, in $pp$ collisions. These proposals, which need relatively reduced additional instrumentation, also serve to maximally exploit the LHC high-energy events in a complementary way. Similar proposals have also been explored for current and future $\ep e^-$ colliders~\cite{Dreyer:2021aqd,Schafer:2022shi,Alimena:2022hfr}. Using T tracks we can extend the lifetime coverage of LLP searches at LHCb, studied for Long and Downstream tracks in Ref.~\cite{Borsato:2021aum}, with no additional instrumentation.

The physics reach of future analyses with T tracks is being explored within the LHCb collaboration. These studies need to consider detailed efficiencies, resolutions, and backgrounds for these events. In this section we will study only the geometrical acceptance of T track events and compare it to Long and Downstream tracks, on which several analyses have already been performed at LHCb~\cite{LHCb:2014jgs,LHCb:2015ujr,LHCb:2016inz,LHCb:2016awg,LHCb:2016buh,LHCb:2017trq,LHCb:2017xxn,LHCb:2019vmc,LHCb:2020akw,LHCb:2021dyu}. Furthermore, the acceptance of LHCb is also compared to that of \textsc{Codex-b}, which envisages a $10\times 10\times 10 \,{\rm m}^3$ instrumented volume to be installed around 30 m away from the LHCb $pp$ collision point, in a space previously occupied by a data-acquisition computing farm. As benchmark channel for this study we will use the $\Bp\to\Kp\chi(\to\mup\mu^-)$ decay (where $\chi$ is the targeted LLP), also proposed in the original paper of \textsc{Codex-b}~\cite{Gligorov:2017nwh}.

\begin{figure} 
	\centering
	\includegraphics[height=0.26\textheight]{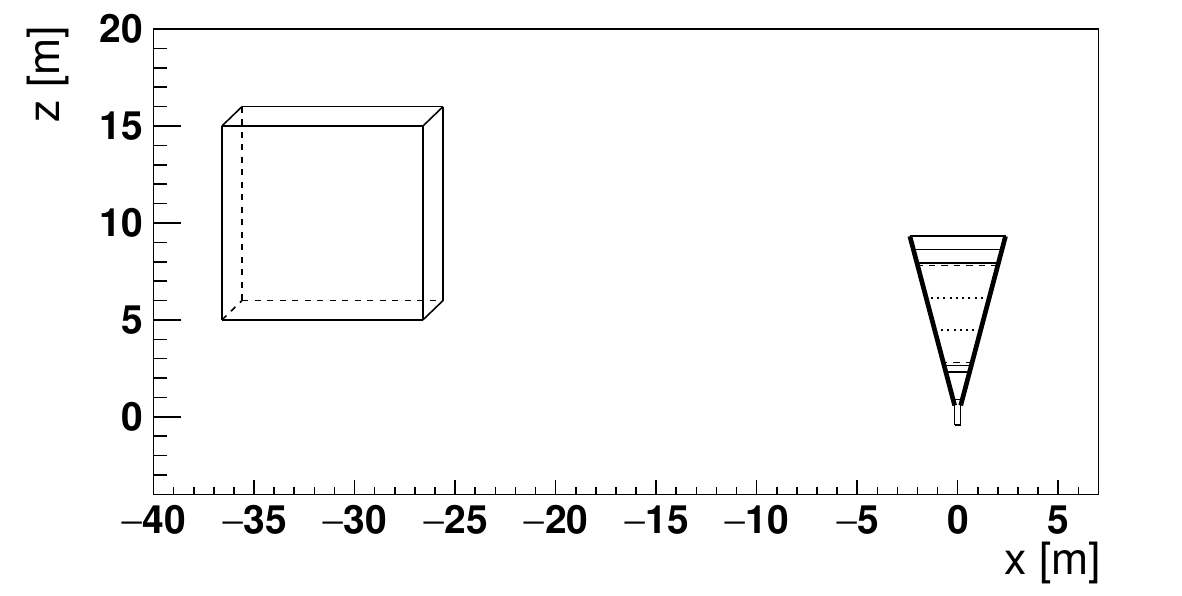}
	\includegraphics[height=0.26\textheight]{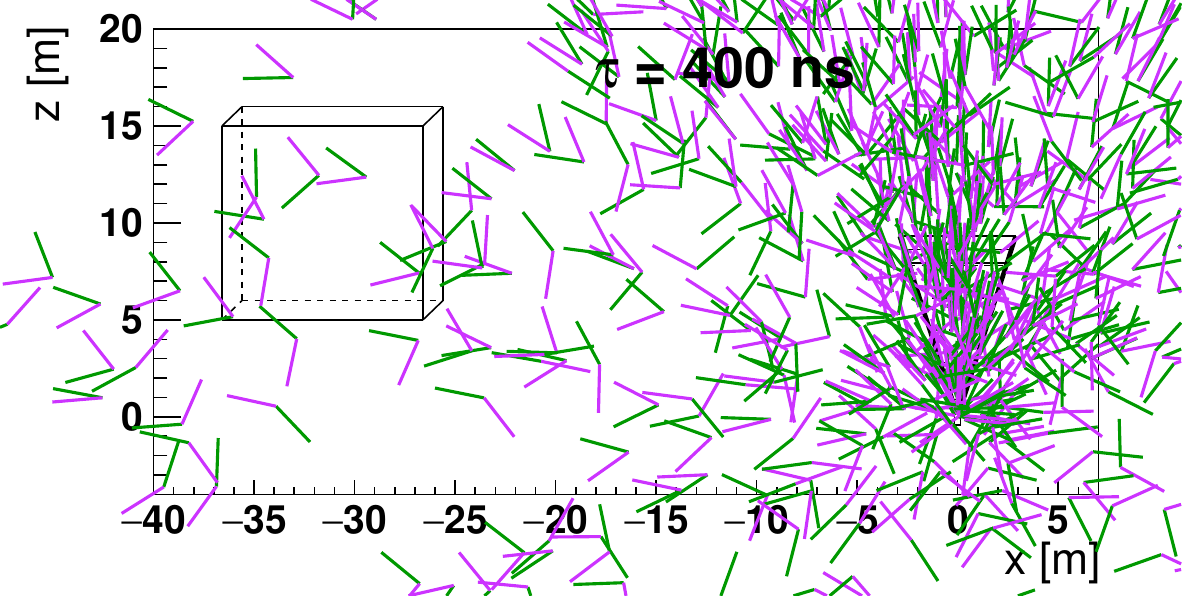}
	\includegraphics[height=0.26\textheight]{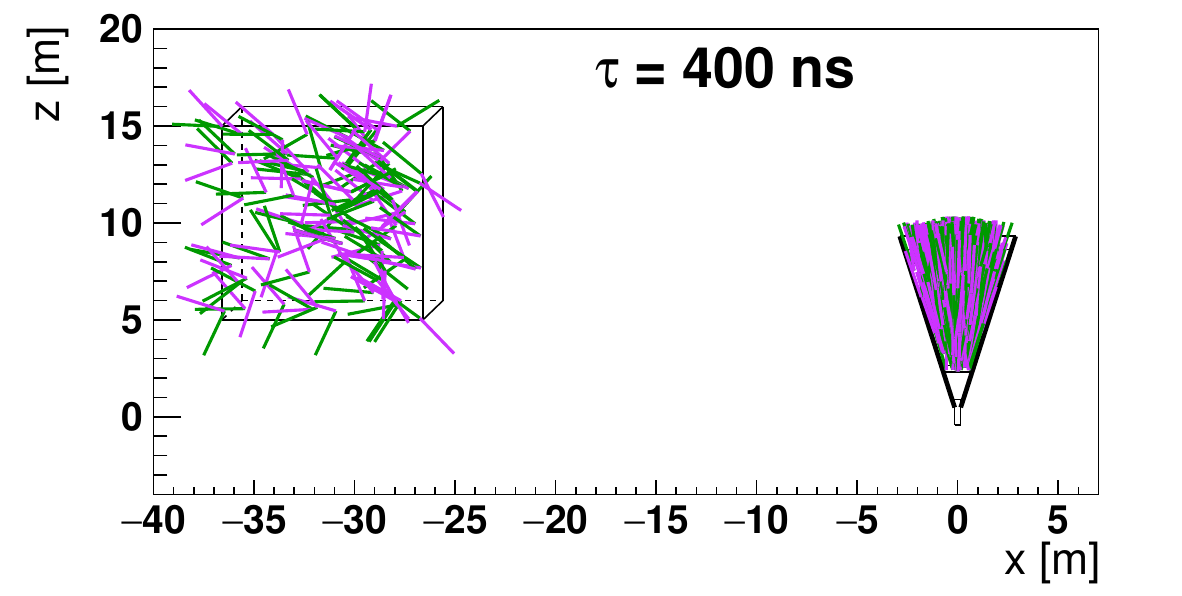}
	\caption{\textbf{Top:} schematic view of the LHCb tracking system (cone on the right) and the \textsc{Codex-b} active volume (cube on the left). \textbf{Middle:} event displays of the muon pair in $\Bp\to\Kp\chi(\to\mup\mu^-)$ events, overlayed on the detector geometries. \textbf{Bottom:} selected particles within Codex-b and T track acceptance for $\chi$ lifetime $\tau=400\,\ns$. This interface was built to validate the selections used in Figure~\ref{fig:acceptance}. }
	\label{fig:LLP_CodexbLHCb}
\end{figure}

\begin{figure}
	\centering
	\includegraphics[width=0.45\linewidth]{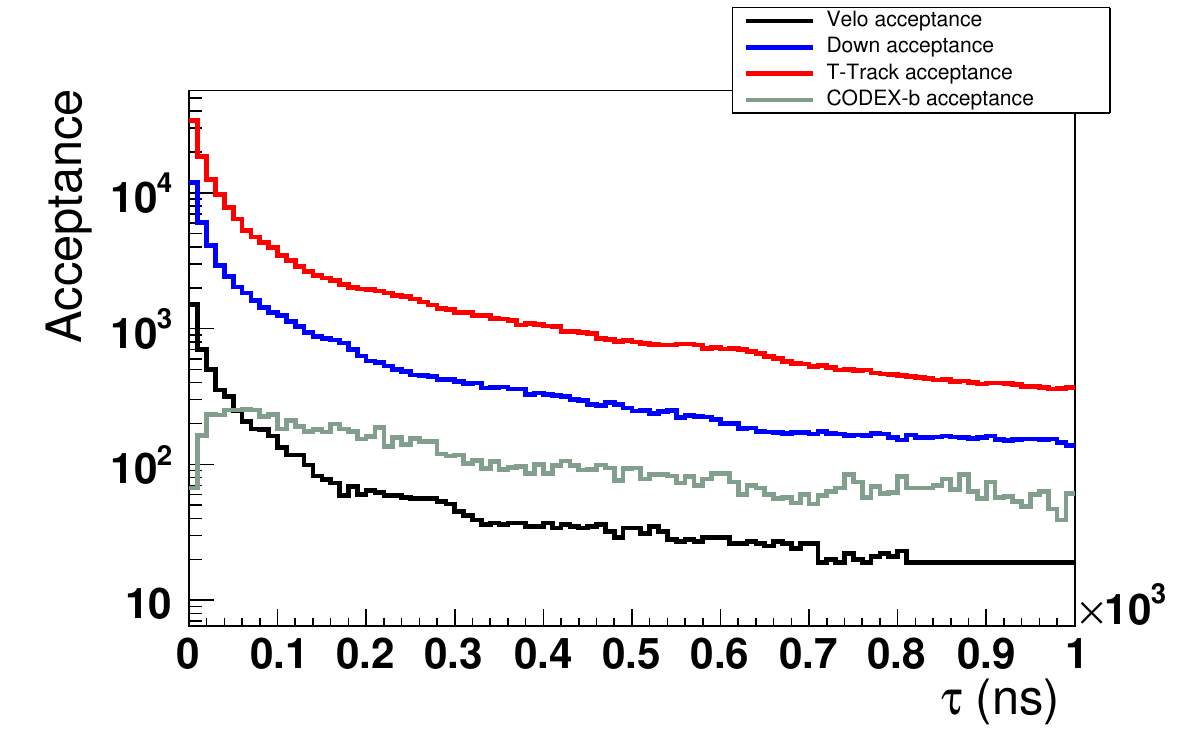}
	\includegraphics[width=0.45\linewidth]{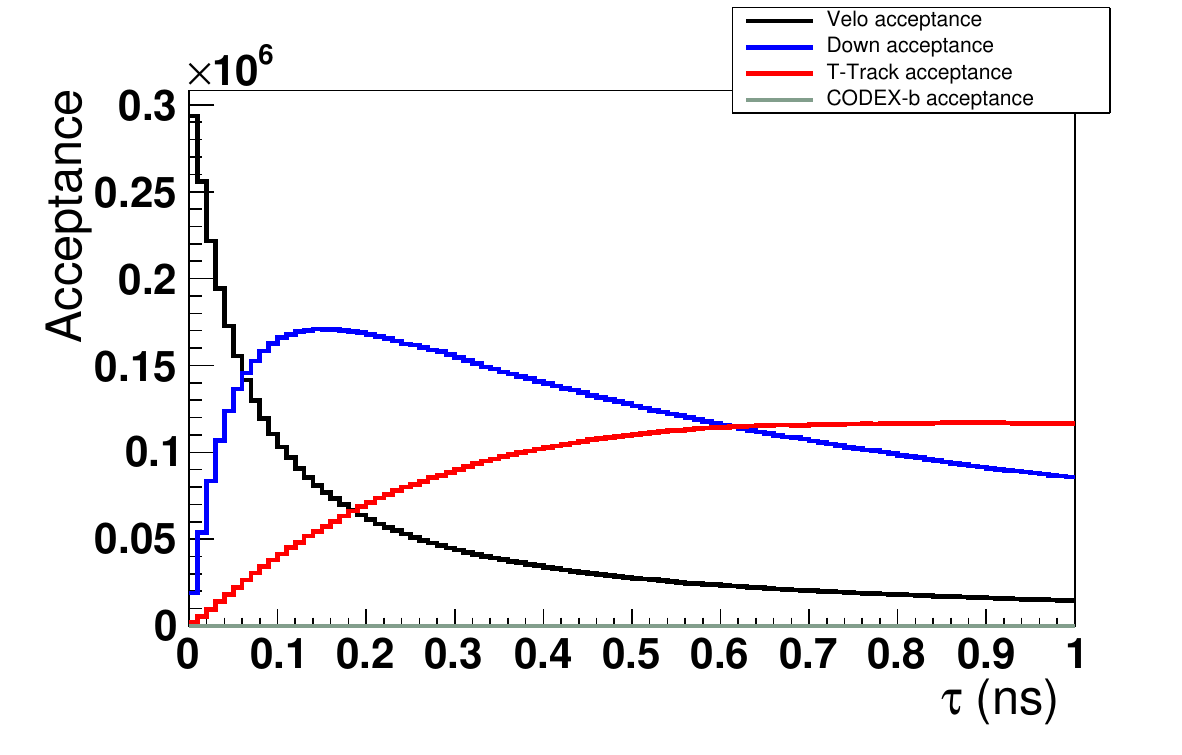}
	\caption{Number of events in acceptance to be reconstructed with Long, Downsteam, or T-Tracks at LHCb, compared to those in the \textsc{Codex-b} detector, using the benchmark chanel presented in the original Codex-b proposal~\cite{Gligorov:2017nwh}, $\Bp\to\Kp\chi(\to\mup\mu^-)$. The efficiency can be obtained dividing by $10^6$, the number of generated events.}
	\label{fig:acceptance}
\end{figure}

First\footnote{The results shown here were already presented in an internal meeting of the LHCb collaboration~\cite{internalmeetingLLP} as part of a preliminary study on the physics reach of T tracks for LLPs, which is not reproduced here.}, simulations of the benchmark channel $\Bp\to\Kp\chi(\to\mup\mu^-)$ with $m_\chi=2\,\gev$ were produced using \pythia and \evtgen. While no specific LHCb reconstruction was performed, the initial \evtgen configuration file (\texttt{EventType = 12113086}) was taken from the LHCb database, as it was previously used by the analysis in Ref.~\cite{LHCb:2016awg}. This file was adapted to eliminate the generator-level cut \textit{DaughtersInLHCb}, which removes particles outside the LHCb cone with no chance of being reconstructed by the detector, but which can reach \textsc{Codex-b}. To evaluate the geometrical efficiency we employ the standalone LHCb geometrical model already introduced in Secs.~\ref{sec:optimbaryons} and \ref{sec:lambdaedm} and add the \textsc{Codex-b} detector volume, as illustrated in Figure~\ref{fig:LLP_CodexbLHCb} (top). To consider an $\chi\to \mup\mu^-$ event in acceptance we require its decay vertex to be inside the \textsc{Codex-b} volume or, for LHCb, to have both muon tracks crossing the relevant tracking stations for Long, Downstream and T tracks. This selection is illustrated for \textsc{Codex-b} and T tracks in the bottom panel of Figure~\ref{fig:LLP_CodexbLHCb}. The mean lifetime of $\chi$ is scanned between 0 and $1000\,\ns$, shifting the decay point accordingly. The final results are presented in Figure~\ref{fig:acceptance}. The geometrical acceptance for \textsc{Codex-b} is about $10^{-4}$ for $\tau_\chi \approx 30\,\ns$ ($c\tau_\chi\approx10\m$), in rough agreement with Ref.~\cite{Gligorov:2017nwh}. The geometrical acceptance of T tracks is larger than that of \textsc{Codex-b} by a factor $\sim10$ or more. In the final bounds on LLP couplings this difference would at least be partially compensated by the reduced backgrounds of \textsc{Codex-b}, which is positioned behind a 3-m concrete wall and will also include a lead shield to stop SM particles from reaching the detector. In this study we have imposed no requirement on the bachelor $\Kp$ track of the $\Bp\to\Kp\chi$ decay. Due to the initial boost of the \Bp meson, requiring the \Kp to be detected as a Long track would strongly disfavour the \textsc{Codex-b} acceptance, positioned in the very-low pseudo-rapidity region. This effect, however, is specific to this benchmark channel and the situation is different for \eg massive LLPs that are prompt-produced in the high-$p_T$ region. On the other hand, LHCb would have the possibility to measure the mass of the LLP candidate, contrary to the case of \textsc{Codex-b}, which would have no magnetic field in its tracking volume.

\part{Angular analysis of $\Lc\to\Lz \pi^+ \pi^+ \pi^-$ decays at LHCb}

\newcommand{\onlyANA}[1]{}

\chapter{Analysis strategy} \label{ch:strategy}

\begin{figure}
	\centering
	\includegraphics[width=0.7\linewidth]{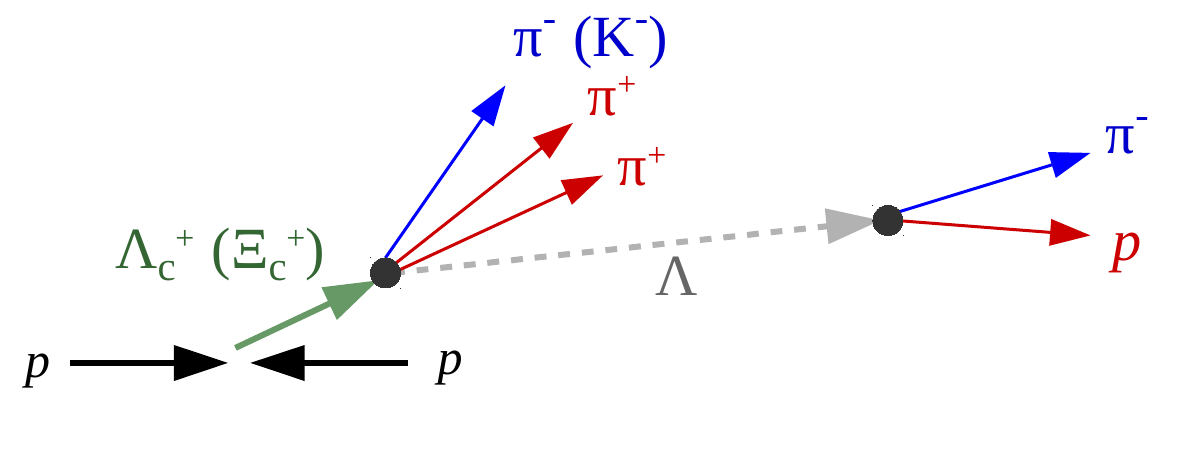}
	\caption{Topology of the \threepi (\kpipi) decay. The position and momentum of the neutral $\Lz$ hyperon is reconstructed with the $p$ and $\pim$ charged tracks originating in its decay. The combination of the \Lz with the $3\pi^\pm$ ($\Km\pip\pip$) system gives the \Lc (\Xicp) 4-momentum.}
	\label{fig:topology4body}
\end{figure}

Motivated by the possibility to perform spin-precession measurements with \Lz hyperons at LHCb (in Section~\ref{sec:lambdaedm}), decay channels of charm baryons into \Lz hyperons are being analysed within the LHCb collaboration. The analysis of these decay modes, in Table~\ref{tab:LambdaChannels}, also presents several opportunities to measure production and decay parameters of multi hadronic charm baryon decays, largely unexplored to date. The list of potential observables with these channels, in Section~\ref{sec:potentialobservables}, includes tests of \CP violation, which was recently observed by LHCb in charm meson decays~\cite{LHCb:2019hro}, and which interpretation in terms of NP is being debated (see \eg~\cite{Chala:2019fdb,Dery:2019ysp}).

In particular, in Part II of this thesis we exploit real LHCb data to analyse \threepi decays.
Given the similarity with the \kpipi decay (Figure~\ref{fig:topology4body}), this decay mode has been partially studied within the same analysis framework. However, for the sake of clarity, we will restrict the discussion and present the complete analysis chain only for \threepi decays.



The first selections to separate the interesting events from the huge amount of data collected by the LHCb experiment are described in Chapter~\ref{ch:dataprep} (data preparation). The resulting samples, which are still large, are further reduced with the offline selection, in Chapter~\ref{ch:selection}, where the agreement between data and Monte Carlo simulations is studied, and machine-learning techniques are employed to separate signal from background events. The fit to extract the \Lz polarization is performed in Chapter~\ref{ch:fit}, where the sources of systematic errors are described. There is usually more than one way to carry on each of the many steps in LHC data analysis. The main objectives and strategies of this analysis are described in the rest of this chapter, which presents an overview of Part II of this thesis.

Many terms and variables appearing recurrently along Part II, and in LHCb analyses in general, are defined in Appendix~\ref{app:analysisterms}.

\section{Objectives} \label{sec:objectives}

The first goal of this analysis is to measure the spin-polarization vector \spol of the \Lz hyperon from \threepi decays, denoted hereafter as \pol. Being a multibody decay, the relative momentum of the decay products is not fixed and we can also study the polarization $\pol(q^2)$ as a function of the momentum transfer $q^2\equiv(p_{\Lc}-p_\Lz)^2$. In the SM, the \Lc decay occurs through the transition $c\to s W^+$. 
On the theory side, however, factorizing the short-distance electroweak amplitude from the hadronic effects is very challenging, especially in multihadronic decays where several possible decay paths, including strong resonances, are possible.\footnote{A more feasible treatment is possible for semileptonic decays. The analogue treatment in this case would be to assume the $3\pipm$ system as originating in the hadronization of the $q\bar{q}'$ pair coming from the $W^+$ decay, which is however not justified.  }

Among many possible combinations, the decay of interest can take place through the production and strong decay of $\mathit\Sigmapm^*$ excited baryons such as $\Lc\to\Sigmacp^*(\Lz \pip) \pip \pim$. The portion of events occurring in this way can be studied by plotting the invariant mass of two or three final-state particles in the Dalitz plot of the decay. Another goal of our study is to qualitatively describe the \textit{resonant structures} in the Dalitz plot, which has never been studied in this decay with event samples of comparable size. The quantitative analysis of the resonant structures requires an \textit{add hoc} model of the decay through a full amplitude analysis. This is left for future studies, together with other interesting measurements listed in Section~\ref{sec:potentialobservables}.

\section{Datasets} \label{sec:datasets}

At the time our interest in these channels started, the LHC Run II was already in motion. The first step was to implement \hlttwo trigger lines for each decay mode in Table~\ref{tab:LambdaChannels}. This allowed saving the information of $pp$ collision events in which the requirements contained in these lines were met. However, these new lines have been only active from the end of the 2017 data-taking period. Recovering some of the interesting events from data-taking periods without dedicated trigger lines is still possible thanks to other generic or dedicated \hlttwo lines for similar decay topologies\footnote{Determining the set of trigger selections that we can use for this period will be the object of Section~\ref{sec:trigger}.}. The information recorded during the data-taking periods is saved on disk, and it is only\footnote{Exceptions to this are the small samples used for calibration or, notably at the LHCb, the Turbo stream of the trigger system.} accessed in a centralized manner through the \textit{stripping} process. In this process, the full reconstruction algorithms are run over the \textit{raw} data. The reconstructed information is compared against the set of selections defined by the \textit{stripping lines}, and the event is saved to be accessed \textit{offline} if it meets the selection criteria of at least one of the stripping lines.
This process is performed regularly (every few years) in the restripping campaigns. Thus, even if a decay mode was initially not targeted by any analysis, like in our case, the recorded events containing these decays can be recovered by defining a stripping line that will collect the relevant data in the next stripping campaign. It is important to remark that stripping and trigger selections are independent, and they are used at different moments in the data flow.

The final objective in this analysis is to use the full Run II dataset for this measurement. The same stripping selection will be consistently used, but the different trigger selections will necessarily split the full Run II data into two big datasets:

\begin{itemize}
	\item \textbf{2015 + 2016 + 2017, independent triggers}
	
	No dedicated \hlttwo trigger line was implemented in the 2015 and 2016 data-taking periods. The first dedicated trigger was only active in the last months of 2017, but its performance was suboptimal (it was designed essentially without Monte Carlo simulations).
	
	\item \textbf{2018, dedicated trigger} 
	
	An optimized \hlttwo trigger line was in place from the start of 2018 which is in fact the year with maximum recorded luminosity at LHCb. The latest stripping line has not run yet in a restripping campaign for 2018 data. This means that we will have to wait for the next restripping campaign or exploit the data with a suboptimal stripping line (that is already available), but with the dedicated trigger.
	
\end{itemize}

Additionally, depending on the flight distance of the \Lz hyperon, both trigger and stripping lines are defined separately in two categories:

\begin{itemize}
	\item \textbf{\lLL sample}
	
	The proton and pion from the \Lz decay are reconstructed as long tracks, \textit{i.e.} they are detected at least in the VELO and T-stations (see Figure~\ref{fig:tracktype}).
	
	\item \textbf{\lDD sample}
	
	The proton and pion from the \Lz decay are reconstructed as Downstream tracks, \textit{i.e.} they are detected in the TT and T-stations only.
	
\end{itemize}

In this thesis, we will develop the analysis chain with the 2016 \lDD data sample. The same treatment can be directly applied to 2015 and 2017 \lDD data (with the appropriate consistency checks). The analysis of the \lLL sample needs slightly different selections and a different treatment of the PID recalibration, as it will be pointed out in Chapter~\ref{ch:selection}\footnote{Similar to the \kpipi decay channel, many steps have also been done for the \lLL sample of the \threepi decay. However, for the sake of clarity, we will restrict the discussion in this thesis to the \lDD sample.}. The extension to 2018 data will need to account for the differences in the trigger selections, but it can be done with the same analysis framework developed for this thesis.

\section{Selection} \label{sec:selection}

In the selection process, the signal events are separated from the background by using the \textit{variables} of the event. For example, a reconstructed \Lc candidate may comply with all kinematical requirements while being reconstructed with a \Lz and three \pipm tracks from the primary vertex (PV) of the $pp$ collision. By \textit{applying a cut} on the \Lc flight distance from below (\ie removing all candidates with a smaller value than that specified by the cut), we require the $\pipm$ tracks and \Lz direction to be consistent with a displaced vertex, removing backgrounds from particles produced at the PV. 
Four different sets of cuts are applied sequentially in this analysis. We will distinguish between those for data preparation (Chapter~\ref{ch:dataprep}) and for offline selection (Chapter~\ref{ch:selection}). The purpose of these selections are listed in the following:

\begin{enumerate}
	\item \textbf{Stripping} (data preparation)
	
	As introduced above, the stripping has the purpose of separating an initial dataset from the huge amount of information recorded on disk by the LHCb experiment. The stripping selection is discussed in Section~\ref{sec:stripping} and the set of cuts given in Table~\ref{tab:stripDD}. These correspond to the \verb|Hc2V03H_Lambdac2Lambda3PiDDLine| line at its version active during the \verb|s28r2| restripping campaign (of 2016 data). 
	
	\item \textbf{Trigger} (data preparation)
	
	The trigger is used \textit{online} to decide which events should be recorded. Since we have no dedicated \hlttwo trigger line for our dataset, we will have to select \textit{offline} the most efficient trigger lines that are compatible with the stripping requirements. For this reason, we list it in second place. The best lines at all trigger levels (L0, \hltone, \hlttwo) are inspired by the rates on Monte Carlo (MC) and tested on real data (in Section~\ref{sec:trigger}).

	\item \textbf{Preselection} (offline selection)
	
	After the data preparation, the percentage of signal events within the whole data sample is still very low. We would like to directly use a multivariate classifier (defined below) to obtain the best signal/background separation. However, before that, it is essential to assess the agreement between real data and MC simulations (Section~\ref{sec:corrMCData}) for which we need a minimal signal purity and the use of the sWeights\footnote{See definition in Appendix~\ref{app:analysisterms}.}. This is achieved by applying a set of cuts in the preselection (Section~\ref{sec:preselection}), with minimum loss in the signal efficiency.
	
	\item \textbf{Multivariate classifier} (offline selection)
	
	The previous selections used strict cuts on the variables, determining if the events were kept or removed from data based on single variable thresholds (often called \textit{rectangular cuts}). However, these variables are often related to each other, and signal and background events can be better discriminated by considering the multidimensional space of variables. A multivariate classifier can find the optimal boundaries between signal and background events in this multidimensional space. The last step of our selection is based on \textit{boosted decision trees} (BDT), a type of multivariate classifier described in Section~\ref{sec:BDT}. 
	
\end{enumerate}

\section{Angular fit} \label{sec:fitintro}

\begin{figure}[ht]
	\centering
	\subcaptionbox{Laboratory frame.}{ \includegraphics[width=0.3\linewidth]{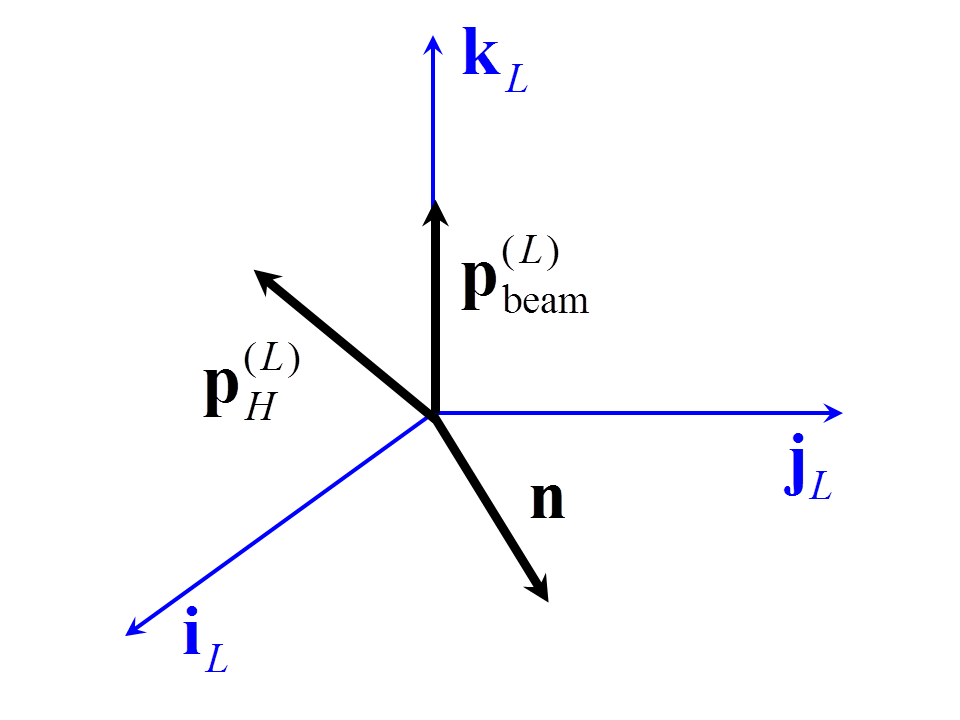} }
	\subcaptionbox{H (mother of \lz) rest frame.}{ \includegraphics[width=0.3\linewidth]{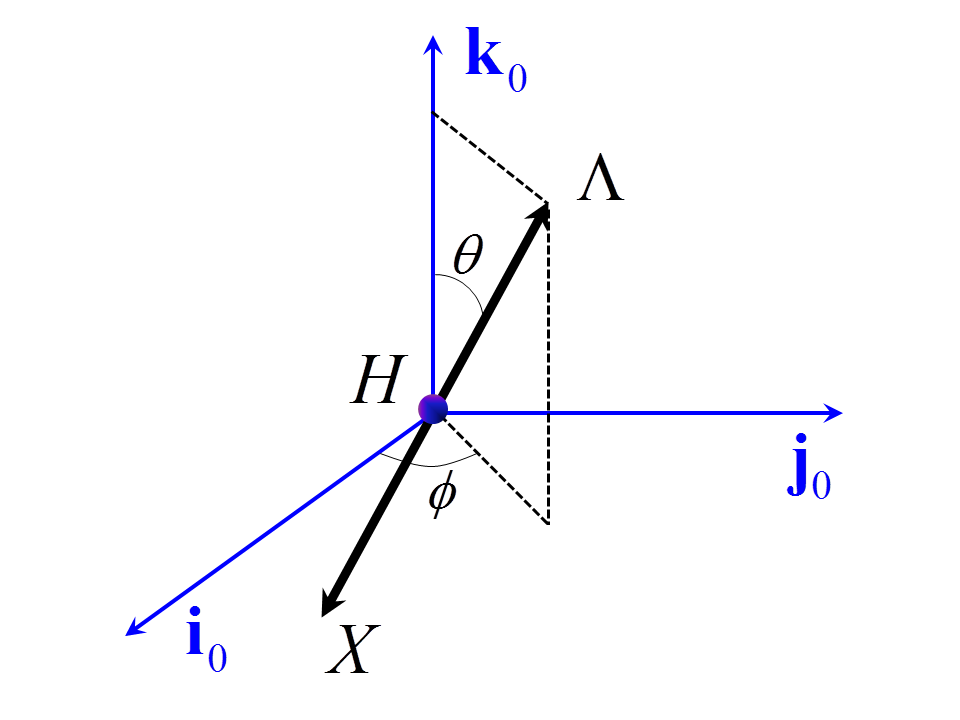} }
	\subcaptionbox{\lz rest frame.}{ \includegraphics[width=0.3\linewidth]{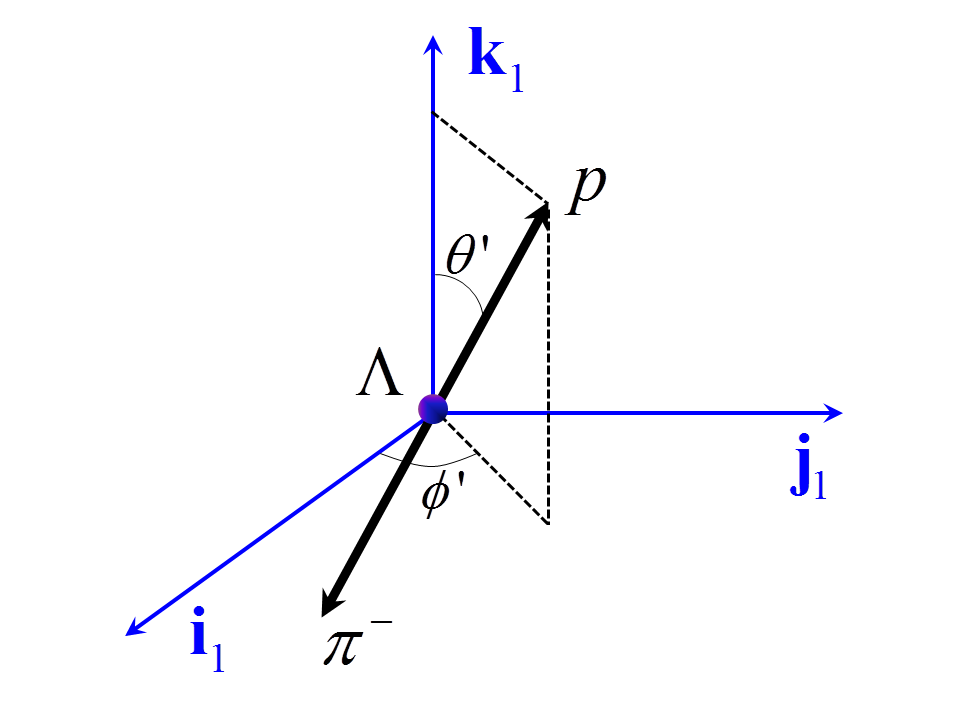} }
	\caption{Definition of the helicity angles and coordinates of the different frames.  }
	\label{fig:frames}
\end{figure}

\subsubsection{Proton angular distribution}

The \Lz polarization vector \pol indicates the preferred direction of the proton $\mathbf{k}$ in the \Lz rest frame. In this frame, the angular distribution of the proton helicity angles $(\theta_p, \phi_p)$ is given by
\begin{equation} \label{eq:angdist}
\frac{d W}{dcos\theta_p\phi_p} = \frac{1}{4\pi}\left[ 1 + \alpha_{\Lambda} \mathbf P_{\Lambda} \cdot \mathbf k \right],
\end{equation}
where $\mathbf k=(\cos\phi_p\sin\phi_p,\sin\phi_p\sin\phi_p,\cos\theta_p)$ and $\mathbf P_{\Lambda} = (P_{\Lambda,x},P_{\Lambda,y},P_{\Lambda,z})$. As we can see, the probability is maximal for $\mathbf{k}$ parallel to \pol.

To reach the helicity frame of the \Lz, one must consider the intermediate rest frame of the mother baryon $H$, in this case \Lc. Denoting the momentum of a particle $X$ in the frame $F$ as $\p{X}{F}$, the three needed coordinate systems (in Figure~\ref{fig:frames}) are defined as follows~\cite{RICHMAN,LEADER}:

\begin{itemize}

	\item \textbf{Frame L:} in the laboratory frame, the $z$ axis is defined in the direction of the beam ${\bf k}_L$. Charge-conjugated decays $\overline{\Lz} \to \overline{p} \pip $ have precisely opposite sign for the decay-asymmetry parameter $\alpha_\Lz$ assuming \CP conservation. To fit both datasets together, this vector is taken as ${\bf -k}_L$ for charge-conjugated decays.

	\item \textbf{Frame 0:} In the $H$ rest frame, the ${\bf k}_0$ vector is defined with the normal direction to the production plane ${\bf n}$. The other two basis vectors, ${\bf i}_0$ and ${\bf j}_0$ can be chosen arbitrarily but their definition must be consistent for all events. These are taken as

	\begin{equation*}
	{\bf k}_0 \equiv {\bf n} = \frac{ \p{\text{beam}}{L} \times \p{H}{L} }{|\p{\text{beam}}{L} \times \p{H}{L} | }  ~~,~~~
	{\bf i}_0 = \frac{ \p{\text{beam}}{L} }{|\p{\text{beam}}{L}|} ~~,~~~
	{\bf j}_0 = {\bf k}_0 \times {\bf i}_0 ~.
	\end{equation*}

	\item \textbf{Frame 1:} In the \Lz rest frame, ${\bf k}_1$ is defined as the momentum of the \Lz in the mother rest frame.  The vector  ${\bf j}_1$ is defined by the $H$ direction,

	\begin{equation*}
	\label{frame1}
	{\bf k}_1 = \frac{  \p{\Lambda}{0} }{| \p{\Lambda}{0} | }  ~~,~~~
	{\bf j}_1 = \frac{ {\bf n} \times \p{\Lambda}{0} }{| {\bf n} \times \p{\Lambda}{0} |} ~~,~~~
	{\bf i}_1 = {\bf j}_1 \times {\bf k}_1 ~.
	\end{equation*}

\end{itemize}

The helicity angles used in the angular fit are the polar and azimuthal angles of the proton in Frame 1, as shown in Figure~\ref{fig:frames} (c).

\subsubsection{Fit strategy}

To extract the \Lz polarization, in Chapter~\ref{ch:fit} we will perform an unbinned maximum-likelihood fit of the proton helicity angles to the PDF in Eq~\eqref{eq:angdist}. However, as a previous step, we must make sure that the distribution of helicity angles is not affected by the applied selections. If the variables used in the selection are correlated to the proton angles, the efficiency on signal events may be uneven across the $(\theta_p, \phi_p)$ plane. We will address these effects with the so-called \textit{acceptance corrections}. The efficiency map in $(\theta_p, \phi_p)$ is parameterized with MC simulations generated by assuming zero polarization (\textit{phase space MC}). By applying the same selections on this MC data we can parameterize any deviation from a flat distribution with the use of Legendre polynomials. Then, the helicity angles are fit to the product of the angular distribution in Eq~\eqref{eq:angdist} by the acceptance correction.

The fit result yields a central value of the polarization components together with the \textit{statistical uncertainty}. Additionally, each of the steps in the analysis chain also has \textit{systematic uncertainties}  associated to them. Commonly, these are assessed by defining some alternative methods for these steps and evaluating their effect on the central value of the final fit. For example, the size of the MC samples used for the acceptance correction is one of the leading systematic errors in the (preliminary) analysis presented here. In Section \ref{sec:systematics} we will estimate this and other relevant systematic errors.

\pagebreak

\section{Other potential observables} \label{sec:potentialobservables}

\begin{table}[htb]
	\centering
	\caption{$H_c \to \Lambda X$ multibody decays. The acronyms in the \textit{Suppression} column stand for Cabibbo-Favoured (CF) and Single-Cabibbo Supressed (SCS). The strangeness \textit{S} of the final state is also indicated.}
	\label{tab:channelsOthers}
	\renewcommand{\arraystretch}{1.1}
	\resizebox{0.6\textwidth}{!}{
		\begin{tabular}{cccc}
			\hline \hline
			\#   & Mode                           & $S$ final state & Suppression \\
			\hline 
			1    & $\Lc\to\Lambda\pim\pip\pip$    & $-1$            & CF     \\
			2    & $\Lc\to\Lambda\pim\Kp\pip$     & $0$             & SCS    \\
			3    & $\Xicp\to\Lambda\Km\pip\pip$   & $-2$            & CF     \\
			4    & $\Xicp\to\Lambda\Km\Kp\pip$    & $-1$            & SCS    \\
			\hline
			5    & $\Xicz\to\Lambda\Km\pip$       & $-2$            & CF     \\
			6    & $\Xicz\to\Lambda\Km\Kp$        & $-1$            & SCS    \\
			\hline \hline
		\end{tabular}
	}
\end{table}

Besides measuring the \Lz polarization in \threepi or \kpipi decays, much more information can be extracted from the analysis of these and other similar channels. The data preparation and event selection presented in this thesis is common to all potential measurements with $\Lc\to\lDD\pip\pip\pim$ events and, with minimal modifications, the four-body decay modes presented in Table~\ref{tab:channelsOthers} (1-4) can be studied too. The three-body decays in this table (5,6) are currently under study mainly by the Milano LHCb group.
The fit procedure and evaluation of systematic uncertainties is necessarily different when treating different observables, but largely common for the different channels.

Some of these possible analyses are listed in the following:

\begin{itemize}
	\item \textbf{Decay-asymmetry parameter of $\Lc\to\Lambda$}
	
	If the $3 \pipm$ system were associated to the $\Wp$ in the $c\to s \Wp$ transition then $P_{\Lz,z}$ can be identified with the decay asymmetry parameter of the $\Lc \to \Lambda \Wp$ decay,
	\begin{equation}
	P_{\Lambda,z} (q^2) = \alpha(q^2) \equiv \frac{\left(H_{\frac12 1}|^2+|H_{\frac12 0}|^2\right) - \left(H_{-\frac12 -1}|^2+|H_{-\frac12 0}|^2\right)}
	{\left(H_{\frac12 1}|^2+|H_{\frac12 0}|^2\right) + \left(H_{-\frac12 -1}|^2+|H_{-\frac12 0}|^2\right)}~,
	\end{equation}
	where $H_{\lambda_\Lambda \lambda_W}$ are the helicity amplitudes with $\lambda_\Lambda=\pm\frac{1}{2}$ and $\lambda_W=0,\pm 1$. 
	However, the three pions may also be created through other subprocesses, diluting the direct interpretation of $P_{\Lz,z}$ as $\alpha$. It would be interesting to study this decay with hadronic models to clarify the interpretation of this parameter in terms of more fundamental interactions. Nevertheless, considering the $\Lc \to \Lambda X^+$ decay, an effective $\alpha_{\rm eff}$ parameter can be determined.
	
	\item \textbf{\CP violation from the decay asymmetry}
	
	Independently of its precise parameterization in terms of fundamental operators, a different value of $\alpha_{\rm eff}$ for the decay with particles and antiparticles, \textit{i.e.} ${\alpha(q^2) \neq -\overline\alpha(q^2)}$ would signal \CP violation in the decay,
	\begin{align}
	A_{CP}(q^2)   =  \frac{\alpha(q^2)+\overline\alpha(q^2)}{\alpha(q^2)-\overline\alpha(q^2)} \approx A_{CP,\Lambda} + A_{CP,H_c}(q^2), \\
	\text{where~~} A_{CP,\Lambda}  = \frac{\alpha_{\Lambda}+\alpha_{\bar\Lambda}}{\alpha_{\Lambda}+\alpha_{\bar\Lambda}} \ , \ \ \
	A_{CP,H_c}(q^2) = \frac{\alpha_{H_c}(q^2)+\overline\alpha_{H_c}(q^2)}{\alpha_{H_c}(q^2)-\overline\alpha_{H_c}(q^2)}.
	\end{align}
	
	\item \textbf{Comparison to suppressed channels}
	
	Similarly to the observable $\Delta A_{\CP}$ with which CPV was observed in charm meson decays~\cite{LHCb:2019hro}, it is also possible to measure the difference between direct decay asymmetries of CF and SCS decays, in Table~\ref{tab:channelsOthers}, as
	
	\begin{equation}
	\Delta A_{\CP} = A_{CP}(\rm SCS) - A_{CP}(\rm CF).
	\end{equation} 
	
	\item \textbf{Final state triple products}
	
	In four-body modes like these, there is a way to directly probe \T violation with \T-odd triple product asymmetries. Considering a decay $P\to abcd$, the triple product asymmetries are constructed as (see \textit{e.g.} Ref.~\cite{Gronau:2015gha})
	\begin{equation}
	A_T = \frac{\Gamma(C_T>0) - \Gamma(C_T<0)}{\Gamma(C_T>0) + \Gamma(C_T<0)},~~\text{where~~} C_T = {\bm p_a} \cdot ({\bm p_b} \times {\bm p_c}).
	\end{equation} 
	This observable is particularly interesting for the Cabibbo-Suppressed modes, in which NP effects may be enhanced.
	
	\item \textbf{Complete angular analysis}
	
	The complete \threepi angular distribution can be effectively described through two additional effective decay-asymmetry parameters and three polarization parameters accounting for the polarization state of the mother \Lc baryon. This four-dimensional angular analysis also requires the helicity angles of the \Lz in the frame 0 and would allow a simultaneous determination of the production and decay polarization properties of the \Lc baryon. This analysis is underway in parallel with the preliminary angular analysis presented in this thesis.

	\item \textbf{Full amplitude analysis and spectroscopy of pentaquark states}

	By analysing the Dalitz plot and complete angular distribution of the decays, it is possible to build an amplitude model containing all the information of the intermediate resonances involved in this final state, together with the polarization of the \Lc baryon. This study may shed light on the existence of a compact pentaquark state $\Sigma^*$ with $J^P = \frac{1}{2}^-$ with mass $m(\Sigma^*)\approx1380 \,\mev$. Experimental analyses~\cite{Wu:2009tu,Wu:2009nw} and hadronic model predictions~\cite{Helminen:2000jb,Zhang:2004xt,Gao:2010hy} have studied this possible state, which would appear as a broad structure under the peak of the well-established $\Sigma(1385)^\pm$ resonance, in the $m(\Lz \pipm)$ spectrum.
	This study is being carried out in collaboration with the UCAS LHCb group.

\end{itemize}

\chapter{Data preparation} \label{ch:dataprep}

In this chapter, the trigger and stripping selections are defined for the 2016 sample, which is exploited in the following chapters. No dedicated trigger was present in this period of data taking and we will need to define the most efficient trigger lines among those available, in Section~\ref{sec:trigger}. Before that, to obtain an initial event sample in the restripping campaigns, a stripping selection had to be defined as described in Section~\ref{sec:stripping}. We tried to maximize the signal efficiency while being selective enough to stay within a reasonable bandwidth, that must be compatible with the many other channels being restripped. A similar study was also carried out to define the selection criteria in the dedicated trigger line, for 2018 data, which is nevertheless not used in this thesis and therefore also not presented in this chapter.

\section{Stripping} \label{sec:stripping}

If the stripping selection is the very first step to obtain any initial reconstructed data, how can we design the stripping selection itself? One possibility is to \textit{blindly} define a set of reasonable cuts based on generator-level simulations\footnote{See definition in Appendix~\ref{app:MCtypes}.}, \ie without detector effects. This will certainly provide \textit{some} data and, in cases where the event signature is particularly differentiated from the combinatorial background (for example in direct searches with high-$p_T$ signals), it may be more than enough to start the analysis process. In our case, the event topology contains three pions with relatively low momentum and a (neutral) \Lz that decays between 0.5 and 2.5 meters from the $pp$ collision point. 
A careful selection involving many variables is thus needed to maximize the signal efficiency.

%
%
%

\subsubsection{Signal: Monte-Carlo-matching selection}

Starting from a set of reconstructed signal MC (\texttt{DST} or \texttt{LDST}) that was \textit{flagged}\footnote{See definition in Appendix~\ref{app:MCtypes}.} with a different stripping line, we can supersede this stripping selection by our custom initial selection. We only require that (five) reconstructed tracks with the global topology of our decay are indeed associated to the true simulated signal tracks, \ie the reconstructed objects are \textit{MC-matched}.

Effectively, we obtain a representative sample of \threepi decays that have a chance of being reconstructed at LHCb. With this baseline, we can design cuts that keep as many of these events as possible, while rejecting enough background.

In Tables \ref{tab:stripLL} and \ref{tab:stripDD} this sample will be used to extract the signal efficiency $\varepsilon({\rm MC})$.

\subsubsection{Background: data with the previous stripping} 

For background events, ideally, we would use a sample of real data that contains absolutely anything that can combine into our event topology (the so-called \textit{minimum bias} selection). These samples are available at LHCb, but they require managing huge initial files which, after a few cuts, are left with very few events, usually not enough to continue a precise study of background efficiencies. Thus, we will take a more practical approach: real data obtained with a previous version of the stripping line is used to ensure that, at least, the background passing that selection is reduced as much as possible, while the cuts can be \textit{opened} in other variables.

In Tables \ref{tab:stripLL} and \ref{tab:stripDD} this sample will be used to extract a proxy to the efficiency on background, noted $\varepsilon({\rm Data})$.

\subsubsection{Particle containers}

One of the first checks was to ensure that the stripping efficiency (of the previous line) was reproduced by running the stripping selection, within the LHCb software, and applying the cuts offline on the MC-matched sample. An initial disagreement served to find all the cuts within the stripping lines that are not obvious at first glance. Starting to build a stripping selection, some basic cuts are always applied in each initial track, defined in the \textit{particle containers}. These, in turn, call other containers or functions that may also contain additional cuts. The structure of this initial stripping selection, including all these ``hidden" cuts, is shown in Appendix~\ref{app:strippedstripping}.

Considering all these cuts, we reproduced the stripping efficiencies both by applying the stripping within the LHCb framework and with our custom MC-matched data plus offline cuts present in the stripping line.

\begin{figure}
	\centering
	\includegraphics[width=0.48\linewidth]{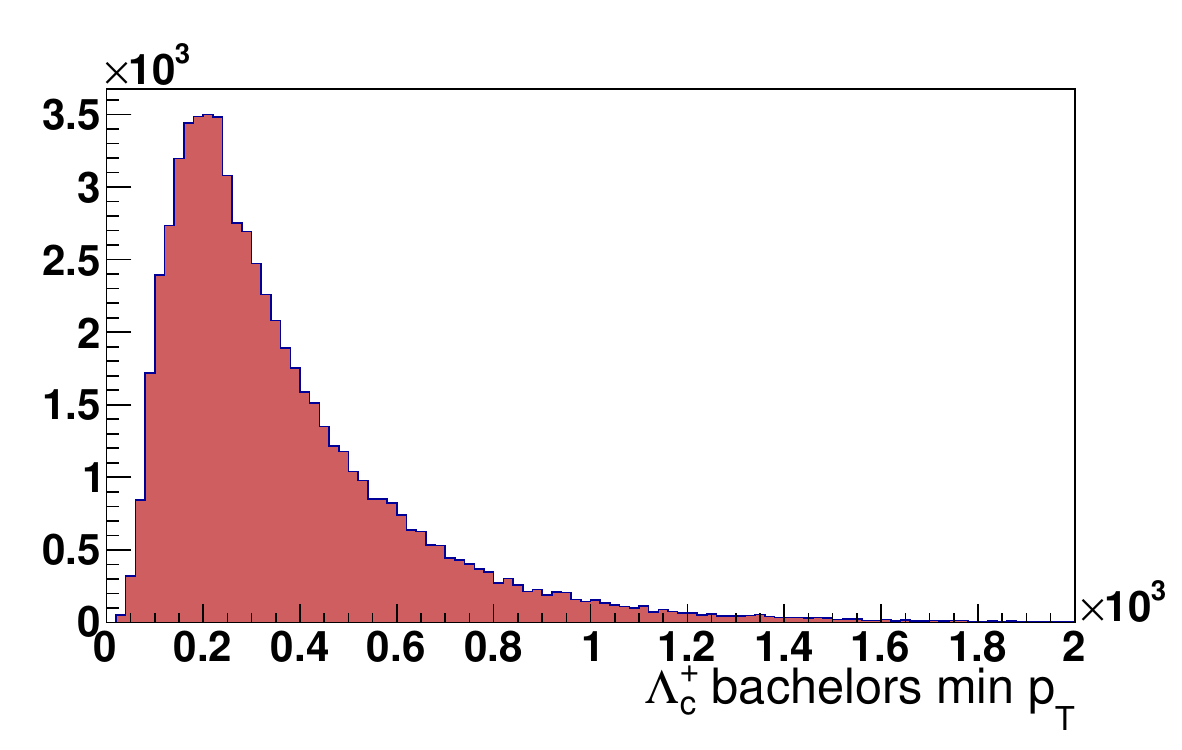}
	\includegraphics[width=0.48\linewidth]{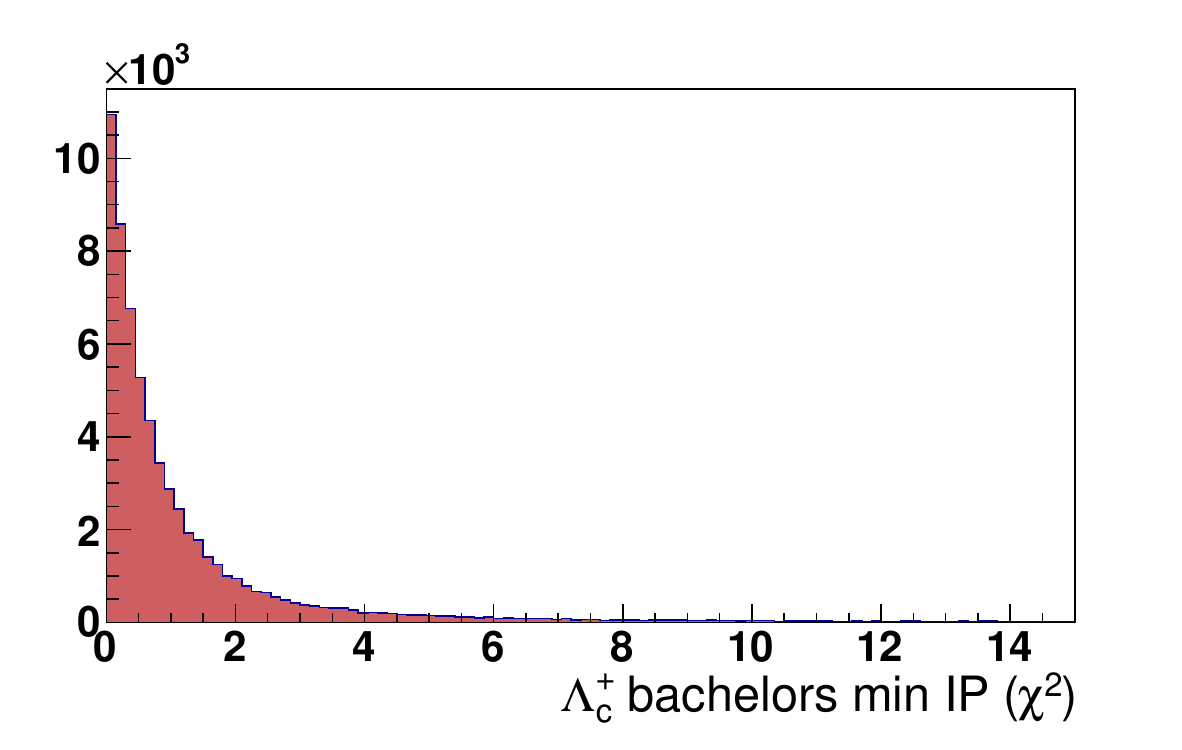}
	\caption{(Left) minimum $p_T$ (in MeV) and (right) $\chi^2_{\rm IP}$ of the 3\pipm(\Lc), for the MC-matched sample. The previous stripping line was applying cuts at $\min p_T >250\,\mev$ and $\min \chi^2_{\rm IP} >4$. The vertical axes indicate the number of events per bin. }
	\label{fig:containersvariables}
\end{figure}

This study served to realise that most of the signal events were being removed already within the particle containers, as shown in Figure~\ref{fig:containersvariables}.  Thus, the first step was to use different particle containers without these cuts and introduce others to keep the rates under control.

\subsubsection{Final stripping selection}

The final stripping selection is reported in Tables~\ref{tab:stripDD} and \ref{tab:stripLL}. The optimized stripping lines have already been used in restripping campaigns of almost all Run II, as reported in Table \ref{tab:yearsdata}. The main changes with respect to the previous selection are summarized in the following.

\begin{table} 
	\centering
	\caption{Run II years of data taking and restripping version with the optimized line \texttt{StrippingHc2V03H\_Lambdac2Lambda3PiDDLine}.	In this thesis we are using s28r2 data.} \label{tab:yearsdata}
	\begin{tabular}{cc}
		\hline \hline
		Year & Stripping version \\ \hline
		2015 & s24r2 \\
		
		2016 & s28r2 \\
		
		2017 & s29r2p1 \\
		
		2018 & - \\
		\hline \hline
	\end{tabular}
\end{table}

In practice, completely removing the cuts of the particle containers was increasing the background to an unmanageable level. First, pions with very low $p_T$ are copiously produced in $pp$ collisions and material interactions of all kinds. Thus, we softened the cut from $\pipm(\Lc) ~p_T\geq 250 \,\mev$ (in \texttt{StdNoPIDPions}) to $\geq 150 \,\mev$. Additionally, due to the low Q-value ($Q=m_\Lc - m_\Lz - 3 m_{\pipm}$) of our decay and the boost of the \Lc baryon, the three charged pions are mostly aligned to the flight direction of the \Lc, which is dominantly produced in the PV. Therefore, the pion tracks are very difficult to separate from the many tracks produced at the $pp$ collision. Thus, a cut on their impact parameter\footnote{See definition in Appendix~\ref{app:analysisterms}.} ($\chi^2_{\rm IP}$) necessarily removes a lot of signal events although it is also necessary to keep the PV backgrounds under control. Instead of requiring the three pions to have $\pipm(\Lc) ~\chi^2_{\rm IP} > 4$ (in \texttt{StdNoPIDPions}), we require one of the pions to have $\chi^2_{\rm IP} > 9$ and another one $\pipm(\Lc)~ \chi^2_{\rm IP} > 1$, with no requirement on the third pion. This solution improves significantly the signal efficiency of the IP requirement. 

Another significant change was implemented on the \Lz selection, for the \lLL sample. The \verb|StdLooseLambdaLL| container was replaced by \verb|StdVeryLooseLambdaLL|, already defined in the selection framework, effectively removing any requirement on the $\pim(\Lz)$ and \pr(\Lz) $p_T$, which before was required to be $p_T\geq 250\,\mev$.

Other cuts, with signal efficiency $\varepsilon({\rm MC})\approx 98\%$, were introduced to reduce the rates, which were severely increased with the previous relaxations of the stripping requirements. In particular we used  \Lz $p$        $>$  10 \,\gev, $\Lambda$ $p_{T}$        $>$    0.5\,\gev, $\Lambda_{c}^{+}$ $\log \chi^2_{\rm IP}$        $<$      5\footnote{This was a very unfortunate decision, since the distribution of  $\log \chi^2_{\rm IP}$ is instrumental to control the number of primary and secondary \Lc particles as shown later in Fig.~\ref{fig:DTFLcLogIPChi2}. In that figure, it is also apparent that this cut does not have a 98\% efficiency. The reason is that MC-matched events are mostly composed of primary \Lc particles peaking to the left side of the distribution. However, after the full selection process, events with non-prompt \Lc particles are strongly preferred.}, and $\Lambda_{c}^{+}$ decay vertex $\chi^{2}$/ndf         $<$      3.

With these modifications, we achieved an improvement in signal efficiency of a factor 2.0 (3.2) for the \lDD (\lLL) sample. Nevertheless, the efficiency on MC-matched events is still very low, of 4.6\% (8.3\%), and different strategies may be considered to improve it for LHC Run III.

\subsubsection{Filtered Monte Carlo productions}

We will need large initial MC samples to define a trigger selection, study the MC/data agreement, train a multivariate classifier and correct for detector efficiencies on the helicity angles before the final fit. To optimize resources for the simulation, we defined generator-level cuts that preselect the events on which the whole reconstruction must be performed, reducing the total time to 19\%. Additionally, these samples were produced with the \textsc{ReDecay} algorithm for fast simulation~\cite{redecay}. 

Besides optimizing online computing resources, in large samples one must also optimize the storage space. Since the optimized stripping line will be the baseline selection for every further study, only the MC events that pass the stripping selection are saved in this \textit{filtered production}, just like the reconstructed real data.


\section{Trigger strategy} \label{sec:trigger}

As already discussed, in the 2015, 2016 and 2017 data-taking periods, there was no dedicated HLT2 trigger line to record \threepi events. Yet, many of these were saved anyway by other trigger lines. Those were recovered with a dedicated stripping line but to have a consistent selection both in real data and MC simulations we should define a set of trigger selections with those lines that keep most of these events.

To do this, we will systematically study the most efficient trigger lines on MC data, in Section~\ref{sec:trigMC}, and, inspired on these results, several combinations of trigger lines will be tested on real data, in Section~\ref{sec:trigData}.

\subsection{Trigger efficiencies on Monte Carlo} \label{sec:trigMC}

\textit{Prescaled} trigger lines are configured to not record every single event in which they fire. This is done to reduce the bandwidth of these lines, without biasing the recorded sample. However, in MC data this prescaling is not applied by default in order to optimize the computing resources (it would not make sense to simulate many events just to randomly throw away most of them at the end). Thus, the trigger efficiencies on MC do not reproduce, by definition, the real data efficiencies in the presence of prescaled lines. Since generally these lines give negligible contributions to the decay of interest, they have been simply commented out from the tables. Note also that the lines that are the composition of many lines such as \texttt{HltxGlobal} or \texttt{HltxPhys} also include some prescaled lines and do not reproduce well the efficiencies on MC. \footnote{To find the prescales one must:
\begin{enumerate}
    \item Define the TCK of the interesting data taking period:
        \begin{itemize}
        	\item \texttt{lb-run -c best Moore/v26r5 TCKsh}
        	\item \texttt{listConfigurations()} (lists of TCK and month in which they were used)
        \end{itemize}
        For \textit{Physics pp July 2016} we took \texttt{0x21361609} (\hlttwo) and \texttt{0x11361609} (\hltone)
    \item Search the prescale of one line \\
    \texttt{\noindent>>listProperties(0x21361609,".*Hlt2Topo2Body.*PreScaler","AcceptFraction")}\\
    
\end{enumerate}

}

To estimate the trigger efficiencies, one must also consider the overlap between different lines. To account for this to first approximation we also provide, for each line, the number of overlapped events with a reference line (one of the most efficient ones), and the relative gain on statistics reached by adding each single line to the reference one.

Last, it is also important to distinguish whether the trigger line was fired by the tracks of the \threepi decay, \ie \textit{trigger on signal} (TOS), or by other particles in the same event, \ie \textit{trigger independent of signal} (TIS).

The efficiencies on the MC sample are systematically listed for all trigger lines separating TIS and TOS decisions. The results are presented for the \elzero~(Tables~\ref{tab:trigMCL0TOS} and \ref{tab:trigMCL0TIS}), \hltone (Tables~\ref{tab:trigMCHLT1TOS} and \ref{tab:trigMCHLT1TIS}) and \hlttwo (Tables~\ref{tab:trigMCHLT2TOS} and \ref{tab:trigMCHLT2TIS}) trigger stages.


\onlyANA{
	\rev{
		I am not sure if I used the preselection also for MC. For sure it was used on data.	
	}
}

\onlyANA{
\rev{
	
\subsection{Combinations to be tested on data}

\textbf{Ideas that came up. But probably we want to already freeze the trigger. Nevertheless, the systematics would have to be studied at some point.}

Combinations
\begin{itemize}
    \item Giorgia:
    \begin{itemize}
        \item L0\_Hadron\_TOS $||$ L0\_Global\_TIS
        \item Hlt1TrackMVADecision\_TOS $||$  Hlt1TwoTrackMVADecision\_TOS
    \end{itemize}
	\item$D^0 \to \pip \pim \piz$  \\ \url{https://indico.cern.ch/event/993170/contributions/4234179/attachments/2191716/3704439/WG_Presentation_2021_3Pi\%283\%29.pdf}
		\begin{itemize}
			\item \verb! Dstr_L0Global_TIS || H1_L0HadronDecision_TOS ! \\ 
			\verb!||H2_L0HadronDecision_TOS || pi0_L0PhotonDecision_TOS!
			\item \verb! TrackMVADecision_TOS  !
		\end{itemize}
\end{itemize}

}

}

\subsection{Trigger efficiencies on data} \label{sec:trigData}

\begin{figure}
	\centering
	\includegraphics[width=0.48\linewidth]{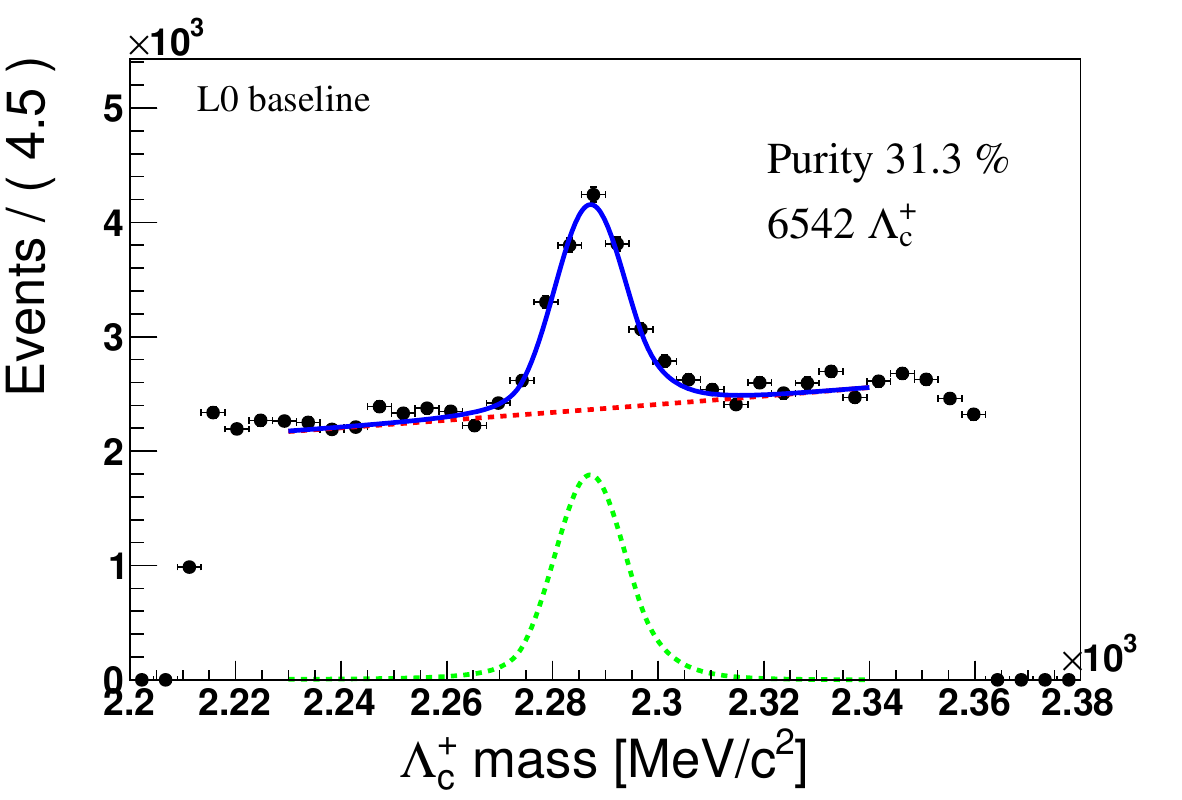}
	\includegraphics[width=0.48\linewidth]{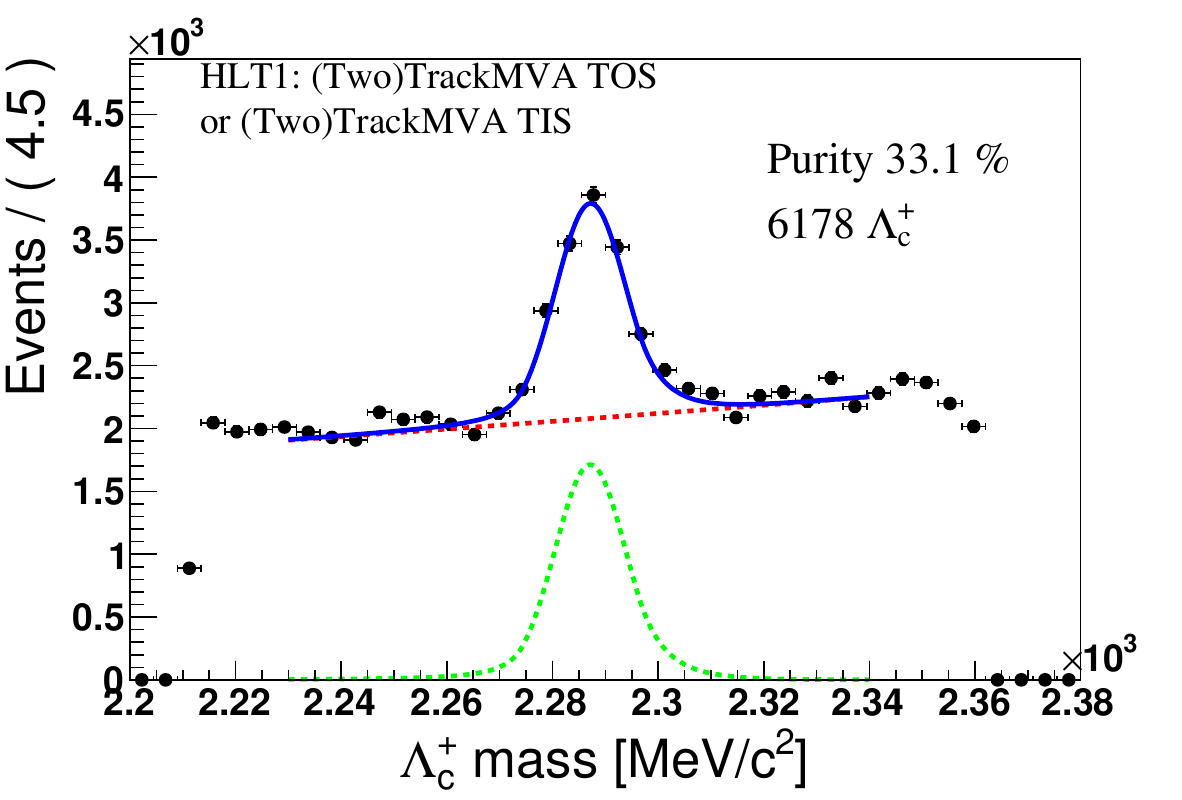}
	\caption{$\Lc$ invariant mass for two example trigger selections: (left) just the \elzero baseline and (right) also requiring \texttt{Hlt1(Two)TrackMVA} TIS or TOS.} 
	\label{fig:yieldsexampleplot}
\end{figure}

After the stripping requirement the real data sample is still overwhelmingly composed by combinatorial background events. In order to compare the yield of signal candidates in each combination of trigger lines, first of all we applied some preselection cuts on this data (given later in Section~\ref{sec:preselection}). With a visible peak on the invariant mass, the number of event candidates is extracted by fitting this distribution with a Gaussian PDF (for the signal) and an exponential function (for the background), as shown in Figure~\ref{fig:yieldsexampleplot}.

In the following, we will test several combination of trigger lines inspired by the efficiencies on MC data. In each trigger stage, we define a set of lines as a baseline to test the following stage.
The yields for \elzero, \hltone and \hlttwo lines are reported in Tables~\ref{tab:trigDataL0}, \ref{tab:trigDataHLT1} and  \ref{tab:trigDataHLT2}, respectively.

\begin{table}[h!]
\caption{Yield on data from different choices of L0 trigger lines. In bold letters, the baseline \elzero requirements. } 
\begin{tabularx}{\textwidth}{{>\raggedright}Xc}
\hline \hline 
             Trigger requirement      &         Yield   \\
\hline
   Only preselection  (and stripping) &  7126  \\
   L0: HadronTOS   &  2544  \\
   L0: HadronTOS or GlobalTIS    &  6838  \\
   {L0: HadronTOS or }{GlobalTIS except HadronTIS}   &  5085  \\
   L0: HadronTOS or ElectronTOS    &  2666  \\
   \textbf{{L0: HadronTOS or }{Had/El/Mu/Phot TIS}}   &  6542  \\
\hline \hline 
\end{tabularx} \label{tab:trigDataL0}
\end{table} 




\begin{table}[h!]
\caption{Yield on data ffor different choices of \hltone trigger lines. In bold letters, the baseline \hltone requirements. } 
\begin{tabularx}{\textwidth}{{>\raggedright}Xc}
\hline \hline 
             Trigger requirement      &         Yield   \\
\hline
   L0 baseline   &  6542  \\
   HLT1: TwoTrackMVA TOS   &  3823  \\
   HLT1: (Two)TrackMVA TOS   &  4477  \\
   {HLT1: (Two)TrackMVA TOS }{or Phys TIS }   &  6225  \\
   \textbf{{HLT1: (Two)TrackMVA TOS }{or (Two)TrackMVA TIS }}   &  6178  \\
   {HLT1: (Two)TrackMVA TOS }{or (Two)TrackMVA / TrackMuon TIS}   &  6246  \\
\hline \hline 
\end{tabularx} \label{tab:trigDataHLT1}
\end{table} 



\begin{table}[h!]
\caption{Yield on data for different choices of HLT2 trigger lines. In bold letters, the baseline \hlttwo requirements. } 
\begin{tabularx}{\textwidth}{{>\raggedright}Xc}
\hline \hline 
             Trigger requirement      &         Yield   \\
\hline
   L0 \&\& HLT1 baseline   &  6246  \\
   HLT2: Phys TOS or Phy TIS   &  3817  \\
   HLT2: Topo[2,3]Body TOS   &  1165  \\
   HLT2: Topo[2,3]Body / Incl[Dst,Sigc] TOS   &  1759  \\
   {HLT2: Topo[2,3]Body / Incl[Dst,Sigc] TOS }{or Phys TIS}   &  3030  \\
   \textbf{{HLT2: Topo[2,3]Body / Incl[Dst,Sigc] TOS or }{Topo(Mu,E)[2,3]Body / Incl[Dst,Sigc] TIS} }  &  2618  \\
   {HLT2: Topo[2,3]Body / Incl[Dst,Sigc] TOS or }{$>$3 \% on MC TIS}   &  2726  \\
   {HLT2: Topo[2,3]Body / Incl[Dst,Sigc] TOS or }{$>$3 \% on MC TIS or Turbo TOS}   &  3410  \\
\hline \hline 
\end{tabularx} \label{tab:trigDataHLT2}
\end{table} 



\onlyANA{
	
Why Phys TIS + TOS take away events. Is it because of the candidates?

\begin{table}[h]
	\caption{Yield on data from different trigger lines, preselecting events with only one candidate. Still removes 38\% events. } 
	\label{tab:caliblines}
	\begin{tabularx}{\textwidth}{Xc} 
		\hline
		Trigger requirement      &         Yield   \\
		\hline
		{L0 \&\& HLT1 baseline.}{ 1 candidate}   &  3986  \\
		{HLT2: Phys TIS or TOS.}{ 1 candidate}   &  2489  \\
		\hline 
	\end{tabularx} 
\end{table} 

The numbers in table \ref{tab:caliblines} suggest that 38\% of the interesting events are triggered \textbf{only} by calibration lines in HLT2.

}

\subsubsection{Discussion} 

In the \elzero and \hltone stages, we could select few lines recovering 91\% and 95\% of the initial signal events, respectively. In the \hlttwo stage, however, only 42\% of the events could be recovered, as shown in Table~\ref{tab:trigDataHLT2}. The bottleneck is clearly identified in the second row of this table. By requiring that either \texttt{Hlt2Phys\_TIS} or \texttt{Hlt2Phys\_TOS} fired in the event, 39\% of the signal events (that are recorded on disk) are directly lost. The \texttt{HltxPhys} lines contain the \textit{or} of all lines used for physics analyses, \ie not for calibration or other purposes. \onlyANA{\rev{In other words, it \textit{seems} that 39\% of signal events are recorded by calibration lines.}}
We have also tested if these low efficiencies may be due to the presence of several signal candidates in the same event. Essentially identical numbers were obtained by preselecting events with only one candidate.
%

Considering now the yields of the \texttt{Hlt2Phys} combo lines as the maximum reachable (with any \hlttwo trigger selection), we can try to get as close as possible with a reduced set of lines. With the lines in bold letters in Table~\ref{tab:trigDataHLT2}, we only recover 71\% of the events triggered by the \texttt{Hlt2Phys} lines (second row). As a test, we also tried adding a large list of TIS lines (adding a 3\% efficiency in MC) that are not prescaled, which in principle would not bias our sample since they fire independently of the signal. In real data, the gain (of 4\%) is approximately reproduced within statistical fluctuations. A significant gain of 30\% with respect to the baseline choice can apparently be achieved by adding all positive decisions on the charm Turbo TOS lines. However, the turbo stream is in principle not available for the stripping (since only reconstructed objects are saved), and these additional events must be fired by a large ensemble of  (non-Turbo) TOS lines. Determining which signal track was used by each TOS line, in each event,  is unfeasible and these lines are almost guaranteed to introduce instabilities and abrupt shapes in several variables of our sample.

\onlyANA{
\begin{table}[h!]
\caption{Yield on data from different choices of Hlt2 Turbo trigger lines } 
\begin{tabularx}{\textwidth}{{>\raggedright}Xc}
\hline
             Trigger requirement      &         Yield   \\
\hline
   {HLT2: Topo3Body TOS }{(no L0,Hlt1)}   &  902  \\
   {HLT2: Topo3Body TOS or }{Dsp2KS0PimPipPip Turbo TOS (no L0,Hlt1)}   &  2032  \\
   {HLT2: Topo3Body TOS or }{Dp2KS0PimPipPip Turbo TOS (no L0,Hlt1)}   &  1752  \\
   {HLT2: Topo3Body TOS or }{Xim2LamPim Turbo TOS (no L0,Hlt1)}   &  1014  \\
\hline 
\end{tabularx} 
\end{table} }

\subsection{Summary}

For the dataset without a dedicated \hlttwo trigger line (analysed in this thesis), we selected a set of trigger lines based on their yields on real data. The goal was to maximize the efficiency while keeping a reduced number of lines. Especially, for the TOS category, different lines may have different cut values on the same variable, which would produce a line shape with step-like functions causing instabilities in the offline selection and final fit. 
The final trigger selection is presented in the following. %
At least one line in each of the trigger stages must be fired to select the event. \onlyANA{This is, we use the binary operator \textit{or} within a trigger stage, and \textit{and} between the stages.}
~\\

\begin{minipage}[t]{0.4\linewidth}
\begin{itemize}
	\item \textbf{L0}
	\begin{itemize}
		\item \verb|Hadron_TOS |
		\item \verb|Hadron_TIS |
		\item \verb|Electron_TIS |
		\item \verb|Muon_TIS |
		\item \verb|Photon_TIS |
	\end{itemize}
	~
	
	\item \textbf{HLT 1}
	\begin{itemize}
		\item \verb|TrackMVA_TOS|
		\item \verb|TwoTrackMVA_TOS|
		\item \verb|TrackMVA_TIS|
		\item \verb|TwoTrackMVA_TIS|
	\end{itemize}

\end{itemize}
\end{minipage}
\begin{minipage}[t]{0.6\linewidth}
\begin{itemize}	

	\item \textbf{HLT 2}
	\begin{itemize}
		\item \verb|CharmHadInclSigc2PiLc2HHXBDT_TOS|
		\item \verb|CharmHadInclDst2PiD02HHXBDT_TOS|
		\item \verb|Topo2Body_TOS|
		\item \verb|Topo3Body_TOS|
		\item \verb|CharmHadInclSigc2PiLc2HHXBDT_TIS|
		\item \verb|CharmHadInclDst2PiD02HHXBDT_TIS|
		\item \verb|Topo2Body_TIS|
		\item \verb|Topo3Body_TIS|
		\item \verb|TopoMu2Body_TIS|
		\item \verb|TopoMu3Body_TIS|
		\item \verb|TopoE2Body_TIS|
		\item \verb|TopoE3Body_TIS|
	\end{itemize}
\end{itemize}
\end{minipage}

\onlyANA{
	\rev{

\subsection{Thresholds on the variables from the combination of TOS lines  - to be investigated} \label{sec:varthresholds}

\begin{itemize}
	\item Only interested in variables that I use for the selection or the BDT. If there are thresholds in others, but they are decorrelated to the interesting variables I do not care.
	\item This peaking thresholds should be reproduced in signal MC? Obviously they appear in the background distribution(because you have more or less bkg to each side of the threshold). They appear on signal if they also remove a significant portion of the signal. 
	\item maxDoca
	\item Problem with the threshold in pip2\_Lc\_ProbNNpi. Probably comming form a TOS trigger line. It should not come from the stripping containers (since all the pions are filtered in the same way, and we do not have the combination of two stripping lines).
	\item bachelors combined pt
\end{itemize}

}}

\onlyANA{
\section{Summary of efficiencies - to be completed in the ANANOTE}

\textbf{TO BE CHANGED! The trigger requiremets are not the final ones, and they were not differentiating TIS and TOS}

	{\small
	\begin{longtable}{lcc}
		\hline
		${}^{\S}$ & LL & DD \\ \hline
		MCMatching~  & $1.3\times 10^{-3}$ &  $1.3\times 10^{-3}$ \\ 
		\& Stripping \textbf{28r2} & $6.9\times 10^{-4}$ &  $6.5\times 10^{-4}$ \\ 
		\& L0Global & $1.9\times 10^{-4}$ &  $2.4\times 10^{-4}$ \\ 
		\& L0Hadron ${}^\ddag$ & $1.2\times 10^{-4}$ &  $1.6\times 10^{-4}$ \\ 
		\& (Hlt1TrackMVA $||$ Hlt1TwoTrackMVA)${}^\ddag$ & $9.3\times 10^{-5}$ &  $8.3\times 10^{-5}$ \\  
		\& \textit{Hlt2FourLines}~${}^\dag$ ${}^\ddag$ & $\mathbf{2.2\times 10^{-5}}$ &  $\mathbf{2.5\times 10^{-5}}$~* \\   \hline 
	\end{longtable} 
}
}
\chapter{Offline selection}  \label{ch:selection}

Once we have an initial dataset with some basic requirements specified in the stripping and trigger selections, we can genuinely start its (offline) analysis.
A soft preselection to improve the purity is defined in Section~\ref{sec:preselection}. The MC samples are recalibrated to better reproduce detector effects, and some treatments are also applied to the data for different purposes. We will refer to these methods as \textit{corrections to MC and data} for short, in Section~\ref{sec:corrMCData}. With these corrections, we will apply the last offline selection by configuring a Boosted Decision Tree, in Section~\ref{sec:BDT}. Lengthy tables and graphics from this chapter are displayed in Appendix~\ref{app:analysistables}.

\section{Preselection} \label{sec:preselection}

\begin{figure}
	\centering
	\includegraphics[width=0.45\linewidth]{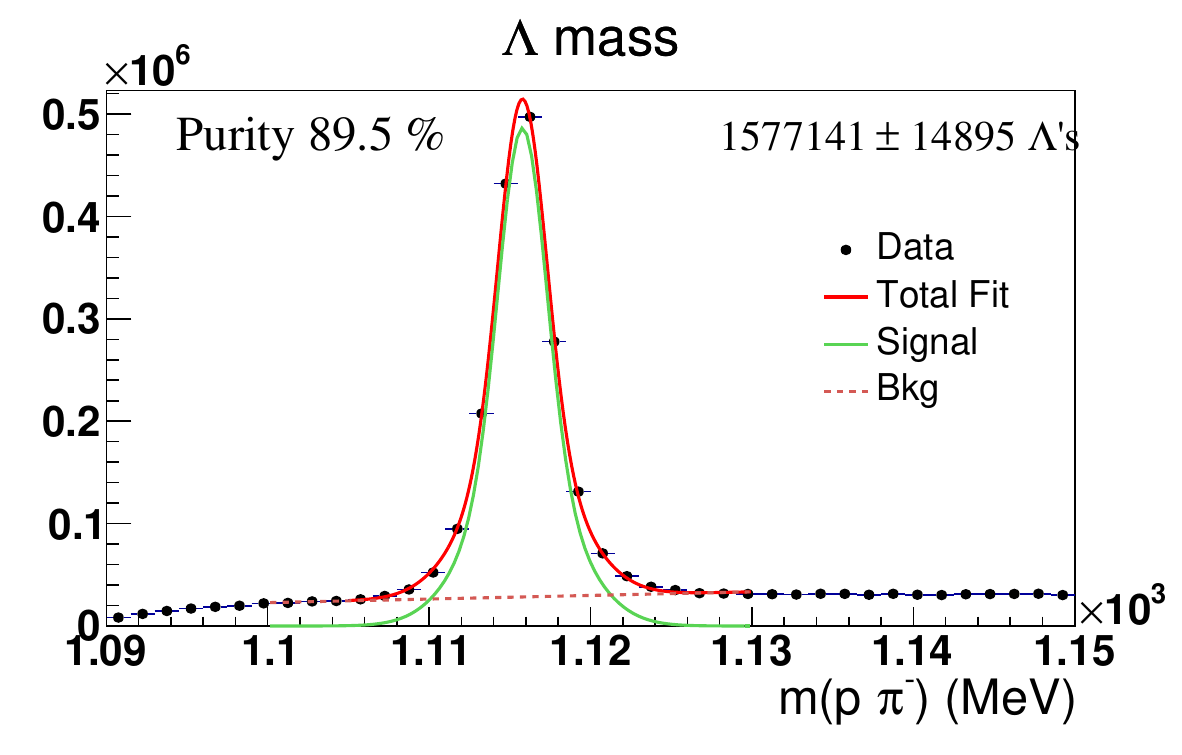}
	\includegraphics[width=0.45\linewidth]{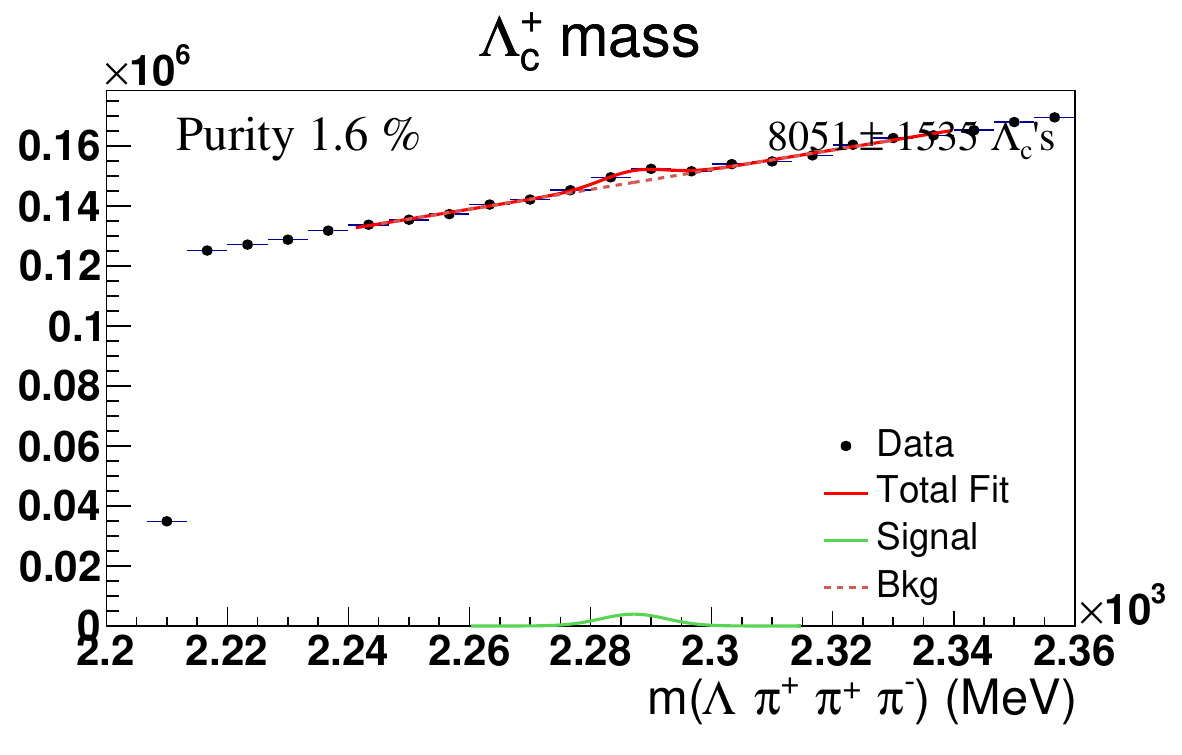}
	\caption{Invariant-mass distribution for (left) $\Lz$ and (right) $\Lc$ after the stripping and trigger requirements. 
		Preliminary binned fit to the \Lz (\Lc) distribution with an exponential function and the mixture of two Gaussian (a single Gaussian) PDFs.
		Subset ($\approx 9\%$) of 2016 data. 
		The vertical axes indicate the number of events per bin.
		\onlyANA{\red{To be replaced with new (TIS and TOS) requirements}}}
	\label{fig:initialPeaks}
\end{figure}

The initial data sample after stripping and trigger requirements is overwhelmingly composed of combinatorial background. While most of the candidates contain real $\Lz$'s, one can barely distinguish any structure in the $\Lc$ invariant mass, as shown in Figure \ref{fig:initialPeaks}. The main goal of this preselection is to reach a signal purity of 30\% (within 2$\sigma$ of the centre of the mass peak), in order to apply the sPlot technique\footnote{See definition in Appendix~\ref{app:analysisterms}.} reliably. The choice for this value is however just a guess, and it is compared with 20\%- and 40\%-purity preselections in Appendix \ref{sec:puritysweight}.

We identified the most discriminating variables by comparing the distribution in MC and real data for about $170$ variables. The first attempts to this selection were done through a combination of soft cuts in many variables
. This approach complicates the description of the selection, preventing an intuitive understanding of the cuts. More importantly, we observed that this strategy reduces the performance of a subsequent multivariate classifier. Thus, we tried to reduce the number of cuts to the bare minimum, keeping similar background rejection and signal efficiencies. In the following, this set of cuts is presented, differentiating between $\Lz$- and \Lc-specific cuts.

\subsubsection{$\Lz$ background}

\begin{figure}
	\centering
	\includegraphics[width=0.45\linewidth]{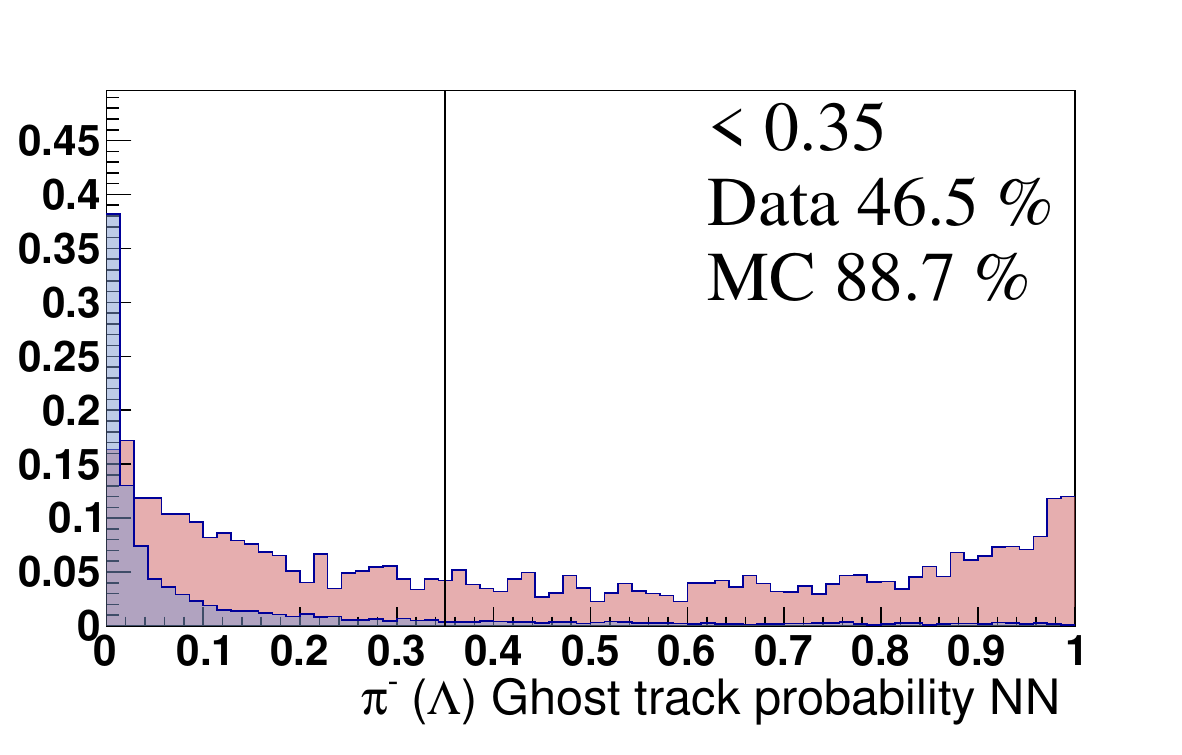} 
	\includegraphics[width=0.45\linewidth]{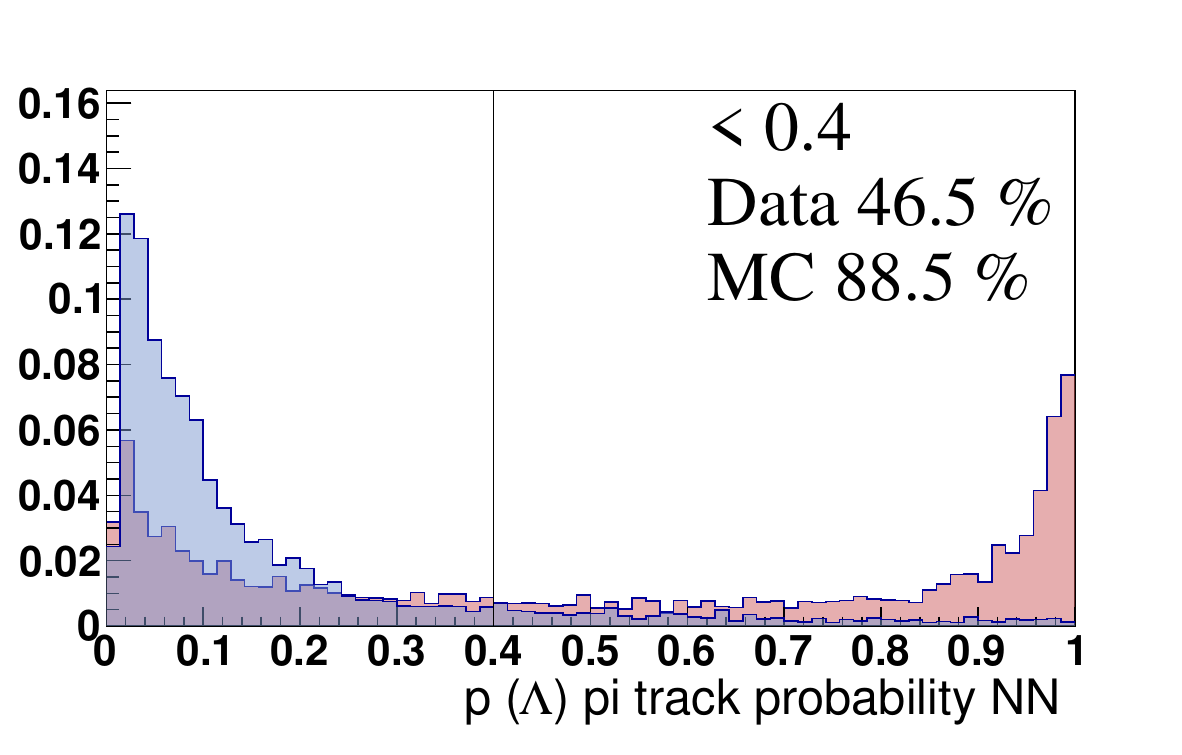}  
	\caption{(Left) ghost probability of the \Lz pion and (right) probability of the proton to be a pion for (red) data and (blue) signal MC distributions, together with the cut value and the corresponding efficiencies. The data distribution contains mostly combinatorial background events. The vertical axes indicate the relative number of events per bin. The histograms are rescaled to allow their comparison. }
	\label{fig:initial_pi_L0_ProbNNghost}
\end{figure}

\begin{figure}
	\centering
	\includegraphics[width=0.45\linewidth]{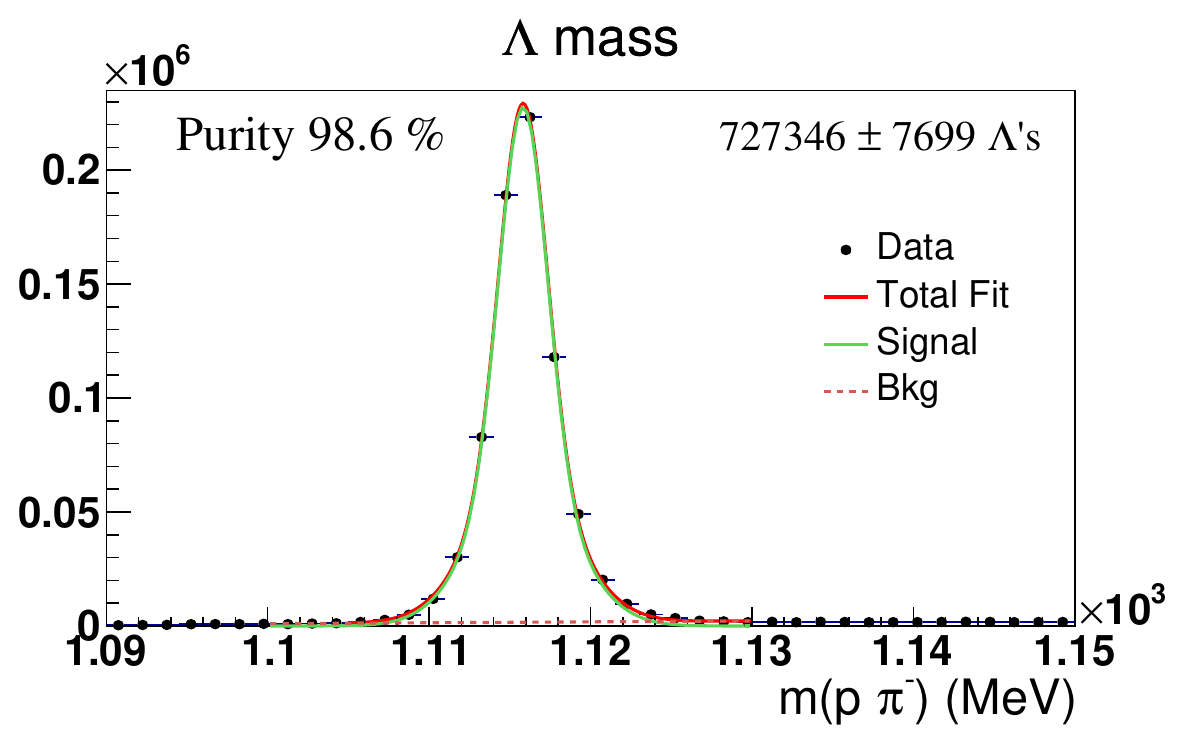}
	\includegraphics[width=0.45\linewidth]{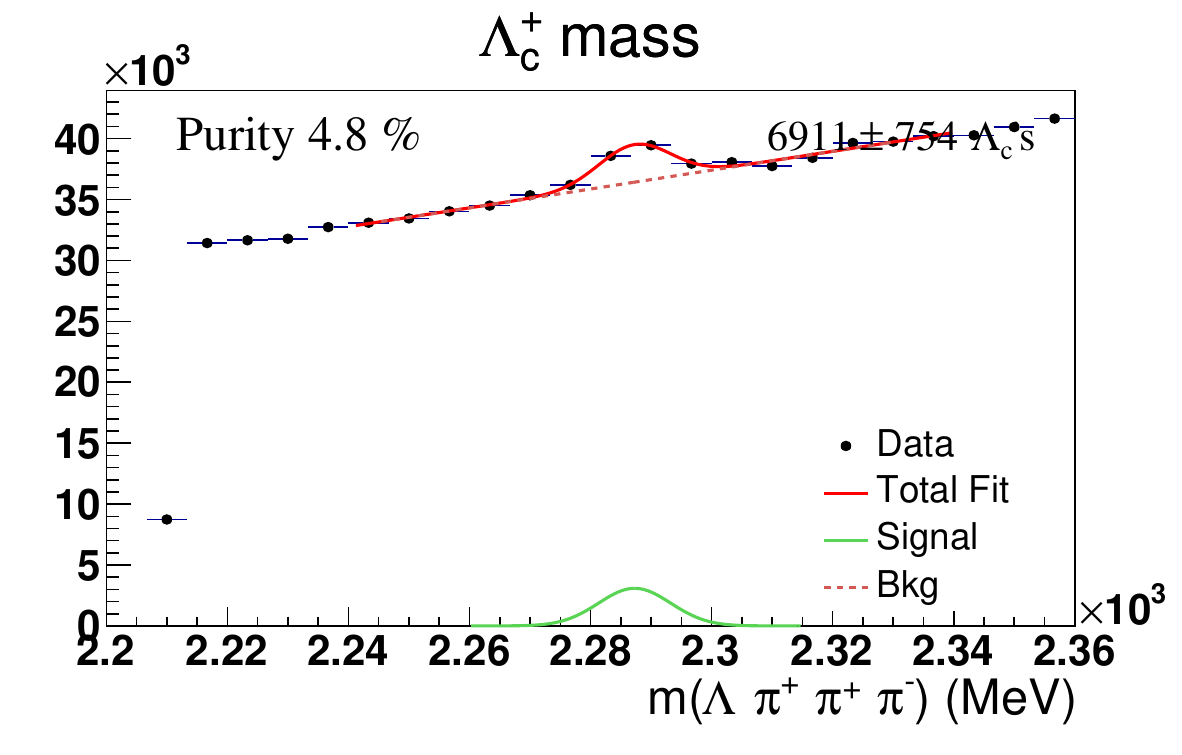}
	\caption{Invariant-mass distribution for (left) $\Lz$ and (right) $\Lc$ after the stripping and trigger requirements. Subset of 2016 data. The vertical axes indicate the number of events per bin. \onlyANA{\red{To be replaced with new (TIS and TOS) requirements}}}
	\label{fig:onlyLambdaCutsPeaks}
\end{figure}

Most of the $\Lz$ background can be removed with cuts on only two variables of the proton and pion. The variable distributions are shown in Fig.~\ref{fig:initial_pi_L0_ProbNNghost}. The efficiency of the combined cuts and the invariant mass distribution are shown in Table \ref{tab:onlyLambdaCuts} and  Fig. \ref{fig:onlyLambdaCutsPeaks}, respectively. These cuts also improve the purity on the $\Lc$ peak by about a factor 3.

%

\begin{longtable}{lccc}
	\caption{Efficiencies of the cuts specific for the $\Lz$ background.} \label{tab:onlyLambdaCuts} \\
	\hline	\hline
	Variable      &           Cut  &    $\varepsilon$(Data)  &      $\varepsilon$(MC)  \\*
	\hline
	p ($\Lambda$) pi track probability NN      &    $<$    0.4 &    46.2\%  &    90.7\%  \\
	$\pi^{-}$ ($\Lambda$) Ghost track probability NN      &    $<$   0.35 &    52.4\%  &    89.4\%  \\
	\hline   
	
	All cuts     &                &  \textbf{ 23.7\%}  &  \textbf{ 80.8\%}  \\
	\hline \hline
\end{longtable}

\pagebreak

\subsubsection{\Lc background}

\begin{figure}
	\centering
	\includegraphics[width=0.45\linewidth]{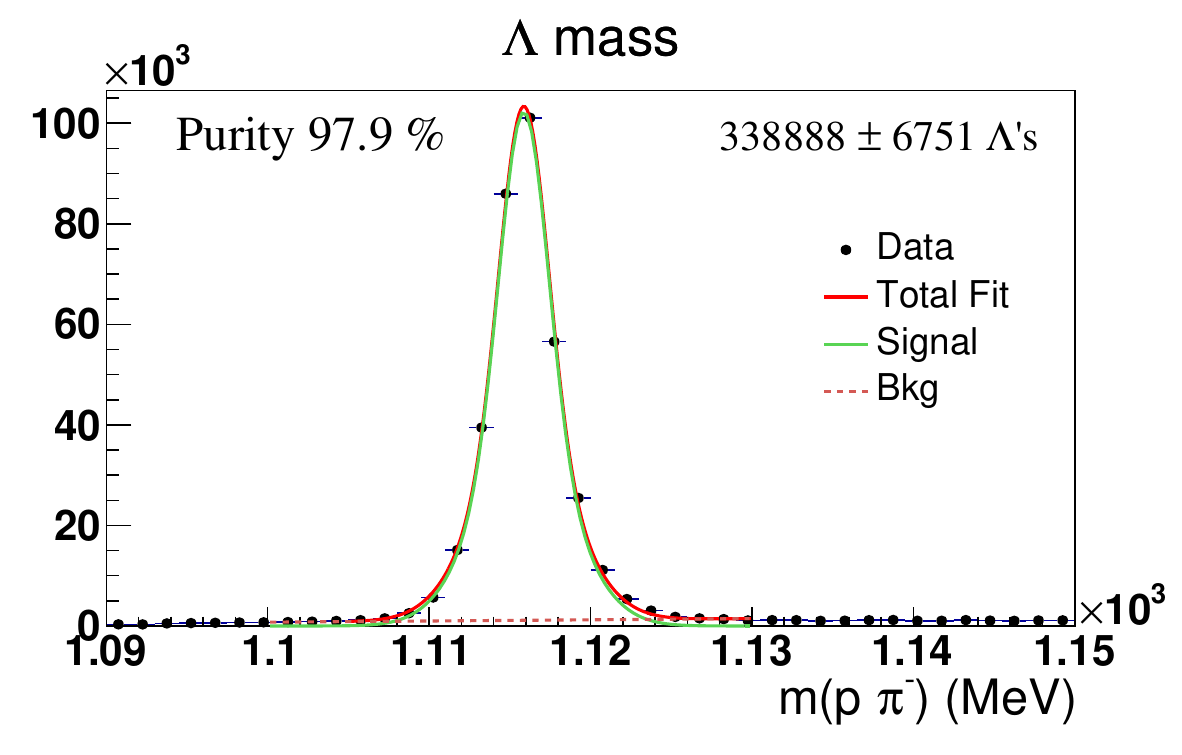}
	\includegraphics[width=0.45\linewidth]{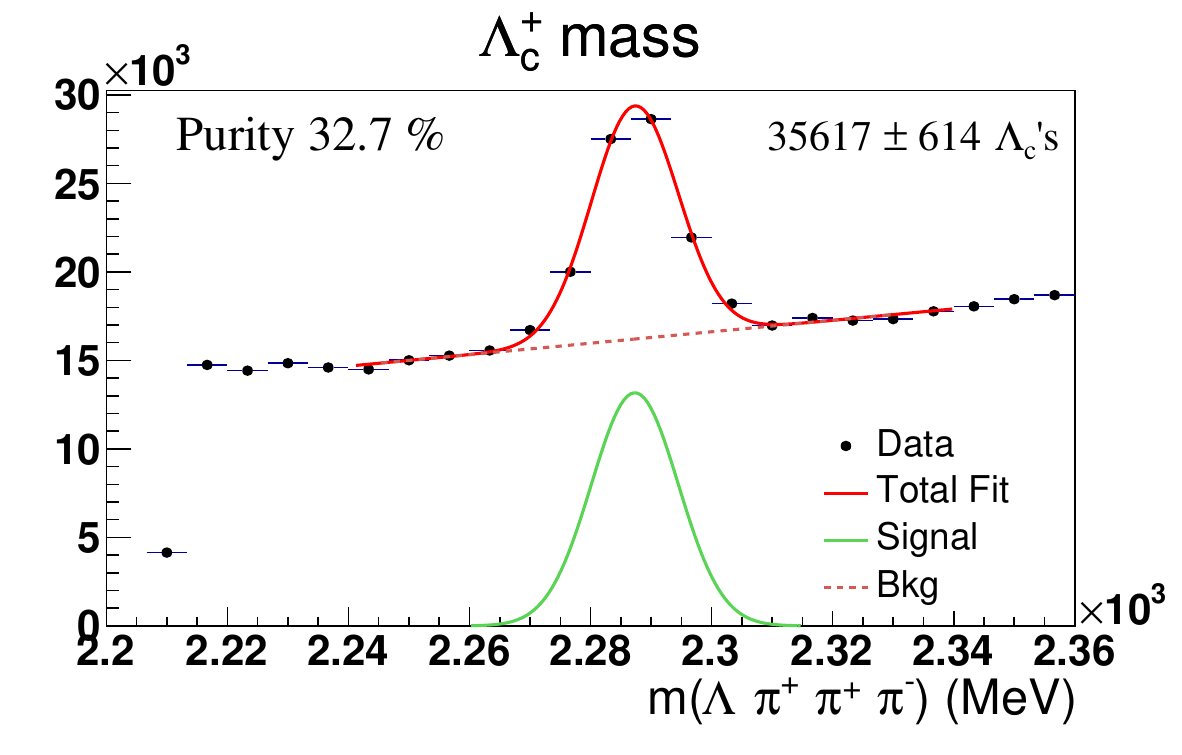}
	\caption{Invariant-mass distribution for (left) $\Lz$ and (right) $\Lc$ after the preselection. Complete 2016 \lDD dataset.
		The vertical axes indicate the number of events per bin.
		 \onlyANA{ \textbf{Showing the full 2016 data set, with the correct trigger (TIS and TOS) requirements}}}
	\label{fig:preselectionPeaks}
\end{figure}

\begin{figure}
	\centering
	\includegraphics[width=0.45\linewidth]{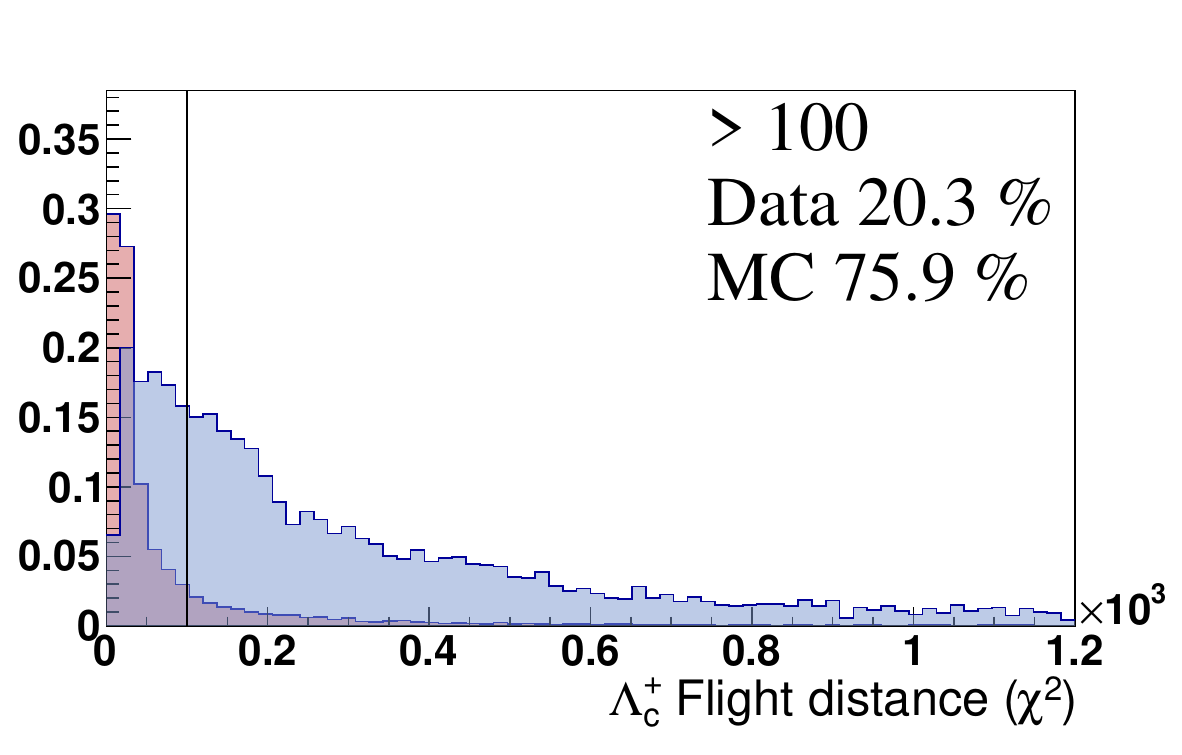}
	\includegraphics[width=0.45\linewidth]{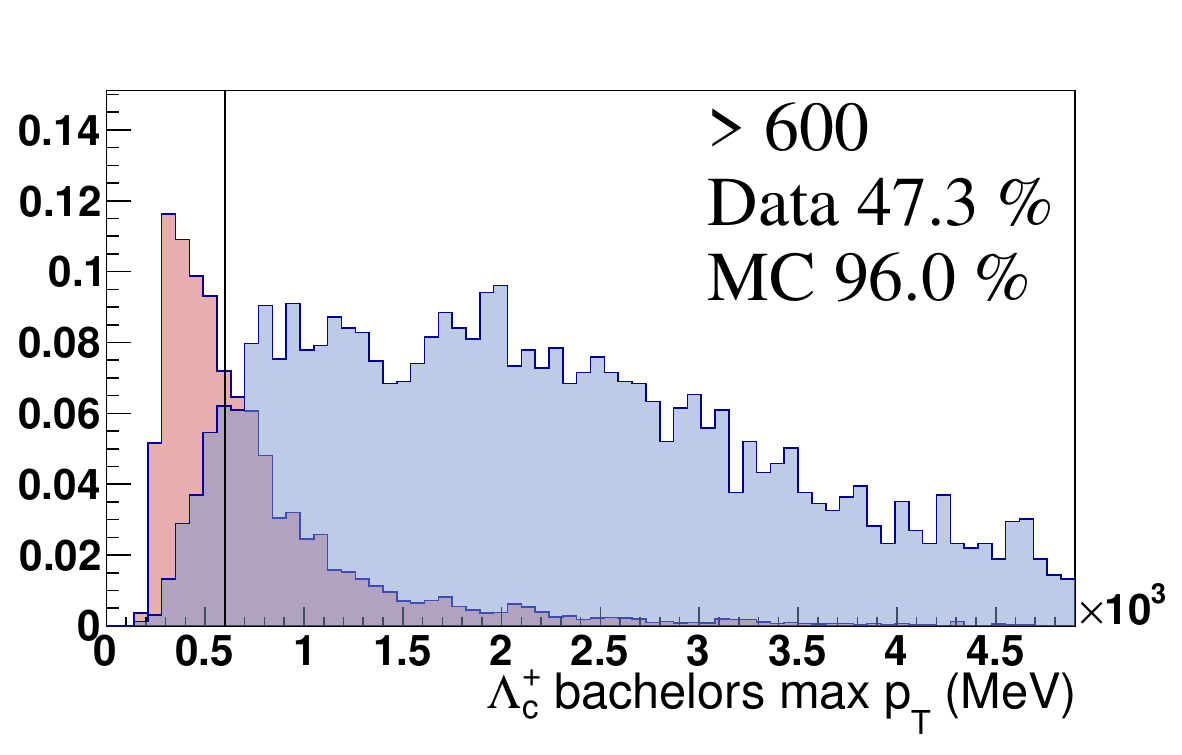}
	\includegraphics[width=0.45\linewidth]{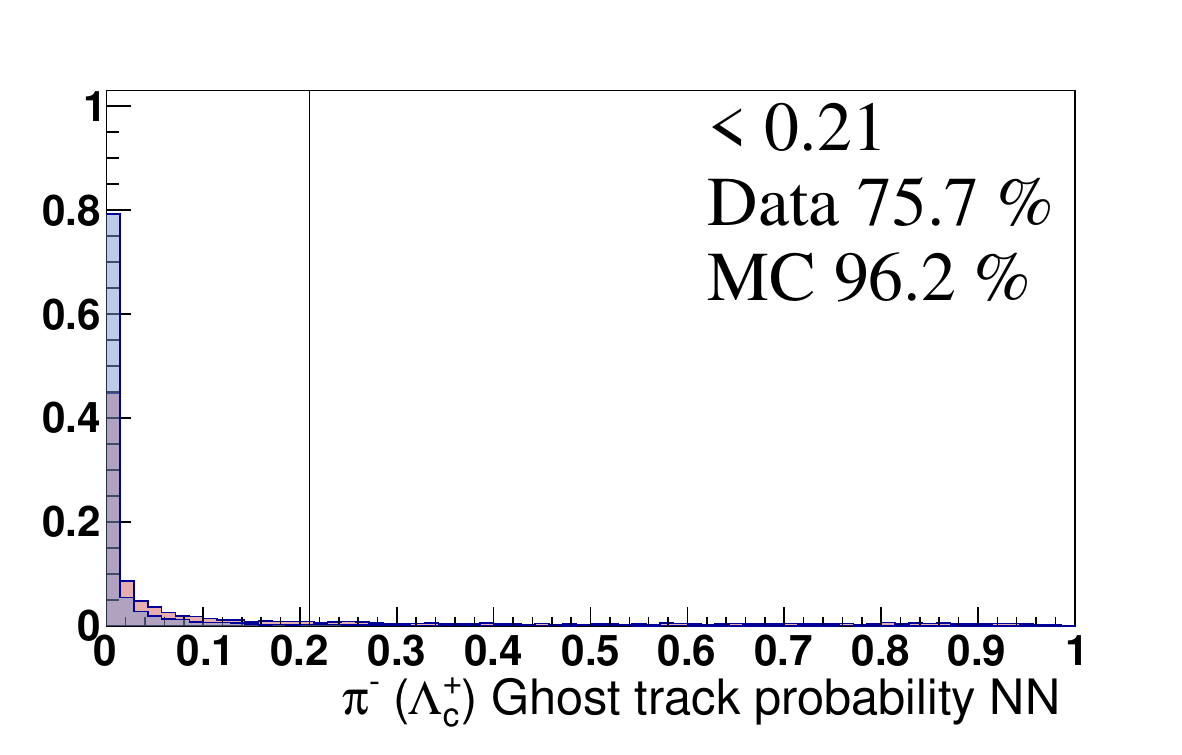}
	\includegraphics[width=0.45\linewidth]{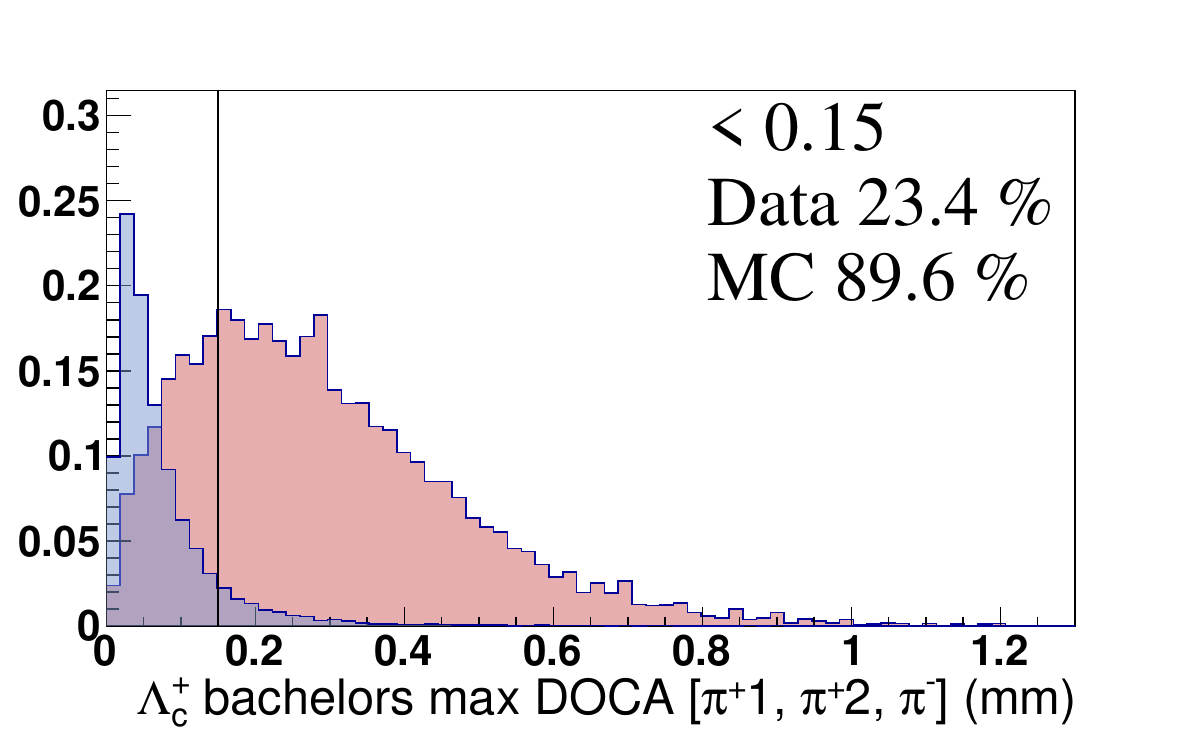}
	\caption{Distribution of variables in the preselection for (red) data and (blue) signal MC, together with the cut value and the corresponding efficiencies. The vertical axes indicate the relative number of events per bin.}
	\label{fig:preselectionVariables}
\end{figure}

The purity goal of 30\% can be reached with four additional cuts, determined with the distributions of  Figure \ref{fig:preselectionVariables}. 
Note that the condition on the \Lc vertex is applied only on the 3\pipm tracks. Thus, there are no further requirements on the \Lz and 3\pipm combination besides those in the stripping selection.
Table \ref{tab:preselectionCuts} contains the final efficiencies of the preselection, and Figure \ref{fig:preselectionPeaks} the resulting invariant mass distribution.

\begin{table}[] 
	\centering
	\caption{Set of cuts and corresponding efficiencies in the nominal preselection. } 
	\begin{tabular}{lccc}
		\hline \hline	
		Variable      &           Cut  &    $\varepsilon$(Data)  &      $\varepsilon$(MC)  \\
		\hline
		p ($\Lambda$) pi track probability NN  Corrected     &    $<$    0.4 &    46.5\%  &    88.5\%  \\
		
		$\pi^{-}$ ($\Lambda$) Ghost track probability NN      &    $<$   0.35 &    46.5\%  &    88.7\%  \\
		\hline   
		$\Lambda_{c}^{+}$ Flight distance ($\chi^{2}$)     &    $>$    100 &    20.3\%  &    75.9\%  \\
		$\Lambda_{c}^{+}$ bachelors max $p_{T}$     &    $>$    600 &    47.3\%  &    96.0\%  \\
		
		$\pi^{+/-}$ ($\Lambda_{c}^{+}$) Ghost track probability NN      &    $<$   0.21 &    75.7\%  &    96.2\%  \\
		
		$\Lambda_{c}^{+}$ bachelors max DOCA [$\pi^{+}$1, $\pi^{+}$2, $\pi^{-}$]     &    $<$   0.15 &    23.4\%  &    89.6\%  \\
		\hline 
		All cuts     &                &   1.1\%  &   48.6\%  \\
		\hline \hline
	\end{tabular}
	\label{tab:preselectionCuts}
\end{table}

\onlyANA{
	
	\subsection{PID of the proton}  \label{sec:ProbNNp}
	
	\red{\textbf{This issue with the proton PID only applies for LL events.}}

	The stripping contains a cut on the proton PID, \verb|p_L0_ProbNNp>0.10|, that is applied in the production of the filtered signal MC. The subsequent correction on the PID variables (here using \verb|PIDCorr|) may lead to threshold effects, thus losing the agreement between real data and MC. For that reason, we apply yet a tighter cut on this variable in the preselection, counteracting the threshold effects due to the PID correction. To determine the safe value for this cut, we evaluate the standard deviation of the PID correction for events below the threshold. A small sample of non-filter MC signal (\textit{LDST} format) is used.
	
	\red{Unfortunately, this sample contains only 1M generated events. An alternative would be to re-strip the sample with 5M events in MDST format. This is possible in principle, see \href{https://twiki.cern.ch/twiki/bin/view/LHCb/RestrippingMDST}{RestrippingMDST} twiki page. However, after much work on that, I could not get the restripping to work on MDST.}
	
	\pagebreak
	
	With the available statistics, only $\sim 5\%$ of the events go above the original threshold at 0.10, and none goes above 0.30. Thus, we include another cut in the preselection (with PID corrected MC) at \verb|p_L0_ProbNNp>0.50|.
	
	\begin{figure}
		\centering
		\includegraphics[width=0.45\linewidth]{content/figs/preselection/checkDispProbNNp/ProbNNp}
		\includegraphics[width=0.45\linewidth]{content/figs/preselection/checkDispProbNNp/ProbNNpCorr} \\
		\includegraphics[width=0.45\linewidth]{content/figs/preselection/checkDispProbNNp/ProbNNp2D}
		\caption{Proton ProbNNp for events below 0.10 on this variable. Distribution for (left) original MC signal, (right) PID-correctected MC, and (bottom) correlation of the two.}
		\label{fig:probnnpcorr}
	\end{figure}
	
}

\subsubsection{\Lz flight}

To speed up the MC production, we used some generator-level cuts, which remove events before the reconstruction step of the simulation is applied. This set of cuts removes events that would not pass the stripping selection (\textit{e.g.} very-low momentum tracks) and others that are directly outside the LHCb acceptance (\textit{e.g.} low or high pseudorapidity). Among the latter, the \Lz decay (end) vertex along the $z$ axis was required to be below 2400 mm\footnote{See the Decfiles in the \href{https://gitlab.cern.ch/lhcb-datapkg/Gen/DecFiles/-/commit/f753a692df245c7eda1a2ad760612b68f5ce968c}{LHCb GitLab}.}, since the first TT station is positioned at 2350 mm~\cite{Gassner:2004yda}.

However, this generator-level cut was applied by mistake before the smearing of the PV, which reproduces the distribution of $pp$ collisions along the beam, happening approximately in a range of 10\,\cm. The \Lz decay vertex is repositioned according to this smearing, and some of the removed events would have ended up within the acceptance. Then, to avoid possible threshold effects, we apply a tighter cut on this variable, $z(\Lz \text{ decay})<2300 \mm$.

\begin{figure}[h!]
	\centering
	\includegraphics[width=0.45\linewidth]{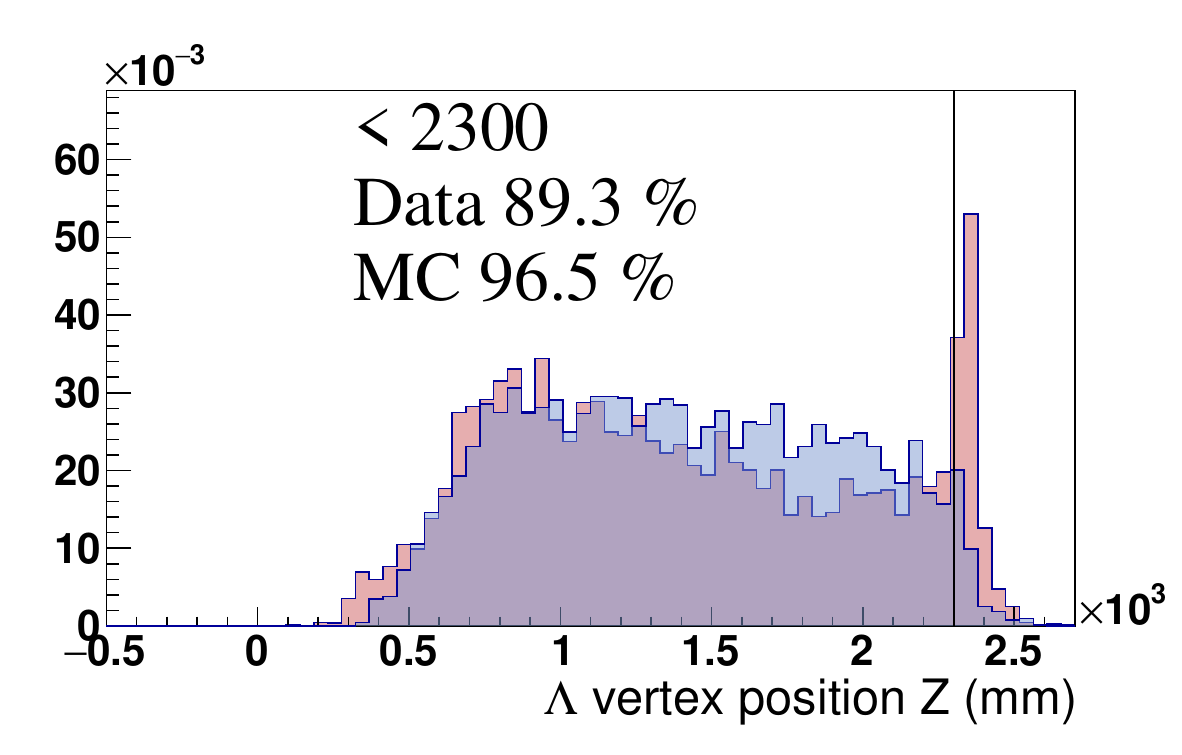}
	\caption{End vertex $z$ position of the $\Lz$ baryon for (red) data and (blue) signal MC. The events between $500$ and $1000 \mm$ correspond to $\Lz$ candidates that decay inside the VELO but are reconstructed as DD (and not as LL). The further away from the interaction point, the less likely to be reconstructed by the VELO modules.
		The peak at 2350\,\mm corresponds to random $p$ and $\pim$ tracks generated in material interactions at the first TT module that were combined contributing to the \Lz combinatorial background. By applying the displayed cut (vertical line) we are also removing these events (although the majority of them were already removed by the preselection). The distributions are normalized to unity, with the vertical axis indicating the normalized number of events per bin.
		%
	}
	\label{fig:L0_FD_Z}
\end{figure}

Part of the \Lc combinatorial background comes from material interactions in the first TT module. This produces a peaking structure in $z(\Lz \text{ decay})$ (shown in Figure \ref{fig:L0_FD_Z}), meaning that some \Lz candidates are originated in this plane, and matched \textit{accidentaly} to a three-track vertex.

\subsubsection{Negative lifetime and $\chi^2$ variables}

Some events have negative values for variables that are defined positive while having perfectly normal values for other variables. These effects are, in general, known within the experiment. Since we would like to use all the available information (variables) in the multivariate classifier (Section~\ref{sec:BDT}), for the time being, we will cut out these events, which represent less than 1\% of the sample. The events with negative \Lz flight distance $\chi^2$ represent the 0.2\% (0.06\%) of the real data (signal MC) sample. The events with negative \Lz lifetime or uncertainty-normalized lifetime ($\chi^2$) represent the 0.1\% (0.07\%). Finally, the events with negative \Lc lifetime, uncertainty-normalized lifetime, or lifetime-fit $\chi^2$ amount to 0.4\% (0.03\%). All of these events together, accounting for overlaps, compose the 0.7\% (0.2\%) of the preselected samples. \onlyANA{There are also 3 events, on real data, that have \verb|Lc\_minDOCACHI2| and \verb|Lc_minBachDOCACHI2| $<$0, which are removed as well. The summary with the number of eliminated events, and their e

{\scriptsize
	\begin{longtable}{lccccc}
		\centering
	\caption{Variables defined positive, in principle, for which there are events with negative values. Among the events with negative values (second and fourth column) there are some with specific error code values (third and fifth columns), identifying failed fits in the offline reconstruction. The rest may be due to numerical precision in the data processing. The analyzed events have passed the preselection described above. "Real Data" includes background from the side bands.} \label{tab:negativevalues} \\
	\hline
	& \multicolumn{2}{c}{ Real Data } & \multicolumn{2}{c}{ Signal MC } &  Error value \\ 
	condition & events &  with error & events &  with error  &   \\ \hline
	all events  & 388042 & - & 31019 & - &  \\ 
	L0\_FD\_CHI2\_OWNPV (L0\_BPVVDCHI2) $<0$  & 716 & - & 11 & - &  - \\ 
	L0\_FD\_CHI2\_ORIVX $<0$  & 482 & - & 13 & - &  - \\ 
	L0\_FD\_CHI2\_OWNPV $<0$ $|$  L0\_FD\_CHI2\_ORIVX $<0$  & 904 & - & 19 & - & - \\ \hline
	L0\_TAU $<0$  & 205 & 156 & 9 & 8 &  $-10^{2}$ \\ 
	L0\_TAUCHI2 $<0$  & 304 & - & 23 & - & - \\ 
	L0\_BPVLTIME $<0$  & 276 & 190 & 17 & 13 &  $-10^{10}$ \\ 
	L0\_TAU $<0$ $|$ L0\_TAUCHI2 $<0$ $|$ L0\_BPVLTIME $<0$ & 517 & - & 23 & - & - \\ \hline
	Lc\_TAU $<0$  & 1515 & 14 & 10 & 1 &  $-10^{2}$ \\ 
	Lc\_BPVLTIME $<0$  & 1515 & 14 & 10 & 1 &  $-10^{10}$ \\ 
	Lc\_BPVLTSIGNCHI2 $<0$  & 1515 & 14 & 10 & 1 &  $-10^{10}$ \\ 
	Lc\_TAUCHI2 $<0$  & 14 & 14 & 1 & 1 &  $-10^{2}$ \\ 
	Lc\_BPVLTCHI2 $<0$  & 14 & 14 & 1 & 1 &  $-10^{10}$ \\ 
	Lc\_BPVLTFITCHI2 $<0$ & 14 & 14 & 1 & 1 &  $-10^{10}$ \\ 
	any of the Lc variables & 1515 & - & 10 & - &  $-10^{2}$ \\ \hline
\end{longtable}
}

}

\subsubsection{Angular distribution distorsion}

\begin{figure}
	\centering
	\includegraphics[width=0.45\linewidth]{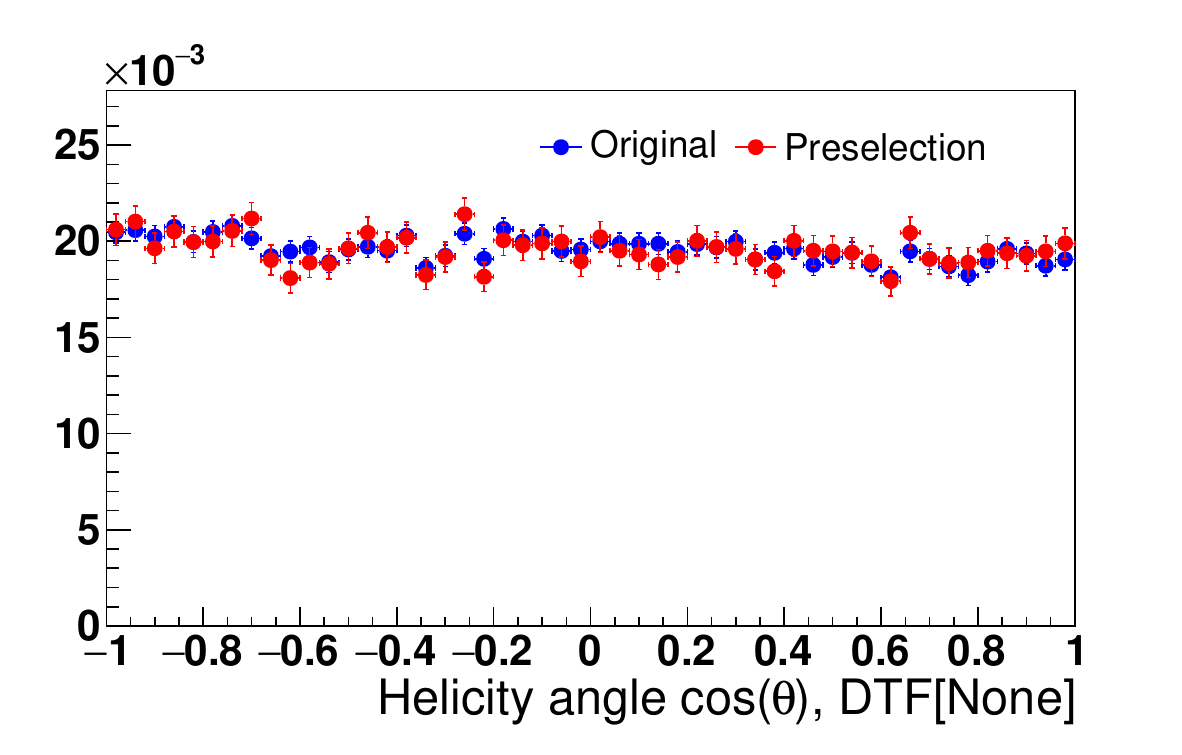}
	\includegraphics[width=0.45\linewidth]{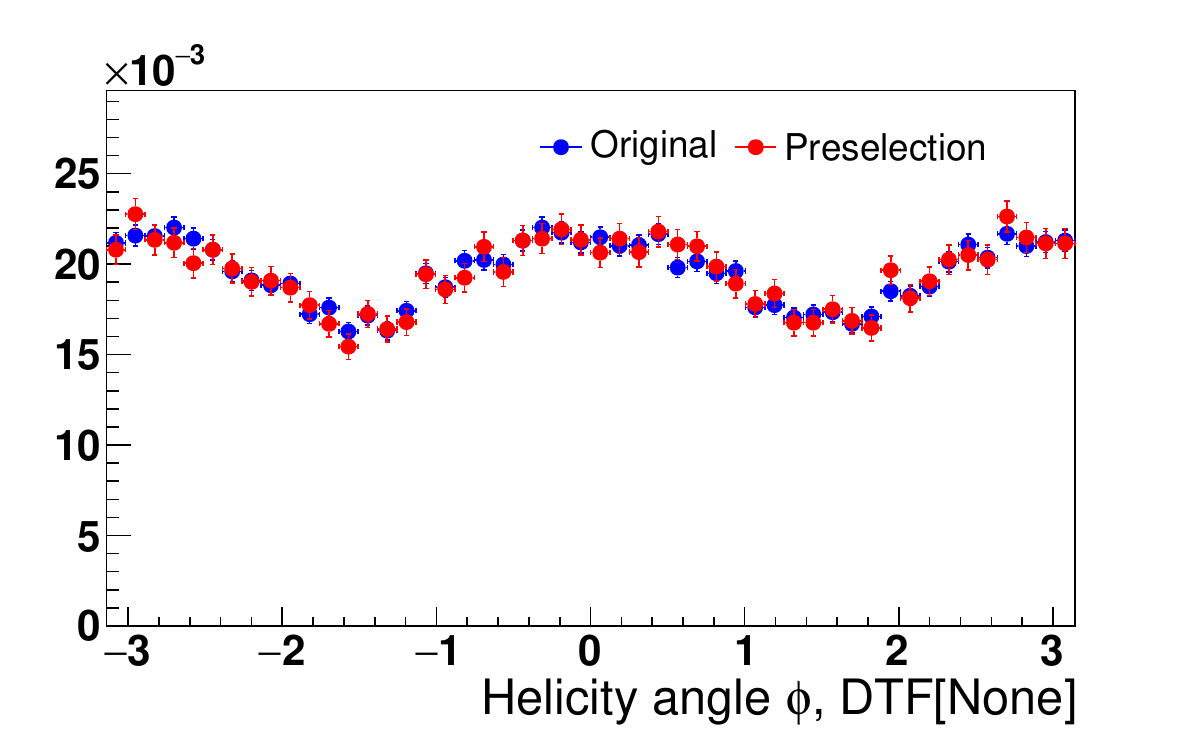}
	\caption{Distribution of the helicity angles in signal MC before and after the preselection. These variables are calculated from refitted momenta (DTF) but no mass or vertex constraints. The distributions are normalized to unity, with the vertical axes indicating the normalized number of events per bin.}
	\label{fig:anglesPreselection}
\end{figure}

To check that the preselection cuts are not biasing the final measurement, we compare the distribution of the angular variables before and after these cuts. Even though up to 50\% of signal events were removed with the preselection, these distributions are practically identical before and after the cuts, as shown in Figure \ref{fig:anglesPreselection}.

\section{Corrections to Monte Carlo and data} \label{sec:corrMCData}

To improve the agreement between signal MC and real-data signal events, several corrections are applied to the simulation. Some treatments are also applied to data to recalibrate the momentum reconstruction (momentum scaling), to simultaneously use the complete vertex and momentum information (Decay Tree Fitter) 
and to reweight the events mimicking a pure signal sample (sWeights).
Although some of these data treatments are not strictly corrections (there was nothing \textit{wrong} in first place), we will refer to these methods as \textit{corrections to MC and data}, summarized in Table~\ref{tab:correctionssummary}.

\begin{table}[ht!]
	\centering
	\caption{Corrections to Monte Carlo and data. These corrections are either generic($^*$), \ie defined from calibration samples or using generic reconstruction algorithms, or specific($^\dag$) to the dataset.}
	\label{tab:correctionssummary}
	\resizebox{1\linewidth}{!}{
		\begin{tabular}{ p{5cm} p{4cm} p{4cm} }
			\hline \hline
			& Data & Monte Carlo \\ \hline
			&  &   \\  [-0.6em]
			Variable resampling & Mom. scaling$^*$	\newline Decay Tree Fitter$^*$ & Mom. smearing$^*$ \newline Decay Tree Fitter$^*$ \newline PIDCalib$^*$ \\ [1.0em]
			Event reweighting & sWeights$^\dag$ & GBReweighter$^\dag$ \newline L0 Hadron$^*$ \\ [1.0em]
			\hline \hline
	\end{tabular} }
\end{table}

\subsection{Momentum scaling and smearing}

On data, the track momentum is rescaled introducing a correction at the per-mille level. This recalibration is done using large samples of $J/\psi\to\mup\mu{-}$ decays, and \textit{forcing} the resulting average mass to the central value of the well-known $J/\psi$ mass. This correction is configured within the tool \verb|TrackScaleState|\footnote{The \texttt{TrackScaleState} tool can be used as a \texttt{DaVinci().UserAlgorithms} before the processing of \texttt{DecayTreeTuple}. Within the \textit{modern selection framework}, this tool can be applied directly on the \texttt{AutomaticData} with the \texttt{MomentumScaling} wraper. This is partially explained in the \href{https://twiki.cern.ch/twiki/bin/view/LHCb/RefitTracksFromDST\#Momentum_Scale_correction}{main documentation}.}, 
which has been thoroughly validated with many other resonances.

The momentum smearing corrects the MC for a slightly underestimated detector resolution. It can be called with \texttt{TrackSmearState}.\onlyANA{\footnote{\rev{Louis Henry sent me this. Only God knows if this is documented anywhere.}}}

Both of these corrections account for small (per mille or lower) effects and using them makes sense only in analyses of large data samples (like ours), where the statistical uncertainty is also very small, and with the most optimal reconstruction techniques. In particular, the tools implementing them can only be used together with Decay Tree Fitter, studied next.

\subsection{Decay Tree Fitter} \label{sec:DTF}

Decay Tree Fitter (DTF)~\cite{Hulsbergen:2005pu} is a method to refit the kinematics of the event by simultaneously using the vertex and momentum information and introducing constraints either on the invariant mass of unstable particles and/or on the primary vertex (PV) position. This method improves the resolution on momenta, which propagates to the invariant masses (when they are not fixed) and the angular variables. In this section, the effect of DTF on these variables is evaluated in order to determine the optimal DTF configuration for this decay.

\subsubsection{Resolution on mass and helicity angles}

To extract the resolution on the helicity angles, $\cos\theta_p$ and $\phi_p$, the truth-matched MC sample is used after applying the preselection and trigger requirements according to Sections \ref{sec:trigger} and \ref{sec:preselection}. The distribution of $\cos\theta_{p,{\rm DTF}}^{\rm reco} - \cos\theta_p^{\rm true}$, and analogously for $\phi_p$, is fitted with the mixture of two Gaussian PDFs. The variance of this PDF is the weighted sum of the two widths,

\begin{equation}
\sigma^2 = w_{\rm core} \sigma_{\rm core}^2 + w_{\rm tail} \sigma_{\rm tail}^2 ,
\end{equation}
where $w_{\rm core(tail)}$ is the fraction of the narrower (wider) Gaussian. This distribution does not fit well the most central region, but it can be used as a figure of merit to compare the performance of each DTF configuration. The results of these fits are shown in Figure \ref{fig:DTFAngularResolution}.

\begin{figure}[h!]
	\centering
	\includegraphics[width=0.45\linewidth]{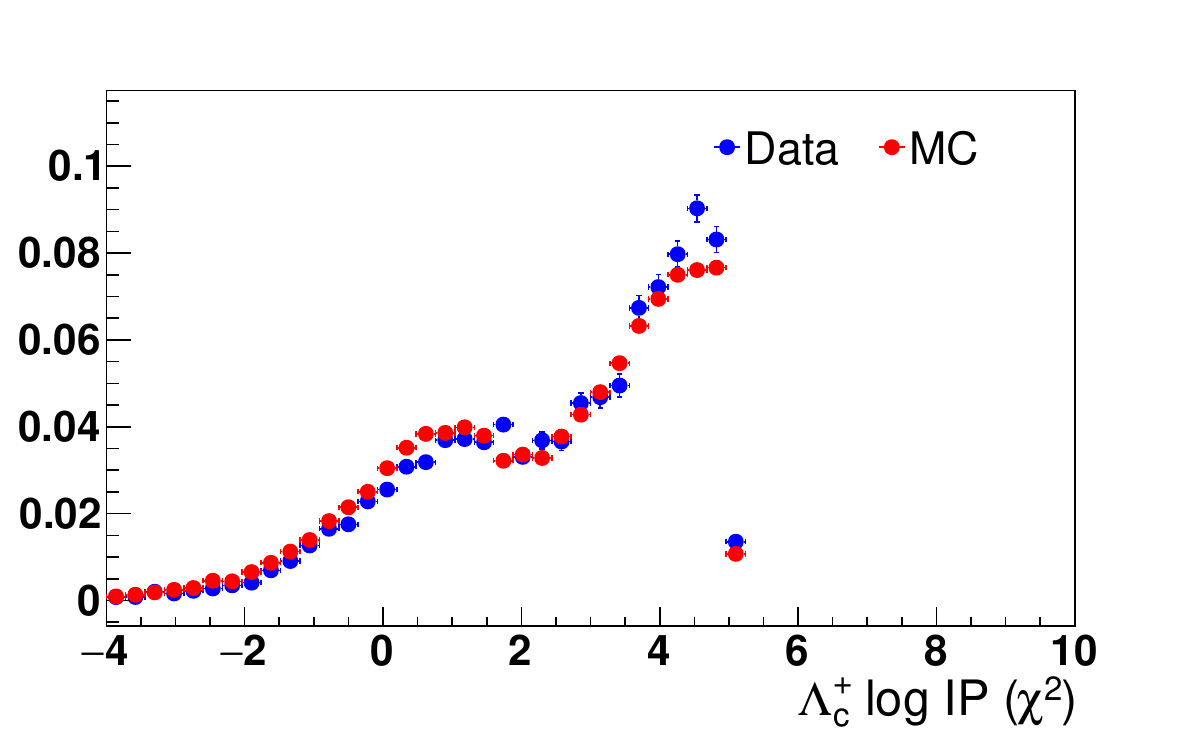}
	\caption{Impact parameter (IP) of the \Lc with respect to the PV divided by the estimated uncertainty. The two visible structures contain primary and secondary \Lc's, with smaller and larger IP values, respectively. The stripping cut at $\log({\rm IP} \chi^2)<5$ is very effective in removing background candidates. The distributions are normalized to unity, with the vertical axis indicating the normalized number of events per bin.}
	\label{fig:DTFLcLogIPChi2}
\end{figure}

The width and central value of the $\Lc$ invariant mass are evaluated on real data for the different DTF configurations. The corresponding fits are shown in Figure \ref{fig:DTFMassFits}, except for the two cases in which the constraint is applied on the \Lc mass itself. Table \ref{tab:DTFResolutions} summarizes the systematic comparison between all the DTF configurations. The resolution on the helicity angles improves by about 10\% when the masses of the \Lz or \Lz and \Lc are constrained to their PDG value~\cite{PDG}. In turn, the angular resolution worsens when the $\Lc$ direction is forced to intersect with the PV. This effect is associated to the large amount of \Lc candidates produced in secondary decays of heavier particles, which do not come from the PV. The worsened resolution of these configurations is reproduced on real data when fitting the \Lc invariant mass. The $\sigma$ of the signal PDF (double-sided Crystal Ball) increases by about 12\% in this case, similar to the angular resolution on signal MC. To support the hypothesis of the secondary \Lc's, there should be a comparable amount of primary and secondary \Lc particles on both signal MC and real data. This can be tested by comparing the distribution of the $\log (\chi_{\rm IP}^2)$ on sWeighted data and signal MC, in Figure \ref{fig:DTFLcLogIPChi2}. The two distributions are in remarkable agreement.

On another note, we evaluated the number of events for which the DTF did not converge, on signal MC. Even with no additional constraint, around 2.2\% of candidates are \textit{lost} when the fit is rerun (see Table \ref{tab:DTFFailedFits}). Instead, when the PV is constrained the fit fails only for 1.5\% of the candidates. We found no direct explanation for this effect, which needs further investigation.

In conclusion, the optimal resolution for our data sample is obtained by fixing the \Lz or the \Lz and \Lc invariant masses. We will use the \Lz constraint in the nominal fit and the other one to estimate the systematic uncertainty of this choice.
\begin{table}
	\ centring
	\caption{Comparison between the different DTF configurations. Results of the fit to the \Lc invariant mass using a double-sided Crystal Ball PDF~\cite{Skwarnicki:1986xj} (on data), and resolution on the helicity angles, fitted with a mixture of two Gaussian PDFs (on signal MC).} 
	\label{tab:DTFResolutions}
	\resizebox{1\linewidth}{!}{
		\begin{tabular}{lcccc}
			\hline \hline
			DTF constraint & $\mu (m_\Lc)$  & $\sigma (m_\Lc)$ & $\sigma (\cos\theta)$ &  $\sigma (\phi)$ \\ \hline
			None   &  $ 2287.04 \pm  0.08 $  &  $ 6.79 \pm  0.09 $  &  $0.0239 \pm 0.0001$  &  $0.0438 \pm 0.0004$ \\ 
			\Lz mass   &  $ 2286.90 \pm  0.07 $  &  $ 6.37 \pm  0.08 $  &  $0.0214 \pm 0.0001$  &  $0.0402 \pm 0.0003$ \\ 
			\Lz and \Lc mass   &  $ - $  &  $ - $  &  $0.0213 \pm 0.0001$  &  $0.0392 \pm 0.0003$ \\ 
			PV   &  $ 2287.63 \pm  0.10 $  &  $ 7.83 \pm  0.12 $  &  $0.0271 \pm 0.0002$  &  $0.0480 \pm 0.0004$ \\ 
			PV and \Lz mass   &  $ 2287.51 \pm  0.09 $  &  $ 7.41 \pm  0.11 $  &  $0.0248 \pm 0.0001$  &  $0.0452 \pm 0.0004$ \\ 
			PV, \Lz, and \Lc mass   &  $ - $  &  $ - $  &  $0.0247 \pm 0.0001$  &  $0.0442 \pm 0.0004$ \\ \hline \hline
		\end{tabular}
	}
\end{table}

\begin{table}
	\centering
	\caption{Portion of signal MC events in which the DTF did not converge. The refit with no additional constraints (first row) also looses candidates. } \label{tab:DTFFailedFits}
	\resizebox{0.55\linewidth}{!}{
		\begin{tabular}{lc}
			\\ \hline \hline
			DTF constraint  & Fraction of failed refits  \\ \hline
			None   &  2.22\%  \\
			\Lz mass   &  6.14\%  \\
			\Lz and \Lc mass   &  6.06\%  \\
			PV   &  1.49\%  \\
			PV and \Lz mass   &  5.74\%  \\
			PV, \Lz, and \Lc mass   &  5.69\%  \\ \hline \hline
		\end{tabular}
	}
\end{table}

\subsubsection{Background shapes of helicity angles}

\begin{figure}
	\centering
	\includegraphics[width=0.45\linewidth]{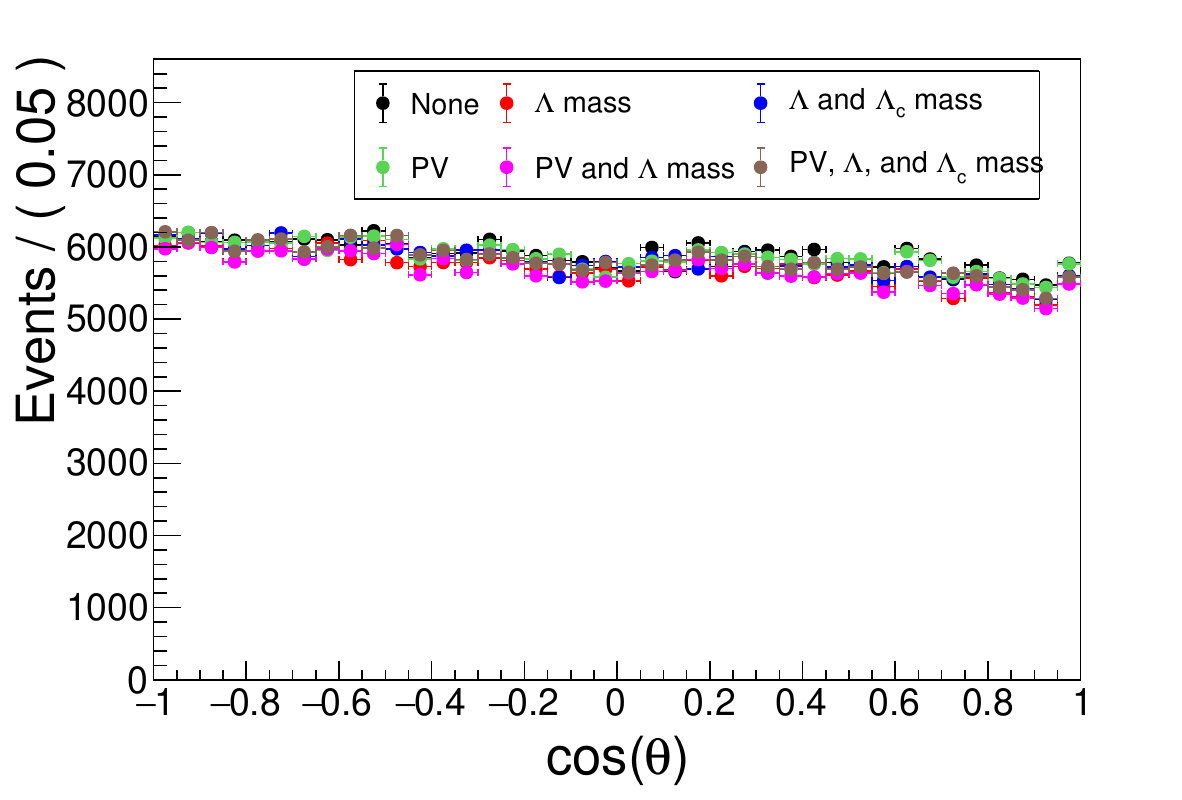}
	\includegraphics[width=0.45\linewidth]{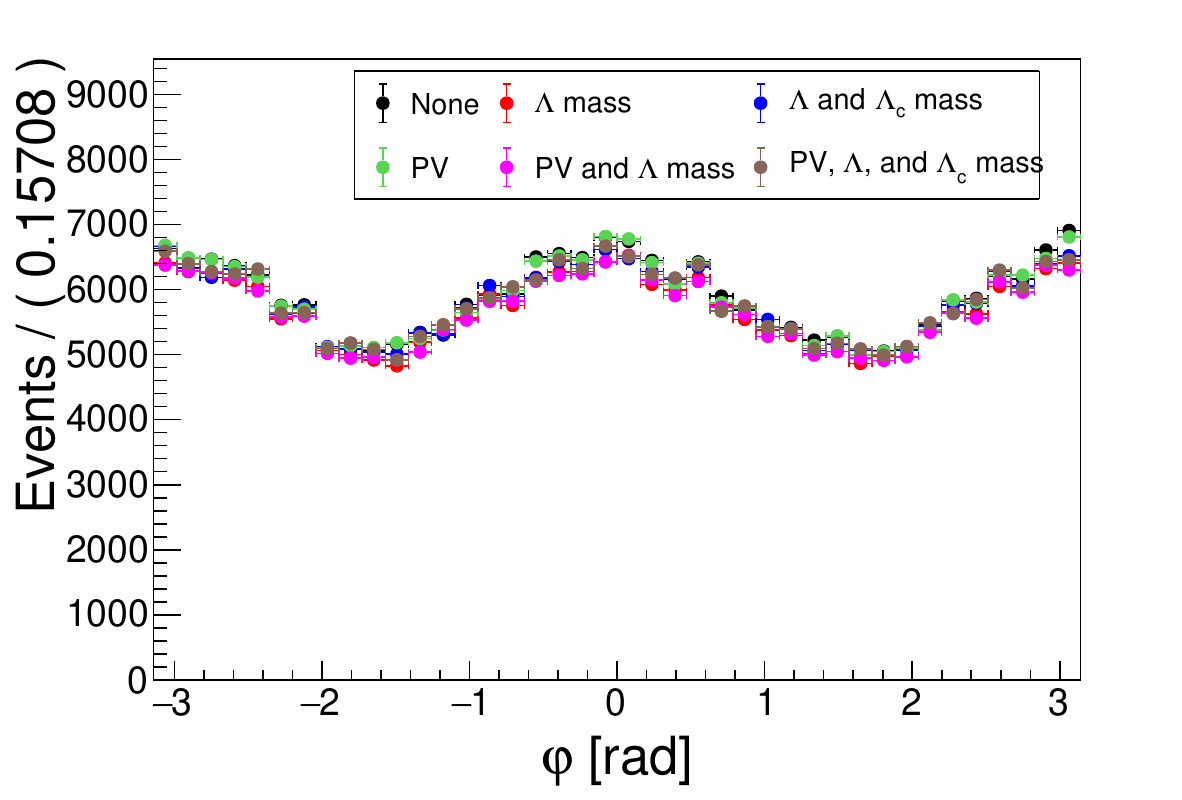}
	\caption{Angular distribution of background candidates for different DTF configurations, for (left) $\cos\theta_p$ and (right) $\phi_p$. Both lower and higher sidebands are included through the selection cut $|m(\Lc)^{\rm reco} - m(\Lc)^{\rm PDG} | > 25 \mev$. }
	\label{fig:DTFbkgAngles}
\end{figure}

\begin{figure}
	\centering
	\includegraphics[width=0.45\linewidth]{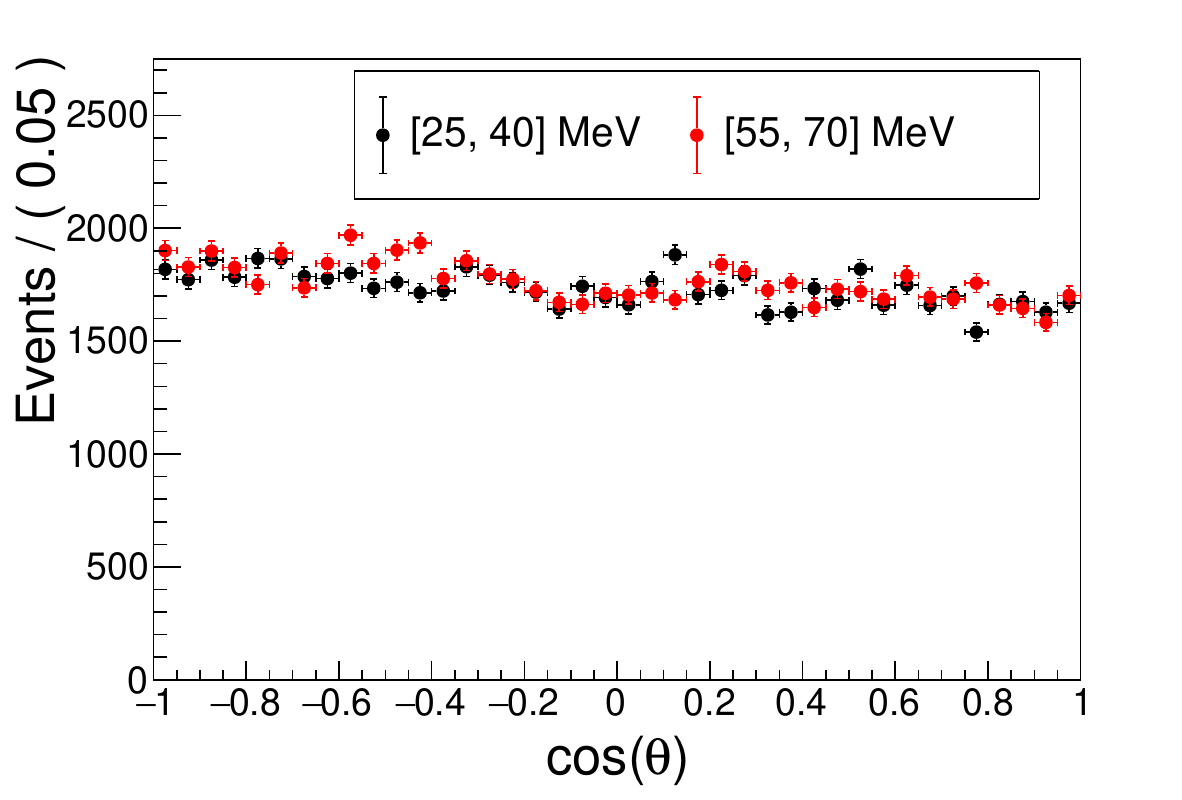}
	\includegraphics[width=0.45\linewidth]{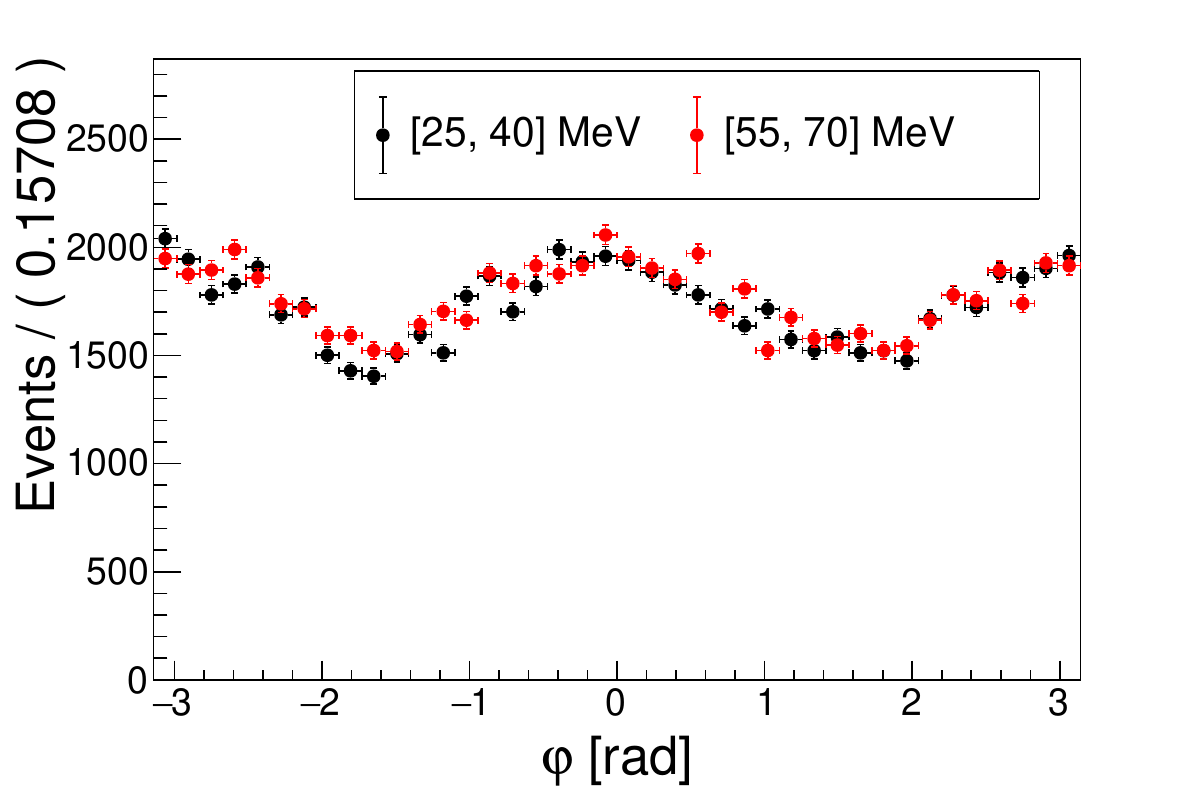}
	\caption{Angular distribution of background candidates for the DTF configuration with constraints on the \Lz and \Lc mass, for (left) $\cos\theta_p$ and (right) $\phi_p$. Two background regions, at different distances of the real $\Lc$ mass are compared. The considered events are in different ranges of $|m(\Lc)^{\rm reco} - m(\Lc)^{\rm PDG} | $, as shown in the legend. }
	\label{fig:DTFbkgLcAngles}
\end{figure}

Even though the angular resolution improves for signal events, the angular distribution of irreducible background candidates may be affected by these constraints. In Figure \ref{fig:DTFbkgAngles}, the angular distribution of the background component is compared for the different DTF configurations. The refit does not introduce any distortion. While the constraint on the \Lz mass is only applied to real \Lz candidates (the purity in the \Lz invariant mass is close to 100\%), the constraint on the \Lc mass forces also the kinematics of background events to produce the \Lc invariant mass, which may introduce more prominent distortions on the angular distributions. We evaluate this effect for sideband candidates (which will be cut out in the final fit) in Figure \ref{fig:DTFbkgLcAngles}. Again, no evident distortions are seen.

\onlyANA{
\subsubsection{Background shapes on \Lc invariant  mass}

\rev{
	\textbf{TO-DO (maybe)}
	
	The DTF constraints affect the shape of the background on the invariant mass distribution. Specially, when constraining the $\Lz$ mass in the sample without any offline selection (besides the stripping), we get a distorted background as seen in Figure X1. By selecting events on the $\Lz$ mass window $|m(p \pim) - m_{\Lz}^{\rm PDG} | < xx \mev$, the mass constrain is closer to the actual measured mass of the $\Lz$ candidate and the background on the $\Lc$ invariant mass is back to an exponential (Figure X2).

Once the preselection (presented in Section X) is applied, the $\Lz$ candidates have basically a 100\% purity and this cut is unnecessary to do the checks presented in this section.

\begin{itemize}
	\item Fig X1: \Lc mass with standard reco and L0 constrained; after stripping.
	\item Fig X2: same as above, after mass cut on lambda. Caption:events on the lambda mass window such that $|m(p \pim) - m_{\Lz}^{\rm PDG} | < xx \mev$. 
	\item Fig X3: same, but after whole preselection.
\end{itemize}

}

}

\pagebreak

\pagebreak

\subsection{PID correction} \label{sec:PIDcorrection}

\onlyANA{\red{Re-do this comparison plots eliminating the green distribution (previous to the MC reweighting)}
	
	\red{Ideally, use only one plot per variable: three-fold comparison with sweighted data / default MC / PIDCorr. Probably have to modify the code slightly.}
}

The MC simulation of the RICH detector response is especially challenging since there are many charged tracks in each event producing Cherenkov photons which can overlap in the RICH photomultipliers and, in general, cannot be associated to single tracks. Thus, any LHCb analysis using PID information and MC simulations (almost all of them) must apply offline PID corrections to the MC samples.

The PIDCalib package provides two alternatives to apply corrections to the PID variables, based on calibration data:
\begin{itemize}
	\item \verb|PIDGen.py| recalculates the PID variables from scratch based on large calibration tables and the relevant information of the event, without simulating the RICH response.
	\item \verb|PIDCorr.py| transforms the initially simulated PID variables, keeping the correlation among the PID variables of the same track.
\end{itemize}
We opted for the second option, as these variables are later used in the multivariate classifier.

To compute the new PID of each track, \verb|PIDCorr.py| takes two variables for each track (momentum and transverse momentum) and the number of tracks in each event (also called multiplicity). The multiplicity, however, is also poorly reproduced in MC and we rescale the number of tracks in each event by a factor 1.11 to improve the data/MC agreement previously to the correction.

As shown in Figure~\ref{fig:PIDCompBach} the PID distributions of the 3\pipm from the \Lc improves slightly for \verb|ProbNNp|, but it has no apparent effect on \verb|ProbNNpi| or \verb|ProbNNk|. For the $p(\Lambda)$ and $\pi^-(\Lambda)$, in Figures.~\ref{fig:PIDCompLambda1} and \ref{fig:PIDCompLambda2}, the data/MC agreement worsens with \verb|PIDCorr|. The reason is that the calibration samples from real data, in the default configuration, consider only tracks originating in the central VELO region, and the RICH signature of downstream tracks is naturally different. In the final selection, only the 3\pipm(\Lc) PID variables will be used.

\onlyANA{\rev{
		
		BUT: we already cut on the proton PID in the preselection.
		
		The solution was suggested in the Charm WG: use PIDGen or PIDCalib (don't remember which); and delve into the code to use calibration samples with \Lz or Ks.
		
		People working with downstream lambdas (in real data analysis; not in ethereal reconstruction ideas) must know how.

	}
	
}

The \texttt{ProbNN} PID variables peak at zero or/and one. To illustrate the (dis)agreement of sWeighted data and signal MC, we applied some prior transformations to these variables, which keep the variable domain between 0 and 1 as well:

\begin{itemize}
	\item Variables peaking at zero (misID):
	$$
	\texttt{ProbNNx} \to (\texttt{ProbNNx})^{0.3} 
	$$
	
	\item Variables peaking at one (PID of the particle itself):
	$$
	\texttt{ProbNNx} \to 1-(1-\texttt{ProbNNx})^{0.3}
	$$	
\end{itemize}

\onlyANA{
	\rev{
		
		\subsubsection{Discussion}
		
		~\\
		$\Lambda$ variables
		\begin{itemize}
			\item The correction sample of the proton is based on $\Lambda_{b}\to \Lambda_{c}^+ \pi^-$
			\item $\pi^-(\Lambda)$ is \textbf{not well reproduced}
			\item $p(\Lambda)$ and $\pi^-(\Lambda)$ Data/MC agreement \textbf{worsens wtih} \verb|PIDCorr|
			\item $p(\Lambda)$ and $\pi^-(\Lambda)$ are also \textbf{not so important (?)} ($97\%$ purity on $\Lambda$'s after preselection) 
		\end{itemize}
		
		~\\
		$\Lambda_c^+$ variables
		\begin{itemize}
			\item $\pi^{\pm}(\Lambda_c^+)$ are relatively \textbf{well reproduced}
			\item Agreement \textbf{improves slightly} after \verb|PIDCorr|
		\end{itemize}

		\begin{itemize}
			\item Problem with the threshold in pip2\_Lc\_ProbNNpi. Probably coming from a TOS trigger line. It should not come from the stripping containers (since all the pions are filtered in the same way, and we do not have the combination of two stripping lines).
		\end{itemize}
		
	}
}

\subsection{Reweighting of MC events} \label{sec:reweightMC}

In the following, the signal MC sample is corrected by computing event weights such that the distributions of the \Lc $p$ and $p_T$ match the sWeighted data. Conceptually, the weights are obtained by dividing this two-dimensional space in bins, and finding a scale factor for each bin of MC data such that the two samples match. However, since making evenly spaced bins in two (or more) dimensions would require too many events, the space of variables is split via decision trees: the parameter space is recursively bisected into regions that maximize the deficit or surplus of MC events with respect to the data. After a few splits (depth of the tree), the weights are calculated for each region. The process is repeated several times (number of trees), taking into account the event weights accumulated by previous iterations. The fact that the events are reweighted after each iteration, and this acquired \textit{knowledge} is used by subsequent trees is known as a \textit{boosting}. We used the tool \verb|GBReweight| in the \verb|hep-ml| package for python. We will use decision trees again in Section~\ref{sec:BDT}, to finish the selection process with a multivariate classifier.

\begin{figure}[h!]
	\centering
	\includegraphics[width=0.45\linewidth]{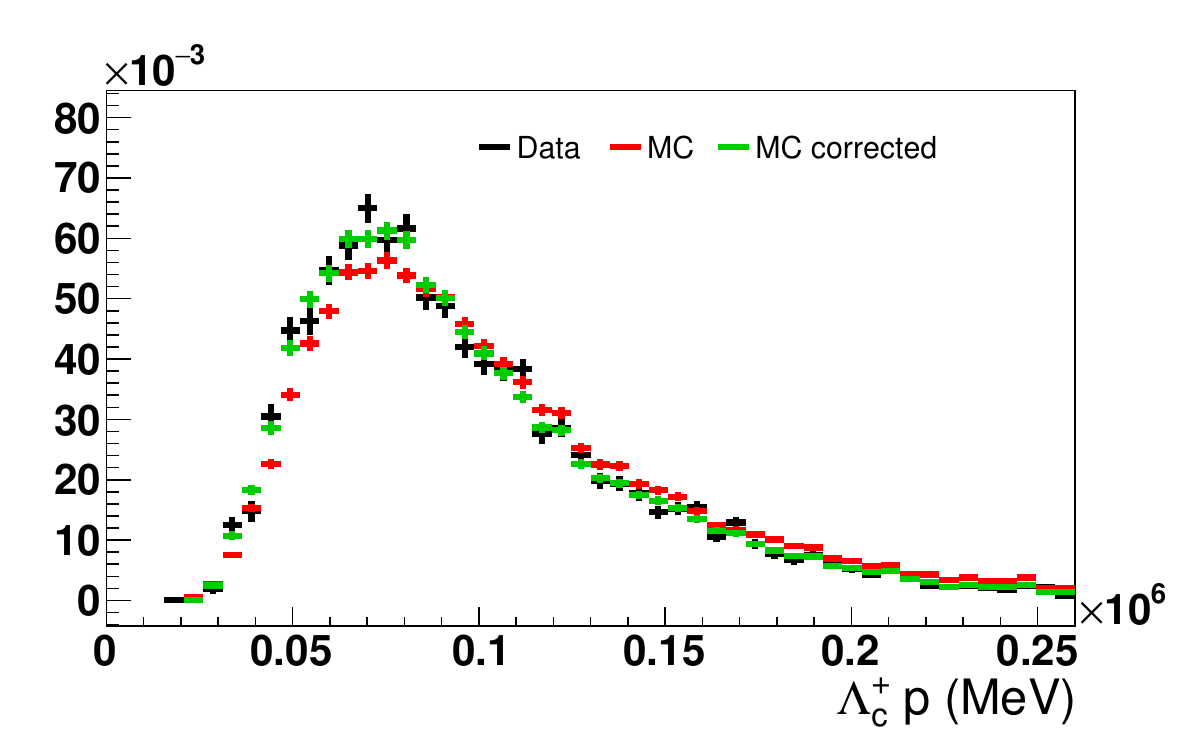}
	\includegraphics[width=0.45\linewidth]{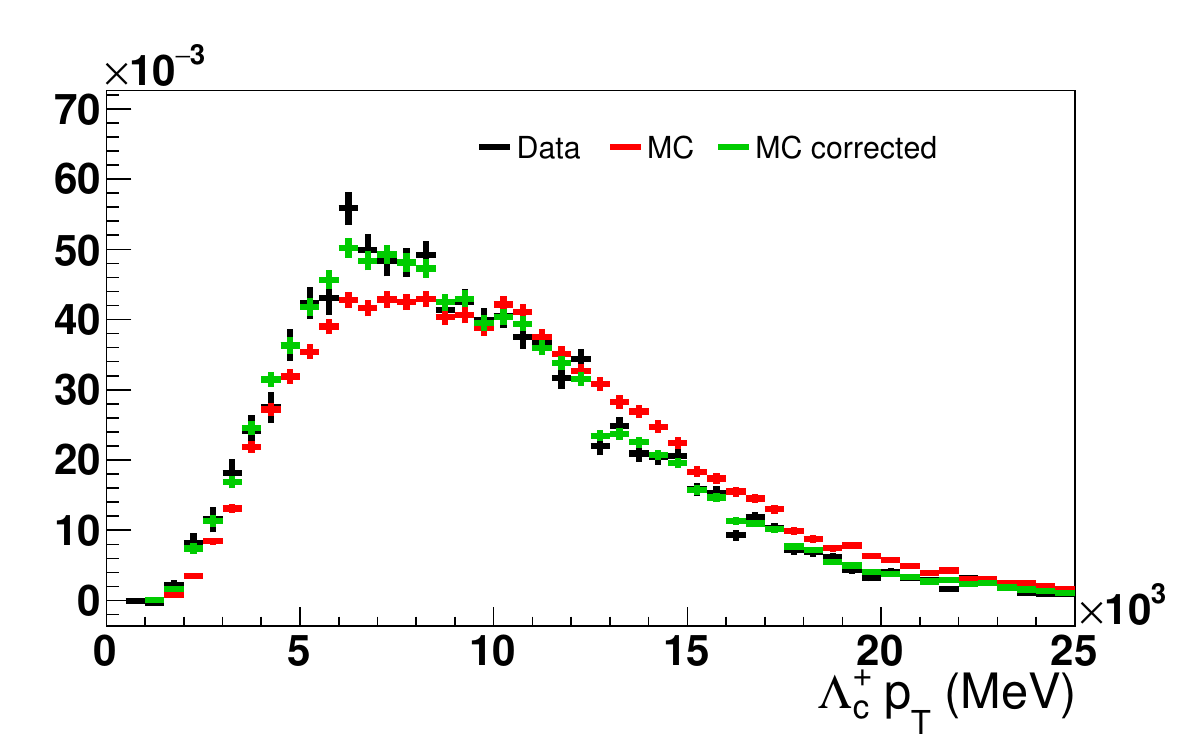}
	\caption{Variables used for reweighting the signal MC sample. The "MC corrected" distribution is forced to match the sWeighted "Data". The distributions are normalized to unity, with the vertical axes indicating the normalized number of events per bin.}
	\label{fig:reweightLcPPT}
\end{figure}

The reweighted distributions are shown for the reference variables, \Lc $p$ and $p_T$, in Figure \ref{fig:reweightLcPPT}. This correction improves the agreement in the distributions of other variables as well, which is crucial since we will rely on MC events to train a multivariate classifier in Section~\ref{sec:BDT}.

The data/MC agreement for the 14 variables that will be used in the final configuration of the classifier is shown in Figs.~\ref{fig:BDTvars} and \ref{fig:BDTvars2}.


\subsection{Reweighting of L0 Hadron trigger}

As introduced in Section~\ref{sec:LHCb}, \texttt{L0} is the first level of the LHCb trigger system, which is based on hardware and uses only the signals in the muon chambers and calorimeters. A positive signal in the hadronic calorimeter is recorded through the \texttt{L0Hadron} line. Nevertheless, it is well known that the efficiency of the \texttt{L0Hadron} line is not well reproduced in MC simulations of the detector response. The disagreement has been parameterized depending on the type of particle (\pipm, \piz, \Kpm, $p$), and its position at the HCAL\footnote{This tool is documented in the \href{https://twiki.cern.ch/twiki/bin/viewauth/LHCbPhysics/CalorimeterObjectsToolsGroupDOC}{LHCb twiki pages}.}. This correction must be considered when the particle triggering the calorimeter is one of the signal particles (\ie affects the \texttt{L0Hadron\_TOS}, not the \texttt{TIS}). In our multihadronic decay, about 30\% of candidates are recorded with the \texttt{L0Hadron\_TOS} line. This correction, however, has not been applied yet in the preliminary analysis presented in this thesis.

\onlyANA{
	
	To apply this correction one needs the \verb|L0Hadron_TOS| decision for single tracks and the point where the track hit the HCAL, provided by \verb|TupleToolL0Calo|. 
	
	{\ colour {BrickRed} We did not apply this correction. How to do it, and its estimated importance for our analysis is still unclear:

		\begin{itemize}
			\item The changes might be very small (compared to other systematic errors). But the systematic error associated to this correction will need to be computed anyway.
			\item This only corrects the candidates triggered by the L0 Hadron \textit{TOS} line (38\% of the signal candidates), but not the L0 Hadron \textit{TIS} (also about 40\%). If the overlap needs to be subtracted, \textit{i.e.} it only applies to L0 Hadron \textit{TOS not TIS}, it would affect around 30\% of the MC sample.
			\item Patrick Robbe told me he is preparing new scripts to simplify the application of the correction (although seeing the previous scripts it does not look any \textit{simple}).
		\end{itemize}
		
	}

}


\onlyANA{

\section{Data/MC comparison (BDT variables)} \label{sec:compDataMC}

Plot the MC helicity angles with the correction.

\pagebreak

\pagebreak

\subsection{PID variables}

\red{Include here some plots after the MC reweighting (before we only showed some distribution with and without PID corr). Showing how this correction does not help for the agreement of the PID variables.}

\subsection{Other variables with bad Data/MC agreement} \label{sec:badCompDataMC}

Variables that have been discarded due to very bad data/MC agreement.
\begin{itemize}
	\item 3\pipm system combined mass. 
	\item Reconstructed track $\chi^2$
	\item Global event variables (number of tracks, clusters, PVs, candidates)
	\item DTF $\chi^2$ for fits with constrained \Lc mass
	\item RHO and PV\_PCHI2 of constrained vertex
\end{itemize}

}

\onlyANA{

\begin{figure}[ht!]
	\centering
	\includegraphics[width=0.45\linewidth]{content/figs/compDataMC/CompAllVarsDataCorrMCPur30Plus3DD/Variables/Lc_bachelors_COMBM}
	\includegraphics[width=0.45\linewidth]{content/figs/compDataMC/CompAllVarsDataCorrMCPur30Plus3DD/Variables/pim_Lc_TRACK_CHI2NDOF}
	\includegraphics[width=0.45\linewidth]{content/figs/compDataMC/CompAllVarsDataCorrMCPur30Plus3DD/Variables/pim_Lc_TRACK_PCHI2}
	\includegraphics[width=0.45\linewidth]{content/figs/compDataMC/CompAllVarsDataCorrMCPur30Plus3DD/Variables/nLongTracks}
	\includegraphics[width=0.45\linewidth]{content/figs/compDataMC/CompAllVarsDataCorrMCPur30Plus3DD/Variables/nTTracks}
	\includegraphics[width=0.45\linewidth]{content/figs/compDataMC/CompAllVarsDataCorrMCPur30Plus3DD/Variables/nPVs}
	\includegraphics[width=0.45\linewidth]{content/figs/compDataMC/CompAllVarsDataCorrMCPur30Plus3DD/Variables/p_L0_OWNPV_CHI2}
	\includegraphics[width=0.45\linewidth]{content/figs/compDataMC/CompAllVarsDataCorrMCPur30Plus3DD/Variables/Lc_BPV_RHO}
	\caption{Examples of variables with poor Data/MC agreement. Among the considered 179 variables, all of those with Data/MC differences are represented by or closely related to the shown distributions. \textit{E.g.} the track $\chi^2$ disagrees for all tracks (illustrated here by the \pim(\Lc) track $\chi^2$); the number of (any type of) tracks and clusters disagrees in general; the $\chi^2$ of the best PV compatible with the tracks also disagrees; etc.}
	\label{fig:compDataMCNoAgreement}
\end{figure}

\begin{figure}[ht!]
	\centering
	\includegraphics[width=0.45\linewidth]{content/figs/compDataMC/CompAllVarsDataCorrMCPur30Plus3DD/Variables/log(Lc_DTF_None_CHI2)}
	\includegraphics[width=0.45\linewidth]{content/figs/compDataMC/CompAllVarsDataCorrMCPur30Plus3DD/Variables/log(Lc_DTF_L0C_CHI2)}
	\includegraphics[width=0.45\linewidth]{content/figs/compDataMC/CompAllVarsDataCorrMCPur30Plus3DD/Variables/log(Lc_DTF_L0LcC_CHI2)}
	\includegraphics[width=0.45\linewidth]{content/figs/compDataMC/CompAllVarsDataCorrMCPur30Plus3DD/Variables/log(Lc_DTF_PVC_CHI2)}
	\includegraphics[width=0.45\linewidth]{content/figs/compDataMC/CompAllVarsDataCorrMCPur30Plus3DD/Variables/log(Lc_DTF_PVL0C_CHI2)}
	\includegraphics[width=0.45\linewidth]{content/figs/compDataMC/CompAllVarsDataCorrMCPur30Plus3DD/Variables/log(Lc_DTF_PVL0LcC_CHI2)}
	\caption{Goodness of fit ($\chi^2$) for the different DTF configurations. The background events (\textit{removed} here via the sweights) populate the distributions at higher values. However, when the \Lc mass is constrained, the $\chi^2$ is heavily correlated to the measured mass, and the negative sweights (from the sidebands) appear always at larger $\chi^2$ values. For that reason, the "Data" distribution has negative counts (hardly visible with the shown y-range). }
	\label{fig:compDataMCDTFCHI2}
\end{figure}


}


\section{Multivariate selection} \label{sec:BDT}

Having recalibrated the MC sample to ensure the agreement between data and MC, we can move to the last selection step, a multivariate classifier. The discussion is restricted to the boosted decision tree algorithm (BDT), with \textit{Adaptive Boost}. In some initial tests, this type of classifier was found to be the best performing algorithm among those included in the TMVA package~\cite{Hocker:2007ht} of the \root toolkit. An updated study to compare the different algorithms with the full set of MC corrections would be interesting but the potential improvement is expected to be marginal. 

Compared to Neural Networks, BDTs are very transparent algorithms. To some extent, BDTs allow to retrieve the \textit{inner} working of the trained classifier since their structure is relatively simple. In this section, we will introduce the general algorithm, describe its configuration for our analysis and optimize the final BDT selection.

Some basic terms appearing recurrently in multivariate classifiers for this type of data analyses are defined in Appendix \ref{app:BDTterms}.

\subsection{Boosted Decision Trees} \label{sec:BDTintroduce}

\begin{figure}
	\centering
	\includegraphics[width=0.4\linewidth]{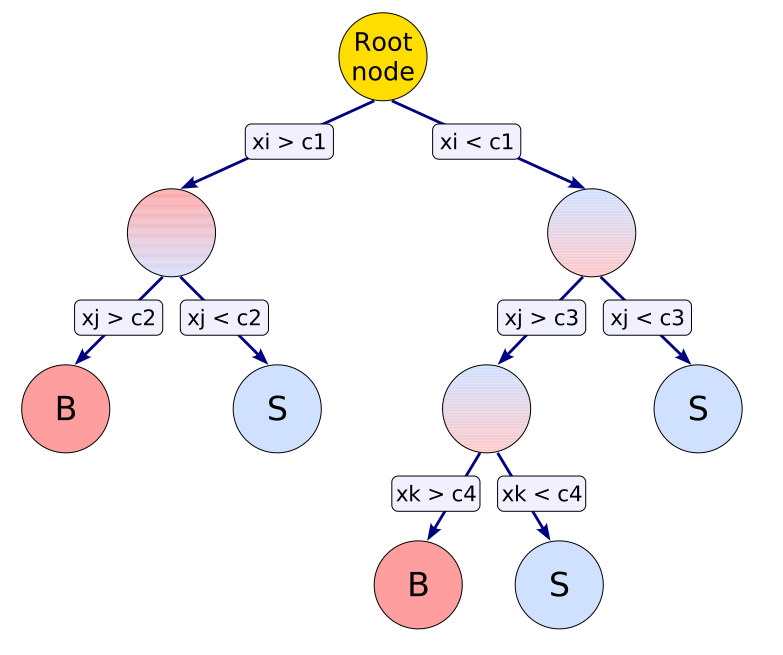}
	\caption{From Ref.~\cite{Hocker:2007ht}. Structure of a (single) decision tree. The initial data sample is sequentially split with binary selection cuts on any of the variables that maximize the discrimination between signal and background. }
	\label{fig:BDTscheme}
\end{figure}

A decision tree consists of a set of cuts applied sequentially over a dataset giving rise to a tree structure, as shown in Figure~\ref{fig:BDTscheme}. In each \textit{node}, the most discriminating variable and corresponding cut are determined to split the data in two. The datasets at the end of the tree (the \textit{leaves}) are labelled as signal or background on a majority basis. In this sense, a single decision tree is no different to a selection with rectangular cuts (like our preselection), in which the final dataset is considered just signal (that we want to keep). The breakthrough in performance emerges when, instead of one, many decision trees are used in the selection, and each of them is built using the gathered information by the previous ones (\textit{boosting}). 


More specifically, the misclassified events (\textit{e.g.} background events in a leaf of signal majority) will have an increased importance in the next trees. This is achieved by reweighting these events by the \textit{boost weight}
\begin{equation}
\alpha=\frac{1-\varepsilon}{\varepsilon},
\end{equation}
where $\varepsilon$ is the misclassification error of the complete tree, which is equal to the percentage of events that ended up in leaves of opposite categories, \textit{i.e.} background (signal) events in signal- (background-)majority leaves. Thus, the misclassified events in a highly discriminant tree will get an enormous weight, \textit{e.g.} if there are $\varepsilon$ = 4\% of misclassified events in a tree, these are reweighted by a factor $\alpha=0.96/0.04=24$. The performance of the full classifier is often improved by reducing this factor. In the \textit{Adaptive Boost} technique, which we will use in this thesis, the weight is transformed as $\alpha \to \alpha^\beta$, where the parameter $\beta \in [0,1]$ must be configured for the data sample.

The configuration of the algorithm consists of two steps: determining the set of event variables to be analysed by the BDT and the values of the \textit{hyperparameters}, which control the intrinsic configuration of the algorithm. There are four hyperparameters in our BDT:

\begin{itemize}
	\item \textbf{Maximum depth:}{ Maximum number of successive cuts in a tree.}
	\item \textbf{Minimum node size:} { Minimum number of events allowed in a final leaf. A tree will not apply further cuts if one of the two resulting datasets contains less than \textit{e.g.} 2\% of the initial sample.}
	\item \textbf{Number of trees:} Total number of decision trees.
	\item \textbf{Learning rate $\beta$}: Regulates the importance of the boost weight, $\alpha\to \alpha^\beta$, as described above.
\end{itemize}

These are optimized in Section~\ref{sec:hyperparameters}, extracting the uncertainties with the k-fold technique. 

\subsection{Variables} 

We will take a systematic approach to define the set of event variables used by the BDT. We start with 179 variables that are identically defined in the signal MC and data samples. These include kinematical, topological, global event information, PID variables and others. Some of these variables are transformed with a logarithm to avoid very long tails in unevenly spread distributions, which can reduce the performance of the BDT.\footnote{The domain of each variable is automatically determined by TMVA and divided into \texttt{nCuts=80} segments to find the most discriminating cut value. If most of the events are accumulated within the first segment all the discrimination power is directly lost. }

\subsubsection{Prerequisites}

The following variables are excluded:\footnote{See definitions in Appendix~\ref{app:analysisvariables}.}

\begin{itemize}
	
	\item Global-event variables (number of tracks, clusters, PVs, and decay candidates) that are stored as integers in the ntuple. The MC/data agreement on these is typically bad.
	
	\item Variables with poor data/MC agreement.
	
	\item \textit{Difference Log-Likelihood} (DLL) PID variables. We will only keep \verb|ProbNN| variables.
	\item Estimated reconstruction uncertainties: \verb|TAUERR|, \verb|MMERR|.
	
	\item $\chi_{\rm DTF}^2$ for DTF configurations with constrained \Lc mass. 
	\item The probability to be a muon track, \verb|x_ProbNNmu|. Most of the events have error code values, since it is only calculated on tracks with potentially matching signals on the muon chambers (positive \texttt{isMuon} decision). 
	\item Redundant variables: keeping \verb|log[Lc(L0)_IPCHI2]| (excluding \verb|Lc(L0)_IPCHI2|); \verb|BPVLTIME| (\verb|TAU|); \verb|FDCHI2_OWNPV| (\verb|BPVVDCHI2|); \verb|BPVLTCHI2| (\verb|BPVLTSIGNCHI2|).
	
	\item Finally, we also exclude some events from data and MC ($\approx 6\%$ of signal MC) on which some DTF configuration failed to refit the data, as shown in Table \ref{tab:DTFFailedFits}. 
	
\end{itemize}

After this initial reduction, we are left with 111 variables, many of which are highly correlated to each other. We configured a BDT with this set of variables, finding a high discriminating power with AUC = 0.924,  optimal significance on real data of $S/\sqrt{S+B} = 128.7$, signal purity $53.9\%$, and signal efficiency $\varepsilon_{\rm sig} = 97\%$.

\subsubsection{Optimization}

The TMVA libraries of \root include a tool to evaluate the importance of each variable in the BDT (\textit{specific method}), which is related to the number of times each variable was found to be the most discriminant one, \textit{i.e.} in how many splitting nodes it was used. If there are two highly correlated variables that essentially have the same discriminating power, each will be used approximately half the time, both of them appearing lower in the ranking. For the same reason, correlated variables introduce instabilities in the BDT performance. Depending on which of the two correlated variables is slightly more discriminating in a given node, the rest of the tree may change substantially. These instabilities may be seen in the curve of signal efficiency against background rejection.
Thus, guided by the variable ranking of TMVA and the correlation matrix, we will reduce the number of variables to the bare minimum, always making sure that the performance is not affected. 
The data/MC agreement is also considered and, if the BDT performance was not affected, any variable with suboptimal data/MC agreement was also removed.
As a figure of merit, we used the optimal significance on data. The AUC, however, will be slightly reduced but not the final significance or purity of the sample. 
%
%

\begin{figure}
	\centering
	\includegraphics[width=0.95\linewidth]{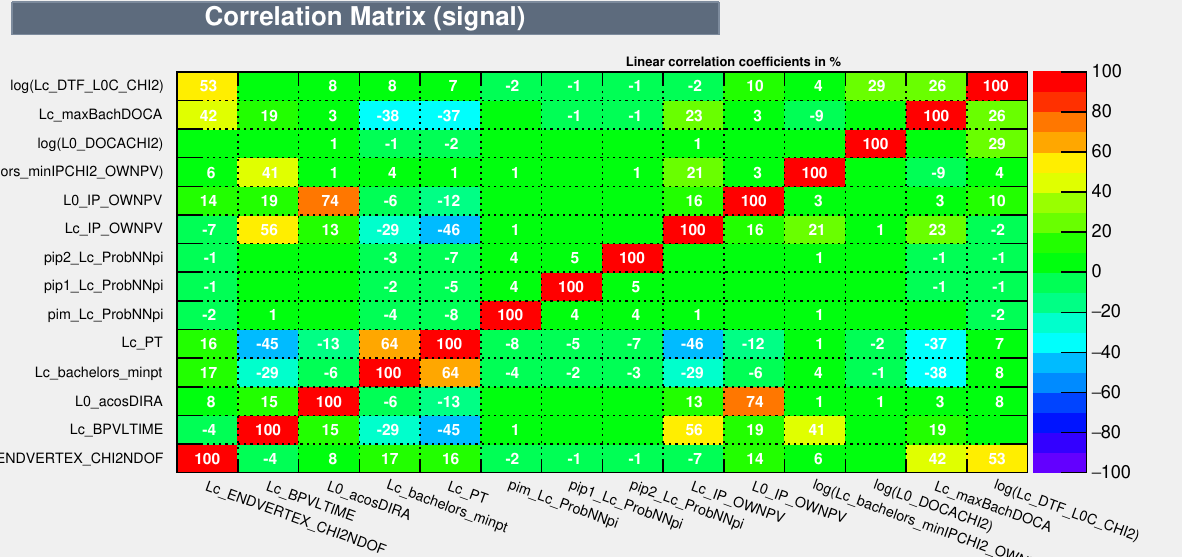}
	\caption{Linear correlation coefficients between the BDT variables, for the signal MC sample.}
	\label{fig:corr}
\end{figure}

The number of variables was reduced to just 14 variables, yielding essentially the same BDT performance. These are listed in Table~\ref{tab:BDTvars}, and their correlation shown in Figure~\ref{fig:corr}.

\begin{table}[ht!]
	\centering
	\footnotesize
	\caption{Final set of BDT variables.}
	\begin{tabular}{ll} 
		\hline \hline
		Name in the ntuple & Readable  name\\ 
		\hline 
		\verb|Lc_PT| & $\Lambda_{c}^{+}$ $p_{T}$ (MeV) \\ 
		\verb|Lc_IP_OWNPV| & $\Lambda_{c}^{+}$ IP (mm)  \\ 
		\verb|Lc_BPVLTIME| & $\Lambda_{c}^{+}$ lifetime, BPVLTIME (ns) \\ 
		\verb|Lc_ENDVERTEX_CHI2NDOF| & $\Lambda_{c}^{+}$ decay vertex $\chi^{2}/ndf$ \\ 
		\verb|L0_acosDIRA| & $\Lambda$ DIRA angle (mrad) \\ 
		\verb|L0_IP_OWNPV| & $\Lambda$ IP (mm)  \\ 
		\verb|log(L0_DOCACHI2)| & $\Lambda$ vertex log DOCA ($\chi^{2}$) [p, $\pi^{-}$] \\  
		\verb|pim_Lc_ProbNNpi| & $\pi^{-} (\Lambda_{c}^{+}$) pi track probability NN  \\ 
		\verb|pip1_Lc_ProbNNpi| & $\pi^{+} 1 (\Lambda_{c}^{+})$ pi track probability NN \\ 
		\verb|pip2_Lc_ProbNNpi| & $\pi^{+} 2 (\Lambda_{c}^{+})$ pi track probability NN \\ 
		\verb|Lc_maxBachDOCA| & $\Lambda_{c}^{+}$ bachelors max DOCA [$\pi^{+}1$, $\pi^{+}2$, $\pi^{-}$] (mm) \\ 
		\verb|log(Lc_bachelors_minIPCHI2_OWNPV)| & $\Lambda_{c}^{+}$ bachelors min log IP ($\chi^{2}$) \\
		\verb|Lc_bachelors_minpt| & $\Lambda_{c}^{+}$ bachelors min $p_{T}$ (MeV) \\ 
		\verb|log(Lc_DTF_L0C_CHI2)| & DTF [$\Lambda$] log $\chi^{2}$ \\ 
		\hline \hline
	\end{tabular} 
	\label{tab:BDTvars}
\end{table}

\subsection{Hyperparameters} \label{sec:hyperparameters}

A scan over the BDT hyperparameters was carried out comparing the performance of the BDT through the AUC measure. These scans were performed in one dimension at a time, fixing the other hyperparameters to the best value and repeating the process iteratively. In practice, studying the AUC in these scans was not the only factor to determine the optimal values. To ensure that there was no overtraining we had to continuously monitor the BDT response distribution.\footnote{Ideally, this could be achieved by plotting together the AUC measure calculated with the testing and training sample, as a function of the hyperparameters. In the presence of over-training, these two measures of performance diverge since the training sample yields overly optimistic discrimination power and the testing sample stays constant or decreases in performance. However, this test is not implemented in the \textit{out-of-the-box} TMVA libraries, as discussed in the \href{https://root-forum.cern.ch/t/roc-integral-from-test-and-training-sample/42026}{\root forum}.}.
The (four) hyperparameters are necessarily correlated and this optimization should, in principle, be carried out in four dimensions. However, identifying where the over-training starts without a direct measure is significantly more complex in two or more dimensions and, in any case, we observed large plateau regions (without overtraining) in the one-dimensional scan and a multidimensional optimization is not expected to improve the performance substantially.

\begin{figure}
	\centering
	\includegraphics[width=0.41\linewidth]{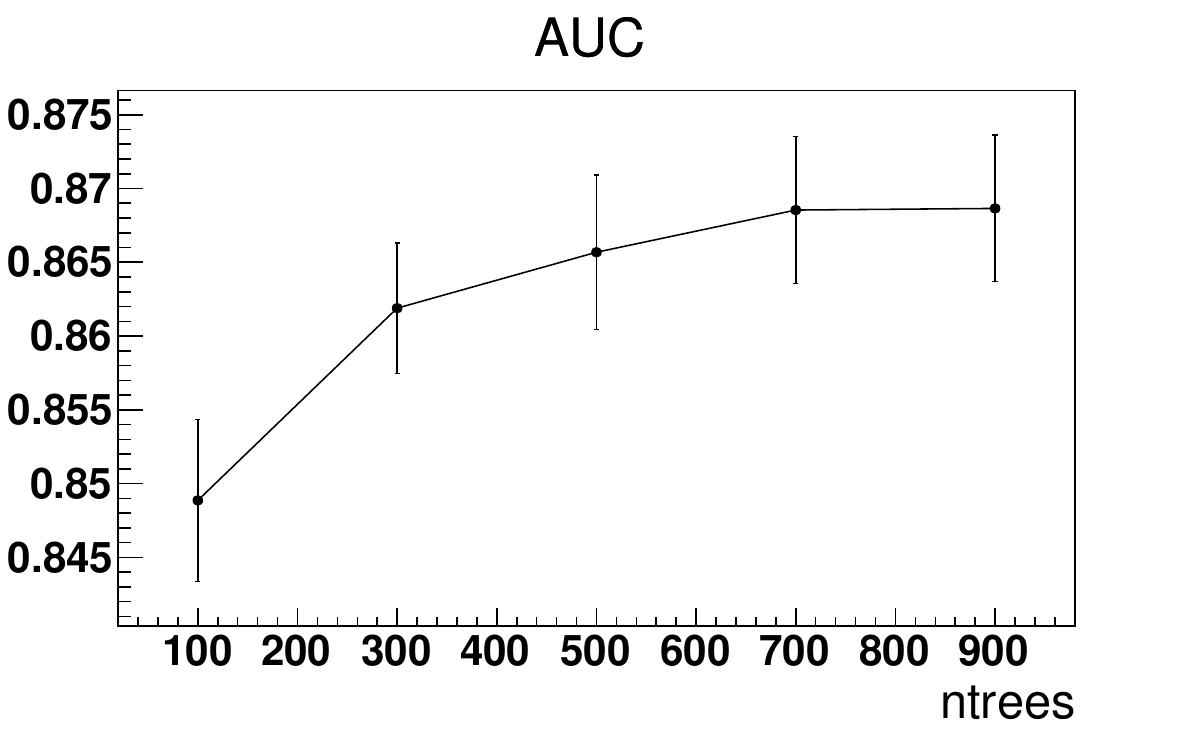}
	\includegraphics[width=0.41\linewidth]{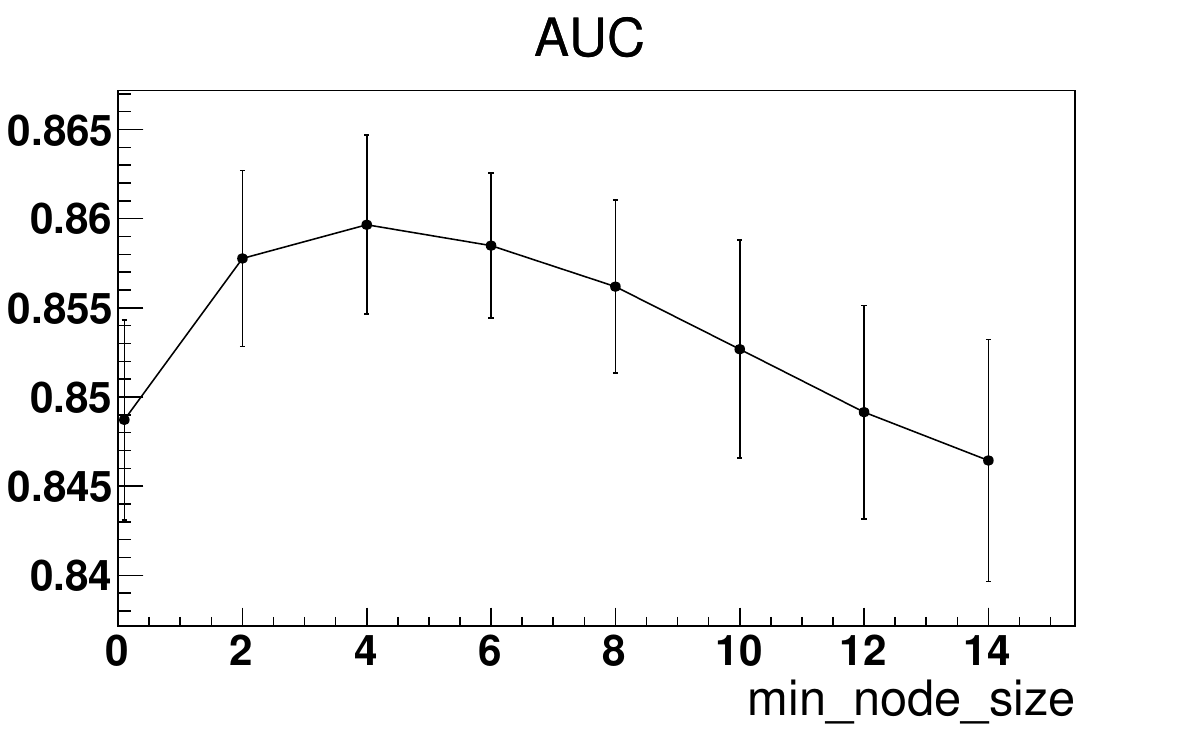}
	\includegraphics[width=0.41\linewidth]{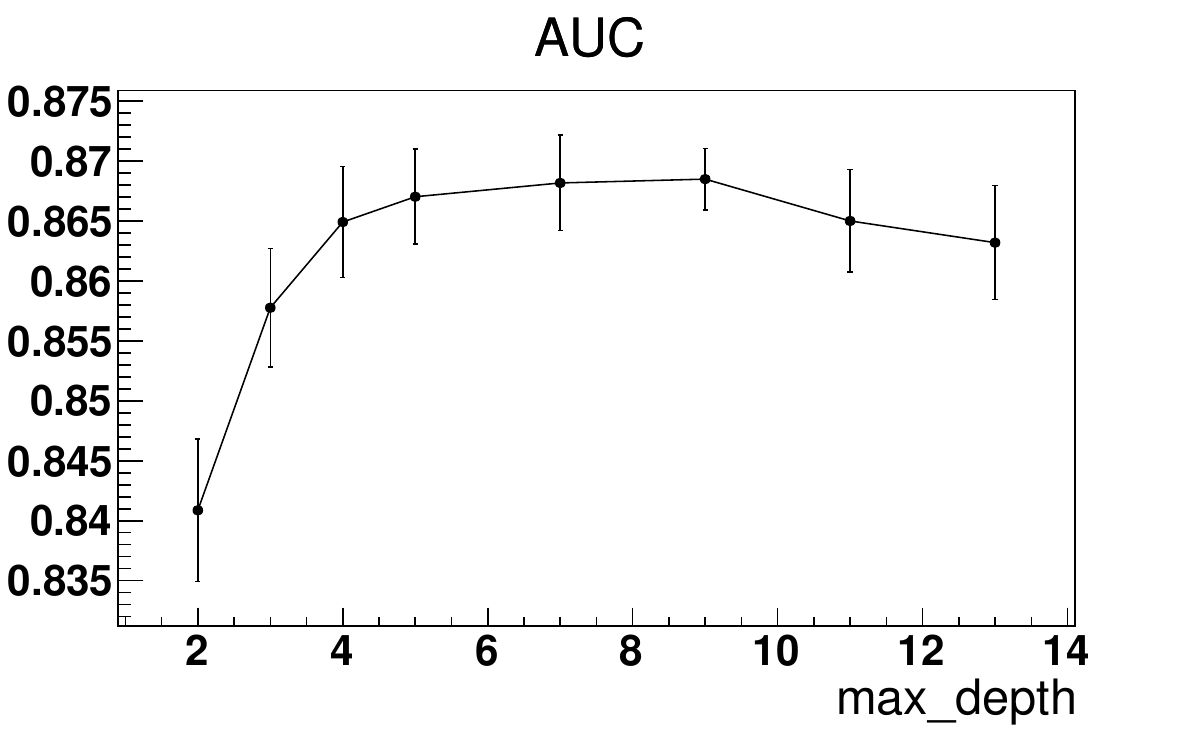}
	\includegraphics[width=0.41\linewidth]{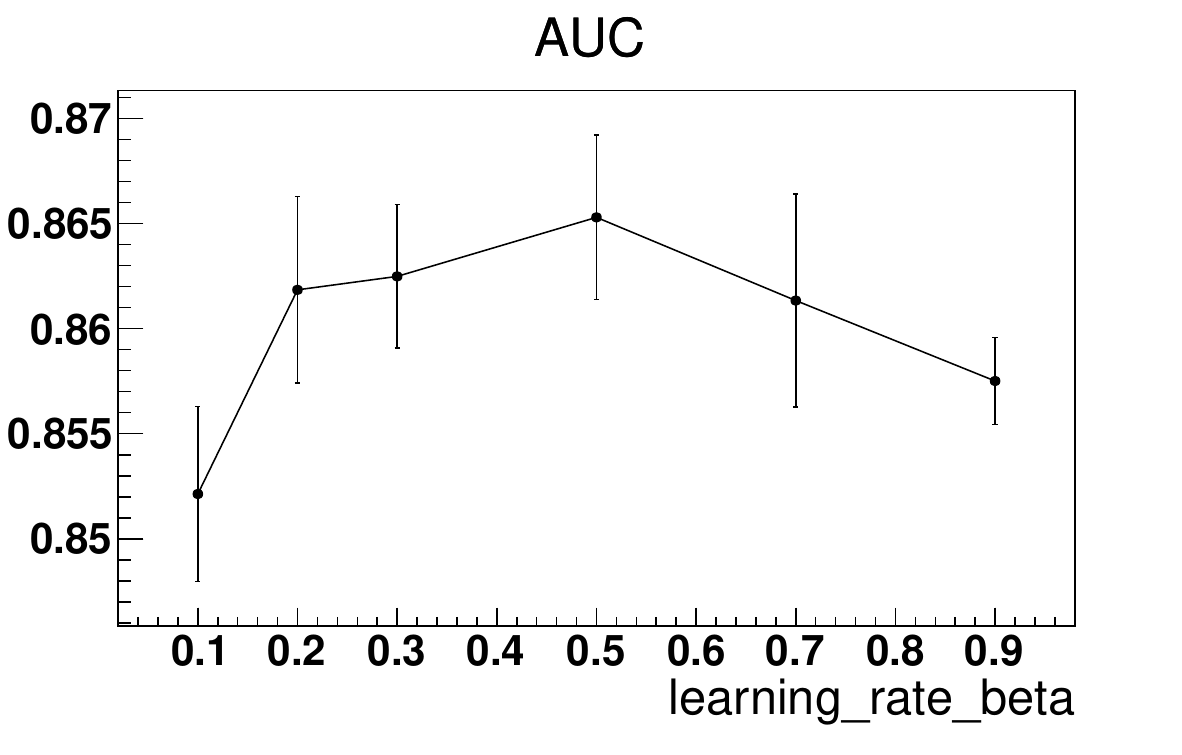}
	\caption{BDT performance (AUC values) for one-dimensional scans of the BDT hyperparameters: number of trees, minimum node size, maximum depth of the tree, and learning rate $\beta$. The error bars represent the standard deviation of five independent trainings of the BDT as using the k-fold technique. }
	\label{fig:hyperparameters}
\end{figure}

To account for statistical fluctuations, each point of the hyperparameter scan was independently evaluated five times, obtaining the standard deviation of the resulting AUC values. To obtain independent tests of the performance without reducing the training dataset by a factor of 5 we used the k-fold technique, which separates the training sample into five subsamples and each time uses four of them for training and one for testing. The result of the one-dimensional scans is shown in Figure~\ref{fig:hyperparameters}. The optimal values are

\begin{center}
	
	\begin{tabular}{lclc} 
		{number of trees} & 900, & {minimum node size} & 2\%, \\
		{maximum tree depth} & 4, & {learning rate} $\beta$ & 0.15. \\ 
	\end{tabular}
\end{center}

\subsection{Results} \label{sec:bdtresults}

\begin{figure}[ht!]
	\centering
	\includegraphics[width=0.47\columnwidth]{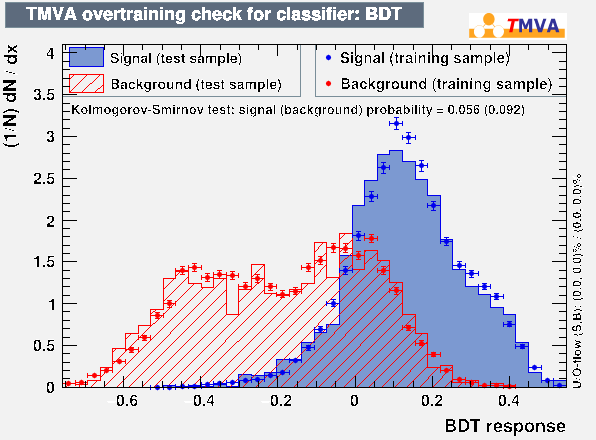}
	\includegraphics[width=0.47\columnwidth]{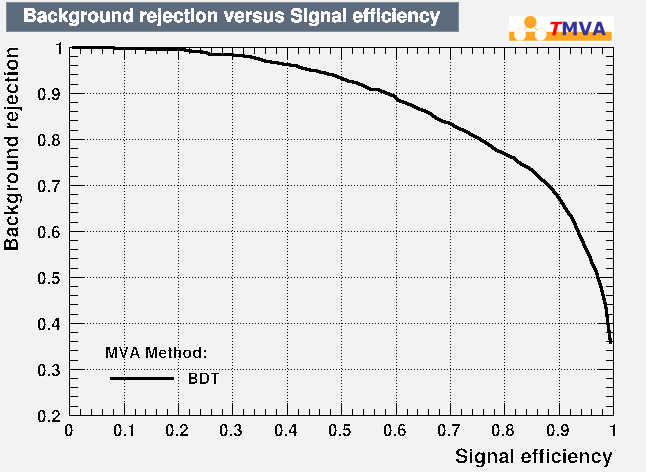}
	\caption{(Left) distribution of the BDT response on signal and background events. The TMVA software~\cite{Hocker:2007ht} automatically  applies cuts over this distribution and evaluates the background rejection and signal effiency at each point. Then, these are plotted together in the so-called ROC curve (right).}
	\label{fig:BDTresponse}
\end{figure}

\begin{figure}
	\centering
	\includegraphics[width=0.45\columnwidth]{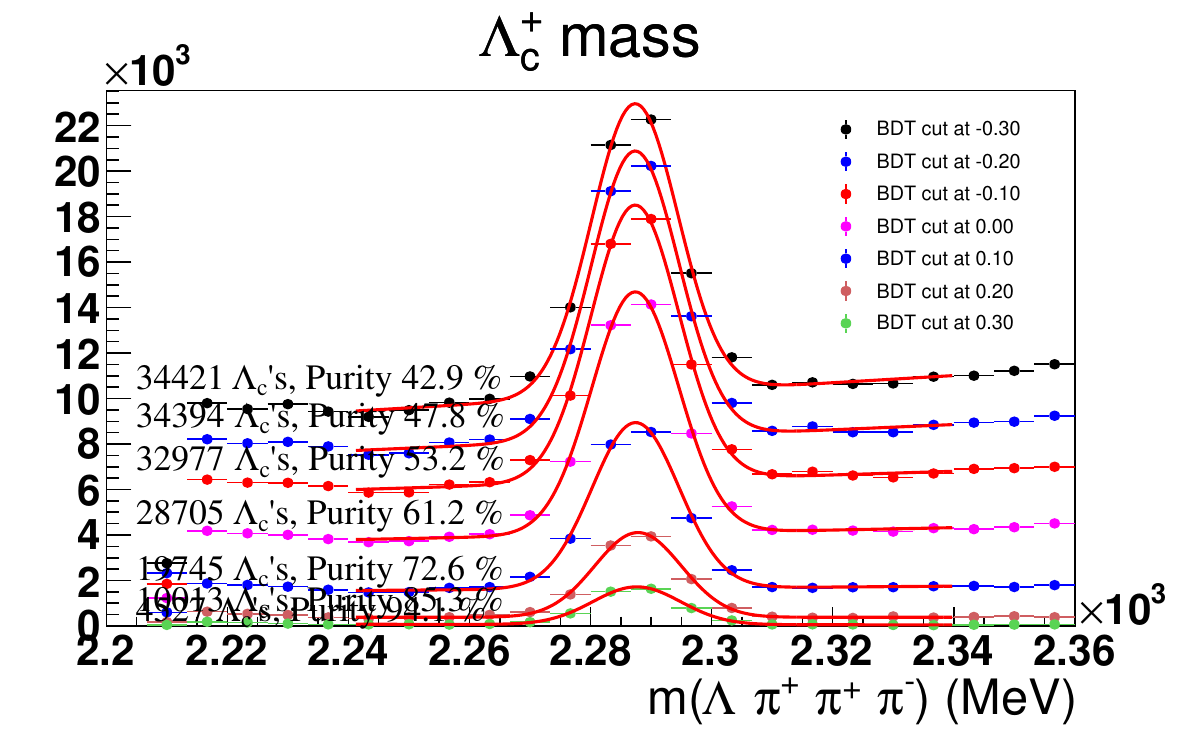}
	\includegraphics[width=0.45\columnwidth]{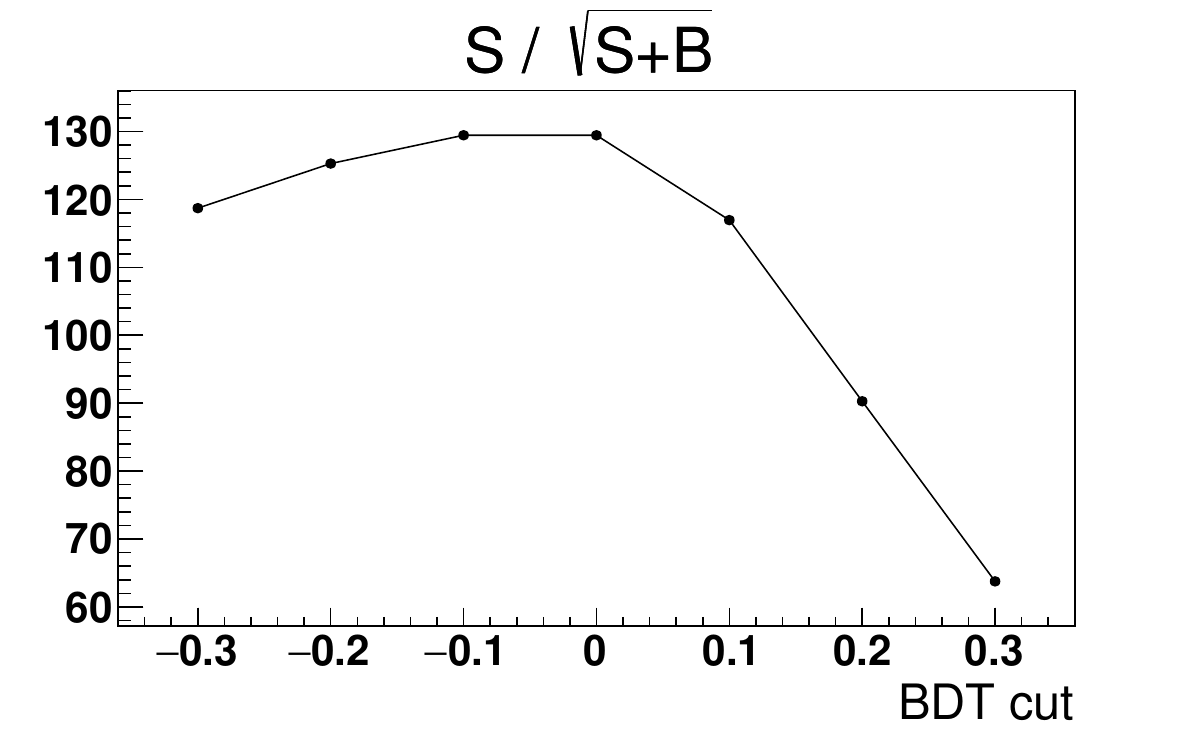}
	\caption{(Left) \Lc mass distributions on the data sample after applying different cuts on the BDT response. The vertical axis indicates the number of events per bin. (Right) the resulting significance is plotted to determine the optimal cut.}
	\label{fig:BDToptimsignificance}
\end{figure}

\begin{figure}
	\centering
	\includegraphics[width=0.47\columnwidth]{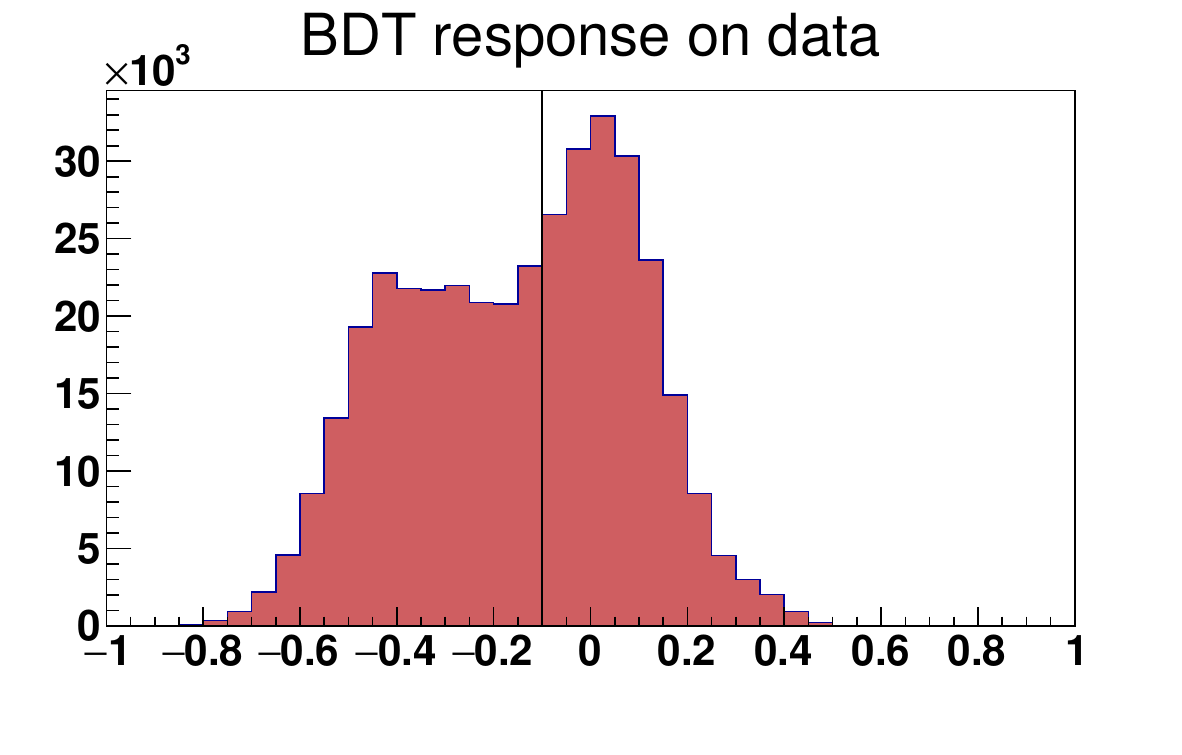}
	\includegraphics[width=0.47\columnwidth]{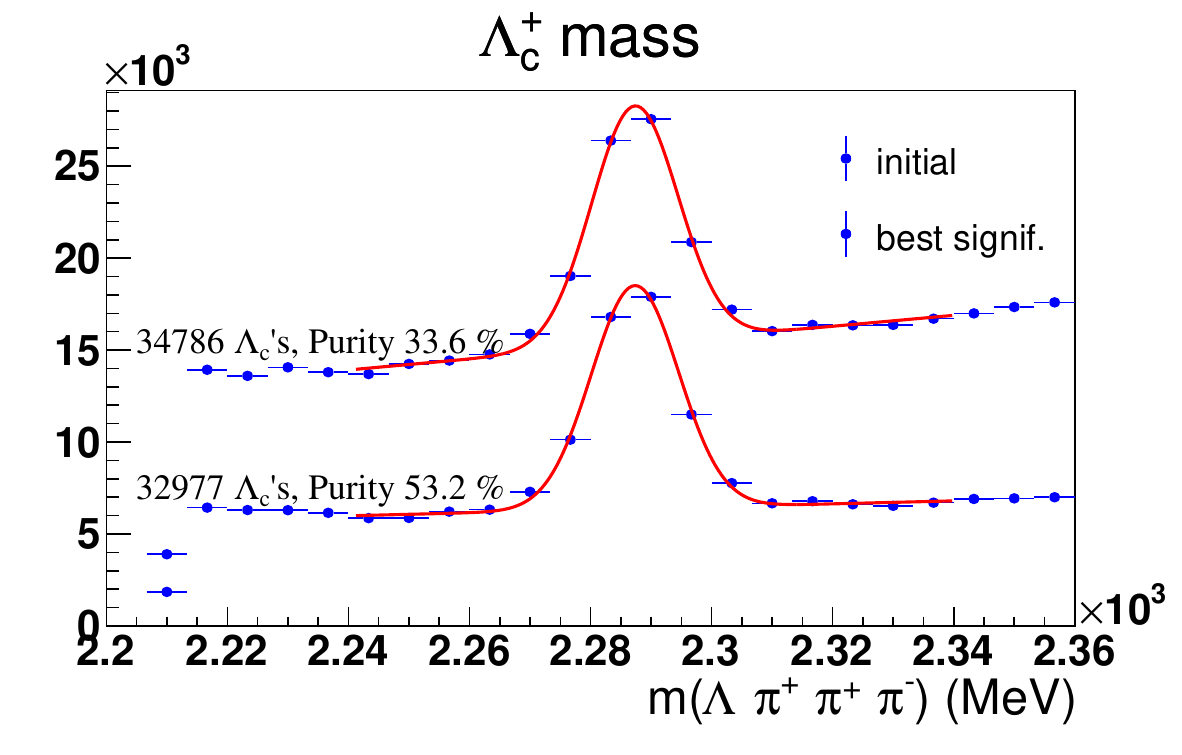}
	\caption{(Left) BDT response on real data (with mixed signal and background, unlike Figure~\ref{fig:BDTresponse}) with the optimal cut value ($>-0.1$). (Right) \Lc mass distributions on data before and after the BDT selection.  The vertical axes indicate the number of events per bin.}
	\label{fig:BDTfinalcut}
\end{figure}

The main results of the BDT are shown in Figures.~\ref{fig:BDTresponse}, \ref{fig:BDToptimsignificance} and \ref{fig:BDTfinalcut}. The purity of signal candidates in the signal region (at 2$\sigma$ from the centre of the invariant-mass peak) improves from 34 to 53\% retaining most of the initial signal with an efficiency of approximately 94\%. The optimal significance is $S/\sqrt{S+B} = 129.4$, where S (B) are the number of signal (background) events in the signal region, as evaluated on real data. The BDT configured with the reduced set of variables is in fact performing slightly worse than the first one, with 111 variables. Specifically, the AUC changes from 0.92 to 0.88. However, at the optimal cut value, the signal significance, as we have seen, is essentially identical for both configurations.
%
%

\begin{figure}
	\centering
	~~~~~~~~~~\includegraphics[width=0.70\columnwidth]{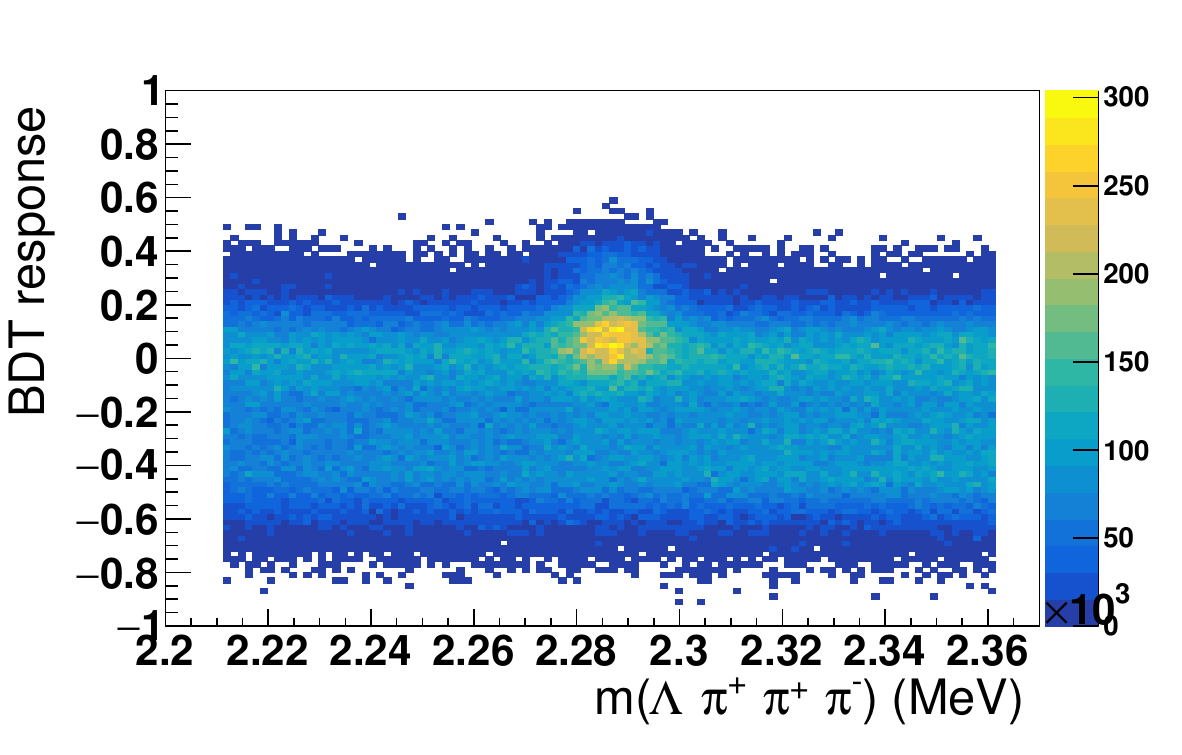}
	\caption{Distribution of the BDT response against the \Lc invariant mass. The hot spot in the middle corresponds to the signal events in the real data sample, which are naturally centered at high values of the BDT response. In the sidebands there is no correlation between the two variables and we can discard a possible mass-dependent BDT response.}
	\label{fig:massdiff}
\end{figure}

The BDT cannot base its decision on whether or not the invariant mass of the \Lc is within the signal region since it does not have the necessary kinematic information to recover the \Lc mass. However, there may be indirect correlations with other variables and it is always recommended to check for correlations of the BDT response with the mass. The two-dimensional histogram is shown in Figure~\ref{fig:massdiff}, finding no problematic structures.

\chapter{Angular fit} \label{ch:fit}

The final event sample for the preliminary analysis of 2016 data has been obtained in the previous section. With the information given to the multivariate classifier, the algorithm has determined the optimal discrimination between signal and background. The number of signal \Lc candidates is about 53\% of the total in the signal region, as obtained by a preliminary mass fit that served to obtain a figure of merit to optimize the selection.

In this section, the \Lc invariant-mass distribution will be fitted with a more complex parameterization. Besides extracting the number of signal candidates, this parameterization will also serve to weight the amount of signal and background at each mass value. The angular fit performs best when this information is included as we will see in Section \ref{sec:angfit}. Estimates of the systematic errors are obtained in Section~\ref{sec:systematics} and preliminary results on the \Lz polarization are presented in Section~\ref{sec:angfitresults}. To finalize, the Dalitz plots are qualitatively analysed in Section~\ref{sec:dalitz}.

\section{Mass fit} \label{sec:massfit}

The combinatorial background below the signal peak is parameterized with a simple exponential line shape,
\begin{equation}
M_b(\mLc) \propto \exp(C_{\rm exp} \mLc),
\end{equation}
where the dependence of $M_b(\mLc)$ with the fit parameter $C_{\rm exp}$ is implicit.
For the signal, in the binned fit of the previous chapter we used a Gaussian line shape, 
\begin{equation}
M_s^{\rm Gauss}(\mLc) \propto \exp\left( - \frac{(\mLc - \mu)^2}{2\sigma^2}\right),
\end{equation}
which we will compare to an asymmetric parameterization of the mass peak with a Crystal Ball function\footnote{From ROOT v6.24, it is implemented by default in the \roofit class \texttt{RooCrystalBall}. } $M_s^{\rm CB}$, named after the Crystal Ball collaboration~\cite{Skwarnicki:1986xj}. This piecewise function is composed of a central Gaussian core and two exponential tails that take over at $\alpha_R$ and $\alpha_L$ number of sigmas from the center of the peak. It is given by
\begin{align}
M_s^{\rm CB} (\mLc)&=\left\lbrace \begin{array}{llc}
A_L \cdot (B_L - \frac{m-\mu}{\sigma})^{-n_L} & \text{ for } & -\infty < \frac{m-\mu}{\sigma} < -\alpha_L \\
\exp\left( - \frac{1}{2} \left( \frac{m-\mu}{\sigma} \right)^2\right)  & \text{ for } & -\alpha_L \leq \frac{m-\mu}{\sigma} < \alpha_R \\
A_R \cdot (B_R - \frac{m-\mu}{\sigma})^{-n_R} & \text{ for } & \alpha_R \leq \frac{m-\mu}{\sigma} < \infty \\
\end{array} \right. , \\
~ & \nonumber  \\
\text{where~} & A_i = \left( \frac{n_i}{|\alpha_i|}\right)^{n_i} \exp \left(-\frac{\alpha_i^2}{2}\right)  \text{~~~and~~~}  B_i = \frac{n_i}{|\alpha_i|} - |\alpha_i|~. \nonumber
\end{align}

\begin{figure}[h]
	\centering
	\includegraphics[width=0.75\linewidth]{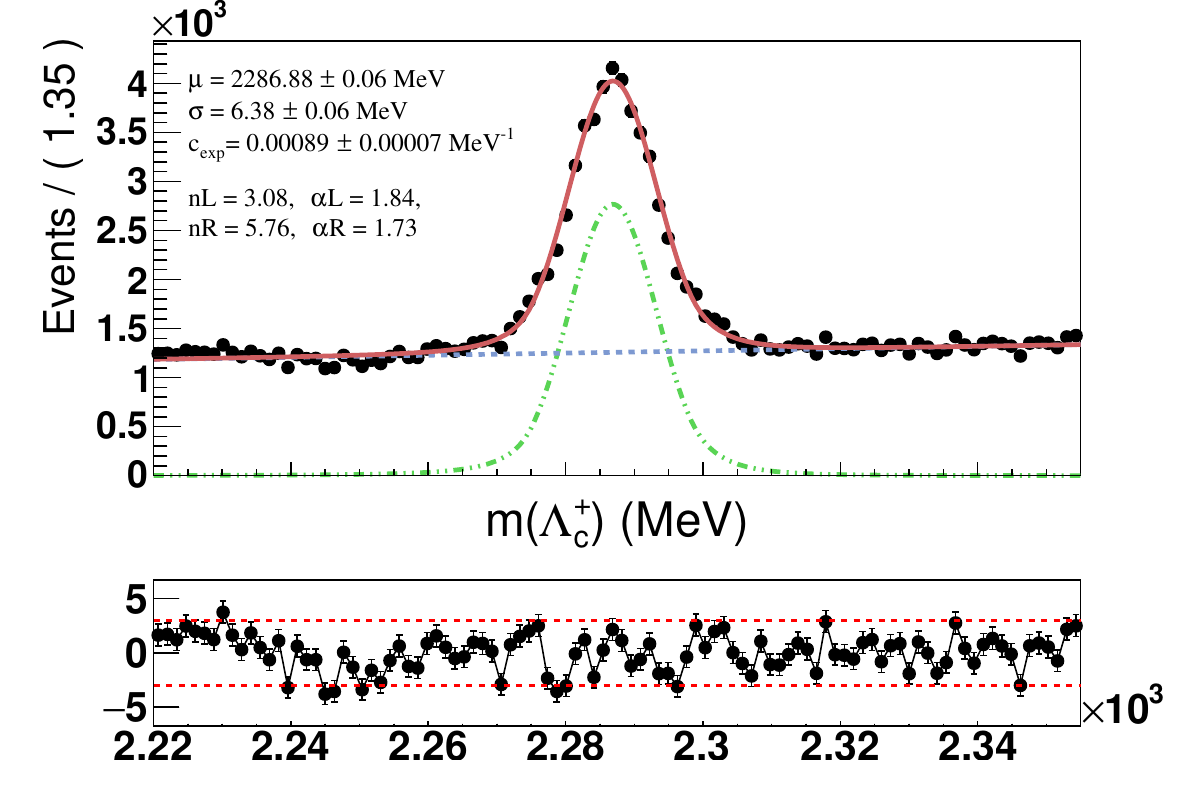}
	\caption{Fit to \Lc invariant-mass distribution modelling the signal with a double-sided Crystal Ball PDF (green dashed line) and the background with an exponential function  (blue dashed line). The bottom panel shows the normalized residuals between the data points and the fitted total PDF. The signal purity, selecting the region within $2\sigma$ of the centre of the peak, is 56.8\% and the number of \Lc baryons 31282.  }
	\label{fig:massfit}
\end{figure}

We perform an unbinned maximum-likelihood fit to the \mLc distribution with the PDF given by
\begin{align}
P^{\rm CB} (m_\Lc) = 	n_s M_s^{\rm CB}(\mLc)~+~ n_b M_b(\mLc),
\end{align}
where $n_s$ and $n_b$ represent the signal and background yields.
The result of the fit is shown in Figure~\ref{fig:massfit}.

\section{Angular fit description} \label{sec:angfit}

\newcommand{\cospr}{\ensuremath{\cos\theta_p\xspace}}
\newcommand{\phipr}{\ensuremath{\phi_p\xspace}}

\newcommand{\costp}{\ensuremath{\cos\theta'_p\xspace}}
\newcommand{\phip}{\ensuremath{\phi'_p\xspace}}

Before fitting the helicity angles we first need to correct for possible distortions of their distribution due to the reconstruction and selection process. The MC sample that we have used was generated assuming the distribution of decay-product directions to be flat in phase space. Then, evaluating the deviation from a uniform distribution in the MC sample allows us to account for these acceptance effects.

The efficiency will be parametrised as a function of the helicity angles using Legendre polynomials. First, the $\phi_p$ angle is rescaled to take values in the range $\phi'_p\in[-1,1]$ in which the Legendre polynomials are orthogonal. For homogeneity, we will denote $\costp = \cospr$. The efficiency parametrisation reads
\begin{equation}
\varepsilon(\cos\theta'_p, \phi'_p) = \sum_{i,j} c_{i,j} {\rm Leg}(\cos \theta'_p , i) {\rm Leg}(\phi'_p , j) ,
\end{equation}
where $ {\rm Leg}(x , i)$ is the Legendre polynomial of order $i$ on the variable $x$ and the coefficients are
\begin{equation}
c_{i,j} = M_{i,j} (2i+1)(2j+1),
\end{equation}
where $M_{i,j}$ is obtained using the method of moments
\begin{equation}
M_{i,j} = \frac{1}{N} \sum_{\rm evs} {\rm Leg}(\cos \theta'_p , i) {\rm Leg}(\phi'_p , j),
\end{equation}
where $N$ is the number of MC events and the sum goes over all events (each with different \costp~and \phip). Events with \Lc and $\bar{\Lambda}_c^-$ are assumed to have no difference in reconstruction efficiency and are simultaneously considered. We are using the normalization
\begin{equation}
\int_{-1}^{1} {\rm Leg}(x , m)  {\rm Leg}(x , n) dx = \frac{2}{2n+1} \delta_{mn},
\end{equation}
where $\delta_{mn}$ is the Kronecker delta. The allowed maximum order of the polynomials is varied until we find a reasonable agreement with the MC data. In our case, we use Legendre polynomials of up to order two for $\cos\theta'_p$ and eight for $\phi'_p$.

The signal PDF for the angular fit is the product of the acceptance function $\varepsilon(\cos\theta_p, \phi_p)$\footnote{We use the same notation $\varepsilon(\cos\theta^{(\prime)}_p, \phi^{(\prime)}_p)$ for the acceptance efficiency on original and transformed angular variables.} with the physical angular-distribution function (also in Eq.~\eqref{eq:angdist}),
\begin{equation} \label{eq:angdistcompleteangles} 
F(\cos\theta_p, \phi_p) = 1 + \alpha_{\lz} [ P_{\Lambda,z} \cos \theta_p + (P_{\Lambda,x} \cos \phi_p + P_{\Lambda,y} \sin \phi_p)\sin\theta_p],
\end{equation}
where $\alpha_\lz$ is fixed to its known value~\cite{PDG}.
Altogether, the signal PDF for the angular distribution reads
\begin{equation} \label{eq:angidstsignal}
A_s(\cos\theta_p, \phi_p) = \varepsilon(\cos\theta_p, \phi_p) ~F(\cos\theta_p, \phi_p).
\end{equation}

The combinatorial background is dominated by real \Lz baryons, which purity in the \Lz invariant-mass signal region is around 98\%, as shown in Figure~\ref{fig:preselectionPeaks}. The polarization of these \Lz baryons is expected to be zero due to (1) the strong dilution introduced by the (randomly defined) quantization frame of the (non-physical) \Lc candidate and (2) the dominant \Lz production in prompt $pp$ interactions, for which the polarization is known to be negligible at the LHC~\cite{ATLAS:2014ona}.

In any case, we do not need to assume a flat distribution on the helicity angles $(\cos\theta_p, \phi_p)$ of the background. Since this distribution is also affected by the reconstruction and selection cuts, we will parametrise the background angular distribution directly using real data events from the sidebands. \roofit includes by default two methods to do this. On the one hand, \texttt{RooHistPdf} separates the $(\cos \theta_p, \phi_p)$ space in bins, freezes their relative content, and allows to fit this distribution to a dataset extracting a global scale factor. On the other hand, \texttt{RooNDKeysPDF} saves all the events of the input data and is independent of binning choices. For the size of our dataset, \texttt{RooHistPdf} is faster than \texttt{RooNDKeysPDF} approximately by a factor 150~\footnote{This difference could be substantially reduced with some optimization of the \href{https://root-forum.cern.ch/t/roondkeyspdf-is-very-slow-for-print/46929}{\texttt{RooNDKeysPDF}}  class for large datasets.}, with comparable results. 
For clarity, we will note the background PDF for the angular fit as
\begin{equation}
A_b(\cos\theta_p, \phi_p) = \texttt{RooHistPDF}(\text{\mLc sideband}).
\end{equation}
We are in a position to join the signal and background PDFs and perform the fit. We have compared two possible approaches:
\begin{enumerate}
	\item[\textbf{I}] \textbf{Sequential}: after the mass fit, the events in the \mLc signal region are selected within 2$\sigma$ of the centre of the peak, the yields $(n_s,~n_b)$ in this region are fixed, and the $ (\cos\theta_p, \phi_p)$ distribution of these events is fitted to the total PDF
	\begin{equation}
	A_{\rm I} (\cos\theta_p, \phi_p) = n_s A_s(\cos\theta_p, \phi_p)~+~ n_b A_b(\cos\theta_p, \phi_p).
	\end{equation}
	\item[\textbf{II}] \textbf{Simultaneous}: combining the mass fit with the angular fit by building a three-dimensional PDF that shares the yields of signal and background as
	\begin{equation} \label{fig:pdfsimul}
	A_{\rm II} (m_\Lc , \cos\theta_p, \phi_p) = 
	n_s M_s(m_\Lc) A_s(\cos\theta_p, \phi_p)~+~ n_b M_b(m_\Lc) A_b(\cos\theta_p, \phi_p).
	\end{equation}
\end{enumerate}

In the simultaneous fit, the relative importance of the events is considered also within the signal regions (\textit{i.e.} tails have a smaller weight), whereas in the sequential fit all the events in the signal region count the same. Moreover, defining the signal region to fix the yields has a systematic error associated that is completely avoided in the simultaneous fit, simplifying the propagation of systematic uncertainties from previous steps in the analysis chain. On the other hand, the simultaneous fit is performed on many more events, as the sidebands need to be included to fit the exponential background in \mLc. For the same reason, as most of the events are background, the projection of the angular helicity angles is almost fully diluted and the shape of signal events cannot be easily visualized.

The correlation between the variables \mLc and $(\cos\theta_p, \phi_p)$ has been verified to be negligible, both for signal and background, as assumed in Eq.~\ref{fig:pdfsimul}.

The results of the fits are shown in Table~\ref{tab:comparefits}, where we observe that the error of most of the parameters decreases in the simultaneous fit. For example, the uncertainty of the main polarization component, $P_{\Lz, z}$, is reduced by 4\%. Nevertheless, the $\sigma$ parameters of the mass signal PDF deviate, a difference that can be accounted for by the different systematics in the two procedures.
In these fits the parameters $(\alpha_L,~\alpha_R,~n_L,~n_R)$ of the $M_s^{\rm CB}(m_\Lc)$ function have been fixed to the values obtained in a mass fit to the MC sample, which are displayed in Figure~\ref{fig:massfit}.

{\scriptsize

	\begin{table}[] 
		\centering
		\caption{Results of the fit with sequential and simultaneous strategies regarding the mass and angular fit. The rule in the middle separates the results of the two fits in the sequential strategy.}
		\renewcommand{\arraystretch}{1.1}
		\begin{tabular}{lr p{0.2cm} lrcl}
			\hline  \hline
			&  \multicolumn{3}{c}{\textbf{ I: Sequential}}  & \multicolumn{3}{c}{\textbf{ II: Simultaneous}}  \\ \hline
			$C_{\rm exp}$            & (8.79   & $\pm$ & 0.73)$\,\times 10^{-4}$ & (8.90  & $\pm$ & 0.73)$\,\times 10^{-4}$ \\
			$n_b$    & (1.266  & $\pm$ & 0.004)$\,\times 10^{5}$ & (1.259  & $\pm$ & 0.004)$\,\times 10^{5}$ \\
			$\mu$   & (2.28689   & $\pm$ & 0.00006)$\,\times 10^{3}$ & (2.28688  & $\pm$ & 0.00006)$\,\times 10^{3}$ \\
			$\sigma$   & (6.26   & $\pm$ & 0.06) & (6.38  & $\pm$ & 0.06) \\
			$n_s$ & (3.35   & $\pm$ & 0.03)$\,\times 10^{4}$ & (3.43  & $\pm$ & 0.03)$\,\times 10^{4}$ \\ \cline{1-4}
			$P_{\Lz,x}$              & (1.97   & $\pm$ & 1.69)$\,\times 10^{-2}$ & (0.97 & $\pm$ &  1.61)$\,\times 10^{-2}$ \\
			$P_{\Lz,y}$              & (3.19   & $\pm$ & 1.81)$\,\times 10^{-2}$ & (3.94  & $\pm$ & 1.74)$\,\times 10^{-2}$ \\
			$P_{\Lz,z}$             & ($-23.7$  & $\pm$ & 1.71)$\,\times 10^{-2}$ & ($-24.2$ & $\pm$ & 1.64)$\,\times 10^{-2}$ \\ \hline  \hline
		\end{tabular}
		\label{tab:comparefits}
	\end{table}

}

\section{Systematic uncertainties} \label{sec:systematics}

The best fit values of the previous section were obtained together with an uncertainty that accounts for the statistical fluctuations of the data sample. Additionally, also the different choices in each step of the analysis chain introduce systematic uncertainties in the fit values. Estimating the genuine systematic uncertainties isolating the statistical fluctuations, and without double counting them is not an easy task. In this section, we provide preliminary estimates of some of the most important systematic uncertainties along with cross-checks of the nominal fit with alternative configurations.

The systematic uncertainties shown in Table~\ref{tab:systematics} were obtained as follows:

\begin{itemize}
	\item \textbf{Limited MC statistics}\\
	The parametrisation of the acceptance function $\varepsilon(\cospr,\phipr)$ with a finite-size MC sample represents a dominant source of systematic uncertainty. This was estimated by rerunning the fit to the data 100 times, each of them doing the acceptance corrections anew with a different (mimicked) MC sample, and evaluating the standard deviation of the fit results through a Gaussian fit. Each of the acceptance corrections was itself constructed with the procedure outlined in Section~\ref{sec:angfit} using a $(\cospr,\phipr)$ binning and fluctuating the bin contents according to Poisson errors. 
	
	\item \textbf{Angular acceptance parametrisation}\\
	An alternative parametrisation of the angular acceptance was employed reducing in one order the Legendre polynomials for the two proton angles. The difference between the central values of this fit and the nominal fit is taken as an estimate of the systematic uncertainty, in Table~\ref{tab:systematics}. 
	\item \textbf{Background angular distribution}
	\begin{itemize}
		\item Different PDF: the background PDF for the angular fit was obtained from the same sideband events, but using \texttt{RooNDKeysPDF}. In the nominal fit we use a binned approach with \texttt{RooHistPDF}, as explained in the previous section.
		\item Different sideband: instead of selecting all sideband events positioned at least at $2\sigma$ from the centre of mass peak, the fit is redone by (1) selecting those that start at $3\sigma$ from the peak and (2) removing the events at $m_\Lc \leq 2.24 \,\gev$, where a small bump is observed on the mass distribution (Figure~\ref{fig:massfit}).
	\end{itemize}
	
	\item \textbf{External input $\mathbf{\alpha_\Lz}$}\\
	The decay-asymmetry parameter, fixed to its central value in the nominal fit, is taken at the extreme values within its uncertainty.
	
	\item \textbf{Reweighting of MC events}\\
	The event weights extracted in Section~\ref{sec:reweightMC} to improve the data/MC agreement in the multivariate classifier are used to calculate the angular acceptance.
	
	\item \textbf{Mass parametrisation}\\
	The Crystal Ball PDF to describe the mass distribution is replaced by a simple Gaussian and a Johnson function.
	The former gives the largest effect and is taken as systematic uncertainty.
\end{itemize}

Other potentially relevant sources of systematic uncertainty not estimated yet include the PID correction (Section~\ref{sec:PIDcorrection}), the trigger selection (Section~\ref{sec:trigger}), or the DTF configuration.

The fit was redone with different BDT selections, determining the background PDF with only one sideband, and separating the data collected with different polarities of the LHCb dipole magnet. The polarization results are shown in Table~\ref{tab:crosschecks}. The different fit variations are affected by statistical differences, yet to be evaluated. The test with different magnet polarities shows large differences on the $P_{\Lz,x}$ component and needs further investigation.

%

%
%

\begin{table}
	\caption{Estimated systematic uncertainties. The signed contributions were obtained as the difference between the nominal fit value and the alternative method. All values are in percentage (\%).}
	\centering
	\begin{tabular}{lccc}
		\hline \hline
		\multicolumn{4}{c}{Systematic uncertainties  }       \\
		Source   &  $P_{\Lz,x}$  & $P_{\Lz,y}$ & $P_{\Lz,z}$  \\
		\hline
		Limited MC statistics (on efficiency)                      & $\phantom{+}1.300 $    & $\phantom{+}1.100 $   & $\phantom{+}0.800 $     \\
		Angular acceptance parametrisation                         & $+0.181$    & $-0.444$   & $-0.014$      \\
		Background angular distribution (\texttt{RooNDKeysPdf})    & $+0.035$    & $+0.155$   & $+0.739$      \\
		Background angular distribution (3$\sigma$)       & $+0.454  $  & $-0.280  $ & $+0.016  $    \\
		Background angular distribution (cut sideband)    & $+0.306  $  & $+0.824  $ & $+0.442  $    \\
		External input $\alpha_\Lz=0.750\pm 0.014$        & $\pm0.019$  & $\pm0.075$ & $\pm0.460$      \\
		Reweighting of MC events (on efficiency)          & $-0.094  $  & $+0.152  $ & $+0.234  $    \\
		$\Lc$ mass parametrisation                        & $+0.765  $  & $-0.110  $ & $+0.167  $    \\
		\hline 
		Sum in quadrature                                 & 1.43     & 1.49    & 1.21      \\
		\hline \hline
	\end{tabular}
	\label{tab:systematics}
\end{table}


\begin{table}
	\caption{Cross checks of the polarization results. All values are in percentage (\%).}
	\centering
	\resizebox{1\linewidth}{!}{
		\begin{tabular}{lccc}
		\hline \hline
		\multicolumn{4}{c}{Cross-checks  }       \\
		    Fit configuration   &  $P_{\Lz,x}$  & $P_{\Lz,y}$ & $P_{\Lz,z}$   \\
			\hline
			Nominal     & 0.97 $\pm$ 1.61  & 3.94 $\pm$ 1.74 & $-24.12$ $\pm$ 1.64  \\
			Alternative BDT (with 111 variables)                                   & 1.28 $\pm$ 1.61    & 4.04 $\pm$ 1.74   & $-24.56$ $\pm$ 1.64      \\
			Looser BDT cut (-0.2)                             & 2.68 $\pm$ 1.66    & 3.38 $\pm$ 1.79   & $-26.43$ $\pm$ 1.69      \\
			Tighter BDT cut (0.0)                             & 0.97 $\pm$ 1.61    & 3.94 $\pm$ 1.74   & $-24.19$ $\pm$ 1.64      \\
			Background angular dist. from upper sideband      & 0.41 $\pm$ 1.58    & 1.97 $\pm$ 1.69   & $-25.61$ $\pm$ 1.61      \\
			Background angular dist. from lower sideband      & 1.43 $\pm$ 1.57    & 5.77 $\pm$ 1.70   & $-21.43$ $\pm$ 1.60      \\
			Dipole magnet polarity MagUp                      & $-7.42$ $\pm$ 2.37$\phantom{+}$    & 2.90 $\pm$ 2.56   & $-21.52$ $\pm$ 2.41      \\
			Dipole magnet polarity 	MagDown                   & 8.01 $\pm$ 2.14    & 4.69 $\pm$ 2.30   & $-24.90$ $\pm$ 2.17      \\
			\hline \hline
	\end{tabular}}
	\label{tab:crosschecks}
\end{table}

\section{Results} \label{sec:angfitresults}

The \Lz polarization obtained in this preliminary analysis of \threepi decays collected by the LHCb experiment during 2016, and using only the \lDD sample is
\begin{equation}
\begin{array}{lcc} \label{eq:finalpol}
P_{\Lz,x} &=& (1.0\pm 1.6\pm 1.4)\%, \\
P_{\Lz,y} &=& (4.0\pm 1.7 \pm 1.5)\%,  \\
P_{\Lz,z} &=& (-24.1\pm 1.6 \pm 1.2)\%,
\end{array}
\end{equation}
where the first uncertainty is statistical and the second systematic. 

\begin{figure}
	\centering
	\includegraphics[width=0.4\linewidth]{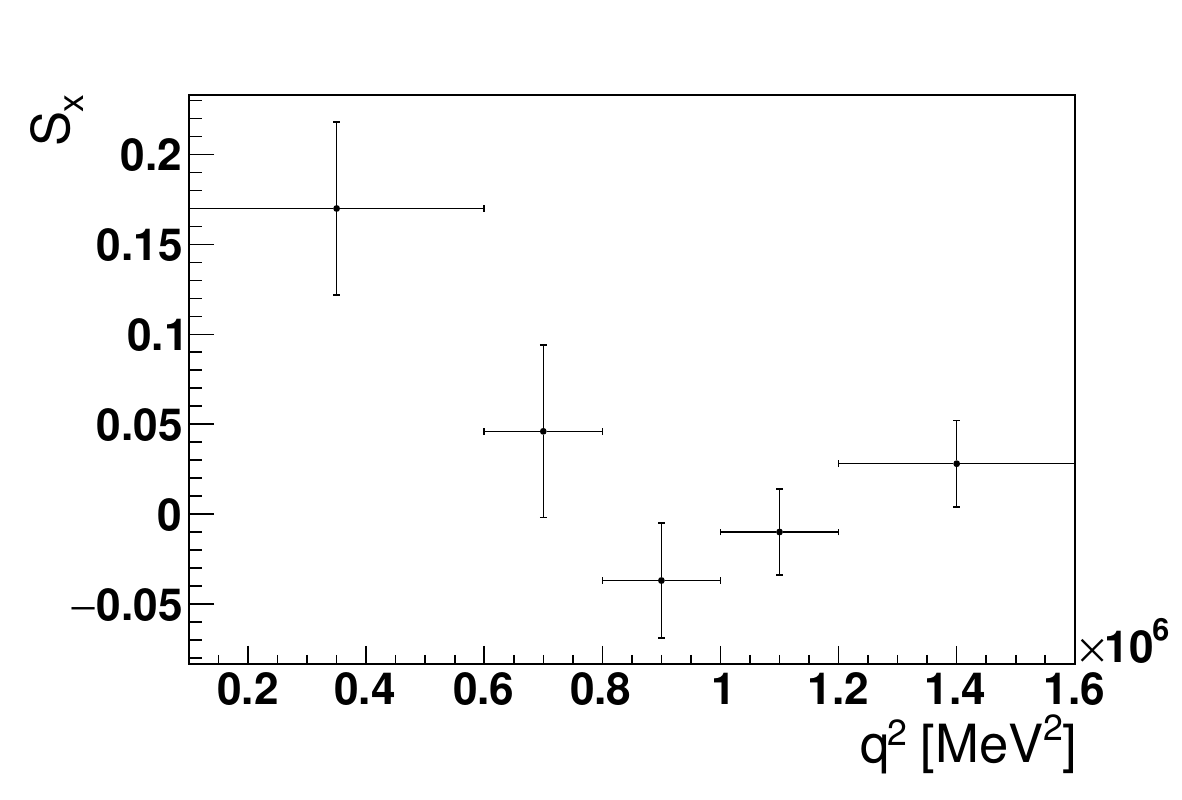}
	\includegraphics[width=0.4\linewidth]{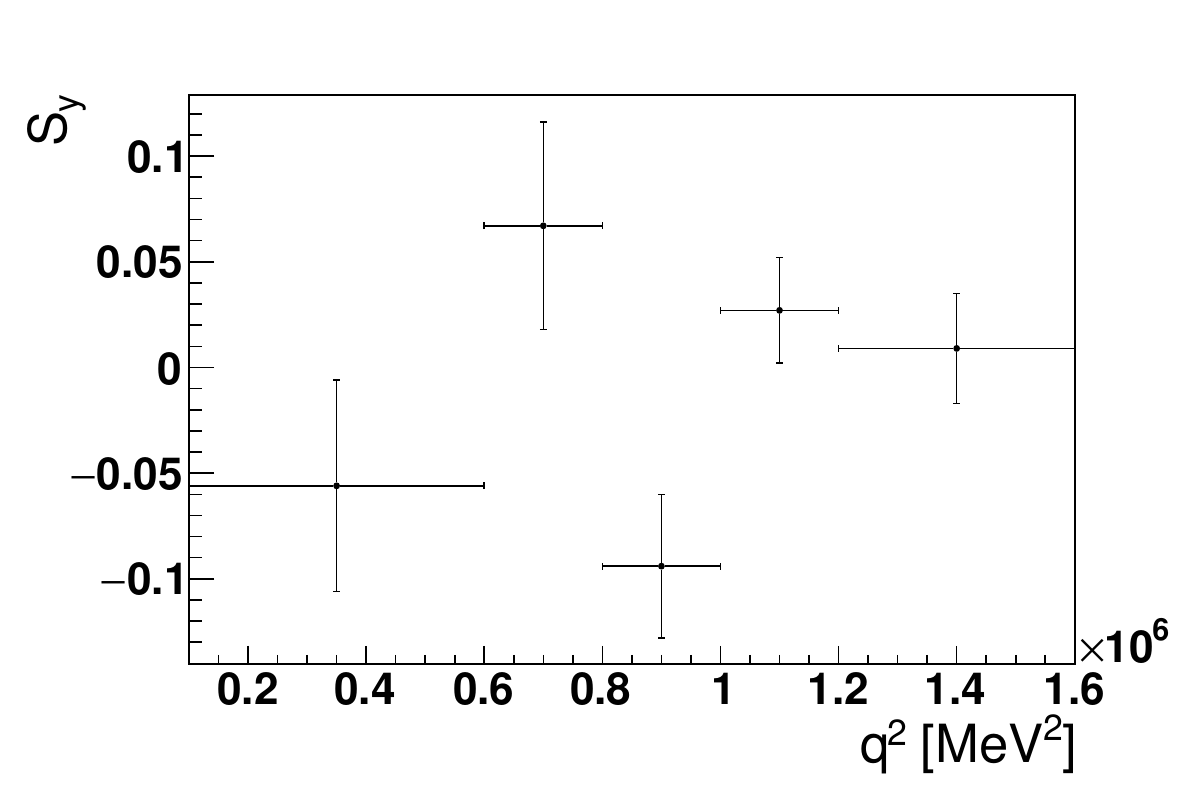}
	\includegraphics[width=0.4\linewidth]{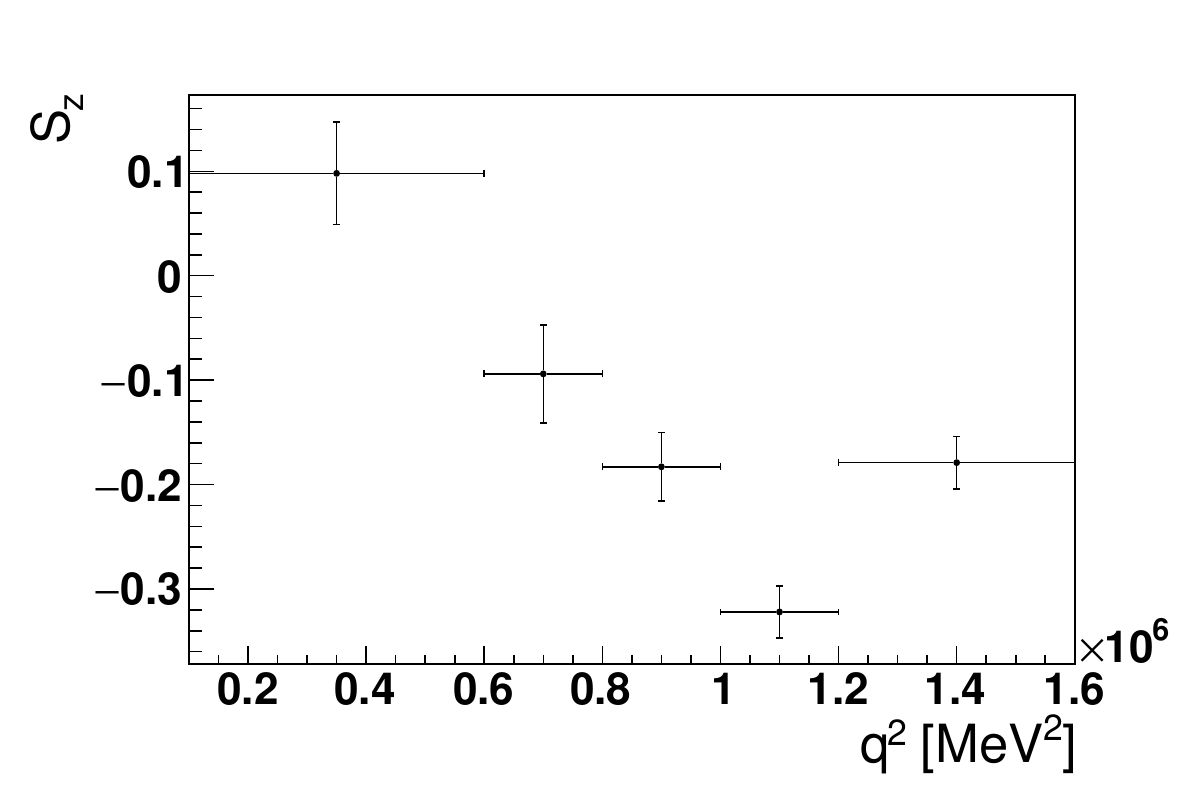}
	\caption{Polarization values for different bins of $q^2$. In these plots the polarization components  $(P_{\Lz,x},P_{\Lz,y},P_{\Lz,z})$ are noted as $(s_x,s_y,s_z)$. The vertical error bars represent only the statistical uncertainty. }
	\label{fig:polq2}
\end{figure}

As discussed in Chapter~\ref{ch:strategy}, it is also very interesting to study the dependence of the polarization as a function of $q^2 = p_\Lc^2 - p_\Lc^2$.
Separating the dataset in five bins of $q^2$, the fitting process is repeated for each of them including the acceptance correction. The results are shown in Figure~\ref{fig:polq2}.

\section{Dalitz plot} \label{sec:dalitz}

In all of the previous experimental analyses with \threepi decays that we are aware of~\cite{BESIII:2015bjk,FOCUS:2005sye,CLEO:1990unu,Anjos:1989tc,ACCMOR:1990gke,ARGUS:1988hly}, the number of signal candidates was at least a factor $30$ smaller than in our preliminary analysis of part of the LHCb data, with only the 2016 \lDD sample. This large amount of signal candidates opens the possibility to study in detail the dynamics of strong intermediate resonances in the decay. In the following, we provide only a qualitative description of the relevant invariant mass distributions.

\begin{figure}
	\centering
	\includegraphics[width=0.48\linewidth]{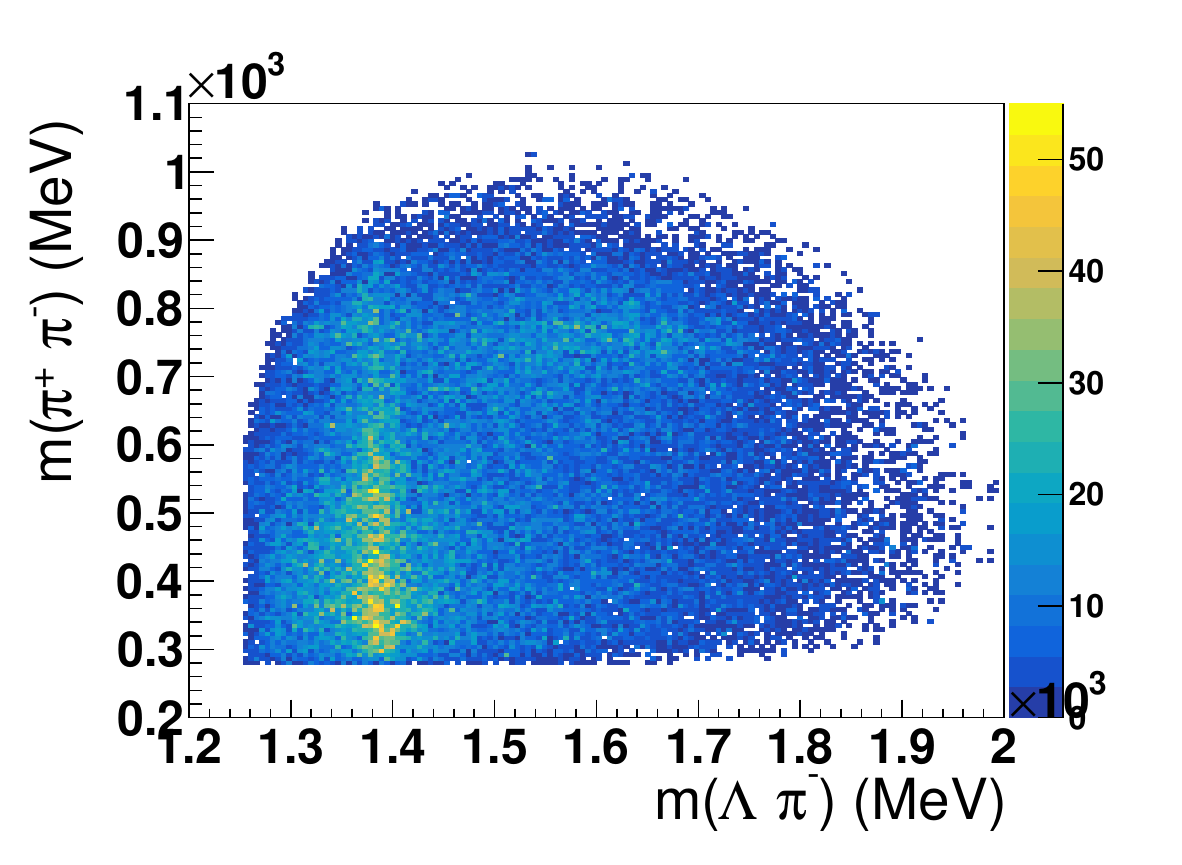}
	\includegraphics[width=0.48\linewidth]{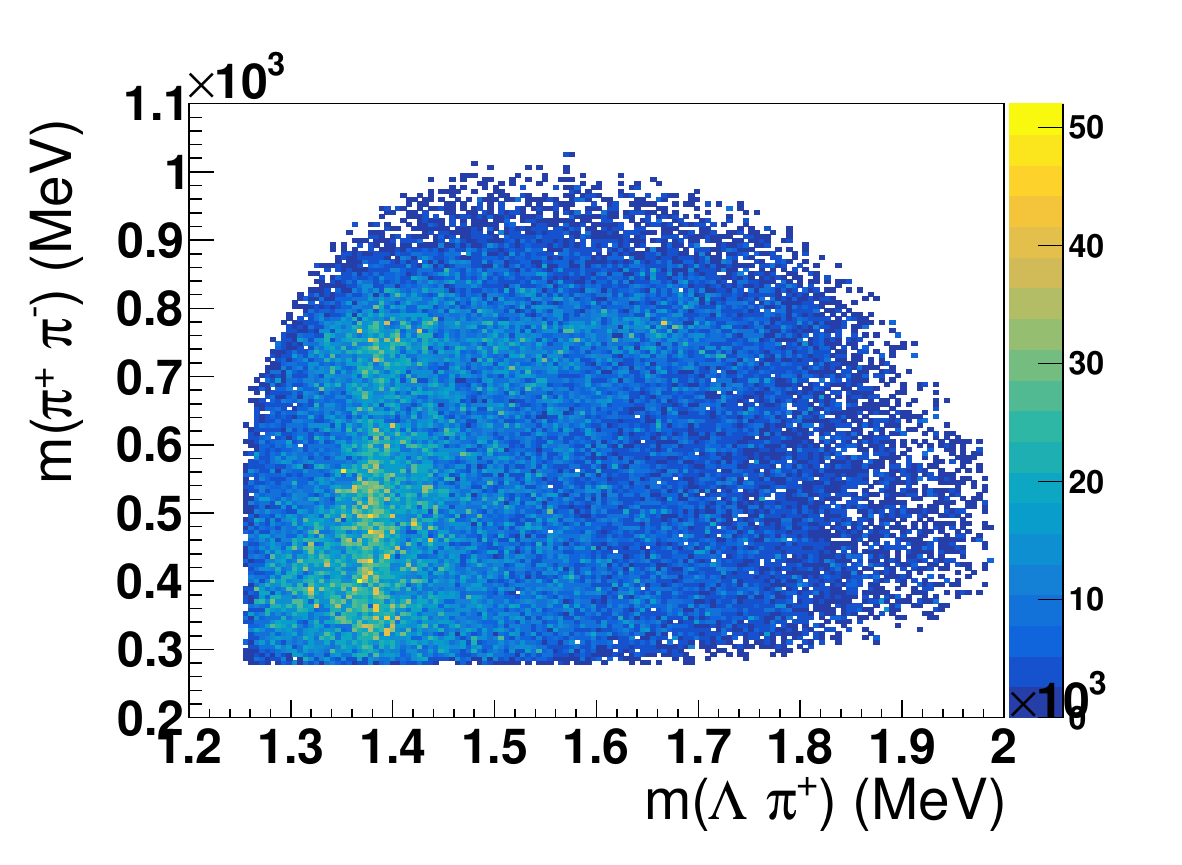}
	\includegraphics[width=0.48\linewidth]{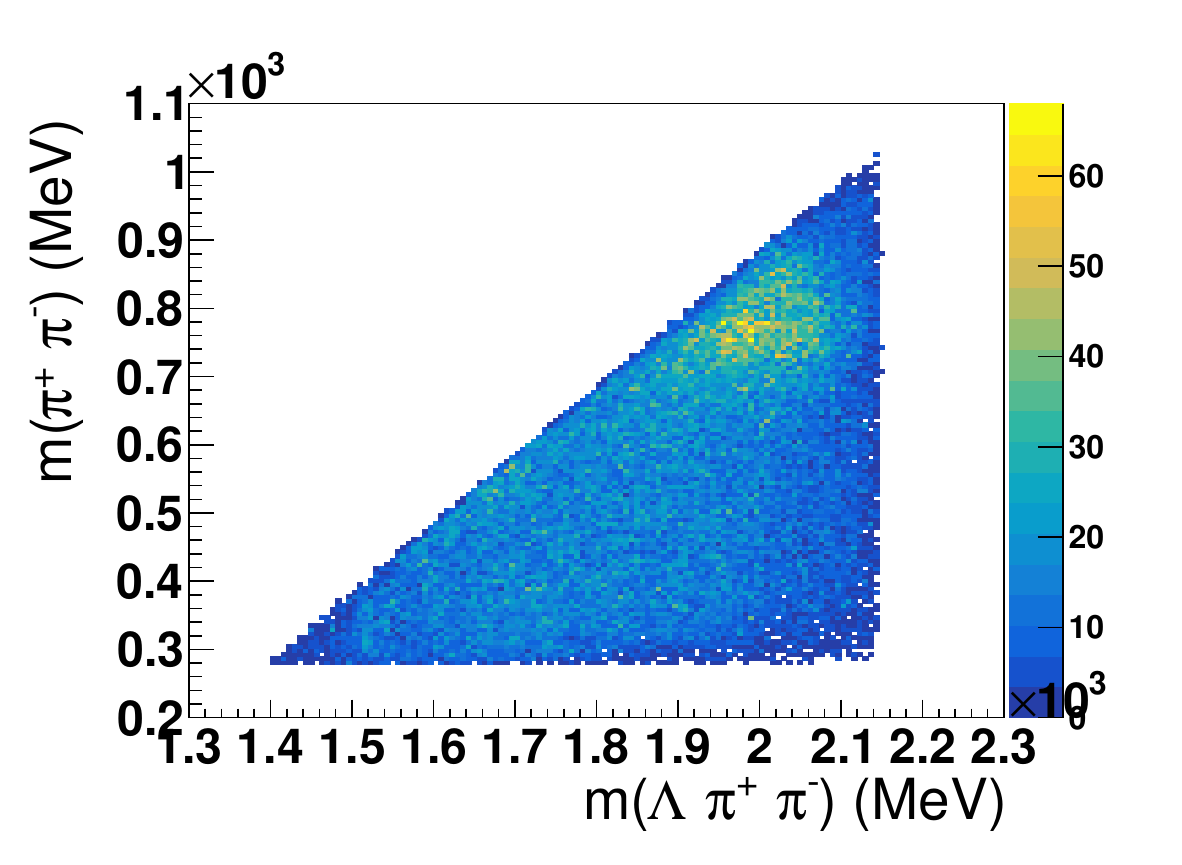}
	\caption{Dalitz plots of the \threepi decay.  Since the two \pip are indistinguishable, the invariant-mass combinations with each of them are superimposed (symmetrized), thus increasing the statistics of the plot by a factor of two.}
	\label{fig:dalitz}
\end{figure}

\begin{figure}
	\centering
	\includegraphics[width=0.45\linewidth]{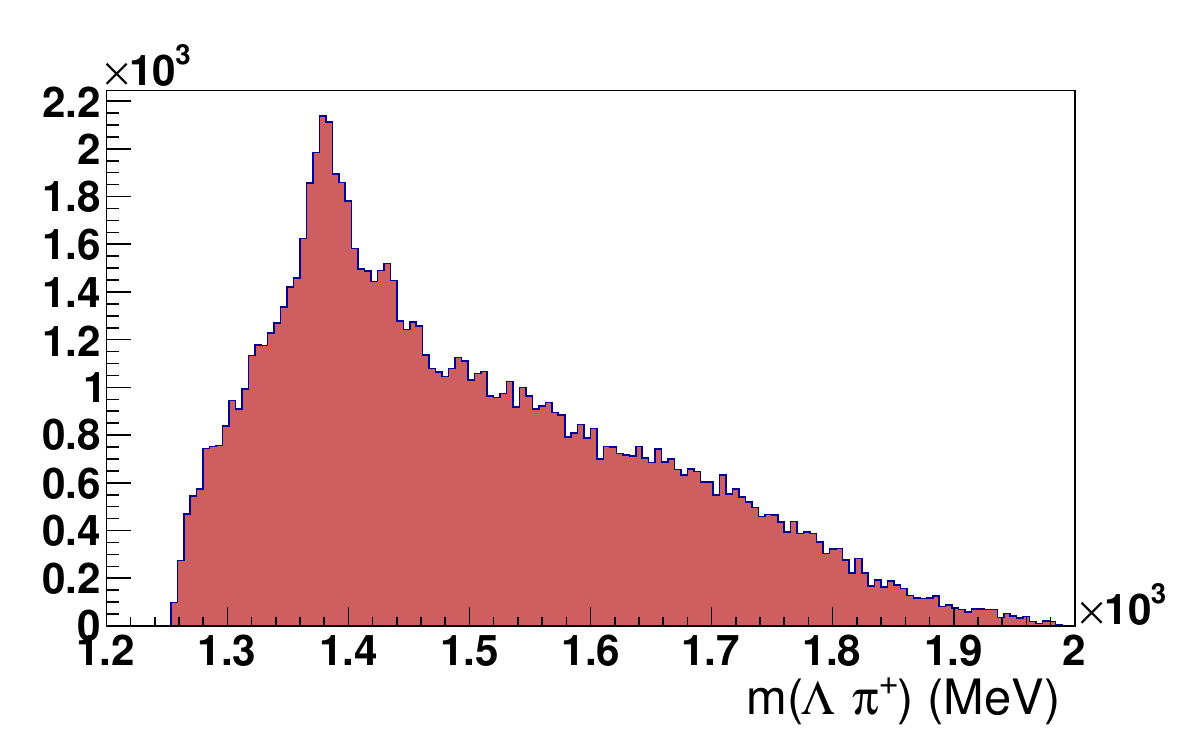}
	\includegraphics[width=0.45\linewidth]{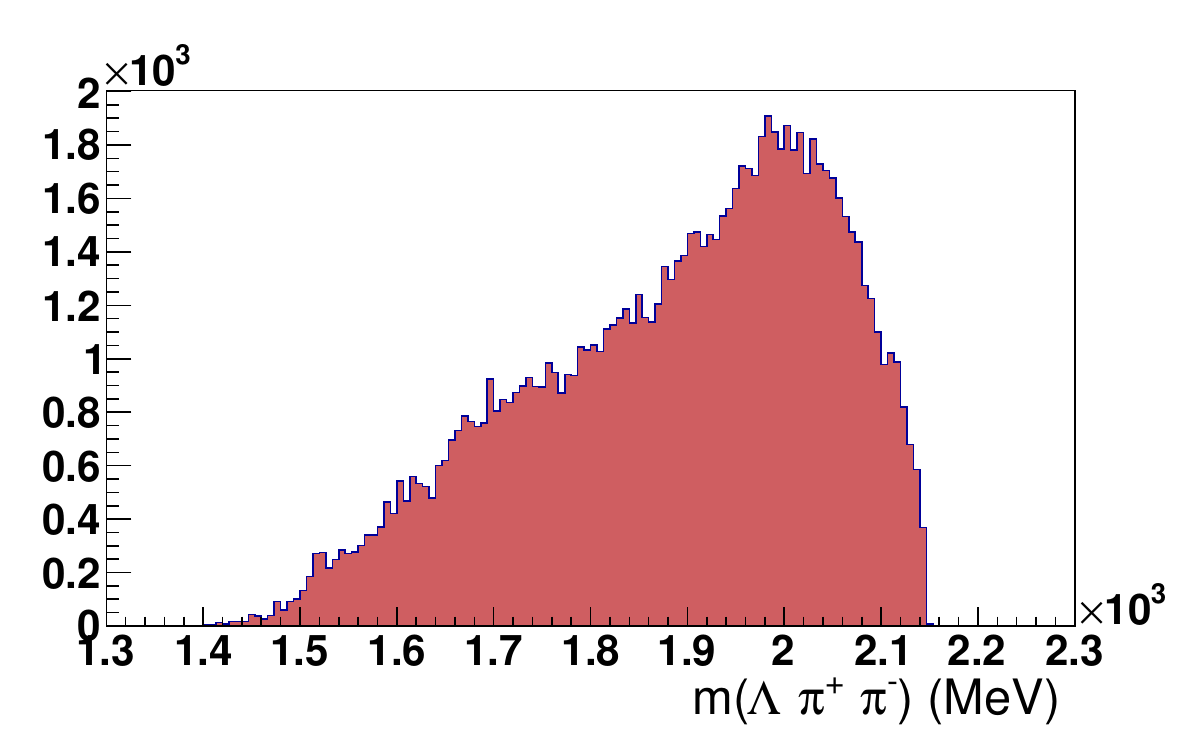}
	\caption{One-dimensional projections of the Dalitz plots, showing the invariant mass of the (\Lz\pip) and (\Lz\pip\pim) systems.}
	\label{fig:dalitz1D}
\end{figure}

In the PDG~\cite{PDG}, the only resonances quoted for this decay are $\Sigma(1385)^\pm \to \Lambda \pipm$ and $\rho(770)\to \pip \pim$. These are unequivocally present in our sample, as shown in Figure~\ref{fig:dalitz}. However, we do not find any other obvious structure. In the one-dimensional projection of the Dalitz plots (Figure~\ref{fig:dalitz1D}) we find some small bumps in the distributions of $m(\Lz \pipm)$ and $m(\Lambda \pip \pim)$, which nevertheless may be due to statistical fluctuations. The only possible structure that matches a known resonance is the $\Lambda(1520) \to \Lz \pip \pim$. Its presence could be clarified with the combination of all Run II data. However, to quantify its contribution an amplitude analysis would be required.
Probably a more appealing test consists on the analysis of the $m(\Lz \pipm)$ distribution, attempting to simultaneously fit the $\Sigma(1385)^\pm$ peak together with a broader $J^P=\frac{1}{2}^+$ structure below it that may correspond to the pentaquark state $\Sigma^*$~\cite{Wu:2009tu,Wu:2009nw,Helminen:2000jb,Zhang:2004xt,Gao:2010hy}, introduced in Section~\ref{sec:potentialobservables}.


\section{Summary}

A preliminary measurement of the \Lz polarization has been obtained with a partial sample of \threepi decays collected by the LHCb experiment in 2016. This polarization corresponds to the effective \P-violating parameter $\alpha_{\rm eff}$ of the \threepi decay. To arrive at this result, in Eq.~\eqref{eq:finalpol}, all the analysis chain has been developed anew: the implementation of an optimized stripping line; the analysis of trigger efficiencies to define a reduced set of L0, \hltone and \hlttwo lines; a soft preselection; several treatments of real and simulated data (\textit{corrections}); the development of a BDT; mass and angular fits; and the estimation of some of the dominant systematic uncertainties.

Besides extending the analysis framework to the full Run II dataset, additional studies must be carried out first. These include evaluating the systematic uncertainties from early steps of the analysis chain, \textit{e.g.} the preselection, \texttt{L0Hadron} efficiencies or PID corrections, and assessing the fit stability and compatibility between different fit strategies. The analysis of the \lLL sample will provide valuable input to study the systematic errors originating directly in the event reconstruction. 

A better control of the systematic errors and the possible additional steps in the analysis chain to refine this measurement will also offer the possibility to probe direct \CP violation via the comparison of the effective asymmetry parameter for \Lc and $\mathit{\bar{\Lambda}_c^-}$ decays. Beyond the \Lz polarization measurement, the angular fit can be extended to include information of the \Lc angles and thereby extract the \Lc polarization. This observable, naturally, is sensitive to the \Lc production mechanism and it would be most interesting to separate the sample in prompt and secondary \Lc particles, \textit{i.e.} being produced directly in the $pp$ collision or as the weak decay product of some heavier state such as $\Lb$. In practice, the best results would be obtained from a simultaneous fit to the decay angles and some variable that can discriminate between prompt and secondary \Lc. Common choices would be the \Lc $\log(\chi^2_{\rm IP})$ or the pseudo-propertime (see definition \eg in Ref.~\cite{LHCb:2017ygo}).

Analogously, the \Lz polarization is also sensitive to the specific production mechanism even within \threepi decays. Depending on the phase space region of the Dalitz plot, the \Lz may be generated in $\Sigma(1385)^\pm\to\Lambda\pip$ strong decays, in which case the rest frame of this resonance should also be accounted for. \textit{Missing} an intermediate rest frame can lead to dilutions of the polarization of a factor 1/3~\cite{Becattini:2016gvu}.

To summarize, the very initial objective of evaluating the \Lz polarization in this channel has been accomplished. This will be an important input to compare the sensitivity of different \Lz production modes in the measurement of its electric and magnetic dipole moment at LHCb.
Several important studies still need to be completed and the analysis methods must be scrutinized by the LHCb collaboration to convert this first layout of the analysis chain into an official LHCb measurement.

\part{Phenomenology of electric dipole moments }

\chapter{Theory introduction } \label{ch:introtheo}

In the last part of this thesis, two phenomenological studies on EDM observables are presented. The first one, in Chapter~\ref{ch:improvedbounds}, arose from a bibliographic search on the physics reach of charm baryon EDM measurements, presented in Part I. All indirect bounds on heavy quark EDMs (to our knowledge) are collected at the beginning of that chapter.\footnote{Previous versions of those summary tables were in fact already presented in the context of the experiment proposal~\cite{internalnote}.} From the study of those limits, we will motivate a simple yet very powerful method to extract new bounds on the quark EDM operators. The implications were studied for different BSM scenarios including the Manohar-Wise model (MW), with new colour octet scalars. Although the new limits were effective in constraining the parameter space of this model, it became clear that only a complete analysis of EDM phenomenology, also considering light quark EDMs, could place the most stringent constraints on the \CP violating parameters of the model. All the relevant contributions of the MW model to hadronic EDMs are obtained in Chapter~\ref{ch:edmsmw}, which are used to explore the constraints on the model parameters.

As preparation, in the remainder of this Chapter~\ref{ch:introtheo}, the theoretical framework for these studies is briefly introduced. First, the main elements of the SM are presented in Section~\ref{sec:SM}. The extension of the SM that will be treated in more detail, the Manohar-Wise model, is briefly described in Section~\ref{sec:MWmodel} along with its motivations.
To study low-energy observables we will make use of effective field theories, which are introduced in Section~\ref{sec:EFT}. In that section, also the relevant effective operators for our analyses of EDMs are provided.

Further concepts and techniques used later are not introduced in this chapter. Some technical details are provided in Appendices~\ref{app:RGEsolution} and \ref{app:loops}, and introductory reviews of some of these topics can be found in Ref.~\cite{Pich:1998xt} (matching EFTs), Ref.~\cite{Buras:1998raa} (renormalization group equations) or Ref.~\cite{Ilisie:2016jta} (tensor integrals and loop calculations). To develop the calculations and obtain the numerical results we have used the software \textsc{Mathematica} and the open-source packages \textsc{FeynCalc}~\cite{Shtabovenko:2020gxv}, \textsc{Package-X}~\cite{Patel:2015tea}, \textsc{FeynArts}~\cite{Hahn:2000kx} and \textsc{FeynRules}~\cite{Alloul:2013bka}.

\section{Standard Model} \label{sec:SM}

The SM is based on the symmetry principle of gauge invariance. In this section we will see an overview of the construction of the SM Lagrangian based on the gauge symmetry group $SU(3)_C\otimes SU(2)_L\otimes U(1)_Y$. 
To this end, we will start with the explicit derivation of quantum electrodynamics (QED), which is based on the group $U(1)_Q$, where $Q$ stands for electric charge. This was in fact the first gauge theory to be formulated~\cite{Tomonaga:1946zz} 
and it satisfactorily describes all electromagnetic phenomena also at high energies. By analogy, we will outline the derivation of quantum chromodynamics (QCD) and the electroweak theory (EW), based on the symmetry groups $SU(3)_C$ and $SU(2)_L\otimes U(1)_Y$, respectively. 
Towards the end of this section, we will briefly present the Higgs mechanism which introduces fermion and gauge boson masses and leads to the emergence of the CKM matrix.

\subsubsection{Quantum electrodynamics}

The starting point will be the Dirac Lagrangian, describing a free fermion with spinor $\psi$,
\begin{equation}
\lag_0=i\overline{\psi}(x)(\gamma^\mu\partial_\mu-m)\psi(x).
\label{eq:lagDirac}
\end{equation}
Applying a global $U(1)_Q$ transformation is equivalent to rotating the value of each spinor-field component in the complex plane by an angle $Q\theta$ (\textit{rephasing}),
\begin{equation}
\psi(x)\rightarrow\psi'(x)=\text{exp}(iQ\theta)\psi(x),
\end{equation}
where $Q$ stands for the (conserved\footnote{The conservation of charge is proven through the Noether theorem. Examples can be found in any QFT textbook, \textit{e.g.} in Ref.~\cite{Peskin:1995ev}.}) electric charge, and can be understood as the single generator of the $U(1)$ group. The Lagrangian in Eq.~\eqref{eq:lagDirac} is clearly invariant under this transformation. However, it is more natural to expect the physical laws to be independent of this phase even if its value is not the same at all points of space-time, \textit{i.e.} $\theta=\theta(x)$. In this case, the mass term is still invariant under the now \textit{local} gauge transformations, but the derivative transforms as
\begin{equation}
\partial_\mu\psi(x)\rightarrow\partial_\mu\psi'(x)=\text{exp}(iQ\theta(x))(\partial_\mu+iQ\partial_\mu\theta(x))\psi(x).
\label{eq:derivTransf}
\end{equation}
To cancel out the extra term, we add the following term to the original Lagrangian
\begin{equation}\label{eq:QEDinteraction}
\lag\supset-eQA_\mu(x)\overline{\psi}(x)\gamma^\mu\psi(x),
\end{equation}
and impose the transformation rule 
\begin{equation}
A_\mu(x)\rightarrow A'_\mu(x)=A_\mu(x)-\frac{1}{e}\partial_\mu\theta.
\end{equation}

This can also be seen as a substitution of the derivative by a covariant derivative $ D_\mu\psi(x)=[\partial_\mu+ieQA_\mu(x)]\psi(x)$ which transforms under local gauge transformations like the usual derivative does under global ones, 
\begin{equation}
D_\mu\psi(x)\rightarrow\psi'(x)=\text{exp}(iQ\theta(x)) D_\mu\psi(x).
\end{equation}
Then we can write the Lagrangian as
\begin{equation}
\lag=i\overline{\psi}(x)(\gamma^\mu D_\mu-m)\psi(x).
\end{equation}
Here we have added the gauge field $A_\mu(x)$ representing the photon, which is necessarily massless as the $U(1)_Q$-invariant Lagrangian does not admit a mass term. To allow this field to propagate while respecting the symmetries we can add the kinetic term
\begin{equation}
\lag_\text{kin}=-\frac{1}{4}F_{\mu\nu}(x)F^{\mu\nu}(x),
\end{equation}
where we have introduced the field strength tensor
\begin{equation} \label{eq:strengthtensorF}
F_{\mu\nu}(x)\equiv\frac{i}{e}[D_\mu,D_\nu]=\partial_\mu A_\nu-\partial_\nu A_\mu~.
\end{equation}

%
%

\subsubsection{Quantum chromodynamics}

The strong interactions are described by QCD, a gauge theory based on the group $SU(3)_C$, where $C$ stands for \textit{colour}, a new quantum number introduced to reconcile the baryon structure of quarks with the Pauli exclusion principle. Its origins lie in the classification and interpretation of the several hadron species (\textit{zoo} of particles) that were being discovered along the 20th century\footnote{See more on the historical development of QCD in \textit{e.g.} Ref.~\cite{Fritzsch:2015jfa}.}.

The group $SU(N)$ is the \textit{special} group of unitary matrices $U\in SU(N)$ with $\det U = 1$, in an $N$-dimensional space, and it is a subgroup of $U(N)$. For $N=3$, these transformations can be expressed as
\begin{equation}
U=\text{exp}\left( i T^a \theta_a\right),
\end{equation}
where $T^a~(a=1,2,...,8)$ are the generators of the $SU(3)$ group which, in its irreducible representation, are $3\times 3$ matrices. These are proportional to the so-called Gell-Mann matrices $\lambda^a = 2 T^a$ (or simply colour matrices).

Each quark spinor $q_f$ with flavor $f$, has three components in colour space, on which the transformations $U$ act. Starting with the Lagrangian for free quarks,
\begin{equation}
\lag_0=\sum_f\overline{q}_f(i\gamma^\mu\partial_\mu -m_f)q_f,
\end{equation}
we can derive the covariant derivative corresponding to local $SU(3)_C$ transformations analogously to the case of QED, obtaining
\begin{equation}
D^\mu q_f=\Big[\partial^\mu + i g_s\frac{\lambda^a}{2}G^\mu_a(x)\Big]q_f\equiv[\partial^\mu + i g_sG^\mu(x)]q_f,
\end{equation}
where we had to introduce eight new gauge boson fields $G^\mu_a(x)$, the gluons, which are massless. Another key difference with respect to QED is that the generators of $SU(3)$ do not commute, exposing the non-Abelian character of QCD. The commutator of the colour matrices reads
\begin{equation}
\left[T^a,T^b\right]=i f^{abc}T^c,
\end{equation}
where $f^{abc}$ are the $SU(3)_C$ structure constants. For this reason, the gluon strength tensor $G^{\mu\nu} (x)$, defined analogously to $F^{\mu\nu}(x)$ in Eq.~\eqref{eq:strengthtensorF}, has an additional term involving two gluon fields,
\begin{equation} \label{eq:strengthtensorG}
G^{\mu\nu}(x)
=
-\frac{i}{g_s}[D^\mu,D^\nu]
=
\partial^\mu G^\nu-\partial^\nu G^\mu + i g_s [G^\mu,G^\nu].
%
\end{equation}
Factoring out the colour matrix $T^a$, one obtains for the gluon fields
\begin{equation} \label{eq:strengthtensorGa}
G^{\mu\nu}_a(x)
=
\partial^\mu G^\nu_a-\partial^\nu G^\mu_a - g_s f^{abc} G^\mu_b G^\nu_c.
\end{equation}
Finally, including also the gluon kinetic term, the complete QCD Lagrangian reads
\begin{equation} \label{eq:lagQCD}
\lag_{\rm QCD}=-\frac{1}{4}G_a^{\mu\nu}G^a_{\mu\nu}+\sum_f\overline{q}_f(i\gamma^\mu D_\mu -m_f)q_f.
\end{equation}

However, there is another term of dimension four that abides by the colour $SU(3)$ symmetry and contains only gluon fields. This is the so-called $\theta$-QCD term, already introduced in Eq.~\eqref{eq:lagrangianEDMs}, 
\begin{equation}
\lag_\theta= \theta \frac{g_s^2}{64\pi^2} \varepsilon^{\mu \nu \sigma \rho} G^a_{\mu\nu} G^a_{\sigma\rho}.
\end{equation}
This term violates the discrete symmetries $P$ and $T$ and would give rise to a large neutron EDM of the order $d_n\sim 10^{-16}\,\ecm$ if $\theta\sim 1$. However, even if QCD does respect $P$ and $T$, and $\theta=0$, the same operator appears from the quark sector via an axial $U(1)_A$ rotation of the quark fields. The experimental limits on the neutron EDM constrain a combination of $\theta$ with the phase of the determinant of the quark mass matrix $M_q$~\cite{Pospelov:2005pr},
\begin{equation}
|\bar\theta| = |\theta - \arg \det M_q| \lesssim 10^{-10}.
\end{equation}
The unnatural level of fine-tuning between these quantities to comply with the experimental results is known as the \textit{strong CP problem}. A family of solutions based on the Peccei-Quinn mechanism~\cite{Peccei:1977hh,Weinberg:1977ma,Wilczek:1977pj}  predict the existence of the axion, which has not been observed to date.

%
%

\subsubsection{Electroweak theory}

The other main pillar of the SM is the theory of electroweak (EW) interactions. An important difference compared to QCD is that EW interactions do not conserve parity based on its formulation through the gauge group $SU(2)_L\otimes U(1)_Y$. The fermion spinors can be decomposed in right- and left-handed parts and we will only require invariance under $SU(2)$ for the left-handed components. Moreover, these transformations act on the combination of the up- and down-type quarks (or charged leptons and neutrinos), which form doublets of $SU(2)_L$, revealing the family structure of the fermion content of the SM.

Considering only the first quark generation, the spinor fields are organized as
\begin{equation} \label{eq:defLRFields}
\psi_1(x)=       \left( {\begin{array}{c}
	u\\
	d \\
	\end{array} } \right)_L,\hspace{1cm}
\psi_2(x)=u_R,\hspace{1cm}
\psi_3(x)=d_R.
\end{equation}
Their transformations under $SU(2)_L\otimes U(1)_Y$ are
\begin{align}
\nonumber\psi_1(x)&\rightarrow\psi_1'=\text{exp}\{iy_1\beta\}\text{exp}\left\{i\frac{\sigma_j}{2}\alpha^j\right\}\psi_1(x),\\
\psi_2(x)&\rightarrow\psi_2'=\text{exp}\{iy_2\beta\}\psi_2(x),\\
\nonumber\psi_3(x)&\rightarrow\psi_3'=\text{exp}\{iy_3\beta\}\psi_3(x),
\end{align}
where $\sigma_j$ ($j=1,2,3$) are the commonly known Pauli matrices, generators of the $SU(2)$ group, and $y_i$ is the hypercharge, which is related to the QED electric charge as we will see. In the Lagrangian, a Dirac mass term connects the left and right components of the fields and, since they transform differently, it will not be invariant under the gauge group. Thus, although the QCD Lagrangian in Eq.~\eqref{eq:lagQCD} can have a mass term in principle, this is forbidden by the EW symmetry group, and the fermion masses will have to be introduced by other means\footnote{The invariance of the Lagrangian under independent rotations of left and right fields in known as chiral symmetry, which will be broken with the introduction of mass terms.}. The starting-point Lagrangian will simply be 
\begin{equation}
\lag_0=\sum_{j=1}^3i\overline{\psi}_j(x)\gamma^\mu\partial_\mu\psi_j(x).
\end{equation}
Requiring invariance under local $SU(2)_L\otimes U(1)_Y$ transformations, the new covariant derivatives read
\begin{align}
\nonumber D_\mu\psi_1(x)&=\big[\partial_\mu+ig\frac{\sigma_j}{2}W^j_\mu(x)+ig'y_1B_\mu(x)\big]\psi_1(x),\\
D_\mu\psi_2(x)&=\big[\partial_\mu+ig'y_2B_\mu(x)\big]\psi_2(x),\\
\nonumber  D_\mu\psi_3(x)&=\big[\partial_\mu+ig'y_3B_\mu(x)\big]\psi_3(x).
\end{align}
The new gauge fields $W_\mu^j$ and $B_\mu$ must transform as 
\begin{align}
B_\mu(x)&\rightarrow B'_\mu(x)=B_\mu(x)-\frac{1}{g'}\partial_\mu\beta(x),\\
\widetilde{W}_\mu(x)&\rightarrow \widetilde{W}'_\mu(x)=U_L(x)\widetilde{W}_\mu U_L^\dagger(x)+\frac{i}{g}\partial_\mu U_L(x)U_L^\dagger,
\end{align}
where the notation $\widetilde{W}_\mu(x)=\frac{\sigma_j}{2}W^j_\mu(x)$ and $U_L(x)=\text{exp}\{i\frac{\sigma_j}{2}\alpha^j(x)\}$ has been introduced.

Analogously to the gauge bosons of QED and QCD, the EW gauge fields also cannot have a mass term, as it would break the symmetry. Their propagation is included through the kinetic term, with the associated strength tensors $B^{\mu\nu}$ and $\widetilde{W}^{\mu\nu}$ (see \textit{e.g.} Ref.~\cite{Pich:2012sx}). 
%
%
Finally, the fundamental EW Lagrangian, invariant under local gauge symmetry $SU(2)_L\otimes U(1)_Y$, is
\begin{equation} \label{eq:lagEW}
\lag_{\rm EW}=i\overline{\psi}_j(x)\gamma^\mu D_\mu\psi^j(x)-\frac{1}{4}B_{\mu\nu}B^{\mu\nu}-\frac{1}{4}W_{\mu\nu}^iW^{\mu\nu}_i.
\end{equation}

At this point, we would like to identify the charged- and neutral-current interactions of the gauge bosons with the fermions, which are implicit in the covariant derivative. Writing explicitly $\widetilde{W}_\mu$ with the Pauli matrices, we find
\begin{equation}
\widetilde{W}_\mu = \frac{\sigma_j}{2}W_\mu^j= \frac{1}{2}\left( {\begin{array}{cc}
	W_\mu^3&\sqrt{2}W^\dagger_\mu\\
	\sqrt{2}W_\mu&-W^3_\mu \\
	\end{array} } \right),
\end{equation}
where we have defined $W_\mu\equiv(W^1_\mu+iW^2_\mu)/\sqrt{2}$. The off-diagonal elements connect the up and down components of the field $\psi_1$ defined in Eq.~\eqref{eq:defLRFields} and give rise to charged current interactions (CC), which we will write in Eq.~\eqref{eq:CC} directly with the CKM matrix. Conversely, $W_\mu^3$ couples similarly to $B_\mu$, without mixing up and down fields. 
The photon field $A_\mu$, with the same coupling of QED, can be identified as a linear combination of these fields through the weak angle $\theta_W$,
\begin{equation}
\left( {\begin{array}{c}
	W^3_\mu\\
	B_\mu \\
	\end{array} } \right)=
\left( {\begin{array}{cc}
	\text{cos}\theta_W& \text{sin}\theta_W\\
	- \text{sin}\theta_W& \text{cos}\theta_W \\
	\end{array} } \right)
\left( {\begin{array}{c}
	Z_\mu\\
	A_\mu \\
	\end{array} } \right), 
\end{equation}
giving rise to the neutral current (NC) interaction terms
\begin{align}\label{eq:NC}
\nonumber{\cal L}_{NC}=-\sum^3_{j=1}\overline{\psi}_j\gamma^\mu\Big\{A_\mu\Big[\underbrace{g\frac{\sigma_3}{2}\text{sin}\theta_W+g'y_j\text{cos}\theta_W}_{e Q}\Big]+
Z_\mu\Big[g\frac{\sigma_3}{2}\text{cos}\theta_W-g'y_j\text{sin}\theta_W\Big]\Big\}\psi_j,
\end{align}
where we identified the QED coupling constant by comparison with the photon QED interaction in Eq.~\eqref{eq:QEDinteraction}. However, in $SU(2)_L$ space, $Q$ must be represented as a diagonal matrix with the quark charges. This also fixes the interpretation of the hypercharge matrix $Y$ as the combination of the electric charge with the third generator of $SU(2)_L$,
\begin{equation}
Y=Q-T_3, ~\text{ and }~ g\sin\theta_W=g'\cos\theta_W=e.
\label{eqn:cond2}
\end{equation}

\subsubsection{Higgs mechanism}


Gluons, photons, and weak gauge bosons emerge naturally in the SM based only on the local gauge symmetry principle. However, the masses of fermions and weak gauge bosons had no place in the SM Lagrangian. Thus, the fundamental symmetry must break in some way. In the SM, this is achieved through the spontaneous symmetry breaking (SSB) in form of the Higgs mechanism~\cite{Higgs:1964pj,Englert:1964et,Guralnik:1964eu,Weinberg:1967tq} which predicts a new scalar boson. The search for this (or a "similar"\footnote{Even if the Higgs mechanism did not describe nature, another mechanism to generate fermion masses would necessarily predict new particles below the TeV scale.
	%
	%
	Before the construction of the LHC, this was known as \textit{no-lose theorem}. 
}) particle was the main motivation to construct a high-energy hadron collider. A new boson was discovered in 2012 at the LHC~\cite{ATLAS:2012yve,CMS:2012qbp} and, to date, all analyses characterizing it yield it consistent with the SM Higgs boson. 
In the following we will only describe the main concepts, while a more detailed explanation can be found \textit{e.g.} in Ref.~\cite{Pich:2012sx}

First, let us consider two additional complex scalar fields that transform as a doublet under $SU(2)_L$,
\begin{equation} \label{eq:higgsfield}
\phi(x) = \begin{pmatrix} \phi^{(+)} (x) \\ \phi^{(0)} (x) \end{pmatrix} .
\end{equation}
Kinetic and potential terms associated to these fields can be added to the Lagrangian while conserving the fundamental gauge symmetries as
\begin{equation}  \label{eq:lagScalar}
\mathcal{L}_S = \left( D_\mu \phi \right)^\dagger D^\mu \phi -V(\phi) \, , \quad V(\phi)= \mu^2 \phi^\dagger \phi + h \left( \phi^\dagger \phi\right)^2\, , 
\end{equation}
where $h>0$ is necessary to find solutions of minimum energy and $\mu^2<0$ leads to a non-zero value for the ground state of $\phi^{(0)}(x)$, with $|\langle 0|\phi^{(0)}|0\rangle|=v/\sqrt{2}$. The constant $v$ is the so-called vacuum expectation value (VEV) of the Higgs field, which can be determined through the Fermi constant as
\begin{equation}
v = \left( \sqrt{2} G_F \right)^{-1/2}=246 \gev\,.
\end{equation}

The scalar doublet may be written as
\begin{equation}
\phi(x) = \text{exp}\left\{i\frac{\sigma_j}{2}\alpha^j\right\} \frac{1}{\sqrt{2}} \begin{pmatrix} 0 \\ v + H(x) \end{pmatrix}\, ,
\end{equation}
where $H(x)$ parametrizes the excitation above the ground state and the exponential represents any possible transformation of $SU(2)_L$. As we can see, if $H(x)=0$ there are infinitely many ground states with the same energy. However, if we choose one of them by setting \textit{e.g.} $\alpha^j=0$ (unitary gauge), the symmetry is spontaneously broken and the scalar Lagrangian in Eq.~\eqref{eq:lagScalar} is no longer invariant under $SU(2)_L$\footnote{The missing Lagrangian degrees of freedom after the gauge fixing appear as the longitudinal degrees of freedom of the gauge bosons $Z$ and $W^\mp$ when they acquire mass. See \textit{e.g.} Ref.~\cite{Pich:2012sx}.}.

The masses of the $Z$ and $W^\pm$ bosons appear from the kinetic term of the scalar doublet, which contains the covariant derivative of $SU(2)_L\otimes U(1)_Y$ and hence the coupling of the scalar doublet to the weak gauge bosons. After the spontaneous symmetry breaking, the kinetic term is written in terms of field $H(x)$ as
\begin{equation}
\left( D_\mu \phi \right)^\dagger D^\mu \phi = \left(\partial_\mu H \right)^\dagger \partial^\mu H \\
+ (1+2/vH+H^2/v^2) \left( \frac{v^2g^2}{4} W_\mu^\dagger W^\mu + \frac{v^2 g^2}{8\cos ^2 \theta_W} Z_\mu Z^\mu \right) \, , \nonumber
\end{equation}
where we can identify the weak boson masses $M_W = v g/2$ and $M_Z = M_W/(\cos^2 \theta_W)$.

In turn, the potential produces the mass term of the Higgs boson, with $M_H = \sqrt{-2\mu^2}$, and also the cubic and quartic self-interaction terms,
\begin{equation}
-V(\phi) = \frac{M_H^2 v^2}{8} -\frac{ M_H^2}{2}H^2 -\frac{ M_H^2}{2 v}H^3-\frac{ M_H^2}{8 v^2}H^4\, .
\end{equation}

As anticipated, the fermion masses are also introduced thanks to the Higgs mechanism. Consider a new interaction term between a left-handed fermion doublet, the scalar doublet and one of the corresponding right-handed singlets,  
$(\bar{u}~ \bar{d})_L \left(\begin{smallmatrix}\phi^{(+)}\\ \phi^{(0)}\end{smallmatrix} \right) u_R$. This is the so-called Yukawa interaction and it is easy to prove that it respects the gauge symmetry: $SU(2)_L$ transformations \textit{cancel out} between $(\bar{u}~ \bar{d})_L$ and $\left(\begin{smallmatrix}\phi^{(+)}\\ \phi^{(0)}\end{smallmatrix} \right)$, and $U(1)_Y$ phases are compensated by the hypercharge values of the three elements ($-1/6$,~$+1/2$ and $-1/3$~\cite{Pich:2012sx}).


After the SSB, the Yukawa interaction for all quark and lepton flavours becomes
\begin{equation}
\mathcal{L}_Y = -  \left(1+\frac{H}{v}\right) \sum_f \left( m_{u_f} \bar{u}_fu_f + m_{d_f} \bar{d}_f{d_f}  + m_{l_f} \bar{l}_f l_f \right), \label{eq:lagYukawaSM}
\end{equation}
granting masses to the fermions. An immediate consequence is that the Yukawa couplings of the fermions to the physical Higgs boson are proportional to the fermion masses. To arrive at Eq.~\eqref{eq:lagYukawaSM}, with the sum over quark and lepton flavours $f$, we had to diagonalize the initial mass matrices in flavour space, inducing a rotation of the flavour basis into the physical fields, \textit{e.g.} for up-type quarks $(u_1,u_2,u_3)\longrightarrow(u'_1,u'_2,u'_3) = (u,c,t)$. However, the CC interactions are not invariant under this change of basis since they mix up and down flavours. For this reason, an additional matrix $\boldsymbol{V}_{ij}$ is introduced in CC interactions, \textit{connecting} up and down flavours from different quark generations. This is the commonly known CKM matrix~\cite{Cabibbo:1963yz,Kobayashi:1973fv}, which leads to the extremely rich phenomenology of flavour physics. The CC interaction term after diagonalization reads
\begin{equation} \label{eq:CC}
\mathcal{L}_\text{CC} = - \frac{g}{2\sqrt{2}} \left( W_\mu^\dagger \left[ \sum_{i,j} \bar{u}_i \gamma^\mu \left( 1 - \gamma_5 \right) \boldsymbol{V}_{ij} d_j + \sum_l \bar{\nu}_l \gamma^\mu \left( 1- \gamma_5 \right) l \right] + h.c. \right)\, .
\end{equation}
The CKM matrix contains, in principle, several imaginary parameters. However, through the appropriate rephasing of the quark fields in the Lagrangian, it is possible to eliminate all complex phases but one. This phase gives rise to \CP-violation in EW interactions\footnote{The PMNS neutrino mixing matrix also contains a phase of \CP violation~\cite{Akhmedov:1999uz}. In the SM the neutrinos are massless and, although measuring and studying neutrino oscillations is completely \textit{standard}, anything having to do with neutrino masses is not quite \textit{Standard Model}. }. 

To illustrate the connection of complex phases in the Lagrangian to \CP violation, we shall consider a transition amplitude $\mathcal{M}(A\to B)$, which is a complex number, with at least two possible subprocesses $\mathcal{M} = \mathcal{M}_1 + \mathcal{M}_2$ with associated phases $\theta_1$ and $\theta_2$. The corresponding \CP conjugate processes with antiparticles $\bar{\mathcal{M}}(\bar A\to \bar B)$ have opposite complex phases and, developing the expressions with the absolute values, it is easy to see that $|\mathcal{M}|-|\bar{\mathcal{M}}| \propto \sin (\theta_1 - \theta_2) \neq 0$ in general.

\subsubsection{Complete SM Lagrangian}

Having described the QCD \eqref{eq:lagQCD} and EW \eqref{eq:lagEW} theories, with the addition of the scalar doublet \eqref{eq:lagScalar} and its Yukawa interaction, given after SSB in Eq.~\eqref{eq:lagYukawaSM}, we can write the complete SM Lagrangian as 
\begin{equation}
\mathcal{L}_\text{SM} = \mathcal{L}_\text{QCD} + \mathcal{L}_\text{EW} + \mathcal{L}_S + \mathcal{L}_Y\, .
\end{equation}

\section{Manohar-Wise model} \label{sec:MWmodel}

Despite its success, we know that the SM is incomplete and other, more fundamental, theories of nature must address its issues. To date, many extensions of the SM have been proposed, some of which (\textit{e.g.} leptoquarks and supersymmetry) will be referenced in Chapter~\ref{ch:improvedbounds}. Among them, special attention will be dedicated to the MW theory, on which the Chapter~\ref{ch:edmsmw} is based.

In this model, a new scalar field
\begin{equation} \label{eq:defS}
S^a (x) = \left( \begin{array}{c}
S^{a,+} (x) \\
S^{a,0} (x)
\end{array} \right) 
\end{equation}
is introduced that transforms as a doublet under $SU(2)_L$ (\eg like the left-handed quarks) and as an octet under the colour $SU(3)$ group (like the gluon fields). In Eq.~\eqref{eq:defS}, $a=1, ..., 8$ are colour indices of the adjoint representation. 
These scalars were first proposed by Manohar and Wise \cite{Manohar:2006ga}, the original motivation being that they are one of the few scalar representations of the SM gauge group that can implement Minimal Flavour Violation (MFV) \cite{Chivukula:1987py,DAmbrosio:2002vsn}. In addition, these scalars emerge naturally with a mass of few TeVs from $SU(4)$, $SU(5)$ or $SO(10)$ unification theories at high energy scales~\cite{Georgi:1974sy,Georgi:1979df,Dorsner:2006dj,FileviezPerez:2013zmv,Perez:2016qbo,FileviezPerez:2019ssf,Bertolini:2013vta}. 
Their interactions with the SM fields also introduce new sources of \CP violation, needed to explain the matter-antimatter imbalance, as introduced in Chapter~\ref{ch:introexp}. 

The transformation properties of these scalars allow the construction of many new gauge-invariant interactions terms~\cite{Manohar:2006ga}. In general, the Lagrangian describing colour-octet scalar interactions can be written as 
\begin{align}
\mathcal{L}_{\text{MW}}\,=\,\mathcal{L}_{\text{SM}}\,+\,\mathcal{L}_{\text{kin}}\,+\,\mathcal{L}_{\text{S}}\,+\,\mathcal{L}_{Y}\,,
\end{align}
where $\mathcal{L}_{\text{SM}}$, $\mathcal{L}_{\text{kin}}$, $\mathcal{L}_{Y}$, and $\mathcal{L}_{\text{S}}$ represent the SM Lagrangian, the colour-octet scalar kinetic term, the scalar sector and the interaction with SM fermions (Yukawa couplings), respectively. 
The kinetic term
\begin{align}
\mathcal{L}_{\text{kin}}=2\,\Tr[(D_\mu S)^\dagger D^\mu S]~,
\label{eq:kin}
\end{align}
introduces interactions with the SM gauge bosons through the covariant derivative $D_\mu S=\partial_\mu S+ i\, g_s\,[G_\mu,S]\,+\,i\,g\, \widetilde{W}_\mu\, S+\frac{i}{2} \,g^\prime\,B_\mu \,S$. The scalar interaction Lagrangian $\mathcal{L}_{\text{S}}$ encodes the self-interaction of the octet scalars, and the interactions with the SM Higgs doublet, in Eq.~\eqref{eq:higgsfield}. These are given by~\cite{Manohar:2006ga} 
\begin{align} 
\mathcal{L}_{\text{S}}&=2\,m_S^2\,\Tr(S^{\dagger i}S_i)+\lambda_1\,\phi^{\dagger i}\phi_i\,\Tr(S^{\dagger j}S_j)+\lambda_2\,\phi^{\dagger i}\phi_j\,\Tr(S^{\dagger j}S_i)\nonumber\\
&+\left[\lambda_3\,\phi^{\dagger i}\phi^{\dagger j}\,\Tr(S_i S_j)+\lambda_4\,\phi^{\dagger i}\, \Tr(S^{\dagger j} S_j S_i)+\lambda_5\,\phi^{\dagger i} \,\Tr(S^{\dagger j} S_i S_j)+\text{h.c.}\right]\label{eq:potential} \\&+\lambda_6\, \Tr(S^{\dagger i}S_iS^{\dagger j}S_j)+\lambda_7\, \Tr (S^{\dagger i}S_jS^{\dagger j}S_i)+\lambda_8\, \Tr(S^{\dagger i}S_i)\Tr( S^{\dagger j}S_j)\nonumber\\&+
\lambda_9 \,\Tr(S^{\dagger i}S_j)\Tr (S^{\dagger j}S_i)+\lambda_{10}\, \Tr( S_iS_j)\,\Tr( S^{\dagger i}S^{\dagger j})+\lambda_{11} \,\Tr(S_iS_j S^{\dagger j}S^{\dagger i})~,\nonumber
\end{align}
where $i$ and $j$ are $SU(2)_L$ indices and all traces are in colour space. 
The parameters $\lambda_{3,4,5}$ contain two phases of \CP violation which can contribute to EDM observables. Nevertheless, the phenomenological studies in this thesis will be limited to EDM observables arising from the \CP violation in the Yukawa couplings, which contribution is enhanced. 
%

Decomposing the complex field of neutral scalars into two real scalars,
\begin{align}
S^{a,0}\,=\,\frac{1}{\sqrt{2}}\,\left(S^{a,0}_R\,+\,i\,S^{a,0}_I\right)~,
\end{align}
one can substitute them in the expression of $\lag_{\rm S}$ together with the Higgs doublet after SSB (with the unitary gauge) and identify the mass terms of the physical fields, yielding
\begin{align} \label{eq:massesMW}
m_{S^\pm}^2=m_S^2+\lambda_1\frac{v^2}{4}~,\hspace{1cm} m_{S^{0}_{R,I}}^2=m_S^2+(\lambda_1+\lambda_2\pm2\,\lambda_3)\,\frac{v^2}{4}~,
\end{align}
where $m_{S^\pm}$ is the mass of the charged scalar, $m_{S^{0}_{R}}$ the mass of the neutral CP-even scalar, $m_{S^{0}_{I}}$ the mass of the CP-odd scalar, and $m_S$ represents the (non-physical) mass of the unbroken scalar doublet, introduced in the first term of Eq.~\eqref{eq:potential}. Note that the mass splitting between the physical fields is a consequence of the non-zero VEV.

Assuming MFV, the Yukawa interaction of the new scalars can be parametrised by two complex numbers, $\eta_U$ and $\eta_D$, in
\begin{equation}\label{eq:yukawa_lagrangian}
\mathcal{L}_{Y}=-\sum^3_{i,j=1}\Big[\eta_D \,Y^d_{ij}\,\overline{Q}_{L_i}\,S\, d_{R_j}+\eta_U\, Y^u_{ij}\, \overline{Q}_{L_i}\, \widetilde{S}\,u_{R_j}+\text{h.c.}\Big]~,
\end{equation}
inducing new CP-violating sources beyond the SM that contribute to hadronic EDMs. In Eq.~\eqref{eq:yukawa_lagrangian}, $Q_{L}$ represents the left-handed quark doublet, and $u_{R}$ and $d_R$ correspond to the right-handed up- and down-quark singlets, respectively. The new scalar fields are written as $S=S^a\,T^a$ with $S^a = (S^{a,+},\, S^{a,0})^T$. In Eq.~\eqref{eq:yukawa_lagrangian}, the shorthand notation $\widetilde{S}=i\, \sigma_2\,S$ is employed, where $\sigma_2$ is the usual Pauli matrix.

\section{Effective theory framework} \label{sec:EFT}

In this section the concept of EFTs is briefly introduced together with the effective Lagrangian that will be used in Chapter~\ref{ch:improvedbounds} and~\ref{ch:edmsmw}.

\subsubsection{Effective Field Theories}


An Effective Field Theory (EFT) is a quantum field theory that can account for all particle dynamics within a certain energy range. If there is an energy gap between the energy scale of these interactions and the mass of some heavier degrees of freedom, the physical amplitudes can be described in terms of the \textit{active} degrees of freedom alone. The strength of their couplings is regulated by the Wilson coefficients in the effective Lagrangian. The values of the Wilson coefficients can be determined using experimental data. Thus, EFTs provide a model-independent theory framework to analyse experimental results without being limited to any extension of the SM. 
In this use of EFTs, all Wilson coefficients are independent. Instead, by \textit{matching} a given fundamental theory to the EFT, the Wilson coefficients can be expressed in terms of the fundamental parameters, and many relations among them may arise at the high-energy scale. Even if we are certain of the fundamental theory we want to study, working with effective theories valid at intermediate energy ranges can immensely simplify the calculations of low-energy observables. This is commonly referred to as \textit{top-down} approach, and we will see an example in Chapter~\ref{ch:edmsmw}.

At any given scale, there are infinitely many effective operators that can be constructed with the active degrees of freedom. Among these, only dimension-four operators have adimensional Wilson coefficients. Beyond that, the Wilson coefficients are suppressed by inverse powers of the energy scale $\Lambda$ which can be connected to the mass of the lightest new particle in the fundamental theory. Thus, it is possible to truncate the chain of operators based on their dimension, the most common choice being to consider operators up to dimension six. Examples of effective field theories are the SM effective field theory (SMEFT) and the low energy effective field theory (LEFT). In the SMEFT, all the particle content of the SM is considered with its range of validity below the electroweak scale, at $v \sim 246\,\gev$. The effective operators must comply with the symmetries of the SM gauge group $SU(3)_C\otimes SU(2)_L\otimes U(1)_Y$. In the LEFT, the energy gap between the bottom quark $m_b\approx 4.18\,\gev$ and the weak bosons $m_{W^\pm} \approx 80.4\,\gev$ is used to define the EFT without the heavy particles of the SM, \ie the top quark, Higgs boson, and weak gauge bosons. At these scales, the electroweak symmetry is already broken and operators can be constructed with just the $SU(3)_C\otimes U(1)_Q$ symmetries, which amounts to the conservation of electric and colour charge. Some of the operators that comply with this reduced symmetry but not with that of the SMEFT are the dipole operators, coupling left- and right-handed fermions with a gauge boson.

\subsubsection{Effective operators for EDM analyses}

To contribute to EDM observables, the effective operators must violate \CP while preserving quark (and lepton) flavour. In Chapters~\ref{ch:improvedbounds} and \ref{ch:edmsmw} we will use the following operators which compose our flavour-diagonal CP-violating effective Lagrangian,\footnote{We adopt the same conventions as in \eg Ref.~\cite{Hisano:2012cc}.}
\begin{align}\label{eq:lagrangian}
\begin{split}
\mathcal{L}_{\text{CPV}}\,=&\,\sum_{q}\,C_1^q(\mu)\,\mathcal{O}^q_1(\mu)\,+\,\sum_{q}\,C_2^q(\mu)\,\mathcal{O}^q_2(\mu)\,+\,C_3(\mu)\,\mathcal{O}_3(\mu)~,
\end{split}
\end{align}
where the sum over $q$ runs for all quark flavours except for the top. The effective operators are defined as
\begin{align}\label{eq:opEFF}
\begin{split}
\mathcal{O}^q_1\,&=\,-\,\frac{i}{2}\,e\,\mathcal{Q}_q\,m_q\,(\bar{q}\,\sigma^{\mu\nu}\gamma_5\,q)\,F_{\mu\nu}~,\\
\mathcal{O}^q_2\,&=\,-\,\frac{i}{2}\,g_s\,m_q\,(\bar{q}\,\sigma^{\mu\nu}\,\gamma_5\,T^a\,q)\,G^a_{\mu\nu}~,\\
\mathcal{O}_3\,&=\,-\,\frac{1}{6}\,g_s\,f^{a b c}\,\epsilon^{\mu\nu\lambda\sigma}\,G^{a}_{\mu\rho}\,G^{b \rho}_{\nu}\,G^{c}_{\lambda\sigma}~.
\end{split}
\end{align}
Here, $F_{\mu\nu}$ and $G_{\mu\nu}^a$ with $a=1,...,8$ are the electromagnetic and gluon field strength tensors, $g_s$ is the strong coupling constant ($\alpha_s\equiv g_s^2/4\pi$), and $\sigma_{\mu\nu}\,=\,\frac{i}{2}[\gamma_\mu,\,\gamma_\nu]$. The matrix $T^a$ represents the generators of the $SU(3)_C$ group with normalisation $\Tr(T^a\,T^b)=\delta^{ab}/2$, and the tensor $f^{abc}$ the structure constant. The charge of up- and down-type quarks is $\mathcal{Q}_q=(2/3,-1/3)$. The expression for the covariant derivative will be relevant to define the $3\times 3$ anomalous dimension matrix later in Eq.~\eqref{eq:anomalousmatrix}. It is defined as $D_\mu\,=\,\partial_\mu\,-\,i\,e\,\mathcal{Q}_q\,A_\mu\,-\,i\,g_s\,G_\mu^a\,T^a$, where $A_\mu$ and $G_\mu^a$ are photon and gluon fields, respectively. Additionally, also dimension-six four-quark operators can contribute to EDMs. We will argue why they can be excluded from our analyses in the next chapters. 

The quark EDM $d_q(\mu)$, chromo-EDM $\widetilde{d}_q(\mu)$, and the usually defined coefficient $w(\mu)$ of the Weinberg operator are related to the Wilson coefficients by
\begin{align} \label{eq:basesEDMs}
d_q(\mu)\:&=\:e\:\mathcal{Q}_q\:m_q(\mu)\:C_1^q(\mu) ~,\nonumber\\
\widetilde{d}_q(\mu)\:&=\:m_q(\mu)\:C_2^q(\mu)~,\\
w(\mu)\:&=\:-\:C_3(\mu)~.\nonumber
\end{align}

\chapter{Improved bounds on heavy quark EDMs} \label{ch:improvedbounds}

\begin{flushright}
	This chapter is based on Ref.~\cite{Gisbert:2019ftm}. \\Tables \ref{tab:charmEDMbounds} and \ref{tab:bottomEDMbounds} were previously presented in Ref.~\cite{internalnote}.\\
\end{flushright}

As we have seen in Part I, direct EDM searches on heavy-flavoured hadrons represent an experimental challenge due to their very small lifetime. On the other hand, heavy-quark EDMs may be largely enhanced in NP models, especially in scenarios with non-trivial flavour structure suggested by the persisting B-anomalies~\cite{Buttazzo:2017ixm,Dekens:2018bci}.
To date, only indirect limits have been set in these quantities.
In this section, a new approach for setting indirect bounds on quark EDM couplings is presented. By exploiting the mixing of operators under the renormalization group and using current constraints on the chromo-EDM of charm and bottom quarks~\cite{Sala:2013osa,Chang:1990jv}, we extract new bounds on their corresponding EDMs that improve the current limits by several orders of magnitude.

\section{Previous bounds}

Attempts to constraint heavy quark EDMs and chromo-EDMs have followed different strategies. All limits in the literature, to our knowledge, are compiled in Tables~\ref{tab:charmEDMbounds} and~\ref{tab:bottomEDMbounds} for charm and bottom quarks, respectively.

Probing the chromo-EDMs is relatively straightforward with the neutron EDM. When a heavy quark is integrated out, its chromo-EDM gives a finite contribution to the Weinberg operator\footnote{We will see this in more detail in the next chapter, in  Eq.~\eqref{eq:thresholdweinberg}.}~\cite{Braaten:1990gq,Chang:1990jv,Boyd:1990bx}, which is strongly constrained from the limits on the neutron EDM. The resulting bounds on the quark chromo-EDMs are~\cite{Sala:2013osa,Chang:1990jv},
\begin{align}
|\dtildec(m_c)| &< \: 1.0 \times 10^{-22}\:\cm~, \nonumber\\
|\dtildeb(m_b)| &< \: 1.1 \times 10^{-21}\:\cm~. \label{eq:bounddtildeq}
\end{align}

\begin{table}[ht]
	\centering
	\caption{Indirect bounds on the charm quark EDM and chromo-EDM extracted from different experimental measurements. Ordered by year of publication.}
	\label{tab:charmEDMbounds}
	\resizebox{1.0\textwidth}{!} {%
		\renewcommand{\arraystretch}{1.1}
		\begin{tabular}{  l  l  p{2.7cm} p{8cm} }
			\hline \hline
			Bound & Ref. & Primary obs. & Method \\  [1ex]
			\hline \hline
			\multicolumn{4}{c}{Charm EDM} \\ 
			&  &  &  \\ [-2ex]\hline
			$|d_c| < 8.9\times 10^{-17} ~e$cm & \cite{Escribano:1993xr} & $\Gamma( Z \to c\overline{c})$ & Measurement at the Z peak (LEP). Weights electic ($d_c$) and weak ($d^{w}_c$) dipole moments through model-dependent relations. \\  \hline
			$|d_c| < 5\times 10^{-17} ~e$cm & \cite{Blinov:2008mu} & $e^+ e^- \to c\overline{c}$ & The total cross section (from the LEP combination~\cite{ALEPH:2006bhb}) is enhanced by the charm EDM vertex $c\overline{c}\gamma$. \\ \hline	
			$|d_c| < 3\times 10^{-16} ~e$cm & \cite{Grozin:2009jq} & electron EDM & Considers contribution of $d_c$ into $d_e$ through light-by-light scattering (three-loop) diagrams.  \\ \hline
			$|d_c| < 1\times 10^{-15} ~e$cm & \cite{Grozin:2009jq} & neutron EDM & Similar approach than Ref. \cite{Sala:2013osa} with different treatment of diverging integrals and more conservative assumptions. \\ \hline
			$|d_c| < 4.4\times 10^{-17} ~e$cm & \cite{Sala:2013osa} & neutron EDM & Considers contribution of $d_c$ into $d_d$ via $W^\pm$ loops. Expressions from Ref.~\cite{CorderoCid:2007uc}.\\ \hline
			$|d_c| < 3.4\times 10^{-16} ~e$cm & \cite{Sala:2013osa} & $\text{BR}(B \to X_s \gamma)$ & Considers contributions of $d_c$ into the Wilson coefficient $C_7$. \\ \hline
			$|d_c| < 1.5\times 10^{-21} ~e$cm & \cite{Gisbert:2019ftm} & neutron EDM \newline (ours)& Renormalization group mixing of $d_c$ into $\dtildec$ (see its bound below). \\ \hline
			$|d_c| < 6\times 10^{-22} ~e$cm & \cite{Ema:2022pmo} & neutron EDM & Contribution of $\dc$ to photon-gluon operators, to neutron EDM. \\ \hline
			$|d_c| < 1.3\times 10^{-20} ~e$cm & \cite{Ema:2022pmo} & electron EDM   & Contribution of $\dc$ to photon-gluon operators, to electron-nucleon operators, to paramagnetic molecule ThO (used for $d_e$). \\ [1ex] \hline \hline
			\multicolumn{4}{c}{Charm chromo-EDM} \\ 
			&  &  &  \\ [-2ex]\hline
			$|\dtildec| < 3\times 10^{-14} ~e$cm & \cite{Kuang:2012wp} & $\psi ' \to J / \psi \pi^+ \pi^-$ & The $\tilde{d_c}$ contributes to the static potential betwen $c$ and $\bar c$ in $\psi ' $ and $J / \psi$. It also affects the dynamical transition amplitudes. \\ \hline
			$|\dtildec| < 1.0\times 10^{-22} ~e$cm & \cite{Sala:2013osa} & neutron EDM & Considers threshold contributions of $d_c$ into the Weinberg operator $w$ and the light quark EDMs $d_{u,d}$. \\ \hline \hline		
		\end{tabular}   
	}
\end{table}

\begin{table}[tt]
	\centering
	\caption{Same as Table~\ref{tab:charmEDMbounds} for bottom quark dipole couplings.}
	\label{tab:bottomEDMbounds}
	\resizebox{1.0\textwidth}{!} {%
		\renewcommand{\arraystretch}{1.1}
		\begin{tabular}{  l  l  p{2.7cm}  p{8cm} }
			\hline \hline
			Bound & Ref. & Primary obs. & Method \\  [1ex]
			\hline \hline
			\multicolumn{4}{c}{Bottom EDM} \\ 
			&  &  &  \\ [-2ex]\hline
			$|d_b| < 8.9\times 10^{-17} ~e$cm & \cite{Escribano:1993xr} & $\Gamma( Z \to b\overline{b})$ & Measurement at the Z peak (LEP). Weights electic ($d_b$) and weak ($d^{w}_b$) dipole moments through model-dependent relations. \\  \hline
			$|d_b| < 1.22\times 10^{-13} ~e$cm & \cite{CorderoCid:2007uc} & neutron EDM & Similar to Ref.~\cite{Grozin:2009jq}, but neglects longitudinal component in the $W$ propagator, thus missing emerging divergences. \\ \hline
			$|d_b| < 2\times 10^{-17} ~e$cm & \cite{Blinov:2008mu} & $e^+ e^- \to b\overline{b}$ &  The total cross section (from the LEP combination~\cite{ALEPH:2006bhb}) is enhanced by the bottom EDM vertex $b\overline{b}\gamma$. \\ \hline
			$|d_b| < 7\times 10^{-15} ~e$cm & \cite{Grozin:2009jq} & electron EDM & Considers contribution of $d_b$ into $d_e$ through light-by-light scattering (three-loop) diagrams. \\ \hline
			$|d_b| < 2\times 10^{-12} ~e$cm & \cite{Grozin:2009jq} & neutron EDM & Considers contribution of $d_b$ into $d_u$ via $W^\pm$ loops.  \\ \hline
			$|d_b| < 1.2\times 10^{-20} ~e$cm & \cite{Gisbert:2019ftm} & neutron EDM \newline (ours) &  Renormalization group mixing of $d_b$ into $\dtildeb$ (see its bound below). \\ \hline
			$|d_b| < 2\times 10^{-20} ~e$cm & \cite{Ema:2022pmo} & neutron EDM & Contribution of $\db$ to photon-gluon operators, to neutron EDM. \\ \hline
			$|d_b| < 7.6\times 10^{-19} ~e$cm & \cite{Ema:2022pmo} & electron EDM   & Contribution of $\db$ to photon-gluon operators, to electron-nucleon operators, to paramagnetic molecule ThO (used for $d_e$). \\ [1ex] \hline \hline
			\multicolumn{4}{c}{Bottom chromo-EDM} \\ 
			&  &  &  \\ [-2ex]\hline
			$|\dtildeb| \lesssim 1.1\times 10^{-21} ~$cm & \cite{Konig:2014iqa}  & neutron EDM & Numerical result based on the the contribution of the bottom CEDM into the Weinberg opperator derived in \cite{Chang:1990jv}. \\ \hline \hline
		\end{tabular}   
	}
\end{table}

In the case of heavy quark EDMs, the situation is quite different as they do not give direct contributions to purely gluonic operators and the connection to other observables often involves flavour-changing suppression factors.
Some attempts have used flavor-mixing contributions into light quark EDMs~\cite{Sala:2013osa,CorderoCid:2007uc,Grozin:2009jq},
$b\to s \gamma$ transitions~\cite{Sala:2013osa}, 
mixing into the electron EDM via light-by-light scattering diagrams~\cite{Grozin:2009jq} and
tree-level contributions to the ${e^+e^-\to q\:\bar{q}}$ total cross section~\cite{Escribano:1993xr,Blinov:2008mu}. The most restrictive (previous) bounds yield~\cite{Sala:2013osa,Blinov:2008mu}
\begin{align}
|\dc(m_c)| &< \: 4.4 \times 10^{-17}\:\ecm~, \nonumber\\
|\db(m_b)| &< \: 2.0 \times 10^{-17}\:\ecm~, \label{eq:prevbounddq}
\end{align}
which are weaker than the corresponding chromo-EDM limits by several orders of magnitude.

In this work we follow a new strategy that relates the EDM and chromo-EDM operators in order to find new limits on \dq from the already available strong bounds on \dtildeq. This relation is done in a model-independent way using the renormalization group equations, which mix the effective operators when the energy scale is changed. 
The relevant diagrams include photon loops which have been neglected in previous works due to their small size compared with pure QCD corrections. Nevertheless, they represent the first non-zero contribution to the mixing we are interested in.\\

Just a few weeks before the publication of this thesis, new limits on the EDM of charm and bottom quarks have been derived in Ref.~\cite{Ema:2022pmo}. These are similar in size to our constraints but have been obtained with an independent method. In this reference, the contributions of the quark EDM to different photon-gluon interaction operators are evaluated directly at the energy scale of the quark mass and the low-energy contribution of these operators to the neutron and ThO-molecule EDM is subsequently evaluated to extract the bounds quoted in Tables~\ref{tab:charmEDMbounds} and \ref{tab:bottomEDMbounds}.

\section{Renormalization group equations}

\begin{figure}[t]
	\centering
	\raisebox{-.5\height}{
		\includegraphics[width=0.35\columnwidth]{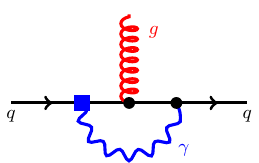}
	}
	\caption{The quark EDM coupling (blue square) induces a chromo-EDM through photon-loop diagrams, which are divergent. Treating these divergences gives rise to the renormalization group evolution of the couplings with the energy scale, encoded in the the anomalous dimension matrix. This diagram contributes to its element $(\gamma_e)_{12}^{(0)}$.}
	\label{fig:feynmandiagram}
\end{figure}

The evolution of the Wilson coefficients is given by the RGEs, 
\begin{align}
\frac{\text{d}}{\text{d}\ln\mu}\:\overrightarrow{C}(\mu)\:=\:\widehat{\gamma}^{\text{T}}\:\overrightarrow{C}(\mu)~,\label{eq:RGE}
\end{align}
where $\overrightarrow{C}\equiv(C_1^q,\:C_2^q,\:C_3)$ and $\widehat{\gamma}$ is the anomalous dimension matrix. This matrix can be expanded in powers of the QCD and QED coupling constants, $\als$ and $\ale$, respectively,
\begin{align}
\widehat{\gamma}\:=\:\frac{\als}{4\:\pi}\:\gamma_s^{(0)}\:+\:\left(\frac{\als}{4\:\pi}\right)^2\:\gamma_s^{(1)}\:+\:\frac{\ale}{4\:\pi}\:\gamma_e^{(0)}\:+\:\cdots~,
\end{align}
where $\gamma_s^{(0)}$ and $\gamma_s^{(1)}$ represent the one- and two-loop QCD corrections, while $\gamma_e^{(0)}$ encodes the one-loop QED correction~\cite{Weinberg:1989dx,Wilczek:1976ry,Braaten:1990gq,Degrassi:2005zd,Jenkins:2017dyc}. At $\mathcal{O}(\als^2)$, the quark EDM does not mix into the chromo-EDM and the first contribution only appears at $\mathcal{O}(\ale)$ from photon-loop diagrams as shown in Figure~\ref{fig:feynmandiagram}. 
The standard procedure for the computation of the anomalous dimension matrix $\hat\gamma$ can be found in \eg Refs.~\cite{Buchalla:1995vs,Buras:1998raa} and our calculation is explicitly shown in Ref.~\cite{GisbertMullor:2019vwg}. We obtain the matrix element $(\gamma_e)_{12}^{(0)}=8$, in agreement with a recent calculation in Ref.~\cite{Jenkins:2017dyc}.

The next step is solving Eq.~\eqref{eq:RGE} including this contribution and accounting also for the QCD leading order. The solution is derived explicitly in Appendix~\ref{app:RGEsolution}, in terms of evolution matrices. Beyond the first order in $\als$ there is no  simple analytical form in terms of the $\hat{\gamma}$ matrix elements. By adding this contribution, the evolution of the charm and bottom chromo-EDMs read
\begin{align}
\dtildec(m_c)\:&=\:-\:0.04\:\frac{\dc(M_{\text{NP}})}{e}\:+\:0.74\:\dtildec(M_{\text{NP}})~,\label{eq:cCEDMmixing}  \\
\dtildeb(m_b)\:&=\:0.08\:\frac{\db(M_{\text{NP}})}{e}\:+\:0.88\:\dtildeb(M_{\text{NP}})~,\label{eq:bCEDMmixing}  
\end{align}
where we have taken $M_{\text{NP}}\sim 1\:$TeV as the scale of NP.
In this result, we have neglected the mixing of the Weinberg operator into the chromo-EDM due to the very strong bounds on $\omega$ from constraints on the neutron EDM~\cite{Pospelov:2005pr,Baker:2006ts}.
The mixing of \dtildeq into itself, described by the second term on the right-hand side of Eqs.~\eqref{eq:cCEDMmixing} and \eqref{eq:bCEDMmixing}, has leading contributions from pure QCD corrections, therefore corrections of \cOale can be safely neglected.

\section{New bounds}
\begin{figure}[ht]
	\centering
	\includegraphics[width=0.45\linewidth]{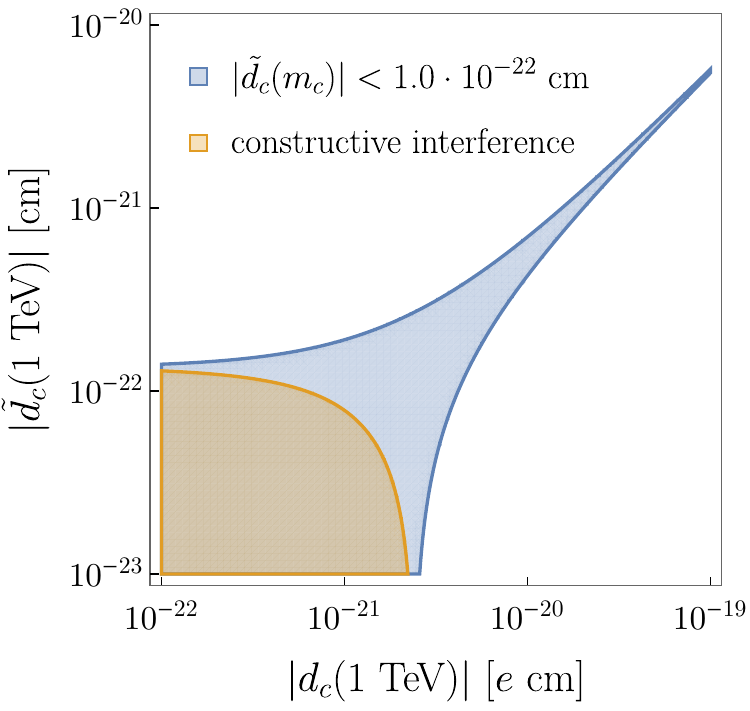}
	\includegraphics[width=0.45\linewidth]{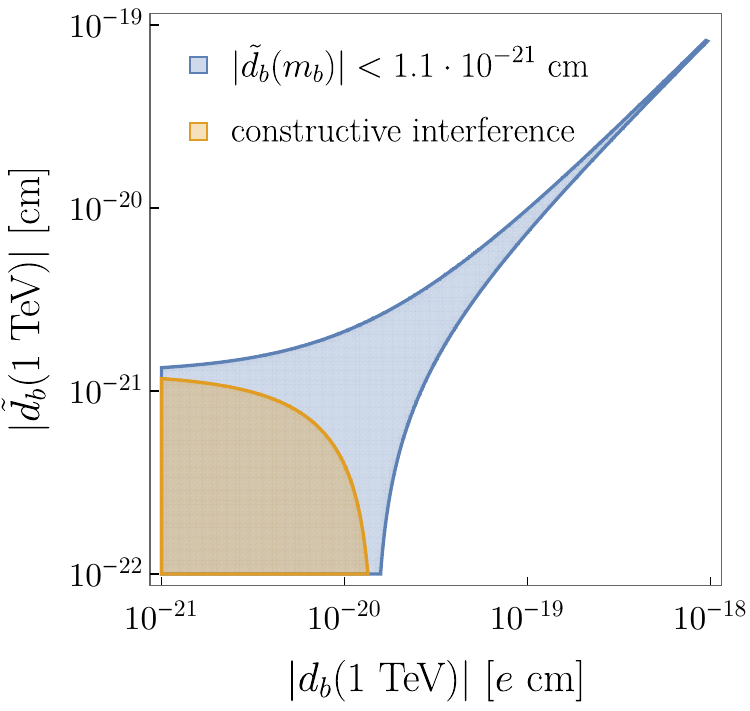}
	\caption{Bounds on the charm (bottom) chromo-EDM constrain the \dtildec-\dc (\dtildeb-\db) plane to the allowed blue area. The narrow region results from strong cancellation effects that are not present in the case of constructive interference, displayed in orange.}
	\label{fig:cEDMvsEDM}   
\end{figure}

Using the bounds on the chromo-EDMs at the low scales quoted in Eq.~\eqref{eq:bounddtildeq}, the parameter space on the \dtildeq-\dq plane is constrained as shown in Figure \ref{fig:cEDMvsEDM}. Strong fine-tuned cancellations between the two pieces of Eqs.~\eqref{eq:cCEDMmixing} and \eqref{eq:bCEDMmixing} result in an allowed region extending along a straight line which is unlikely to be realised in NP models.

Hence, we assume constructive interference between the EDM and chromo-EDM contributions at the NP scale to extract bounds on $\dq(M_{\text{NP}})$. Then, using the evolution of the EDM operator to bring these bounds down to the quark mass scale, the new bounds on the charm and bottom quark EDMs are
\begin{align}
|\dc(m_c)| &< \: 1.5\times 10^{-21}\:\ecm~,\nonumber\\
|\db(m_b)| &< \: 1.2 \times 10^{-20}\:\ecm~,
\label{eq:newlimits}
\end{align}
which improve the previous ones quoted in Eq.~\eqref{eq:prevbounddq} by three and four orders of magnitude, respectively.
This approach does not improve the current bounds on the top quark EDM~\cite{Cirigliano:2016njn,Fuyuto:2017xup} given that the limit on its chromo-EDM is of similar size~\cite{Kamenik:2011dk}.
These results directly depend on the chromo-EDM bounds, in Eq.~\eqref{eq:bounddtildeq}, which are obtained from the neutron EDM by neglecting cancellations between the light quarks (C)EDM and the Weinberg operator. The large uncertainty on the Weinberg operator contribution to the neutron EDM is treated conservatively by taking the smallest value within the confidence interval. 
We should point out that using the mercury EDM provides better bounds by about a factor 2~\cite{Graner:2016ses,Engel:2013lsa}. However, given the additional sources of uncertainty together with the cancellation effects that may arise between the several contributions to the mercury EDM, we consider only the direct experimental bounds on the neutron EDM. 
Note also that higher values of the NP scale yield less conservative results, e.g. a 30\% stronger bounds for $M_{\text{NP}} =10 \:\text{TeV}$.
The inclusion of dimension-six four-quark operators would add extra terms in the right-hand side of Eqs.~\eqref{eq:cCEDMmixing} and \eqref{eq:bCEDMmixing}. The resulting cancellation effects are nevertheless smaller than the self-correction of the chromo-EDM, shown in Figure~\ref{fig:cEDMvsEDM}.

\section{Consequences for New Physics} \label{sec:boundsConsequencesNP}

In the following we evaluate the effect of the new bounds for the charm and bottom quark EDMs on the parameter space of different BSM theories.

In the context of minimal flavour violation (MFV)~\cite{Buras:2000dm}, the EDM of different quarks only differ by the quark mass. When this dependence goes as $d_q \propto m_q$, the strong bounds on the light quark EDMs, ${|d_{u,d}|\lesssim10^{-25}\,\ecm}$, impose stronger constraints than our bounds. However, if the quark EDM scales with larger powers of the quark mass, the heavy quark EDMs are greatly enhanced and may become competitive.

\begin{figure}[ht]
	\centering 
	\includegraphics[height=0.2\columnwidth]{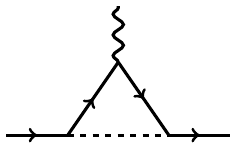} \hspace{0.3cm}    
	\includegraphics[height=0.2\columnwidth]{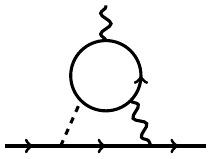}

	\caption{Example diagrams contributing to the quark (C)EDM in BSM theories at (left) one loop and (right) two loops, through Barr-Zee diagrams. Dashed lines represent new scalars, the external wavy lines represent photons (gluons), while the internal one can be either a photon, gluon, or weak gauge bosons.}
	\label{fig:2HDMdiagrams}
\end{figure}

\subsubsection{Two-Higgs doublet model}

This is the case of the Two-Higgs-Doublet model (THDM), which generates fermion EDMs via the Yukawa couplings of new scalars. To avoid flavour-changing neutral-currents at tree level, which are very constrained at the TeV scale, we restrict the discussion to the THDM with flavour alignment~\cite{Pich:2009sp,Penuelas:2017ikk} in which the Yukawa matrices 
$Y_{D,U}$ of the new scalars are proportional to the quark mass matrices,
\begin{align}
Y_{D}\,=\,\varsigma_D\,M_D~,\quad Y_{U}\,=\,\varsigma_U\,M_U~,    
\end{align}
%
%
%
where $\varsigma_{U,D}$ are complex numbers and contain the \CP violation.
In this type of models, the quark EDMs arise at one-loop level mediated by neutral or charged scalars (see Figure~\ref{fig:2HDMdiagrams}), giving contributions proportional to $m_{q}^3/M_{S^0}^2$ or $m_q m_{q'}^2/M_{S^{\pm}}^2 |V_{qq'}|^2$, respectively, where $V$ is the CKM matrix.
These mass factors suppress the light quark EDMs, which are actually dominated by two-loop Barr-Zee contributions, as shown Figure~\ref{fig:2HDMdiagrams}. The EDM of heavy quarks are much larger and, even with weaker experimental bounds, they can be more restrictive.

\subsubsection{Manohar-Wise model}

\begin{figure}[t]
	\centering
	\includegraphics[width=0.45\linewidth]{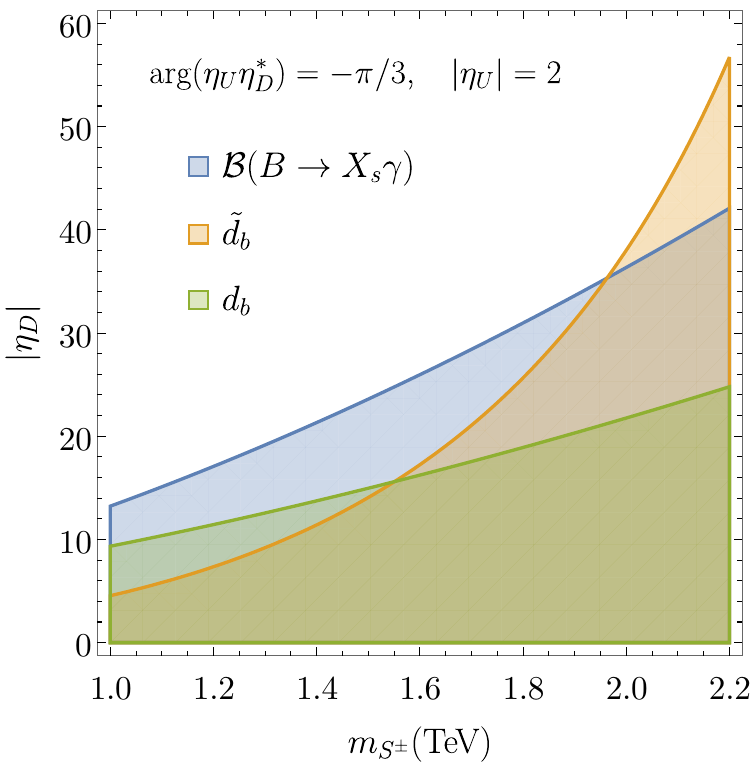}
	\includegraphics[width=0.45\linewidth]{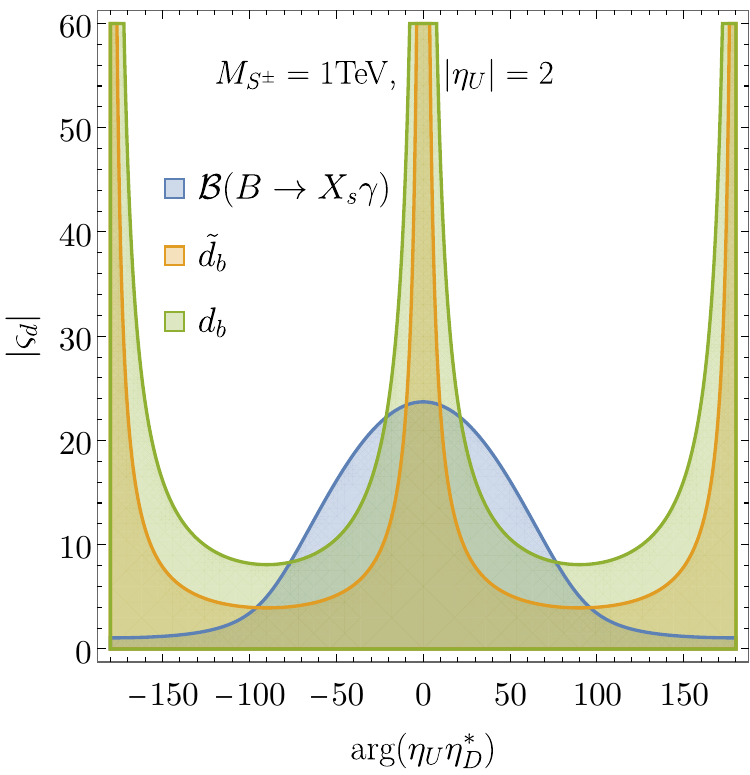}
	\caption{Constraints on the parameter space of the MW model. The shaded areas represent the allowed regions by each obserable. The lower-limit on the mass range follows from~Ref.\cite{Miralles:2019uzg}.}
	\label{fig:MW_zetaD-M}   
\end{figure}

Among these models, we shall consider the contribution to the bottom quark \mbox{({chromo-})} EDM by the colour-octet scalars appearing in the MW model \cite{Manohar:2006ga}, introduced already in Section~\ref{sec:MWmodel}. In the next chapter we will present a comprehensive study of EDM phenomenology within this model. For the moment, let us briefly see the implications of the new bound on $\db$. The relevant one-loop diagrams were originally computed in Ref.~\cite{Martinez:2016fyd} and are dominated by the exchange of a charged scalar with mass $m_{S^\pm}$. 
The constraints on this model from the experimental results on the $B^0_s - \overline{B}^0_s$ mixing or the $B_s^0\to \mu\mu$ and $B^0\to X_s\gamma$ decays were analysed in Ref.~\cite{Cheng:2015lsa}. Among them, the inclusive branching ratio $\mathcal{B}(B\to X_s\gamma)$ dominates the constraints on the $|\eta_U\,\eta_D| - m_{S^\pm}$ plane. As it is shown in Figure~\ref{fig:MW_zetaD-M} (left), the bounds on the bottom EDM derived above are more restrictive than this observable and even surpass the constraining power of the chromo-EDM for $m_{S^\pm} \gtrsim 1.5 \text{~TeV}$\footnote{The fact that the EDM is even more restrictive than the CEDM might seem very surprising, as the EDM bound was derived from the CEDM itself. The explanation is however rather simple: the contributions to $\dtildeb$ cancel out approximately at $m_{S^\pm}\approx2.5\,\tev$, completely diluting the restrictive power of this operator. Well before that, the restrictions from \db are already more stringent.}.
Fixing $m_{S^\pm} = 1\,\tev$ (the current lower limit~\cite{Miralles:2019uzg,Eberhardt:2021ebh}), we can see in Fig.~\ref{fig:MW_zetaD-M} (right) the potential of EDM observables to restrict the Yukawa couplings when their CP-violating phases deviate from zero or, conversely, to restrict these phases for reasonable values of the Yukawa couplings. 

With the current limits on the masses, the two-loop Barr-Zee diagrams largely dominate the EDM of light quarks, which may impose even stronger bounds on the MW model. This was the motivation to delve into this specific model and study its EDM phenomenology in Ref.~\cite{Gisbert:2021htg} (and Chapter~\ref{ch:edmsmw}). There, we will derive all the relevant one- and two-loop contributions and provide more detailed discussions on the interplay of parameters under the EDM bounds. Nevertheless, truly indisputable restrictions on the model will only come from global-fit analyses including the primary EDM observables. To this end, a new study using the \texttt{HEPFit} software~\cite{DeBlas:2019ehy}
%
%
%
is in preparation.

\subsubsection{Leptoquarks}

In recent years, a series of measurements hinting at the violation of lepton flavour universality have motivated new physics extensions with non-universal couplings between the three families. 
When the new \CP violation sources are specific to the quark family, the EDM of each quark carries independent and complementary information that should be used in complete phenomenological analyses of such models.

Examples of family non-universal \CP-violating interaction are found in models with scalar leptoquarks. These models are currently receiving a lot of attention as they are able to explain naturally the deviations in $ b\to c\tau\bar{\nu_\tau}$ transitions
~\cite{Buttazzo:2017ixm,Murgui:2019czp,Becirevic:2016yqi,Cornella:2019hct,Fajfer:2012jt,Hiller:2016kry}.
The additional charged currents contributing to this process are parametrized through the coefficient $g_{S_L}$. For scalar leptoquarks in the representation $(3, 2, 7/6)$ of the SM gauge group, the experimental values of $R_D$ and $R_{D^*}$ result in allowed regions for $g_{S_L}$ away from the real axis~\cite{Becirevic:2018afm} which induce a sizeable charm EDM~\cite{Dekens:2018bci}.
If no signal is observed in the planned neutron EDM experiments with sensitivities
of a few times $10^{-27} \ecm$~\cite{Chupp:2017rkp}, the resulting upper limits on the charm EDM (extracted with the method presented here) will rule out this model as an explanation for the B-anomalies.
In fact, exclusion regions on the model parameters from the charm EDM are already presented in Ref.~\cite{Dekens:2018bci}. These results are nevertheless based on lattice QCD calculations for the strange quark tensor charge, whose translation into the charm quark is highly uncertain.

\subsubsection{Supersymmetry}

The next BSM extension we discuss is the minimal supersymmetric standard model (MSSM). Among the large number of free parameters that it contains, there are many new sources of \CP violation. It is customary to restrict phenomenological analyses to just two sources: the trilinear couplings $A$, and the $\mu$-term (see definitions in Ref.~\cite{Martin:1997ns}). Since the fermion EDMs appear at one-loop level, these parameters are strongly constrained by the neutron and electron EDMs~\cite{Pospelov:2005pr}. Nevertheless, in more general cases the $A$ coupling can be a $3\times3$ matrix which elements are specific to the quark family. In particular, the charm quark EDM accesses the element $A_c$ predominantly via gluino loops \cite{Aydin:2002ie}. Updating the numerical analysis of Ref.~\cite{Aydin:2002ie} by taking into account the LHC restrictions on the masses~\cite{ATLAS:2019jvl}, we still find values of $\dc \sim 10^{-20}\ecm$ for scharm masses $M_{c1}$($M_{c2}$) of $1(2)$ TeV, gluino mass $m_{\tilde{g}}=1.6$ TeV, and  $\arg(A_c)=\pi/4$. These regions of the parameter space are therefore excluded and the new bounds should be included in more detailed analyses of this model.

Beyond the MSSM, there are new \CP-violating sources that can generate contributions to quark EDMs. 
An example of these is the MSSM with gauged baryon and lepton numbers (BLMSSM). Scaling the results of Ref.~\cite{Zhao:2016jcx} accounting for the top quark EDM bounds~\cite{Cirigliano:2016njn,Fuyuto:2017xup} we obtain values of $\dc$ reaching $10^{-19} \ecm$, \textit{i.e.} two orders of magnitude above the new upper limit in Eq.~\eqref{eq:newlimits}. As a consequence, the new heavy quark EDM bounds impose stringent constraints on the additional $\CP$-violating phases of the BLMSSM. 
In the R-parity violating supersymmetry, the EDM of heavy fermions are the only EDM observables that directly access the bilinear combinations of the third quark generation $\text{Im}(\lambda_{i 3 3}{\lambda_{i 3 3}'^*})$, for $i=1,2$~\cite{Yamanaka:2014nba}. Nevertheless, the leading contribution appears in this model at two-loop level~\cite{Yamanaka:2014mda} and it is suppressed in comparison with other supersymmetric extensions. For this reason, the bottom EDM is not yet competitive with other observables when considering the effect of one coupling $\lambda_{ijj}$ at a time,  but it could be used in global analyses to restrict extended regions of fine-tuned cancellations.

In the literature there exist other models giving predictions on heavy quark EDMs at the level of our bound or higher. For example, we found works based on Composite Higgs~\cite{Panico:2016ull} and THDM with non-universal extra dimensions~\cite{Iltan:2004xr}. 

\section{Summary}

We have presented a simple way to access the quark EDM through the corresponding chromo-EDM. The method relies on the inclusion of photon-loop corrections \cOale in the renormalization group equations, which are often overlooked due to their small size. Nevertheless, these corrections provide a new window to access effective operators which are otherwise unconstrained.
We derived new upper limits for the charm and bottom quark EDMs and showed the potential of these operators to constrain the parameters of NP models. 
These limits will provide valuable input for detailed phenomenological analyses of BSM physics.

In fact, since the publication of the new bounds, their implications have been explored for different extensions of the SM, including the BLMSSM~\cite{Yang:2019aao},  U(1)XSSM~\cite{Yan:2020ocy}, and THDM~\cite{Cai:2022xha}.


\chapter{Electric dipole moments from coloured scalars} \label{ch:edmsmw}

\begin{flushright}
	This chapter is based on Ref.~\cite{Gisbert:2021htg}.
\end{flushright}

\begin{figure}[t] 
	\centering
	\vbox{
		\resizebox{0.55\columnwidth}{!}{
			\includegraphics[scale=1]{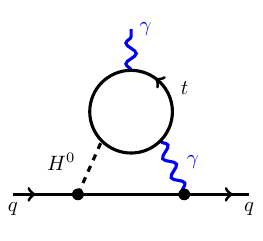}
			\includegraphics[scale=1]{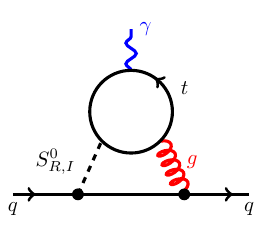}
		}
	}
	\caption{Opposed to the THDM with colourless scalars (left), the leading contribution to the light quark EDM appears in the MW model through gluon exchange (right), enhancing the EDM by a factor $(C_F\,\alpha_s)/\alpha\sim 50$, with the couplings evaluated at the hadronic scale.}
	\label{fig:barr-zee-THDM}
\end{figure}

In Chapter~\ref{ch:introexp} we argued the need for theories beyond the SM to explain the observed baryon asymmetry of the Universe, which differs by several orders of magnitude from the SM prediction. The MW model constitutes one of such theories, bringing new sources of CP violation that can be tested with experimental data. The building blocks of the MW model were already introduced in Section~\ref{sec:MWmodel} along with more details on its different motivations. 
In Section~\ref{sec:boundsConsequencesNP}, we saw the potential of bottom EDM bounds to restrict the parameter space of the MW model when CP-violating phases are allowed. 
However, to derive robust limits on the model accounting for cancellation effects, the direct contributions to the neutron EDM through gluonic or light-quark operators need to be considered. The relevant contributions include two-loop diagrams which, to our knowledge, have not been computed in the literature treating EDM observables within the MW model~\cite{Heo:2008sr,Hisano:2012cc,Fajfer:2014etr,Martinez:2016fyd}. 
The new scalars appearing in this model transform as octets under $SU(3)_C$, differently from those in THDMs, which transform as singlets. As a result, light-quark EDMs are greatly enhanced in this model, as explained in Figure~\ref{fig:barr-zee-THDM}, which are dominated by two-loop Barr-Zee diagrams in the range of masses allowed by LHC searches~\cite{Miralles:2019uzg}. This feature makes hadronic EDMs powerful observables to assess the viability of the MW theory. 
In this chapter we will derive all the relevant contributions to the neutron EDM from the Yukawa couplings of the colour-octet scalars, and make some considerations about the restrictions of these observables on the model parameters.

The chapter is organised as follows. 
We will start by describing the available experimental input on the EDM of hadrons and see its well-studied relation to the quark and gluon effective operators, in Section~\ref{sec:hadronEDMs}. 
The running of these operators from the NP scale to the hadronic scale was shown in Section~\ref{sec:EFT}, complementing the discussion of Chapter~\ref{ch:improvedbounds} by adding the mixing from the Weinberg operator. 
The main results of this study are shown in Sections~\ref{sec:MWEDMcontrib} and \ref{sec:pheno}. First, in Section~\ref{sec:MWEDMcontrib}, we provide analytical expressions for the relevant one- and two-loop contributions to the quark (C)EDM and Weinberg operator. 
The EDM predictions for each quark flavour are compared at the end of this section. In Section~\ref{sec:pheno}, the expected size of the neutron EDM, with all its contributions, is shown for different values of the model parameters. Next, we turn this around and use the current bounds on the neutron EDM to study the allowed values for the model parameters, showing the restrictive power of the EDMs as compared to other observables. 
The intriguing result of the $W$-boson mass measurement by the CDF collaboration~\cite{CDF:2022hxs} has an important impact on the parameter space of this model. Its interplay with the EDM bounds is briefly analysed in Section~\ref{sec:Wmass}. The main results are summarized in Section~\ref{sec:summary}.

\section{Experimental input} \label{sec:hadronEDMs}

To start our analysis of EDMs, we need to know what are the available experimental inputs. The scalars of this model carry colour charge and do not couple to leptons. Thus, we shall focus on hadron EDMs. Current limits and projected sensitivities on the neutron, proton, and mercury EDM are summarized in Table~\ref{tab:bounds}.

\begin{table}[h!]
	\centering
	\resizebox{0.75\linewidth}{!}{\begin{tabular}{lcc}
		\hline
		\hline
		Observable  &  Current bound [$e\cm$] & Future sensitivities [$e\cm$]\\
		\hline
		$d_n$ &  $1.8\,\cdot\,10^{-26}$  & $1.0\,\cdot\,10^{-28}$       \\
		$d_{p}$ & --   & $1.0\,\cdot\,10^{-29}$     \\
		$d_{\text{Hg}}$ & $6.3\,\cdot\,10^{-30}$   & $1.0\,\cdot\,10^{-30}$       \\
		\hline
		\hline
	\end{tabular}}
	\caption{Current experimental limits (at 90\% C.L.) on the absolute value of the electric dipole moments of the neutron $d_n$~\cite{nEDM:2020crw}, and mercury $d_{\text{Hg}}$ \cite{Graner:2016ses}, and the future experimental sensitivity, including the proton $d_p$ in $e\,$cm units.
	}
	\label{tab:bounds}
\end{table}

The relation between the EDM of hadrons and the (CP-violating) quark and gluon operators is computed with non-perturbative techniques of strong interactions at low energies. These computations have undergone much progress in the last decade. 
State-of-the-art coefficients for the CEDMs and Weinberg operator have been obtained in the literature with QCD sum rules~\cite{Pospelov:2000bw,Lebedev:2004va,Hisano:2012sc,Haisch:2019bml} and the quark model~\cite{Yamanaka:2020kjo}, while the contributions of the quark EDMs have been computed in lattice QCD~\cite{Bhattacharya:2015esa,Bhattacharya:2015wna,Bhattacharya:2016zcn,Gupta:2018qil,Gupta:2018lvp}, with significantly smaller errors. Assuming a Peccei-Quinn mechanism, these read~\cite{Dekens:2021bro}
\begin{align}\label{eq:edm_fun}
d_n&=g_T^u\,d_u+g_T^d\,d_d-(0.55\pm0.28)\,e\,\tilde d_u-(1.1\pm0.55)\,e\,\tilde d_d -20\,(1\pm0.5)\,{\rm MeV}\,e\,g_s\,w,\nonumber\\
d_{p}&=g_T^d\,d_u+g_T^u\,d_d+(1.30\pm0.65)\,e\, \tilde d_u+(0.60\pm0.30)\,e\,\tilde d_d +18\,(1\pm0.5)\,{\rm MeV}\,e\,g_s\,w\,,\nonumber\\
d_{\text{Hg}}&=-(2.1\pm 0.5)\cdot 10^{-4}\left[(1.9\pm0.1)d_n +(0.20\pm0.06) d_p \right]\,,
\end{align}
where the contributions to $d_{\rm Hg}$ from pion-nucleon interactions that are compatible with zero within $1\sigma$ have been left out. In these expressions, the tensor charges describing the light quark EDM contributions read $g_T^u = -0.213 \pm 0.012$ and $g_T^d = 0.820 \pm 0.029$.

The current bound on the mercury EDM~\cite{Graner:2016ses} is four orders of magnitude stronger than that of the neutron, which has been recently reported by the nEDM collaboration~\cite{nEDM:2020crw}. However, this difference is compensated by the suppression factor in front of $d_{n,p}$ in its contribution to $d_{\rm Hg}$, as shown in Eq.~\eqref{eq:edm_fun}.
As a result, we obtain almost identical constraints on the model parameters by using the $d_n$ or $d_{\rm Hg}$ experimental limits (within less than a 10\% difference), and in the numerical analysis of Section~\ref{sec:pheno} we will use only the direct limit on $d_n$.

\section{Renormalization group evolution}

The Wilson coefficients in Eq.~\eqref{eq:edm_fun} are evaluated at the hadronic scale $\mu_{\text{had}}\sim1\,$GeV, but the NP predictions will be extracted at the NP scale $\Lambda_{\text{NP}}\sim1\,$TeV. We can determine their evolution with the energy scale through the RGEs,
\begin{align}\label{eq:rge}
\frac{\text{d}\,\overrightarrow{\mathcal{C}}(\mu)}{\text{d}\,\text{ln}\,\mu}\,=\,\widehat{\gamma}^{T}(\mu)\,\overrightarrow{\mathcal{C}}(\mu)\,.
\end{align}
In the previous chapter we used the evolution of EDM and CEDM operators. Since the gluonic Weinberg operator $w$ also has, in principle, a sizeable contribution to the neutron EDM, we consider the three operators in  $\overrightarrow{\mathcal{C}}(\mu)=(C_1^q(\mu),\,C_2^q(\mu),\,C_3(\mu))$. These are related, in Eq.~\eqref{eq:basesEDMs}, to the usually defined coefficients $\dq,~\dtildeq,~w$. 
%
At leading order in $\alpha_s$, the anomalous dimension matrix is~\cite{Degrassi:2005zd,Dai:1989yh,Braaten:1990gq,Shifman:1976de,Brod:2018pli,Yamanaka:2017mef,Hisano:2012cc,deVries:2019nsu}
\begin{align} \label{eq:anomalousmatrix}
\widehat{\gamma}(\mu)\,=\,\frac{\alpha_s(\mu)}{4\pi}\begin{pmatrix}
8\,C_F & 0 & 0 \\
8\,C_F & 16\,C_F\,-4\,N_C& 0\\
0 & 2\,N_C & N_C\,+\,2\,n_f\,+\,\beta_0
\end{pmatrix}\,,
\end{align}
where $C_F=(N_C^2-1)/(2 N_C)$, $\beta_0=(11 N_C -2 n_f)/3$, $N_C=3$ and $n_f$ denotes the number of active flavours. Solving Eq.~\eqref{eq:rge} with the methods of Appendix~\ref{app:RGEsolution} we obtain the scale dependence of the Wilson coefficients $\overrightarrow{\mathcal{C}}(\mu)$ for a theory with constant number of active flavours. Starting at the NP scale $\mu=\Lambda_{\text{NP}}$, close to the top quark mass,\footnote{ The masses of the new scalars are in fact constrained to be above 1 TeV~\cite{Miralles:2019uzg,Eberhardt:2021ebh}. By simultaneously integrating out the new scalars and the top quark, we ignored the running in the range $[1\,\text{TeV},m_t]$. Conservatively, we estimate this effect to be at most of 30\%, which is smaller than the leading systematic error from the hadronic matrix elements in Eq.~\eqref{eq:edm_fun}.
} where the fundamental theory is matched to the effective one, the Wilson coefficients are evolved down to the bottom-quark mass scale with $n_f=5$. At this point, the bottom quark is integrated out, generating a threshold contribution of the bottom CEDM to the Weinberg operator as~\cite{Dai:1989yh,Boyd:1990bx,Dekens:2018bci}
\begin{align}\label{eq:thresholdweinberg}
C_3(\mu_b^-)\,=\,C_3(\mu_b^+)\,+\,\frac{ \alpha_s(\mu_b^+)}{8\,\pi}\,C_2^q(\mu_b^+)~.
\end{align}
Here $\mu_b^+$ and $\mu_b^-$ refer to the scale $\mu_b\sim m_b$ in the theories with $n_f=5$ and $n_f=4$, respectively.  Analogously, also the charm CEDM induces a threshold correction to the Weinberg operator, although it is numerically irrelevant for our study on the MW model. The final running with $n_f=4$ and $n_f=3$ brings the Wilson coefficients down to the hadronic scale $\mu_{\text{had}}\sim 1\,$GeV.

%

Equation~\eqref{eq:thresholdweinberg} shows the key relation to find bounds on the heavy-quark CEDM from the Weinberg operator (and in turn from the neutron EDM). These bounds were used in Chapter~\ref{ch:improvedbounds} to find limits on the heavy-quark EDMs. In the MW model, we have checked that the leading constraints on the Yukawa couplings appear from the light quark (C)EDMs, as we will see in Section~\ref{sec:pheno}, and therefore we neglected $\mathcal{O}(\alpha)$ corrections in Eq.~\eqref{eq:anomalousmatrix} for the sake of simplicity.

\section{Contributions to EDMs}\label{sec:MWEDMcontrib}

In this section, we analyse the different contributions to the neutron EDM from the colour-octet scalars. Namely, we will derive the expressions for the quark (C)EDM at one-loop level and the enhanced contributions at two-loop level, together with the leading two-loop contributions to the Weinberg operator. The constraints on the model using these expressions are discussed in Section~\ref{sec:pheno}.

\subsection{One-loop contributions}\label{sec:oneloop}

\begin{figure}[htp!]
	\centering
	
	\vbox{
		\resizebox{1.0\columnwidth}{!}{
			\subcaptionbox{}{\includegraphics[scale=1]{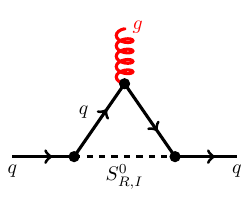}}
			\subcaptionbox{}{\includegraphics[scale=1]{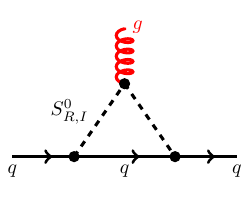}}
			\subcaptionbox{}{\includegraphics[scale=1]{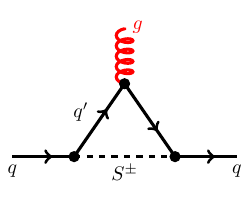}}
			\subcaptionbox{}{\includegraphics[scale=1]{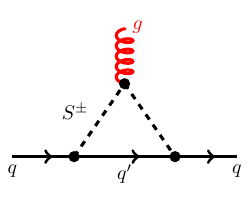}}
		}
	}
	\caption{Neutral $S^0$ and charged $S^\pm$ scalars contributing to the CEDM of a quark $q$.}
	\label{fig:oneloopCEDM}
\end{figure}

\begin{figure}[h!]
	\centering	
	\vbox{
		\resizebox{0.75\columnwidth}{!}{
			\subcaptionbox{}{\includegraphics[scale=1]{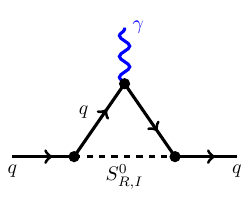}}
			\subcaptionbox{}{\includegraphics[scale=1]{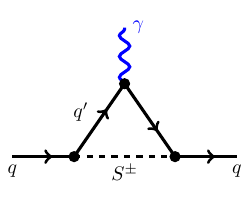}}
			\subcaptionbox{}{\includegraphics[scale=1]{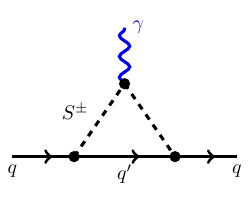}}
		}
	}
	\caption{Neutral $S^0$ and charged $S^\pm$ scalars contributing to the EDM of a quark $q$. }\label{fig:oneloopEDM}
\end{figure}

At one-loop level, the EDM (CEDM) of a quark $q$ receives contributions from neutral and charged scalars, $S^0_{I,R}$ and $S^\pm$, as shown in Figure~\ref{fig:oneloopEDM} (Figure~\ref{fig:oneloopCEDM}). These contributions can be computed using standard techniques, and are finite since the Lagrangian does not contain any tree-level (C)EDM. Our results for the CEDM of a quark $q$ reads
\begin{align}
\begin{split}
\widetilde{d}_q\,=\,\text{sgn}(\mathcal{Q}_q)\left(\mathcal{N}_{(a)}^{qg}\,\widetilde{d}^{\,\text{(a)}}_q+\mathcal{N}_{(b)}^{qg}\,\widetilde{d}^{\,\text{(b)}}_q+\mathcal{N}_{(c)}^{qg}\,\widetilde{d}^{\,\text{(c)}}_q+\mathcal{N}_{(d)}^{qg}\,\widetilde{d}^{\,\text{(d)}}_q\right)\,,
\end{split}
\end{align}
with\footnote{Equations~\eqref{eq:CEDM_oneloop} and \eqref{eq:EDM_oneloop} can be translated to the basis of Ref.~\cite{Martinez:2016fyd} through the following replacements: $d_q\to-\,d_q$ and $\widetilde{d}_q\to-\,\widetilde{d}_q$, as well as $\eta_U\to\eta_U\,\text{e}^{i(\alpha_U+\pi)}$ and $\eta_D\to\eta_D\,\text{e}^{-i\alpha_D}$.} 
\begin{align}\label{eq:CEDM_oneloop}
\begin{split}
\widetilde{d}^{\,\text{(a)}}_{q}\,&=\,-\, \frac{G_F}{\sqrt{2}}\,\frac{m_{q}^3}{4\,\pi^2}\,\text{Re}(\eta_Q)\,\text{Im}(\eta_Q)\,\left(\frac{F_{2,0}(r_{q R})}{m_{S_R^0}^2}\,-\,\frac{F_{2,0}(r_{q I})}{m_{S_I^0}^2}\right)\,,\\
\widetilde{d}^{\,\text{(b)}}_{q}\,&=\,-\,
\frac{G_F}{\sqrt{2}}\,\frac{m_{q}^3}{4\,\pi^2}\,\text{Re}(\eta_Q)\,\text{Im}(\eta_Q)\,\left(\frac{F_{1,1}(r_{q R})}{m_{S_R^0}^2}\,-\,\frac{F_{1,1}(r_{q I})}{m_{S_I^0}^2}\right)\,,\\
\widetilde{d}^{\,\text{(c)}}_{q}\,&=\,\frac{G_F}{\sqrt{2}}\,\frac{m_{q}\,m_{q^\prime}^2\,|V_{q q^\prime}|^2}{4\,\pi^2}\,\bigg(\text{Re}(\eta_Q)\, \text{Im}(\eta_{Q^\prime})-\text{Re}(\eta_{Q^\prime}) \,\text{Im}(\eta_Q)\bigg)\,\left(\frac{G_{2,0}(r_{q},r_{q^\prime})}{m_{S^\pm}^2}\right)\,,\\
\widetilde{d}^{\,\text{(d)}}_{q}\,&=\,\frac{G_F}{\sqrt{2}}\,\frac{m_{q}\,m_{q^\prime}^2\,|V_{q q^\prime}|^2}{4\,\pi^2}\,\bigg(\text{Re}(\eta_Q)\, \text{Im}(\eta_{Q^\prime})-\text{Re}(\eta_{Q^\prime}) \,\text{Im}(\eta_Q) \bigg)\,\left(\frac{G_{1,1}(r_{q},r_{q^\prime})}{m_{S^\pm}^2}\right)\,,
\end{split}
\end{align}
where
\begin{align}\label{eq:colorfactors}
\mathcal{N}_{(a)}^{qg}\,=\,\mathcal{N}_{(c)}^{qg}\,=\,-\frac{C_A - 2 \,C_F}{2}\,=\,-\,\frac{1}{6}\,,\quad
\mathcal{N}_{(b)}^{qg}\,=\,\mathcal{N}_{(d)}^{qg}\,=\,\frac{C_A}{2}\,=\,\frac{3}{2}\,,
\end{align}
are colour factors emerging from the color structures appearing in the Feynman diagrams $T^{b} T^{a} T^{b}=-\frac{(C_A - 2C_F)}{2}\,T^{a}$ and $f^{a b c} \,T^{b} T^{c} = i\,\frac{C_A}{2}\,T^{a}$ with $C_A=3$.\footnote{Sum over repeated (colour) indices is understood here and in the following. } The loop functions $F_{n,m}(r)$ and $G_{n,m}(r_1,r_2)$ are defined in the notation of Ref.~\cite{Martinez:2016fyd} as
\begin{align}
F_{n,m}(r)\,&=\,\int_{0}^1\,\frac{x^n\,(1-x)^m}{1\,-\,x\,+\,r^2\,x^2}\,\text{d}x\,,\\
G_{n,m}(r_1,r_2)\,&=\,\int_{0}^1\,\frac{x^n\,(1-x)^m}{(1-x)(1\,-\,r_1^2\,x)\,+\,r_2^2\,x^2}\,\text{d}x\,,
\end{align}
with $r_q\equiv\frac{m_q}{m_{S^\pm}}$, $r_{qR}\equiv\frac{m_q}{m_{S^0_R}}$, and $r_{qI}\equiv\frac{m_q}{m_{S^0_I}}$.

%
%

While both neutral and charged scalars couple to gluons, only charged scalars couple to photons, making the CEDM receive an extra contribution with respect to the EDM, depicted in Figure~\ref{fig:oneloopCEDM} (b). The contributions to the EDM of a quark $q$, shown in Figure~\ref{fig:oneloopEDM}, share the same loop functions with the CEDM diagrams. Therefore, the results for the EDM can be given in terms of the CEDM expressions (Eqs.~\eqref{eq:CEDM_oneloop}) as

\begin{align}
\begin{split}
d_{q}\,=\,\text{sgn}(\mathcal{Q}_q)\left(\mathcal{N}_{(a)}^{q\gamma}\,d^{\,\text{(a)}}_{q}+\mathcal{N}_{(b)}^{q\gamma}\,d^{\,\text{(b)}}_{q}+\mathcal{N}_{(c)}^{q\gamma}\,d^{\,\text{(c)}}_{q}\right)\,,
\end{split}
\end{align}
where 
\begin{align}\label{eq:EDM_oneloop}
d^{\,\text{(a)}}_{q}\,=\,e\,\mathcal{Q}_q\,\widetilde{d}^{\,\text{(a)}}_{q}\,,\quad
d^{\,\text{(b)}}_{q}\,=\,e\,\mathcal{Q}_{q^\prime}\,\widetilde{d}^{\,\text{(c)}}_{q}\,,\quad
d^{\,\text{(c)}}_{q}\,=\,e\,(\mathcal{Q}_q\,-\,\mathcal{Q}_{q^\prime})\,\widetilde{d}^{\,\text{(d)}}_{q}\,,
\end{align}
and
\begin{align}
\mathcal{N}_{(a)}^{q\gamma}\,=\,\mathcal{N}_{(b)}^{q\gamma}\,=\,\mathcal{N}_{(c)}^{q\gamma}\,=\,C_F\, .
\end{align}
The colour factor $C_F$ emerges from the combination of two colour matrices $ (T^a T^a)_{ij}=C_F\,\delta_{ij}$, each of them provided by one of the two Yukawa couplings in the diagrams of Figure~\ref{fig:oneloopEDM}.

Correcting by colour factors, our results are in good agreement with the literature of the colourless THDM~\cite{Iltan:2001vg,Jung:2013hka}. Additionally, for the diagrams that do not appear in the THDM but emerge through the introduction of colour-octet scalars (see Figure~\ref{fig:oneloopCEDM} (b) and (d)) we found agreement with the previous calculation in Ref.~\cite{Martinez:2016fyd}. However, in this reference, the loop function $F_{2,0}$ in Eqs.~\eqref{eq:CEDM_oneloop} is replaced by $F_{0,0}$, which differs from our results and those of Refs.~\cite{Iltan:2001vg,Jung:2013hka}.

\subsection{Two-loop contributions}\label{sec:twoloop}

In the previous section, we have studied all one-loop contributions to the (C)EDM of the quarks. Since light quark (C)EDMs are heavily suppressed at one-loop level by powers of the quark masses, the leading contributions to the neutron EDM appear at two-loop level. In the following, we derive these two-loop contributions to the quark (C)EDM and the Weinberg operator.

\subsubsection{Barr-Zee diagrams}

\begin{figure}[h]
	\centering
	\vbox{
		\resizebox{0.95\columnwidth}{!}{
			\subcaptionbox{}{\includegraphics[scale=1]{pheno/fEDMsMW/diagrams/barzeeS0gluontphoton}}
			\subcaptionbox{}{\includegraphics[scale=1]{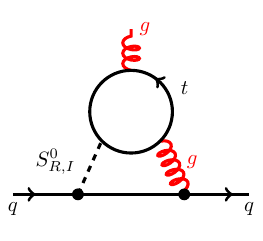}}
			\subcaptionbox{}{\includegraphics[scale=1]{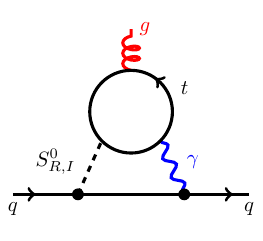}}
			\subcaptionbox{}{\includegraphics[scale=1]{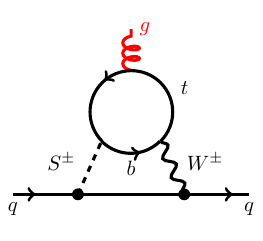}}
		}
	}
	\caption{Barr-Zee type diagrams contributing to the EDM (a) and the CEDM (b, c and d) of quarks in the MW model. 
	}\label{fig:barr_zee}
\end{figure}

Although being suppressed by additional coupling constants and loop factors, the Barr-Zee type diagrams, shown in Figure~\ref{fig:barr_zee}, benefit from the enhancement of the top-quark Yukawa coupling in flavour models. 
Among the contributions to the CEDM, the one depicted in Figure~\ref{fig:barr_zee} (b) largely dominates, since it is enhanced by the strong coupling constant (from the internal gluon propagator) and it is not suppressed by any mass of the electroweak bosons. For this reason, this will be the only diagram that will be included in the expressions of the CEDM. 
Contributing to the quark EDM, only the diagram (a) of Figure~\ref{fig:barr_zee} is numerically relevant.

The Barr-Zee contribution to the CEDM is given by
\begin{align}
\begin{split}
\widetilde{d}_{q}^{\text{BZ}} =\,-\, 2\,\sqrt{2}\,G_F\frac{\alpha_s \,m_q }{(4\pi)^3}\, \mathcal{N}^{qg}_{\text{BZ}}\,
\bigg( 
&\imEtaQ \reEtaU \mathcal{F}^{(1)}\left(r_{tR}\right)
+\reEtaQ\imEtaU\mathcal{F}^{(1)}\left(r_{tI}\right)   \\
+ &\imEtaU \reEtaQ \widetilde{\mathcal{F}}^{(1)}\left(r_{tR}\right)
+\imEtaQ\reEtaU \widetilde{\mathcal{F}}^{(1)}\left(r_{tI}  \right)\bigg),
\end{split}  
\end{align}
where we have defined the loop functions\footnote{$\mathcal{F}^{(1)}$ and $\widetilde{\mathcal{F}}^{(1)}$ are related to $f$ and $g$ from Ref.~\cite{Jung:2013hka}, through $\mathcal{F}^{(1)}(\sqrt{r})=-f(r)$ and $\widetilde{\mathcal{F}}^{(1)}(\sqrt{r})=g(r)$~.
}
\begin{align}
&\mathcal{F}^{(1)}(r)=\frac{r^2}{2}\int_0^1 \text{d}x\, \frac{2x(1-x) -1}{r^2 - x(1-x)} \log \frac{r^2}{x(1-x)}\,,\\
&\widetilde{\mathcal{F}}^{(1)}(r)=\frac{r^2}{2}\int_0^1 \text{d}x \, \frac{1}{r^2 - x(1-x)} \log \frac{r^2}{x(1-x)}\,,
\end{align}
and the colour factor $\mathcal{N}^{qg}_{\text{BZ}}=\frac{N_C^3-N_C-4}{4N_C}$ comes from $\Big(\Tr(T^a T^b T^c)+\Tr(T^b T^a T^c)\Big)T^b T^c=\frac{d^{a b c}}{2} T^b T^c=\frac{N_C^3-N_C-4}{4N_C} T^a$. The first two traces correspond to the current flow of the top quark in the loop, clockwise and counterclockwise. 
Similar to the one-loop case, the EDM depends on the same loop functions as the CEDM and they are related by 
\begin{align}
d_{q}^{\text{BZ}} = e\, \frac{\mathcal{Q}_t\,C_F}{\mathcal{N}^{qg}_{\text{BZ}}} \,
\widetilde{d}_{q}^{\text{BZ}}\,,
\end{align}
with $\mathcal{Q}_t=\frac{2}{3}$ being the charge of the top quark. For details on the computation of these diagrams see Appendix~\ref{app:BarrZee}.

\subsubsection{Weinberg contribution}

\begin{figure}[h!] 
	\centering
	\vbox{
		\resizebox{0.70\columnwidth}{!}{
			\subcaptionbox{}{\includegraphics[scale=1]{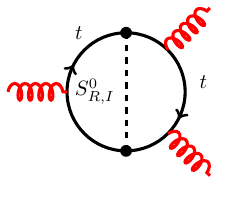}}
			\subcaptionbox{}{\includegraphics[scale=1]{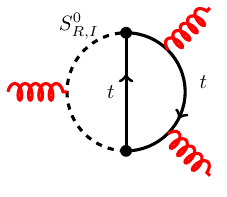}}
			\subcaptionbox{}{\includegraphics[scale=1]{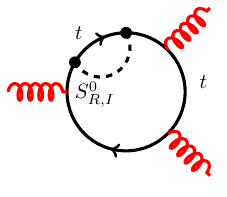}}
		}
	}
	\caption{Neutral scalar contributions to the Weinberg operator. The colour structure of diagram (a) yields a suppression factor of $1/6$ with respect to a THDM with colour-singlet scalars. In turn, diagram (b) is specific to colour-octet scalars, and diagram (c) vanishes. 
	}
	\label{fig:twoloopWeinbergNeut}
\end{figure}

\begin{figure}[h!] 
	\centering
	\vbox{
		\resizebox{0.9\columnwidth}{!}{
			
			\subcaptionbox{}{\includegraphics[scale=1]{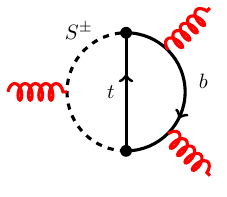}}
			\subcaptionbox{}{\includegraphics[scale=1]{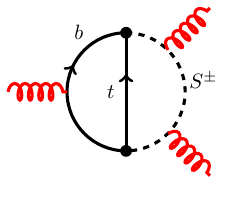}}
			\subcaptionbox{}{\includegraphics[scale=1]{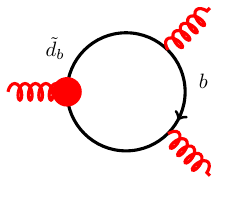}}
			\subcaptionbox{}{\includegraphics[scale=1]{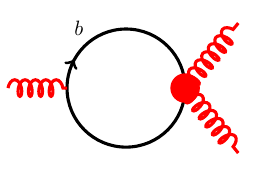}}

		}
	}
	\caption{Charged scalar contributions to the Weinberg operator. Below the top quark mass scale, diagrams (a) and (b) are accounted for through the effective operators of the bottom quark, depicted as red circles in diagrams (c) and (d). These induce a threshold correction to the Weinberg operator at the bottom quark mass scale, as shown in Eq.~\eqref{eq:thresholdweinberg}.
	}
	\label{fig:twoloopWeinbergCharg}
\end{figure}

The Weinberg operator can also play an important role since it contributes directly to the neutron EDM and does not suffer from light quark mass suppressions. Furthermore, it also has an impact on the light quarks' (C)EDM due to the operator mixing in the RGEs. The first-order contribution appears at the two-loop level via the exchange of neutral (Figure~\ref{fig:twoloopWeinbergNeut}) or charged (Figure~\ref{fig:twoloopWeinbergCharg}) scalars. For the diagrams with neutral scalars, the masses of the top quark and new scalars running in the loop are assumed to be of the same order and, therefore, the complete two-loop diagrams must be calculated, yielding\footnote{Notice that Wilson coefficient of the Weinberg operator from Ref.~\cite{Jung:2013hka} can be translated to our basis through $C_W\,=\,-\,g_s\,C_3$.}
\begin{align}\label{eq:loopWeinberg_neutral}
\begin{split}
w^{\,\text{(a)}}\,&=\,4\,\sqrt{2}\,G_F\,\frac{\alpha_s}{(4\pi)^3}
\,\imEtaU\,\reEtaU\,\mathcal{N}_{(a)}^w  \, \big( h(r_{tR}) - h(r_{tI}) \big)\,,\\
w^{\,\text{(b)}}\,&=\,4\,\sqrt{2}\,G_F\,\frac{\alpha_s}{(4\pi)^3}\,\imEtaU\,\reEtaU\,\mathcal{N}_{(b)}^w  \, \big( g(r_{tR}) - g(r_{tI}) \big)\, .
\end{split}
\end{align}
Here $\mathcal{N}_{(a)}^w=\mathcal{N}_{(b)}^w=-\frac{C_A - 2 \,C_F}{2}$ are the colour factors emerging from $ T^{b} T^{a} T^{b}$, similar to Eq.~\eqref{eq:colorfactors}, and
\begin{align}
\begin{split}
h(r)&=\frac{r^4}{4}\int_0^1 \text{d}x \int_0^1 \text{d}y \frac{y^3x^3(1-x)}{[r^2 x (1-y x)+(1-y)(1-x)]^2}\,,\\
g(r)&=\frac{r^4}{4}\int_0^1 \text{d}x \int_0^1 \text{d}y \frac{y^3x^2(1-x)^2}{[r^2 x (1-y x)+(1-y)(1-x)]^2}\,,  
\end{split}
\end{align}
are the corresponding loop functions. In turn, the diagram (c) of Figure~\ref{fig:twoloopWeinbergNeut} vanishes. The details of the calculation are found in Appendix~\ref{app:wein}. Notice that $h(r)$ corresponds to the well-known Weinberg loop function of Ref.~\cite{Weinberg:1989dx} but $g(r)$ is only appearing for coloured scalars since they can couple to gluons. Looking at Eq.~\eqref{eq:loopWeinberg_neutral} we see how, as it happened for the one-loop contribution to the (C)EDMs of the quarks, there is a relative minus sign between the contribution of the CP-even and CP-odd neutral scalars. Therefore, this contribution will be suppressed by the mass splitting of the neutral scalars which, as can be seen in Ref.~\cite{Eberhardt:2021ebh}, is around two orders of magnitude smaller than the mass of the scalars. This suppression is large and, for the numerical analysis in Section~\ref{sec:pheno}, we will assume that all scalar masses are the same, effectively neglecting the contribution of the neutral scalars in the Weinberg operator.

The calculation of the charged scalar contributions, shown in Figure~\ref{fig:twoloopWeinbergCharg}, proceeds differently. Having a bottom quark propagator in the loop, these diagrams will only induce a contribution to the Weinberg operator below the bottom-quark mass scale. At this scale, the NP particles and the top quark have already been integrated out and their information is encoded in the effective vertices shown in diagrams (c) and (d) of Figure~\ref{fig:twoloopWeinbergCharg}. These one-loop diagrams generate a threshold contribution to the Weinberg operator from the bottom CEDM $\tilde d_b$, as shown in Eq.~\eqref{eq:thresholdweinberg}. The main contribution to $\tilde d_b$ comes from the one-loop diagrams 
(c) and (d) of Figure~\ref{fig:oneloopCEDM}, which are not suppressed by any kind of light quark mass or CKM factor.

\subsection{Four-quark contributions}\label{sec:fourquark}

\begin{figure}[h!]
	\centering
	\vbox{
		\includegraphics[scale=1]{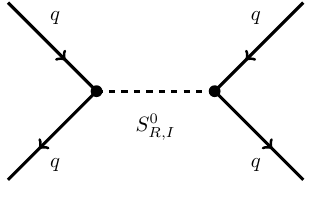}
	}
	\caption{Four-quark contribution from the neutral scalars $S^0_{R,I}$.}\label{fig:4quark}
\end{figure}

Four-quark interactions due to the exchange of colour-octet scalars (Figure~\ref{fig:4quark}) have been studied in detail in Ref.~\cite{Hisano:2012cc}. Using the results from that work, we observe that they represent a sub-leading contribution to the neutron EDM 
\begin{align}
d_n^{(4\,q)}\,\sim\,3\,\cdot\,10^{-29}\,e\,\text{cm}\,\imEtaQ\, \reEtaQ\,\left(\frac{1\,\text{TeV}}{m_{S_{R,I}^0}}\right)^2~,
\end{align}
well below the two-loop contributions presented in Section~\ref{sec:twoloop}. Therefore, we neglect them in the following.

\subsection{Flavours of quark EDM: discussion }

\label{sec:discussionQEDM}

\begin{figure}[h!]
	\centering
	\includegraphics[width=0.45\textwidth]{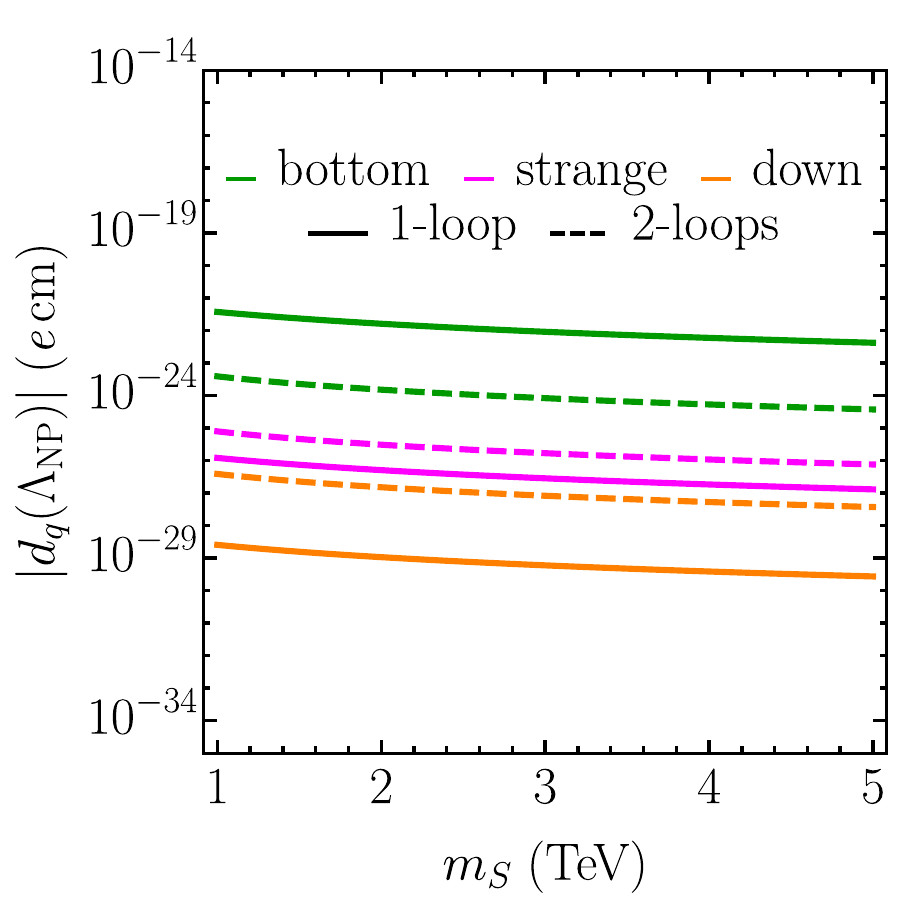} \includegraphics[width=0.45\textwidth]{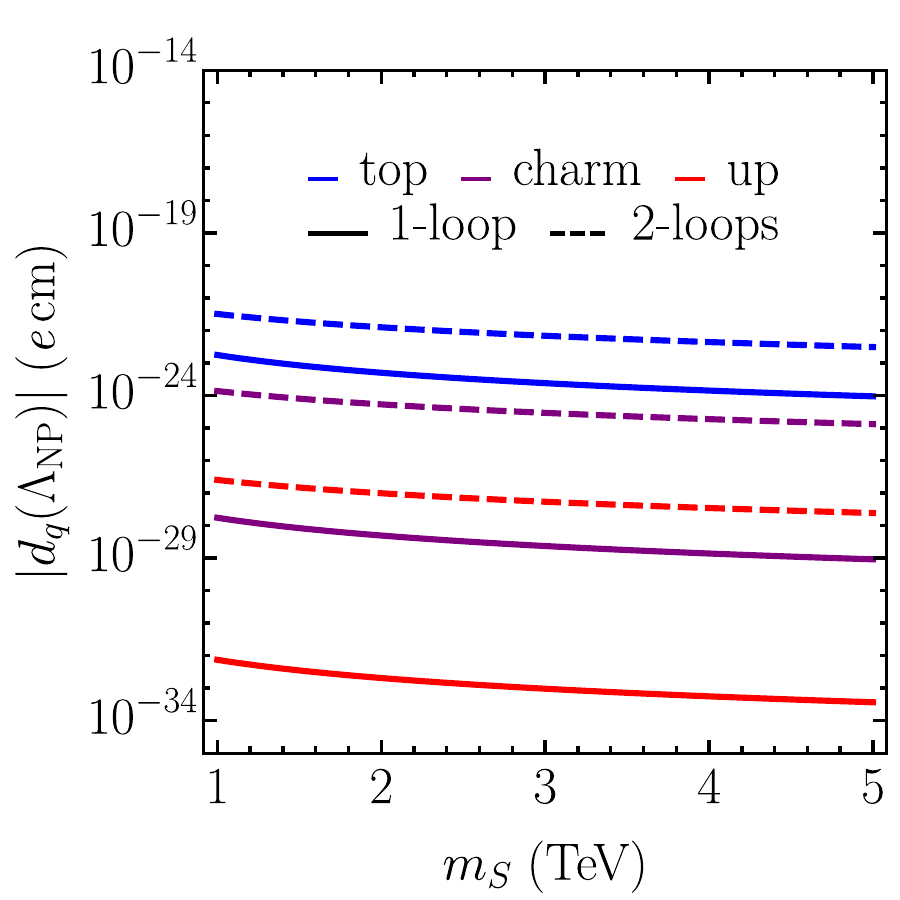}
	\\
	\includegraphics[width=0.45\textwidth]{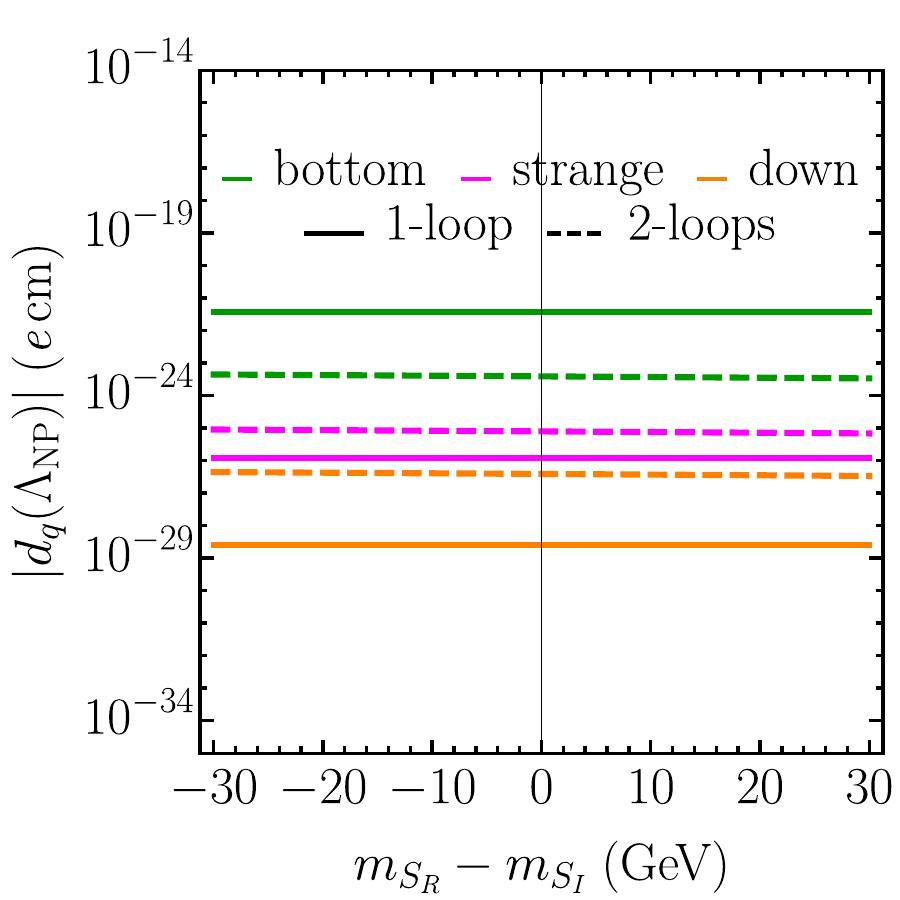}
	\includegraphics[width=0.45\textwidth]{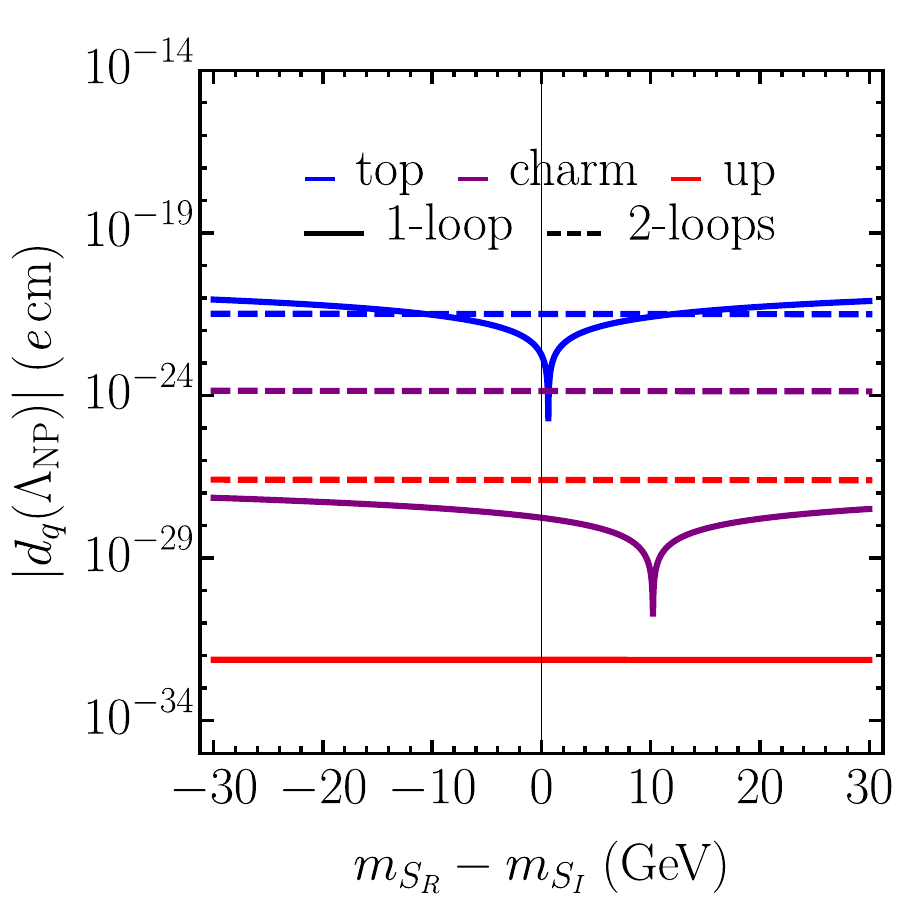}
	\caption{Comparison of the 1- and 2-loop contributions to the EDM of the (left) up-type quarks and (right) down-type quarks, as a function of (top) the mass of the scalars $m_S$ and (bottom) the mass splitting of the neutral scalars $m_{S_R} - m_{S_I}$. The Yukawa couplings have been fixed to $|\eta_U| = |\eta_D|= 1$, $ \arg(\eta_U)=\pi/4$, and $ \arg(\eta_D)=0$. The masses have been fixed to $m_{S^+} = m_{S_R} = m_{S_I} = m_S$, and $m_{S^+} =  m_{S_I} =1\,\text{TeV}$ in the top and bottom panels, respectively.
	}
	\label{fig:EDM_heavy_quarks}
\end{figure}

Before moving on with the phenomenological analysis to compare the model predictions against current experimental upper limits, it is worthwhile to compare the size of the Barr-Zee contributions, obtained above, to the one-loop contributions presented originally in Ref.~\cite{Martinez:2016fyd}.\footnote{Note, however, that our results do not agree with this reference for one of the loop functions. See the discussion above for more details.} 

Due to the strong suppression from powers of the light quark masses, the light quark (C)EDMs are dominated by two-loop Barr-Zee diagrams. Even though these contributions include additional coupling constants and loop suppression factors, the top-quark Yukawa coupling, proportional to $m_t$, makes the two-loop diagrams the dominant contribution for light quark (C)EDMs. With the same argument, one would expect heavy quark (C)EDMs to be dominated by one-loop diagrams, as they are not suppressed neither by light quark masses nor by loop suppression factors. This hierarchy of contributions is illustrated in Figure \ref{fig:EDM_heavy_quarks} (top left), where we see that $d_b^{\rm 1-loop} > d_b^{\rm 2-loop}$, and vice-versa for the strange and down quark, $d_s^{\rm 1-loop} < d_s^{\rm 2-loop}$ and $d_d^{\rm 1-loop} < d_d^{\rm 2-loop}$.

This pattern is generally expected in models where the Yukawa couplings of new scalars are proportional to the quark mass. However, in the MW model, this hierarchy of contributions is apparently not respected for up-type quarks, as shown in Figure \ref{fig:EDM_heavy_quarks} (top right), where also the 2-loop contribution dominates for the top quark EDM, $d_t^{\rm 1-loop} < d_t^{\rm 2-loop}$. This counter-intuitive behaviour can be explained by inspecting the expressions of $d_q$ at one-loop level. There, the CP-odd and CP-even neutral scalar contributions to the (C)EDMs have opposite signs, cancelling each other to a large extent, since the mass splitting $|m_{S_R^0} - m_{S_I^0} | \leq 30\,\text{GeV}$ as established by unitarity bounds \cite{Eberhardt:2021ebh}. In the limit where the masses are degenerate, $m_{S_R^0} = m_{S_I^0}$, only the charged scalar contribution is relevant at one-loop level, being $d_q^{\rm 1-loop}\propto m_q m_{q^\prime}^2 |V_{q q^\prime}|^2$. As a consequence, the top quark EDM $d_q^{\rm1-loop}$ is suppressed by the bottom quark mass, while only top-quark masses appear in $d_t^{\rm2-loop}$. The dependence on the mass splitting is explicitly shown in Figure \ref{fig:EDM_heavy_quarks} (bottom right), where we see that as soon as $|m_{S_R^0} - m_{S_I^0} |$ deviates from zero, the neutral scalar contribution dominates and the expected hierarchy with $d_t^{\rm 1-loop} > d_t^{\rm 2-loop}$ is recovered. Nevertheless, within the allowed range of  $|m_{S_R^0} - m_{S_I^0} |$, note that the one- and two-loop level contributions to the top quark are of similar size, at least for some regions of the parameter space. In this figure, the dip on $d_q^{\rm 1-loop}$ away from $|m_{S_R^0} - m_{S_I^0} | = 0$ is due to the cancellation of the neutral and charged scalar contributions.

This feature is not reproduced for down-type quarks, in Figure \ref{fig:EDM_heavy_quarks} (bottom left), since the charged-scalar contribution dominates in all the range of masses. The reason for this is the enhancement from the heavier mass of the quark running in the loop. Namely, $d_b^{\rm 1-loop}\propto m_t^2$ and $d_s^{\rm 1-loop} \propto m_c^2$. As a consequence, at one-loop level, the bottom and strange quark EDMs are in fact larger than their up-type partners.

\section{Phenomenological analysis} \label{sec:pheno}

With all the relevant contributions of the coloured scalars to the EDM of hadrons obtained, we can study the constraints that these observables impose on the model. Currently, there is no direct limit on the EDM of the proton and it will not be used for this analysis. Furthermore, as mentioned previously, the implications from the neutron and mercury EDM on the MW model are extremely similar. Therefore, in order to provide clearer limits, we will only consider the direct limit on the neutron EDM in the numerical analysis. The values for the hadronic and nuclear matrix elements of Eq.~\eqref{eq:edm_fun} have been set to the central values. Of course, with only one observable and a total of seven parameters,
$$
|\eta_U|,\, |\eta_D|,\, \text{arg}(\eta_U),\,\text{arg}(\eta_D),\, m_{S_R},\, m_{S_I}, \text{ and } m_{S^\pm},
$$
we do not provide a complete phenomenological analysis here, but just a brief study showing the potential of EDM observables to constrain the parameter space of the model. To this end, the interplay of the model parameters appearing in the neutron EDM is discussed, comparing its limits to those imposed by other powerful observables studied in the literature. A global fit with all the relevant observables for the CP-violating MW model, although interesting, is beyond the scope of this work. 

To reduce the total number of free parameters and provide some sensible plots, we will first assume that the mass of the scalars is degenerate, $m_{S_R}=m_{S_I}=m_{S^\pm}=m_S$.~\footnote{This assumption is reasonable given the constraints on the scalar mass splitting found from unitarity bounds in Ref.~\cite{Eberhardt:2021ebh}. However, as we will see in Section~\ref{sec:Wmass}, this assumption is incompatible with the recent measurement of the $W$ mass by the CDF collaboration.} In addition, we will show here the results for $|\eta_U|=1$, which is below the maximum allowed value found in  Ref.~\cite{Eberhardt:2021ebh} for masses of the coloured scalars of a few TeVs. If $|\eta_U|=0$, the contributions with the top quark Yukawa coupling studied here vanish, and only the suppressed diagrams with bottom-quark propagators running in the loops contribute.

\newpage
\subsection{Neutron EDM predictions}

\begin{figure}[h!]
	\centering
	\includegraphics[width=0.6\textwidth]{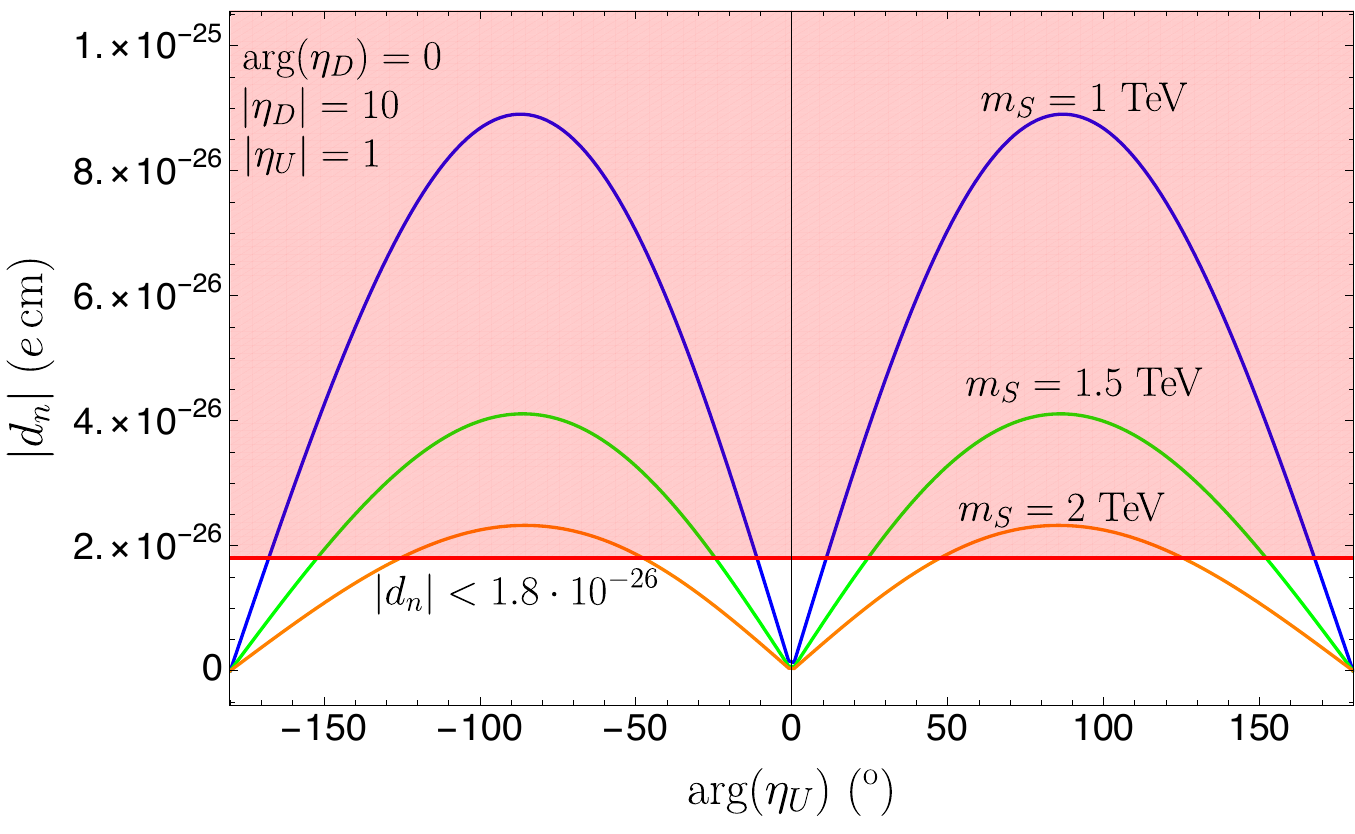}
	\caption{Electric dipole moment of the neutron as a function of the complex phase of $\eta_U$. The shaded region is excluded by the current experimental limit. }
	\label{fig:1D_EDM_alphaU}
\end{figure}

\begin{figure}[h!]
	\centering
	\includegraphics[width=0.65\textwidth]{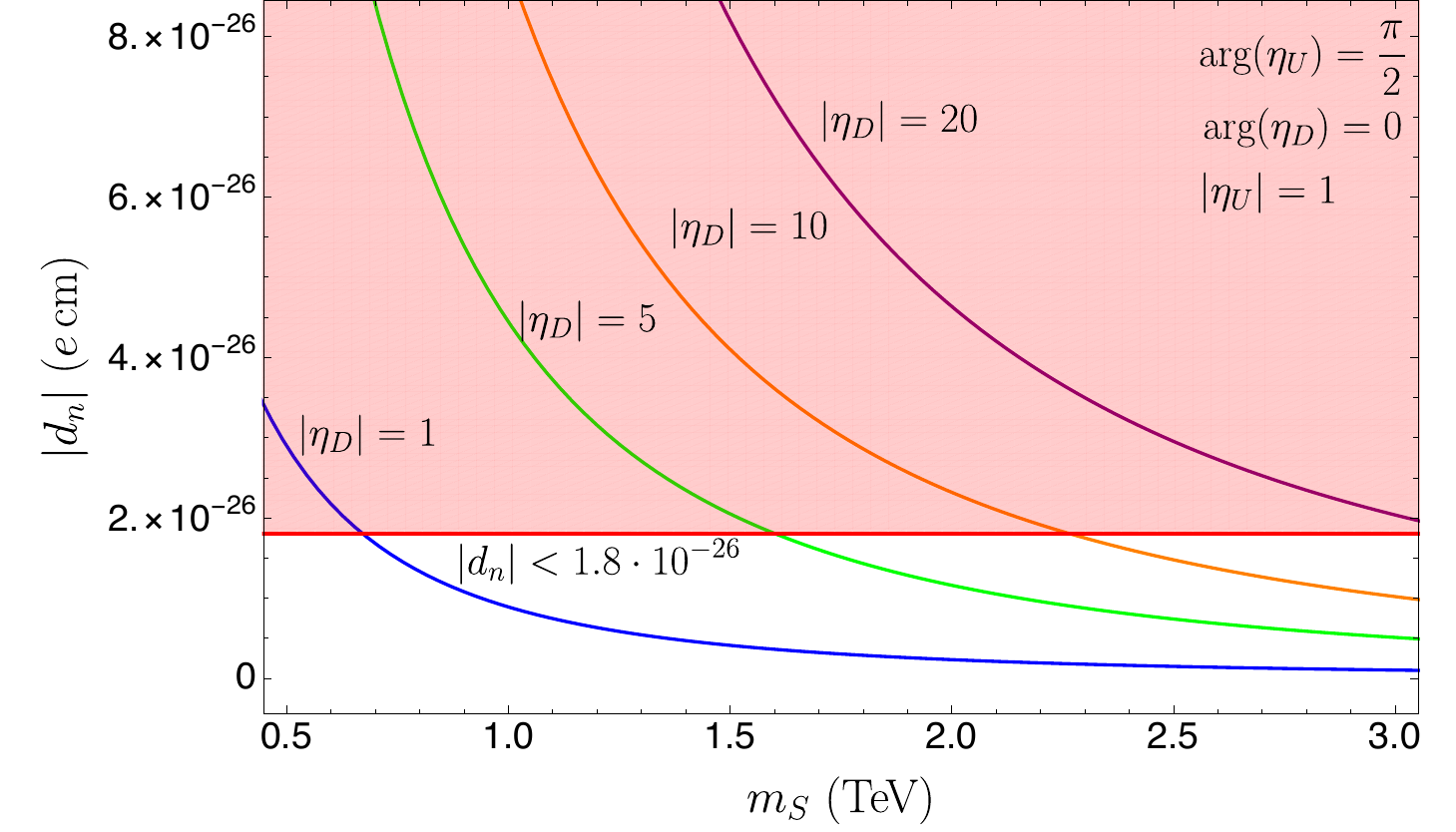}
	\caption{Electric dipole moment of the neutron as a function of the mass of the coloured scalars, $m_S$ (all scalar masses are fixed to the same value). The shaded region is excluded by the current experimental limit.}
	\label{fig:1D_EDM_mS}
\end{figure}

An intuitive way to see the effect of the neutron EDM bound on the MW model is to compare its prediction, as a function of the model parameters, with the current experimental limit. Besides giving clues on the allowed size of the model parameters (studied in more depth in Section~\ref{sec:phenoparams}), this also allows to evaluate the effect of future neutron EDM limits on this model.
In Figure~\ref{fig:1D_EDM_alphaU} we show the size of the neutron EDM as a function of $\text{arg}(\eta_U)$, fixing $|\eta_U|=1$, $|\eta_D|=10$ and $\text{arg}(\eta_D)=0$, for different values of the mass of the coloured scalars. Here we see how a strong constraint for $\text{arg}(\eta_U)$ can be obtained even for masses of the scalars around 1.5 TeV, using the current experimental limits for the neutron EDM. It is also interesting to look at the possible constraints on the scalar masses, obtained by fixing the other parameters. This can be seen in Figure~\ref{fig:1D_EDM_mS} where we have fixed $\text{arg}(\eta_U)=\pi/2$ (which gives the strongest contribution to the neutron EDM), $|\eta_U|=1$ and $\text{arg}(\eta_D)=0$, and we have varied $|\eta_D|$. Here we can see how for reasonable values of $|\eta_D|$, the mass of the scalars could be constrained to be higher than 3 TeV, far beyond the current experimental limit from direct searches.

\subsection{Constraints on the model parameters } \label{sec:phenoparams}

\begin{figure}[h!]
	\centering
	\includegraphics[width=0.45\textwidth]{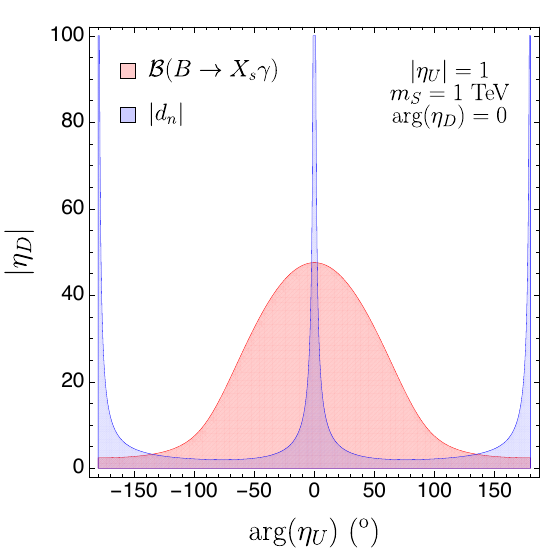}
	\includegraphics[width=0.45\textwidth]{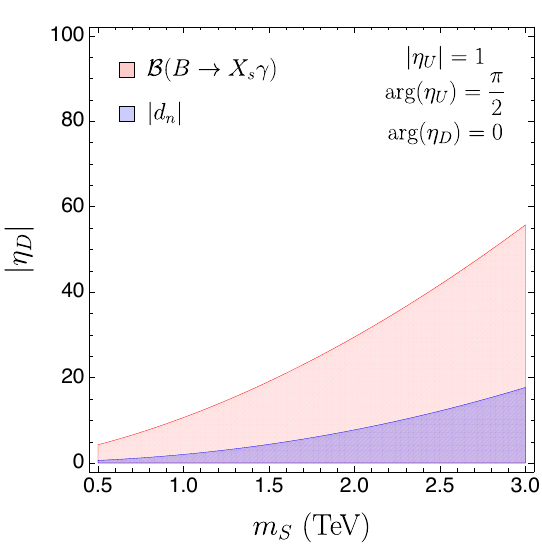}\\
	\includegraphics[width=0.45\textwidth]{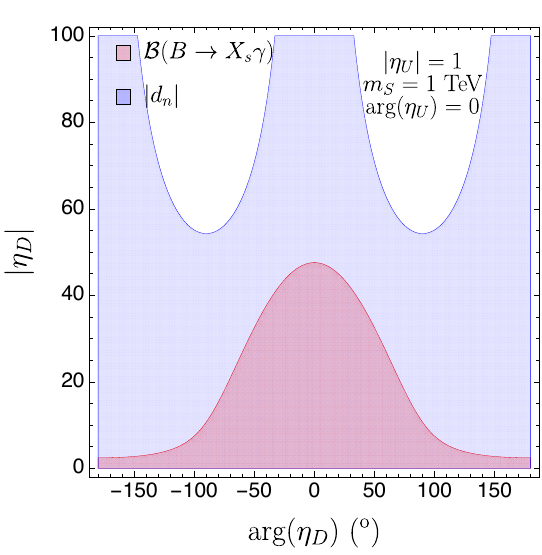}
	\includegraphics[width=0.45\textwidth]{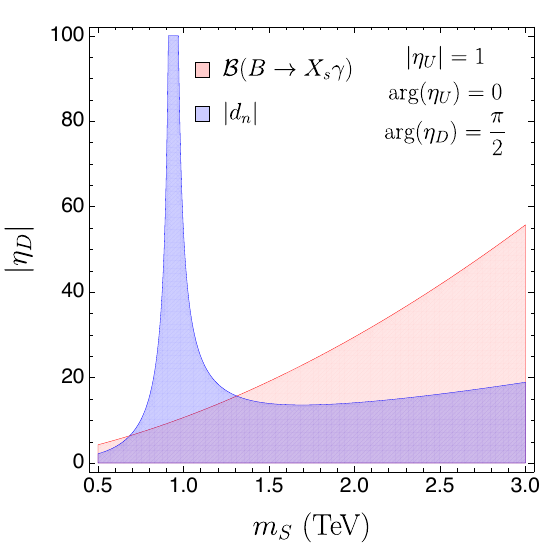}
	\caption{ Constraints on the parameter space of the MW model from the limits on the neutron EDM (blue) compared to those from $\mathcal{B}(B\rightarrow X_s\gamma)$ (red), calculated in Ref.~\cite{Cheng:2015lsa}. The coloured areas represent the allowed regions of parameter space. In the top panels one can see the large impact of the new EDM bounds on some regions of the parameter space. However, its restrictive power is diminished either when the absolute value of the Yukawa couplings is small enough, when the phases are $\arg(\eta_{U,D}) \approx0,\pm\pi$, or in the presence of cancellation effects as in $m_S \approx 1\,\text{TeV}$ (see text for details).
	}
	\label{fig:2D_RegionPlot_EDM}
\end{figure}

To assess the restrictive power of the neutron EDM bounds, we can compare them to the most restrictive observables on the same planes of the parameter space. Following the global-fit analysis of Ref.~\cite{Cheng:2015lsa}, we chose the observable $\mathcal{B}(B\rightarrow X_s\gamma)$ as a benchmark to study the EDM restrictions.
This comparison is done in Figure~\ref{fig:2D_RegionPlot_EDM}, where the region of the parameter space allowed by each observable is shown. In the following we describe these results, pointing out the main patterns in this figure.

We have fixed the phases $\text{arg}(\eta_U)=0$ ($\text{arg}(\eta_D)=0$) in the top (bottom) panels in order to study the effect of CP violation as coming from the down-type (up-type) Yukawa couplings. First, looking at the $|\eta_D|-\text{arg}(\eta_U)$ plane (top left panel) we can see that the constraints from the neutron EDM are stronger than those of $\mathcal{B}(B\rightarrow X_s\gamma)$. The only exceptions lie in the vicinity of the values $\text{arg}(\eta_U) = 0,\pm\,\pi$, where $d_n$ vanishes and it cannot impose any restriction on the model parameters. Fortunately, an excellent experimental precision on $\mathcal{B}(B\rightarrow X_s\gamma)$, which is sensitive to both CP-violating and CP-conserving interactions, allows to restrict those directions even for these limiting values of the phases. This feature shows the power of combining the stringent experimental limits on EDMs with the complementary information from flavour observables.  Nonetheless, as in any interaction beyond the SM, when the absolute value of the new coupling is small enough, restrictions on other model parameters cannot be found. In our case and with the current experimental precision of the neutron EDM, this feature appears as an horizontal band at $|\eta_D|\lesssim1$, spanning along all the domain of $\text{arg}(\eta_U)$ in the top-left panel.

Regarding the $|\eta_D|-m_S$ plane, the constraints from $d_n$ are much stronger than those from $\mathcal{B}(B\rightarrow X_s\gamma)$ when $\text{arg}(\eta_U)=\pi/2$ and $\text{arg}(\eta_D)=0$, in the top right panel of Figure~\ref{fig:2D_RegionPlot_EDM}. However, exchanging the values of the phases, $\text{arg}(\eta_U)=0$ and $\text{arg}(\eta_D)=\pi/2$, an unconstrained direction appears for masses around $m_S\sim 0.9$ TeV (bottom right). 
In this specific region of the parameter space, different contributions to the neutron EDM cancel out. In particular, the light quark (C)EDMs (through Barr-Zee diagrams) and the Weinberg operator (through the threshold contribution proportional to the bottom CEDM, $\tilde{d}_b$) have similar sizes. For $\text{arg}(\eta_U)=\pi/2$ and $\text{arg}(\eta_D)=0$, both contributions interfere constructively resulting in very stringent limits (top right). Conversely, when the values of the phases are switched, $\text{arg}(\eta_U)=0$ and $\text{arg}(\eta_D)=\pi/2$, the Weinberg contribution flips sign, and the interference becomes destructive, preventing any constrain from the neutron EDM. 
To illustrate this dilution of the constraints, we fixed the mass value close to where the destructive interference is produced, $m_S= 1$ TeV, and plotted the allowed regions in the $|\eta_D|-\text{arg}(\eta_D)$ plane (bottom left), keeping $\text{arg}(\eta_U)=0$. 
%



~\\

~\\

\section{Interplay with the CDF $W$-boson mass} 
\label{sec:Wmass}

\begin{flushright}
	This section is based on Ref.~\cite{Gisbert:2022lao}.
\end{flushright}

In April 2022, the CDF collaboration published a new measurement of the $W$-boson mass in tension with the SM prediction~\cite{CDF:2022hxs},
\begin{equation}
m_{W^\pm}^{\rm CDF} = 80.4335 \pm 0.0094\,\gev~,
~m_{W^\pm}^{\rm SM} = 80.357 \pm 0.006\,\gev~.
\end{equation}
This result, published a decade after the last data-taking period in TeVatron, is the most precise single measurement of the $W$ mass. It was achieved thanks to the reduced uncertainties from parton distribution functions and new Monte Carlo techniques.
In principle, the MW model can account for an anomalous value of the $W$ mass via loop corrections. However, accommodating this result has direct consequences on the mass splittings of the new scalars which, as we have seen in Section~\ref{sec:discussionQEDM}, determine the hierarchy of quark EDM contributions. In the following we will briefly explore this connection and outline some consequences of the CDF measurement on the MW model.

	Electroweak global fits taking the new world-average of the $W$ mass as an input find new minima for the Peskin-Takeuchi oblique parameters $S,~T,~U,$~\cite{deBlas:2022hdk,Lu:2022bgw,Asadi:2022xiy}.
	These parameters play an intermediate role between the experimental data and NP phenomenology. Plenty of electroweak precision observables (EWPO) are simultaneously fitted to determine the \textit{experimental values} of $S,~T\text{ and }U$. The results, in turn, are compared to the NP-model prediction without the need to fit all the EWPO directly to the model parameters. 
	The combination of couplings and masses in the construction of these parameters has no contribution from the SM, but their value may deviate from zero in NP models.	And, indeed, 
	with the CDF $W$-mass value, they are in tension with zero by few sigmas.	 Notably, the $U$ parameter gets a larger value than $S$ and $T$, while the opposite is expected from the dimensionality  of the couplings involved.
	For this reason, we consider the case $U=0$ and take the $S$ and $T$ values fitted either with the PDG average of the $W$ mass $(S=0.05\pm0.08,~T=0.09\pm0.07,~\rho_{TS}=0.92)$~\cite{Lu:2022bgw}, or with the new average including the CDF result $(S=0.100\pm0.073 ,~T=0.202\pm0.056,~\rho_{TS}=0.93)$~\cite{deBlas:2022hdk}.

\begin{figure}[t]
		\centering
		\includegraphics[width=0.48\linewidth]{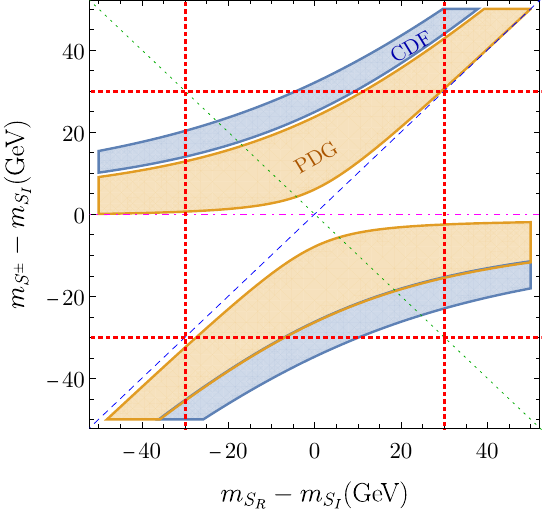}
		\includegraphics[width=0.48\linewidth]{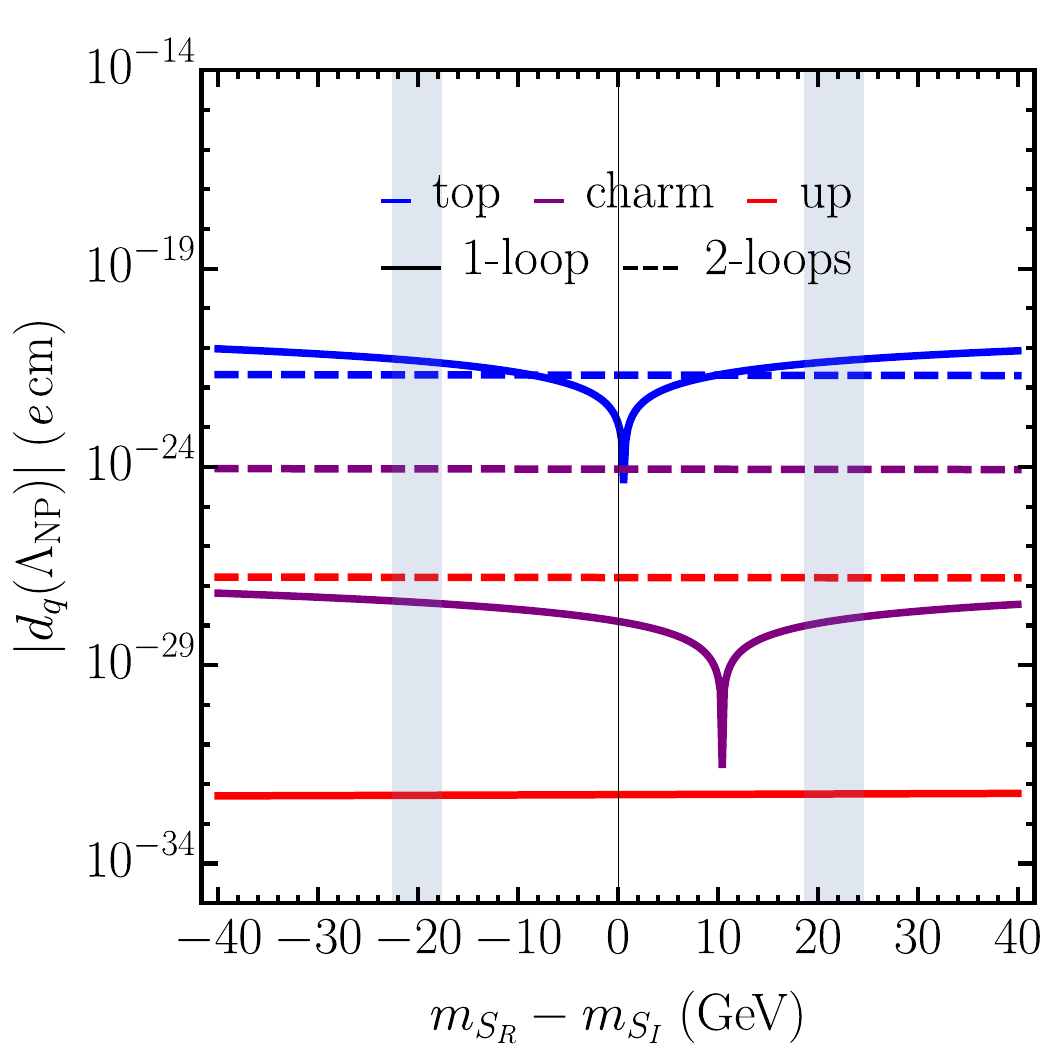}
		\caption{\textbf{Left:} allowed regions at 1$\sigma$ for the mass splittings of the new scalars imposed by the electroweak-fit values of $S$ and $T$ using the PDG average~\cite{Lu:2022bgw} (orange shaded region), and including the CDF result~\cite{deBlas:2022hdk} (blue shaded region). 
			The CDF value of the $W$ mass strongly disfavours (by almost $\sim 6\,\sigma$) a total mass degeneracy at $(0,\,0)$, that is $m_{S^\pm}\approx m_{S_I}\approx m_{S_R}$. Unitarity bounds on the MW model, on the other hand, impose ${|m_{S_i} - m_{S_j}|\lesssim30~\text{GeV}}$ (dotted red lines)~\cite{Eberhardt:2021ebh}. In this plot, $m_{S_I} \sim m_S
			\sim 1~\text{TeV}$. 
			\textbf{Right:}  EDMs of the up-type quarks as a function of the neutral-scalar mass splitting. The gray bands, ${m_{S_R}-m_{S_I}=-20.1\pm 2.5\,\text{GeV}}$ and ${-21.6\pm 3.0\,\text{GeV}}$, represent the preferred values by the CDF $W$ mass measurement when the two mass splittings are assumed to be related as $ (m_{S_R} - m_{S_I}) = \beta \,(m_{S^\pm} - m_{S_I})$ with $\beta=-1$ (left plot, green line). Some other $\beta$ configurations, such as $\beta=0$ and $\beta=1$ (magenta and blue lines, respectively) have no overlap with the blue region, \textit{i.e.} are not compatible with the CDF measurement. In this (right) plot, the Yukawa couplings have been fixed to $|\eta_U| = |\eta_D|= 1$, $ \arg(\eta_U)=\pi/4$, and $ \arg(\eta_D)=0$; and the CP-odd scalar mass to ${m_{S_I} = 1~\text{TeV}}$. 
		}
		\label{fig:Wmass_QEDM}
	\end{figure}

	In the MW model, these parameters are specially sensitive to the mass difference between the new scalar degrees of freedom (with masses $m_{S^\pm}$, $m_{S^0_R}$, and $m_{S^0_I}$), while the parameter $S$ is also sensitive to the non-physical parameter $m_S$, representing the mass of the unbroken scalar doublet (see Eq.~\eqref{eq:massesMW}). 
	To better illustrate the dependency on the scalar mass splittings ${\delta m_{\pm I}\,=\,m_{S^\pm}-m_{S^0_I}}$ and ${\delta m_{RI}\,=\,m_{S^0_R}-m_{S^0_I}}$, we show here the expressions for $T$ and $S$ expanded to leading order on the mass differences,
	\begin{align}\label{eq:ST}
	T \approx -\frac{2\,\delta m_{\pm I}\,\delta m_{RI}}{3\,\pi\,\text{sin}^2\theta_W\,\text{cos}^2\theta_W\,m_{Z}^2}\,,\quad S\approx\frac{2\,m_{S_I}\,(\delta m_{RI}\,-\,2\,\delta m_{\pm I})}{3\,\pi\,m_{S}^2}~,
	\end{align}
	whereas the full expression (in Ref.~\cite{Burgess:2009wm}) is used later in Figure~\ref{fig:Wmass_QEDM}\footnote{A critical reader  could be surprised by the fact that $T$, in Eq.~\eqref{eq:ST}, seems to be independent of the absolute scalar mass. Indirectly, however, the mass splitting is proportional to the parameters of the potential and to the inverse power of the scalar mass.}. 
	Compared to the two-higgs-doublet model, the MW theory has an additional colour factor in the oblique parameters (${\rm dim} = 8$, in the notation of Ref.~\cite{Manohar:2006ga}), increasing their sensitivity to the model parameters.
	In Eq.~\eqref{eq:ST}, the Weinberg angle $\theta_W$ is fixed by the electroweak input parameters $G_F$, $m_{Z}$, and $\alpha_e$~\cite{Brivio:2021yjb}. 
	
	The allowed region of mass splittings is shown in Figure~\ref{fig:Wmass_QEDM} (left), where we see that the case of totally degenerate masses is strongly disfavoured (at $6\sigma$ level) with the CDF measurement of the $W$ mass.
	In addition to the experimental constraints from the oblique parameters, restrictions on the mass splittings of the colour-octet scalars are also obtained by imposing perturbative unitarity and renormalization-group stability \cite{He:2013tla,Cheng:2018mkc,Eberhardt:2021ebh,Cao:2013wqa}. Indeed, using these theoretical constraints the mass splitting is reduced to be smaller than 30 GeV for masses of the scalars of around 1 TeV, as shown in Ref.~\cite{Eberhardt:2021ebh}.
	The combination of the oblique parameters with the theoretical restrictions reduce the parameter space of the theory to the blue regions (oblique parameters) between the red lines (theoretical constraints)\footnote{In an extension of this work, using the \texttt{HEPfit} package to simultaneously include the oblique parameters, theory constraints, and LHC direct searches, we found a much smaller region for the mass splittings~\cite{Miralles:2022jnv}, showing a strong tension between the CDF $W$-boson mass and the unitarity constraints on the model.}. In particular, the mass difference $|m_{S^\pm} - m_{S_R}| \approx  [12,30]~\text{GeV}$, for masses of the coloured scalars of around 1 TeV.

	As shown in Figure~\ref{fig:Wmass_QEDM} (right), taking the benchmark
	$ (m_{S_R} - m_{S_I}) = -(m_{S^\pm} - m_{S_I})$
	inspired by the CDF $W$-mass value,
	the one-loop contribution of the top-quark EDM is greatly enhanced and dominates over the Barr-Zee contribution (see caption of Figure~\ref{fig:Wmass_QEDM}). This enhanced value of the top-quark EDM lies close to its experimental bound, $|d_t| \lesssim 10^{-20}\,e\,\text{cm}$~\cite{Cirigliano:2016njn}, and the phenomenological consequences of this observable together with the $W$ mass should be studied further within this model.

\section{Summary}
\label{sec:summary}

In this chapter we have analysed the relevant contributions to the neutron EDM in the MW model. Expressions for the quark (C)EDM and Weinberg operator have been obtained, which can easily be generalised to other models with colour-octet scalars through the appropriate relations between the coupling constants.
In the case of the Weinberg operator, the neutral scalar contributions turn out to be irrelevant due to the cancellation between CP-odd and CP-even scalars, with the charged scalar contribution being completely dominant for this operator. In turn, only the neutral scalars produce sizable effects in the (C)EDM of light quarks through Barr-Zee type diagrams.

Using the current experimental limits on the neutron EDM, we found new stringent limits on the parameter space of the MW model when the Yukawa CP-violating phases are different from zero. Additionally, in the presence of strong cancellations between the contributions to the neutron EDM, or when the Yukawa phases are zero, we found a valuable complementarity of the neutron EDM with other flavour observables. In future works, the combination of these observables in a global-fit analysis will lead to the most stringent limits on the general CP-violating MW model.

\pagebreak

\renewcommand\chaptername{Appendix}
\appendix

%
%
%


\part*{Appendices} \addcontentsline{toc}{part}{Appendices}

\chapter{Channeling with crystal lenses}

\begin{figure}[H]
	\centering
	\includegraphics[height=0.18\linewidth]{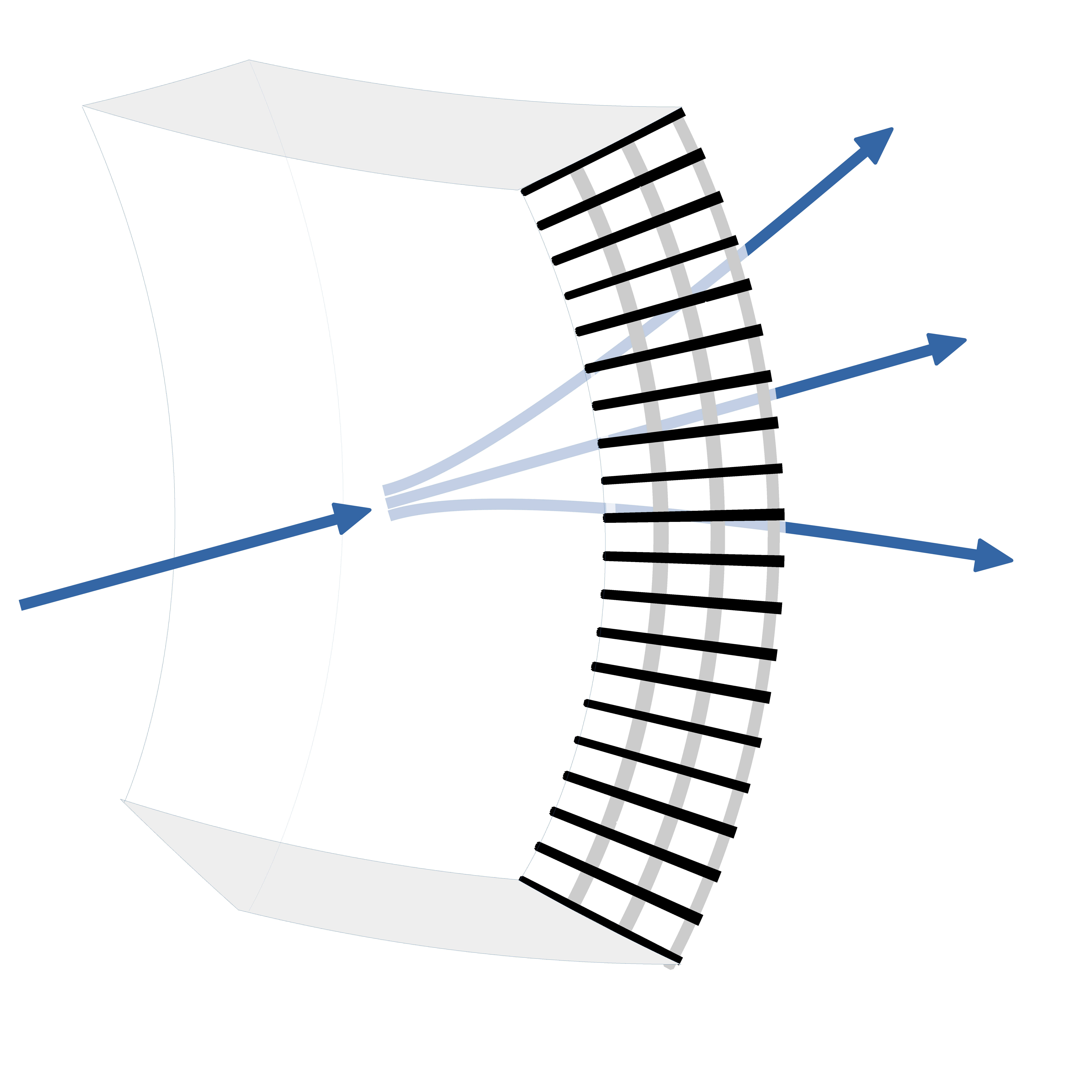}
	\includegraphics[height=0.18\linewidth]{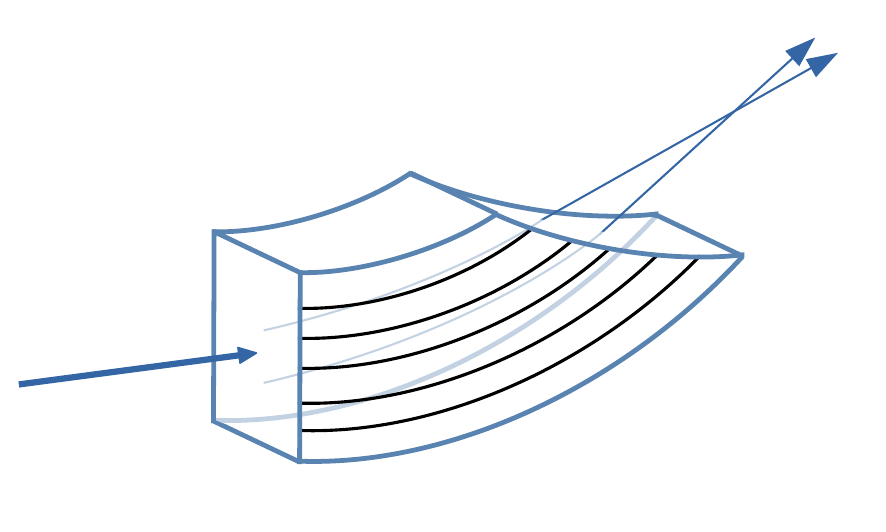}
	\caption{Proposed schemes for (un)focusing parallel beams with crystal lenses.}
	\label{fig:lensmembranev2}
\end{figure}

In the 1970s various techniques were proposed to (un)focus a beam of parallel particles with bent crystals. For instance, by using a bent crystal membrane, as shown in Figure~\ref{fig:lensmembranev2} (left), it should be possible to exploit the deformation of atomic planes in the transverse direction~\cite{Andreev1}\footnote{We have not been able to access the original article, Ref.~\cite{Andreev1}. This is quoted in Ref.~\cite{Biryukov1997}  as Ref.~[112] and in Ref.~\cite{Denisov1} as Ref.~[5].}. 
%
%
%
To our knowledge, the only technique that was successfully tested consists on shaping one of the crystal faces such that the (extended) atomic plane directions are parallel in one end and focused onto a point in the other one, as shown in Figure~\ref{fig:lensmembranev2} (right). These type of crystals were first tested in Ref.~\cite{Denisov1,Denison:1991vf}. With this geometry, the particle beam can be focused and steered outside of the impinging beam at the same time allowing to separate the focused and unfocused beam.

In Section~\ref{sec:optimfocusing} we have presented the possible application of crystal lenses for spin-precession experiments with short-lived particles. 
%
%
In this appendix we derive the geometrical condition for trapping of a particle by a crystal lens. This is a simple trigonometric exercise which result can be used in Monte Carlo simulations of channeling by crystal lenses or in analytical estimates, as discussed in the main text of this thesis. To our knowledge, these formulas are not obtained in previous literature (besides our Ref.~\cite{Biryukov:2021cml}).

\pagebreak

\section{Trapping condition with crystal lenses} \label{app:lensestrapping}

\begin{figure}
	\centering
	\subcaptionbox{}{\includegraphics[width=0.6\columnwidth]{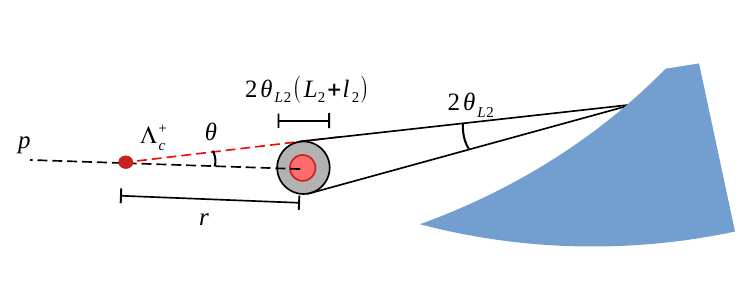}} \\
	\subcaptionbox{}{\includegraphics[width=0.6\columnwidth]{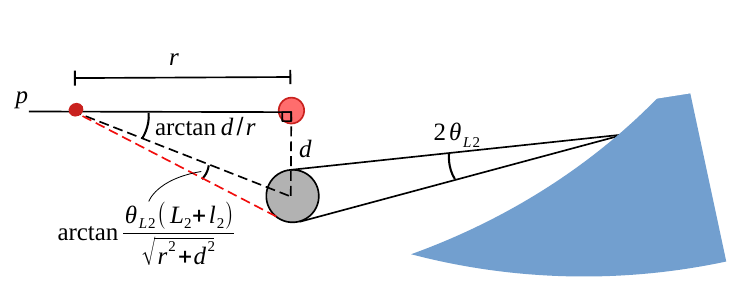}}
	\caption{If we imagine the crystal face to be equidistant to the focal point, \textit{i.e} $l_2 = 0$, then the Lindhard angle projected from every entry point of the crystal defines a circumference (in grey). The extended trajectory of the \Lc particles must overlap this circumference (cylinder in 3D) to be trapped in the crystal. For coinciding focal points (a), the trapping condition depends simultaneously on the \Lc aperture angle, $\theta$, and the distance between the production and the focal point of the first crystal lens, $r$ (both projected onto the bending plane). To define the trapping condition for vertically displaced focal points (b), we assumed the impinging proton to be horizontal, as deviations of the proton trajectory for positive and negative angles approximatley cancel out. This allows to take a right angle and extract the trapping condition with only one additional parameter: the distance $d$ between the two foci. }
	\label{fig:channelingconditions}
\end{figure}

First, we shall consider the case in which the focal points of both crystal lenses are perfectly aligned. This case is shown in Figure~\ref{fig:channelingconditions} (a), and it imposes a condition to trap the \Lc that simultaneously depends on the \Lc angle and its production point. It reads
$$
|r \tan(\theta)| \leq \lindtwo ( \Ltwo + l_2)~,
$$
where $\theta$ is the \Lc angle with respect to the impinging proton, and $r$ is the signed distance between its production and the focal point, with positive values to the right of the focal point, and viceversa.
Taking a small $\theta$ angle approximation and a conservative value of $l_2=0$, we can compare this expression, $|\theta| \leq \lindtwo (\Ltwo/|r|) $, to the trapping condition in the plain-crystal scheme, $\theta \leq \lindtwo$, which is independent of the production point. The gain factor of the double-lens scheme is therefore $\Ltwo/|r|$, which is always greater than one and tends to infinity when $|r|\to0$. This is in perfect agreement with the geometrical view presented in the main text by which all \Lc produced at the focal point are in acceptance. Taking the average value $\langle |r| \rangle = 0.5 \cm$ (for a 2-\cm target), the factor $\Ltwo/\langle |r| \rangle \approx 6$ already provides an order-of-magnitude estimation of the gain with respect to the plain-crystal scheme. This is, nevertheless, only a lower limit to the real gain, which grows rapidly for small $r$.

Since the focal windows $w_{F1}$ and $w_{F2}$ are $\order(1\mum)$ we should consider also the possible misalignment of the two crystal lenses. From Figure \ref{fig:crystalParameters}, one can see that a displacement of the two focal points in the horizontal direction would not have a great effect on the trapping effiency, as the impinging protons and produced particles travel close to the horizontal direction. A displacement in the vertical direction is however much more critical, and it may have a large effect on the trapping efficiency. With the help of Figure \ref{fig:channelingconditions} (b) we can define the trapping condition for two focal points misaligned in the vertical direction by a distance $d$,\footnote{The trapping condition for the plain-crystal scheme, $|\theta|<\theta_{L2}$, can be recovered when the focal length tends to infinity and $r\approx \Ltwo \gg d$.}
\begin{equation}\label{eq:conditionDisplaced}
|\theta - \arctan\frac{d}{r} |\leq 
\arctan \frac{\theta_{L2}(\Ltwo+l_2)}{\sqrt{d^2 + r^2}} ~,
\end{equation}
where $r$ is in this case the signed distance between the proton interaction and the focal point of the first crystal lens. Note that the previous definition of $r$ is just a limiting case for coinciding focal points, as shown in Figs.~\ref{fig:channelingconditions} (a) and (b).

%
%
%
%
%
%
%
%
%
%
%
%
%
%
%

\chapter{Channels for \Lz dipole moments} \label{app:channelsLambda}

\newcommand{\LcLp}{\Lambda_{c}^+\to\Lambda\pi^+}
\newcommand{\LcLppp}{\Lambda_{c}^+\to\Lambda\pi^+\pi^+\pi^-}
\newcommand{\LcLK}{\Lambda_{c}^+\to\Lambda K^+}
\newcommand{\LcLKK}{\Lambda_{c}^+\to\Lambda K^+ \bar{K^0}}

\newcommand{\XicplusLKp}{\Xi_{c}^+\to\Lambda \bar{K^0} \pi^+}
\newcommand{\XicplusLKpp}{\Xi_{c}^+\to\Lambda K^- \pi^+\pi^+}

\newcommand{\XiczeroLK}{\Xi_{c}^0\to\Lambda K_{S}^0}
\newcommand{\XiczeroLKp}{\Xi_{c}^0\to\Lambda K^- \pi^+}
\newcommand{\XiczeroLKK}{\Xi_{c}^0\to\Lambda K^+ K^-}
\newcommand{\XiczeroLphi}{\Xi_{c}^0\to\Lambda \phi(K^+ K^-)}

\newcommand{\LbLJpsi}{\Lambda_{b}^0\to\Lambda J /\psi}
\newcommand{\LbLmm}{\Lambda_{b}^0\to\Lambda\mu^+\mu^-}

\newcommand{\XizeroLg}{\Xi^0\to\Lambda \gamma(e^+ e^-)}
\newcommand{\XiLp}{\Xi^-\to\Lambda \pi^-}
\newcommand{\OLK}{\Omega^-\to\Lambda K^-}

\newcommand{\XibLcK}{\Xi_{b}^0\to\Lambda_{c}^+ K^-}
\newcommand{\LbLcp}{\Lambda_{b}^0\to\Lambda_{c}^+ \pi^-}
\newcommand{\LbLcK}{\Lambda_{b}^0\to\Lambda_{c}^+ K^-}
\newcommand{\LbLcppp}{\Lambda_{b}^0\to\Lambda_{c}^+ \pi^+\pi^-\pi^-}

\newcommand{\ObOJpsi}{\Omega_{b}^-\to\Omega^- J /\psi}
\newcommand{\OcOp}{\Omega_{c}^0\to\Omega^- \pi^+}
\newcommand{\OcOppp}{\Omega_{c}^0\to\Omega^- \pi^+\pi^+\pi^-}
\newcommand{\OcOK}{\Omega_{c}^0\to\Omega^- K^+}
\newcommand{\OcOKp}{\Omega_{c}^0\to\Omega^- K^+ \pi^+}

\newcommand{\OcXizeroKp}{\Omega_{c}^0\to\Xi^0 K^- \pi+}
\newcommand{\XicXizerop}{\Xi_{c}^+\to\Xi^0 \pi+}
\newcommand{\XicXizeroppp}{\Xi_{c}^+\to\Xi^0 \pi^+\pi^+\pi^-}
\newcommand{\LcXizeroK}{\Lambda_{c}^+\to\Xi^0 K^+}
\newcommand{\OXizerop}{\Omega^-\to\Xi^0 \pi^-}

\newcommand{\XibXiJpsi}{\Xi_{b}^-\to\Xi^- J /\psi}
\newcommand{\OcXiKpp}{\Omega_{c}^0\to\Xi^- K^-\pi^+\pi^+}

\newcommand{\XiczeroXizerop}{\Xi_{c}^0\to\Xi^- \pi^+} 					
\newcommand{\XiczeroXizeroppp}{\Xi_{c}^0\to\Xi^- \pi^+\pi^+\pi^-}		
\newcommand{\XiczeroXizeroK}{\Xi_{c}^0\to\Xi^- K^+}						

\newcommand{\XiczeroXip}{\Xi_{c}^0\to\Xi^- \pi^+}						
\newcommand{\XiczeroXippp}{\Xi_{c}^0\to\Xi^- \pi^+\pi^+\pi^-}			
\newcommand{\XiczeroXiK}{\Xi_{c}^0\to\Xi^- K^+}							

\newcommand{\XicXipp}{\Xi_{c}^+\to\Xi^- \pi^+\pi^+}
\newcommand{\LcXiKp}{\Lambda_{c}^+\to\Xi^- K^+\pi^+}
\newcommand{\OXipp}{\Omega^-\to\Xi^- \pi^+\pi^-}

\newcommand{\Jpsimm}{J/\psi \to \mu^+ \mu^-}
\newcommand{\Kspp}{K_S \to \pi^+ \pi^-}


All possible \Lz production channels from weak decays are systematically listed with the criteria presented in Section~\ref{sec:lambdaedm}. Only the most abundant modes are included in the main text (Table~\ref{tab:LambdaChannels}). The complete list of decays can be found in Tables~\ref{tab:AllChShort} and \ref{tab:AllChLong}, where the number of produced \Lz particles are computed for 5\,\invfb of data with the branching ratios quoted in the 2016 version of the PDG booklet~\cite{PDG}, and the cross sections and fragmentation fractions as described in the main text.

\begin{table}[H] 
	\centering
	\caption{Production channels of $\Lz$ particles with only short-lived particle ancestors, noted as SL in the main text. The yields marked with an asterisk (*) are computed imposing the sum of all measured branching rations equal to 100\%. } \label{tab:AllChShort}  
	\renewcommand{\arraystretch}{1.1}
	\begin{tabular}{llcr}
		\hline \hline
		\multicolumn{2}{c}{ Channel } &  $N_{\Lambda}$  & \\ [0ex] \cline{1-4}
		&  &   & \\  [-2ex]
		$\XiczeroLKp$ &  & 3.70$\times 10^{11}$ & * \\ 
		$\LcLppp$ &  & 1.49 $ \times 10^{11}$ &  \\ 
		$\XicplusLKp$ &  & 1.01 $ \times 10^{11}$ & * \\ 
		$\XicplusLKpp$ &  & 9.47 $ \times 10^{10}$ & * \\ 
		$\LcLp$ &  & 6.14 $ \times 10^{10}$ &  \\ 
		$\XiczeroLK$ &  & 2.51 $ \times 10^{10}$ & * \\ 
		$\XiczeroLKK$ &  & 1.00 $ \times 10^{10}$ & * \\ 
		$\LcLKK$ &  & 9.33 $ \times 10^{9}$ &  \\ 
		$\XiczeroLphi$ &  & 5.75 $ \times 10^{9}$ & * \\ 
		$\LcLK$ &  & 2.87 $ \times 10^{9}$ &  \\ 
		$\LbLcppp$ & $\LcLppp$ & 4.69 $ \times 10^{7}$ &  \\ 
		$\LbLcp$ & $\LcLppp$ & 3.34 $ \times 10^{7}$ &  \\ 
		$\LbLcppp$ & $\LcLp$ & 1.93 $ \times 10^{7}$ &  \\ 
		$\LbLcp$ & $\LcLp$ & 1.37 $ \times 10^{7}$ &  \\ 
		$\LbLJpsi$ &  & 1.10 $ \times 10^{7}$ &  \\ 
		$\LbLcppp$ & $\LcLKK$ & 2.93 $ \times 10^{6}$ &  \\ 
		$\LbLcK$ & $\LcLppp$ & 2.46 $ \times 10^{6}$ &  \\ 
		$\LbLcp$ & $\LcLKK$ & 2.09 $ \times 10^{6}$ &  \\ 
		$\LbLcK$ & $\LcLp$ & 1.01 $ \times 10^{6}$ &  \\ 
		$\LbLcppp$ & $\LcLK$ & 9.02 $ \times 10^{5}$ &  \\ 
		$\LbLcp$ & $\LcLK$ & 6.42 $ \times 10^{5}$ &  \\ 
		$\LbLmm$ &  & 2.43 $ \times 10^{5}$ &  \\ 
		$\LbLcK$ & $\LcLKK$ & 1.54 $ \times 10^{5}$ &  \\ 
		$\XibLcK$ & $\LcLppp$ & 6.65 $ \times 10^{4}$ &  \\ 
		$\LbLcK$ & $\LcLK$ & 4.73 $ \times 10^{4}$ &  \\ 
		$\XibLcK$ & $\LcLp$ & 2.73 $ \times 10^{4}$ &  \\ 
		$\XibLcK$ & $\LcLKK$ & 4.16 $ \times 10^{3}$ &  \\ 
		$\XibLcK$ & $\LcLK$ & 1.28 $ \times 10^{3}$ &  \\ 
		\hline \hline
	\end{tabular}
\end{table}

\begin{table}[H]
	\centering
	\caption{ Same as Table~\ref{tab:AllChShort} but for LL modes, \ie with long-living ($c\tau >1\,\cm$) ancestors of the \Lz particle. In the last rows, the channels with a prompt-produced strange baryon are included factorizing their production rate in $pp$ collisions.
	}  \label{tab:AllChLong}
	\renewcommand{\arraystretch}{1.1}
	\begin{tabular}{llcl}
		\hline \hline
		\multicolumn{2}{c}{ Decays } &  $N_{\Lambda}$  & \\ [0ex] \cline{1-4}
		&  &   & \\  [-2ex]
		%
		%
		$\XiczeroXippp$ & $\XiLp$ & 1.14 $ \times 10^{12}$ & * \\ 
		$\XiczeroXip$ & $\XiLp$ & 3.45 $ \times 10^{11}$ & * \\ 
		$\XicXipp$ & $\XiLp$ & 2.93 $ \times 10^{11}$ & * \\ 
		$\LcXiKp$ & $\XiLp$ & 2.92 $ \times 10^{10}$ &  \\ 
		$\XiczeroXiK$ & $\XiLp$ & 9.67 $ \times 10^{9}$ & * \\ 
		$\XicXizeroppp$ & $\XizeroLg$ & 3.79 $ \times 10^{6}$ & * \\ 
		$\XibXiJpsi$ & $\XiLp$ & 1.94 $ \times 10^{6}$ &  \\ 
		$\XicXizerop$ & $\XizeroLg$ & 1.23 $ \times 10^{6}$ & * \\ 
		$\ObOJpsi$ & $\OLK$ & 3.74 $ \times 10^{5}$ &  \\ 
		$\LcXizeroK$ & $\XizeroLg$ & 1.70 $ \times 10^{5}$ &  \\ 
		\hline
		$\XiLp$ &  & 6.38 $ \times 10^{9}$ & \scriptsize{$\times  \, \sigma_{pp\to\Xi^-}\left[\mu b\right]$ }~ \\ 
		$\OLK$ &  & 4.33 $ \times 10^{9}$ & \scriptsize{$\times  \, \sigma_{pp\to\Omega^-}\left[\mu b\right]$ }~ \\ 
		$\XizeroLg$ &  & 4.86 $ \times 10^{4}$ & \scriptsize{$\times  \, \sigma_{pp\to{\Xi^0}}\left[\mu b\right]$ }~ \\ 
		\hline
		$\OXipp$ & $\XiLp$ & 2.36 $ \times 10^{6}$ & \scriptsize{$\times  \, \sigma_{pp\to\Omega^-}\left[\mu b\right]$ } \\ 
		$\OXizerop$ & $\XizeroLg$ & 1.15 $ \times 10^{4}$ & \scriptsize{$\times  \, \sigma_{pp\to\Omega^-}\left[\mu b\right]$ } \\ 
		\hline \hline
	\end{tabular}
\end{table}


\chapter{LHCb analysis: glossary of definitions} \label{app:analysisterms}

In this appendix, definitions for commonly-used terms in Part II are provided. These are very short definitions, considering the context of this type of data analyses, without pretensions to be exhaustive or general. Some common terms are also defined in the main text, but are repeated here for completeness.

\section{Miscellaneous} \label{app:analysistermsmisc}

\newcommand{\term}[2]{\subsubsection{#1}   #2 }

\newcommand{\var}[2]{\subsubsection{#1}   #2 }

\term{Combinatorial background}{Background of the invariant-mass distribution associated to random combinations of tracks which pass the selection criteria. They are distributed smoothly in the mass and are commonly parametrized with an exponential or polynomial function.}

\term{Physical background}{Background of the invariant-mass distribution with peaking structures. It is associated to the (partial) reconstruction of other decays, that nevertheless meet all the selection criteria.}

\term{sPlot and sWeights}{Given an initial dataset mixed with signal and background, the sPlot technique~\cite{Pivk:2004ty} extracts a weight for each event so that the \textit{sWeighted} dataset resembles a pure sample of signal. While events under the peak have weights larger than one, sideband events receive negative weights, effectively subtracting the background distributions from the initial dataset. It can be used to compare the signal data and MC distributions. To have better control of the systematic uncertainties it is sometimes recommended to avoid it in the fitting procedure (in our case, the angular fit). However, if the mass distribution has many peaking backgrounds, this method has advantages over more complex fitting PDFs.}

\term{Stripping}{It is the process of classifying recorded collision events according to the selection criteria defined by analysts in the stripping lines. This process is run centrally on the same data every few years during the (re)stripping campaigns. On MC samples, the stripping can be rerun at any time if the simulation was not filtered, and if sufficient information is saved on the initial data file. For example, it is always possible to run a different stripping selection on \texttt{(L)DST} files and it may or may not be possible on \texttt{MDST} files (see definition below).}

\term{Trigger}{The trigger system is in charge of deciding whether or not the information of each collision event should be recorded for permanent storage and offline analysis. Roughly speaking, at LHCb the L0 trigger level identifies hard collision events; \hltone puts requirements on basic reconstructed objects like charged tracks, photons, electrons, etc.; and \hlttwo studies precisely the combinations of these objects. At \hlttwo there are dedicated lines, defined for specific analyses, and generic lines that \eg identify some general decay topologies. A tiny portion of events is always recorded regardless of the trigger configuration in the \textit{minimum bias} samples. In MC samples, the trigger decisions of all lines in the three trigger stages are simulated and saved. }

\section{Types of Monte Carlo samples} \label{app:MCtypes}

\term{Generator level}{The kinematics of the $pp$ collisions and resulting particles are simulated by \pythia~\cite{Sjostrand:2007gs}. When the mother particle of the interesting decay (for us \Lc) is produced, \evtgen~\cite{Lange:2001uf} takes over and makes it decay into a specified final state. A decay model (including decay asymmetries or intermediate resonances if needed) can be also configured in \evtgen. Commonly, the information of the particles forming the signal decay is saved along with global event information. But every simulated particle (of the rest of the $pp$ collision) is accessible in principle. }

\term{Reconstructed}{From the generator level MC, the signal of each particle in each of the subdetectors is simulated. With the detector responses, the full reconstruction algorithm is run on simulated data, and the reconstructed objects and their combinations are saved.}

\term{Flagged}{If the reconstructed signal candidate meets the criteria of some specified stripping lines the event is flagged. The information on the signal candidate is saved to a separate location within the \texttt{DST}/\texttt{LDST}/\texttt{MDST} file for faster access.  }

\term{Filtered}{Only the candidate events passing the stripping selection are saved. Filtered simulations are designed to save disk storage space in large MC productions.}

\term{ReDecay}{Many particles are simulated and reconstructed for each $pp$ collision, usually with only one signal candidate among them. The fast simulation technique \textsc{ReDecay}~\cite{redecay} re-uses most of these reconstructed objects in several events. Only the detector responses produced by the signal particles are simulated again with a newly generated decay of the mother particle (into the same final state).  }

\term{Phase space}{At the generator level, the kinematics of the decay products is simulated according to a flat distribution in the available phase space of the decay, \ie without decay asymmetries, polarization, or intermediate (interfering) resonances. All of these possibilities for decay models are not listed here. Phase space (reconstructed) MC is commonly used even for highly complex amplitude analyses.}

\term{File formats}{In \texttt{(L)DST} files the complete event information is saved, also with \textit{raw} data. It is possible to rerun the full-event reconstruction. In \texttt{MDST} files only the \textit{hits} (in the subdetectors) and reconstructed objects associated to the signal candidate are saved, also with some configurable list of global event information. Generator-level MC produced with the \textsc{Gauss} simulation framework~\cite{Clemencic_2011} has the file format \texttt{xgen}.} 

\term{\textsc{ROOT} ntuples}{ All the mentioned file formats can be converted into \textsc{ROOT} ntuples for "offline" analysis with the \textsc{DaVinci} framework. Usually, only a small fraction of the information in \texttt{(L)DST} or \texttt{MDST} files is saved in the final ntuples. Which of the information is to be saved can be configured in the \textit{DaVinci option files}. }

\term{Combinations of the above}{These types of MC simulations are not exclusive. In fact, most of them can be combined, \eg both flagged and filtered are by definition reconstructed samples which may be saved in any file format, and have been generated with \textsc{ReDecay} using a phase space model at the generator level. }

\section{Multivariate classifiers} \label{app:BDTterms}

\term{Working phases}{ 
	\begin{itemize}
		\item Training: the classifier \textit{learns} to discriminate any given event as signal or background. It does it essentially by trial and error. In each attempt (in our case, each decision tree), the classifier takes a mixed sample of events, each labeled as signal or background, applies different cuts, and saves the resulting percentage of signal/background events. At the end of the training phase, all the gathered information is combined producing a (proxy to the) probability of any given event to be signal or background (the BDT response).
		
		\item Test: Statistically independent events, not used in the training phase, are used to test the performance of the trained classifier. Each of these events must also be labeled as signal or background.
		\item Application: A mixed sample of (non-labelled) signal and background events is given to the classifier to be separated. In our case, this is done by applying a cut on the BDT response. The optimal separation is determined with the significance of the resulting sample or other figure of merit.
	\end{itemize}
}

\term{Area under the curve (AUC)}{ The best performing classifier is the one that rejects as most as possible background and as little as possible signal. This can be visualized the plot of the signal efficiency against background rejection. The area under this curve (AUC) is a common measure of the classifier performance. }

\term{Labeled signal/background samples}{To train the classifier, it needs two pure samples of signal and background events. In our analysis, the pure signal sample is just the simulated signal MC and the pure background sample is extracted from the side bands of the \Lc invariant-mass distribution. We used both upper and lower side band. It is important that the classifier does not have the full kinematic information to reconstruct the mass of the particle. In that case, it will manage to discriminate signal and background events exclusively based on the mass value.}

\section{Variables} \label{app:analysisvariables}

\var{Bachelor pions}{ In our analysis it refers to the three charged pions in the \threepi decay, \eg \texttt{Lc\_maxBachDOCA} is the maximum DOCA (see definition below) computed for every combination of the three charged pion tracks. The word \textit{bachelor} is commonly used to refer to single tracks associated to a decay vertex (not our case, with three charged tracks).}

\var{DOCA}{Distance of closest approach between two tracks or particle trajectories. It is useful to put requirements on the common origin vertex of different particles. DOCA is always defined for pairs of particles. Vertices with three or more particles can use the maximum or minimum of the different combinations.}

\var{Impact parameter (IP)}{ Distance between the extended particle trajectory (line) and a given vertex (point). It is used to identify particles (in our analysis, the \textit{bachelor} pions) that were not produced in the primary vertex.   } 

\var{DIRA angle}{Applies to unstable particles. It refers to the angle between the momentum vector (as the combination of daughter-particles momenta) and the flight direction (as defined by the production and decay vertices).}


\var{$\chi^2$ variables ($\chi^2_{\rm IP}$, $\chi^2_{\rm FD}$, ...)}{ Can be interpreted as the value of the variable divided by the per-event estimated uncertainty, squared. For instance, $\chi^2_{\rm IP}$ is the difference in the vertex-fit $\chi^2$ of the PV candidate reconstructed with and without the considered particle, thus a particle with $\chi^2_{\rm IP} = 9$ is separated from the PV by 3$\sigma$.}

\var{Lifetime}{Time that an unstable particle lived before decaying. Calculated taking into account its relativistic boost.}

\var{Flight distance}{Applies to unstable particles. It refers to the distance between the origin and end (decay) vertex of the particle (\texttt{FD\_ORIVX}). Instead of the origin vertex possibly within a decay chain, it is also defined with respect to the PV that has a smaller IP for the particle trajectory (\texttt{FD\_OWNPV}).}

\var{DTF $\chi^2$}{Gives the $\chi^2$ fit quality of the decay chain refit performed by Decay Tree Fitter~\cite{Hulsbergen:2005pu}. It provides some general information on the reconstruction quality of the signal candidates. Configurations with mass constraints (in our analysis noted \eg "DTF [$\Lambda$,~$\Lambda_c^+$]") show a strong dependence of $\chi^2_{\rm DTF}$ with the originally reconstructed mass.}

\var{Vertex $\chi^2$}{ Quality of the vertex fit. Accounts for all associated DOCA and uncertainties on the track directions.}

\var{PID: \texttt{x\_ProbNNy}}{Probability of particle \texttt{x} to be of type \texttt{y}. Applies to charged tracks (\texttt{y} =[$\pip, \Kp, p, e^+, \mu^+$]) and is mostly based on the information collected by the RICH detectors combined with the reconstructed track momentum.}

\var{PID: DLL (x-y)}{Difference between the log-likelihood of a charged track to be of type x and type y. By default it is computed with y = $\pip$, but any comparison of particle species can be obtained by subtracting the different DLL variables (canceling out the pion log-likelihood). }

\var{Ghost probability}{Probability for a reconstructed track not to correspond to a physical charged particle. Ghost tracks are reconstructed from the combination of random (aligned) hits. These hits (or clusters) may be caused by different physical particles of the event or by persisting electronic noise from previous collision events (spillover). The ghost track probability includes information on the track fit quality and missing hits (in the trajectory of the track). }

\var{Multiplicity / number of tracks}{Total number of reconstructed charged tracks in the event. It is also computed separately for every track category (\eg downstream tracks, T tracks, etc.).}

\chapter{LHCb analysis: tables and plots} \label{app:analysistables}

Supplementary material to the LHCb analysis in Part II including tables and large lists of plots that are not essential to the main discussion are presented in this appendix.

\section{Structure of stripping lines} \label{app:strippedstripping}

In Table~\ref{tab:containers}, the schematic structure of the stripping line  \texttt{HcV03H\_Lambdac2Lambda3Pi[LL,DD]} is shown in its version previous to the optimization described in Section~\ref{sec:stripping}\footnote{Find the source code of this line, for the stripping s28r1p1 version, in the \href{https://gitlab.cern.ch/lhcb/Stripping/-/blob/Stripping-s28r1p1/Phys/StrippingSelections/python/StrippingSelections/StrippingCharm/StrippingHc2V03H.py}{GitLab of LHCb}.} to illustrate the different levels of particle containers, as discussed in the main text.

%
%

\pagebreak

\begin{table}
	\caption{Structure of the stripping line before our optimization, described in Section~\ref{sec:stripping}. The particle containers include several cuts. The symbols are defined as: } \label{tab:containers}
\end{table}

\vspace*{-1.6cm}

\begin{center}
	{\color{black}\Large \Rightscissors}~~cut ~~~~~
	\firstcont~container ~~~~
	\firstfun function
\end{center}


{\small
	\begin{minipage}[t]{0.5\linewidth}
	
	\textbf{Main selection}
	
	~
	
	\firstitem \Lc cuts
	\begin{itemize}
		\item combCuts
		\begin{itemize}
			\itcut ADAMASS
			\itcut ADOCA(1,2), (1,3), (1,4)
		\end{itemize}
		\item lambdacCuts
		\begin{itemize}
			\itcut VCHI2/VDOF ~~ (vertex $\chi^2$)
			\itcut BPVVDCHI2 ~~ (dist. to PV $\chi^2$)
			\itcut BPVDIRA
			\itcut ADMASS
			\itcut CHILD(VZ) - VZ ~~ ($\Lambda^0$ FD)
			
		\end{itemize}
		\item comb12/comb123Cuts
		\begin{itemize}
			\itcut DAMASS 
			\itcut ADOCA(1,2), (1,3)
		\end{itemize}
		\itfun DaVinci\_\_N4BodyDecays()
	\end{itemize}

	\firstitem \lz cuts
	\begin{itemize}
		\item lambda0Cuts
		\begin{itemize}
			\itcut P, PT
			\itcut BPVVDCHI2 ~~ (FD $\chi^2$)
			\itcut VCHI2/VDOF ~~ (vertex $\chi^2$)
		\end{itemize}
		\item LambdaList
		\begin{itemize}
			\itcut PROBNNp
			\itcont StdLooseLambda
		\end{itemize}
		
	\end{itemize}

	\firstitem \textit{Bachelor} $\pi^{\pm}$ cuts
	\begin{itemize}
		\itcut P
		\itcont StdNoPIDsPions
	\end{itemize}

	\vfill\null
	
\end{minipage}
\begin{minipage}[t]{0.5\linewidth}

\textbf{Containers - first level}

~

	\firstcont StdLooseLambda
	\begin{itemize}
		\itcut p($\pi$) , p($p$)
		\itcut MIPCHI2DV(PRIMARY)
		\itcut ADAMASS , ADOCACHI2CUT 
		\itcut ADMASS , CHI2VX
		\item LL
		\begin{itemize}
			\itcont StdLoosePions
			\itcont StdLooseProtons 
		\end{itemize}
		\item DD
		\begin{itemize}
			\itcont StdNoPIDsDownPions
			\itcont StdNoPIDsDownProtons 
		\end{itemize}
	\end{itemize}

	\firstcont StdNoPIDsPions 
	\begin{itemize}
		\itfun defaultCuts()
		\itcont StdAllNoPIDsPions
	\end{itemize}

~

\textbf{ Containers - deeper levels}

~

\firstcont StdLoosePions
\begin{itemize}
	\itfun defaultCuts()
	\itcont StdAllLoosePions
\end{itemize}

\firstcont StdLooseProtons
\begin{itemize}
	\itfun defaultCuts()
	\itcont StdAllLooseProtons
\end{itemize}

	\vfill\null
	
\end{minipage}
\begin{minipage}[t]{0.5\linewidth}

	\firstcont StdNoPIDsDownPions
	\begin{itemize}
		\itcut Track Chi2Cut
		\itfun NoPIDsParticleMaker()
		\itfun trackSelector()
	\end{itemize}

	\firstcont StdNoPIDsDownProtons
	\begin{itemize}
		\itcut Track Chi2Cut
		\itfun NoPIDsParticleMaker()
		\itfun trackSelector()
	\end{itemize}
	
	\firstcont StdAllNoPIDsPions
	\begin{itemize}
		\itfun NoPIDsParticleMaker()
		\itfun trackSelector()
	\end{itemize}

	\firstcont StdAllLoosePions
\begin{itemize}
	\itfun CombinedParticleMaker()
	\itfun ProtoParticleCALOFilter
	\itfun trackSelector()
\end{itemize}
~\\
	
	\vfill\null
	
\end{minipage}
\begin{minipage}[t]{0.5\linewidth}

	\firstcont StdAllLooseProtons
	\begin{itemize}
		\itcut CombDLL(p-pi)
		\itfun CombinedParticleMaker()
		\itfun ProtoParticleCALOFilter
		\itfun trackSelector()
	\end{itemize}

	\firstfun defaultCuts()
	\begin{itemize}
		\itcut PT $>$ 250 MeV
		\itcut MIPCHI2DV(PRIMARY)  $>$ 4
	\end{itemize}
	
	\vfill 
	
\end{minipage}
}

\section{Stripping cuts}

The final set of cuts of the optimized stripping selection is presented in Tables \ref{tab:stripLL} and \ref{tab:stripDD}. In these tables, also the variables included in the dedicated-trigger selections (present in 2018 data) are listed for reference,
as well as the variables used in the previous particle containers, quoting the new values of the cuts (if any).

\begin{table}[H]
	\centering
\caption{Final set of stripping cuts for \lLL line. Relative eficiencies for real data (2016 - str28r1p1 - \texttt{HcLamXLines} - no trigger) and MC (MC-Matching - loose containers). Number of events: Data 23862, MC 36700.}
\label{tab:stripLL}
	{\footnotesize
	\begin{tabular}{|l|c|c|c|} 
	\hline
	Variable      &           Cut  &    $\varepsilon$(Data) [\%]  &      $\varepsilon$(MC) [\%]  \\
	\hline
	$\Lambda_{c}^{+}$ bachelors min p     &    $>$   2000 &   100.0  &    95.4  \\
	p ($\Lambda$) proton probability (NN)     &    $>$    0.1 &   100.0  &    96.3  \\
	$\Lambda$ p     &    $>$  10000 &    99.6  &    99.3  \\
	$\Lambda$ $p_{T}$     &    $>$    500 &   100.0  &    95.5  \\
	$\Lambda$ Flight distance ($\chi^{2}$)     &     -  &   100.0  &    99.9  \\
	$\Lambda$ Flight distance  ($\chi^{2}$) from PV     &    $>$     49 &    99.2  &    98.9  \\
	$\Lambda$ Flight distance     &    $>$     25 &   100.0  &    98.0  \\
	$\Lambda$ decay vertex $\chi^{2}$/ndf     &    $<$     12 &   100.0  &    96.9  \\
	$\Lambda_{c}^{+}$ ADAMASS     &    $<$     90 &    98.5  &    99.5  \\
	$\Lambda_{c}^{+}$ ADMASS     &    $<$     75 &    83.2  &    99.3  \\
	$\Lambda_{c}^{+}$ vertex max DOCA [$\Lambda$, 3$\pi$]     &    $<$    0.5 &    96.6  &    88.9  \\
	$\Lambda_{c}^{+}$ decay vertex $\chi^{2}$/ndf     &    $<$      3 &    60.8  &    90.3  \\
	$\Lambda_{c}^{+}$ DIRA angle (mrad)     &    $<$    140 &   100.0  &    75.9  \\
	$\Lambda_{c}^{+}$ flight distance ($\chi^{2}$), BPVVDCHI2      &    $>$     16 &   100.0  &    24.3  \\\hline  
	\verb|StdLooseLambdaLL| and therein (new cut values) & & &  \\
	p ($\Lambda$) IP ($\chi^{2}$)     &    $>$      9 &   100.0  &   100.0  \\
	$\pi^{-}$ ($\Lambda$) IP ($\chi^{2}$)     &    $>$      9 &   100.0  &   100.0  \\
	p ($\Lambda$) p      &    $>$   2000 &   100.0  &   100.0  \\
	$\pi^{-}$ ($\Lambda$) p      &    $>$   2000 &   100.0  &   100.0  \\
	$\Lambda$ vertex DOCA ($\chi^{2}$) [p, $\pi^{-}$]     &    $<$     30 &   100.0  &   100.0  \\
	$\Lambda$ ADAMASS     &    $<$     50 &   100.0  &   100.0  \\
	$\Lambda$ ADMASS     &    $<$     35 &   100.0  &   100.0  \\
	p ($\Lambda$) $p_{T}$     &     -  &   100.0  &   100.0  \\
	$\pi^{-}$ ($\Lambda$) $p_{T}$     &     -  &   100.0  &   100.0  \\
	p ($\Lambda$) track $\chi^{2}$/ndf      &     -  &   100.0  &   100.0  \\
	$\pi^{-}$ ($\Lambda$) track $\chi^{2}$/ndf      &     -  &   100.0  &   100.0  \\\hline  
	\verb|StdNoPIDsPions| and therein (new cut values) & & & \\
	$\Lambda_{c}^{+}$ bachelors min $p_{T}$     &    $>$    150 &   100.0  &    81.8  \\
	$\Lambda_{c}^{+}$ bachelors min IP ($\chi^{2}$)      &    $>$      1 &   100.0  &    33.6  \\
	$\Lambda_{c}^{+}$ bachelors max IP ($\chi^{2}$)      &    $>$      9 &    95.5  &    32.2  \\
	$\Lambda_{c}^{+}$ log IP ($\chi^{2}$)     &    $<$      5 &    69.6  &    98.9  \\
	$\Lambda_{c}^{+}$ bachelors max DOCA [$\pi^{+}$1, $\pi^{+}$2, $\pi^{-}$]     &     -  &   100.0  &   100.0  \\
	$\Lambda_{c}^{+}$ bachelors max DOCA ($\chi^{2}$) [ $\pi^{+}$1, $\pi^{+}$2, $\pi^{-}$]     &     -  &   100.0  &   100.0  \\\hline  
	\textit{Only in trigger} & & & \\
	$\Lambda_{c}^{+}$ lifetime, BPVLTIME (ns)     &     -  &   100.0  &   100.0  \\
	$\Lambda_{c}^{+}$ vertex max DOCA ($\chi^{2}$) [$\Lambda$, 3$\pi$]     &     -  &   100.0  &   100.0  \\
	$\pi^{-}$ ($\Lambda_{c}^{+}$) DLL K-pi      &     -  &   100.0  &   100.0  \\
	$\pi^{+}$ 1 ($\Lambda_{c}^{+}$) DLL K-pi      &     -  &   100.0  &   100.0  \\
	$\pi^{+}$ 2 ($\Lambda_{c}^{+}$) DLL K-pi     &     -  &   100.0  &   100.0  \\
	$\Lambda_{c}^{+}$ bachelors max track $\chi^{2}$/ndf      &     -  &   100.0  &   100.0  \\
	$\pi^{-}$ ($\Lambda_{c}^{+}$) track GhostProb      &     -  &   100.0  &   100.0  \\
	$\pi^{+}$ 1 ($\Lambda_{c}^{+}$) track GhostProb     &     -  &   100.0  &   100.0  \\
	$\pi^{+}$ 2 ($\Lambda_{c}^{+}$) track GhostProb     &     -  &   100.0  &   100.0  \\
	$\Lambda$ lifetime, BPVLTIME (ns)     &     -  &   100.0  &   100.0  \\
	\hline 
	All cuts     &                &   29.5  &   8.3  \\
	\hline 
\end{tabular} }
\end{table}

\begin{table}[H]
	\centering
	\caption{Final set of stripping cuts for \lDD line. Relative eficiencies for real data (2016 - str28r1p1 - \texttt{HcLamXLines} - no trigger) and MC (MC-Matching - loose containers). Number of events: Data 23862, MC 36700.}
	\label{tab:stripDD}
		{\footnotesize
\begin{tabular}{|l|c|c|c|}
	\hline
	Variable      &           Cut  &    $\varepsilon$(Data) [\%] &      $\varepsilon$(MC) [\%] \\
	\hline
	$\Lambda_{c}^{+}$ bachelors min p     &    $>$   2000 &   100.0  &    95.4  \\
	p ($\Lambda$) proton probability (NN)     &    $>$      0 &   100.0  &   100.0  \\
	$\Lambda$ p     &    $>$  10000 &    99.3  &    99.7  \\
	$\Lambda$ $p_{T}$     &    $>$    500 &    95.3  &    98.8  \\
	$\Lambda$ Flight distance ($\chi^{2}$)     &     -  &    99.8  &    99.8  \\
	$\Lambda$ Flight distance  ($\chi^{2}$) from PV     &    $>$     49 &    99.4  &    94.6  \\
	$\Lambda$ Flight distance     &    $>$    200 &    98.1  &    99.8  \\
	$\Lambda$ decay vertex $\chi^{2}$/ndf     &    $<$     12 &   100.0  &    98.8  \\
	$\Lambda_{c}^{+}$ ADAMASS     &    $<$     90 &    98.6  &    99.4  \\
	$\Lambda_{c}^{+}$ ADMASS     &    $<$     75 &    83.0  &    99.3  \\
	$\Lambda_{c}^{+}$ vertex max DOCA [$\Lambda$, 3$\pi$]     &    $<$    0.5 &    95.4  &    30.7  \\
	$\Lambda_{c}^{+}$ decay vertex $\chi^{2}$/ndf     &    $<$      3 &    78.0  &    90.5  \\
	$\Lambda_{c}^{+}$ DIRA angle (mrad)     &    $<$     45 &   100.0  &    60.5  \\
	$\Lambda_{c}^{+}$ flight distance ($\chi^{2}$), BPVVDCHI2      &    $>$      9 &   100.0  &    37.0  \\\hline  
	\verb|StdLooseLambdaDD| and therein (new cuts) & & &  \\
	p ($\Lambda$) IP ($\chi^{2}$)     &    $>$      4 &   100.0  &   100.0  \\
	$\pi^{-}$ ($\Lambda$) IP ($\chi^{2}$)     &    $>$      4 &   100.0  &   100.0  \\
	p ($\Lambda$) p      &    $>$   2000 &   100.0  &   100.0  \\
	$\pi^{-}$ ($\Lambda$) p      &    $>$   2000 &   100.0  &   100.0  \\
	$\Lambda$ vertex DOCA ($\chi^{2}$) [p, $\pi^{-}$]     &    $<$     25 &   100.0  &   100.0  \\
	$\Lambda$ ADAMASS     &    $<$     80 &   100.0  &   100.0  \\
	$\Lambda$ ADMASS     &    $<$     64 &   100.0  &   100.0  \\
	p ($\Lambda$) $p_{T}$     &     -  &   100.0  &   100.0  \\
	$\pi^{-}$ ($\Lambda$) $p_{T}$     &     -  &   100.0  &   100.0  \\
	p ($\Lambda$) track $\chi^{2}$/ndf      &    $<$     10 &   100.0  &   100.0  \\
	$\pi^{-}$ ($\Lambda$) track $\chi^{2}$/ndf      &    $<$     10 &   100.0  &   100.0  \\\hline  
	\verb|StdNoPIDDownPions| (new cuts) & & & \\
	$\Lambda_{c}^{+}$ bachelors min $p_{T}$     &    $>$    150 &   100.0  &    85.2  \\
	$\Lambda_{c}^{+}$ bachelors min IP ($\chi^{2}$)      &    $>$      1 &   100.0  &    36.2  \\
	$\Lambda_{c}^{+}$ bachelors max IP ($\chi^{2}$)      &    $>$      9 &    94.0  &    34.8  \\
	$\Lambda_{c}^{+}$ log IP ($\chi^{2}$)     &    $<$      5 &    77.0  &    98.9  \\
	$\Lambda_{c}^{+}$ bachelors max DOCA [$\pi^{+}$1, $\pi^{+}$2, $\pi^{-}$]     &     -  &   100.0  &   100.0  \\
	$\Lambda_{c}^{+}$ bachelors max DOCA ($\chi^{2}$) [ $\pi^{+}$1, $\pi^{+}$2, $\pi^{-}$]     &     -  &   100.0  &   100.0  \\\hline  
	\textit{Only in trigger} & & & \\
	$\Lambda_{c}^{+}$ lifetime, BPVLTIME (ns)     &     -  &   100.0  &   100.0  \\
	$\Lambda_{c}^{+}$ vertex max DOCA ($\chi^{2}$) [$\Lambda$, 3$\pi$]     &     -  &   100.0  &   100.0  \\
	$\pi^{-}$ ($\Lambda_{c}^{+}$) DLL K-pi      &     -  &   100.0  &   100.0  \\
	$\pi^{+}$ 1 ($\Lambda_{c}^{+}$) DLL K-pi      &     -  &   100.0  &   100.0  \\
	$\pi^{+}$ 2 ($\Lambda_{c}^{+}$) DLL K-pi     &     -  &   100.0  &   100.0  \\
	$\Lambda_{c}^{+}$ bachelors max track $\chi^{2}$/ndf      &     -  &   100.0  &   100.0  \\
	$\pi^{-}$ ($\Lambda_{c}^{+}$) track GhostProb      &     -  &   100.0  &   100.0  \\
	$\pi^{+}$ 1 ($\Lambda_{c}^{+}$) track GhostProb     &     -  &   100.0  &   100.0  \\
	$\pi^{+}$ 2 ($\Lambda_{c}^{+}$) track GhostProb     &     -  &   100.0  &   100.0  \\
	$\Lambda$ lifetime, BPVLTIME (ns)     &     -  &   100.0  &   100.0  \\
	\hline 
	All cuts     &                &   40.7  &   4.6  \\
	\hline 
\end{tabular} 
}
\end{table}

\section{Trigger efficiencies in MC}

The efficiencies of different \elzero, \hltone and \hlttwo lines are evaluated on MC. This sample only contains the initial stripping selection. Each trigger stage is separated in two tables, with the trigger lines fired by the tracks of the \threepi decay, \ie \textit{on signal} (TOS), or \textit{independent of signal} (TIS).

\subsubsection{L0 trigger}
%
%
\begin{table}[H]
	\centering
\caption{Number of events fired by each L0 line on the DD sample.} 
\begin{tabular}{lrrr} 
\hline \hline 
   L0 line    &   times fired   &  overlapped evs & added stats [\%]\\
\hline
                                                Lc\_L0Global\_TOS    &  39906 &  35878  &  11.23  \\
                                        Lc\_L0Hadron \_TOS    &  35878 &  35878  &   0.00  \\
                                      Lc\_L0Electron \_TOS    &   6119 &   3491  &   7.32  \\
                                          Lc\_L0Muon \_TOS    &   1435 &    472  &   2.68  \\
                                        Lc\_L0Photon \_TOS    &   1228 &    781  &   1.25  \\
                                        Lc\_L0MuonEW \_TOS    &    261 &     77  &   0.51  \\
                                         Lc\_L0JetEl \_TOS    &    195 &    177  &   0.05  \\
                                        Lc\_L0DiMuon \_TOS    &     88 &     35  &   0.15  \\
                                         Lc\_L0JetPh \_TOS    &     62 &     59  &   0.01  \\
\hline \hline 
\end{tabular} 
\label{tab:trigMCL0TOS}
\end{table}
\begin{table}[H]
	\centering
	\caption{Number of events fired by each L0 line on the DD sample.} 
	\label{tab:trigMCL0TIS}
\begin{tabular}{lrrr} 
\hline \hline 
   L0 line    &   times fired   &  overlapped evs & added stats [\%]\\
\hline
                                                Lc\_L0Global\_TIS    &  56015 &  31123  &  79.98  \\
                                        Lc\_L0Hadron \_TIS    &  31123 &  31123  &   0.00  \\
                                      Lc\_L0Electron \_TIS    &  20466 &   8845  &  37.34  \\
                                          Lc\_L0Muon \_TIS    &  13569 &   3510  &  32.32  \\
                                        Lc\_L0Photon \_TIS    &   8872 &   3880  &  16.04  \\
                                        Lc\_L0MuonEW \_TIS    &   3458 &    892  &   8.24  \\
                                        Lc\_L0DiMuon \_TIS    &   2914 &    762  &   6.91  \\
                                         Lc\_L0JetEl \_TIS    &    817 &    739  &   0.25  \\
                                         Lc\_L0JetPh \_TIS    &    313 &    277  &   0.12  \\
\hline \hline 
\end{tabular} 
\end{table}

\subsubsection{HLT1 trigger}
\begin{table}[H]
\centering
\caption{Number of events fired by each Hlt1 line on the DD sample.} 
\begin{tabular}{lrrr}
\hline \hline 
   Hlt1 line    &   times fired   &  overlapped evs & added stats [\%]\\
\hline
                                                Lc\_Hlt1Phys\_TOS    &  77308 &  36808  &  110.03  \\
                                 Lc\_Hlt1TwoTrackMVA \_TOS    &  36808 &  36808  &   0.00  \\
                                    Lc\_Hlt1TrackMVA \_TOS    &  28533 &  21354  &  19.50  \\
                                      Lc\_Hlt1IncPhi \_TOS    &   5276 &   4316  &   2.61  \\
                                   Lc\_Hlt1TrackMuon \_TOS    &    604 &    418  &   0.51  \\
                               Lc\_Hlt1DiMuonLowMass \_TOS    &     83 &     37  &   0.12  \\
                                  Lc\_Hlt1DiMuonNoL0 \_TOS    &     50 &     11  &   0.11  \\
                          Lc\_Hlt1SingleElectronNoIP \_TOS    &     19 &     16  &   0.01  \\
                             Lc\_Hlt1CalibTrackingKK \_TOS    &     11 &      7  &   0.01  \\
                            Lc\_Hlt1SingleMuonHighPT \_TOS    &     10 &      7  &   0.01  \\
                            Lc\_Hlt1B2PhiGamma\_LTUNB \_TOS    &      5 &      4  &   0.00  \\
                                Lc\_Hlt1B2GammaGamma \_TOS    &      2 &      2  &   0.00  \\
                                    Lc\_Hlt1DiProton \_TOS    &      2 &      2  &   0.00  \\
                                  Lc\_Hlt1DiMuonNoIP \_TOS    &      2 &      2  &   0.00  \\
\hline \hline 
 \label{tab:trigMCHLT1TOS}
\end{tabular} 
\end{table}

\begin{table}[H]
\centering
 \caption{Number of events fired by each Hlt1 line on the DD sample.} 
\begin{tabular}{lrrr}
\hline \hline 
   Hlt1 line    &   times fired   &  overlapped evs & added stats [\%]\\
\hline
                                                Lc\_Hlt1Phys\_TIS    &  77868 &  16767  &  364.41  \\
                                    Lc\_Hlt1TrackMVA \_TIS    &  25397 &  13223  &  72.61  \\
                                 Lc\_Hlt1TwoTrackMVA \_TIS    &  16767 &  16767  &   0.00  \\
                                   Lc\_Hlt1TrackMuon \_TIS    &   3912 &   1983  &  11.50  \\
                             Lc\_Hlt1CalibTrackingKK \_TIS    &   1399 &    959  &   2.62  \\
                                      Lc\_Hlt1IncPhi \_TIS    &   1023 &    870  &   0.91  \\
                           Lc\_Hlt1CalibTrackingPiPi \_TIS    &    762 &    498  &   1.57  \\
                                    Lc\_Hlt1DiProton \_TIS    &    620 &    287  &   1.99  \\
                            Lc\_Hlt1SingleMuonHighPT \_TIS    &    494 &    260  &   1.40  \\
                              Lc\_Hlt1B2HH\_LTUNB\_KPi \_TIS    &    468 &    305  &   0.97  \\
                               Lc\_Hlt1DiMuonLowMass \_TIS    &    451 &    286  &   0.98  \\
                              Lc\_Hlt1DiMuonHighMass \_TIS    &    420 &    206  &   1.28  \\
                               Lc\_Hlt1B2HH\_LTUNB\_KK \_TIS    &    414 &    278  &   0.81  \\
                             Lc\_Hlt1B2HH\_LTUNB\_PiPi \_TIS    &    358 &    246  &   0.67  \\
                          Lc\_Hlt1SingleElectronNoIP \_TIS    &    262 &    104  &   0.94  \\
                                  Lc\_Hlt1DiMuonNoL0 \_TIS    &    229 &    145  &   0.50  \\
\hline \hline 
\label{tab:trigMCHLT1TIS}
\end{tabular} 
\end{table}

\subsubsection{HLT2 trigger}

{\scriptsize

\begin{longtable}{lrrr} 
\caption{Number of events fired by each Hlt2 line on the DD sample.} \\
\hline \hline 
   Hlt2 line    &   times fired   &  ovrlp. evs. & added st. [\%]\\
\hline
                                                Lc\_Hlt2Phys\_TOS    &  23792 &   4953  &  380.36  \\
         Lc\_Hlt2CharmHadDsp2KS0PimPipPip\_KS0DDTurbo \_TOS    &  13225 &   1635  &  234.00  \\
          Lc\_Hlt2CharmHadDp2KS0PimPipPip\_KS0DDTurbo \_TOS    &  10871 &   1151  &  196.24  \\
                                   Lc\_Hlt2Topo3Body \_TOS    &   4953 &   4953  &   0.00  \\
                                   Lc\_Hlt2Topo2Body \_TOS    &   4503 &   1820  &  54.17  \\
                Lc\_Hlt2CharmHadInclSigc2PiLc2HHXBDT \_TOS    &   3400 &    450  &  59.56  \\
                 Lc\_Hlt2CharmHadInclDst2PiD02HHXBDT \_TOS    &   2473 &    376  &  42.34  \\
                                   Lc\_Hlt2Topo4Body \_TOS    &   1600 &   1478  &   2.46  \\
                 Lc\_Hlt2CharmHadXim2LamPim\_DDLTurbo \_TOS    &    785 &    105  &  13.73  \\
                   Lc\_Hlt2CharmHadDp2KS0KS0PipTurbo \_TOS    &    506 &     68  &   8.84  \\
                  Lc\_Hlt2CharmHadDsp2KS0KS0PipTurbo \_TOS    &    424 &     36  &   7.83  \\
              Lc\_Hlt2CharmHadDp2EtapKp\_Etap2PimPipG \_TOS    &    214 &     58  &   3.15  \\
          Lc\_Hlt2CharmHadDsp2KS0KpPimPip\_KS0DDTurbo \_TOS    &    133 &     16  &   2.36  \\
\hline \hline 
\label{tab:trigMCHLT2TOS}
\end{longtable} 
	
\begin{longtable}{lrrr} 
\caption{Number of events fired by each Hlt2 line on the DD sample.} \\
\hline \hline 
   Hlt2 line    &   times fired   &  ovrlp. evs. & added st. [\%]\\
\hline
                                                Lc\_Hlt2Phys\_TIS    &  49260 &   2760  &  1684.78  \\
                                   Lc\_Hlt2Topo2Body \_TIS    &   3421 &   1579  &  66.74  \\
                                   Lc\_Hlt2Topo3Body \_TIS    &   2760 &   2760  &   0.00  \\
                                   Lc\_Hlt2Topo4Body \_TIS    &   1185 &   1007  &   6.45  \\
                                 Lc\_Hlt2TopoMu2Body \_TIS    &   1095 &    378  &  25.98  \\
                Lc\_Hlt2CharmHadInclSigc2PiLc2HHXBDT \_TIS    &   1044 &    279  &  27.72  \\
                                  Lc\_Hlt2TopoE2Body \_TIS    &    923 &    359  &  20.43  \\
                 Lc\_Hlt2CharmHadInclDst2PiD02HHXBDT \_TIS    &    886 &    221  &  24.09  \\
                                  Lc\_Hlt2TopoE3Body \_TIS    &    796 &    506  &  10.51  \\
                                 Lc\_Hlt2TopoMu3Body \_TIS    &    635 &    402  &   8.44  \\
                 Lc\_Hlt2CharmHadXim2LamPim\_DDDTurbo \_TIS    &    382 &     15  &  13.30  \\
                                  Lc\_Hlt2TopoE4Body \_TIS    &    357 &    275  &   2.97  \\
                        Lc\_Hlt2RadiativeIncHHHGamma \_TIS    &    346 &    162  &   6.67  \\
                         Lc\_Hlt2RadiativeIncHHGamma \_TIS    &    335 &     86  &   9.02  \\
                       Lc\_Hlt2CharmHadD02KmPipTurbo \_TIS    &    267 &     78  &   6.85  \\
                                 Lc\_Hlt2TopoMu4Body \_TIS    &    223 &    174  &   1.78  \\
                  Lc\_Hlt2CharmHadDspToKpPimPipTurbo \_TIS    &    197 &     99  &   3.55  \\
                   Lc\_Hlt2CharmHadDpToKpPimPipTurbo \_TIS    &    174 &     76  &   3.55  \\
                           Lc\_Hlt2XcMuXForTauB2XcMu \_TIS    &    170 &     58  &   4.06  \\
                   Lc\_Hlt2CharmHadDpToKmPipPipTurbo \_TIS    &    167 &     52  &   4.17  \\
                         Lc\_Hlt2DiMuonDetachedHeavy \_TIS    &    160 &     32  &   4.64  \\
                  Lc\_Hlt2CharmHadDpToPimPipPipTurbo \_TIS    &    158 &     95  &   2.28  \\
                 Lc\_Hlt2CharmHadDspToPimPipPipTurbo \_TIS    &    157 &     90  &   2.43  \\
                              Lc\_Hlt2SingleMuonRare \_TIS    &    155 &     37  &   4.28  \\
                             Lc\_Hlt2DiMuonJPsiHighPT \_TIS    &    134 &     19  &   4.17  \\
                                   Lc\_Hlt2PhiIncPhi \_TIS    &    134 &     35  &   3.59  \\
                            Lc\_Hlt2SingleMuonHighPT \_TIS    &    133 &     18  &   4.17  \\
     Lc\_Hlt2CharmHadDstp2D0Pip\_D02PimPimPipPipTurbo \_TIS    &    127 &     30  &   3.51  \\
                                Lc\_Hlt2DiMuonBTurbo \_TIS    &    123 &     16  &   3.88  \\
                                     Lc\_Hlt2DiMuonB \_TIS    &    123 &     16  &   3.88  \\
                   Lc\_Hlt2CharmHadDspToKmKpPipTurbo \_TIS    &    114 &     37  &   2.79  \\
                        Lc\_Hlt2CharmHadDspToKmKpPip \_TIS    &    114 &     37  &   2.79  \\
                                Lc\_Hlt2TopoMuE3Body \_TIS    &    113 &     70  &   1.56  \\
      Lc\_Hlt2CharmHadDstp2D0Pip\_D02KmPimPipPipTurbo \_TIS    &    111 &     28  &   3.01  \\
                                Lc\_Hlt2TopoMuE2Body \_TIS    &    110 &     44  &   2.39  \\
                  Lc\_Hlt2CharmHadDp2EtaKp\_Eta2EmEpG \_TIS    &    110 &     36  &   2.68  \\
                 Lc\_Hlt2CharmHadXim2LamPim\_LLLTurbo \_TIS    &    104 &     21  &   3.01  \\
      Lc\_Hlt2CharmHadDstp2D0Pip\_D02KpPimPimPipTurbo \_TIS    &    102 &     21  &   2.93  \\
                               Lc\_Hlt2TopoMuMu2Body \_TIS    &     96 &     39  &   2.07  \\
              Lc\_Hlt2CharmHadDp2EtapKp\_Etap2PimPipG \_TIS    &     89 &     26  &   2.28  \\
                 Lc\_Hlt2CharmHadDp2EtaPip\_Eta2EmEpG \_TIS    &     87 &     27  &   2.17  \\
       Lc\_Hlt2CharmHadDp2EtapKp\_Etap2EtaPimPip\_EtaR \_TIS    &     84 &     35  &   1.78  \\
                  Lc\_Hlt2CharmHadLcpToPpPimPipTurbo \_TIS    &     82 &     35  &   1.70  \\
                               Lc\_Hlt2TopoMuMu3Body \_TIS    &     80 &     59  &   0.76  \\
                                 Lc\_Hlt2TopoEE3Body \_TIS    &     79 &     55  &   0.87  \\
                  Lc\_Hlt2CharmHadDspToKmPipPipTurbo \_TIS    &     78 &     37  &   1.49  \\
                 Lc\_Hlt2CharmHadXim2LamPim\_DDLTurbo \_TIS    &     74 &      7  &   2.43  \\
                    Lc\_Hlt2CharmHadDpToKmKpPipTurbo \_TIS    &     71 &     27  &   1.59  \\
         Lc\_Hlt2CharmHadDp2EtaKp\_Eta2PimPipPi0\_Pi0R \_TIS    &     70 &     33  &   1.34  \\
         Lc\_Hlt2CharmHadDsp2KS0PimPipPip\_KS0LLTurbo \_TIS    &     65 &     30  &   1.27  \\
      Lc\_Hlt2CharmHadDp2EtapPip\_Etap2EtaPimPip\_EtaR \_TIS    &     64 &     27  &   1.34  \\
                  Lc\_Hlt2CharmHadDsp2KS0KS0PipTurbo \_TIS    &     64 &     17  &   1.70  \\
                                 Lc\_Hlt2ExoticaRHNu \_TIS    &     63 &     15  &   1.74  \\
                   Lc\_Hlt2CharmHadDp2KS0KS0PipTurbo \_TIS    &     62 &     17  &   1.63  \\
             Lc\_Hlt2CharmHadDstp2D0Pip\_D02GG\_G2EmEp \_TIS    &     62 &      6  &   2.03  \\
     Lc\_Hlt2CharmHadDstp2D0Pip\_D02PimPip\_LTUNBTurbo \_TIS    &     61 &     13  &   1.74  \\
\hline \hline 
\label{tab:trigMCHLT2TIS}
\end{longtable} 
	
}

\section{Purity goal for sWeights} \label{sec:puritysweight}

\onlyANA{\red{L0logIPChi2 does is missing 'log' in the the label}}

\begin{figure}[h!]
	\centering
	\includegraphics[width=0.4\linewidth]{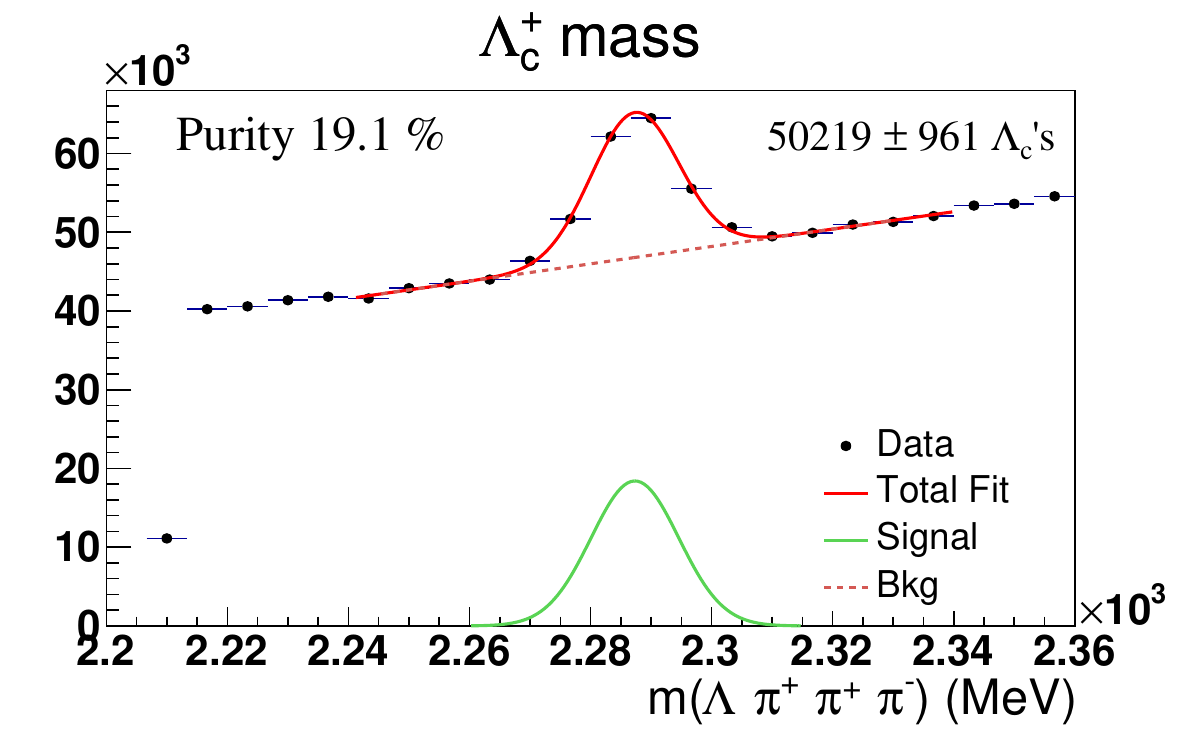}
	\includegraphics[width=0.4\linewidth]{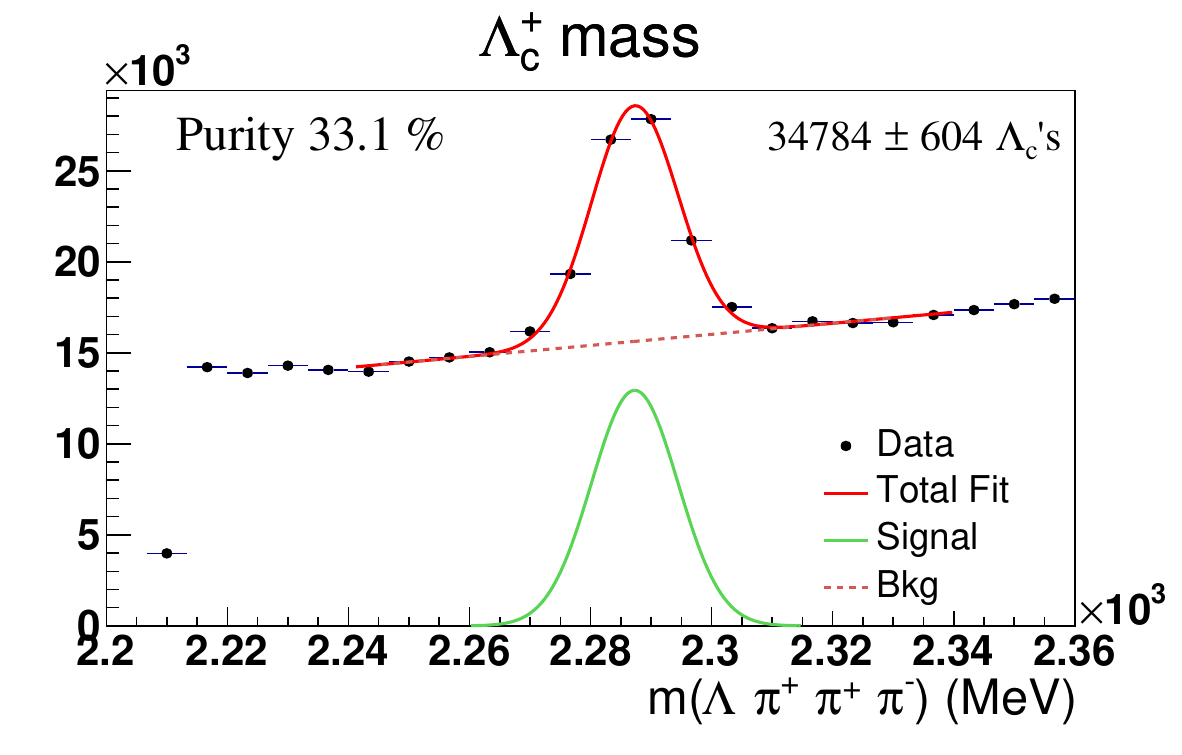}
	\includegraphics[width=0.4\linewidth]{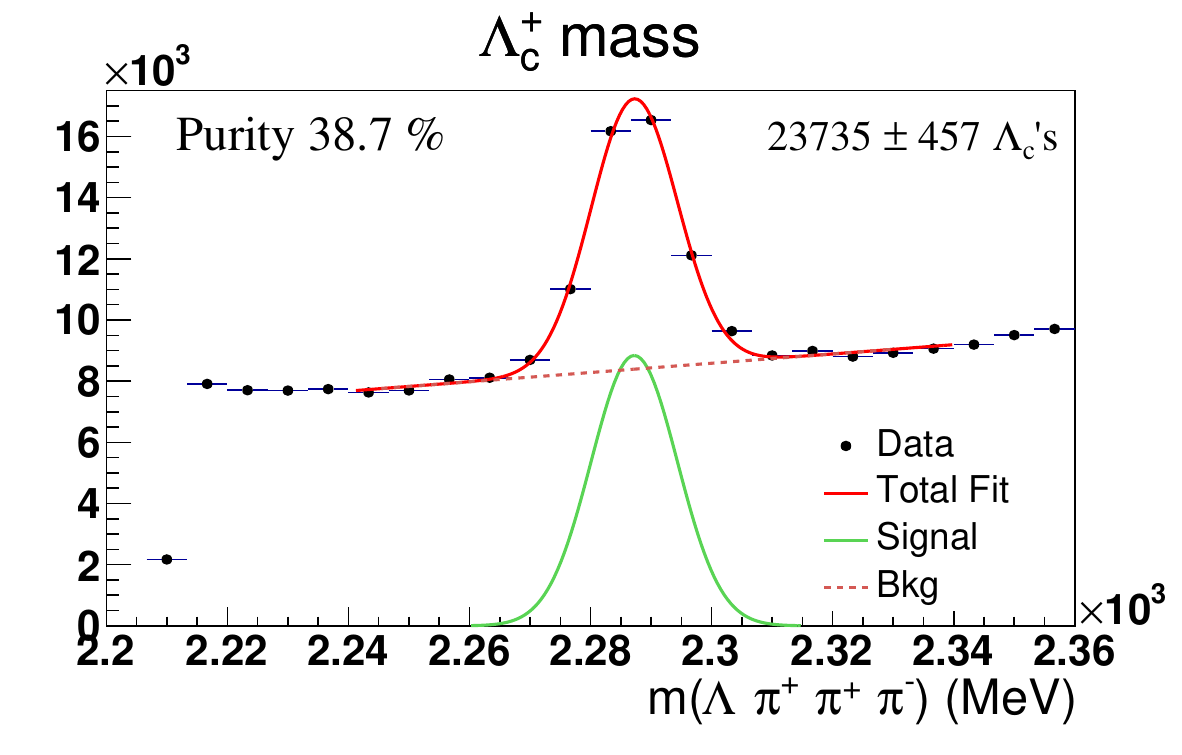}
	\caption{Invariant mass distribution of the $\Lc$ candidates for three different preselections.}
	\label{fig:threePreselectionsPeaks}
\end{figure}

In the multivariate classifier we rely on the similarity between simulated signal MC events and true signal events on data. To ensure this agreement, we correct the MC sample (in Section~\ref{sec:reweightMC}) by forcing some of its variables to match the sWeighted data (see definition of sPlot in Appendix~\ref{app:analysistermsmisc}). However, how is the agreement between the sWeighted real data and the actual signal events (``inside" the peak)? This may depend strongly on the initial purity of the sample used to calculate the sWeights. To study this dependence and eventually associate a systematic uncertainty to this effect, we designed three different preselections such that the resulting purity is $\approx$20, 30 and 40\%. The signal peaks and cut values for each set of cuts are shown in Figure \ref{fig:threePreselectionsPeaks} and Table \ref{tab:threePreselections}, respectively.

\begin{figure}
		\centering
		\includegraphics[width=0.45\linewidth]{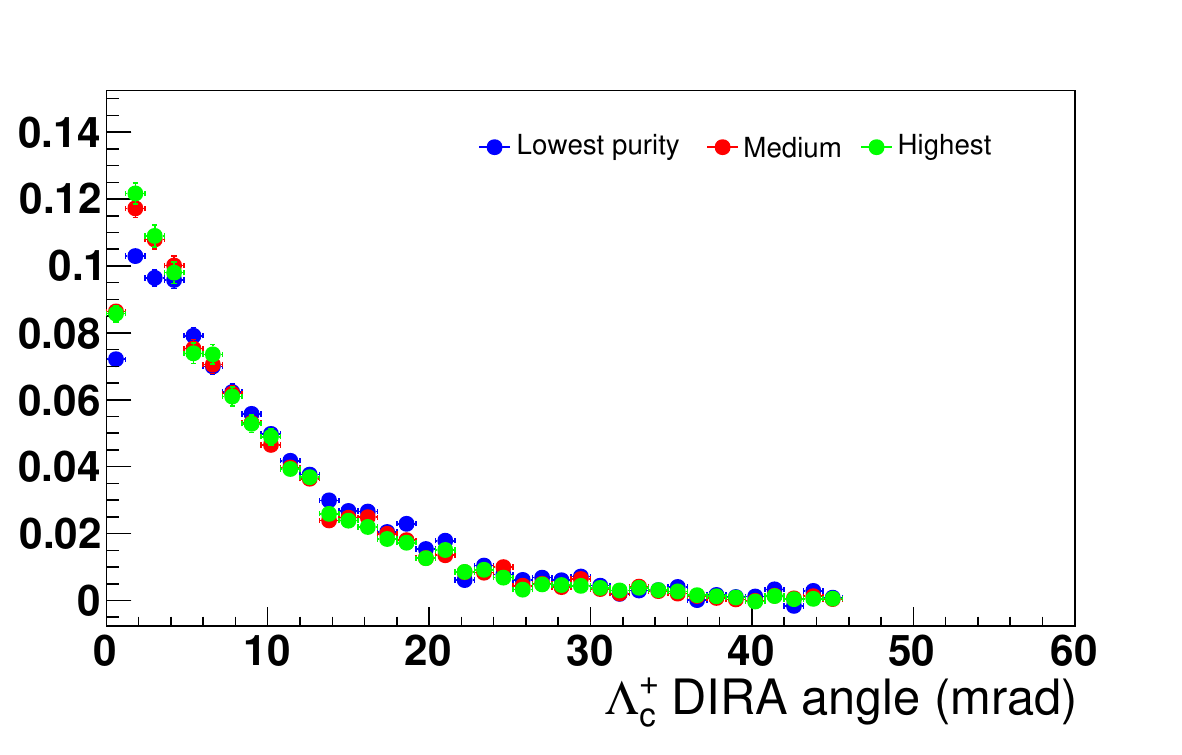}
		\includegraphics[width=0.45\linewidth]{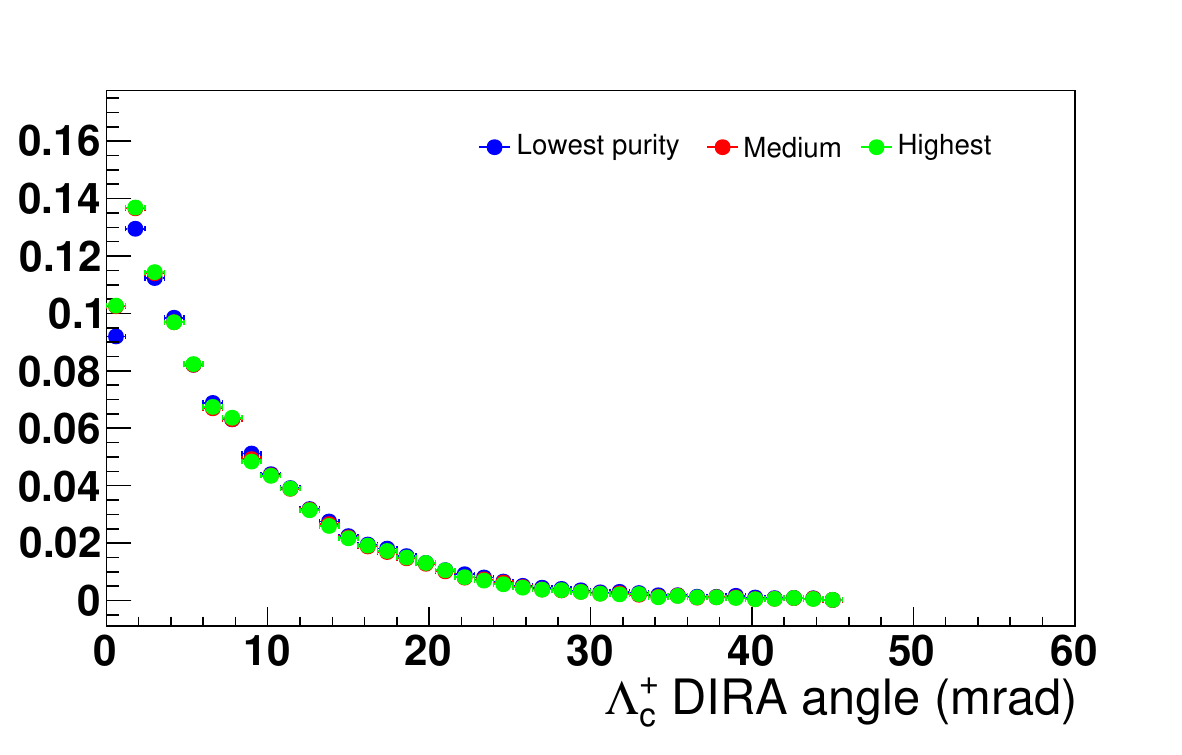}
		\includegraphics[width=0.45\linewidth]{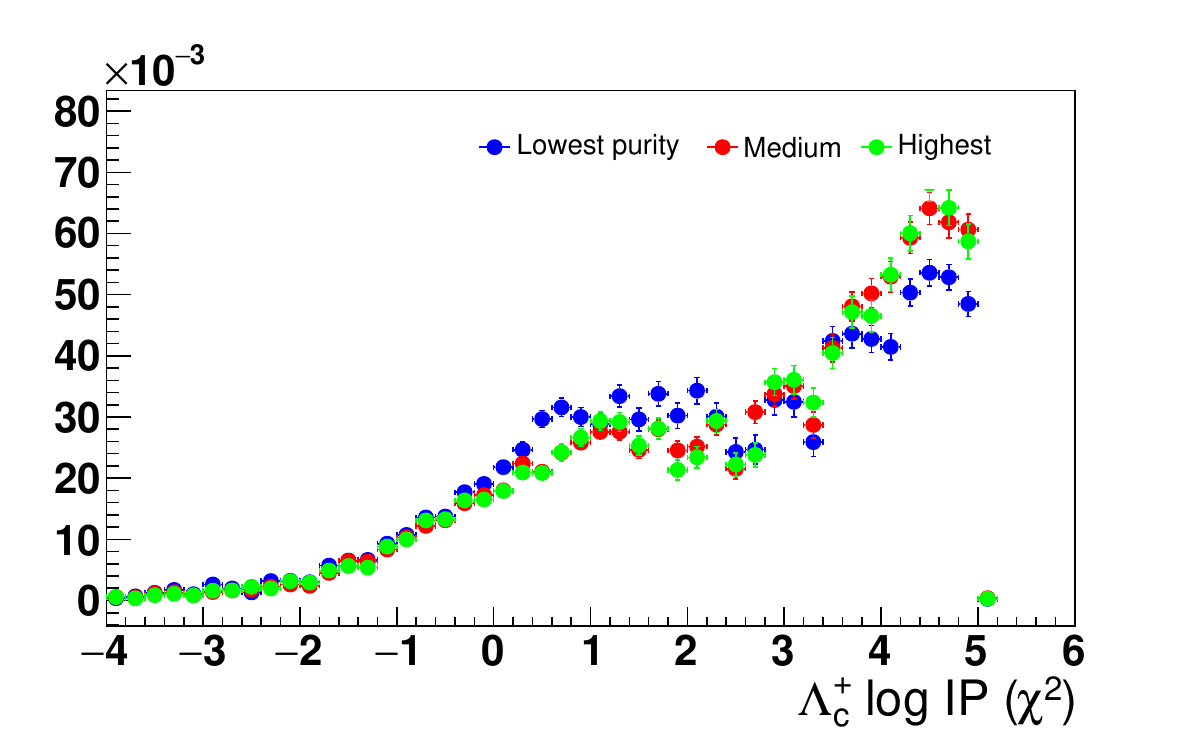}
		\includegraphics[width=0.45\linewidth]{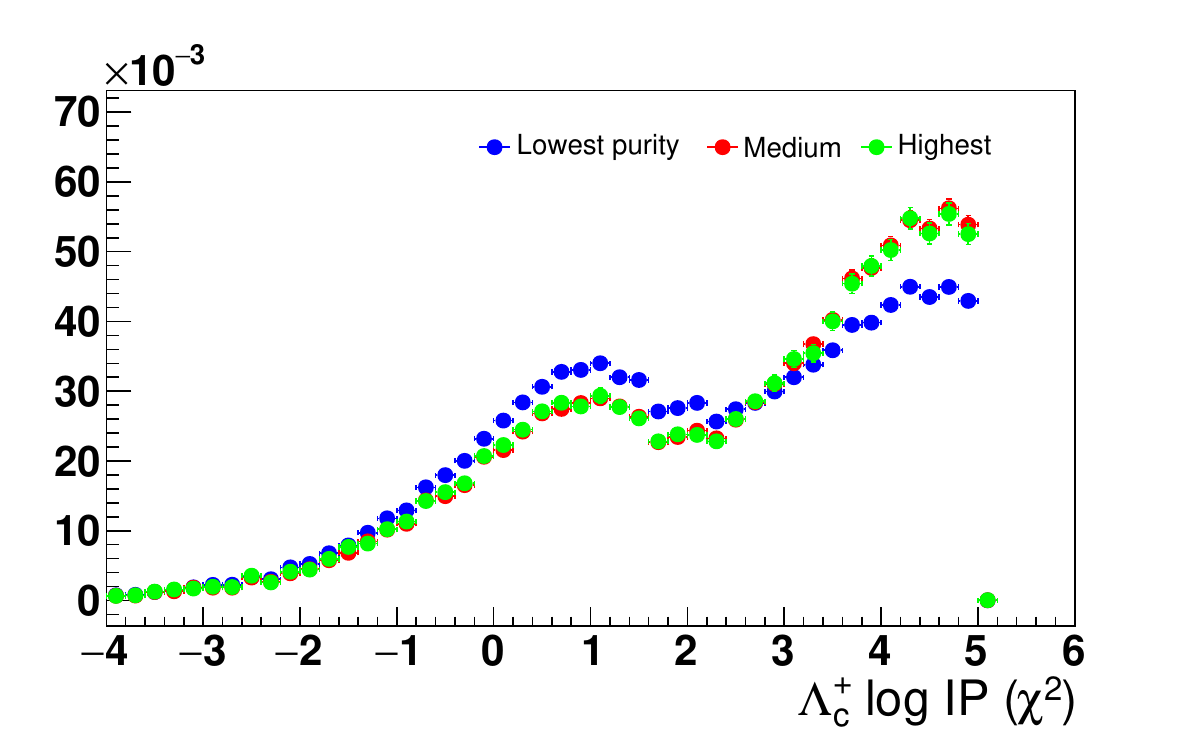}
		\includegraphics[width=0.45\linewidth]{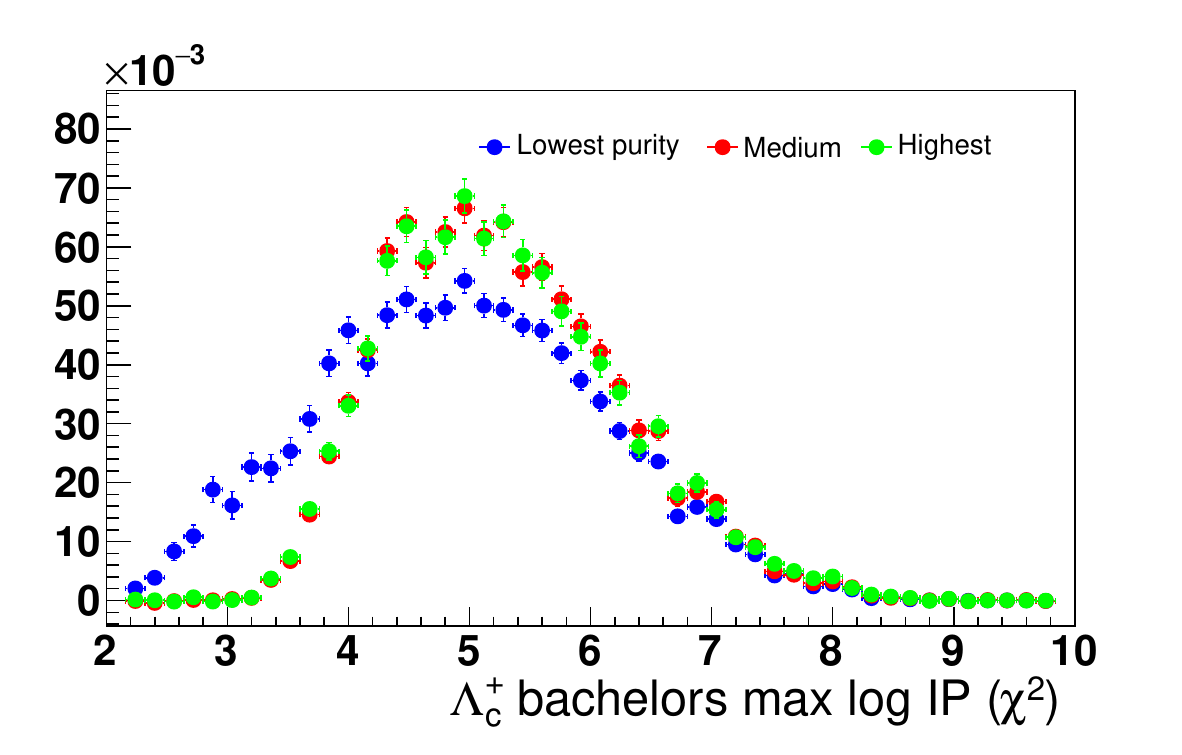}
		\includegraphics[width=0.45\linewidth]{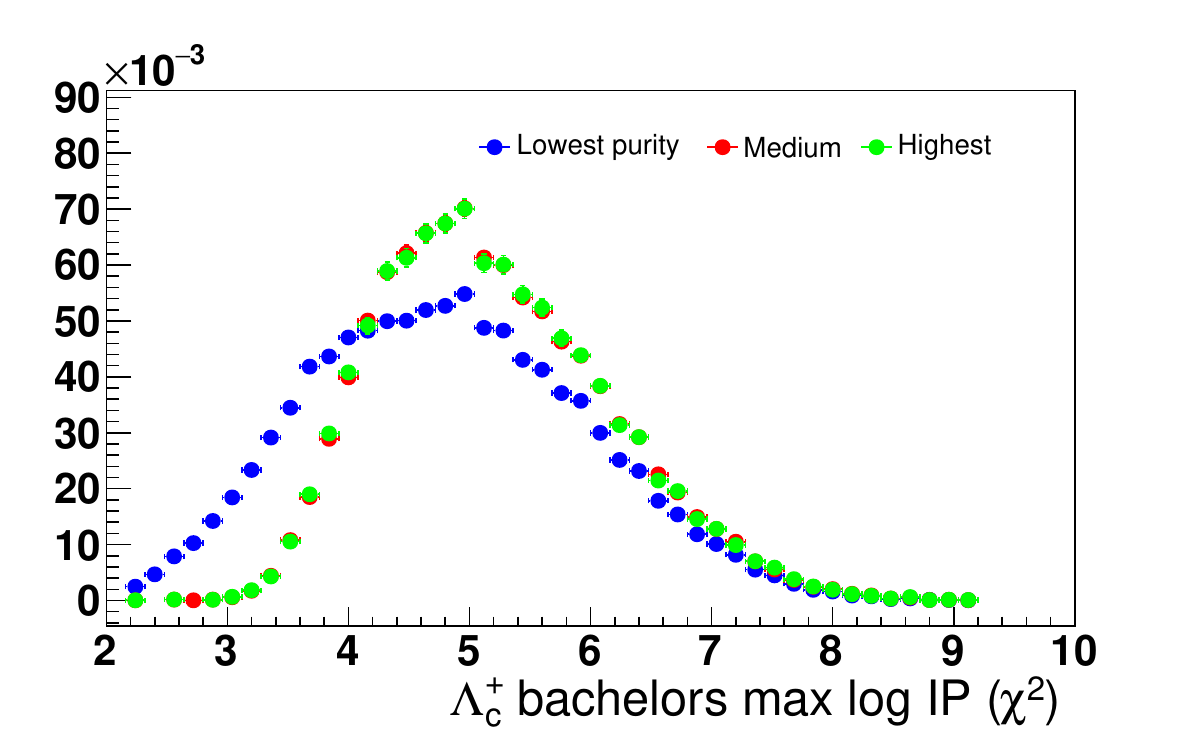}
		\includegraphics[width=0.45\linewidth]{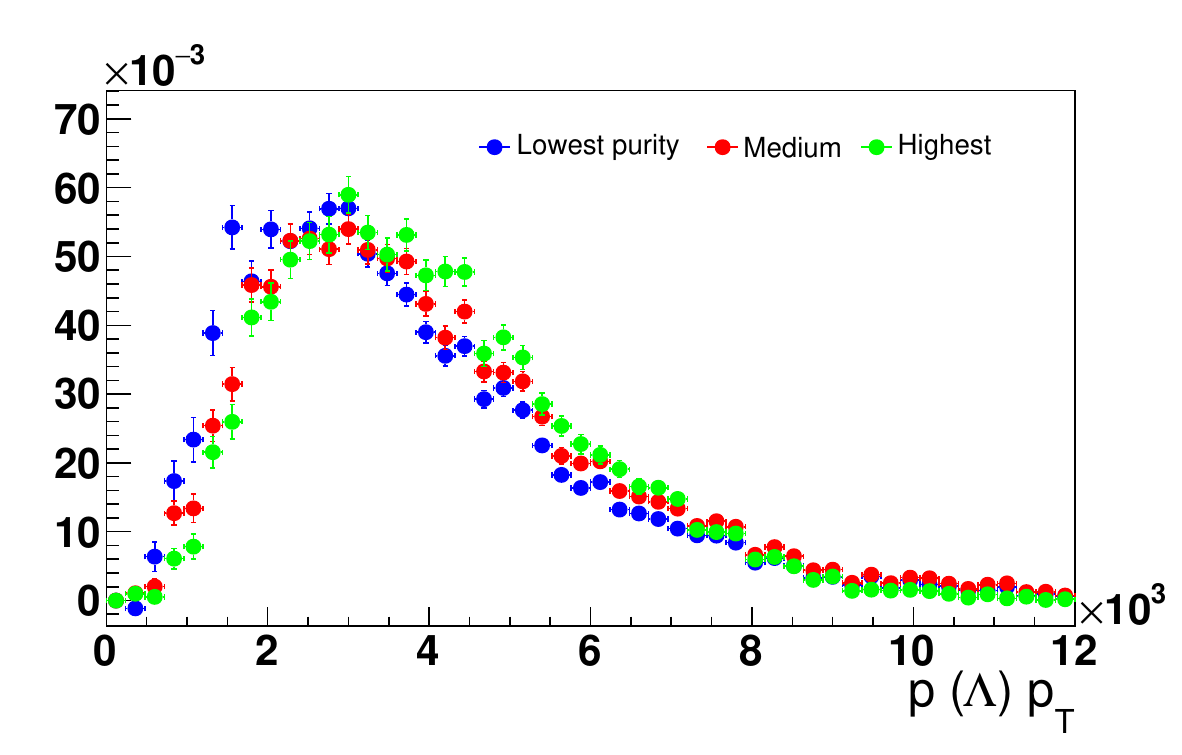}
		\includegraphics[width=0.45\linewidth]{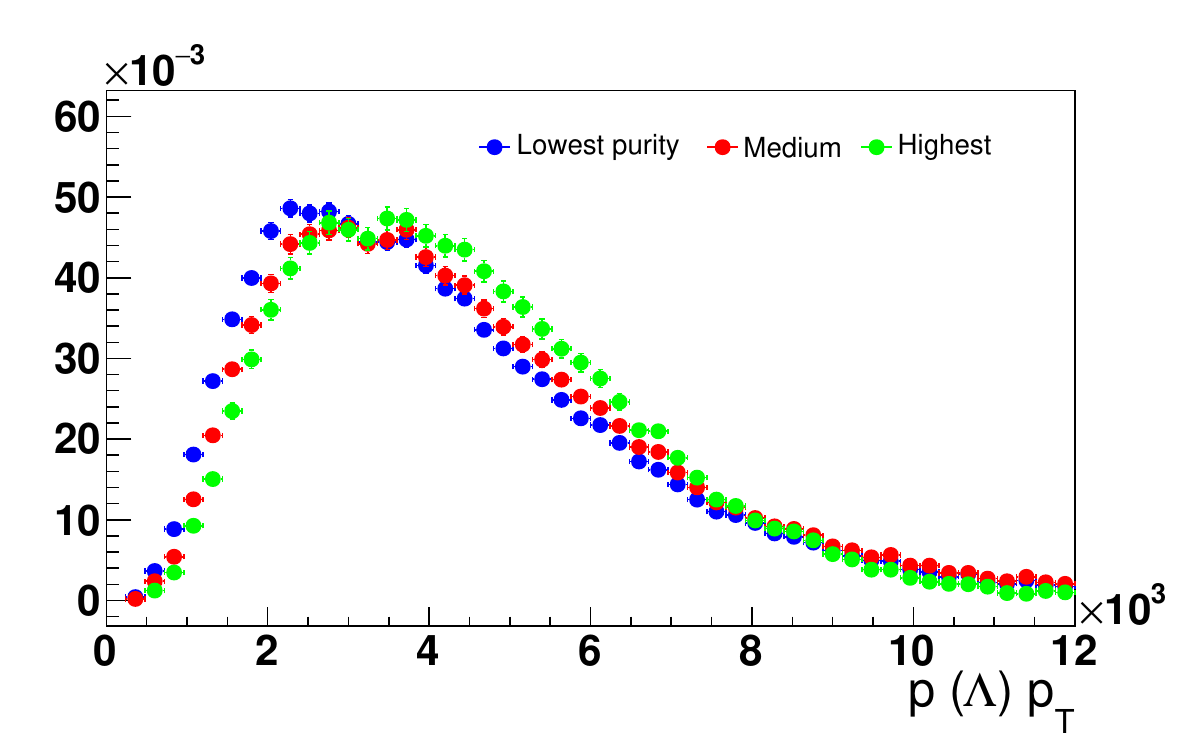}
		\caption{Comparison of the variable distributions for (left) sWeighted real data and (right) signal MC with the three preselections presented in Table \ref{tab:threePreselections}. Since the main differences between the three distributions are reproduced in signal MC, these are independent of the computation of sWeights. The distributions are normalized to unity, and the vertical axes represent the normalized number of events per bin. }
		\label{fig:threePreselectionsDistributions1}
	\end{figure}

\begin{figure}
		\centering
		\includegraphics[width=0.45\linewidth]{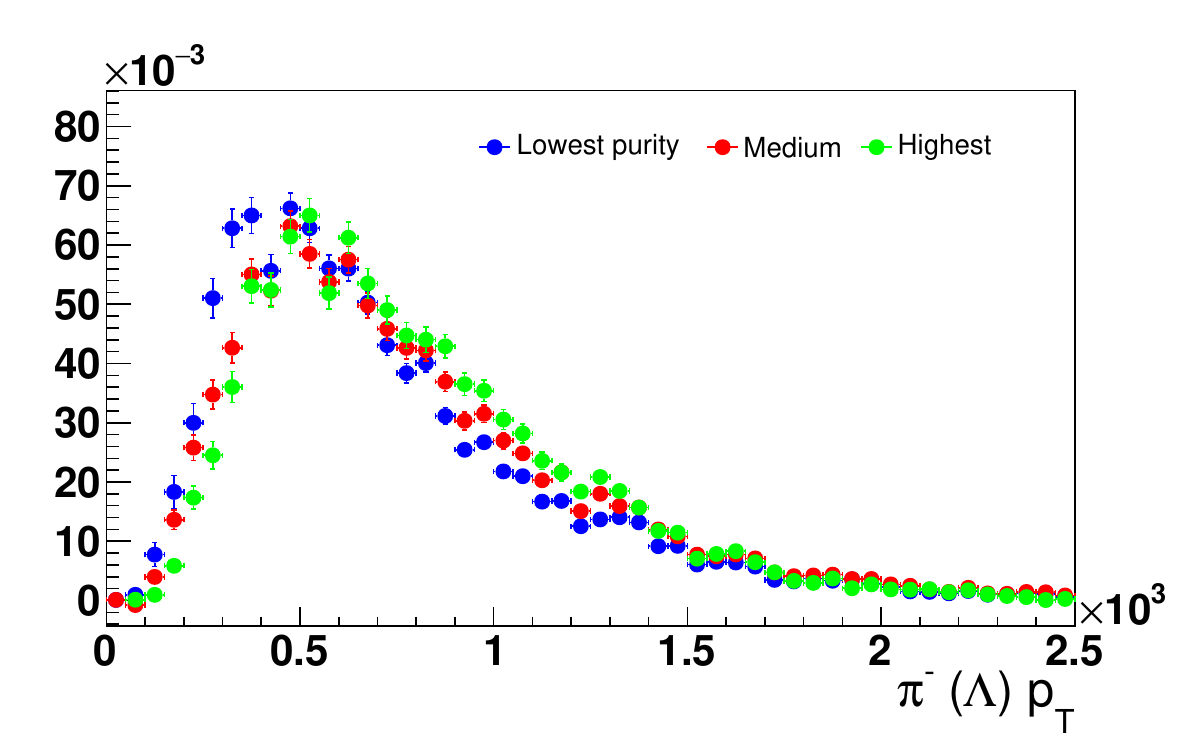}
		\includegraphics[width=0.45\linewidth]{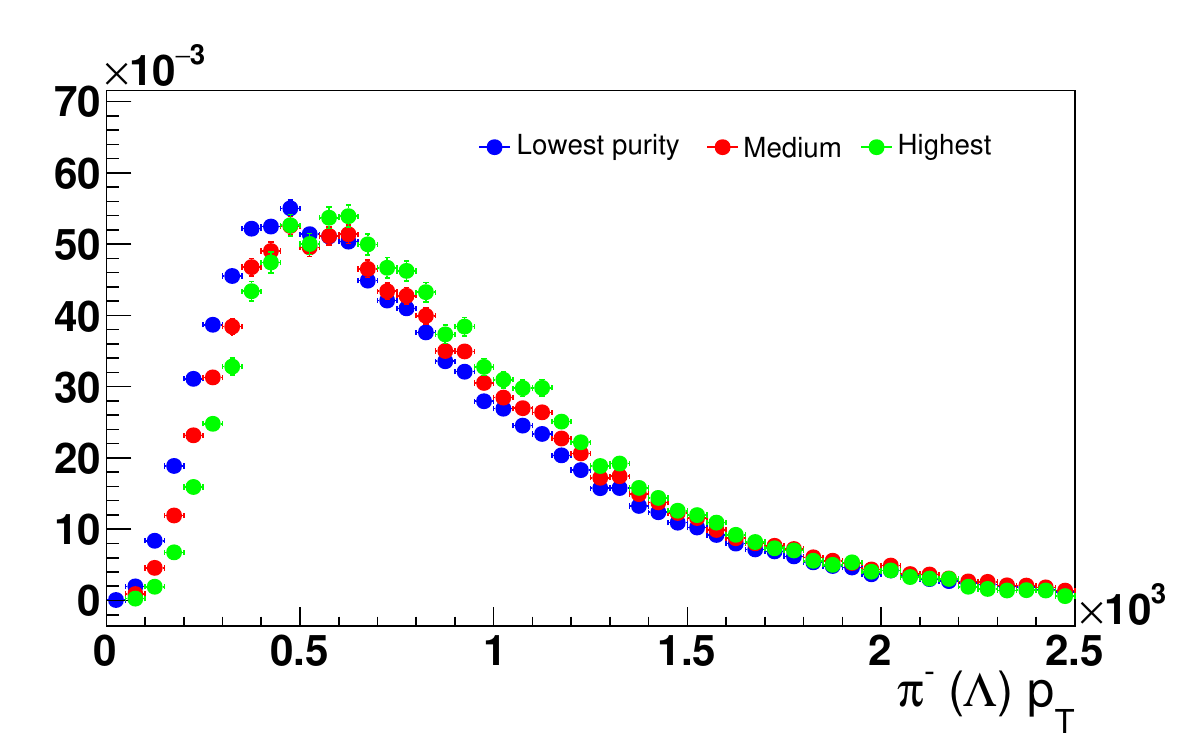}
		\includegraphics[width=0.45\linewidth]{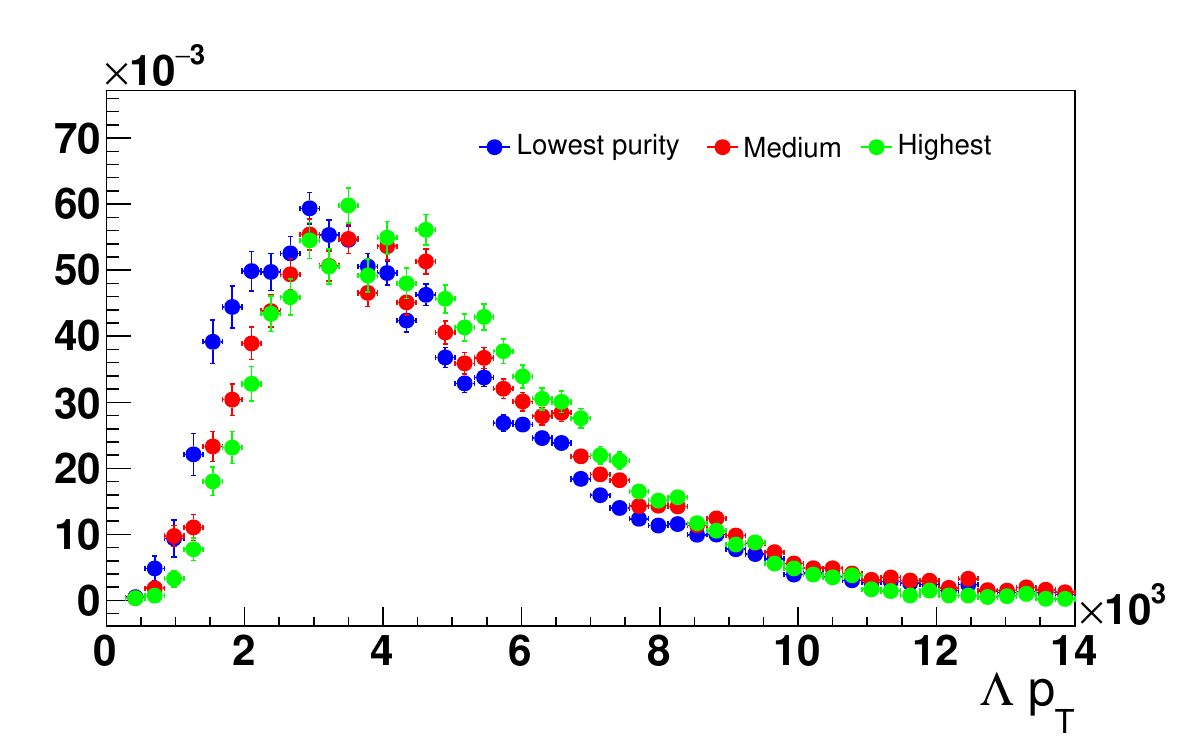}
		\includegraphics[width=0.45\linewidth]{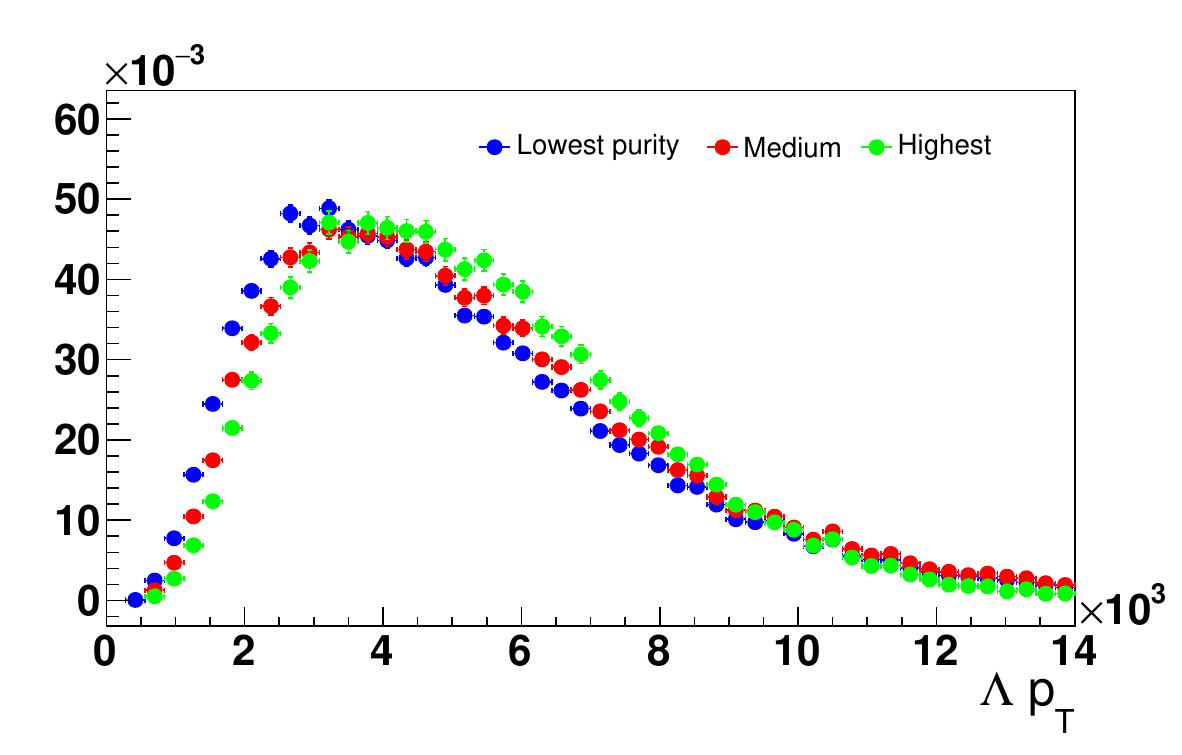}
		\includegraphics[width=0.45\linewidth]{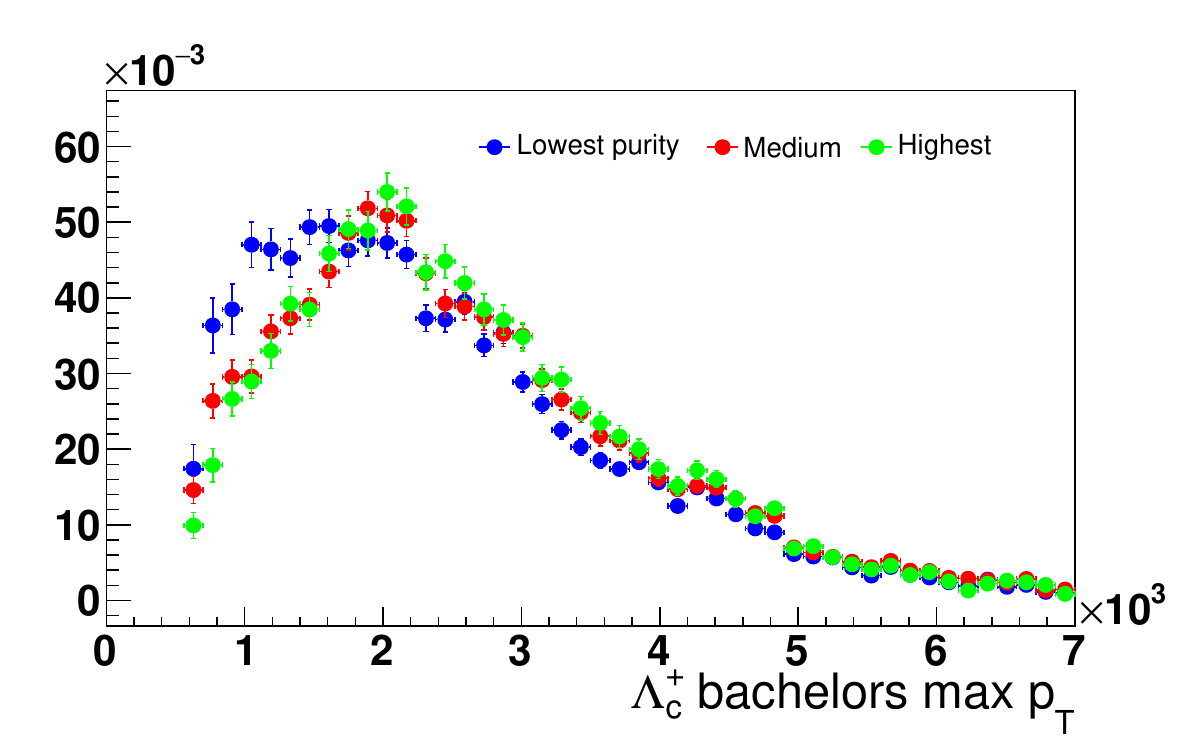}
		\includegraphics[width=0.45\linewidth]{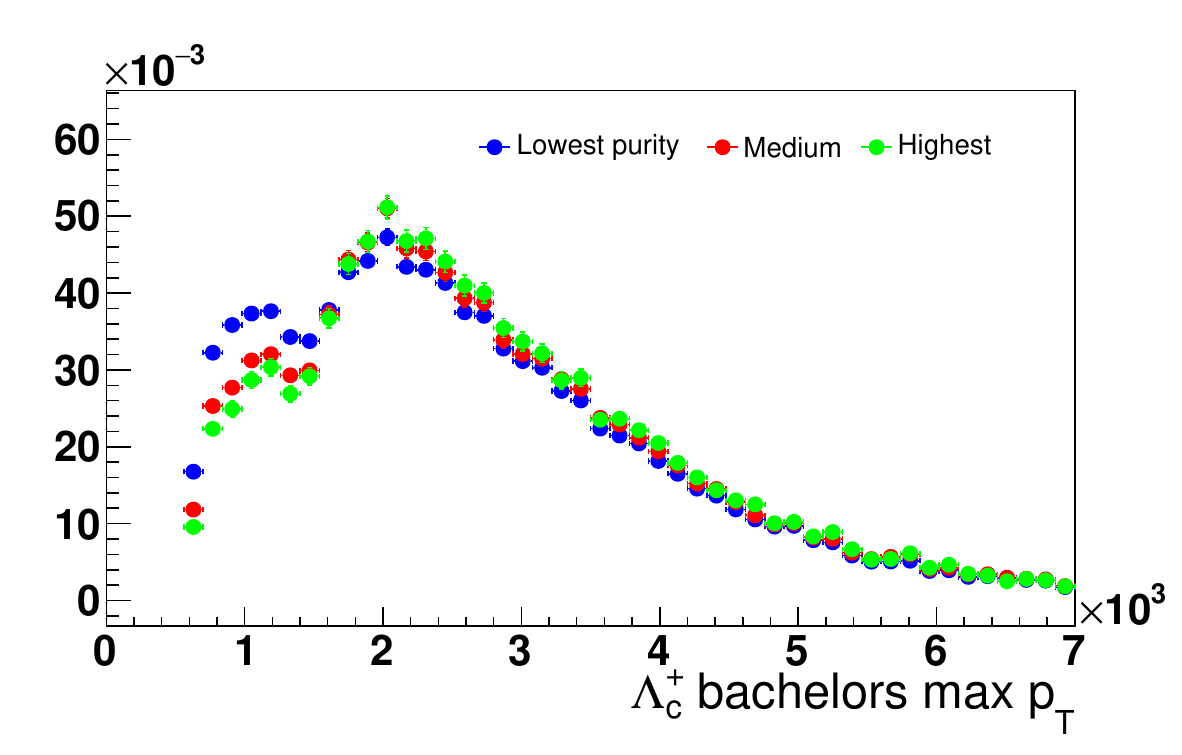}
		\includegraphics[width=0.45\linewidth]{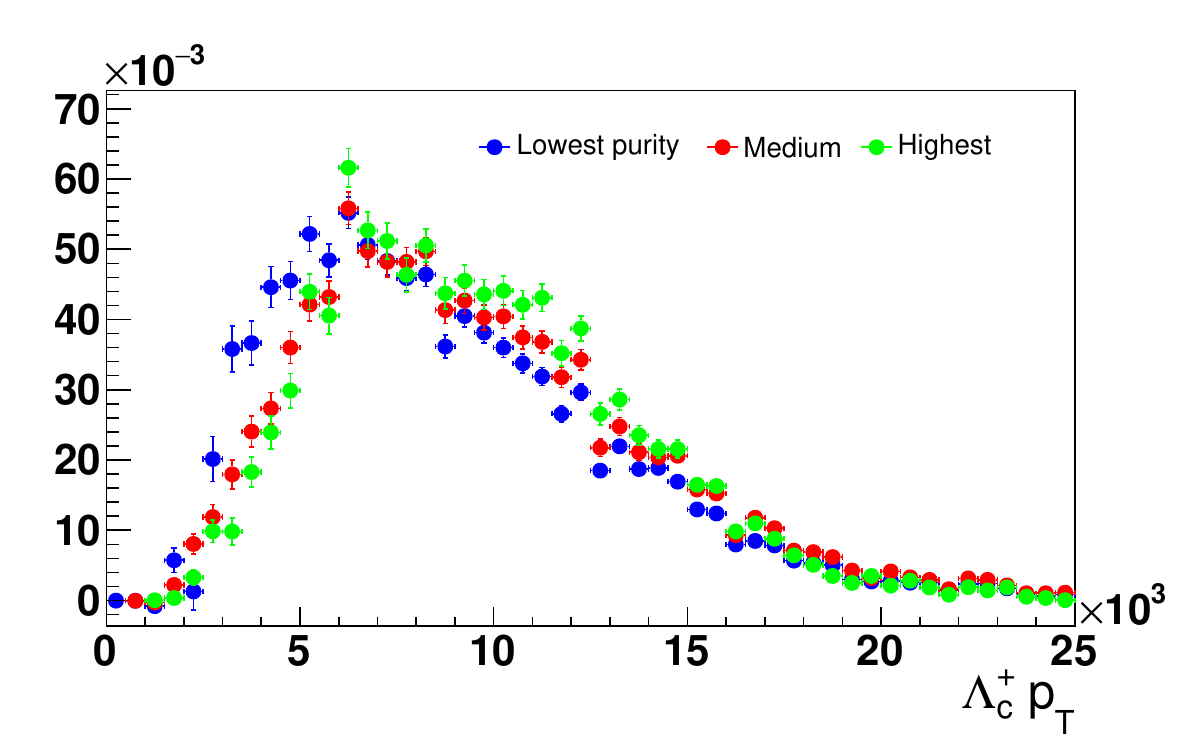}
		\includegraphics[width=0.45\linewidth]{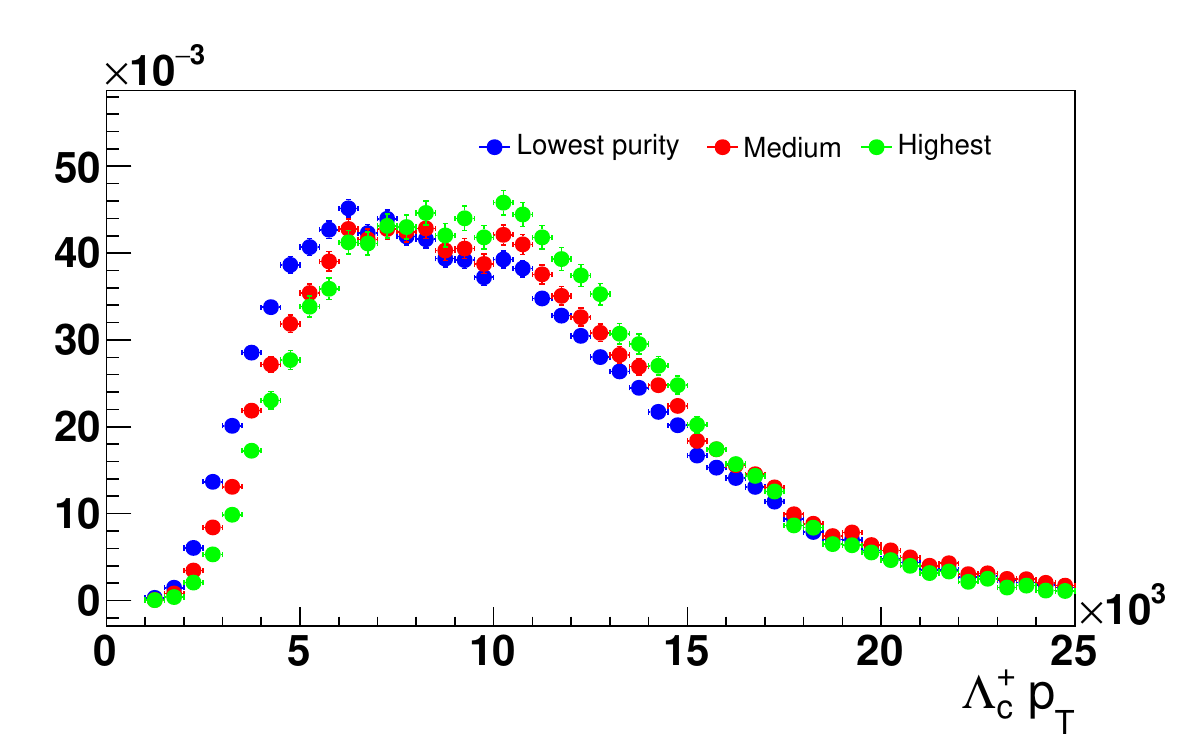}
		\caption{Same as Figure~\ref{fig:threePreselectionsDistributions1} for other variables.}
		\label{fig:threePreselectionsDistributions2}
	\end{figure}

Among the distribution of some relevant variables we observed significant differences between the three datasets. However, we need to rule out if these differences come from the correlation of the shown distributions to the cutting variables. Instead of comparing each dataset to the corresponding signal MC (as the data/MC agreement is in general not good at this stage), we plot the same three-fold comparison with the signal MC, in Figures \ref{fig:threePreselectionsDistributions1} and \ref{fig:threePreselectionsDistributions2}. As we can see, the main differences appearing on real data are reproduced in the signal MC.

We can conclude that a preselection with 20\% purity is equally good to calculate the sWeights reliably. It has the advantage of a higher signal efficiency (70 against 50\%). Thus, in future versions of the analysis this preselection will be softened.

	\begin{table}
			\caption{Cuts and efficiency on signal MC from three different preselections. The initial number of events is 63716.}
	\resizebox{1\linewidth}{!}{	\begin{tabular}{lcccccc}
		\hline 
		Variable      &           Cut  &  $\varepsilon$(MC)   &           Cut  &  $\varepsilon$(MC)  &           Cut  &  $\varepsilon$(MC) \\
		\hline
		$\pi^{-}$ ($\Lambda$) Ghost track probability NN             &    $<$    0.7   &    96.0\% &    $<$    0.35   &    88.7\% &    $<$    0.15   &   78.3\%  \\
		p ($\Lambda$) pi track probability NN  Corrected   &    $<$    0.4   &    88.5\% &    $<$    0.4   &    88.5\% &    $<$    0.2   &    75.2\%  \\ \hline
		$\Lambda_{c}^{+}$ Flight distance ($\chi^{2}$)     &    $>$     20   &    97.7\%  &    $>$     100   &    75.9\% &    $>$     100   &    75.9\%  \\
		$\Lambda_{c}^{+}$ bachelors max $p_{T}$            &    $>$    600  &    96.0\% &    $>$    600  &    96.0\% &    $>$    600  &    96.0\% \\
		$\pi^{-}$ ($\Lambda_{c}^{+}$) Ghost track probability NN     &    $<$   0.21  &    96.2\%  &    $<$   0.21  &    96.2\%  &    $<$   0.12  &    93.9\%  \\
		$\Lambda_{c}^{+}$ bachelors max DOCA [$\pi^{+}$1, $\pi^{+}$2, $\pi^{-}$]     &    $<$    0.2 &      94.3\% &    $<$    0.15 &      89.6\% &    $<$    0.15 &      89.6\%  \\ \hline
		$\Lambda$ vertex position Z                        &    $<$   2300 &      96.5\% &    $<$   2300 &      96.5\% &    $<$   2300 &      96.5\%  \\
		\hline 
		All cuts     &             &   67.4\% & &   48.7\% & &   35.8\%   \\
		\hline 
		\end{tabular}}
		 \label{tab:threePreselections}
	\end{table} 

%
%


	\section{Comparison of DTF configurations}
	
\begin{figure}[h!]
			\centering
			\includegraphics[width=0.45\linewidth]{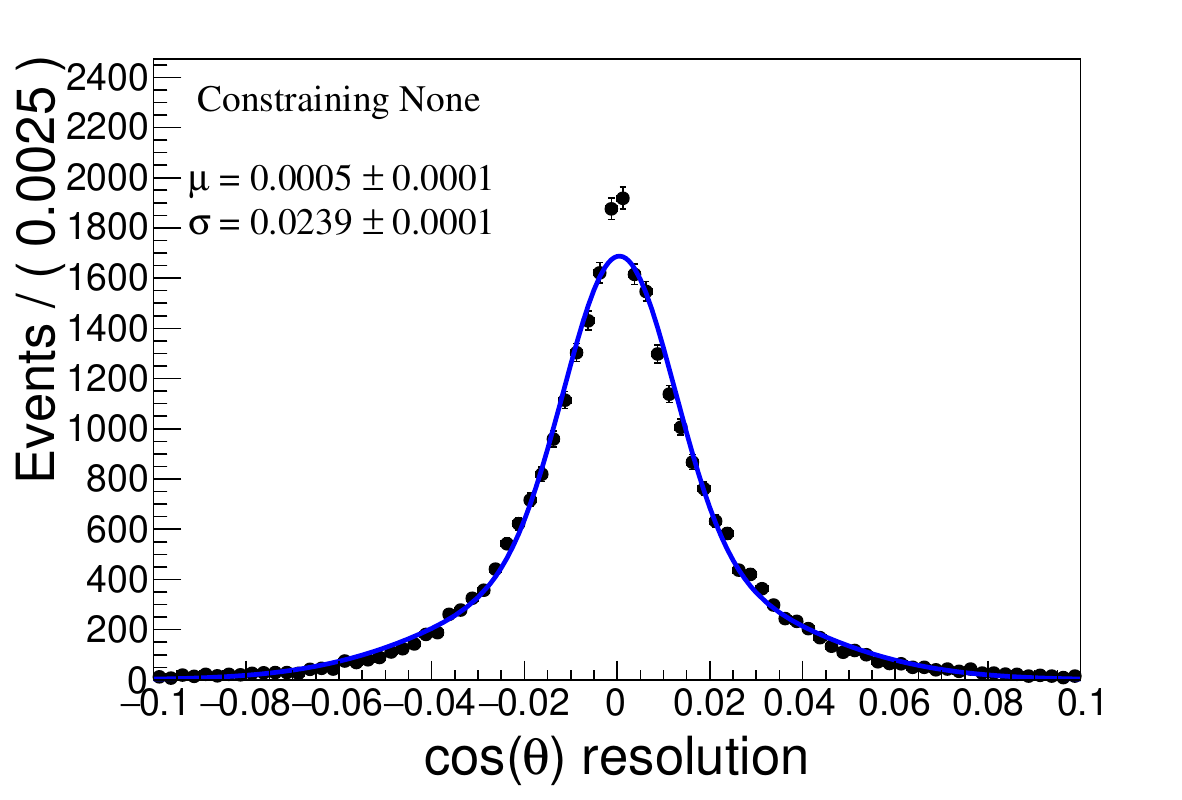}
			\includegraphics[width=0.45\linewidth]{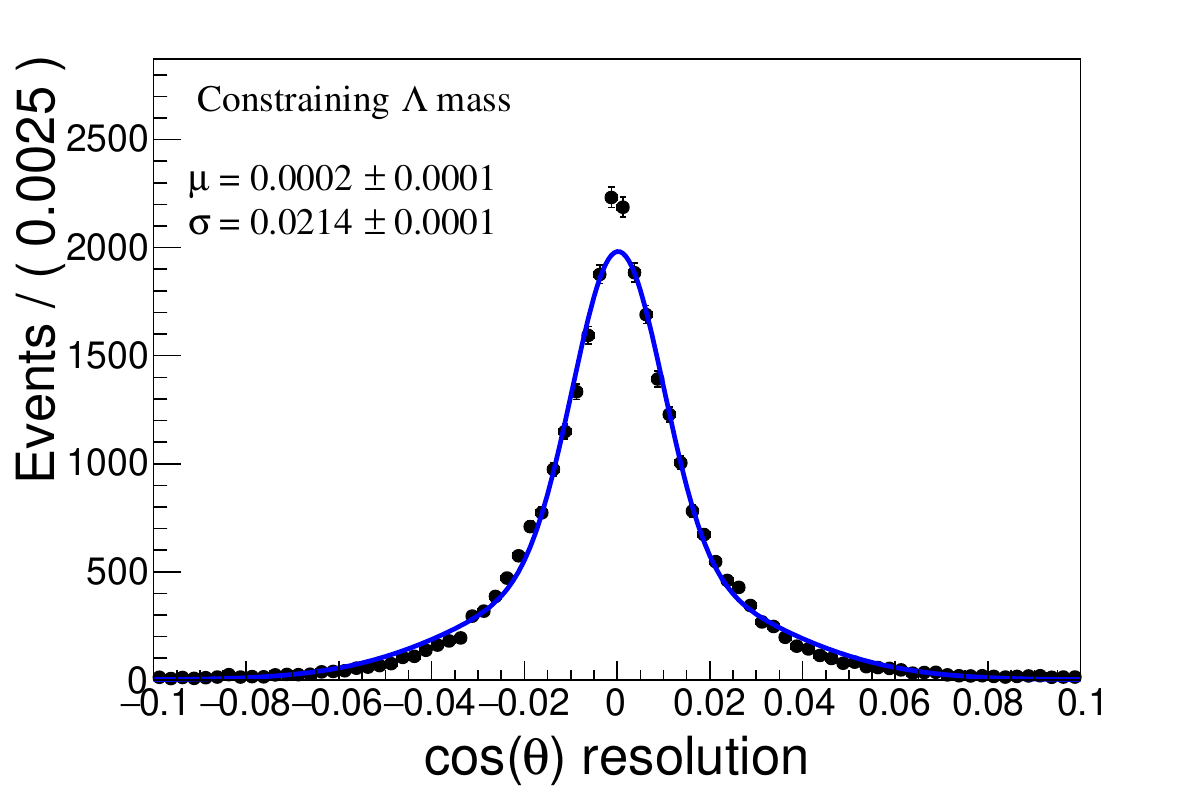}
			\includegraphics[width=0.45\linewidth]{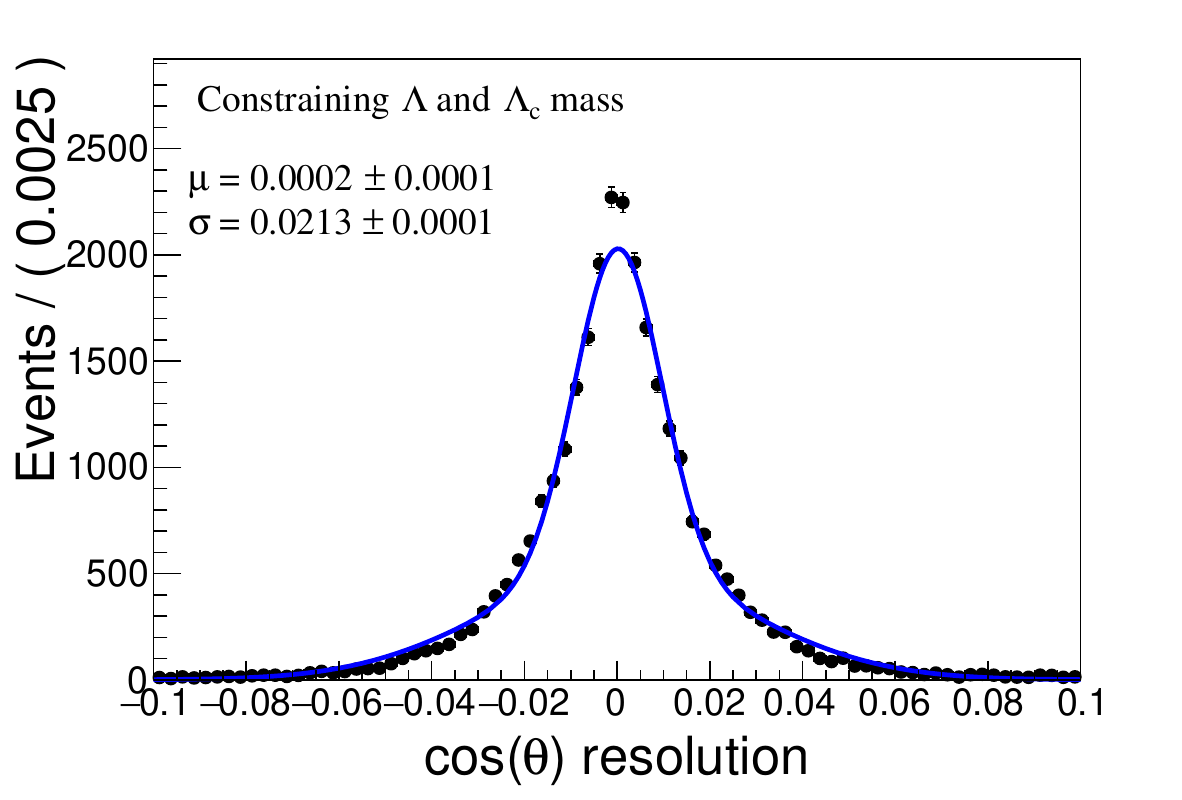}
			\includegraphics[width=0.45\linewidth]{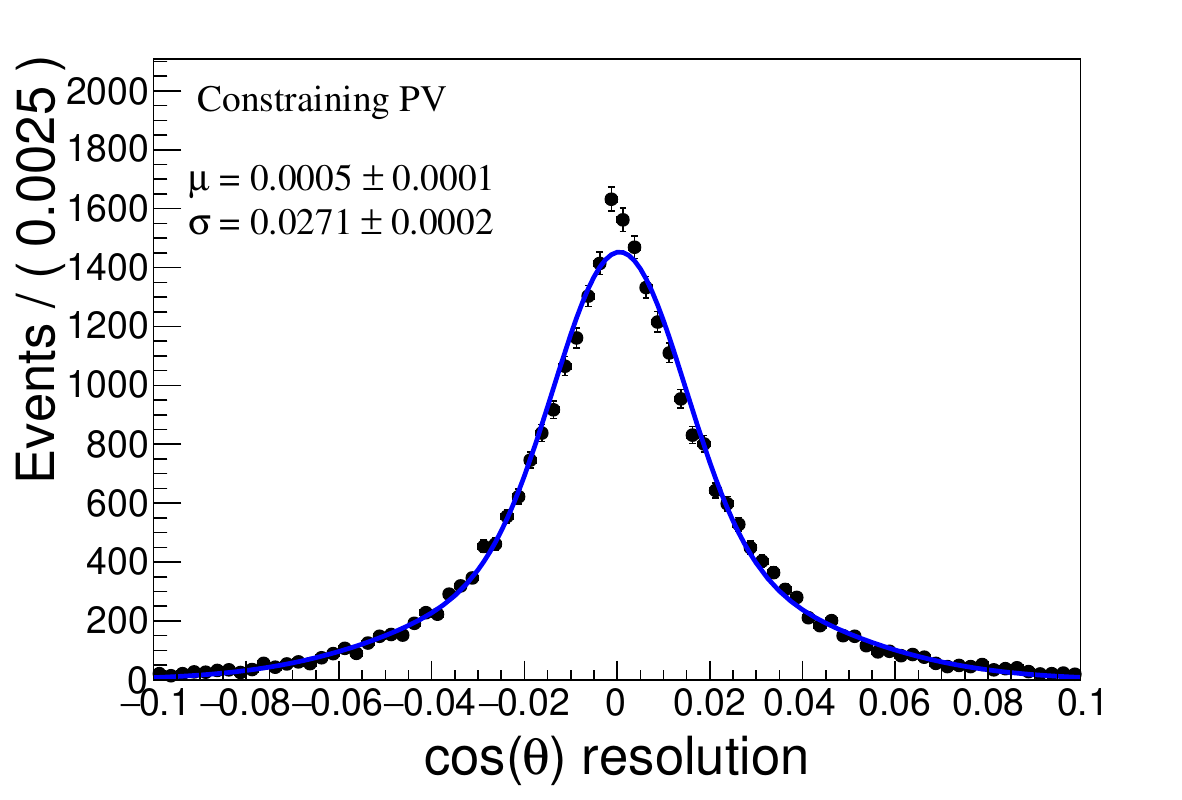}
			\includegraphics[width=0.45\linewidth]{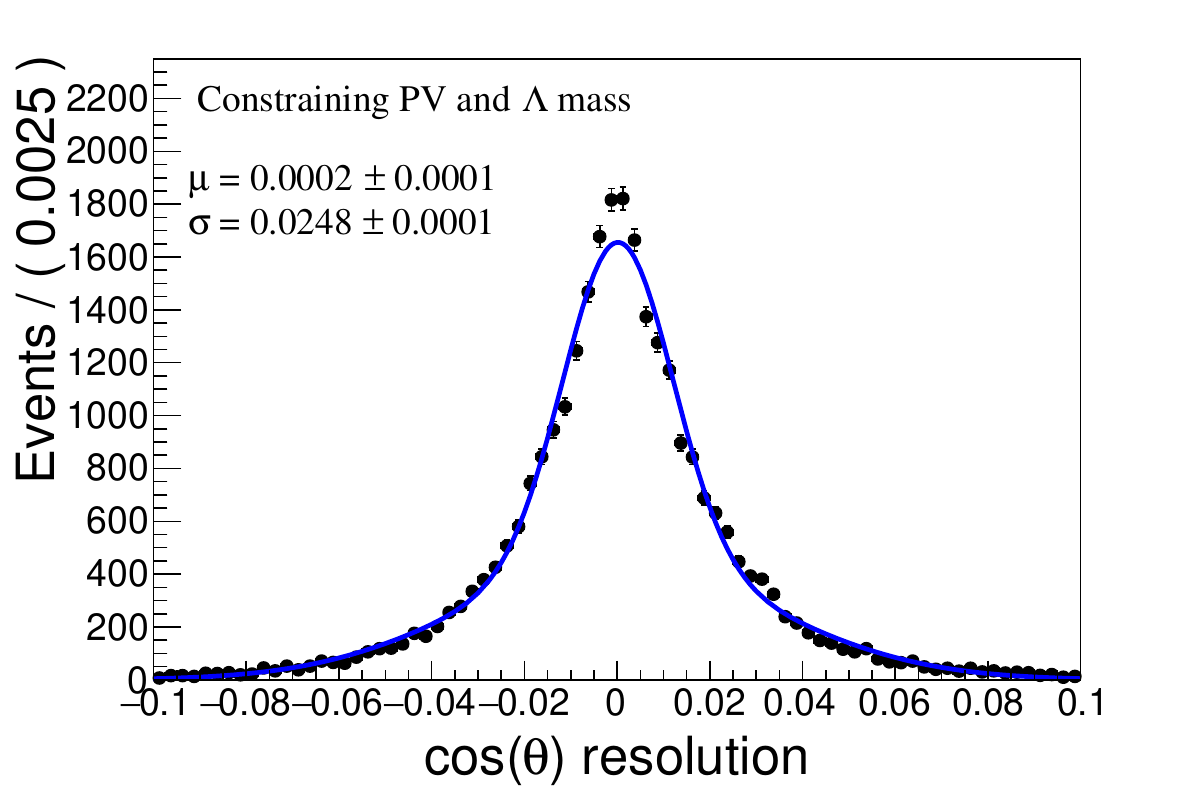}
			\includegraphics[width=0.45\linewidth]{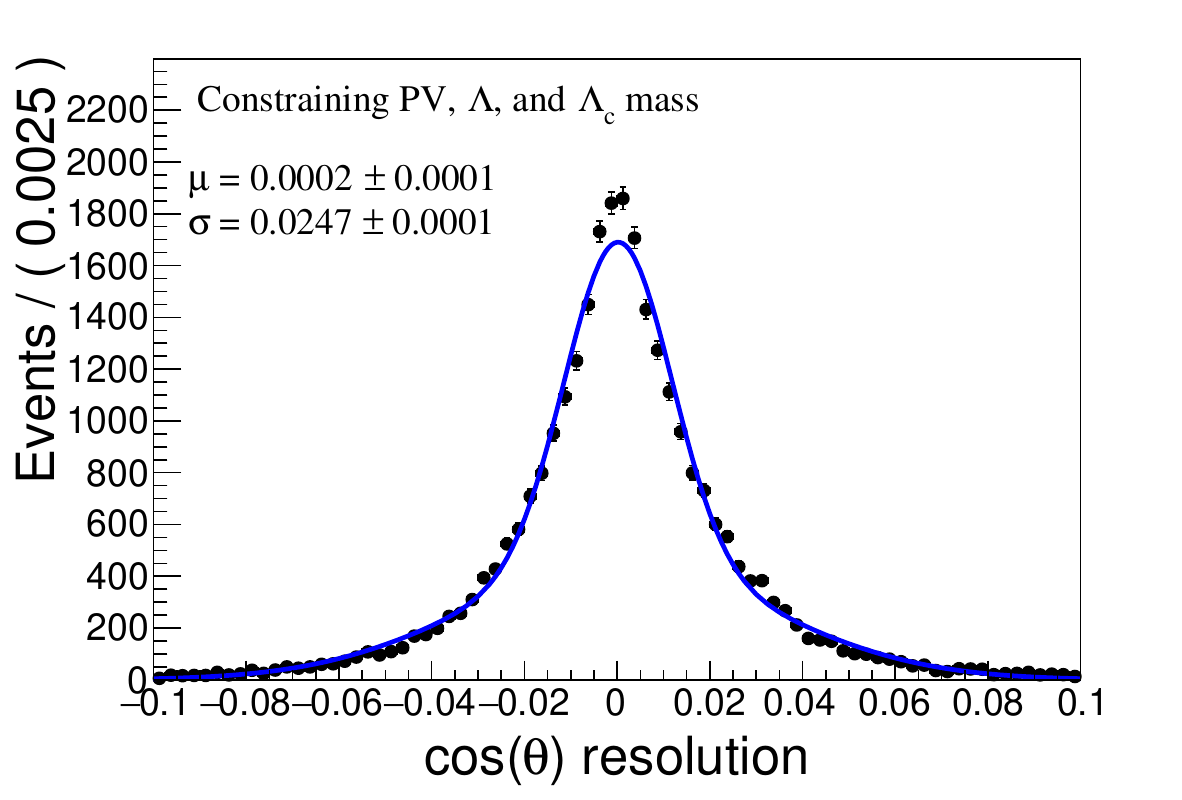}
			\caption{Resolution of $\cos \theta_p$ for the different DTF configurations (labeled on the top left corner of each plot), using MC data. The points are fitted with a double-Gaussian PDF, which does not reproduce well the center of the peak but that, nevertheless, provides a figure of merit to compare the different configurations. }
			\label{fig:DTFAngularResolution}
		\end{figure}

	The fits described in Section~\ref{sec:DTF} to compare the DTF configurations are shown in Figures~\ref{fig:DTFAngularResolution} and \ref{fig:DTFMassFits}.
	
		\onlyANA{\begin{figure}[h!]
			\centering
			\includegraphics[width=0.45\linewidth]{content/figs/DTFConf/resultsDTF2/resPhiNonedata}
			\includegraphics[width=0.45\linewidth]{content/figs/DTFConf/resultsDTF2/resPhiL0Cdata}
			\includegraphics[width=0.45\linewidth]{content/figs/DTFConf/resultsDTF2/resPhiL0LcCdata}
			\includegraphics[width=0.45\linewidth]{content/figs/DTFConf/resultsDTF2/resPhiPVCdata}
			\includegraphics[width=0.45\linewidth]{content/figs/DTFConf/resultsDTF2/resPhiPVL0Cdata}
			\includegraphics[width=0.45\linewidth]{content/figs/DTFConf/resultsDTF2/resPhiPVL0LcCdata}
			\caption{Same as Figure~\ref{fig:DTFAngularResolution}, but with the $\phi_p$ angle.  \red{the resolution on $\phi$ peaks in [-$\pi$,0,$\pi$]. Only showing one third. How to solve this?} }
			\label{fig:DTFAngularResolutionPhi}
		\end{figure}
	}
		
		\begin{figure}[h!]
			\centering
			\includegraphics[width=0.49\linewidth]{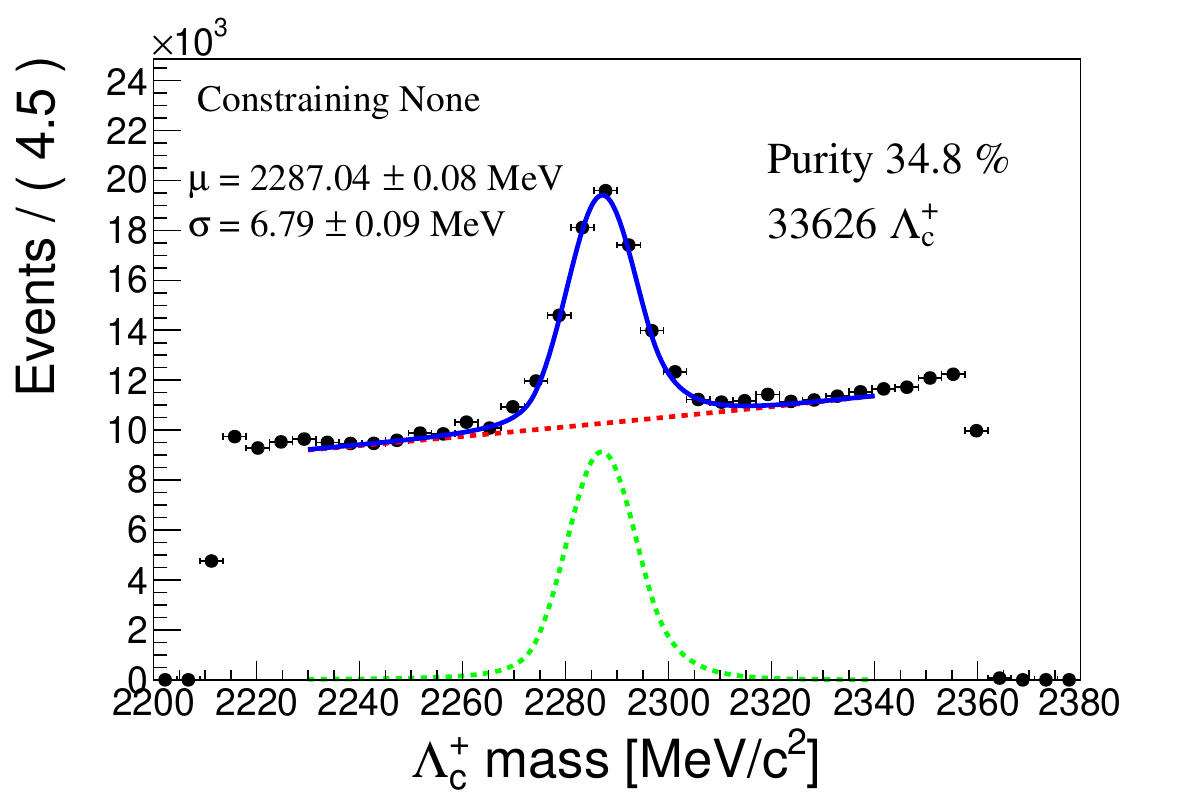}
			\includegraphics[width=0.49\linewidth]{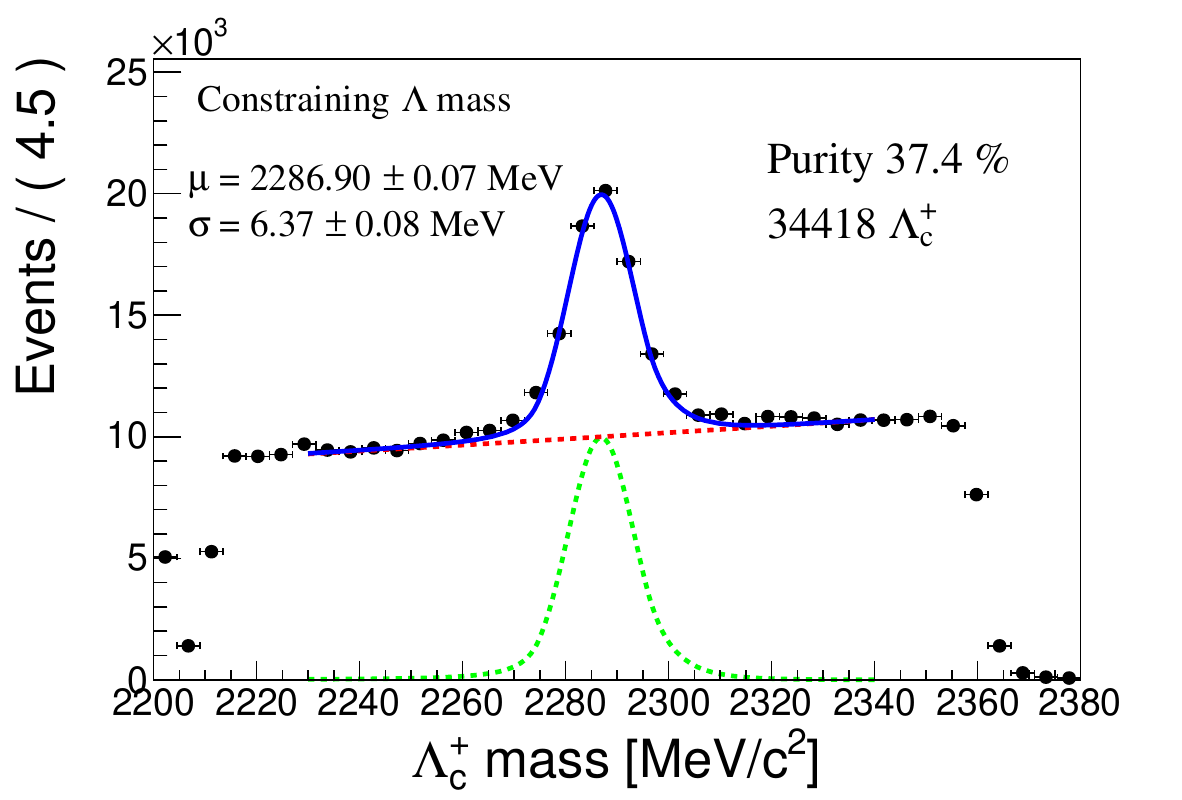}
			\includegraphics[width=0.49\linewidth]{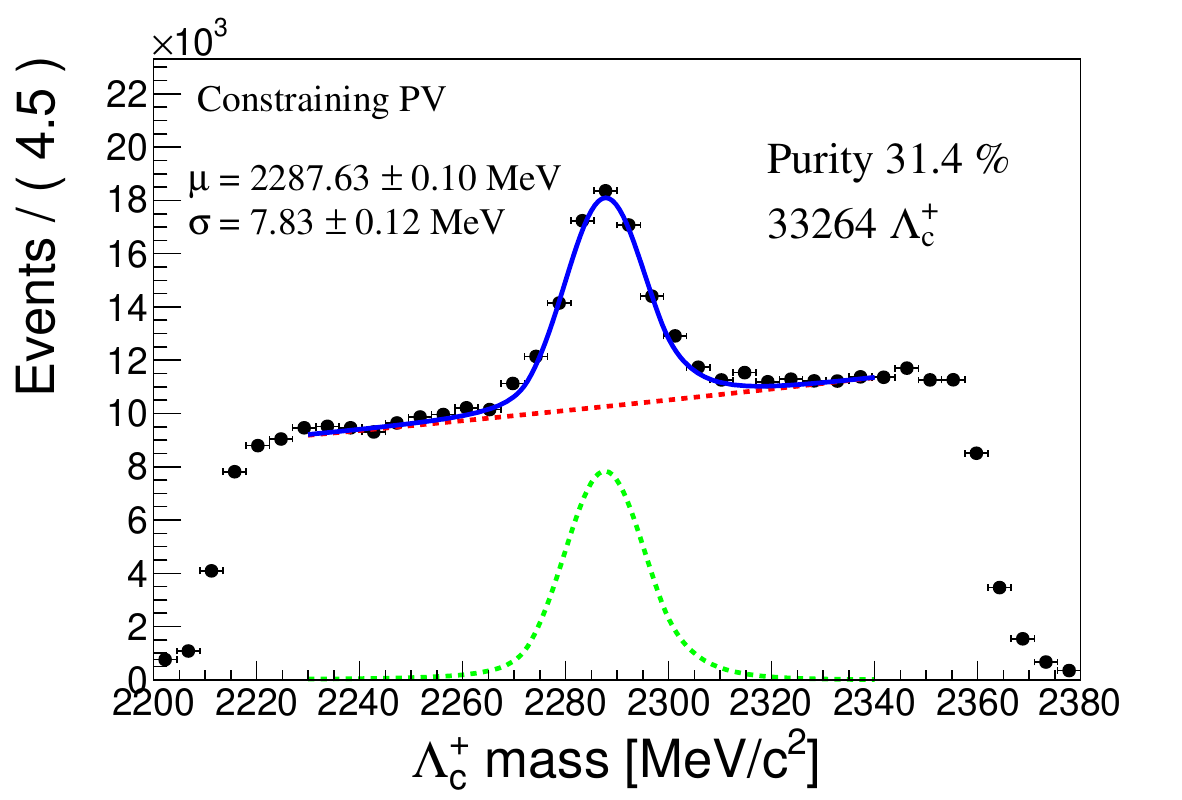}
			\includegraphics[width=0.49\linewidth]{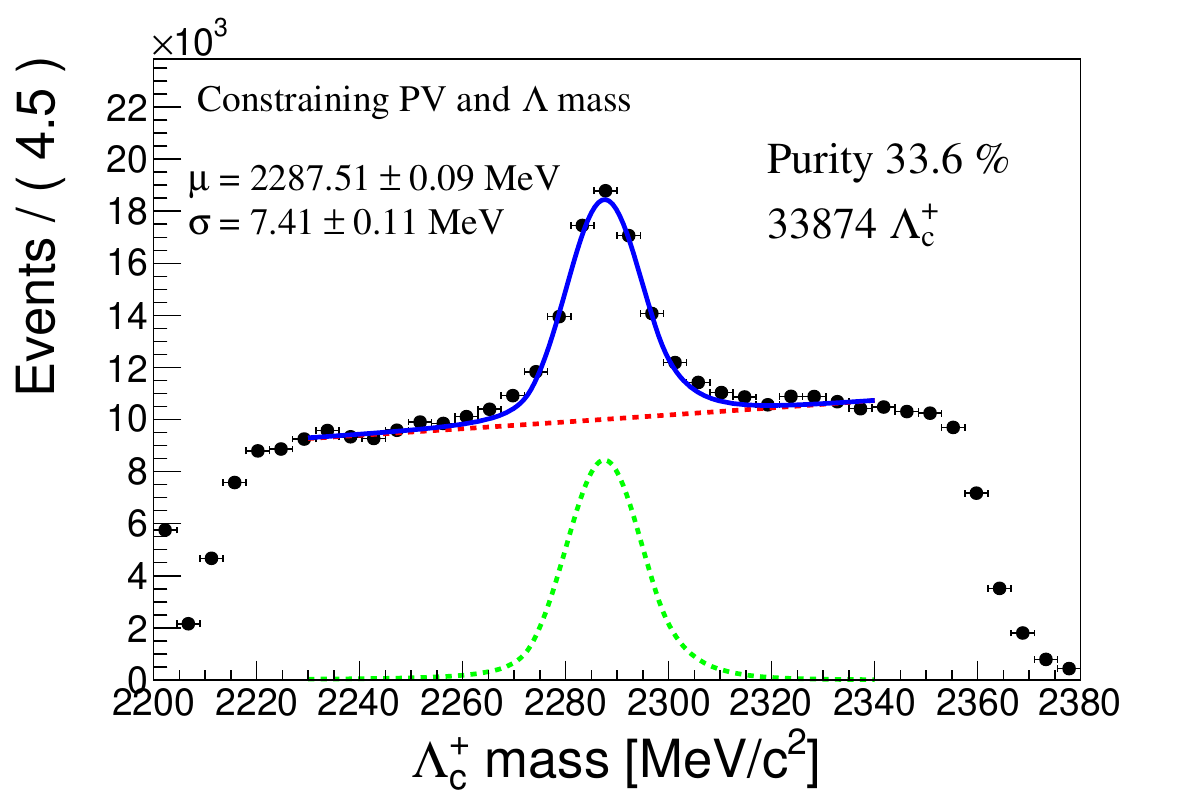}
			\caption{Invariant mass distribution of the \Lc candidates for the different DTF configurations.}
			\label{fig:DTFMassFits}
		\end{figure}

		\section{PID Corrections}
		
\begin{figure}[h!]
					\centering
					
					\includegraphics[width=0.45\linewidth]{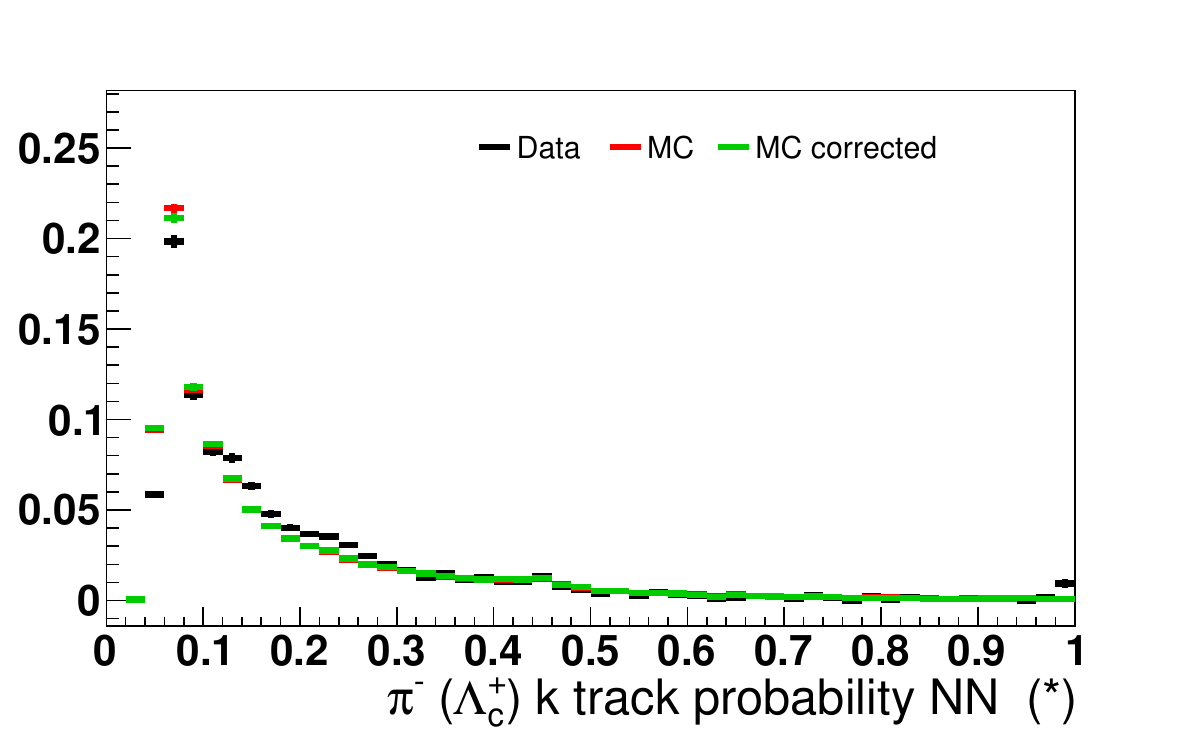}
					\includegraphics[width=0.45\linewidth]{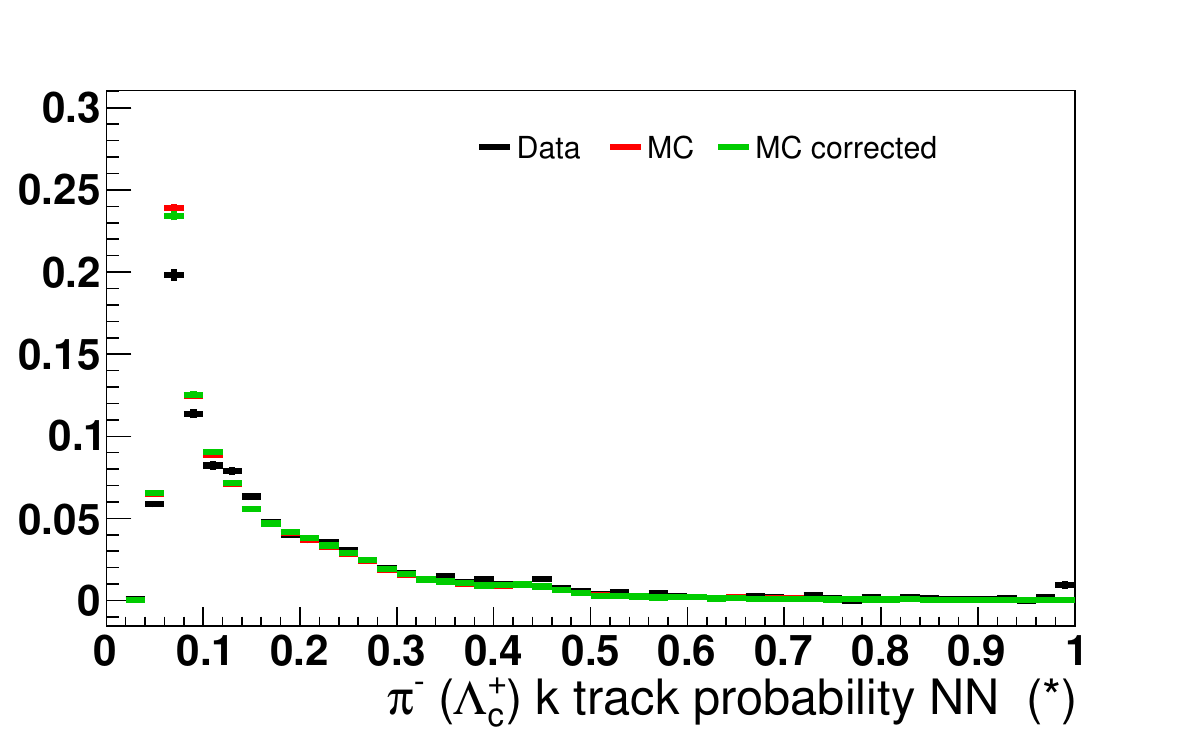}
					
			\includegraphics[width=0.45\linewidth]{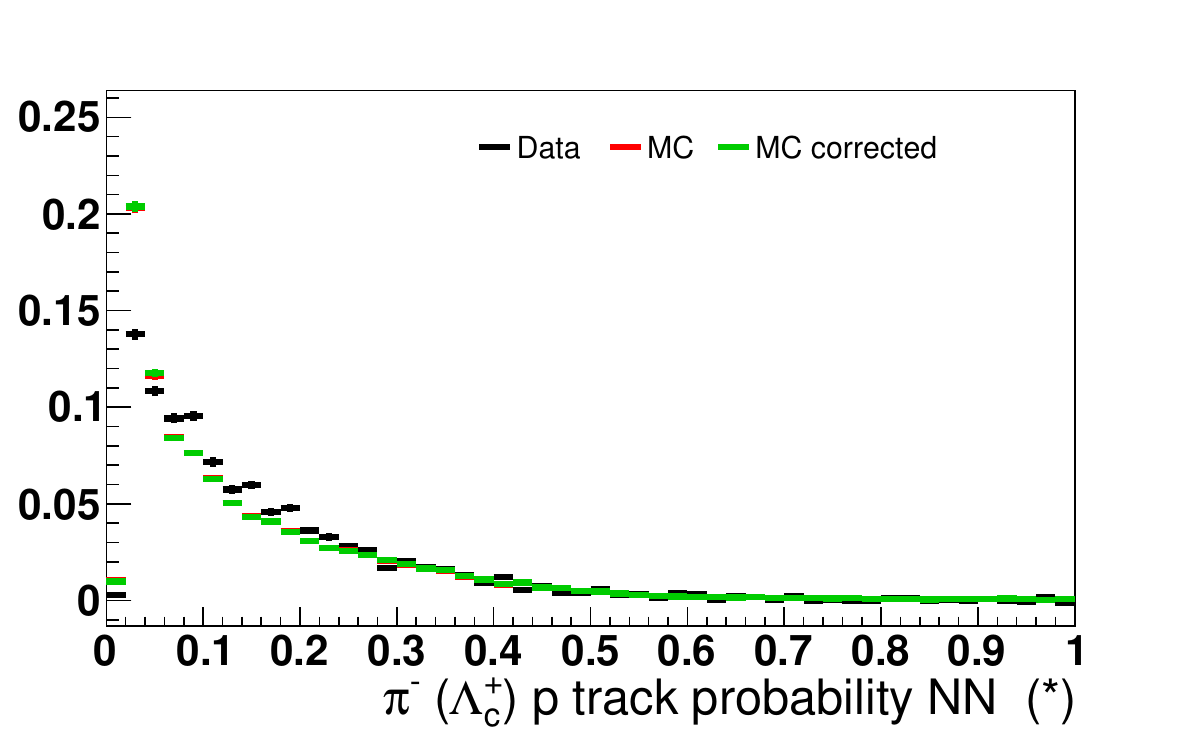}
			\includegraphics[width=0.45\linewidth]{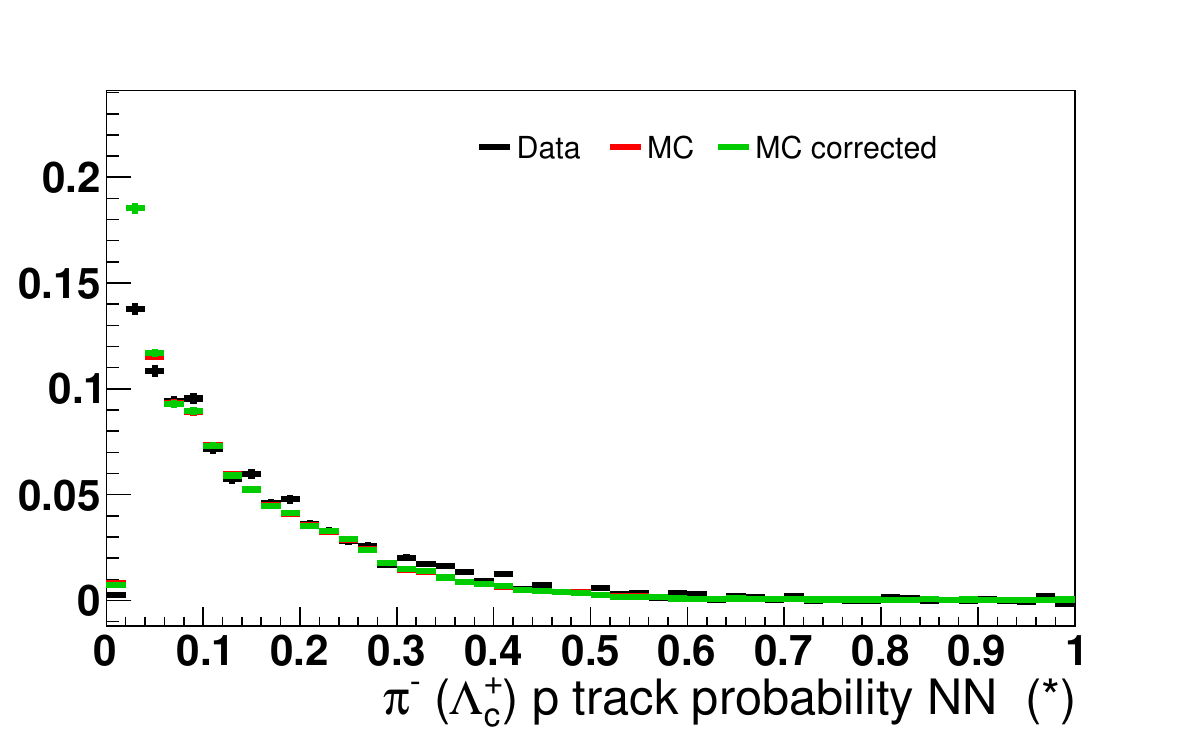}

			\includegraphics[width=0.45\linewidth]{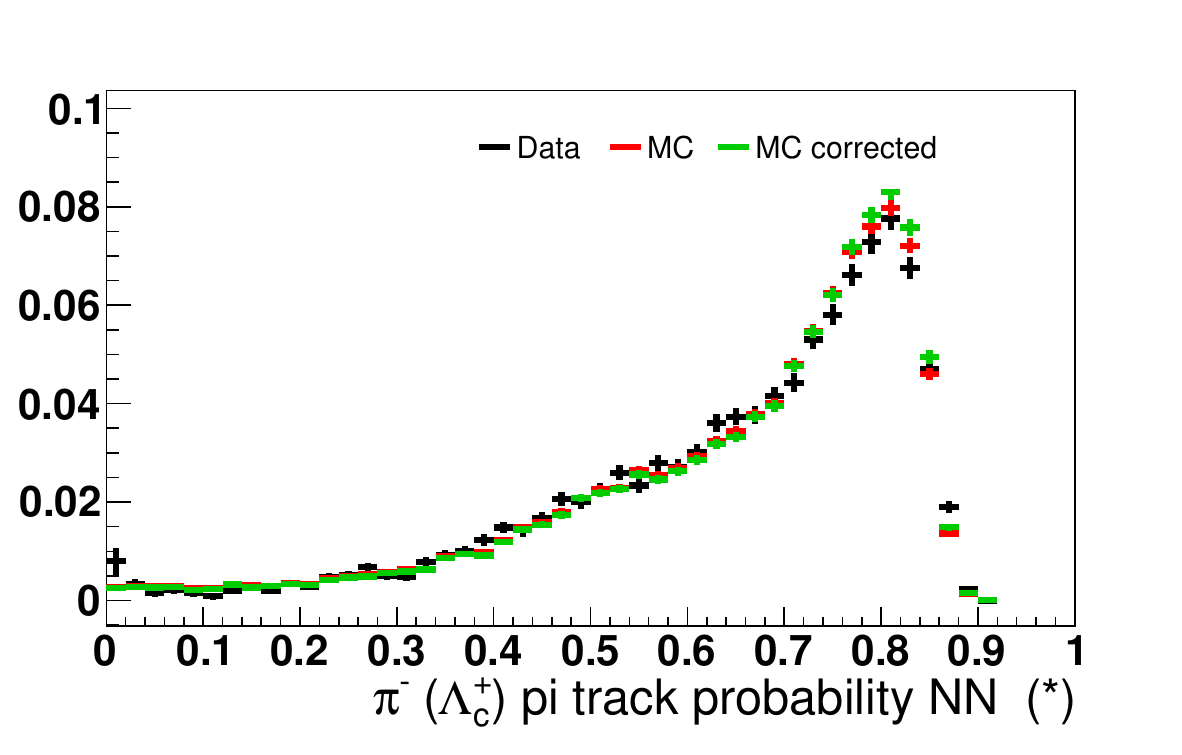}
			\includegraphics[width=0.45\linewidth]{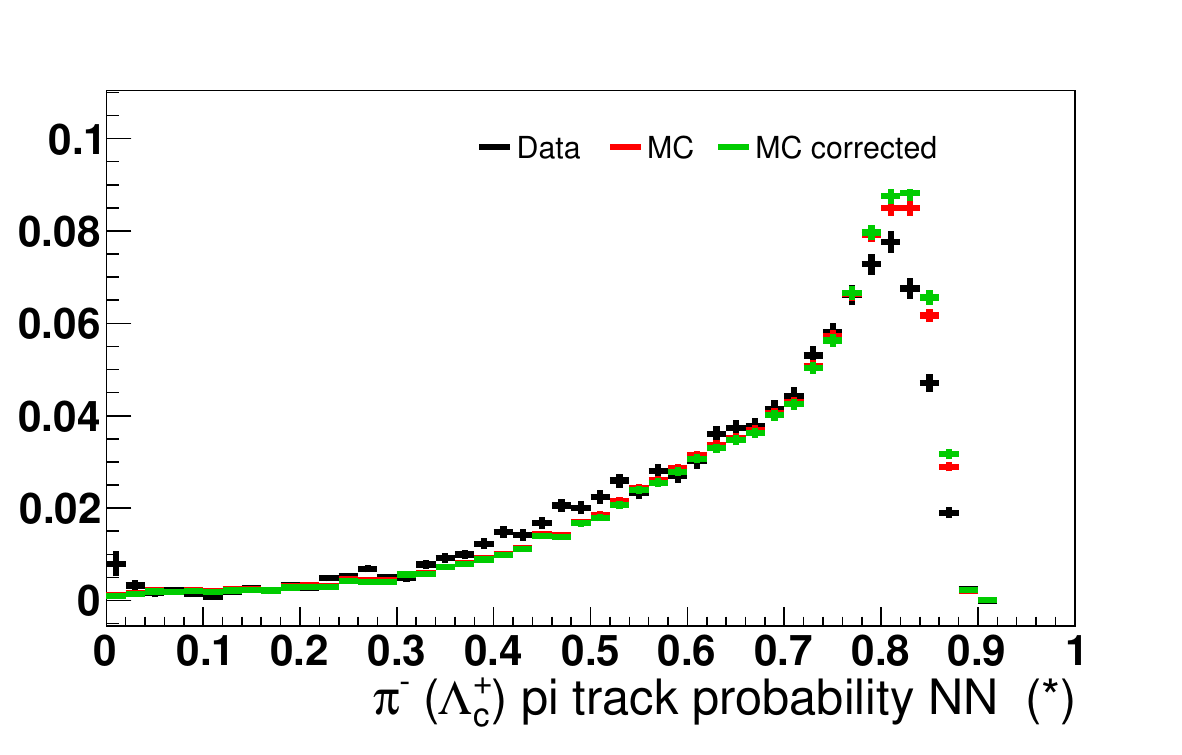}

					\caption{PID variables of the negatively charged pion originated in the \Lc decay. The legend "MC corrected" refers to the reweighting of MC events, which is not related to the PID variables. This figure and the following ones have the objective of comparing the agreement of PID variables in data and MC for the (left) default PID and the (right) \texttt{PIDCorr}-corrected one. The distributions are normalized to unity and the vertical axes represent the normalized number of events per bin. \onlyANA{\rev{Duplicated (below I show one PID variable in each figure, for the three pions)}}} 
					\label{fig:PIDCompBach}
				\end{figure}

		Data/Mc agreement before and after the PID correction is shown in Figs.~\ref{fig:PIDCompBach}, \ref{fig:PIDCompLambda1} and \ref{fig:PIDCompLambda2}.
		

\onlyANA{
			
		\begin{figure}[h!]
			\centering
			
			\includegraphics[width=0.45\linewidth]{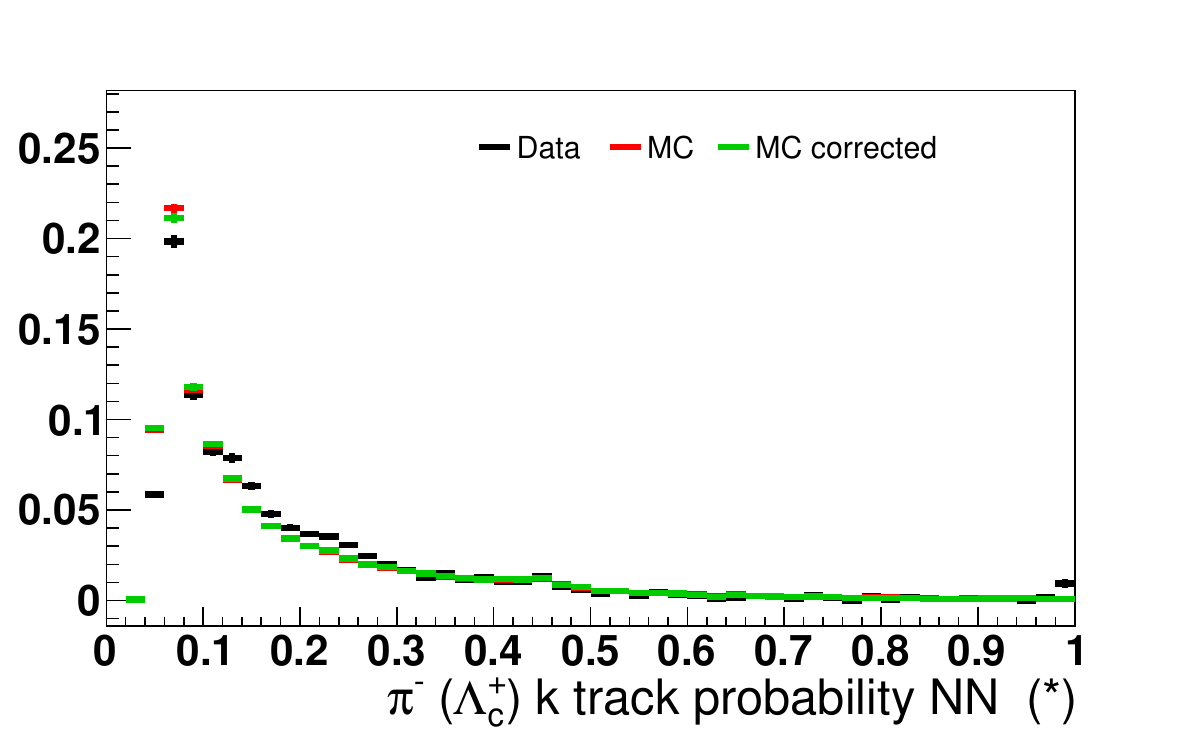}
			\includegraphics[width=0.45\linewidth]{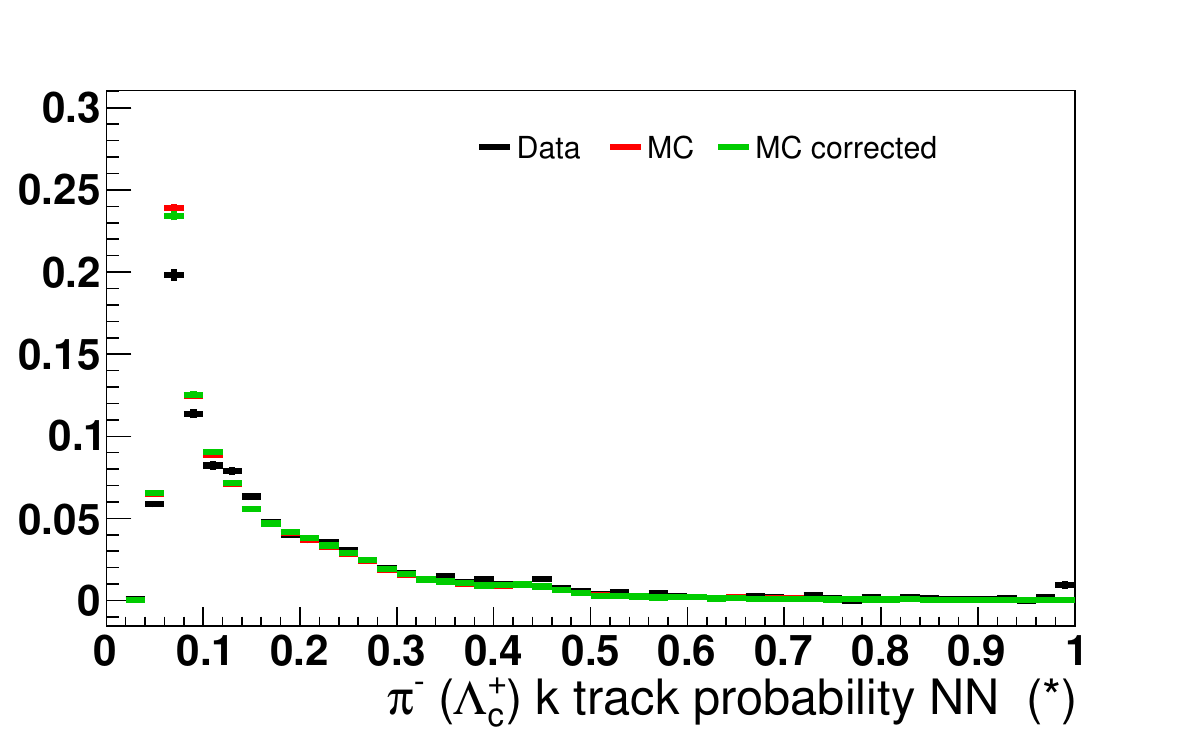}
			
			\includegraphics[width=0.45\linewidth]{content/figs/compDataMC/PIDNotCorrectedPlus3DD/Variables/(pip1_Lc_ProbNNk)03.pdf}
			\includegraphics[width=0.45\linewidth]{content/figs/compDataMC/PIDCorrectedPlus3DD/Variables/(pip1_Lc_ProbNNk)03.pdf}
			
			\includegraphics[width=0.45\linewidth]{content/figs/compDataMC/PIDNotCorrectedPlus3DD/Variables/(pip2_Lc_ProbNNk)03.pdf}
			\includegraphics[width=0.45\linewidth]{content/figs/compDataMC/PIDCorrectedPlus3DD/Variables/(pip2_Lc_ProbNNk)03.pdf}
			\caption{PID variables of the three pions originated in the \Lc decay. The legend "MC corrected" refers to the reweighting of MC events, which is not related to the PID variables. This (and following) Figures have the objective of comparing the agreement of PID variables in data and MC for the default PID (left) and the \texttt{PIDCorr}-corrected one (right). } 
			\label{fig:PIDComp1ana}
		\end{figure}
		
		\begin{figure} 
			\centering

			\includegraphics[width=0.45\linewidth]{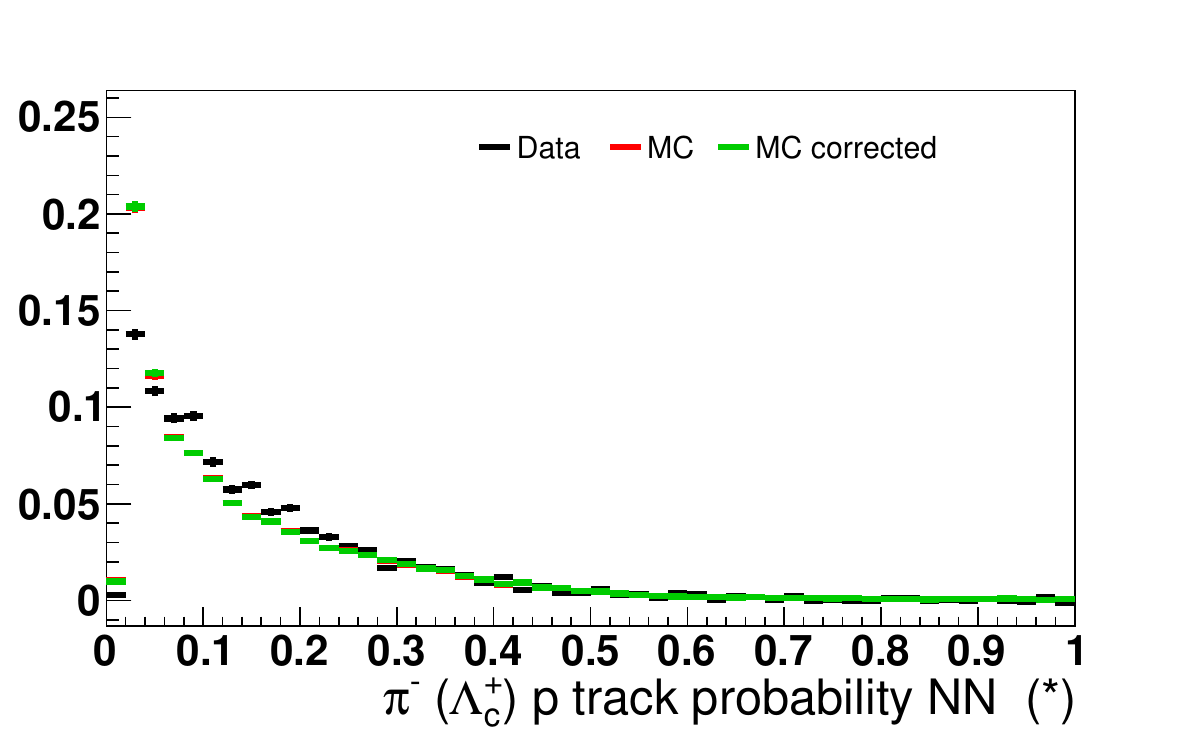}
			\includegraphics[width=0.45\linewidth]{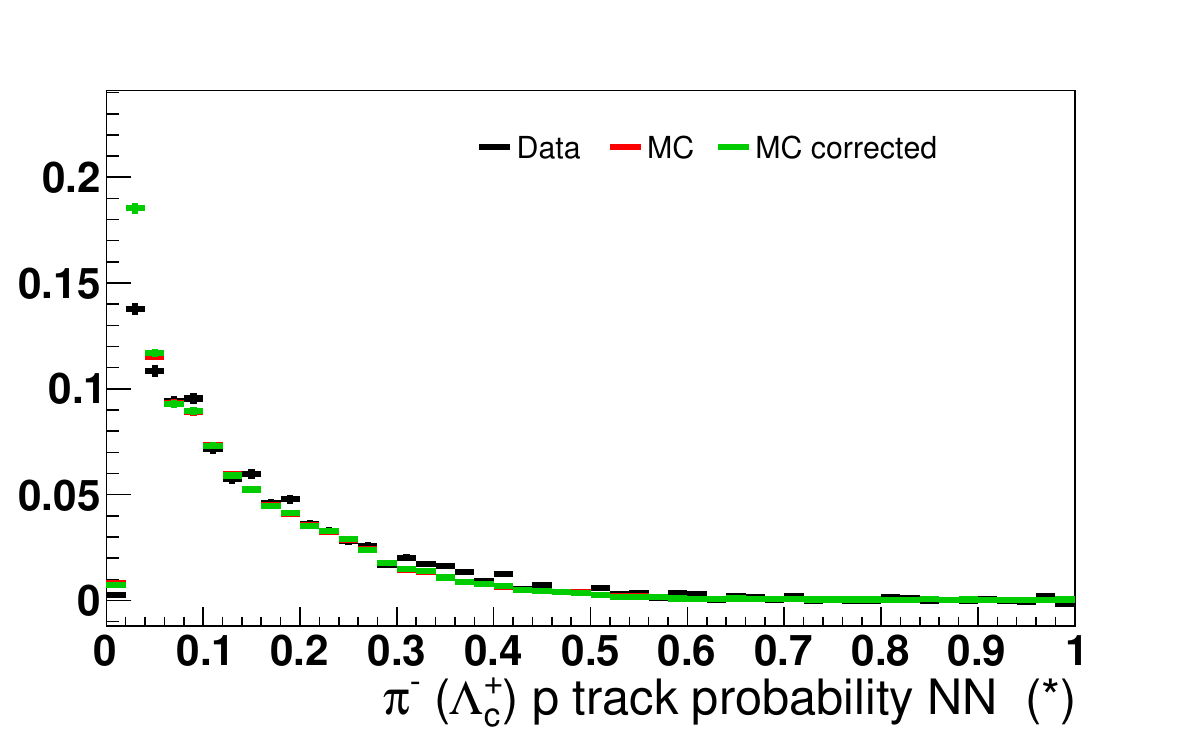}
			
			\includegraphics[width=0.45\linewidth]{content/figs/compDataMC/PIDNotCorrectedPlus3DD/Variables/(pip1_Lc_ProbNNp)03.pdf}
			\includegraphics[width=0.45\linewidth]{content/figs/compDataMC/PIDCorrectedPlus3DD/Variables/(pip1_Lc_ProbNNp)03.pdf}
			
			\includegraphics[width=0.45\linewidth]{content/figs/compDataMC/PIDNotCorrectedPlus3DD/Variables/(pip2_Lc_ProbNNp)03.pdf}
			\includegraphics[width=0.45\linewidth]{content/figs/compDataMC/PIDCorrectedPlus3DD/Variables/(pip2_Lc_ProbNNp)03.pdf}
			
			\caption{Same as Figure~\ref{fig:PIDComp1}. Default PID (left); corrected PID (right). } 
				\label{fig:PIDComp2}
		\end{figure}
		
		\begin{figure} 
			\centering

			\includegraphics[width=0.45\linewidth]{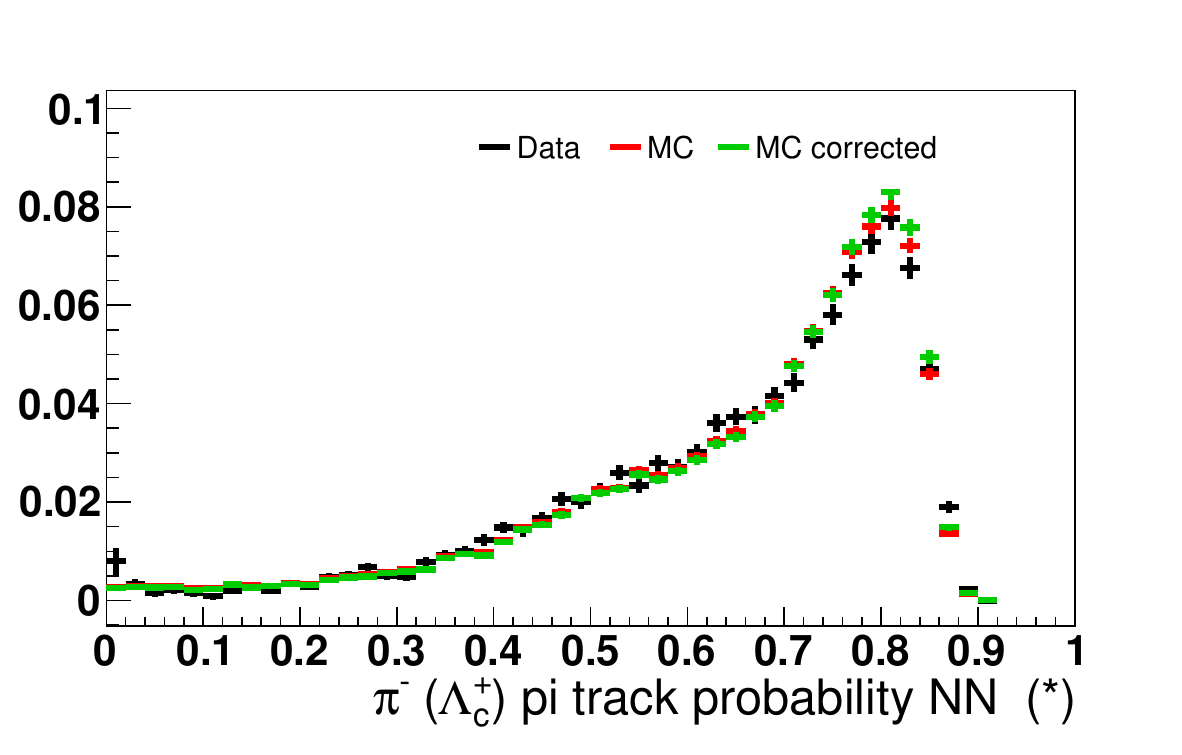}
			\includegraphics[width=0.45\linewidth]{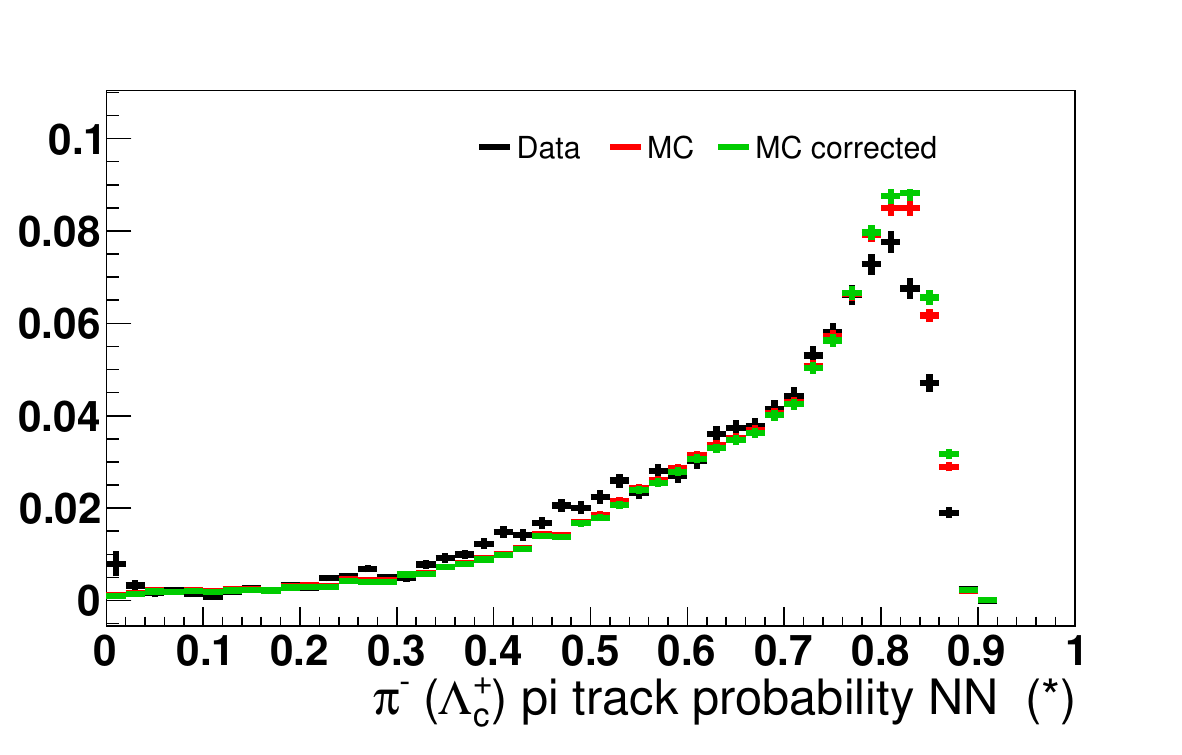}
			
			\includegraphics[width=0.45\linewidth]{content/figs/compDataMC/PIDNotCorrectedPlus3DD/Variables/1-(1-pip1_Lc_ProbNNpi)03.pdf}
			\includegraphics[width=0.45\linewidth]{content/figs/compDataMC/PIDCorrectedPlus3DD/Variables/1-(1-pip1_Lc_ProbNNpi)03.pdf}
			
			\includegraphics[width=0.45\linewidth]{content/figs/compDataMC/PIDNotCorrectedPlus3DD/Variables/1-(1-pip2_Lc_ProbNNpi)03.pdf}
			\includegraphics[width=0.45\linewidth]{content/figs/compDataMC/PIDCorrectedPlus3DD/Variables/1-(1-pip2_Lc_ProbNNpi)03.pdf}
			
			\caption{Same as Figure~\ref{fig:PIDCompBach}. Default PID (left); corrected PID (right). } 
				\label{fig:PIDComp3}
		\end{figure}

}


		\begin{figure}[ht]
			\centering
			
			\includegraphics[width=0.45\linewidth]{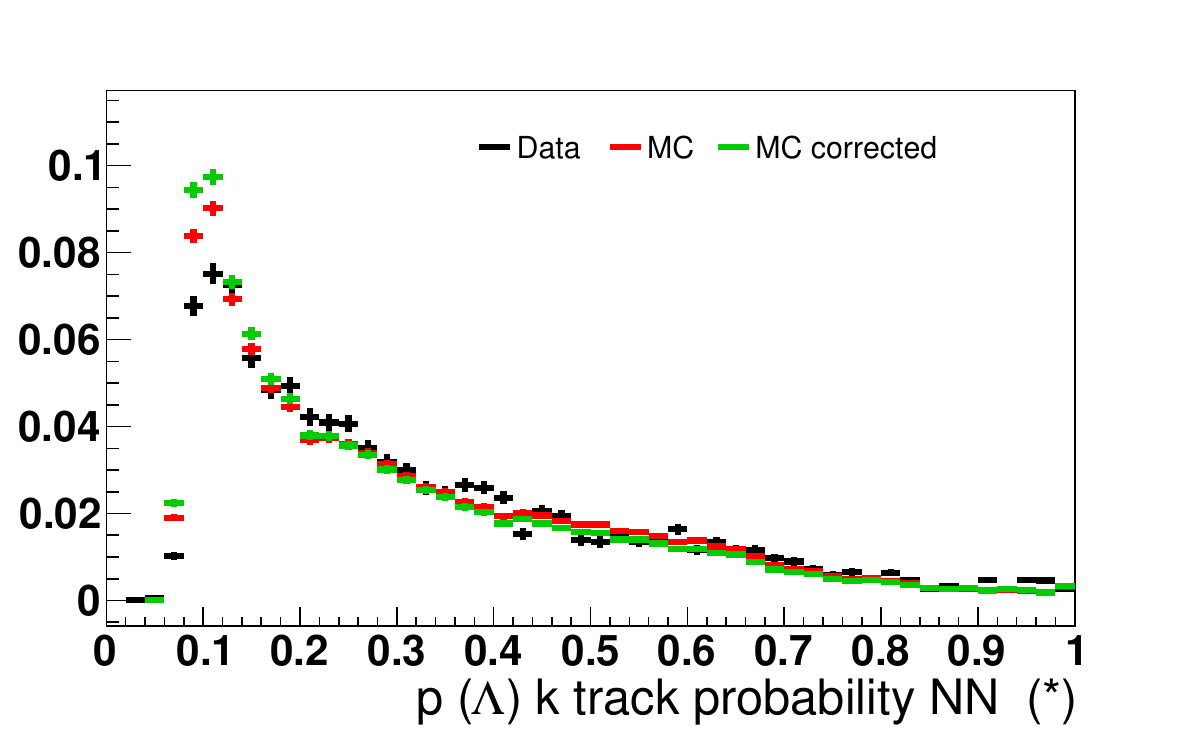}
			\includegraphics[width=0.45\linewidth]{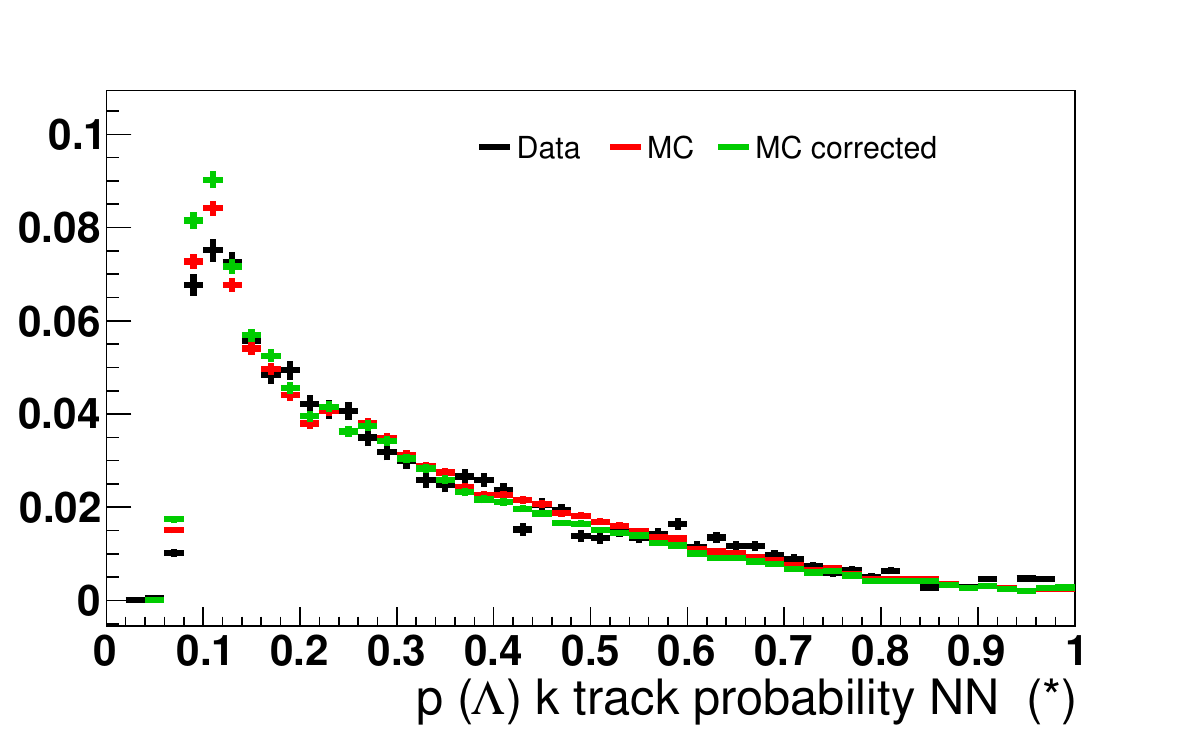}

			\includegraphics[width=0.45\linewidth]{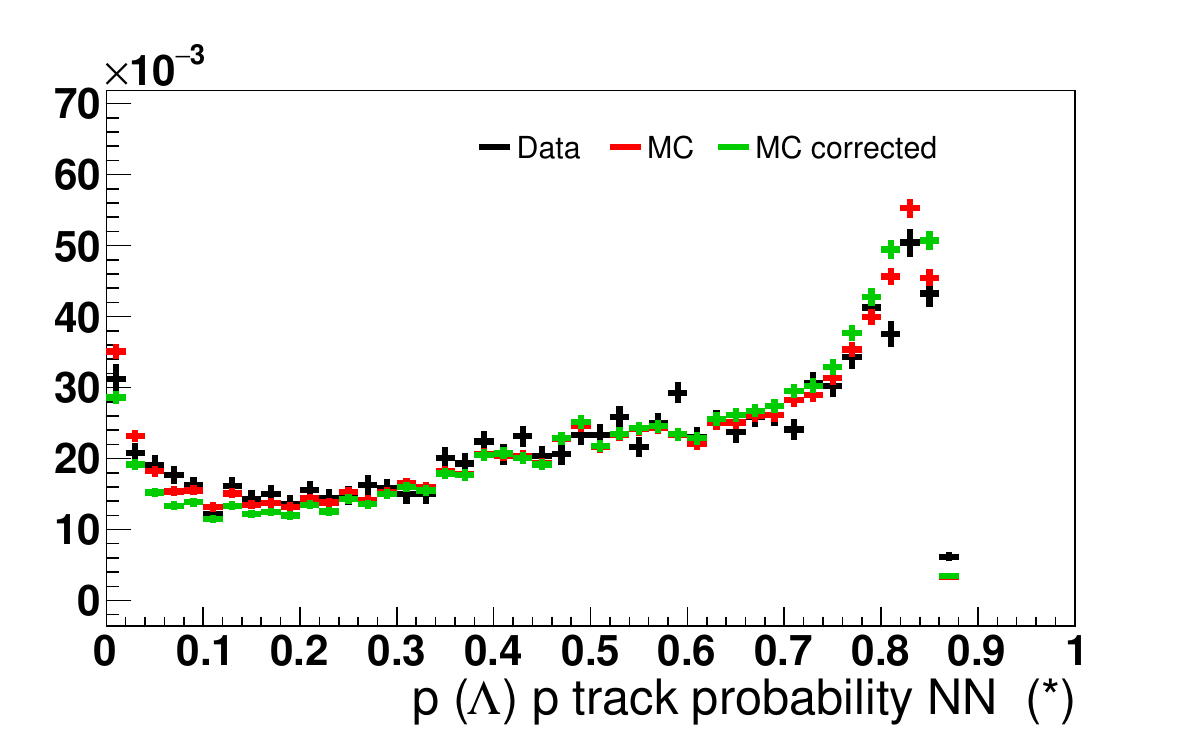}
			\includegraphics[width=0.45\linewidth]{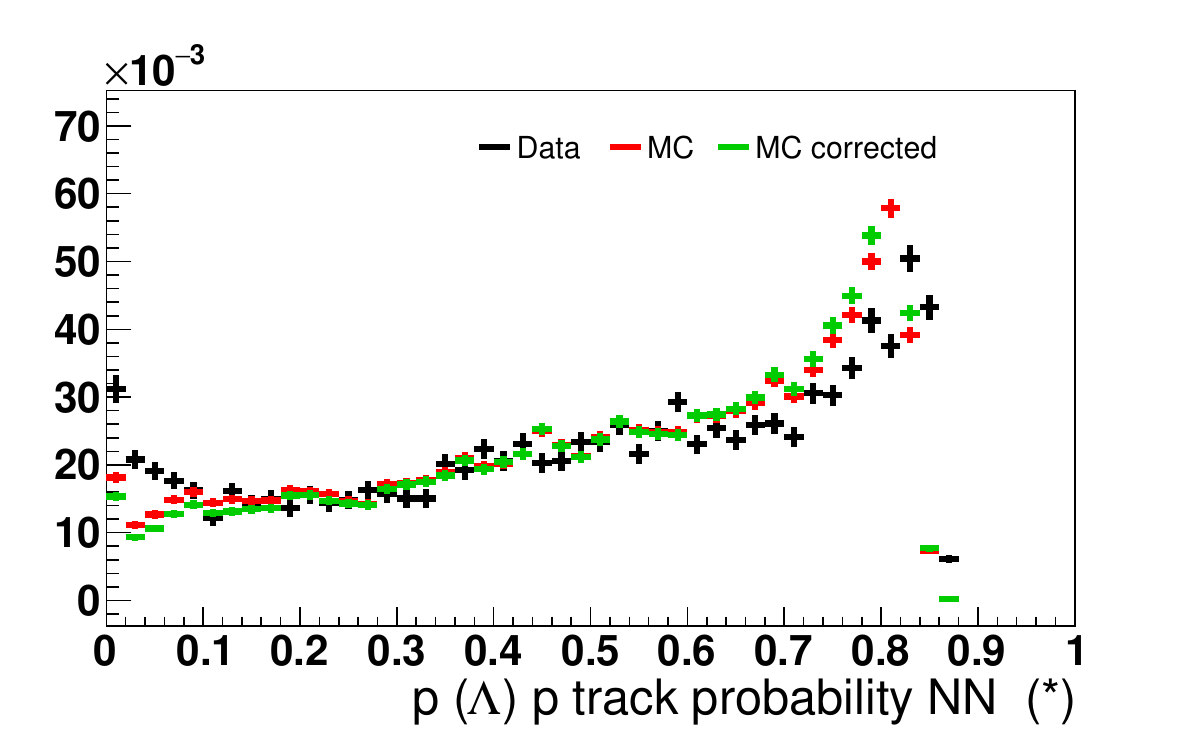}

			\includegraphics[width=0.45\linewidth]{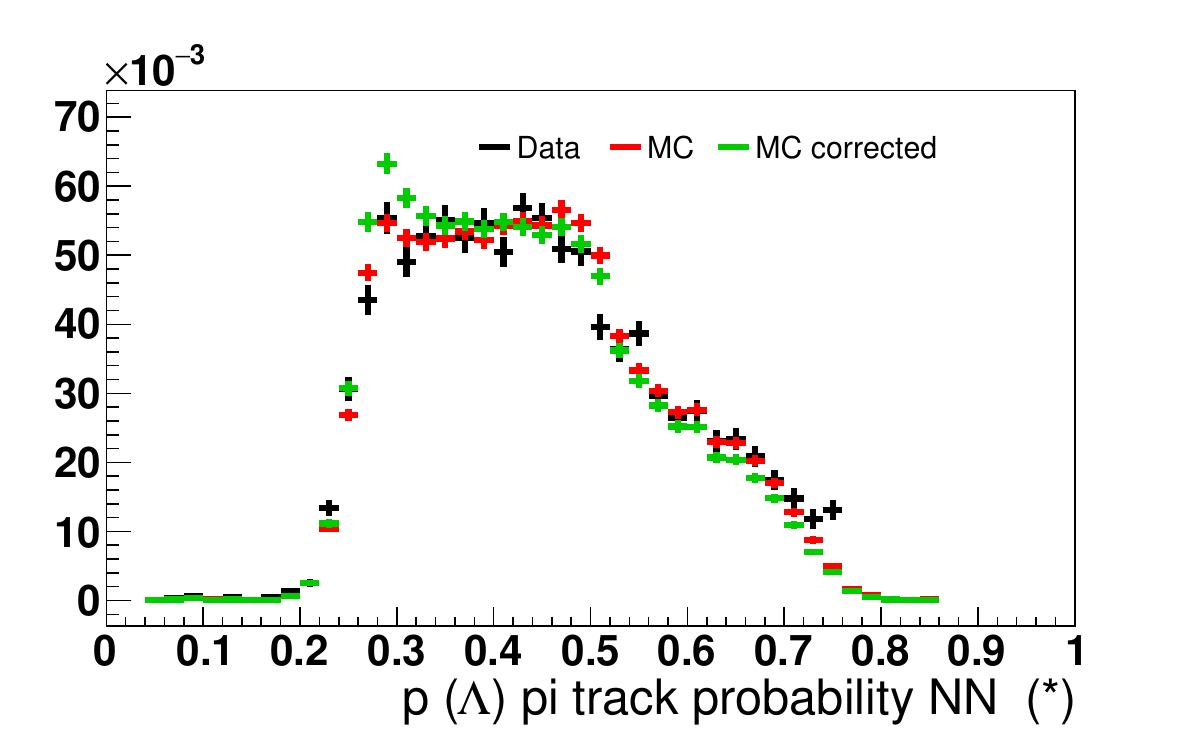}
			\includegraphics[width=0.45\linewidth]{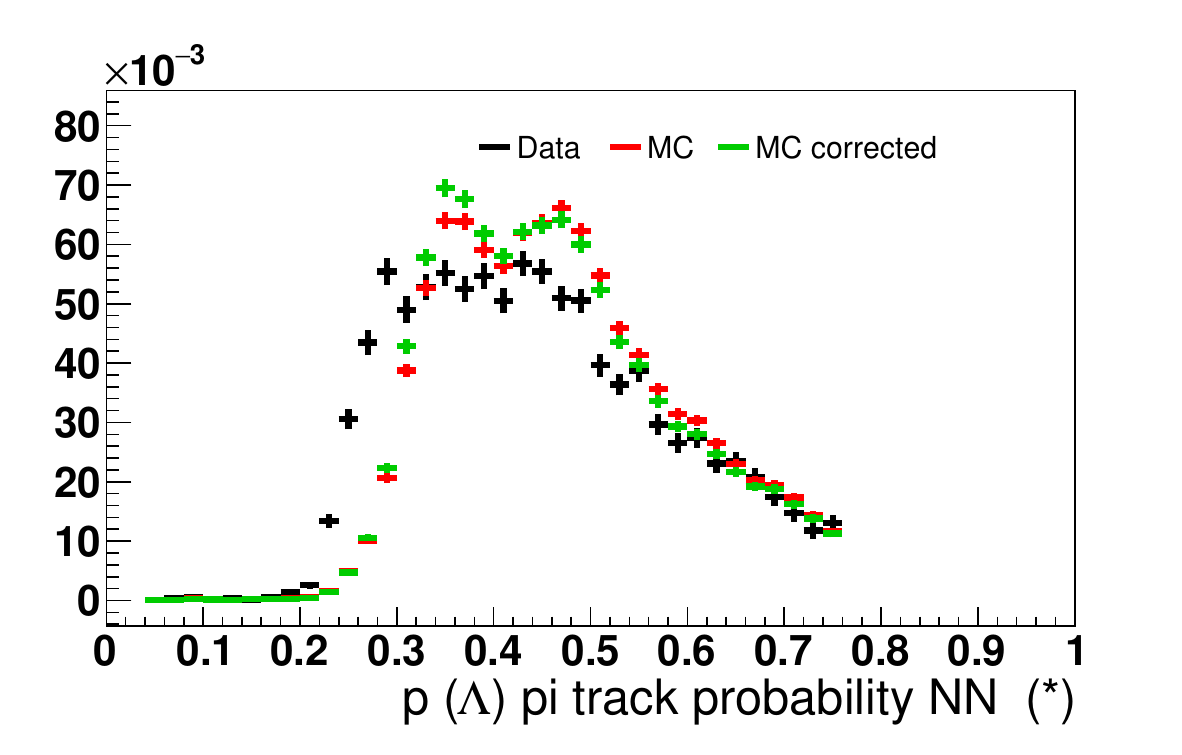}

			\caption{Same as Figure~\ref{fig:PIDCompBach}, for the proton of the \Lz decay. (Left) default PID; (right) corrected PID. } 
				\label{fig:PIDCompLambda1}
		\end{figure}
		
		\begin{figure}[ht!]
			\centering

			\includegraphics[width=0.45\linewidth]{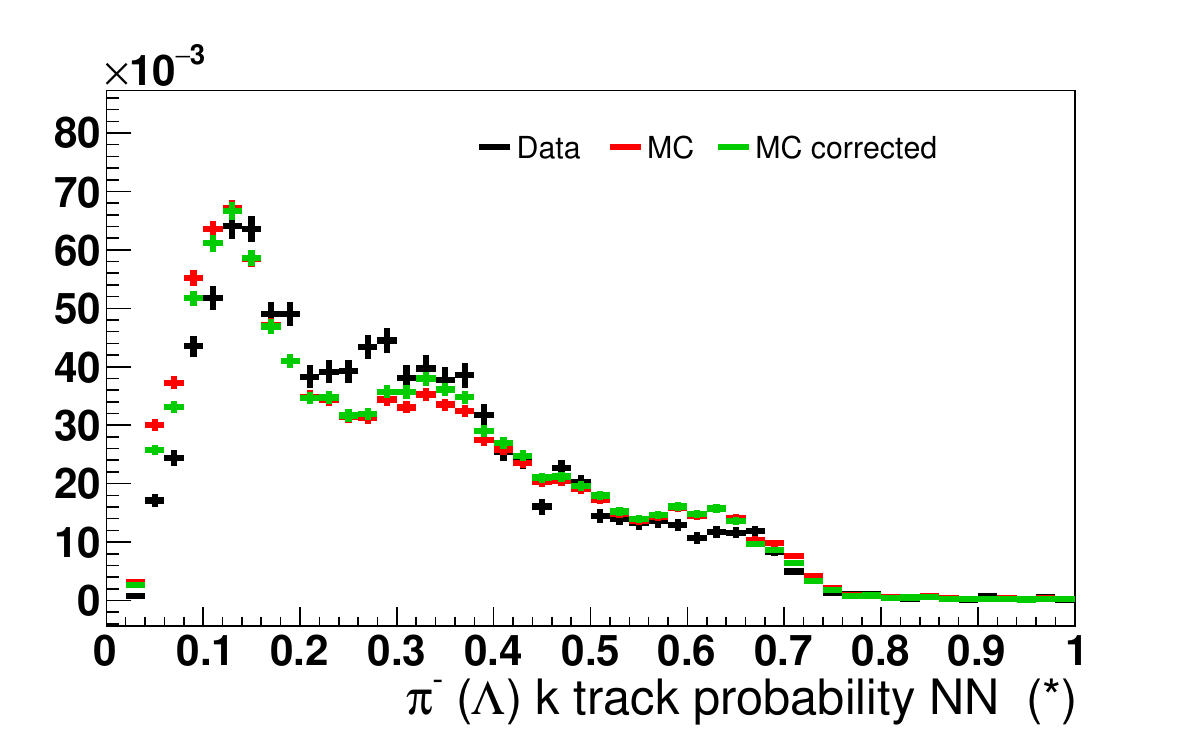}
			\includegraphics[width=0.45\linewidth]{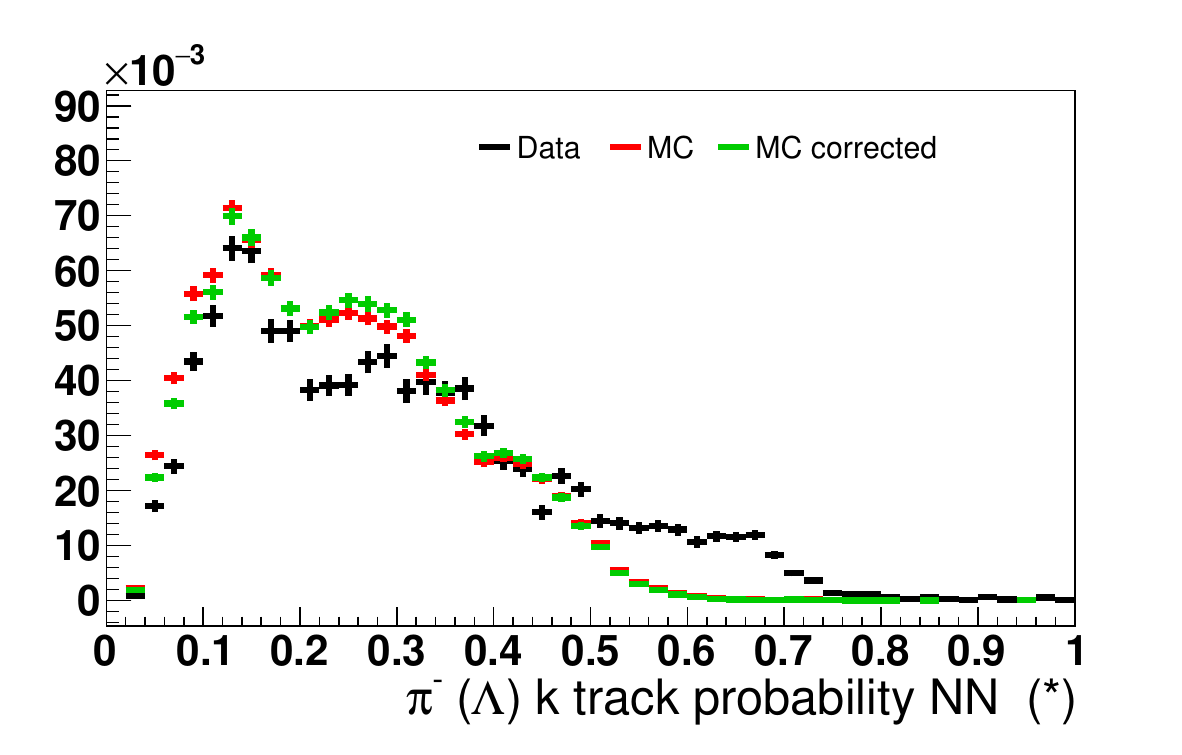}

			\includegraphics[width=0.45\linewidth]{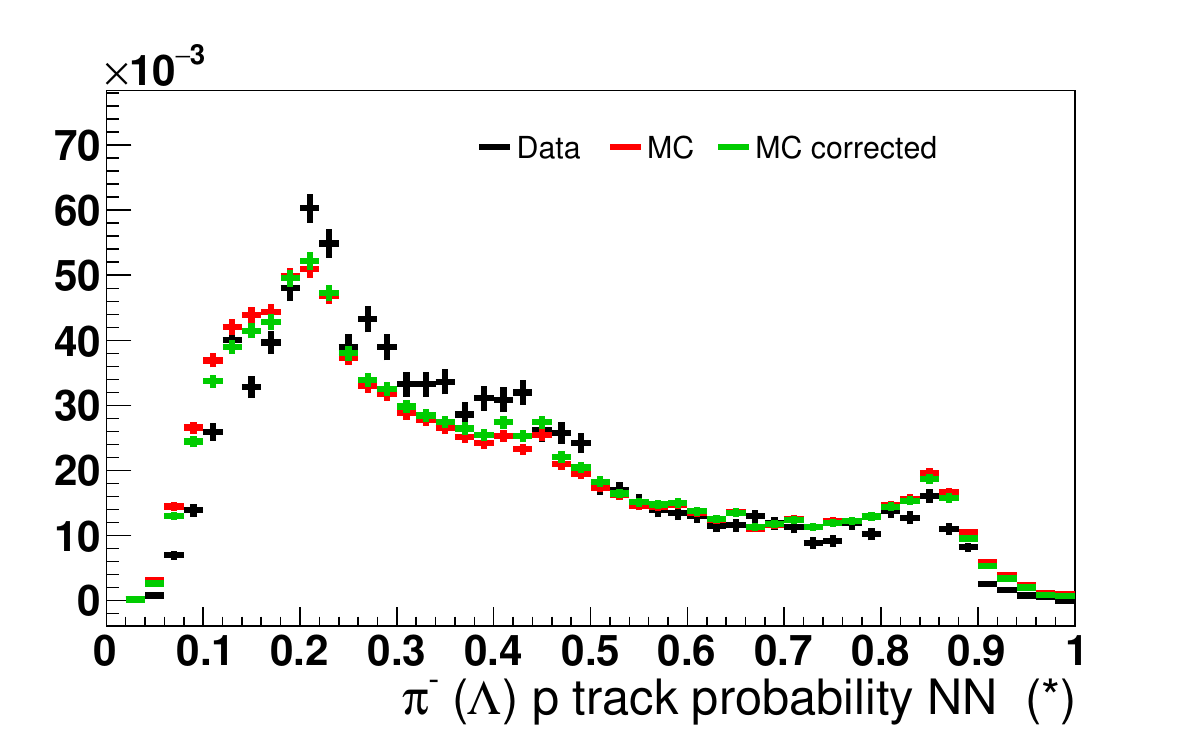}
			\includegraphics[width=0.45\linewidth]{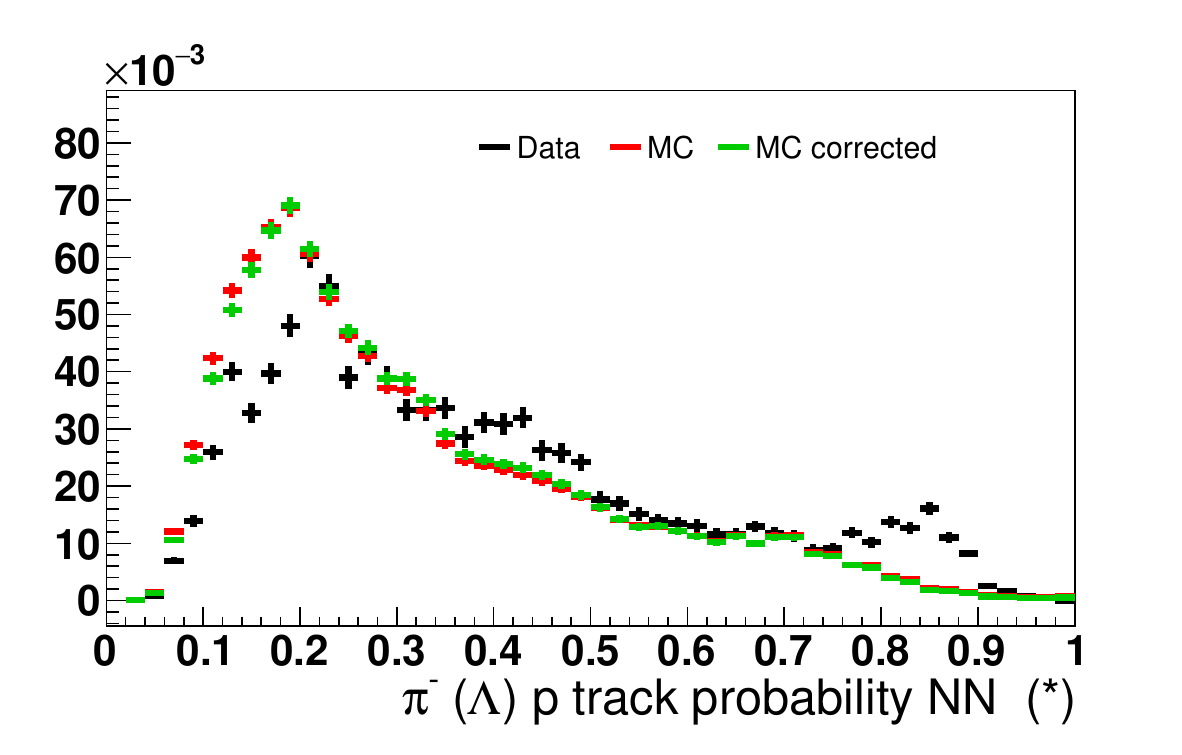}

			\includegraphics[width=0.45\linewidth]{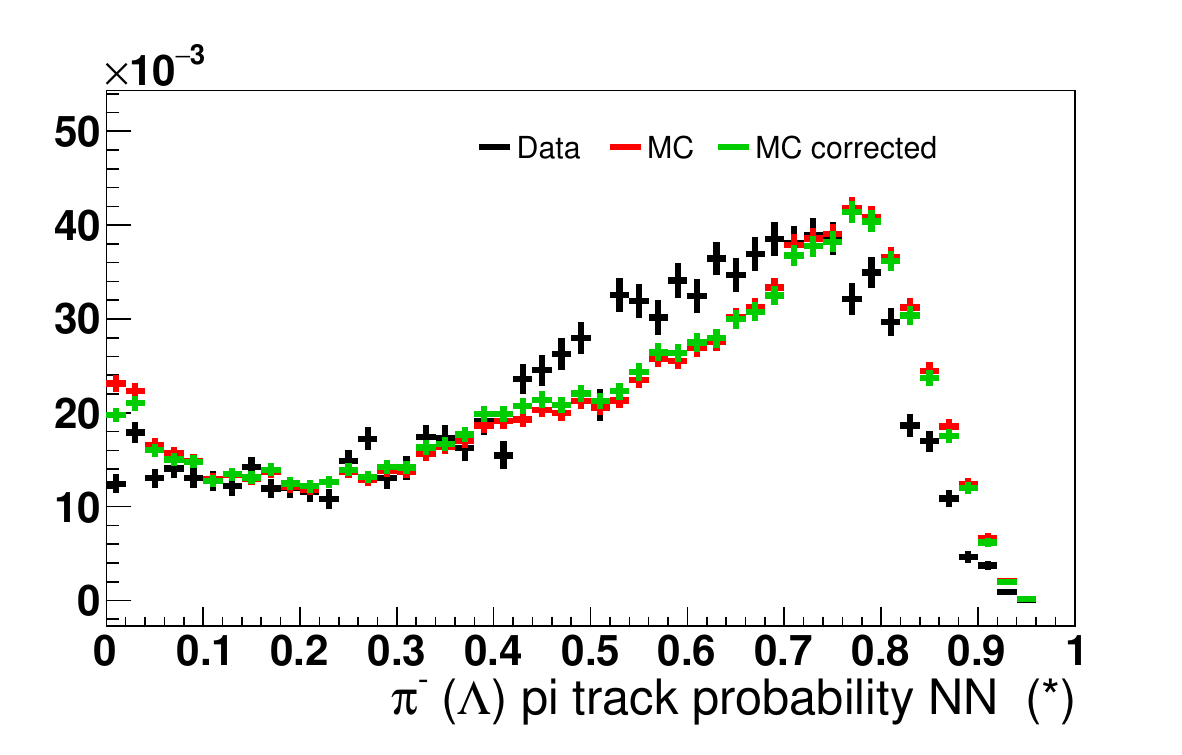}
			\includegraphics[width=0.45\linewidth]{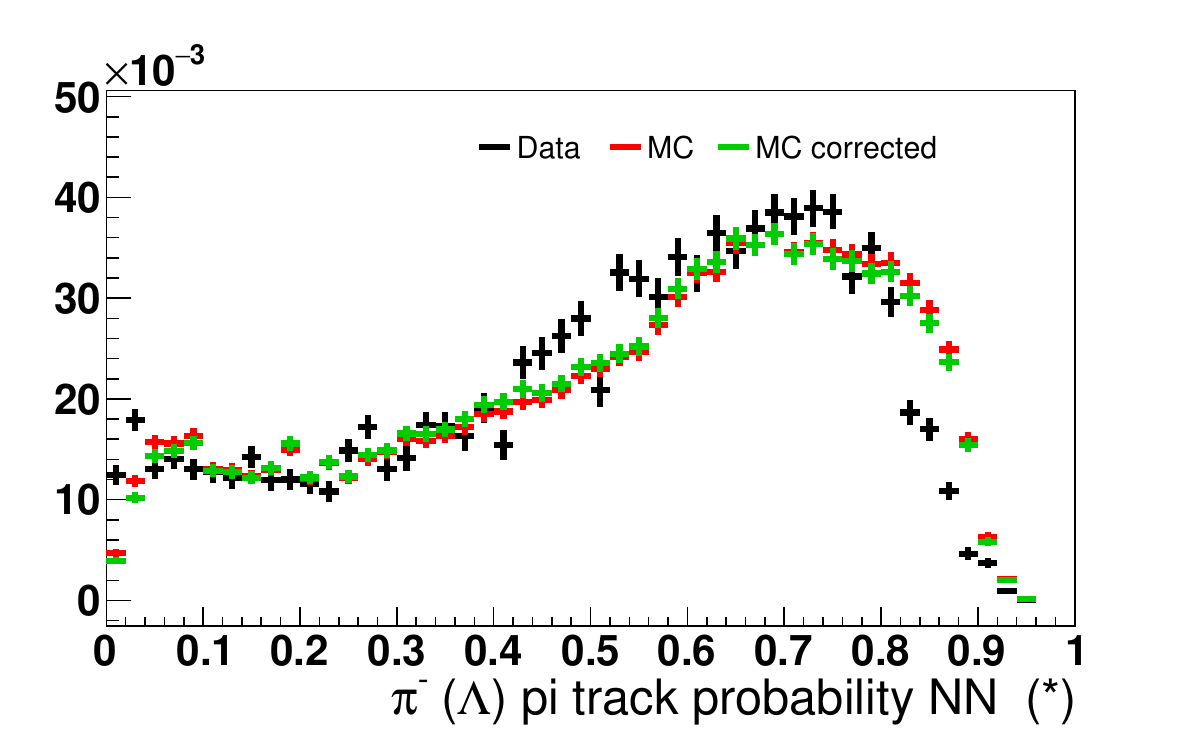}

			\caption{Same as Figure~\ref{fig:PIDCompBach}, for the pion of the \Lz decay. (Left) default PID; (right) corrected PID. } 
		    	\label{fig:PIDCompLambda2}
		\end{figure}
		
		\onlyANA{
		
		\subsubsection{Ghost probability}
		
		\begin{figure} 
			\centering

			\includegraphics[width=0.45\linewidth]{content/figs/compDataMC/PIDNotCorrectedPlus3DD/Variables/(pim_Lc_ProbNNghost)03.pdf}
			\includegraphics[width=0.45\linewidth]{content/figs/compDataMC/PIDNotCorrectedPlus3DD/Variables/(pip1_Lc_ProbNNghost)03.pdf}
			\includegraphics[width=0.45\linewidth]{content/figs/compDataMC/PIDNotCorrectedPlus3DD/Variables/(pip2_Lc_ProbNNghost)03.pdf}
			
			\includegraphics[width=0.45\linewidth]{content/figs/compDataMC/PIDNotCorrectedPlus3DD/Variables/(p_L0_ProbNNghost)03.pdf}
			\includegraphics[width=0.45\linewidth]{content/figs/compDataMC/PIDNotCorrectedPlus3DD/Variables/(pi_L0_ProbNNghost)03.pdf}
			
			\caption{ProbNNGhost. The Ghosts are not corrected. } 
		\end{figure}
		
		\begin{figure} 
			\centering

			\includegraphics[width=0.45\linewidth]{content/figs/compDataMC/CompAllVarsDataCorrMCPur30Plus3DD/Variables/log(pim_Lc_TRACK_GhostProb).pdf}
			\includegraphics[width=0.45\linewidth]{content/figs/compDataMC/CompAllVarsDataCorrMCPur30Plus3DD/Variables/log(pip1_Lc_TRACK_GhostProb).pdf}
			\includegraphics[width=0.45\linewidth]{content/figs/compDataMC/CompAllVarsDataCorrMCPur30Plus3DD/Variables/log(pip2_Lc_TRACK_GhostProb).pdf}
			
			\includegraphics[width=0.45\linewidth]{content/figs/compDataMC/CompAllVarsDataCorrMCPur30Plus3DD/Variables/log(p_L0_TRACK_GhostProb).pdf}
			\includegraphics[width=0.45\linewidth]{content/figs/compDataMC/CompAllVarsDataCorrMCPur30Plus3DD/Variables/log(pi_L0_TRACK_GhostProb).pdf}
			
			\caption{Ghost track probability. The Ghosts are not corrected. } 
		\end{figure}
		
	}

\section{Data/MC agreement after MC reweighting}

\begin{figure}[ht!]
	\centering
	\includegraphics[width=0.45\linewidth]{content/figs/compDataMC/CompAllVarsDataCorrMCPur30Plus3DD/Variables/Lc_PT}
	\includegraphics[width=0.45\linewidth]{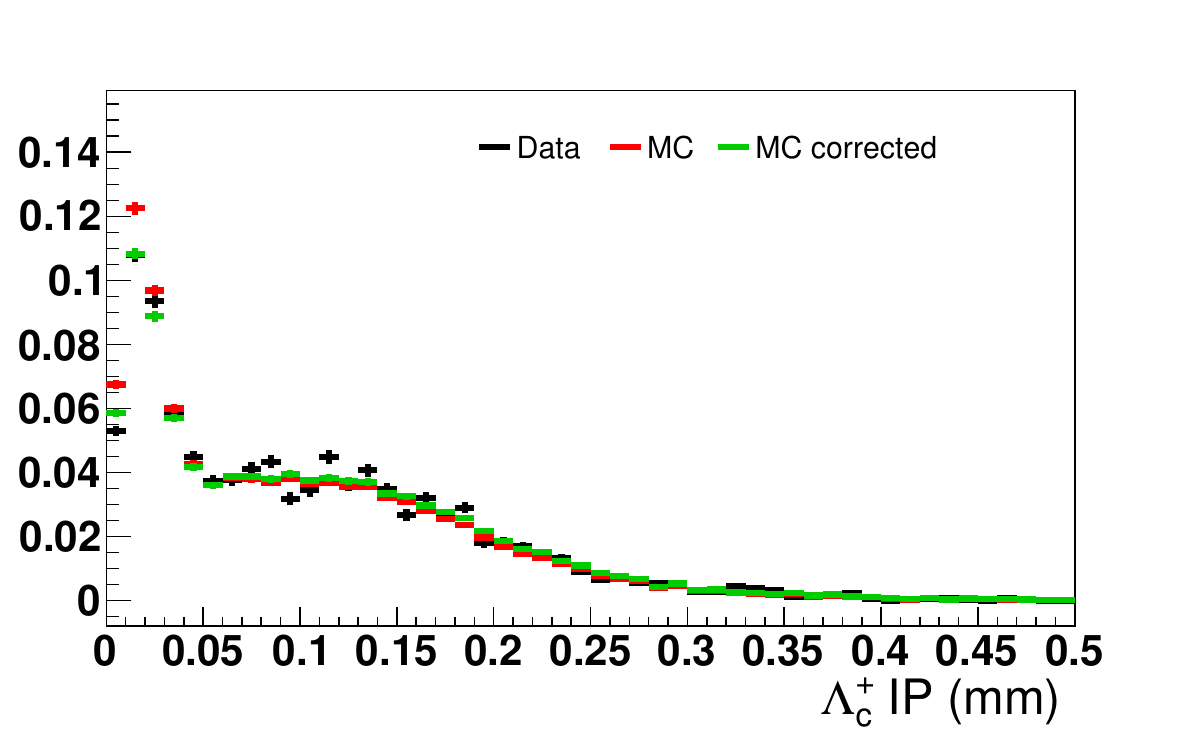}
	\includegraphics[width=0.45\linewidth]{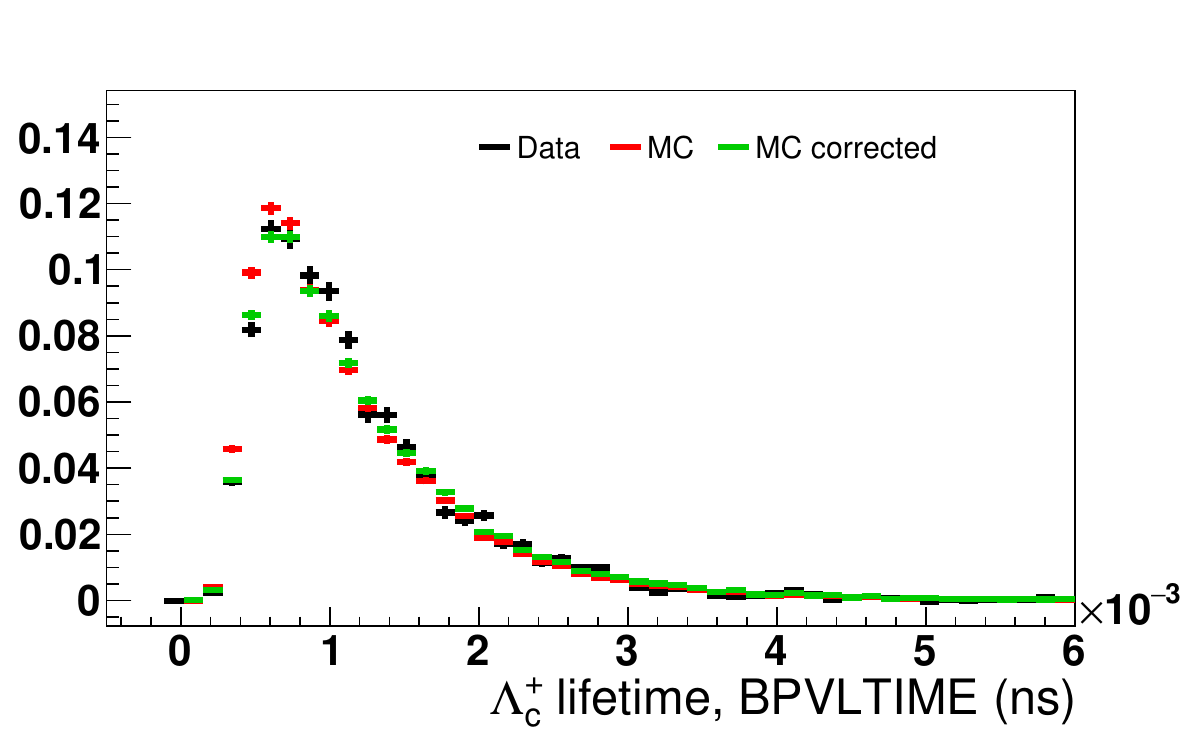}
	\includegraphics[width=0.45\linewidth]{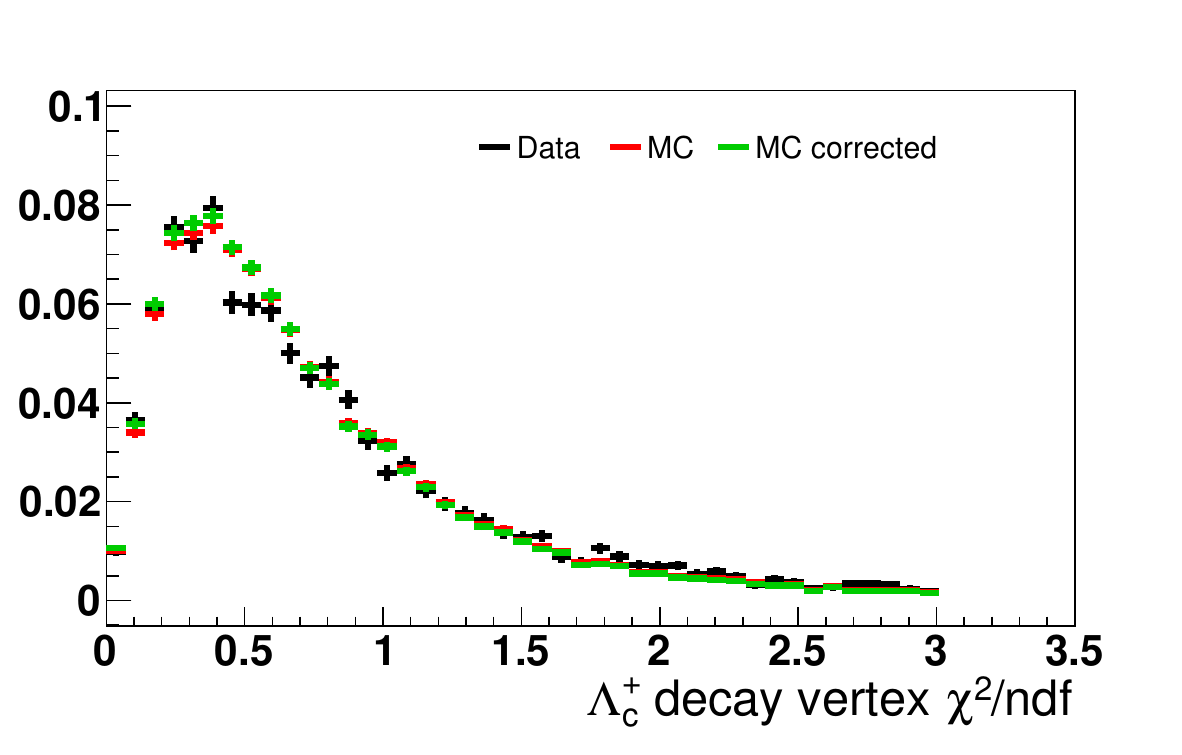}
	\includegraphics[width=0.45\linewidth]{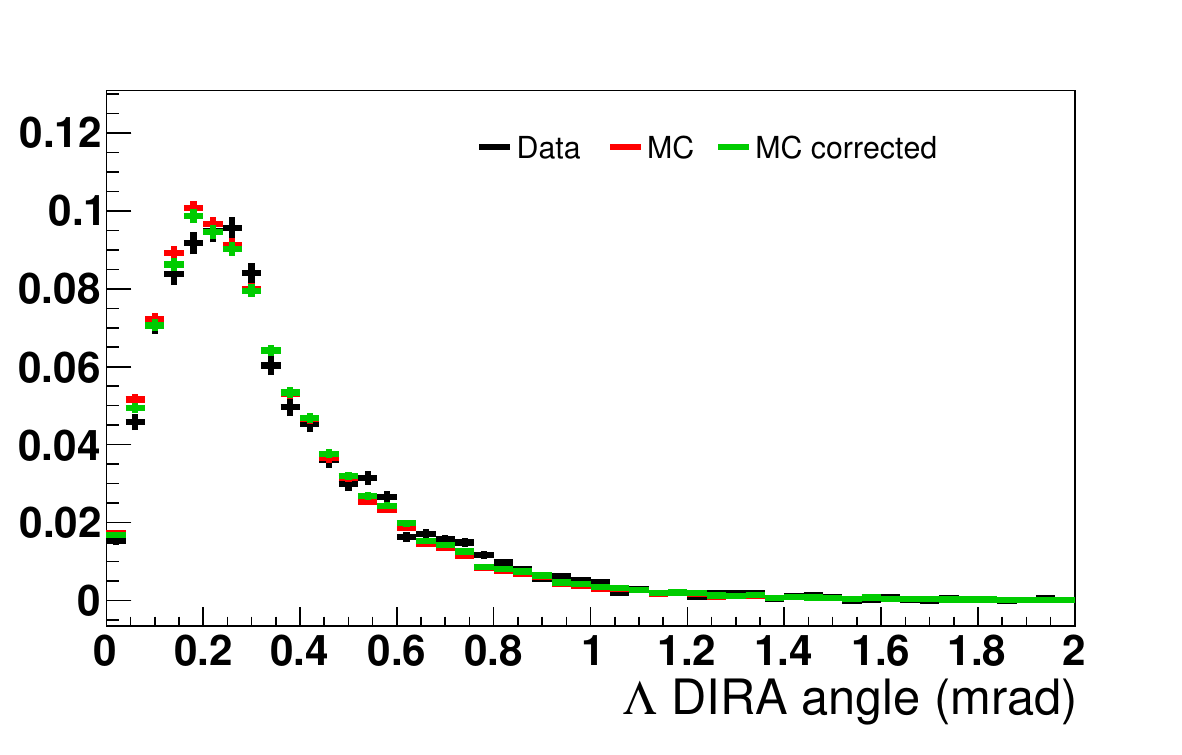}
	\includegraphics[width=0.45\linewidth]{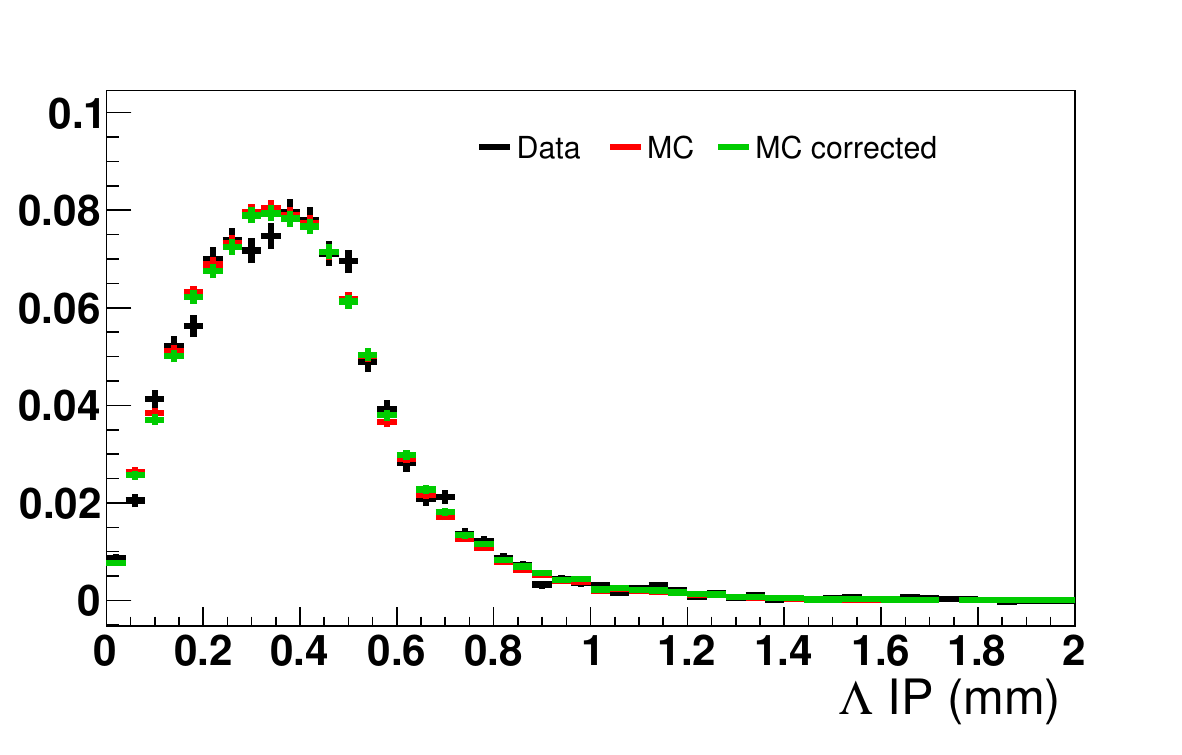}
	\includegraphics[width=0.45\linewidth]{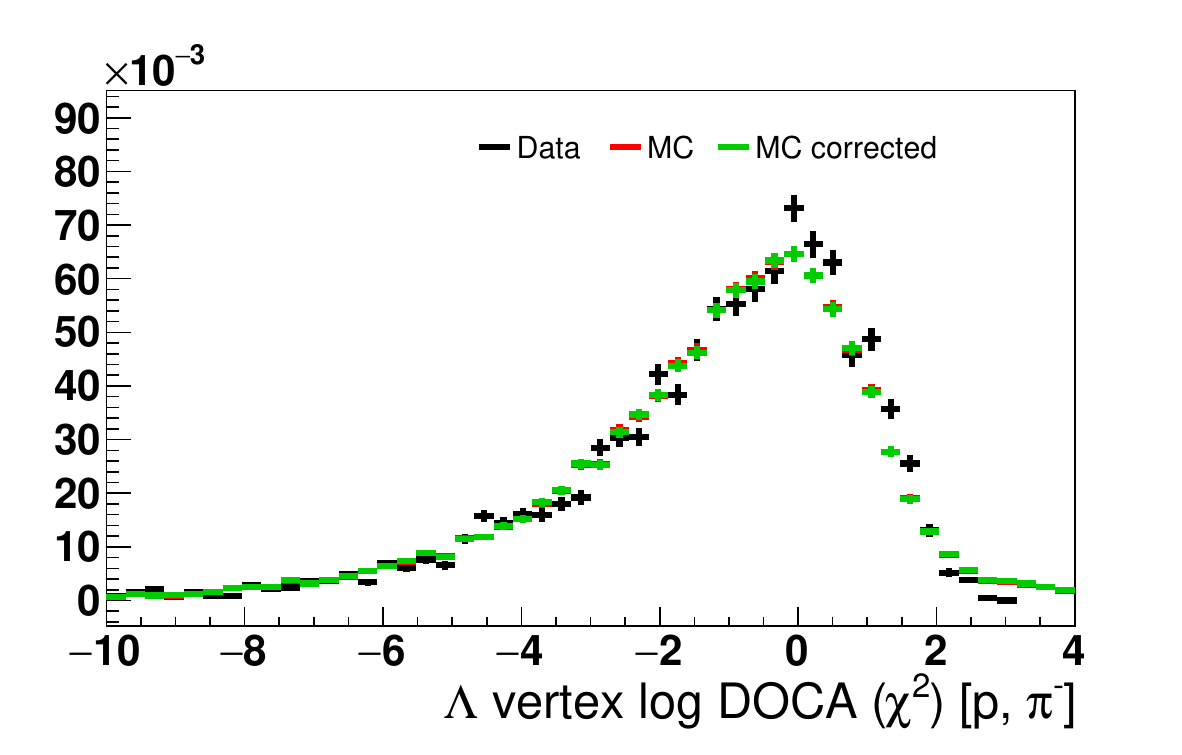}
	\includegraphics[width=0.45\linewidth]{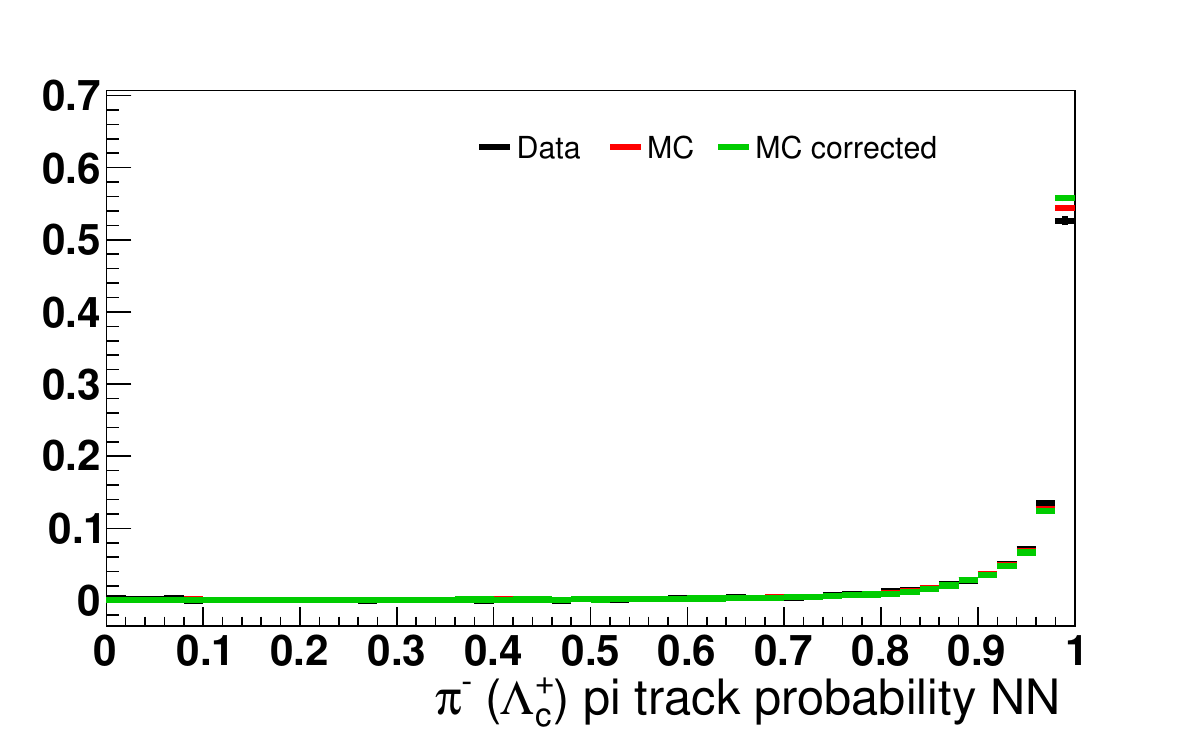}

	\caption{Data/MC agreement for the variables used in the final BDT configuration. The MC distributions are shown before (red) and after (green) the event reweighting introduced to match the $p(\Lc)$ and $p_T(\Lc)$ distributions of sWeighted data. The distributions are normalized to unity and the vertical axes represent the normalized number of events per bin.}
	\label{fig:BDTvars}
\end{figure}

\begin{figure}[ht!]
	\centering
	\includegraphics[width=0.45\linewidth]{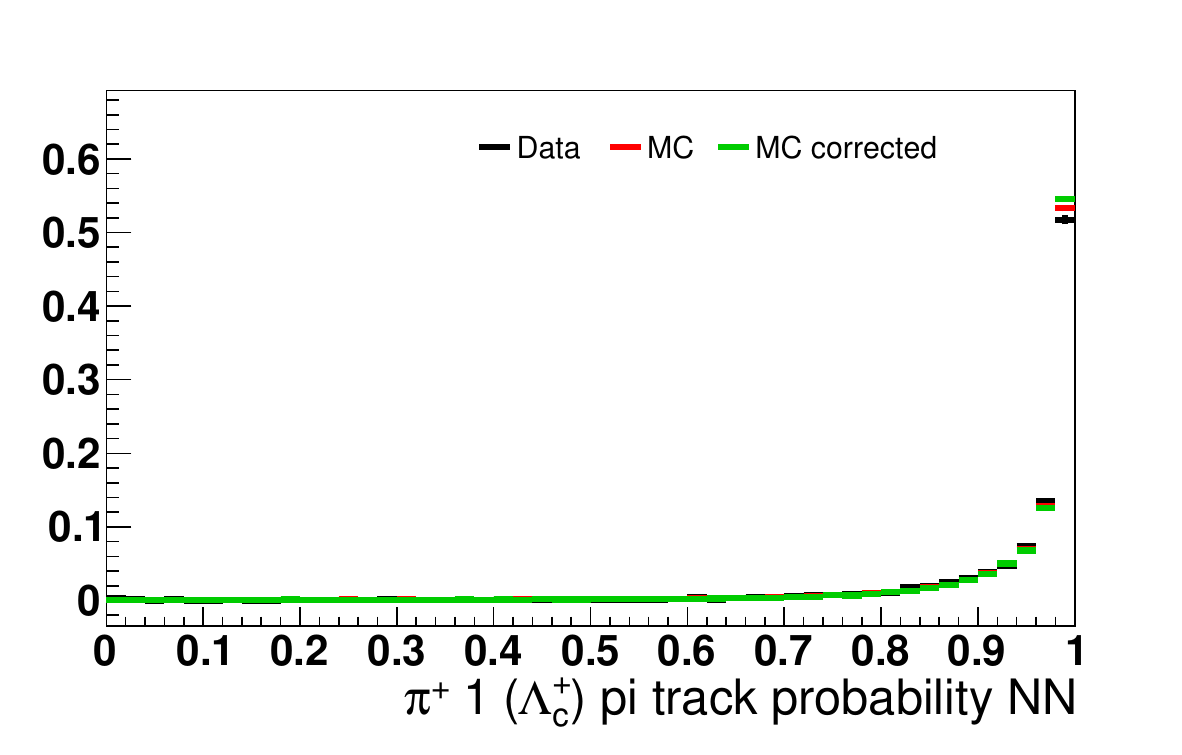}
	\includegraphics[width=0.45\linewidth]{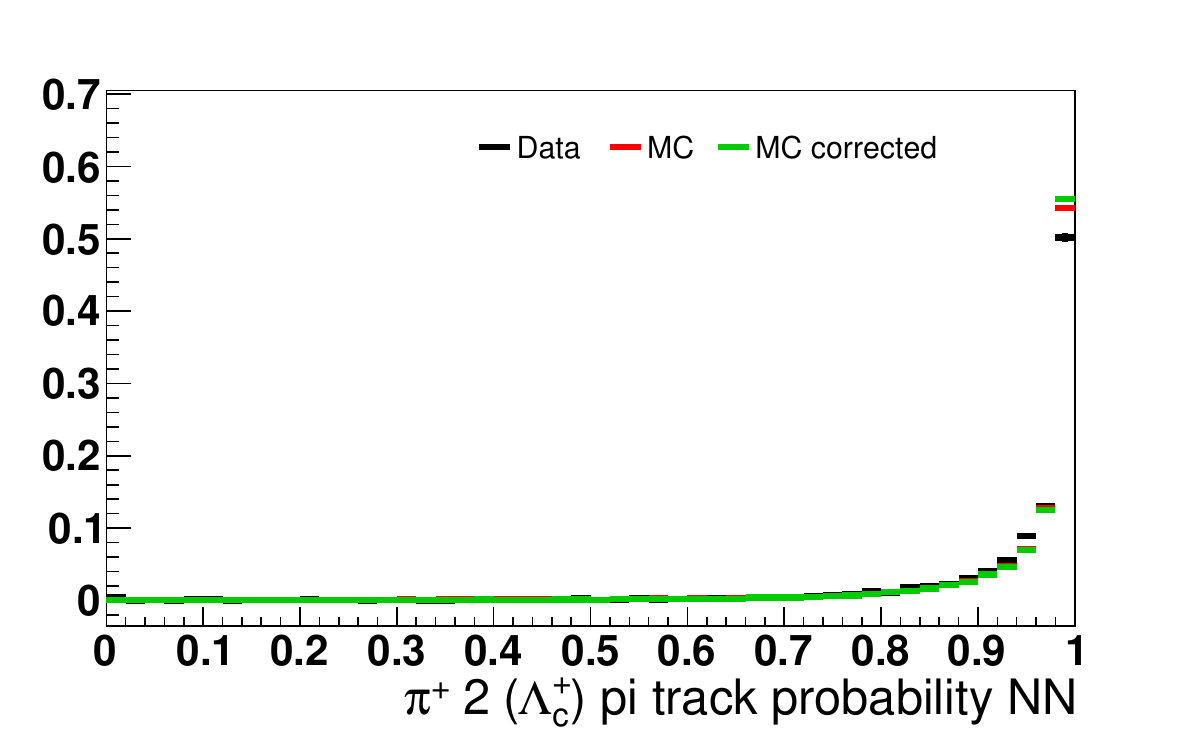}
	\includegraphics[width=0.45\linewidth]{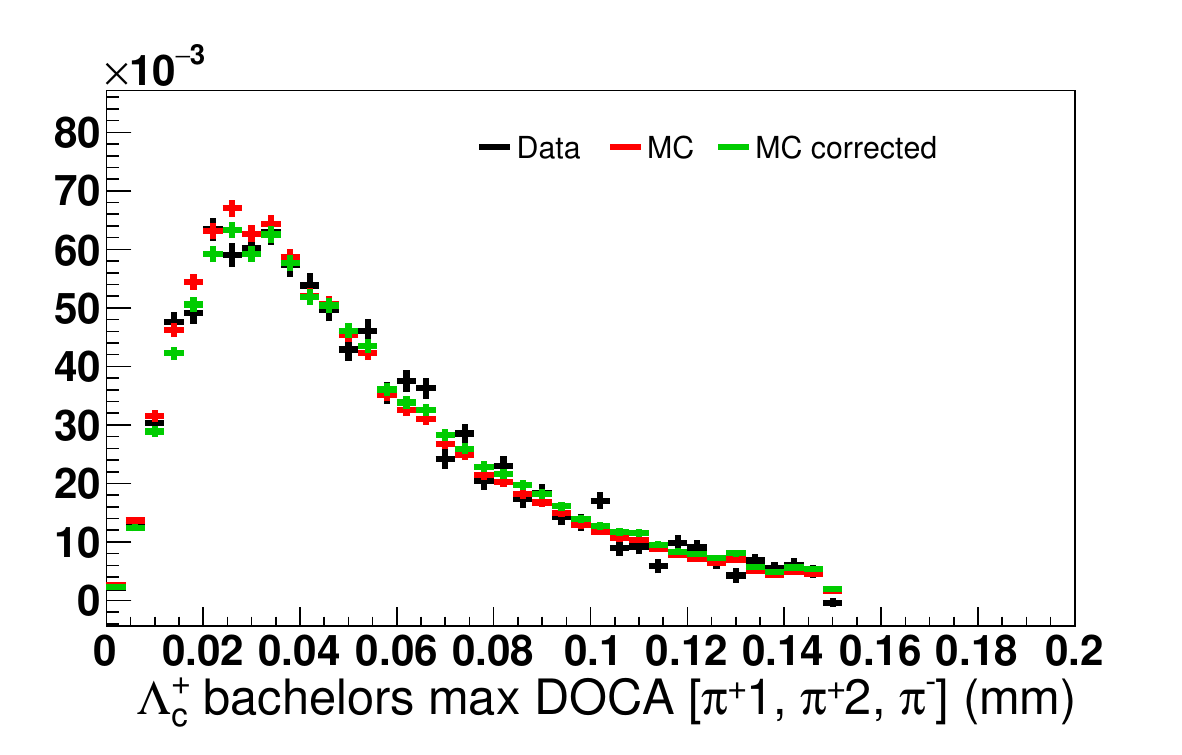}
	\includegraphics[width=0.45\linewidth]{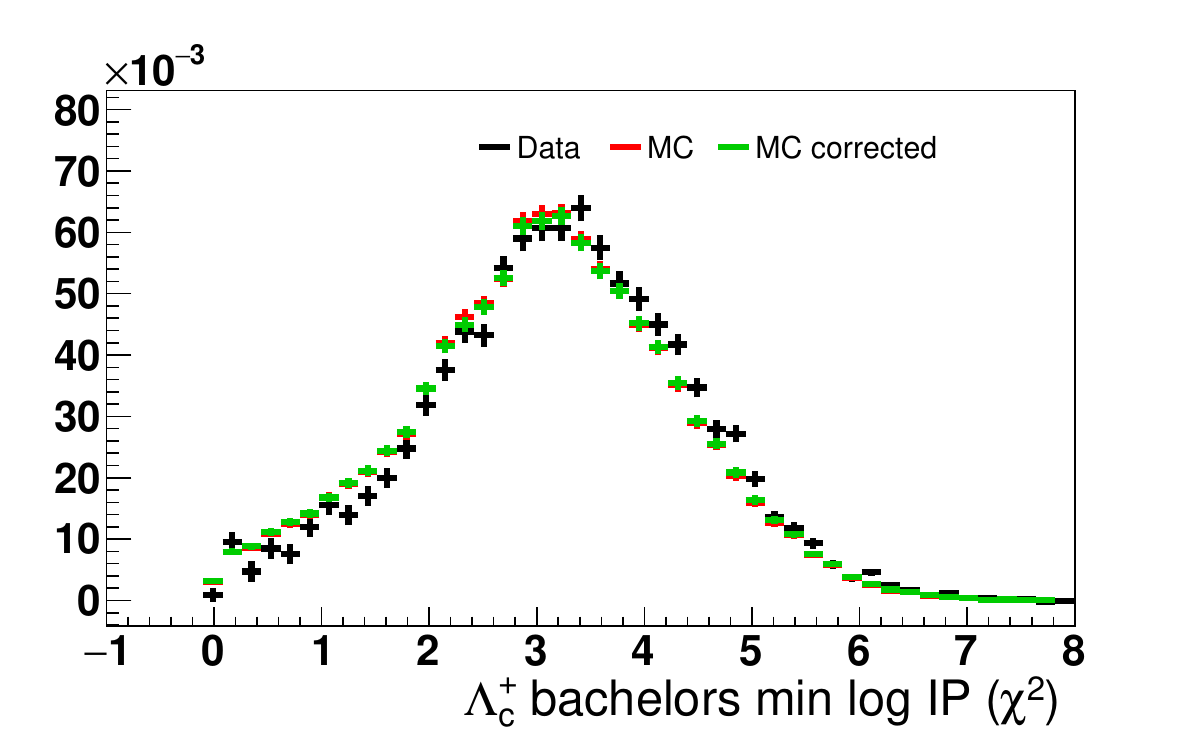}
	\includegraphics[width=0.45\linewidth]{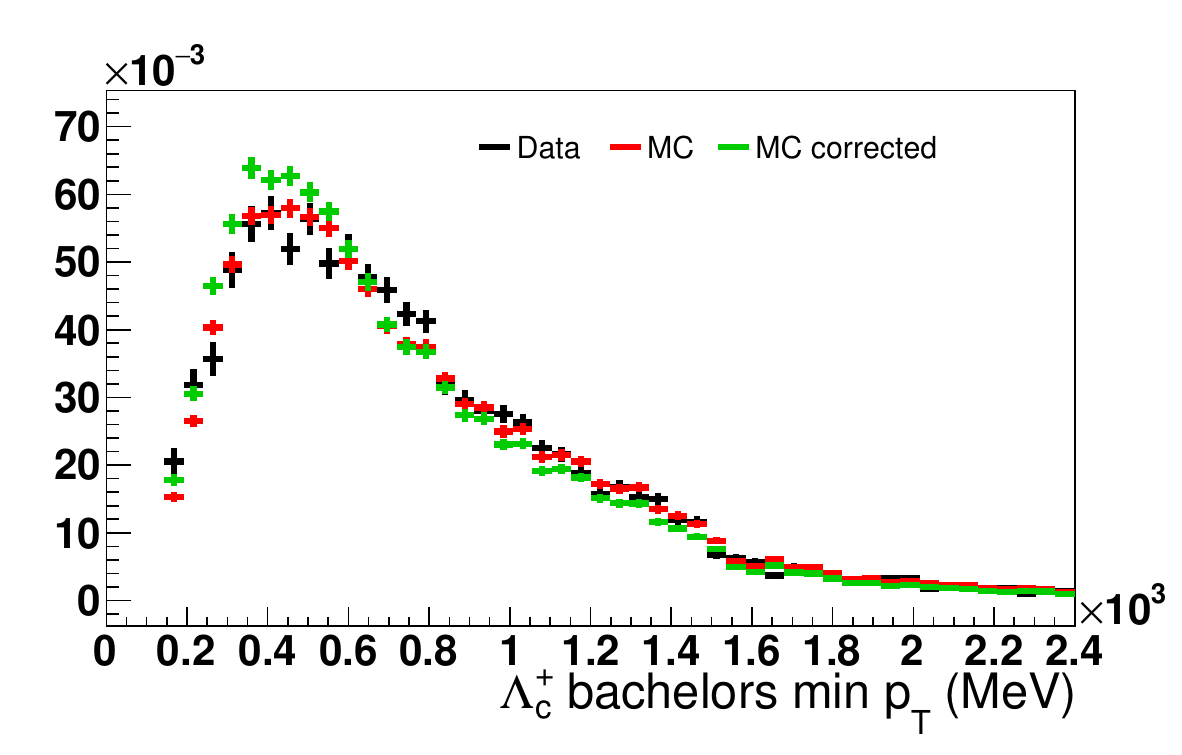}
	\includegraphics[width=0.45\linewidth]{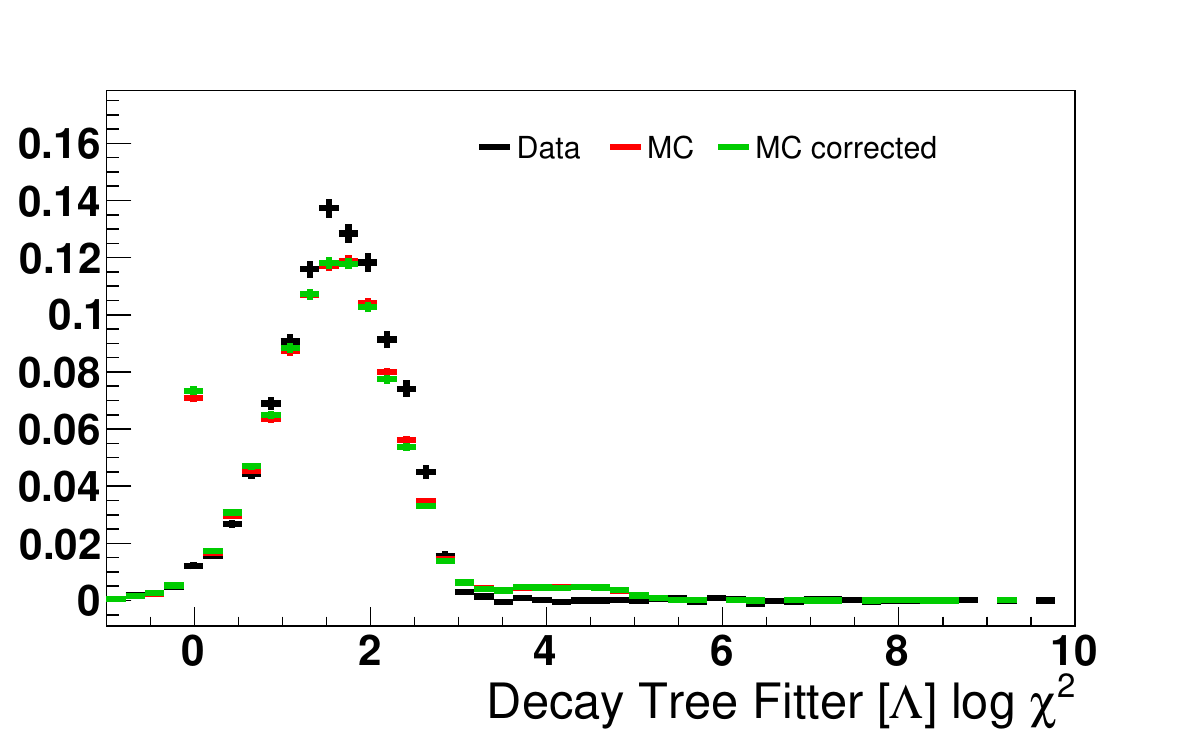}
	
	\caption{Same as Figure~\ref{fig:BDTvars} for other variables.}
	\label{fig:BDTvars2}
\end{figure}

The distributions for the BDT variables are shown in Figs. \ref{fig:BDTvars} and \ref{fig:BDTvars2}.

\chapter{Solution to RGEs}
\label{app:RGEsolution}

The dependence of the Wilson coefficients $\vec C$ on the energy scale $\mu$ is given by the renormalization group equations (RGE) as
\begin{equation}
    \mu \dfrac{d ~{\vec C} (\mu)}{d \mu} = \gamma^T \vec C (\mu).
\end{equation}
The anomalous dimension matrix $\gamma$ can be expanded in powers of the coupling constants \als and \ale as
\begin{equation} \label{eq:expanomalous}
\gamma(\mu, \als, \ale) = \gamma_s(\mu, \als) + \frac{\ale}{4\pi} \Gamma(\mu, \als).
\end{equation}
Introducing an ansatz for the general solution as
\begin{equation}
    {\vec C} (\mu) 
    = \left( \Uqcd(\mu, m_2) + \frac{\ale}{4\pi}\: \Rqed(\mu, m_2) \right) \vec C (m_2),
\end{equation}
we can rewrite the equation as
\be
\begin{array}{c}
     \mu \dfrac{d }{d \mu} \left( \Uqcd(\mu, m_2) + \frac{\ale}{4\pi} \Rqed(\mu, m_2) \right) \vec C (m_2)  \\
     = \left( \gamma_s^T(\mu, \als) + \frac{\ale}{4\pi} \Gamma^T(\mu, \als) \right) \left( \Uqcd(\mu, m_2) + \frac{\ale}{4\pi} \Rqed(\mu, m_2) \right) \vec C (m_2) .
\end{array}
\ee
Collecting terms proportional to 1, $\frac{\ale}{4\pi}$ and $\left(\frac{\ale}{4\pi}\right)^2$ we obtain three equations

\be
\begin{array}{lc} \label{eq:threesolutions}
     \propto 1 : &  \mu \dfrac{d ~\Uqcd (\mu, m_2)}{d \mu} = \gamma_s^T (\mu) \Uqcd(\mu, m_2),\\
     \propto \dfrac{\ale}{4\pi} :&  \mu \dfrac{d ~\Rqed (\mu, m_2)}{d \mu} = \gamma_s^T(\mu) \Rqed(\mu, m_2) + \Gamma^T (\mu) \Uqcd(\mu, m_2),\\
     \propto \left(\dfrac{\ale}{4\pi}\right)^2 :&  \hat 0 = \left(\dfrac{\ale}{4\pi}\right)^2 \Gamma^T (\mu) \Rqed(\mu, m_2) \vec C (m_2).\\
\end{array}
\ee

\noindent The third equation is just consistent with neglecting terms $\order(\ale^2)$ in Eq.~\eqref{eq:expanomalous}.

\section{Solution for  \texorpdfstring{$\Uqcd$}{Uqcd}}
The first equation leads to the result for $\Uqcd$,
\be \label{eq:LOQCD}
\Uqcd(m_1, m_2) = \exp \int_{g(m_2)}^{g(m_1)} dg' \dfrac{\gamma_s^T(g')}{\beta(g')},
\ee
where the beta function
\be
 \beta (g(\mu)) = \dfrac{d g(\mu)}{d\log\mu }
\ee 
describes the dependence of the coupling $g$ with the energy scale and is commonly expanded in powers of the couplings as well.

In Eq.~\eqref{eq:LOQCD}, $\gamma_s$ and $\beta$ can be substituted finding the well-known analytical solution to the RGEs at leading order in QCD. For instance, the self-correction of the first operator going from a high energy scale $M$ to a lower one $m$ is 
\be
C_1 (m) = \left( \frac{\als(M)}{\als(m)}\right)^{\frac{\gamma_{s}(1,1)}{2\beta_0}} C_1 (M) .
\ee

\section{Solution for \texorpdfstring{$\Rqed$}{Rqed}}

The second line in Eq.~\eqref{eq:threesolutions} leads to the solution for $\Rqed(m_1, m_2)$,
\be
\Rqed (m_1, m_2) = \displaystyle\int_{g(m_2)}^{g(m_1)} dg' \dfrac{\Uqcd(m_1, \mu') \Gamma^T (\mu') \Uqcd(\mu', m_2)}{\beta (g')} .
\ee
We provide an outline of the proof for these less-used photon corrections:

~

1. Starting with the second line in Eq.~\eqref{eq:threesolutions}, multiply on the left by $\Uqcd(m_1, \mu) $,
\begin{align}
&\mu ~ \Uqcd(m_1, \mu) \dfrac{d ~\Rqed (\mu, m_2)}{d \mu} - \Uqcd(m_1, \mu) \gamma_s^T(\mu) \Rqed(\mu, m_2) \\
&= \Uqcd(m_1, \mu) \Gamma^T (\mu) \Uqcd(\mu, m_2).
\end{align}

2. Apply the relation 

\be
- \Uqcd(m_1, m_2) \gamma_s^T(m_2) = m_2 \dfrac{d\Uqcd(m_1, m_2)}{d m_2}.
\ee

This relation can be demonstrated (not here) by using Leibniz's integral rule on 

\be
\begin{array}{ll}
     \dfrac{d\Uqcd(m_1, m_2)}{d m_2} &= \dfrac{d}{d m_2} \exp \displaystyle\int_{g(m_2)}^{g(m_1)} dg' \dfrac{\gamma_s^T(g')}{\beta(g')} \\
     &= \Uqcd(m_1, m_2) \underbrace{\dfrac{d}{d m_2} \int_{g(m_2)}^{g(m_1)} dg' \dfrac{\gamma_s^T(g')}{\beta(g')}}_{\text{ use Leibniz rule}} .
\end{array}
\ee
%
%
%
%
%
%

3. On the left-hand side we can identify the result of differentiation by parts,
\be
\begin{array}{rl}
&\mu ~ \Uqcd(m_1, \mu) \dfrac{d ~\Rqed (\mu, m_2)}{d \mu} + \mu  \dfrac{d\Uqcd(m_1, \mu)}{d \mu} \Rqed(\mu, m_2) \\
&= \mu ~ \dfrac{d}{d \mu} \left[ \Uqcd(m_1, \mu) ~\Rqed (\mu, m_2) \right]. \\
\end{array}
\ee

4. Combine with the right-hand side using $ d\log\mu = \dfrac{d g(\mu)}{\beta (g(\mu))}$,

\be
\begin{array}{l}
  d \left[ \Uqcd(m_1, \mu) ~\Rqed (\mu, m_2) \right]=  \dfrac{\Uqcd(m_1, \mu) \Gamma^T (\mu) \Uqcd(\mu, m_2)}{\beta (g(\mu))} d g(\mu).
\end{array}
\ee

5. Integrate in the range $[m_2, m_1]$,

\be
\begin{array}{rl}
\underbrace{\Uqcd(m_1, m_1)}_{=\hat 1} ~\Rqed (m_1, m_2) - &\Uqcd(m_1, m_2) ~\underbrace{\Rqed (m_2, m_2)}_{= \hat 0} = \\
&  \displaystyle\int_{g(m_2)}^{g(m_1)} dg' \dfrac{\Uqcd(m_1, \mu') \Gamma^T (\mu') \Uqcd(\mu', m_2)}{\beta (g')} ,
\end{array}
\ee

where we used the boundary conditions for $\hat U$ and $\Uqcd$

\be
\underbrace{\hat U (m_1, m_1, \ale)}_{\hat 1} = \underbrace{\Uqcd(m_1, m_1)}_{\hat 1} +  \frac{\ale}{4\pi} \Rqed (m_1, m_1) ~~~\rightarrow ~~ \Rqed (m_1, m_1)={\hat 0}.
\ee

Finally,

\be
\Rqed (m_1, m_2) = \displaystyle\int_{g(m_2)}^{g(m_1)} dg' \dfrac{\Uqcd(m_1, \mu') \Gamma^T (\mu') \Uqcd(\mu', m_2)}{\beta (g')} .
\ee

\begin{flushright}
\textbf{Q.E.D.} ~~~~~~~~~~~
\end{flushright}

\chapter{Loop calculations} \label{app:loops}

\section{Weinberg diagrams: straightforward calculation}
\label{app:wein}

In July 2021, the great physicist Steven Weinberg passed away, leaving behind a vast amount of contributions to physics and science. We owe to him 
the Electroweak theory, a keystone of modern particle physics, as well as many other outstanding contributions across many areas of theoretical physics. 
We see some of his footprints in our analysis of EDMs as well: he formulated the CP-odd three-gluon operator, in Eq.~\eqref{eq:opEFF}, and its two-loop leading contribution through the exchange of a scalar field. 
In his original paper~\cite{Weinberg:1989dx}, he provided the $h(r)$ loop function, while the full expression was obtained by D. Dicus in Ref.~\cite{Dicus:1989va}, who referred to the computation of this diagram as \textit{straightforward}.

At various points in the calculation we found, however, that equally justified choices of parameterisation or approximations can take the parametric integral off the \textit{straight} track towards this simple analytic expression. To our knowledge, these technical details are not found in the literature in a comprehensive summary. To facilitate the reproducibility of this analytical shape, we describe these details in the following. Following the same procedure, we arrived to the expression of $g(r)$, which is surprisingly simple as well, and to the cancellation of the diagram in Figure~\ref{fig:twoloopWeinbergNeut}~(c).

\begin{figure}[h]
	\centering
		\centerline{ \includegraphics[scale=1]{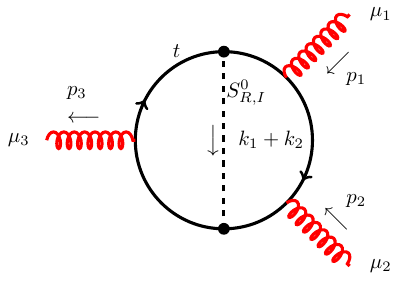} }    
	\caption{Diagram contributing to the Weinberg operator proportional to the loop function $h(r)$. The momentum directions are shown to facilitate the reproducibility of our results.}
	\label{fig:WeinbergMomentumTags}
\end{figure}

The Dirac trace of this two-loop amplitude contains up to eight $\gamma^\mu$, and two $\gamma_5$ matrices. Since the final result is finite, the traces with $\gamma_5$ can be solved with the usual relation ${\rm Tr}(\gamma^\mu \gamma^\nu \gamma^\rho \gamma^\sigma \gamma_5) = 4 i \varepsilon^{\mu \nu \rho \sigma}$. The \CP-violating parts of this amplitude are proportional only to the index structures with Levi-Civita tensors 
$$
\varepsilon^{\mu_1 \mu_2 \sigma \rho},~ \varepsilon^{\mu_2 \mu_3 \sigma \rho},~\varepsilon^{\mu_1 \mu_3 \sigma \rho} ,~\text{and}~\varepsilon^{\mu_1 \mu_2 \mu_3 \sigma},
$$
where the indices $\sigma$ and $\rho$ are contracted with external momenta.
To ease the calculation, it is convenient to select only one of these linearly-independent structures, as the final result is independent of this choice. The directions of the internal loop momenta were chosen as in Figure~\ref{fig:WeinbergMomentumTags}. With this, only three propagator denominators contain external momenta, which are small compared to the heavy mass $M$. Thus, we can expand these denominators in powers of $(p^2/M^2)$ with
\begin{align}
\frac{1}{(k_i+p)^2-M^2}=&
\frac{1}{k_i^2-M^2}
\left[
1-\frac{p^2 +2(p\cdot k_i) }{k_i^2-M^2}+\frac{4 (p \cdot k_i)^2 }{(k_i^2-M^2)^2}
\right] 
+{\cal O}(p^4/M^4)~,
\end{align}
and carefully removing higher-order terms after the expansions. Once the denominator is free of external momenta $p$, the tensor integrals with an odd number of open indices vanish, and, for the rest, we can apply the identity $k_i^\mu k_j^\nu \to (k_i \cdot k_j /D) g^{\mu\nu}$. The resulting master integrals have the shape
\begin{align}\label{eq:masterIntegralWeinberg}
\mathcal{W}_{\left\lbrace 00;~10;~11 \right\rbrace}(\alpha,\beta,\gamma;a,b,c)\equiv \int \frac{\text{d}^D k_1}{(2\pi)^D}\,\int \frac{\text{d}^D k_2}{(2\pi)^D}\,\frac{\left\lbrace 1;~k_1\cdot k_1;~k_1\cdot k_2\right\rbrace }{ \left(k_1^2 - a^2\right)^\alpha \left(k_2^2 - b^2\right)^\beta \left((k_1 \pm k_2)^2 - c^2\right)^\gamma  }~.
\end{align}
To re-express the $\mathcal{W}$ functions in terms of Feynman parameters, one must use Feynman parameterisation of two denominators at a time~\cite{Ilisie:2016jta}, 
\begin{align}
\frac{1}{A^\alpha\,B^\beta}\,=\,\frac{\Gamma(\alpha+\beta)}{\Gamma(\alpha)\,\Gamma(\beta)}\,\int_0^1\,\text{d}x\,\frac{x^{\alpha-1}\,(1-x)^{\beta-1}}{\left[A\,x\,+\,B\,(1-x)\right]^{\alpha+\beta}}~,
\end{align}
and the standard Wick rotation to sequentially integrate over the loop momenta $k_1$ and $k_2$,

	\begin{align}
	\int\frac{\text{d}^D k_2}{(2\pi)^D}\,\frac{(k^2)^\alpha}{(k^2-a^2+i\,\epsilon)^\beta}=\frac{i\,(-1)^{\alpha-\beta}}{(4 \pi)^{D/2}}\,(a^2)^{D/2+\alpha-\beta}\,\frac{\Gamma(\beta-\alpha-D/2)\,\Gamma(\alpha+D/2)}{\Gamma(\beta)\Gamma(D/2)}~.
	\end{align}

The solution to the master integral that leads to the desired analytical shape of $h(r)$ reads 
\begin{align}
    \mathcal{W}_{\delta,\epsilon}(\alpha,\beta,\gamma;a,b,c)=\frac{(-1)^{D/2-\alpha-\gamma+\delta}}{(4\pi)^D}\frac{\Gamma(\alpha+\beta+\gamma-D-\delta)\Gamma(\delta+D/2)}{\Gamma(\alpha)\Gamma(\beta)\Gamma(\gamma)\Gamma(D/2)}w_{\delta\epsilon}(\alpha,\beta,\gamma;a,b,c),
\end{align}
where
\begin{align}
    w_{\delta\epsilon}(\alpha,\beta,\gamma;a,b,c)=&\int_0^1\,\text{d}x\,\int_0^1\,\text{d}y\;x^{D/2-\gamma-1}(1-x)^{\beta+\epsilon+1-D/2}y^{\alpha-1}(1-y)^{\beta+\gamma-D/2-1}\\
    \nonumber&\times \left(a^2y+(1-y)\frac{b^2x+(1-x)c^2}{x(1-x)}^{D-\alpha-\beta-\gamma+\delta}\right).
\end{align}

Selecting the tensor structure $\varepsilon^{\mu_1 \mu_2 \sigma \rho}$, only the function $\mathcal{W}_{11}$ contributes to this amplitude in the diagram of Figure~\ref{fig:WeinbergMomentumTags}. Finally, the three permutations of this diagram, obtained by rotating the internal scalar propagator by $120^o$ at a time, are obtained from the first diagram by renaming the indices and external momenta. This step is crucial to analytically cancel the divergences of different parametric integrals. To obtain the Wilson coefficient, the fundamental amplitude must be matched onto the effective one, that reads~\cite{Dicus:1989va}
\begin{align} \label{eq:effectiveAmplitudeWeinberg}
i\mathcal{M}_{\rm eff} = &- \frac{2}{3}\, f_{abc}\, g_s\, w \,\varepsilon^{\mu_1} (p_1) \,\varepsilon^{\mu_2} (p_2)\, \varepsilon^{\mu_3} (-p_1 - p_2) \\ \nonumber
&\left[ (p_1 - p_2)_{\mu_3}\, \varepsilon_{\mu_1 \mu_2 \sigma \rho} + 2\, (p_{1~\mu_2}\, \varepsilon_{\mu_1 \mu_3 \sigma \rho} + p_{2~\mu_1}\, \varepsilon_{\mu_2 \mu_3 \sigma \rho})
\right] \,p_1^\sigma\, p_2^\rho~.
\end{align}
To develop these expressions we used the help of the open-source packages \textsc{FeynArts} and \textsc{FeynCalc}~\cite{Shtabovenko:2020gxv,Kublbeck:1990xc,Shtabovenko:2016sxi}.

\section{Barr-Zee diagrams}
\label{app:BarrZee}

To simplify the calculation of the Barr-Zee diagrams, the two loops can be computed sequentially. The loop attached to the external photon (gluon) shall be obtained first. The result, in terms of Feynman integrals, can be written in the shape~\cite{Ilisie:2015tra}
\begin{equation}\label{eq:effvert}
i \Gamma^{\mu\nu}_{g,\gamma}=i (g^{\mu\nu} k\cdot q - k^\mu q^\nu) S_{g,\gamma} + i \epsilon^{\mu\nu\alpha\beta} k_\alpha q_\beta \widetilde{S}_{g,\gamma}~,
\end{equation}
where $q$ is the momentum of the external photon (gluon) and $k$ that of the off-shell gauge boson. The scalar functions $S$ and $\widetilde{S}$ encode all the relevant information of the different diagrams. The effective vertex of the dominant contributions to the (C)EDM, and the corresponding scalar form factors, reads

\vspace{0.4cm}

\mpdiag{ \includegraphics[scale=1]{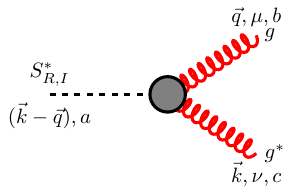}}{

\begin{align*}
S_g=&g_s^2\frac{m_t^2}{16\pi^2 v} d^{a b c} \text{Re}(\eta_U) \int_0^1 {\rm d} x \frac{2x^2-2x+1}{k^2(x-1)x+m_t^2}~,\\
\widetilde{S}_g=&-g_s^2\frac{m_t^2}{16\pi^2 v} d^{a b c}\text{Im}(\eta_U) \int_0^1 {\rm d} x \frac{1}{k^2(x-1)x+m_t^2}~,
\end{align*}

}

\mpdiag{ \includegraphics[scale=1]{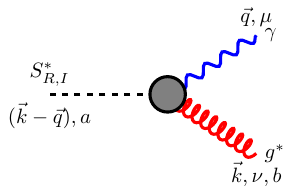}}{
	
\begin{align*}
S_\gamma=&e\,g_s\frac{m_t^2}{12\pi^2 v } \delta^{a b} \text{Re(}\eta_U\text{)}  \int_0^1 {\rm d} x \frac{  2 x^2- 2 x +1}{k^2 (x-1) x  + m_t^2}~,\\
\widetilde{S}_\gamma=&-e\,g_s\frac{m_t^2}{12\pi^2 v } \delta^{a b} \text{Im(}\eta_U\text{)}  \int_0^1 {\rm d} x\frac{  1}{ k^2 (x-1) x  + m_t^2}~.
\end{align*}

	}

\vspace{0.4cm}
	
\begin{figure}[h]
	\centering
	\subcaptionbox{}{\includegraphics[scale=1]{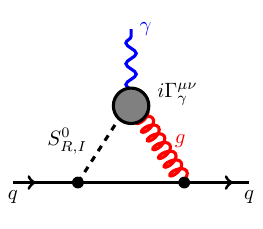}}
	\subcaptionbox{}{\includegraphics[scale=1]{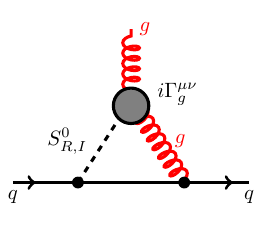}}
	\caption{Second loop of the Barr-Zee contributions to the quark EDM (a) and CEDM~(b).}
	\label{fig:secondloop}
\end{figure}

Only the top quark Yukawa coupling gives a sizeable contribution to this vertex. Thus, we only considered top quarks running in the \textit{inner} loop, as shown in Figure~\ref{fig:barr_zee}.
To arrive at this result, we use Feynman parametrisation in the shape of Eqs.~(6.23) and (6.25) of the detailed guide for loop calculations in Ref.~\cite{Ilisie:2016jta}. Furthermore, the photon (gluon) is assumed to be \textit{soft}, \textit{i.e.} $k\cdot p \to 0$, following the arguments of Ref.~\cite{Ilisie:2015tra}. 
Once the expressions for the first loop are parametrised as in Eq.~\eqref{eq:effvert}, this effective vertex is plugged in the second loop (Figure~\ref{fig:secondloop}), rewriting the denominator $k^2 (x-1)x + m_t^2$ as another propagator with momentum $k$. 
Then, the integrals over $k$ can be identified in terms of Passarino-Veltman functions. Expanding the result in powers of $(m_q/M)$, where $M$ is a heavy mass and $q={u,d}$, only the first term is numerically relevant. 
In this way, we obtained the loop functions $\mathcal{F}$ and $\widetilde{\mathcal{F}}$, in terms of the Feynman parameter $x$, which comes from the \textit{inner loop}. To match the fundamental amplitude to the effective (C)EDM operator, it is convenient to express the Levi-Civita tensor in terms of products of gamma matrices, through the Chisholm identity.

\pagebreak

~ \pagebreak

\def\titol {Resum en valencià}
\chapter*{\titol} \addcontentsline{toc}{chapter}{\titol} 

\fancyhead[LO]{}
\fancyhead[RE]{\titol }

\renewcommand\thefigure{R.\arabic{figure}}    
\setcounter{figure}{0}  

\renewcommand\theequation{R.\arabic{equation}}    
\setcounter{equation}{0}  


\begin{center}

{\Large \textbf{Experiments i fenomenologia dels\\
		moments dipolars elèctrics}}

~\\

~\\

\end{center}


El model estàndard (SM, per les seues sigles en anglès) és la millor descripció que tenim de les partícules fonamentals i les seues interaccions. A partir d'esta teoria tots els fenòmens del món macroscòpic (excepte la gravetat) es poden explicar. A més, fins avui, el SM ha predit amb èxit tots els resultats experimentals dels acceleradors de partícules a la Terra.
No obstant, observacions cosmològiques mostren un gran desequilibri entre la quantitat de matèria i antimatèria a l'Univers, molts ordres de magnitud per damunt de la predicció del SM. 
Per poder explicar estes observacions, han d'existir noves interaccions més enllà del SM que no respecten la simetria de càrrega-paritat (\CP). Estes interaccions, al mateix temps, induirien moments dipolars elèctrics (EDM) en les partícules conegudes, els quals no han segut observats fins l'actualitat.
En la Part I d'aquesta tesi es proposa ampliar el programa experimental de cerques d'EDMs per a barions amb quarks \textit{charm} i \textit{bottom}, leptons $\tau$, i hiperons \Lz. Això també permetria mesurar els seus corresponents moments dipolars magnètics (MDM).
L'EDM i MDM de partícules de molt curta vida es podria mesurar amb un experiment de cristalls corbats que utilitze el feix de protons del Gran Col·lisionador d'Hadrons (LHC), mentre que les partícules \Lz, de vida mitja més llarga, es poden mesurar a l'experiment LHCb sense instrumentació addicional. En la Part II de la tesi es presenta una anàlisi de dades de l'LHCb per mesurar la polarització de la partícula \Lz en desintegracions \threepi, ingredient essencial per a l'experiment proposat a la Part I. En l'última part de la tesi, Part III, es deriven nous límits indirectes en l'EDM de quarks \textit{charm} i \textit{bottom} amb dades ja disponibles de l'EDM del neutró, i s'exploren les implicacions fenomenològiques d'aquests (i altres) observables en models de nova física, amb especial èmfasi en extensions del SM amb noves partícules escalars que són octets de color.

\newpage

\section*{Introducció}

En física estem acostumats a fer preguntes sobre el món natural, com ara \textit{de què està feta la matèria?}, \textit{com funciona aquest fenomen?}, \textit{què són l'espai i el temps?}, ... Pas a pas, fem \textit{teories} o \textit{models} que poden descriure aquests fenòmens i, successivament, trobem teories més i més generals que descriuen simultàniament alguns d'aquests fenòmens i altres que inicialment no es podien explicar. En aquest sentit, una teoria és més \textit{fonamental} si conté l'explicació subjacent a més fenòmens. Avui, al final de la cadena de \textit{fonamentalitat} trobem el model estàndard (SM) de la física de partícules, i la teoria de la relativitat general d'Einstein.

\begin{figure}[h]
	\centering
	\includegraphics[width=0.7\linewidth]{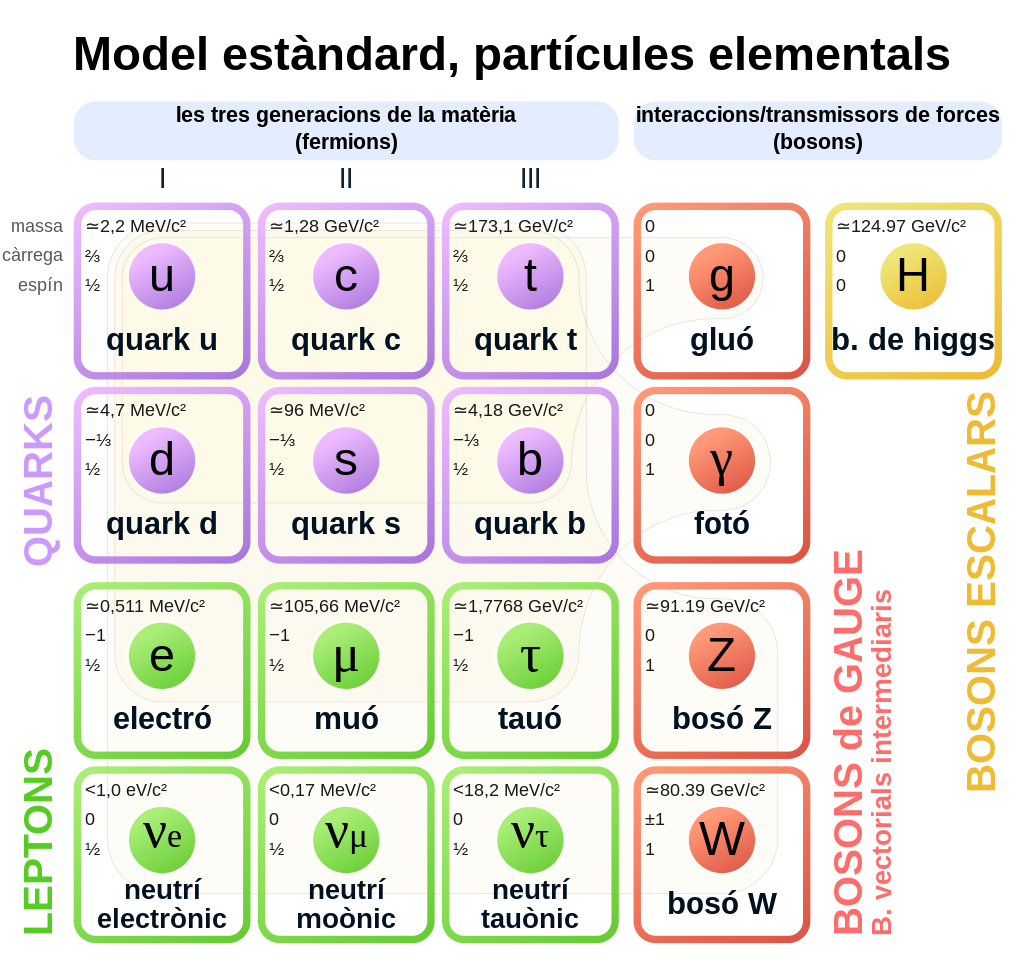}
	\caption{Figura presa de la referència \cite{wikipediaSM}. Contingut de partícules del model estàndard. Tota la matèria ordinaària està composada per quarks \textit{up} i \textit{down} (u i d a la figura), electrons, i bosons \textit{gauge} que mantenen els estats lligats: gluons, que lliguen els quarks$^1$ dins els protons i neutrons (i indirectament els protons i neutrons dins el nucli), i fotons, que mantenen l'estructura dels àtoms i molècules transmetent la força electromagnètica.
		No obstant, la naturalesa a altes energies és molt més rica que allò que ens queda a l'Univers actual de baixa densitat i temperatura. Tenim fins a tres famílies de quarks i leptons; altres dos bosons gauge, $Z$ i $W^\pm$, que medien la força electrofeble$^2$; i el bosó de Higgs, associat al trencament espontani de la simetria gauge del SM, la qual cosa permet donar massa als fermions i bosons $Z$ i $W^\pm$.
	}
	\label{fig:SM_ResumValencia}
\end{figure}

El SM, breument introduït en la Figura~\ref{fig:SM_ResumValencia}, va nàixer en la dècada del 1970 i ha tingut un gran èxit explicant i predient els resultats d'experiments passats i presents de física de partícules. En última instància, este acord entre les prediccions de la teoria i l'evidència experimental és l'única cosa que compta en jutjar l'èxit d'una teoria. Tanmateix, altre aspecte de les teories fonamentals destaca quan s'estudia el SM. Weinberg ho va anomenar \textit{inevitabilitat} i es refereix al fet que tots els elements de la teoria es deriven de molt pocs supòsits inicials o \textit{principis}, que no es poden modificar.

Malgrat el seu èxit, hi ha algunes observacions experimentals que el SM no pot encaixar i altres interrogants (més discutibles) de naturalesa teòrica. Per abordar aquests problemes es necessiten noves teories \textit{més enllà} del SM (BSM). Estes teories contenen noves partícules fonamentals que interaccionen amb les partícules del SM i, per tant, els seus efectes s'haurien de poder observar en experiments de física de partícules d'altes energies. Malauradament, no hi ha cap garantia de trobar aquestes noves partícules dins del rang de masses accessible pels acceleradors actuals o futurs. No obstant això, moltes de les teories més interessants que, a banda dels problemes experimentals, també aborden alguns dels problemes teòrics del SM prediuen que el rang de masses d'aquestes noves partícules ha d'estar al voltant del {\tev}, que requereix energies per produir-les, en principi, a l'abast de la tecnologia actual.

A més de buscar senyals d'aquestes partícules quan es produeixen directament en la seua \textit{capa màsica} (\textit{on shell}, en anglès), també podem reduir la llista de teories candidates amb mesures de precisió que són sensibles als efectes d'aquestes partícules quan apareixen virtualment (\textit{off shell}) en processos on totes les partícules externes són de l'SM.
Per descriure tots aquests possibles efectes de manera sistemàtica, i independent de la teoria de nova física, és possible treballar amb teories efectives (EFT) que són vàlides per escales d'energia davall de la massa d'aquestes partícules. A més a més, les EFTs poden simplificar enormement el càlcul de les prediccions per observables de baixa energia (com els EDMs) a partir de teories fonamentals a alta energia.

En esta tesi, diversos observables de baixa energia es tractaran des de diferents punts de vista. Estos inclouen dues propostes experimentals per mesurar moments dipolars elèctrics (EDM) i magnètics (MDM) de partícules inestables utilitzant cristalls corbats i amb l'imant dipolar convencional del detector LHCb (Part I); una anàlisi experimental de dades de l'LHCb amb desintegracions multihadròniques de barions amb quarks \textit{charm} (d'ací en davant, \textit{barions charm}) (Part II); i dos treballs més fenomenològics al voltant de l'observable EDM (Part III) que utilitzen tant un enfocament independent del model (Capítol \ref{ch:improvedbounds}) i una teoria BSM específica (Capítol \ref{ch:edmsmw}).
Malgrat que alguns d'aquests projectes es troben en marcs de recerca bastant diferents (experimental i teòric), tots van sorgir de forma natural els uns dels altres i el fil conductor de la tesi es veurà de seguida en introduir cada tema.

\newcommand\blfootnote[1]{%
	\begingroup
	\renewcommand\thefootnote{}\footnote{#1}%
	\addtocounter{footnote}{-1}%
	\endgroup
}

\blfootnote{$^1$~Els quarks no poden existir lliurement a la natura degut al confinament quàntic en la interacció forta. Per tant, solament els podem detectar com hadrons, formats per tres quarks (barions) o per una parella de quark-antiquark (mesons).
	Al resum de la Part III tornarem a treballar directament amb quarks i bosons gauge. En les Parts I i II, de física experimental, treballarem sobretot amb barions (\Lc, \Lz, $p$ i altres); mesons (\pipm, \Kpm, \jpsi i altres); i leptons ($\tau$ i $\mu$).}

\blfootnote{$^2$~Els bosons $W^\pm$ interaccionen amb els quarks, transformant-los en altre \textit{sabor} (\textit{u, d, c, s, t} ó \textit{b}), o amb els leptons, intercanviant leptons carregats ($e^\pm$, $\mu^\pm$, $\tau^\pm$) per neutrins ($\nu_e$, $\nu_\mu$, $\nu_\tau$). Estes interaccions fonamentals donen lloc a una ampla gamma de possibles desintegracions d'hadrons, que són l'objecte de la \textit{física del sabor}, en la qual s'especialitza el detector LHCb.}

\section*{Part I: Experiments amb EDMs}

\begin{figure}[th]
	\centering
	\includegraphics[width=0.99\linewidth]{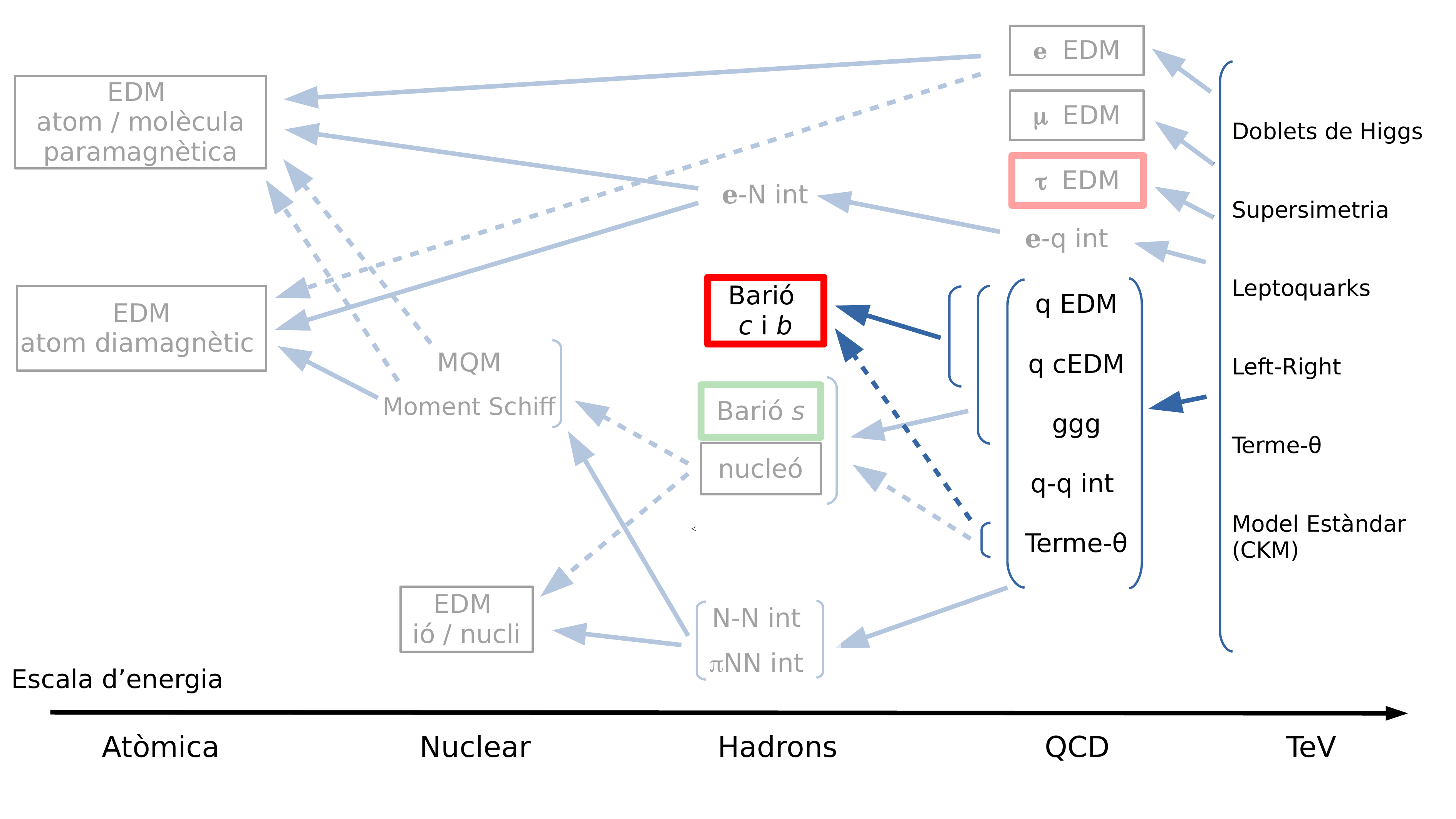}
	\caption{
		Tipus d'observables d'EDM (caixes) i les seues contribucions (fletxes) des d'altes escales d'energia. Les fonts de violació de \CP en les teories fonamentals (\eg doblets de Higgs, supersimetria, model estàndard) donen contribucions als operadors efectius, a l'escala de "QCD", i estos, a l'EDM dels barions charm i bottom. 
	}
	\label{fig:contribResum}
\end{figure}

%
%
%


\subsection*{Motivació teòrica i objectius}

El camp actual de cerques d'EDMs està proporcionant alguns dels resultats més rellevants en física de partícules per estudiar la viabilitat dels models de nova física. Fins l'actualitat, tots estos experiments que han mesurat l'EDM de partícules han trobat un resultat consistent amb zero. La incertesa d'aquest resultat és el que determina la cota superior en l'EDM de la partícula. Des de l'altre costat, amb els models teòrics podem obtenir una expressió per aquests EDMs en funció dels paràmetres lliures del model (masses de les noves partícules i els seus acoblaments). De connectar els models teòrics amb els fenòmens experimentals s'encarrega la \textit{fenomenologia} de partícules, i en la Part III en veurem alguns exemples explícits.

De moment, anem a introduir un esquema general d'aquesta connexió, identificant les fonts potencials de l'EDM dels barions pesats, un dels observables principals en la proposta experimental de la Part I. Açò també ens servirà per a motivar l'experiment i definir clarament el seu potencial per restringir models de nova física.

En termes de diagrames de Feynman, l'EDM del barió s'associa a qualsevol procés que implique un fotó extern enganxat a la línia fermiònica del barió, de forma que la interacció viole la simetria \CP.
Este procés es pot investigar des de diferents escales d'energia, amb teories efectives o fonamentals, i totes estes descripcions s'han de relacionar entre si. Per exemple, el Lagrangià de la teoria que descriga estes interaccions es pot construir per davall de l'escala hadrònica ($\lesssim 0,7 \gev$), amb camps de barions i mesons; per damunt de l'escala de ruptura quiral ($\gtrsim 1.2 \gev$), amb quarks i gluons; i a l'escala de nova física ($\gtrsim200 \gev$), amb el contingut complet de partícules de la teoria fonamental.

Per davall de $\sim 80 \gev$, on s'han integrat les partícules (graus de llibertat) més pesades del SM i més enllà, els operadors efectius no han de respectar la simetria de gauge completa del SM. Les fonts d'EDM bariònic a esta escala estan contingudes en el següent Lagrangià efectiu:

\vspace*{-1.0cm}
\begin{align} \label{eq:lagrangianEDMsresumvalencia}
\mathcal{L}^{\not\text{P}\not\text{T}}_{\text{eff}} =
&-\frac{i}{2} \sum_{q={u,d,s,c,b}}\left. d_q \, \bar q \sigma^{\mu\nu} \gamma_5 q\, F_{\mu\nu} \right. &
\includegraphics[width=0.15\textwidth]{./crystals/fIntro/qEDM.pdf} &
~~~ \text{qEDM} ~~~~ &\nonumber \\
& -\frac{i}{2} \, \sum_{q={u,d,s,c,b}} \left. \tilde d_q\, \bar q \sigma^{\mu\nu} \gamma_5 T_a q\, G^a_{\mu\nu} \right. &
\includegraphics[width=0.15\textwidth]{./crystals/fIntro/qCEDM.pdf} &
~~~ \text{qCEDM} ~~~~ &\nonumber \\
& + \sum_{i,j,k,l = {u,d,s,c,b}}C_{ijkl}\, \bar q_i \Gamma q_j \, \bar q_k \Gamma^\prime q_l\, &
\includegraphics[width=0.10\textwidth]{./crystals/fIntro/4q.pdf} &
~~~ \text{4q int} ~~~~ & \\
& + \frac{w}{6} f_{abc}\varepsilon^{\mu\nu\alpha\beta}G^a_{\alpha\beta}G^b_{\mu\rho}G_{\nu}^{c \, \rho} &
\includegraphics[width=0.12\textwidth]{./crystals/fIntro/gCEDM.pdf} &
~~~ \text{ggg (Weinberg op.)} ~~~~ &\nonumber \\
&- \bar{\theta} \frac{g^2}{64\pi^2}\epsilon^{\mu\nu\alpha\beta} G^a_{\mu \nu}G^a_{\alpha \beta} \quad &
\includegraphics[width=0.12\textwidth]{./crystals/fIntro/theta-T.pdf} &
~~~ \theta\text{-QCD term} , ~~~~ & \nonumber
\end{align}
on $d_q$ i \dtildeq són l'EDM i chromo-EDM (CEDM) del quark, i $F^{\mu\nu}$ i $G^{\mu\nu}$ els camps de fotons i gluons, respectivament. A continuació, tenim molts operadors de contacte de quatre quarks (4q) on les diferents estructures de Dirac estan representades per $\Gamma$ i $\Gamma'$ i els sabors de quarks pels índexs $i,j,k,l$. Finalment, tenim l'operador de Weinberg amb tres gluons (ggg) i el terme $\theta$ de QCD.
Estos operadors també s'inclouen a la Figura \ref{fig:contribResum} a l'escala ``QCD", i juguen un paper intermediari entre les teories de nova física (a alta energia) i l'observable experimental de l'EDM del barió.

Determinar la contribució d'aquests operadors a l'EDM hadrònic requereix tècniques no pertorbatives de la interacció forta a baixa energia. Hi ha diferents possibilitats com ara les teories quirals, les regles de suma de QCD ó el càlcul numèric dels processos en la \textit{retícula} (\textit{lattice QCD}).
La fiabilitat d'estes tècniques es pot avaluar experimentalment a través d'observables hadrònics de baixa energia, entre els quals el moment magnètic dels barions charm, també mesurables amb el nostre experiment, podria jugar un paper rellevant.

\subsection*{Experiment}

Quan estudiem el comportament de les partícules en el règim quàntic no podem predir exactament el resultat d'una mesura, i solament podem obtenir informació estadística de l'experiment. Simplificant, la mitja de moltes mesures d'alguna quantitat ens dona el seu \textit{valor esperat}. Quan tractem amb l'espí d'una partícula, el seu valor esperat és proporcional al vector de polarització $\bm s \equiv \langle \bm{\hat S} \rangle / (\hbar/2)$, on $\bm{\hat S}$ és l'operador d'espí. Si la partícula, a més a més, té un EDM (${\bm \delta}$) i/o MDM (${\bm \mu}$), aquests interaccionen amb el camp elèctric i magnètic (\Evec i \Bvec), canviant la direcció de la polarització en el fenomen anomenat \textit{precessió d'espí}. La precessió d'espí es descriu per l'equació de moviment
\begin{equation}
\label{eq:precesionClassical}
\frac{d \bm s}{d \tau} = \bm \mu \times \bm B^* + \bm \delta \times \bm E^* ~,
\end{equation}
on $\tau$ és el temps propi de la partícula. Aquesta equació s'obté del Hamiltonià clàssic $H=-\bm \delta \cdot \bm E^* -\bm \mu \cdot \bm B^*$, on $\bm E^*$ i $\bm B^*$ són els camps externs en el sistema de referència de la partícula. L'expressió completa d'aquesta equació de moviment ha d'incloure el terme de Thomas per partícules carregades i a més es pot expressar independentment del sistema de referència, de forma covariant. Tot junt, tenim l'equació de Thomas-Bargman-Michel-Telegdi (TBMT), introduïda al text principal en la Secció~\ref{sec:spinprecess}.

Qualsevol configuració experimental per mesurar el fenòmen de precessió d'espí es basa en tres elements principals, que resumim pel nostre cas a continuació.

\begin{figure}
	\centering
	\includegraphics[width=0.99\linewidth]{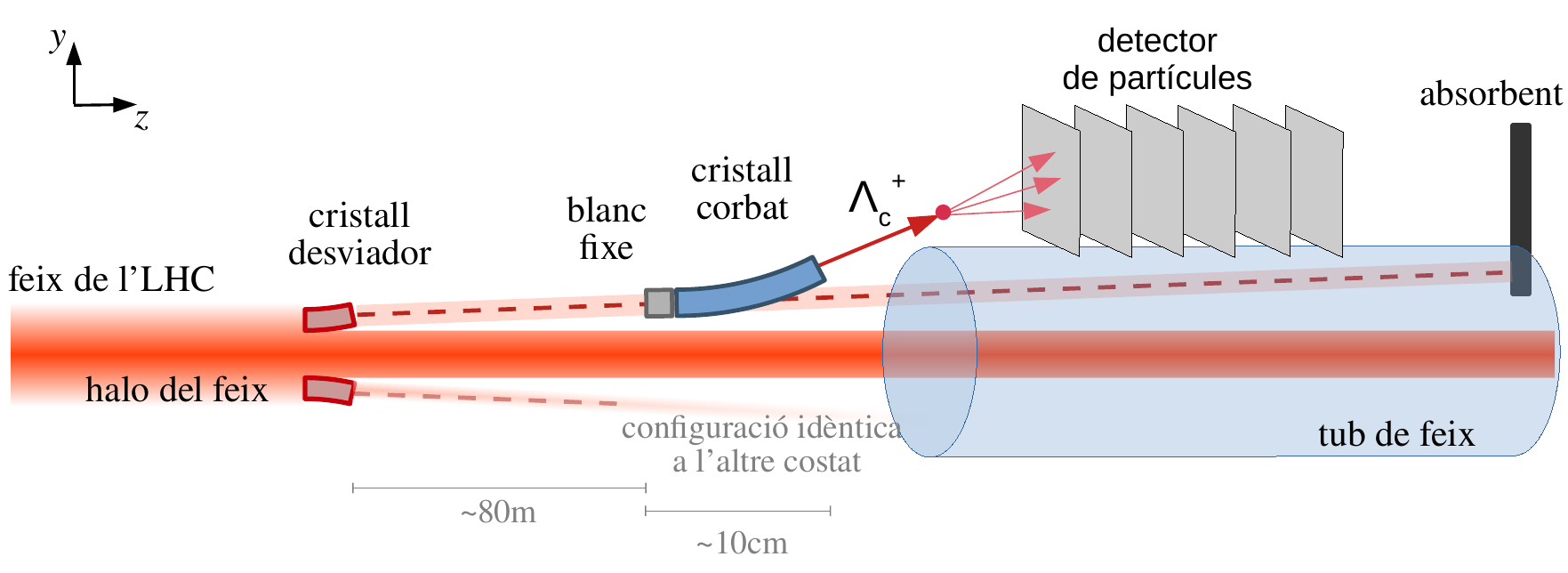}
	\caption{Disposició de l'experiment proposat. Al feix de l'accelerador de partícules LHC, part dels protons en la part externa del feix són desviats amb un cristall corbat (en roig), generant un feix secundari paral·lel al feix principal. Estos protons interaccionen amb un blanc fixe de wolframi, generant partícules noves, entre elles barions charm \Lc (fletxa roja). Una xicoteta part d'estos barions ($10^{-3}$) són produïts molt cap endavant i entren a un cristall corbat paral·lelament als plans atòmics del cristall. La repulsió elèctrica amb els àtoms manté la partícula \textit{atrapada} entre dos plans atòmics, que estan corbats també. En travessar el cristall complet, la partícula \Lc es desintegra en el procés $\LcpKpi$. Els productes de la desintegració es detecten, reconstruint la seua trajectòria. Analitzant la direcció relativa d'estes partícules en el sistema de referència on la \Lc està en repòs, obtenim informació sobre el vector de polarització de la partícula. Finalment, la polarització s'analitza en funció de l'energia de la partícula (\textit{boost} de Lorenz $\gamma$) i s'extrauen els moments dipolars (elèctric i magnètic) del barió \Lc. }
	\label{fig:FT}
\end{figure}

\begin{enumerate}
	\item \textbf{Font de partícules polaritzades}\\
	Part dels protons LHC a la regió externa del feix (\textit{beam halo}) arriben a un blanc fixe de wolframi, on interaccionen amb els protons i neutrons dels nuclis atòmics generant barions charm amb polarització transversal al plà de producció. Veure Fig.~\ref{fig:FT}.
	
	\item \textbf{Camp electromagnètic per induir la precessió d'espí} \\
	La curta vida d'aquestes partícules s'estén en gran mesura en el sistema de referència del laboratori a causa d'efectes relativistes, fent que viatgen uns quants centímetres després del blanc. Les partícules amb càrrega positiva que entren al cristall són repel·lides pels plans atòmics amb càrrega positiva i queden \textit{atrapades} en l'estructura ordenada d'àtoms del cristall. Aquestes partícules són canalitzades al llarg d'un camí corbat, sotmeses al camp electromagnètic dels àtoms carregats, que indueix la rotació del vector de polarització o precessió d'espí.
	
	\item \textbf{Analitzador del vector de polarització final}\\
	Després de la sortida del cristall, les partícules que han sobreviscut han sigut desviades de la seua trajectòria inicial. Aquestes partícules es desintegren i els productes de desintegració poden ser reconstruïts en un detector posicionat fora del tub de feix. Amb una reconstrucció precisa de les seues direccions, la polarització es pot extraure de forma estadística.
\end{enumerate}

El fenomen de precessió d'espí mai s'ha observat en partícules de vida molt curta, que es desintegren en $\sim10^{-13}\,\s$, ja que aquestes presenten grans complicacions respecte a altres sistemes (meta)estables. Avui, amb l'última tecnologia de cristall corbats i el potent feix de protons de l'LHC, tenim una oportunitat única per mesurar estos observables tan evasius.
%

\subsubsection*{Tres disposicions dels cristalls}

\begin{figure}
	\centering
	\includegraphics[width=0.99\linewidth]{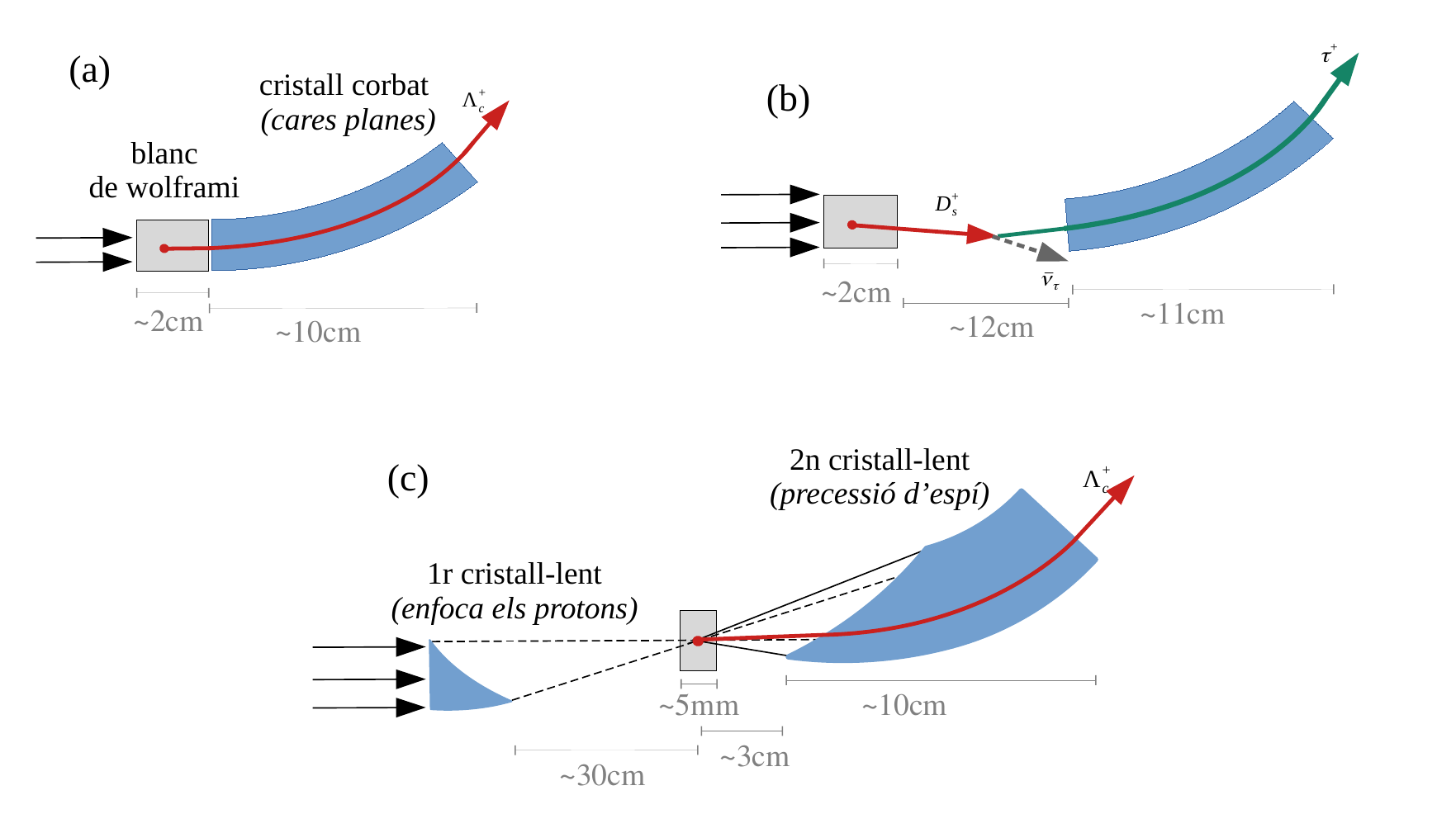}
	\caption{La resta d'elements en la configuració experimental proposada estan inclosos a l'esquema de la Figura~\ref{fig:FT}.    }
	\label{fig:threesetupsresumvalencia}
\end{figure}

Depenent de la partícula que volem mesurar i la geometria del cristall, s'han estudiat les següents configuracions, representades en la Figura \ref{fig:threesetupsresumvalencia}. La resta de la disposició general (cristall desviador, absorbent, posició vertical dins del tub de feix) és conceptualment idèntica a la presentada en la Figura \ref{fig:FT}.

\begin{enumerate}
	\item[a.] \textbf{Barions pesats}
	
	Per mesurar l'EDM i MDM del barió charm utilitzem una configuració on el cristall i el blanc fixe estan junts. Açò minimitza la pèrdua d'esdeveniments degut a la desintegració exponencial de la \Lc, que només vola uns pocs centímetres.
	
	\item[b.] \textbf{Leptó $\tau$}
	
	Donat que el MDM del leptó $\tau$ és molt xicotet, amb $(g-2)_{\tau} \approx 10^{-3}$, i la seua tasa de producció és relativament baixa, sent dominada pel canal de desintegració $\Dsp\to\taup\nu_\tau$, per mesurar la primera xifra significativa del $(g-2)_{\tau}$ necessitaríem períodes extensos de presa de dades. En la configuració experimental, necessitem incorporar una separació entre el blanc i el cristall de $\approx 12 \cm$ per permetre el vol i desintegració del mesó $\Dsp$.
	
	\item[c.] \textbf{Cristalls-lent}
	
	La probabilitat de canalització dels barions en la disposició nominal (a) és molt baixa, $\order(10^{-4})$. El coll de botella d'aquesta eficiència està en el xicotet marge en l'angle de la partícula per que aquesta siga atrapada entre els plans atòmics, anomenat angle de Lindhard. Aquest angle no es pot augmentar, però la geometria general de la configuració sí que es pot canviar per que les partícules siguen atrapades en un rang més ample de direccions. Aquesta idea es basa en l'ús de cristalls-lent, que tenen una geometria un poc diferent als cristalls corbats amb cares planes. En la configuració amb cristalls-lent tots els plans atòmics a l'entrada del cristall apunten cap al blanc fixe, on es produeixen les partícules.

\end{enumerate}

\subsection*{Resultats i sensitivitat de la mesura}

En aquesta tesi em realitzat les simulacions adients de l'experiment proposat per estimar la incertesa final en la mesura de l'EDM i el MDM. 

Per barions charm, amb la disposició nominal (a), obtenim que amb $1.37\times 10^{13}$ protons impactant en el blanc fixe, integrats en dos anys de presa de dades aproximadament, amb un flux de protons de $10^6\,p/s$, la sensibilitat assolible en el MDM (EDM) de la \Lc és de $2\times 10^{-2}\ {\mu_{ N}}$ ($ 3\times 10^{-16}~e\cm$). Amb esta incertesa, el moment magnètic dels barions charm es podria mesurar per primera vegada amb una precisió del 2\%, molt per davall de les incerteses de les prediccions teòriques. Per l'EDM, els límits indirectes en l'EDM del quark charm, derivats també en aquesta tesi, al Capítol \ref{ch:improvedbounds}, són més restrictius que la incertesa projectada de la mesura per un factor $\sim10^{5}$. No obstant, per obtenir aquestes cotes indirectes hem fet diverses aproximacions i, en general, no s'ha de menysprear el valor intrínsec d'una mesura directa, lliure de consideracions teòriques.

Amb la disposició (c) dels cristalls-lent, podem augmentar el nombre de partícules \Lc mesurades en un factor $\approx 20$, que es tradueix en una disminució de la incertesa de la mesura d'un factor $\sqrt{20}\approx4.5$. Malgrat això, aquesta disposició presenta complicacions addicionals en la fabricació dels cristalls corbats i el seu posicionament dins del tub de feix, que ha de controlar-se amb un nivell de precisió de $\lesssim 1 \mum$, amb un cristall de $\approx 10\,\cm$ de llarg. No obstant és important mencionar que la tecnologia per obtenir aquesta precisió ja existeix i s'ha provat amb cristalls desviadors de cares planes al feix de l'LHC i de l'accelerador RHIC en Nova York.

Pels leptons $\tau$, la predicció del SM pel MDM del \Ptau es podria verificar experimentalment amb {\CHb una mostra} d'al voltant de {\CHc $10^{17}$ protons}, fent possible la cerca del seu EDM amb una precisió al nivell de $10^{-17}~e\cm$.
{\CHc Això requeriria un 10\% dels protons emmagatzemats durant una dècada d'operació de l'\lhc.}

\section*{Part II: Anàlisi de dades a l'LHCb}

\subsection*{Motivació i objectius}

\begin{figure}
	\centering
	\includegraphics[width=0.8\linewidth]{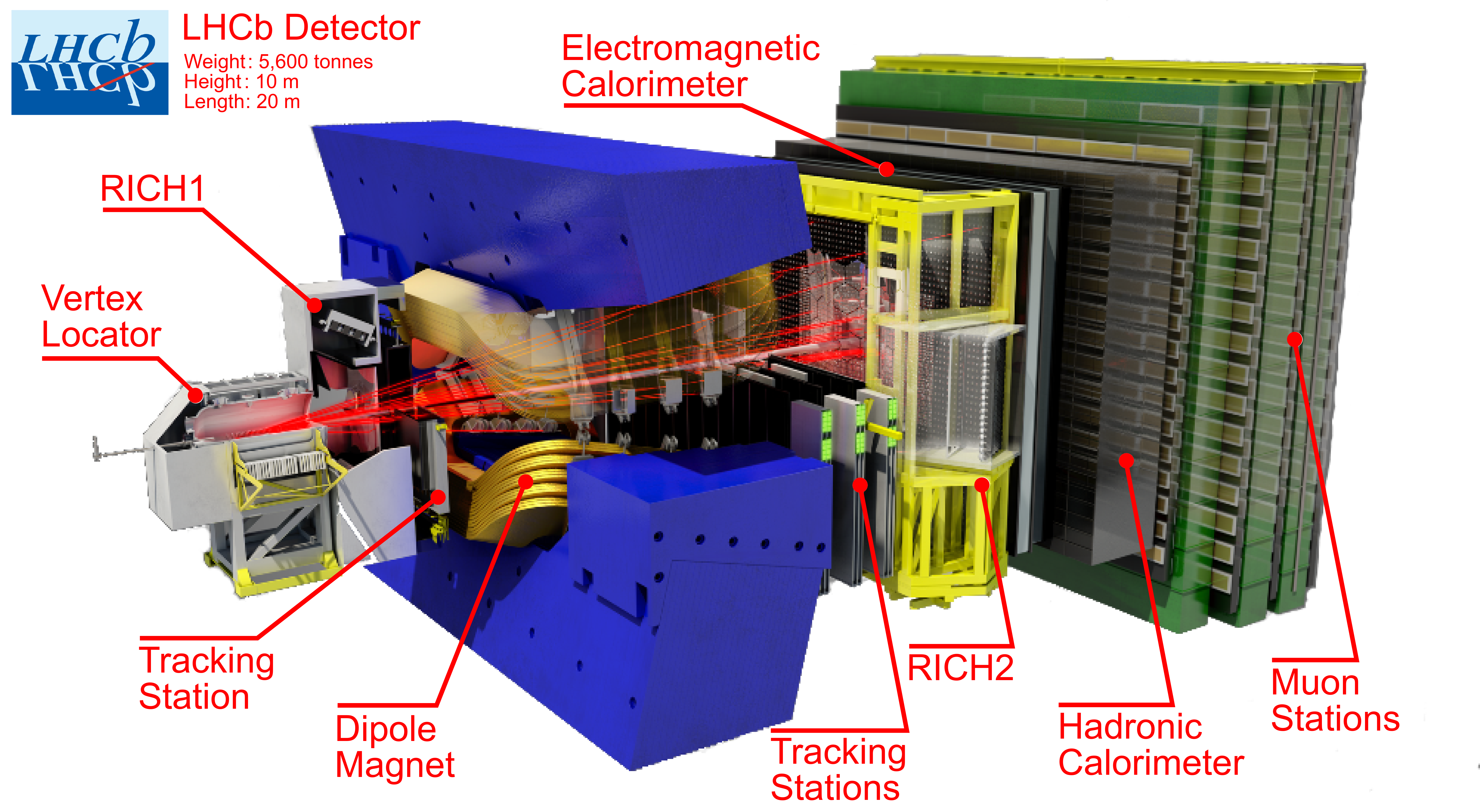}
	\caption{Figura presa de la referència \cite{Koppenburg:2015pca}. Detector LHCb actual. La seua geometria permet reconstruir amb gran precisió les desintegracions de partícules amb quarks pesats que es produeixen típicament en un con d'uns 300\,\mrad al voltant del feix de protons. En aquesta tesi hem utilitzat dades de la versió anterior d'aquest detector, que va estar operatiu des del començament de l'LHC, al 2010, fins el final del Run II, al 2018.}
	\label{fig:lhcb}
\end{figure}

Per a barions \Lz amb un quark \textit{strange}, que tenen un temps de vida més llarg que els barions charm o leptons $\Ptau$, l'experiment LHCb (Figura~\ref{fig:lhcb}) ofereix una oportunitat diferent per mesurar els seus moments dipolars elèctrics i magnètics, aquesta vegada sense instrumentació addicional. Reconstruint esdeveniments en els quals els hiperons \Lz es desintegren abans i després de l'imant de l'LHCb és possible comparar la seua polarització abans i després del camp magnètic, extraient els moments dipolars. L'imant dipolar de l'LHCb s'utilitza per desviar la trajectòria de les partícules carregades i mesurar el seu moment. Tanmateix, si en aquesta proposta de mesura volem utilitzar el cap magnètic per induir la precessió d'espí, com anem a mesurar el moment dels productes de desintegració? La reconstrucció d'estos esdeveniments suposa, efectivament, un repte per a l'experiment, però als últims anys hem aconseguit avanços molt significatius en aquest front, alguns explicats al Capítol~\ref{ch:lambdasLHCb} (de la Part I).
No obstant això, tenir accés a este tipus d'esdeveniments pot expandir el programa de física de l'experiment LHCb a través de mesures directes de moments dipolars electromagnètics. D'altra banda, la reconstrucció d'aquests esdeveniments augmenta enormement el rang de vida mitja en les cerques de partícules predites per teories més enllà del SM.

A banda de les millores en reconstrucció, per fer aquesta mesura necessitem saber la polarització inicial de l'hiperó \Lz (abans d'entrar al camp magnètic). En concret, volem saber \textit{quin tipus de \Lz} tenen major polarització o, dit d'altra manera, quins són els canals de desintegració donant lloc a partícules \Lz que produeixen major polarització. L'única manera de saber-ho és mesurant la polarització de la partícula \Lz en tots estos canals, resumits en la Taula~\ref{tab:LambdaChannels} del text principal. Amb aquest objectiu, començarem pel canal de desintegració \threepi amb les dades completes del Run II (2015-2018). En la Part II de la tesi, hem construit tota la cadena d'anàlisi usant les dades del 2016, proporcionant una mesura preliminar de la polarització.

Gran part d'aquesta anàlisi de dades es pot ``reutilitzar" per obtenir altres mesures que són interessants per si mateixes. Un exemple seria la cerca d'un estat lligat amb cinc quarks, un \textit{pentaquark} $\Sigma^*$, que podria intermediar aquesta desintegració en el procés $\Lc\to\Sigma^*(\to\Lz \pip) \pip \pim$.

\subsection*{Preparació de les dades }

En el moment en què va començar el nostre interès per aquests canals, el \textit{Run II} de l'LHC ja estava en marxa. El primer pas va ser implementar línies de \textit{trigger} per registrar els esdeveniments d'interés. No obstant això, inclús si prèviament no existien triggers dedicats pel nostre canal de desintegració, podem recuperar part d'estos esdeveniments que van ser registrats amb triggers independents del nostre canal ó genèrics. Per fer açò necessitem incloure el nostre canal en el procés d'\textit{stripping} de l'LHCb. En aquest procés, s'accedeix de forma centralitzada a totes les dades ja registrades per l'LHCb, executant de nou els algorismes de reconstrucció. La informació reconstruïda es compara amb el conjunt de criteris de selecció definits per les línies d'stripping i l'esdeveniment es guarda si compleix amb els requeriments d'alguna de les línies.

\subsection*{Selecció \textit{offline}}

Una vegada tenim el primer conjunt de dades del 2016 obtingut a l'stripping, que ocupa al voltant de 1\,Tb, volem reduir aquestes dades per aïllar la senyal del fons, composat per combinacions aleatòries de partícules que s'assemblen a aquesta desintegració però que no tenen cap interés físic. Ho farem en dos pasos:

\begin{enumerate}
	
	\item \textbf{Preselecció}
	
	Una cosa que sabem de les partícules involucrades en la nostra desintegració és la seua massa. Per tant, si traiem l'histograma amb la massa invariant de la \Lc per a tots els esdeveniments, hauríem de veure una acumulació d'esdeveniments al voltant del valor real de la seua massa, en la Figura~\ref{fig:initialPeaks} (dreta). Com veiem, aquesta acumulació d'esdeveniments és quasi inapreciable i la gran majoria d'esdeveniments són fons. Per eliminar aquest fons, posem requeriments en altres variables de l'esdeveniment que, idealment, no eliminarien cap candidat de senyal. Tanmateix, aquest ideal està lluny de la realitat i amb aquesta \textit{preselecció} hem eliminat al voltant del 50\% de la senyal, però també ens em desfet del 98.9\% del fons, com veiem a la Taula~\ref{tab:preselectionCuts}. El pic resultant està a la Figura~\ref{fig:preselectionPeaks}.
	
	
	~\\
	
	\item \textbf{Classificador multivariant}
	
	No ens conformem amb els resultats de la preselecció i volem augmentar la puresa de la senyal el màxim possible. Per fer açò utilitzem un algorisme d'aprenentatge automàtic que és capaç de trobar la separació òptima entre esdeveniments de senyal i de fons, basat en les variables de l'esdeveniment que posem a la seua disposició. Per resumir, aquest algorisme pot \textit{veure} les relacions entre totes aquestes variables en un espai multidimensional (amb tantes dimensions com nombre de variables), i trobar les diferències entre els esdeveniments de senyal i fons. En concret, hem fet servir \textit{arbres de decisió potenciats} (\textit{boosted decision trees}) amb el programari TMVA del marc d'anàlisi de dades \root, desenvolupat al CERN precisament per fer aquest tipus d'anàlisi de dades.
	
\end{enumerate}

\subsection*{Ajust angular}

Estadísticament, la direcció del protó en la desintegració $\Lprp$ segueix una distribució de probabilitat que depén de la polarització de la partícula mare, ${\bm P_{\Lz}}$ (abans hem utilitzat la notació $\spol$).
Per extreure aquesta polarització hem realitzat un ajust de màxima versemblança (\textit{maximum likelihood fit}) de la direcció del protó en els esdeveniments seleccionats.
Finalment, obtenim les tres components de la polarització,

\begin{equation}
\begin{array}{lcc}
P_{\Lz,x} &=& (1.0\pm 1.6\pm 1.4)\%, \\
P_{\Lz,y} &=& (4.0\pm 1.7 \pm 1.5)\%, \\
P_{\Lz,z} &=& (-24.1\pm 1.6 \pm 1.2)\%,
\end{array}
\end{equation}
on la primera incertesa és estadística (associada a les fluctuacions estadístiques en una mostra finita d'esdeveniments) i la segona és sistemàtica (associada al tipus de tractament de les dades que hem fet, i que hem avaluat utilitzant mètodes alternatius en alguns dels passos de l'anàlisi).

\subsection*{Conclusions}

En resum, s'ha assolit l'objectiu inicial d'avaluar la polarització de l'hiperó \Lz en el canal de desintegració \threepi. Aquest resultat és rellevant per comparar la sensibilitat de la mesura de l'EDM i MDM de la \Lz en diferents canals de producció.
Diversos estudis importants encara han de completar-se i els mètodes d'anàlisi han de ser examinats meticulosament per la col·laboració per convertir aquest primer esborrany de la cadena d'anàlisi en una mesura oficial de l'LHCb.

\section*{Part III: Fenomenologia d'EDMs}

\subsection*{Noves cotes a l'EDM dels quarks pesats}

\begin{figure}[t]
	\centering
	\raisebox{-.5\height}{
		\subcaptionbox{}{\includegraphics[width=0.35\columnwidth]{pheno/fBounds/photonCorrection.pdf}}
		\subcaptionbox{}{\raisebox{0.23cm}{\includegraphics[scale=1.2]{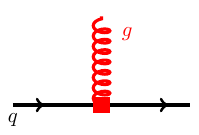}}}
	}
	\caption{L'EDM del quark (quadrat blau) indueix un chromo-EDM del quark (quadrat roig) a través d'un bucle tancat de fotons (línia blava). Calculant aquesta contribució obtenim un resultat infinit, que podem \textit{renormalitzar}. Malgrat que el paràmetre de l'operador EDM i CEDM és adimensional (en certa base d'operadors), aquesta renormalització dona lloc a una dependència ``anòmala" de l'EDM i chromo-EDM amb l'escala d'energia. Aquesta evolució amb l'escala d'energia és descrita per les \textit{equacions del grup de renormalització}, on la informació essencial està inclosa en la \textit{matriu de dimensió anòmala}. Aquest diagrama (a) contribueix al seu element $(\gamma_e)_{12}^{(0)}$.}
	\label{fig:loopfoto}
\end{figure}

Mesurar l'EDM dels barions pesats ens dona informació directa de l'EDM i el CEDM dels quarks pesats. A banda, també podem explotar al màxim els resultats d'altres experiments per treure informació indirecta sobre aquestes quantitats. Els intents de restringir l'EDM i CEDM de quarks pesats han seguit diferents estratègies. Totes les cotes que s'han obtés a la literatura científica, fins on sabem, estan compilades en les Taules \ref{tab:charmEDMbounds} i \ref{tab:bottomEDMbounds} del text principal per a quarks charm i bottom, respectivament.

En aquesta tesi hem seguit una nova estratègia que justament relaciona aquests dos operadors (EDM i CEDM) i ens permet treure noves cotes en l'EDM a partir del CEDM. Aquesta relació es fa d'una manera independent del model, utilitzant les equacions del grup de renormalització, que \textit{mesclen} els operadors efectius quan es canvia l'escala d'energia.
Els diagrames rellevants inclouen bucles de fotons, en la Figura~\ref{fig:loopfoto}, que s'han passat per alt en altres estudis a causa de la seua xicoteta contribució en comparació amb les correccions provinents de bucles de gluons. No obstant això, estos bucles de fotons representen la primera contribució no nul·la a la mescla d'operadors que ens interessa.

Assumint una interferència constructiva entre l'EDM i CEDM a l'escala de nova física, $M_{\text{NP}}\sim 1\,\tev$, podem treure límits en aquest EDM, $\dq(M_{\text{NP}})$, basats en les cotes al CEDM a l'escala de la massa dels quarks. Els nous límits en l'EDM dels quarks charm i bottom són
\begin{align}
|\dc(m_c)| &< \: 1.5\times 10^{-21}\:\ecm~,\nonumber\\
|\db(m_b)| &< \: 1.2 \times 10^{-20}\:\ecm~,
\label{eq:newlimits}
\end{align}
els quals milloren els anteriors límits, en l'Equació \eqref{eq:prevbounddq} del text principal, per tres i quatre ordres de magnitud, respectivament.

\subsection*{EDMs en models de nova física amb escalars de color}

Les noves cotes a l'EDM dels quarks pesats, en la secció anterior, tenen implicacions per models de nova física que tracten de donar resposta a alguns dels problemes (experimentals i teòrics) del model estàndard. Un d'estos models és l'anomenat model de Manohar-Wise, que prediu l'existència de noves partícules escalars semblants al bosó de Higgs però que, a més, tenen càrrega de color com els gluons. Aquestes ``implicacions" es materialitzen com nous límits en els paràmetres lliures del model. En concret, aquests límits s'apliquen sobre la combinació de paràmetres que apareix en la predicció de l'EDM del quark en aquest model. 

En avaluar aquestes noves restriccions a través de l'EDM dels quarks pesats va quedar clar que només una anàlisi completa d'observables d'EDM en aquest model podria posar les cotes més restrictives a l'espai de paràmetres del model. Tot i que els resultats dels experiments (en concret, de l'EDM del neutró) ja estaven disponibles, les prediccions del model per aquests observables no s'havien calculat encara. Totes les contribucions rellevants a l'EDM del neutró han sigut obtingudes al Capítol~\ref{ch:edmsmw}, i podem donar ací un esquema de tot el procés.

\begin{figure}[t]
	\centering
	\vbox{
		\resizebox{0.85\columnwidth}{!}{
			\subcaptionbox{}{\raisebox{0.3cm}{\includegraphics[scale=1.2]{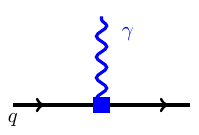}}}
			\subcaptionbox{}{\includegraphics[scale=1.1]{pheno/fEDMsMW/diagrams/oneloopqS+qprimephoton}}
			\subcaptionbox{}{\includegraphics[scale=1]{pheno/fEDMsMW/diagrams/barzeeS0gluontphoton}}
			
		}
	}
	\caption{L'operador efectiu de l'EDM del quark (a) rep contribucions, dels nous escalars $S^\pm$ i $S^0_{R,I}$ en el model de Manhoar-Wise. Estes contribucions es poden representar amb diagrames de Feynman, que són extremadament útils per realitzar els càlculs ja que cada línia i vèrtex d'estos diagrames té una definició matemàtica precisa. Les contribucions amb un bucle (b) dominen per l'EDM de quarks pesats, mentre que els diagrames de \textit{Barr-Zee} amb dos bucles (c) donen la màxima contribució per l'EDM dels quarks lleugers.}\label{fig:diagrames}
\end{figure}

Com hem introduït al començament del resum de la tesi, connectar les teories de nova física amb els observables a baixa energia no és senzill ni directe\footnote{Tot i que en aquests càlculs a sovint se sent la paraula \textit{straightforward}. Veure l'Apèndix~\ref{app:loops}.}. Aquests càlculs tenen quatre parts:

\begin{enumerate}
	\item \textbf{Definir la teoria efectiva}
	
	Necessitem definir una teoria efectiva (EFT) vàlida a escales d'energia intermèdies. En el nostre cas, utilitzem el Lagrangià de l'Equació~\eqref{eq:lagrangianEDMsresumvalencia}.
	
	\item \textbf{Càlcul en el model fonamental}
	
	Hem de relacionar els paràmetres lliures d'aquesta teoria efectiva (\textit{coeficients de Wilson}) amb els de la teoria fonamental. Entre aquests coeficients de Wilson es troba l'EDM del quark $\dq$, a la Figura~\ref{fig:diagrames} (a). Aquesta és la part més complicada, especialment si tenim contribucions amb dos bucles o més\footnote{Els càlculs amb un bucle tampoc són especialment ``fàcils" però, per la part més crítica del càlcul, la integral en el quadrimoment del bucle, ja existeixen solucions tabulades. En concret, la parametrització d'aquestes integrals amb funcions de Passarino-Veltman està implementada i automatitzada en programaris com \textsc{FeynCalc}, per \textsc{Mathematica}. }, com es mostra a la Figura~\ref{fig:diagrames} (c).
	
	\item \textbf{Evolucionar els coeficients amb l'escala d'energia}
	
	Els coeficients de Wilson de la teoria efectiva depenen de l'escala d'energia com hem explicat en la Figura~\ref{fig:loopfoto}. A partir del seu valor a alta energia ($\sim \tev$) hem de calcular el seu valor a l'escala hadrònica ($\sim \gev$). Açò ho podem fer amb les equacions del grup de renormalització, que hem solucionat a l'apèndix~\ref{app:RGEsolution}.
	
	~\\
	
	\item \textbf{Relacionar la teoria efectiva amb l'observable}
	
	Aquesta part del càlcul competeix en complexitat amb el punt (2.) i, en casos com el de l'EDM del neutró, aquesta és amb diferència la part més complicada. Per obtenir resultats fiables, molts grups estan activament revisant aquests càlculs i refent-los amb diferents tècniques de física hadrònica a baixa energia. Afortunadament per nosaltres, podem senzillament prendre els seus resultats, que tenen associada una incertesa que haurem de tenir en compte.
	
\end{enumerate}

Amb tots estos ingredients, hem obtingut alguns resultats fenomenològics del model: la comparació de l'EDM per a diferents sabors de quarks (Figura~\ref{fig:EDM_heavy_quarks} del text principal), les prediccions de l'EDM del neutró en funció dels paràmetres del model (Figures~\ref{fig:1D_EDM_alphaU} i \ref{fig:1D_EDM_mS}), i noves restriccions en l'espai de paràmetres del model (Figura~\ref{fig:2D_RegionPlot_EDM}).

\newpage
\thispagestyle{plain}

\fancyhead[LE,RO]{\textbf{\thepage}} 
\fancyhead[LO]{}
\fancyhead[RE]{\nouppercase{\leftmark}}

\bibliographystyle{utphysCUSTOM}
\bibliography{newthesis.bib,focusing.bib,internalnote.bib,bibTFM.bib,phenonotes.bib,EDMsMW.bib,proceedingsW.bib}

\providecommand{\href}[2]{#2}\begingroup\raggedright\begin{thebibliography}{100}

\bibitem{Botella:2016ksl}
F.~J. Botella, L.~M. Garcia~Martin, D.~Marangotto, F.~Martinez~Vidal, A.~Merli,
  N.~Neri, A.~Oyanguren, and J.~Ruiz~Vidal, ``{On the search for the electric
  dipole moment of strange and charm baryons at LHC}'',
  \href{http://dx.doi.org/10.1140/epjc/s10052-017-4679-y}{{\em Eur. Phys. J. C}
  {\bfseries 77} no.~3, (2017) 181},
  \href{http://arxiv.org/abs/1612.06769}{{\ttfamily arXiv:1612.06769
  [hep-ex]}}.

\bibitem{internalnote}
L.~Henry, D.~Marangotto, F.~Martinez~Vidal, A.~Merli, N.~Neri, P.~Robbe, and
  J.~Ruiz~Vidal, ``{Proposal to search for baryon EDMs with bent crystals at
  LHCb}'', {\em {\rm
  \href{https://cds.cern.ch/record/2265593/}{LHCb-INT-2017-011} (restricted
  access)}} .

\bibitem{Bagli:2017foe}
E.~Bagli {\em et~al.}, ``{Electromagnetic dipole moments of charged baryons
  with bent crystals at the LHC}'',
  \href{http://dx.doi.org/10.1140/epjc/s10052-017-5400-x}{{\em Eur. Phys. J. C}
  {\bfseries 77} no.~12, (2017) 828},
  \href{http://arxiv.org/abs/1708.08483}{{\ttfamily arXiv:1708.08483
  [hep-ex]}}. [Erratum: Eur.Phys.J.C 80, 680 (2020)].

\bibitem{Fu:2019utm}
J.~Fu, M.~A. Giorgi, L.~Henry, D.~Marangotto, F.~M. Vidal, A.~Merli, N.~Neri,
  and J.~Ruiz~Vidal, ``{Novel Method for the Direct Measurement of the
  \ensuremath{\tau} Lepton Dipole Moments}'',
  \href{http://dx.doi.org/10.1103/PhysRevLett.123.011801}{{\em Phys. Rev.
  Lett.} {\bfseries 123} no.~1, (2019) 011801},
  \href{http://arxiv.org/abs/1901.04003}{{\ttfamily arXiv:1901.04003
  [hep-ex]}}.

\bibitem{Aiola:2020yam}
S.~Aiola {\em et~al.}, ``{Progress towards the first measurement of charm
  baryon dipole moments}'',
  \href{http://dx.doi.org/10.1103/PhysRevD.103.072003}{{\em Phys. Rev. D}
  {\bfseries 103} no.~7, (2021) 072003},
  \href{http://arxiv.org/abs/2010.11902}{{\ttfamily arXiv:2010.11902
  [hep-ex]}}.

\bibitem{Biryukov:2021cml}
V.~M. Biryukov and J.~Ruiz~Vidal, ``{Improved experimental layout for dipole
  moment measurements at the LHC}'',
  \href{http://dx.doi.org/10.1140/epjc/s10052-022-10114-5}{{\em Eur. Phys. J.
  C} {\bfseries 82} no.~2, (2022) 149},
  \href{http://arxiv.org/abs/2110.00845}{{\ttfamily arXiv:2110.00845
  [hep-ex]}}.

\bibitem{DEMONSTRATOR}
{LHCb} collaboration, ``{Long-lived particle reconstruction downstream of the
  \lhcb magnet (to appear)}'', 2022.
\newblock {CERN-LHCb-DP-2022-001}.

\bibitem{Gisbert:2019ftm}
H.~Gisbert and J.~Ruiz~Vidal, ``{Improved bounds on heavy quark electric dipole
  moments}'', \href{http://dx.doi.org/10.1103/PhysRevD.101.115010}{{\em Phys.
  Rev. D} {\bfseries 101} no.~11, (2020) 115010},
  \href{http://arxiv.org/abs/1905.02513}{{\ttfamily arXiv:1905.02513
  [hep-ph]}}.

\bibitem{Gisbert:2021htg}
H.~Gisbert, V.~Miralles, and J.~Ruiz-Vidal, ``{Electric dipole moments from
  colour-octet scalars}'',
  \href{http://dx.doi.org/10.1007/JHEP04(2022)077}{{\em JHEP} {\bfseries 04}
  (2022) 077}, \href{http://arxiv.org/abs/2111.09397}{{\ttfamily
  arXiv:2111.09397 [hep-ph]}}.

\bibitem{Merli:2019hyz}
A.~Merli, {\em {Search for $CP$ violation in the angular distribution of
  $\Lambda_b^0 \to p \pi^+\pi^-\pi^+$ baryon decays and a proposal for the
  search of heavy baryon EDM with bent crystal at LHCb}}.
\newblock PhD thesis, Milan University, 2019.

\bibitem{Marangotto:2020tzf}
D.~Marangotto, {\em {Amplitude Analysis and Polarisation Measurement of the
  $\Lambda_c$ Baryon in \pr\Km\pip Final State for Electromagnetic Dipole
  Moment Experiment}}.
\newblock PhD thesis, Milan University, 2020.

\bibitem{GisbertMullor:2019vwg}
H.~Gisbert~Mullor, {\em {Phenomenological applications in CP-violating systems
  in the SM and beyond}}.
\newblock PhD thesis, Universitat de Val\`encia, 2019.

\bibitem{ThesisVictor}
V.~Miralles, {\em {Hunting for new physics in the LHC era}}.
\newblock PhD thesis, Universitat de Val\`encia, 2021.

\bibitem{Weinberg:1992nd}
S.~Weinberg, {\em {Dreams of a final theory: The Search for the fundamental
  laws of nature}}.
\newblock 1992.

\bibitem{Pendlebury:2000an}
J.~M. Pendlebury and E.~A. Hinds, ``{Particle electric dipole moments}'',
  \href{http://dx.doi.org/10.1016/S0168-9002(99)01023-2}{{\em Nucl. Instrum.
  Meth. A} {\bfseries 440} (2000) 471--478}.

\bibitem{Smith:1957ht}
J.~H. Smith, E.~M. Purcell, and N.~F. Ramsey, ``{Experimental limit to the
  electric dipole moment of the neutron}'',
  \href{http://dx.doi.org/10.1103/PhysRev.108.120}{{\em Phys. Rev.} {\bfseries
  108} (1957) 120--122}.

\bibitem{Wu:1957my}
C.~S. Wu, E.~Ambler, R.~W. Hayward, D.~D. Hoppes, and R.~P. Hudson,
  ``{Experimental Test of Parity Conservation in $\beta$ Decay}'',
  \href{http://dx.doi.org/10.1103/PhysRev.105.1413}{{\em Phys. Rev.} {\bfseries
  105} (1957) 1413--1414}.

\bibitem{SternGerlach1}
W.~Gerlach and O.~Stern, ``{Der experimentelle Nachweis des magnetischen
  Moments des Silberatoms}'', {\em Zeitschrift für Physik} {\bfseries 8}
  (1921) 110--111.

\bibitem{SternGerlach2}
O.~Stern, ``{Ein Weg zur experimentellen Prüfung der Richtungsquantelung}'',
  {\em Zeitschrift für Physik} {\bfseries 7} (1921) 249--253.

\bibitem{SternGerlach3}
W.~Gerlach and O.~Stern, ``{Der experimentelle Nachweis der Richtungsquantelung
  im Magnetfeld}'', {\em Zeitschrift für Physik} {\bfseries 9} (1922b)
  349--352.

\bibitem{GoudsmitUhlenbeck}
G.~E. Uhlenbeck and S.~Goudsmit, ``{Ersetzung der Hypothese vom unmechanischen
  Zwang durch eine Forderung bezüglich des inneren Verhaltens jedes einzelnen
  Elektrons}'', {\em Die Naturwissenschaften} {\bfseries 13} (1925) 953--954.

\bibitem{Dirac:1928hu}
P.~A.~M. Dirac, ``{The quantum theory of the electron}'',
  \href{http://dx.doi.org/10.1098/rspa.1928.0023}{{\em Proc. Roy. Soc. Lond. A}
  {\bfseries 117} (1928) 610--624}.

\bibitem{Schwinger:1948iu}
J.~S. Schwinger, ``{On Quantum electrodynamics and the magnetic moment of the
  electron}'', \href{http://dx.doi.org/10.1103/PhysRev.73.416}{{\em Phys. Rev.}
  {\bfseries 73} (1948) 416--417}.

\bibitem{Kusch:1948mvb}
P.~Kusch and H.~M. Foley, ``{The Magnetic Moment of the Electron}'',
  \href{http://dx.doi.org/10.1103/PhysRev.74.250}{{\em Phys. Rev.} {\bfseries
  74} no.~3, (1948) 250}.

\bibitem{Fritzsch:2015jfa}
H.~Fritzsch and M.~Gell-Mann, eds., \href{http://dx.doi.org/10.1142/9249}{{\em
  {50 years of quarks}}}.
\newblock World Scientific, Hackensack, 2015.

\bibitem{Guth:1980zm}
A.~H. Guth, ``{The Inflationary Universe: A Possible Solution to the Horizon
  and Flatness Problems}'',
  \href{http://dx.doi.org/10.1103/PhysRevD.23.347}{{\em Phys. Rev. D}
  {\bfseries 23} (1981) 347--356}.

\bibitem{Iocco:2008va}
F.~Iocco, G.~Mangano, G.~Miele, O.~Pisanti, and P.~D. Serpico, ``{Primordial
  Nucleosynthesis: from precision cosmology to fundamental physics}'',
  \href{http://dx.doi.org/10.1016/j.physrep.2009.02.002}{{\em Phys. Rept.}
  {\bfseries 472} (2009) 1--76},
  \href{http://arxiv.org/abs/0809.0631}{{\ttfamily arXiv:0809.0631
  [astro-ph]}}.

\bibitem{Sakharov:1967dj}
A.~D. Sakharov, ``{Violation of \CP Invariance, c Asymmetry, and Baryon
  Asymmetry of the Universe}'',
  \href{http://dx.doi.org/10.1070/PU1991v034n05ABEH002497}{{\em Pisma Zh. Eksp.
  Teor. Fiz.} {\bfseries 5} (1967) 32--35}.
[Usp. Fiz. Nauk161,61(1991)].

\bibitem{Bennett:2012zja}
{WMAP} collaboration, C.~L. Bennett {\em et~al.}, ``{Nine-Year Wilkinson
  Microwave Anisotropy Probe (WMAP) Observations: Final Maps and Results}'',
  \href{http://dx.doi.org/10.1088/0067-0049/208/2/20}{{\em Astrophys. J.
  Suppl.} {\bfseries 208} (2013) 20},
\href{http://arxiv.org/abs/1212.5225}{{\ttfamily arXiv:1212.5225
  [astro-ph.CO]}}.

\bibitem{Angelopoulos:1998dv}
{CPLEAR} collaboration, A.~Angelopoulos {\em et~al.}, ``{First direct
  observation of time reversal noninvariance in the neutral kaon system}'',
\href{http://dx.doi.org/10.1016/S0370-2693(98)01356-2}{{\em Phys. Lett.}
  {\bfseries B444} (1998) 43--51}.

\bibitem{Lees:2012uka}
{BaBar} collaboration, J.~P. Lees {\em et~al.}, ``{Observation of Time Reversal
  Violation in the \Bz Meson System}'',
  \href{http://dx.doi.org/10.1103/PhysRevLett.109.211801}{{\em Phys. Rev.
  Lett.} {\bfseries 109} (2012) 211801},
\href{http://arxiv.org/abs/1207.5832}{{\ttfamily arXiv:1207.5832 [hep-ex]}}.

\bibitem{Bernabeu:2014mva}
J.~Bernabeu and F.~Martinez-Vidal, ``{Colloquium: Time-reversal violation with
  quantum-entangled B mesons}'',
  \href{http://dx.doi.org/10.1103/RevModPhys.87.165}{{\em Rev. Mod. Phys.}
  {\bfseries 87} (2015) 165},
\href{http://arxiv.org/abs/1410.1742}{{\ttfamily arXiv:1410.1742 [hep-ph]}}.

\bibitem{Pospelov:2005pr}
M.~Pospelov and A.~Ritz, ``{Electric dipole moments as probes of new
  physics}'', \href{http://dx.doi.org/10.1016/j.aop.2005.04.002}{{\em Annals
  Phys.} {\bfseries 318} (2005) 119--169},
  \href{http://arxiv.org/abs/hep-ph/0504231}{{\ttfamily arXiv:hep-ph/0504231}}.

\bibitem{Nowakowski:2004cv}
M.~Nowakowski, E.~A. Paschos, and J.~M. Rodriguez, ``{All electromagnetic
  form-factors}'', \href{http://dx.doi.org/10.1088/0143-0807/26/4/001}{{\em
  Eur. J. Phys.} {\bfseries 26} (2005) 545--560},
\href{http://arxiv.org/abs/physics/0402058}{{\ttfamily arXiv:physics/0402058
  [physics]}}.

\bibitem{Guo:2012vf}
F.-K. Guo and U.-G. Meissner, ``{Baryon electric dipole moments from strong \CP
  violation}'', \href{http://dx.doi.org/10.1007/JHEP12(2012)097}{{\em JHEP}
  {\bfseries 12} (2012) 097},
\href{http://arxiv.org/abs/1210.5887}{{\ttfamily arXiv:1210.5887 [hep-ph]}}.

\bibitem{Wirzba:2014mka}
A.~Wirzba, ``{Electric dipole moments of the nucleon and light nuclei}'',
  \href{http://dx.doi.org/10.1016/j.nuclphysa.2014.04.003}{{\em Nucl. Phys.}
  {\bfseries A928} (2014) 116--127},
\href{http://arxiv.org/abs/1404.6131}{{\ttfamily arXiv:1404.6131 [hep-ph]}}.

\bibitem{Yamanaka:2014mda}
N.~Yamanaka, \href{http://dx.doi.org/10.1007/978-4-431-54544-6}{{\em {Analysis
  of the Electric Dipole Moment in the R-parity Violating Supersymmetric
  Standard Model}}}.
\newblock PhD thesis, Osaka U., Res. Ctr. Nucl. Phys., 2013.
\newblock
\url{http://www.springer.com/gp/book/9784431545439}.
\newblock

\bibitem{Baron:2013eja}
{ACME} collaboration, J.~Baron {\em et~al.}, ``{Order of Magnitude Smaller
  Limit on the Electric Dipole Moment of the Electron}'',
  \href{http://dx.doi.org/10.1126/science.1248213}{{\em Science} {\bfseries
  343} (2014) 269--272},
\href{http://arxiv.org/abs/1310.7534}{{\ttfamily arXiv:1310.7534
  [physics.atom-ph]}}.

\bibitem{Afach:2015sja}
J.~M. Pendlebury {\em et~al.}, ``{Revised experimental upper limit on the
  electric dipole moment of the neutron}'',
  \href{http://dx.doi.org/10.1103/PhysRevD.92.092003}{{\em Phys. Rev.}
  {\bfseries D92} no.~9, (2015) 092003},
\href{http://arxiv.org/abs/1509.04411}{{\ttfamily arXiv:1509.04411 [hep-ex]}}.

\bibitem{Mannel:2012qk}
T.~Mannel and N.~Uraltsev, ``{Loop-Less Electric Dipole Moment of the Nucleon
  in the Standard Model}'',
  \href{http://dx.doi.org/10.1103/PhysRevD.85.096002}{{\em Phys. Rev.}
  {\bfseries D85} (2012) 096002},
\href{http://arxiv.org/abs/1202.6270}{{\ttfamily arXiv:1202.6270 [hep-ph]}}.

\bibitem{Pondrom:1981gu}
L.~Pondrom, R.~Handler, M.~Sheaff, P.~T. Cox, J.~Dworkin, O.~E. Overseth,
  T.~Devlin, L.~Schachinger, and K.~J. Heller, ``{New Limit on the Electric
  Dipole Moment of the $\Lambda$ Hyperon}'',
\href{http://dx.doi.org/10.1103/PhysRevD.23.814}{{\em Phys. Rev.} {\bfseries
  D23} (1981) 814--816}.

\bibitem{Flambaum:2014jta}
V.~V. Flambaum, D.~DeMille, and M.~G. Kozlov, ``{Time-reversal symmetry
  violation in molecules induced by nuclear magnetic quadrupole moments}'',
  \href{http://dx.doi.org/10.1103/PhysRevLett.113.103003}{{\em Phys. Rev.
  Lett.} {\bfseries 113} (2014) 103003},
\href{http://arxiv.org/abs/1406.6479}{{\ttfamily arXiv:1406.6479
  [physics.atom-ph]}}.

\bibitem{Anastassopoulos:2015ura}
V.~Anastassopoulos {\em et~al.}, ``{A Storage Ring Experiment to Detect a
  Proton Electric Dipole Moment}'',
  \href{http://dx.doi.org/10.1063/1.4967465}{{\em Rev. Sci. Instrum.}
  {\bfseries 87} no.~11, (2016) 115116},
  \href{http://arxiv.org/abs/1502.04317}{{\ttfamily arXiv:1502.04317
  [physics.acc-ph]}}.

\bibitem{CPEDM:2019nwp}
{CPEDM} collaboration, F.~Abusaif {\em et~al.},
  \href{http://dx.doi.org/10.23731/CYRM-2021-003}{{\em {Storage Ring to Search
  for Electric Dipole Moments of Charged Particles -- Feasibility Study}}}.
\newblock CERN, Geneva, 6, 2021.
\newblock \href{http://arxiv.org/abs/1912.07881}{{\ttfamily arXiv:1912.07881
  [hep-ex]}}.

\bibitem{Muong-2:2008ebm}
{Muon (g-2)} collaboration, G.~W. Bennett {\em et~al.}, ``{An Improved Limit on
  the Muon Electric Dipole Moment}'',
  \href{http://dx.doi.org/10.1103/PhysRevD.80.052008}{{\em Phys. Rev. D}
  {\bfseries 80} (2009) 052008},
  \href{http://arxiv.org/abs/0811.1207}{{\ttfamily arXiv:0811.1207 [hep-ex]}}.

\bibitem{Chislett:2016jau}
{Muon g-2} collaboration, R.~Chislett, ``{The muon EDM in the g-2 experiment at
  Fermilab}'', \href{http://dx.doi.org/10.1051/epjconf/201611801005}{{\em EPJ
  Web Conf.} {\bfseries 118} (2016) 01005}.

\bibitem{Abe:2019thb}
M.~Abe {\em et~al.}, ``{A New Approach for Measuring the Muon Anomalous
  Magnetic Moment and Electric Dipole Moment}'',
  \href{http://dx.doi.org/10.1093/ptep/ptz030}{{\em PTEP} {\bfseries 2019}
  no.~5, (2019) 053C02}, \href{http://arxiv.org/abs/1901.03047}{{\ttfamily
  arXiv:1901.03047 [physics.ins-det]}}.

\bibitem{Adelmann:2021udj}
A.~Adelmann {\em et~al.}, ``{Search for a muon EDM using the frozen-spin
  technique}'', \href{http://arxiv.org/abs/2102.08838}{{\ttfamily
  arXiv:2102.08838 [hep-ex]}}.

\bibitem{Dekens:2014jka}
W.~Dekens, J.~de~Vries, J.~Bsaisou, W.~Bernreuther, C.~Hanhart, U.-G. Meißner,
  A.~Nogga, and A.~Wirzba, ``{Unraveling models of \CP violation through
  electric dipole moments of light nuclei}'',
  \href{http://dx.doi.org/10.1007/JHEP07(2014)069}{{\em JHEP} {\bfseries 07}
  (2014) 069},
\href{http://arxiv.org/abs/1404.6082}{{\ttfamily arXiv:1404.6082 [hep-ph]}}.

\bibitem{Sharma:2010vv}
N.~Sharma, H.~Dahiya, P.~K. Chatley, and M.~Gupta, ``{Spin $1/2^+$, spin
  $3/2^+$ and transition magnetic moments of low lying and charmed baryons}'',
  \href{http://dx.doi.org/10.1103/PhysRevD.81.073001}{{\em Phys. Rev.}
  {\bfseries D81} (2010) 073001},
\href{http://arxiv.org/abs/1003.4338}{{\ttfamily arXiv:1003.4338 [hep-ph]}}.

\bibitem{Unal:2020ezc}
Y.~\"Unal and U.-G. Mei\ss{}ner, ``{Strong CP violation in spin-1/2 singly
  charmed baryons}'', \href{http://dx.doi.org/10.1007/JHEP01(2021)115}{{\em
  JHEP} {\bfseries 01} (2021) 115},
  \href{http://arxiv.org/abs/2008.01371}{{\ttfamily arXiv:2008.01371
  [hep-ph]}}.

\bibitem{Unal:2021lhb}
Y.~\"Unal, D.~Severt, J.~de~Vries, C.~Hanhart, and U.-G. Mei\ss{}ner,
  ``{Electric dipole moments of baryons with bottom quarks}'',
  \href{http://arxiv.org/abs/2111.13000}{{\ttfamily arXiv:2111.13000
  [hep-ph]}}.

\bibitem{Georgi:1992dw}
H.~Georgi, ``{Generalized dimensional analysis}'',
  \href{http://dx.doi.org/10.1016/0370-2693(93)91728-6}{{\em Phys. Lett.}
  {\bfseries B298} (1993) 187--189},
\href{http://arxiv.org/abs/hep-ph/9207278}{{\ttfamily arXiv:hep-ph/9207278
  [hep-ph]}}.

\bibitem{LHCb:2019hro}
{LHCb} collaboration, R.~Aaij {\em et~al.}, ``{Observation of CP Violation in
  Charm Decays}'', \href{http://dx.doi.org/10.1103/PhysRevLett.122.211803}{{\em
  Phys. Rev. Lett.} {\bfseries 122} no.~21, (2019) 211803},
  \href{http://arxiv.org/abs/1903.08726}{{\ttfamily arXiv:1903.08726
  [hep-ex]}}.

\bibitem{Schacht:2021jaz}
S.~Schacht and A.~Soni, ``{Enhancement of charm CP violation due to nearby
  resonances}'', \href{http://dx.doi.org/10.1016/j.physletb.2021.136855}{{\em
  Phys. Lett. B} {\bfseries 825} (2022) 136855},
  \href{http://arxiv.org/abs/2110.07619}{{\ttfamily arXiv:2110.07619
  [hep-ph]}}.

\bibitem{Bediaga:2022sxw}
I.~Bediaga, T.~Frederico, and P.~Magalhaes, ``{Enhanced charm CP asymmetries
  from final state interactions}'',
  \href{http://arxiv.org/abs/2203.04056}{{\ttfamily arXiv:2203.04056
  [hep-ph]}}.

\bibitem{Pich:1991fq}
A.~Pich and E.~de~Rafael, ``{Strong \CP violation in an effective chiral
  Lagrangian approach}'',
\href{http://dx.doi.org/10.1016/0550-3213(91)90019-T}{{\em Nucl. Phys.}
  {\bfseries B367} (1991) 313--333}.

\bibitem{Borasoy:2000pq}
B.~Borasoy, ``{The electric dipole moment of the neutron in chiral perturbation
  theory}'', \href{http://dx.doi.org/10.1103/PhysRevD.61.114017}{{\em Phys.
  Rev.} {\bfseries D61} (2000) 114017},
\href{http://arxiv.org/abs/hep-ph/0004011}{{\ttfamily arXiv:hep-ph/0004011
  [hep-ph]}}.

\bibitem{Faessler:2006at}
A.~Faessler, T.~Gutsche, S.~Kovalenko, and V.~E. Lyubovitskij, ``{Implications
  of R-parity violating supersymmetry for atomic and hadronic EDMs}'',
  \href{http://dx.doi.org/10.1103/PhysRevD.74.074013}{{\em Phys. Rev.}
  {\bfseries D74} (2006) 074013},
\href{http://arxiv.org/abs/hep-ph/0607269}{{\ttfamily arXiv:hep-ph/0607269
  [hep-ph]}}.

\bibitem{Anselm:1978vu}
A.~A. Anselm and D.~Diakonov, ``{On Weinberg's Model of \CP Violation in Gauge
  Theories}'',
\href{http://dx.doi.org/10.1016/0550-3213(78)90425-X}{{\em Nucl. Phys.}
  {\bfseries B145} (1978) 271--284}.

\bibitem{MartinCamalich:2010nab}
J.~Martin~Camalich, {\em {Properties of the lowest-lying baryons in chiral
  perturbation theory}}.
\newblock PhD thesis, Universitat de Val\`encia, 2010.

\bibitem{Dhir:2013nka}
R.~Dhir, C.~S. Kim, and R.~C. Verma, ``{Magnetic Moments of Bottom Baryons:
  Effective mass and Screened Charge}'',
  \href{http://dx.doi.org/10.1103/PhysRevD.88.094002}{{\em Phys. Rev.}
  {\bfseries D88} (2013) 094002},
\href{http://arxiv.org/abs/1309.4057}{{\ttfamily arXiv:1309.4057 [hep-ph]}}.

\bibitem{Fomin:2019wuw}
A.~S. Fomin, S.~Barsuk, A.~Y. Korchin, V.~A. Kovalchuk, E.~Kou, M.~Liul,
  A.~Natochii, E.~Niel, P.~Robbe, and A.~Stocchi, ``{The prospect of charm
  quark magnetic moment determination}'',
  \href{http://dx.doi.org/10.1140/epjc/s10052-020-7891-0}{{\em Eur. Phys. J. C}
  {\bfseries 80} no.~5, (2020) 358},
  \href{http://arxiv.org/abs/1909.04654}{{\ttfamily arXiv:1909.04654
  [hep-ph]}}.

\bibitem{Nagahama:2017eqh}
{BASE} collaboration, H.~Nagahama {\em et~al.}, ``{Sixfold improved single
  particle measurement of the magnetic moment of the antiproton}'',
\href{http://dx.doi.org/10.1038/ncomms14084}{{\em Nature Commun.} {\bfseries 8}
  (2017) 14084}.

\bibitem{DiSciacca:2013hya}
{ATRAP} collaboration, J.~DiSciacca {\em et~al.}, ``{One-Particle Measurement
  of the Antiproton Magnetic Moment}'',
  \href{http://dx.doi.org/10.1103/PhysRevLett.110.130801}{{\em Phys. Rev.
  Lett.} {\bfseries 110} no.~13, (2013) 130801},
\href{http://arxiv.org/abs/1301.6310}{{\ttfamily arXiv:1301.6310
  [physics.atom-ph]}}.

\bibitem{VanDyck:1987ay}
R.~S. Van~Dyck, P.~B. Schwinberg, and H.~G. Dehmelt, ``{New high-precision
  comparison of electron and positron \textit{g} factors}'',
\href{http://dx.doi.org/10.1103/PhysRevLett.59.26}{{\em Phys. Rev. Lett.}
  {\bfseries 59} (1987) 26--29}.

\bibitem{Bennett:2004pv}
{Muon g-2} collaboration, G.~W. Bennett {\em et~al.}, ``{Measurement of the
  negative muon anomalous magnetic moment to 0.7 ppm}'',
  \href{http://dx.doi.org/10.1103/PhysRevLett.92.161802}{{\em Phys. Rev. Lett.}
  {\bfseries 92} (2004) 161802},
\href{http://arxiv.org/abs/hep-ex/0401008}{{\ttfamily arXiv:hep-ex/0401008
  [hep-ex]}}.

\bibitem{QCDWorkingGroup:2019dyv}
{QCD Working Group} collaboration, A.~Dainese {\em et~al.}, ``{Physics Beyond
  Colliders: QCD Working Group Report}'',
  \href{http://arxiv.org/abs/1901.04482}{{\ttfamily arXiv:1901.04482
  [hep-ex]}}.

\bibitem{Aoyama:2020ynm}
T.~Aoyama {\em et~al.}, ``{The anomalous magnetic moment of the muon in the
  Standard Model}'',
  \href{http://dx.doi.org/10.1016/j.physrep.2020.07.006}{{\em Phys. Rept.}
  {\bfseries 887} (2020) 1--166},
  \href{http://arxiv.org/abs/2006.04822}{{\ttfamily arXiv:2006.04822
  [hep-ph]}}.

\bibitem{Miller:2007kk}
J.~P. Miller, E.~de~Rafael, and B.~L. Roberts, ``{Muon (g-2): Experiment and
  theory}'', \href{http://dx.doi.org/10.1088/0034-4885/70/5/R03}{{\em Rept.
  Prog. Phys.} {\bfseries 70} (2007) 795},
\href{http://arxiv.org/abs/hep-ph/0703049}{{\ttfamily arXiv:hep-ph/0703049
  [hep-ph]}}.

\bibitem{Muong-2:2021ojo}
{Muon g-2} collaboration, B.~Abi {\em et~al.}, ``{Measurement of the Positive
  Muon Anomalous Magnetic Moment to 0.46 ppm}'',
  \href{http://dx.doi.org/10.1103/PhysRevLett.126.141801}{{\em Phys. Rev.
  Lett.} {\bfseries 126} no.~14, (2021) 141801},
  \href{http://arxiv.org/abs/2104.03281}{{\ttfamily arXiv:2104.03281
  [hep-ex]}}.

\bibitem{Mohr:2015ccw}
P.~J. Mohr, D.~B. Newell, and B.~N. Taylor, ``{CODATA Recommended Values of the
  Fundamental Physical Constants: 2014}'',
  \href{http://dx.doi.org/10.1103/RevModPhys.88.035009}{{\em Rev. Mod. Phys.}
  {\bfseries 88} no.~3, (2016) 035009},
  \href{http://arxiv.org/abs/1507.07956}{{\ttfamily arXiv:1507.07956
  [physics.atom-ph]}}.

\bibitem{Muong-2:2006rrc}
{Muon g-2} collaboration, G.~W. Bennett {\em et~al.}, ``{Final Report of the
  Muon E821 Anomalous Magnetic Moment Measurement at BNL}'',
  \href{http://dx.doi.org/10.1103/PhysRevD.73.072003}{{\em Phys. Rev. D}
  {\bfseries 73} (2006) 072003},
  \href{http://arxiv.org/abs/hep-ex/0602035}{{\ttfamily arXiv:hep-ex/0602035}}.

\bibitem{DELPHI:2003nah}
{DELPHI} collaboration, J.~Abdallah {\em et~al.}, ``{Study of tau-pair
  production in photon-photon collisions at LEP and limits on the anomalous
  electromagnetic moments of the tau lepton}'',
  \href{http://dx.doi.org/10.1140/epjc/s2004-01852-y}{{\em Eur. Phys. J. C}
  {\bfseries 35} (2004) 159--170},
  \href{http://arxiv.org/abs/hep-ex/0406010}{{\ttfamily arXiv:hep-ex/0406010}}.

\bibitem{Eidelman:2007sb}
S.~Eidelman and M.~Passera, ``{Theory of the tau lepton anomalous magnetic
  moment}'', \href{http://dx.doi.org/10.1142/S0217732307022694}{{\em Mod. Phys.
  Lett. A} {\bfseries 22} (2007) 159--179},
  \href{http://arxiv.org/abs/hep-ph/0701260}{{\ttfamily arXiv:hep-ph/0701260}}.

\bibitem{Belle:2021ybo}
{Belle} collaboration, K.~Inami {\em et~al.}, ``{An improved search for the
  electric dipole moment of the $\tau$ lepton}'',
  \href{http://dx.doi.org/10.1007/JHEP04(2022)110}{{\em JHEP} {\bfseries 04}
  (2022) 110}, \href{http://arxiv.org/abs/2108.11543}{{\ttfamily
  arXiv:2108.11543 [hep-ex]}}.

\bibitem{Belle:2002nla}
{Belle} collaboration, K.~Inami {\em et~al.}, ``{Search for the electric dipole
  moment of the tau lepton}'',
  \href{http://dx.doi.org/10.1016/S0370-2693(02)02984-2}{{\em Phys. Lett. B}
  {\bfseries 551} (2003) 16--26},
  \href{http://arxiv.org/abs/hep-ex/0210066}{{\ttfamily arXiv:hep-ex/0210066}}.

\bibitem{Blinov:2008mu}
A.~E. Blinov and A.~S. Rudenko, ``{Upper Limits on Electric and Weak Dipole
  Moments of tau-Lepton and Heavy Quarks from $e^+ e^-$ Annihilation}'',
  \href{http://dx.doi.org/10.1016/j.nuclphysbps.2009.03.043}{{\em Nucl. Phys.
  Proc. Suppl.} {\bfseries 189} (2009) 257--259},
\href{http://arxiv.org/abs/0811.2380}{{\ttfamily arXiv:0811.2380 [hep-ph]}}.

\bibitem{Grozin:2009jq}
A.~G. Grozin, I.~B. Khriplovich, and A.~S. Rudenko, ``{Upper limits on electric
  dipole moments of tau-lepton, heavy quarks, and W-boson}'',
  \href{http://dx.doi.org/10.1016/j.nuclphysb.2009.06.026}{{\em Nucl. Phys.}
  {\bfseries B821} (2009) 285--290},
\href{http://arxiv.org/abs/0902.3059}{{\ttfamily arXiv:0902.3059 [hep-ph]}}.

\bibitem{Dekens:2018bci}
W.~Dekens, J.~de~Vries, M.~Jung, and K.~K. Vos, ``{The phenomenology of
  electric dipole moments in models of scalar leptoquarks}'',
  \href{http://dx.doi.org/10.1007/JHEP01(2019)069}{{\em JHEP} {\bfseries 01}
  (2019) 069}, \href{http://arxiv.org/abs/1809.09114}{{\ttfamily
  arXiv:1809.09114 [hep-ph]}}.

\bibitem{Gutierrez-Rodriguez:2013eaa}
A.~Gutierrez-Rodriguez, M.~A. Hernandez-Ruiz, and C.~P. Castaneda-Almanza,
  ``{Dipole moments of the tau-lepton and Z-Z-prime mixing angle induced in a
  331 model}'', \href{http://dx.doi.org/10.1088/0954-3899/40/3/035001}{{\em J.
  Phys. G} {\bfseries 40} (2013) 035001}.

\bibitem{Ibrahim:2010va}
T.~Ibrahim and P.~Nath, ``{Large Tau and Tau Neutrino Electric Dipole Moments
  in Models with Vector Like Multiplets}'',
  \href{http://dx.doi.org/10.1103/PhysRevD.81.033007}{{\em Phys. Rev. D}
  {\bfseries 81} no.~3, (2010) 033007},
  \href{http://arxiv.org/abs/1001.0231}{{\ttfamily arXiv:1001.0231 [hep-ph]}}.
  [Erratum: Phys.Rev.D 89, 119902 (2014)].

\bibitem{Gutierrez-Rodriguez:2009weo}
A.~Gutierrez-Rodriguez, ``{Bounding the electromagnetic and weak dipole moments
  of the tau-lepton in a simplest little Higgs model}'',
  \href{http://dx.doi.org/10.1142/S0217732310032238}{{\em Mod. Phys. Lett. A}
  {\bfseries 25} (2010) 703--713},
  \href{http://arxiv.org/abs/0910.4217}{{\ttfamily arXiv:0910.4217 [hep-ph]}}.

\bibitem{Iltan:2005iy}
E.~O. Iltan, ``{Electric dipole moments of charged leptons in the split fermion
  scenario in the two Higgs doublet model}'',
  \href{http://dx.doi.org/10.1140/epjc/s2005-02373-y}{{\em Eur. Phys. J. C}
  {\bfseries 44} (2005) 411--417},
  \href{http://arxiv.org/abs/hep-ph/0503001}{{\ttfamily arXiv:hep-ph/0503001}}.

\bibitem{Pospelov:2013sca}
M.~Pospelov and A.~Ritz, ``{CKM benchmarks for electron electric dipole moment
  experiments}'', \href{http://dx.doi.org/10.1103/PhysRevD.89.056006}{{\em
  Phys. Rev. D} {\bfseries 89} no.~5, (2014) 056006},
  \href{http://arxiv.org/abs/1311.5537}{{\ttfamily arXiv:1311.5537 [hep-ph]}}.

\bibitem{Fomin:2017ltw}
A.~S. Fomin {\em et~al.}, ``{Feasibility of measuring the magnetic dipole
  moments of the charm baryons at the LHC using bent crystals}'',
  \href{http://dx.doi.org/10.1007/JHEP08(2017)120}{{\em JHEP} {\bfseries 08}
  (2017) 120}, \href{http://arxiv.org/abs/1705.03382}{{\ttfamily
  arXiv:1705.03382 [hep-ph]}}.

\bibitem{Fomin:2018ybj}
A.~S. Fomin, A.~Y. Korchin, A.~Stocchi, S.~Barsuk, and P.~Robbe, ``{Feasibility
  of $\tau$ -lepton electromagnetic dipole moments measurement using bent
  crystal at the LHC}'', \href{http://dx.doi.org/10.1007/JHEP03(2019)156}{{\em
  JHEP} {\bfseries 03} (2019) 156},
  \href{http://arxiv.org/abs/1810.06699}{{\ttfamily arXiv:1810.06699
  [hep-ph]}}.

\bibitem{Biryukov:2021phs}
V.~M. Biryukov, ``{Double-lens technique for efficient capture of short-lived
  particles by a crystal}'',
  \href{http://dx.doi.org/10.1016/j.nimb.2021.10.008}{{\em Nucl. Instrum. Meth.
  B} {\bfseries 509} (2021) 34--38},
  \href{http://arxiv.org/abs/2105.13628}{{\ttfamily arXiv:2105.13628
  [physics.ins-det]}}.

\bibitem{Mirarchi:2019vqi}
D.~Mirarchi, A.~S. Fomin, S.~Redaelli, and W.~Scandale, ``{Layouts for
  fixed-target experiments and dipole moment measurements of short-lived
  baryons using bent crystals at the LHC}'',
  \href{http://dx.doi.org/10.1140/epjc/s10052-020-08466-x}{{\em Eur. Phys. J.
  C} {\bfseries 80} no.~10, (2020) 929},
  \href{http://arxiv.org/abs/1906.08551}{{\ttfamily arXiv:1906.08551
  [physics.acc-ph]}}.

\bibitem{Lyuboshits:1979qw}
V.~L. Lyuboshits, ``{The Spin Rotation at Deflection of Relativistic Charged
  Particle in Electric Field}'',
{\em Sov. J. Nucl. Phys.} {\bfseries 31} (1980) 509.

\bibitem{Kim:1982ry}
I.~J. Kim, ``{Magnetic moment measurement of baryons with heavy flavored quarks
  by planar channeling through bent crystal}'',
\href{http://dx.doi.org/10.1016/0550-3213(83)90363-2}{{\em Nucl. Phys.}
  {\bfseries B229} (1983) 251--268}.

\bibitem{Chen:1992wx}
{E761} collaboration, D.~Chen {\em et~al.}, ``{First observation of magnetic
  moment precession of channeled particles in bent crystals}'',
\href{http://dx.doi.org/10.1103/PhysRevLett.69.3286}{{\em Phys. Rev. Lett.}
  {\bfseries 69} (1992) 3286--3289}.

\bibitem{Baublis:1994ku}
V.~V. Baublis {\em et~al.}, ``{Measuring the magnetic moments of short-lived
  particles using channeling in bent crystals}'',
\href{http://dx.doi.org/10.1016/0168-583X(94)95524-7}{{\em Nucl. Instrum.
  Meth.} {\bfseries B90} (1994) 112--118}.

\bibitem{Samsonov:1996ah}
V.~M. Samsonov, ``{On the possibility of measuring charm baryon magnetic
  moments with channeling}'',
\href{http://dx.doi.org/10.1016/0168-583X(96)00348-5}{{\em Nucl. Instrum.
  Meth.} {\bfseries B119} (1996) 271--279}.

\bibitem{Baryshevsky:2016cul}
V.~G. Baryshevsky, ``{The possibility to measure the magnetic moments of
  short-lived particles (charm and beauty baryons) at LHC and FCC energies
  using the phenomenon of spin rotation in crystals}'',
\href{http://dx.doi.org/10.1016/j.physletb.2016.04.025}{{\em Phys. Lett.}
  {\bfseries B757} (2016) 426--429}.

\bibitem{nEDM:2020crw}
{nEDM} collaboration, C.~Abel {\em et~al.}, ``{Measurement of the permanent
  electric dipole moment of the neutron}'',
  \href{http://dx.doi.org/10.1103/PhysRevLett.124.081803}{{\em Phys. Rev.
  Lett.} {\bfseries 124} no.~8, (2020) 081803},
  \href{http://arxiv.org/abs/2001.11966}{{\ttfamily arXiv:2001.11966
  [hep-ex]}}.

\bibitem{Biryukov1997}
V.~M. Biryukov {\em et~al.}, {\em {Crystal Channeling and Its Application at
  High-Energy Accelerators}}.
\newblock Springer-Verlag Berlin Heidelberg, 1997.

\bibitem{Sytov:2019gad}
A.~I. Sytov, V.~V. Tikhomirov, and L.~Bandiera, ``{Simulation code for modeling
  of coherent effects of radiation generation in oriented crystals}'',
  \href{http://dx.doi.org/10.1103/PhysRevAccelBeams.22.064601}{{\em Phys. Rev.
  Accel. Beams} {\bfseries 22} no.~6, (2019) 064601}.

\bibitem{Szwed:1981rr}
J.~Szwed, ``{Hyperon Polarization at High-Energies}'',
\href{http://dx.doi.org/10.1016/0370-2693(81)90788-7}{{\em Phys. Lett.}
  {\bfseries B105} (1981) 403--405}.

\bibitem{Jezabek:1992ke}
M.~Jezabek, K.~Rybicki, and R.~Rylko, ``{Experimental study of spin effects in
  hadroproduction and decay of $\Lambda_c^+$}'',
\href{http://dx.doi.org/10.1016/0370-2693(92)90177-6}{{\em Phys. Lett.}
  {\bfseries B286} (1992) 175--179}.

\bibitem{Aitala:1999uq}
{E791} collaboration, E.~M. Aitala {\em et~al.}, ``{Multidimensional resonance
  analysis of $\Lc \to p K^- \pi^+$}'',
  \href{http://dx.doi.org/10.1016/S0370-2693(99)01397-0}{{\em Phys. Lett.}
  {\bfseries B471} (2000) 449--459},
\href{http://arxiv.org/abs/hep-ex/9912003}{{\ttfamily arXiv:hep-ex/9912003
  [hep-ex]}}.

\bibitem{ACCMOR:1994fcu}
{ACCMOR} collaboration, S.~Barlag {\em et~al.}, ``{A Study of the transverse
  polarization of Lambda0 and Anti-lambda0 hyperons produced in pi- Cu
  interactions at 230-GeV/c}'',
  \href{http://dx.doi.org/10.1016/0370-2693(94)90052-3}{{\em Phys. Lett. B}
  {\bfseries 325} (1994) 531--535}.

\bibitem{LHCb:2014vhh}
{LHCb} collaboration, R.~Aaij {\em et~al.}, ``{Precision luminosity
  measurements at LHCb}'',
  \href{http://dx.doi.org/10.1088/1748-0221/9/12/P12005}{{\em JINST} {\bfseries
  9} no.~12, (2014) P12005}, \href{http://arxiv.org/abs/1410.0149}{{\ttfamily
  arXiv:1410.0149 [hep-ex]}}.

\bibitem{Bursche:2018orf}
A.~Bursche {\em et~al.}, ``{Physics opportunities with the fixed-target program
  of the LHCb experiment using an unpolarized gas target}'', {\em {\rm
  LHCb-PUB-2018-015, CERN-LHCb-PUB-2018-015 }} (2018) .

\bibitem{HERA-B:2006rds}
{HERA-B} collaboration, I.~Abt {\em et~al.}, ``{Polarization of Lambda and
  anti-Lambda in 920-GeV fixed-target proton-nucleus collisions}'',
  \href{http://dx.doi.org/10.1016/j.physletb.2006.05.040}{{\em Phys. Lett. B}
  {\bfseries 638} (2006) 415--421},
  \href{http://arxiv.org/abs/hep-ex/0603047}{{\ttfamily arXiv:hep-ex/0603047}}.

\bibitem{Ramberg:1994tk}
E.~J. Ramberg {\em et~al.}, ``{Polarization of Lambda and anti-Lambda produced
  by 800-GeV protons}'',
  \href{http://dx.doi.org/10.1016/0370-2693(94)91397-8}{{\em Phys. Lett. B}
  {\bfseries 338} (1994) 403--408}.

\bibitem{Fanti:1998px}
V.~Fanti {\em et~al.}, ``{A Measurement of the transverse polarization of
  Lambda hyperons produced in inelastic p N reactions at 450-GeV proton
  energy}'', \href{http://dx.doi.org/10.1007/s100520050337}{{\em Eur. Phys. J.
  C} {\bfseries 6} (1999) 265--269}.

\bibitem{ATLAS:2014ona}
{ATLAS} collaboration, G.~Aad {\em et~al.}, ``{Measurement of the transverse
  polarization of $\Lambda$ and $\bar{\Lambda}$ hyperons produced in
  proton-proton collisions at $\sqrt{s}=7$ TeV using the ATLAS detector}'',
  \href{http://dx.doi.org/10.1103/PhysRevD.91.032004}{{\em Phys. Rev. D}
  {\bfseries 91} no.~3, (2015) 032004},
  \href{http://arxiv.org/abs/1412.1692}{{\ttfamily arXiv:1412.1692 [hep-ex]}}.

\bibitem{NegreSimoMasterThesis}
A.~Negre~Simó, ``{Moments dipolars electromagnètics del leptó $\tau$ a
  l’LHC i a l’SPS}'', {\em Master thesis. Universitat de Valencia, 2019
  {\rm(unpublished)}} .

\bibitem{FRENKEL}
J.~Frenkel, ``{Die Elektrodynamik des rotierenden Elektrons}'',
\href{http://dx.doi.org/10.1007/BF01397099}{{\em Z. Phys.} {\bfseries 37}
  (1926) 243--262}.

\bibitem{Bargmann:1959gz}
V.~Bargmann, L.~Michel, and V.~L. Telegdi, ``Precession of the polarization of
  particles moving in a homogeneous electromagnetic field'',
  \href{http://dx.doi.org/10.1103/PhysRevLett.2.435}{{\em Phys. Rev. Lett.}
  {\bfseries 2} (May, 1959) 435--436}.
  \url{http://link.aps.org/doi/10.1103/PhysRevLett.2.435}.

\bibitem{Thomas:1926dy}
L.~H. Thomas, ``{The motion of a spinning electron}'',
\href{http://dx.doi.org/10.1038/117514a0}{{\em Nature} {\bfseries 117} (1926)
  514}.

\bibitem{Thomas:1927yu}
L.~H. Thomas, ``{The kinematics of an electron with an axis}'',
{\em Phil. Mag.} {\bfseries 3} (1927) 1--21.

\bibitem{Fukuyama:2013ioa}
T.~Fukuyama and A.~J. Silenko, ``{Derivation of Generalized
  Thomas-Bargmann-Michel-Telegdi Equation for a Particle with Electric Dipole
  Moment}'', \href{http://dx.doi.org/10.1142/S0217751X13501479}{{\em Int. J.
  Mod. Phys.} {\bfseries A28} (2013) 1350147},
\href{http://arxiv.org/abs/1308.1580}{{\ttfamily arXiv:1308.1580 [hep-ph]}}.

\bibitem{Jackson:1998nia}
J.~D. Jackson, {\em {Classical Electrodynamics}}.
\newblock Wiley,
1998.
\newblock

\bibitem{Leader2011}
E.~Leader, {\em {Spin in particle physics}}, vol.~15.
\newblock Camb. Monogr. Part. Phys. Nucl. Phys. Cosmol.,
2011.
\newblock

\bibitem{Silenko:2014uca}
A.~J. Silenko, ``{Spin precession of a particle with an electric dipole moment:
  contributions from classical electrodynamics and from the Thomas effect}'',
  \href{http://dx.doi.org/10.1088/0031-8949/90/6/065303}{{\em Phys. Scripta}
  {\bfseries 90} no.~6, (2015) 065303},
\href{http://arxiv.org/abs/1410.6906}{{\ttfamily arXiv:1410.6906 [hep-ph]}}.

\bibitem{Good:1962zza}
R.~H. Good, ``{Classical Equations of Motion for a Polarized Particle in an
  Electromagnetic Field}'',
\href{http://dx.doi.org/10.1103/PhysRev.125.2112}{{\em Phys. Rev.} {\bfseries
  125} (1962) 2112--2115}.

\bibitem{Metodiev:2015gda}
E.~M. Metodiev, ``{Thomas-BMT equation generalized to electric dipole moments
  and field gradients}'', \href{http://arxiv.org/abs/1507.04440}{{\ttfamily
  arXiv:1507.04440 [physics.acc-ph]}}.
2015.

\bibitem{Biryukov:2021gsd}
V.~M. Biryukov, ``{Possibility to make a beam of tau-leptons and charmed
  particles by a channeling crystal}'',
  \href{http://arxiv.org/abs/2101.05085}{{\ttfamily arXiv:2101.05085
  [physics.ins-det]}}.

\bibitem{Alves:2008zz}
{LHCb} collaboration, A.~A. Alves~Jr. {\em et~al.}, ``{The \lhcb detector at
  the LHC}'', \href{http://dx.doi.org/10.1088/1748-0221/3/08/S08005}{{\em
  JINST} {\bfseries 3} (2008) S08005}.

\bibitem{proofofprincipletalk}
{{See for example: F. Mart\'inez Vidal, LHC Crystals, talk given at Physics
  Beyond Colliders General Working Group meeting, 2-3 December 2021, CERN
  (Geneva, Switzerland)}}.
\newblock \url{https://indico.cern.ch/event/1089151/timetable/}.

\bibitem{LHCb-TDR-015}
{LHCb} collaboration, ``{LHCb Tracker Upgrade Technical Design Report}'', 2014.
\newblock {LHCb-TDR-015}.

\bibitem{Apollinari:2017lan}
G.~Apollinari, I.~B\'ejar~Alonso, O.~Br\"uning, P.~Fessia, M.~Lamont, L.~Rossi,
  and L.~Tavian, ``{High-Luminosity Large Hadron Collider (HL-LHC)}: {Technical
  Design Report V. 0.1}'', 2017.
\newblock CERN-2017-007-M.

\bibitem{Scandale:2021zbn}
W.~Scandale {\em et~al.}, ``{Double-crystal measurements at the CERN SPS}'',
  \href{http://dx.doi.org/10.1016/j.nima.2021.165747}{{\em Nucl. Instrum. Meth.
  A} {\bfseries 1015} (2021) 165747},
  \href{http://arxiv.org/abs/2103.14681}{{\ttfamily arXiv:2103.14681
  [physics.acc-ph]}}.

\bibitem{Scandale:2016krl}
W.~Scandale {\em et~al.}, ``{Observation of channeling for 6500~\gevc protons
  in the crystal assisted collimation setup for LHC}'',
\href{http://dx.doi.org/10.1016/j.physletb.2016.05.004}{{\em Phys. Lett.}
  {\bfseries B758} (2016) 129--133}.

\bibitem{PDG}
{Particle Data Group} collaboration, K.~A. Olive {\em et~al.}, ``{Review of
  Particle Physics}'',
\href{http://dx.doi.org/10.1088/1674-1137/38/9/090001}{{\em Chin. Phys.}
  {\bfseries C38} (2014) 090001}.

\bibitem{Davier:1992nw}
M.~Davier, L.~Duflot, F.~Le~Diberder, and A.~Rouge, ``{The Optimal method for
  the measurement of tau polarization}'',
  \href{http://dx.doi.org/10.1016/0370-2693(93)90101-M}{{\em Phys. Lett. B}
  {\bfseries 306} (1993) 411--417}.

\bibitem{Sjostrand:2006za}
T.~Sjostrand, S.~Mrenna, and P.~Z. Skands, ``{PYTHIA 6.4 Physics and Manual}'',
  \href{http://dx.doi.org/10.1088/1126-6708/2006/05/026}{{\em JHEP} {\bfseries
  05} (2006) 026},
\href{http://arxiv.org/abs/hep-ph/0603175}{{\ttfamily arXiv:hep-ph/0603175
  [hep-ph]}}.

\bibitem{Lange:2001uf}
D.~J. Lange, ``{The EvtGen particle decay simulation package}'',
\href{http://dx.doi.org/10.1016/S0168-9002(01)00089-4}{{\em Nucl. Instrum.
  Meth.} {\bfseries A462} (2001) 152--155}.

\bibitem{Verkerke:2003ir}
W.~Verkerke and D.~P. Kirkby, ``{The RooFit toolkit for data modeling}'', {\em
  eConf} {\bfseries C0303241} (2003) MOLT007,
  \href{http://arxiv.org/abs/physics/0306116}{{\ttfamily
  arXiv:physics/0306116}}.

\bibitem{Brun:1997pa}
R.~Brun and F.~Rademakers, ``{ROOT: An object oriented data analysis
  framework}'', \href{http://dx.doi.org/10.1016/S0168-9002(97)00048-X}{{\em
  Nucl. Instrum. Meth. A} {\bfseries 389} (1997) 81--86}.

\bibitem{LHCb-DP-2014-002}
{LHCb} collaboration, R.~Aaij {\em et~al.}, ``{LHCb detector performance}'',
  \href{http://dx.doi.org/10.1142/S0217751X15300227}{{\em Int. J. Mod. Phys.}
  {\bfseries A30} (2015) 1530022},
\href{http://arxiv.org/abs/1412.6352}{{\ttfamily arXiv:1412.6352 [hep-ex]}}.

\bibitem{LHCb-TDR-013}
L.~collaboration, ``{LHCb VELO Upgrade Technical Design Report}'', tech. rep.,
  Nov, 2013.
\newblock \url{http://cds.cern.ch/record/1624070}.

\bibitem{Bezshyyko:2017var}
A.~S. Fomin {\em et~al.}, ``{Feasibility of measuring the magnetic dipole
  moments of the charm baryons at the LHC using bent crystals}'',
  \href{http://dx.doi.org/10.1007/JHEP08(2017)120}{{\em JHEP} {\bfseries 08}
  (2017) 120},
\href{http://arxiv.org/abs/1705.03382}{{\ttfamily arXiv:1705.03382 [hep-ph]}}.

\bibitem{Korner:1991ph}
J.~G. Korner and M.~Kramer, ``{Polarization effects in exclusive semileptonic
  Lambda(c) and Lambda(b) charm and bottom baryon decays}'',
  \href{http://dx.doi.org/10.1016/0370-2693(92)91623-H}{{\em Phys. Lett. B}
  {\bfseries 275} (1992) 495--505}.

\bibitem{Konig:1993wz}
B.~Konig, J.~G. Korner, and M.~Kramer, ``{On the determination of the b
  ---\ensuremath{>} c handedness using nonleptonic lambda(c) decays}'',
  \href{http://dx.doi.org/10.1103/PhysRevD.49.2363}{{\em Phys. Rev. D}
  {\bfseries 49} (1994) 2363--2368},
  \href{http://arxiv.org/abs/hep-ph/9310263}{{\ttfamily arXiv:hep-ph/9310263}}.

\bibitem{Bonvicini:1994mr}
G.~Bonvicini and L.~Randall, ``{Optimized variables for the study of Lambda(b)
  polarization}'', \href{http://dx.doi.org/10.1103/PhysRevLett.73.392}{{\em
  Phys. Rev. Lett.} {\bfseries 73} (1994) 392--395},
  \href{http://arxiv.org/abs/hep-ph/9401299}{{\ttfamily arXiv:hep-ph/9401299}}.

\bibitem{Korner:1995my}
J.~G. Korner, ``{Lambda(b) polarization from its inclusive semileptonic
  decay}'', \href{http://dx.doi.org/10.1016/0920-5632(96)00379-9}{{\em Nucl.
  Phys. B Proc. Suppl.} {\bfseries 50} (1996) 130--134},
  \href{http://arxiv.org/abs/hep-ph/9512322}{{\ttfamily arXiv:hep-ph/9512322}}.

\bibitem{Diaconu:1995mp}
C.~Diaconu, M.~Talby, J.~G. Korner, and D.~Pirjol, ``{Improved variables for
  measuring the Lambda(b) polarization}'',
  \href{http://dx.doi.org/10.1103/PhysRevD.53.6186}{{\em Phys. Rev. D}
  {\bfseries 53} (1996) 6186--6194},
  \href{http://arxiv.org/abs/hep-ph/9512330}{{\ttfamily arXiv:hep-ph/9512330}}.

\bibitem{Korner:1998nc}
J.~G. Korner and D.~Pirjol, ``{Spin momentum correlations in inclusive
  semileptonic decays of polarized Lambda(b) baryons}'',
  \href{http://dx.doi.org/10.1103/PhysRevD.60.014021}{{\em Phys. Rev. D}
  {\bfseries 60} (1999) 014021},
  \href{http://arxiv.org/abs/hep-ph/9810511}{{\ttfamily arXiv:hep-ph/9810511}}.

\bibitem{Halzen:1984mc}
F.~Halzen and A.~D. Martin, {\em {Quarks and leptons: an introductory course in
  modern particle physics}}.
\newblock John Wiley \& Sons, 1984. ISBN:978-0-471-88741-6.

\bibitem{Berestetskii:1982qgu}
V.~B. Berestetskii, E.~M. Lifshitz, and L.~P. Pitaevskii, {\em {QUANTUM
  ELECTRODYNAMICS}}, vol.~4 of {\em Course of Theoretical Physics}.
\newblock Pergamon Press, Oxford, 1982.

\bibitem{Barschel:2020drr}
C.~Barschel {\em et~al.}, \href{http://dx.doi.org/10.23731/CYRM-2020-004}{{\em
  {LHC fixed target experiments : Report from the LHC Fixed Target Working
  Group of the CERN Physics Beyond Colliders Forum}}}, vol.~4/2020 of {\em CERN
  Yellow Reports: Monographs}.
\newblock CERN, Geneva, 2020.

\bibitem{sps-rep}
W.~Scandale and A.~M. Taratin, ``{Channeling and volume reflection of
  high-energy charged particles in short bent crystals. Crystal assisted
  collimation of the accelerator beam halo}'',
  \href{http://dx.doi.org/10.1016/j.physrep.2019.04.003}{{\em Phys. Rept.}
  {\bfseries 815} (2019) 1--107}.

\bibitem{ATLAS:2017fur}
{ATLAS} collaboration, M.~Aaboud {\em et~al.}, ``{Evidence for light-by-light
  scattering in heavy-ion collisions with the ATLAS detector at the LHC}'',
  \href{http://dx.doi.org/10.1038/nphys4208}{{\em Nature Phys.} {\bfseries 13}
  no.~9, (2017) 852--858}, \href{http://arxiv.org/abs/1702.01625}{{\ttfamily
  arXiv:1702.01625 [hep-ex]}}.

\bibitem{LHCb-PAPER-2018-031}
{LHCb} collaboration, R.~Aaij {\em et~al.}, ``{Measurement of antiproton
  production in \proton{}He collisions at $\sqrt{s_{NN}}=110$ GeV}'',
  \href{http://dx.doi.org/10.1103/PhysRevLett.121.222001}{{\em Phys. Rev.
  Lett.} {\bfseries 121} (2018) 222001} {LHCb-PAPER-2018-031 CERN-EP-2018-217},
  \href{http://arxiv.org/abs/1808.06127}{{\ttfamily arXiv:1808.06127
  [hep-ex]}}.

\bibitem{LHCb-PAPER-2018-023}
{LHCb} collaboration, R.~Aaij {\em et~al.}, ``{First measurement of charm
  production in fixed-target configuration at the LHC}'',
  \href{http://dx.doi.org/10.1103/PhysRevLett.122.132002}{{\em Phys. Rev.
  Lett.} {\bfseries 122} (2019) 132002} {LHCb-PAPER-2018-023 CERN-EP-2018-266},
  \href{http://arxiv.org/abs/1810.07907}{{\ttfamily arXiv:1810.07907
  [hep-ex]}}.

\bibitem{JINSTLHCB}
{LHCb} collaboration, A.~A. Alves, Jr. {\em et~al.}, ``{The LHCb Detector at
  the LHC}'',
\href{http://dx.doi.org/10.1088/1748-0221/3/08/S08005}{{\em JINST} {\bfseries
  3} (2008) S08005}.

\bibitem{LHCbupgrade}
{LHCb} collaboration, ``{Framework TDR for the LHCb Upgrade: Technical Design
  Report}'', 2012.
\newblock {CERN-LHCC-2012-007}.

\bibitem{LHCbRICHGroup:2012mgd}
{LHCb RICH Group} collaboration, M.~Adinolfi {\em et~al.}, ``{Performance of
  the LHCb RICH detector at the LHC}'',
  \href{http://dx.doi.org/10.1140/epjc/s10052-013-2431-9}{{\em Eur. Phys. J. C}
  {\bfseries 73} (2013) 2431}, \href{http://arxiv.org/abs/1211.6759}{{\ttfamily
  arXiv:1211.6759 [physics.ins-det]}}.

\bibitem{LHCb-TDR-016}
{LHCb} collaboration, ``{LHCb Trigger and Online Technical Design Report}'',
  2014.
\newblock {LHCb-TDR-016}.

\bibitem{LHCb-TDR-017}
{LHCb} collaboration, ``{LHCb Upgrade Software and Computing}'', 2018.
\newblock {CERN-LHCC-2018-007}.

\bibitem{LHCb-TDR-014}
{LHCb} collaboration, ``{LHCb PID Upgrade Technical Design Report}'', 2013.
\newblock {CERN-LHCC-2013-022}.

\bibitem{Yang:2020rpi}
M.~Yang and P.~Wang, ``{Electromagnetic form factors of octet baryons with the
  nonlocal chiral effective theory}'',
  \href{http://dx.doi.org/10.1103/PhysRevD.102.056024}{{\em Phys. Rev. D}
  {\bfseries 102} no.~5, (2020) 056024},
  \href{http://arxiv.org/abs/2005.11971}{{\ttfamily arXiv:2005.11971
  [hep-ph]}}.

\bibitem{MAGN78}
L.~Schachinger {\em et~al.}, ``{A Precise Measurement of the $\Lambda^0$
  Magnetic Moment}'',
\href{http://dx.doi.org/10.1103/PhysRevLett.41.1348}{{\em Phys. Rev. Lett.}
  {\bfseries 41} (1978) 1348}.

\bibitem{PONDROM}
L.~Pondrom, R.~Handler, M.~Sheaff, P.~T. Cox, J.~Dworkin, O.~E. Overseth,
  T.~Devlin, L.~Schachinger, and K.~J. Heller, ``{New Limit on the Electric
  Dipole Moment of the $\Lambda$ Hyperon}'',
\href{http://dx.doi.org/10.1103/PhysRevD.23.814}{{\em Phys. Rev.} {\bfseries
  D23} (1981) 814--816}.

\bibitem{Hulsbergen:2005pu}
W.~D. Hulsbergen, ``{Decay chain fitting with a Kalman filter}'',
  \href{http://dx.doi.org/10.1016/j.nima.2005.06.078}{{\em Nucl. Instrum.
  Meth.} {\bfseries A552} (2005) 566--575},
  \href{http://arxiv.org/abs/physics/0503191}{{\ttfamily arXiv:physics/0503191
  [physics]}}.

\bibitem{LINK05}
{FOCUS} collaboration, J.~M. Link {\em et~al.}, ``{Study of the decay asymmetry
  parameter and CP violation parameter in the $\Lambda_c^+ \to \Lambda \pi^+$
  decay}'', \href{http://dx.doi.org/10.1016/j.physletb.2006.01.017}{{\em Phys.
  Lett.} {\bfseries B634} (2006) 165--172},
\href{http://arxiv.org/abs/hep-ex/0509042}{{\ttfamily arXiv:hep-ex/0509042
  [hep-ex]}}.

\bibitem{Aaij:2015bpa}
{LHCb} collaboration, R.~Aaij {\em et~al.}, ``{Measurements of prompt charm
  production cross-sections in $pp$ collisions at $ \sqrt{s}=13 $ TeV}'',
  \href{http://dx.doi.org/10.1007/JHEP03(2016)159}{{\em JHEP} {\bfseries 03}
  (2016) 159}, \href{http://arxiv.org/abs/1510.01707}{{\ttfamily
  arXiv:1510.01707 [hep-ex]}}.
[Erratum: JHEP09,013(2016)].

\bibitem{FONLLWEB}
M.~Cacciari, ``{FONLL Heavy Quark Production}.''
  {\url{HTTP://WWW.LPTHE.JUSSIEU.FR/ ~CACCIARI/FONLL/FONLLFORM.HTML}}.
\newblock {Accessed: 17.05.2016}.

\bibitem{Aaij:2010gn}
{LHCb} collaboration, R.~Aaij {\em et~al.}, ``{Measurement of $\sigma(pp \to b
  \bar{b} X)$ at $\sqrt{s}=7~\rm{TeV}$ in the forward region}'',
  \href{http://dx.doi.org/10.1016/j.physletb.2010.10.010}{{\em Phys. Lett.}
  {\bfseries B694} (2010) 209--216},
\href{http://arxiv.org/abs/1009.2731}{{\ttfamily arXiv:1009.2731 [hep-ex]}}.

\bibitem{Aaij:2015rla}
{LHCb} collaboration, R.~Aaij {\em et~al.}, ``{Measurement of forward $J/\psi$
  production cross-sections in $pp$ collisions at $\sqrt{s}=13$ TeV}'',
  \href{http://dx.doi.org/10.1007/JHEP10(2015)172}{{\em JHEP} {\bfseries 10}
  (2015) 172},
\href{http://arxiv.org/abs/1509.00771}{{\ttfamily arXiv:1509.00771 [hep-ex]}}.

\bibitem{Lisovyi:2015uqa}
M.~Lisovyi, A.~Verbytskyi, and O.~Zenaiev, ``{Combined analysis of charm-quark
  fragmentation-fraction measurements}'',
  \href{http://dx.doi.org/10.1140/epjc/s10052-016-4246-y}{{\em Eur. Phys. J.}
  {\bfseries C76} no.~7, (2016) 397},
\href{http://arxiv.org/abs/1509.01061}{{\ttfamily arXiv:1509.01061 [hep-ex]}}.

\bibitem{Gladilin:2014tba}
L.~Gladilin, ``{Fragmentation fractions of $c$ and $b$ quarks into charmed
  hadrons at LEP}'',
  \href{http://dx.doi.org/10.1140/epjc/s10052-014-3250-3}{{\em Eur. Phys. J.}
  {\bfseries C75} no.~1, (2015) 19},
\href{http://arxiv.org/abs/1404.3888}{{\ttfamily arXiv:1404.3888 [hep-ex]}}.

\bibitem{Amhis:2014hma}
{Heavy Flavor Averaging Group (HFAG)} collaboration, Y.~Amhis {\em et~al.},
  ``{Averages of $b$-hadron, $c$-hadron, and $\tau$-lepton properties as of
  summer 2014}'',
\href{http://arxiv.org/abs/1412.7515}{{\ttfamily arXiv:1412.7515 [hep-ex]}}.

\bibitem{Galanti:2015pqa}
M.~Galanti, A.~Giammanco, Y.~Grossman, Y.~Kats, E.~Stamou, and J.~Zupan,
  ``{Heavy baryons as polarimeters at colliders}'',
  \href{http://dx.doi.org/10.1007/JHEP11(2015)067}{{\em JHEP} {\bfseries 11}
  (2015) 067},
\href{http://arxiv.org/abs/1505.02771}{{\ttfamily arXiv:1505.02771 [hep-ph]}}.

\bibitem{Olive:2016xmw}
C.~Patrignani, ``{Review of Particle Physics}'',
\href{http://dx.doi.org/10.1088/1674-1137/40/10/100001}{{\em Chin. Phys.}
  {\bfseries C40} no.~10, (2016) 100001}.

\bibitem{Hicheur:2007jfk}
A.~Hicheur and G.~Conti,
  \href{http://dx.doi.org/10.1109/NSSMIC.2007.4436650}{``{Parameterization of
  the LHCb magnetic field map}'',} in {\em {Proceedings, 2007 IEEE Nuclear
  Science Symposium and Medical Imaging Conference (NSS/MIC 2007): Honolulu,
  Hawaii, October 28-November 3, 2007}}, pp.~2439--2443.
\newblock
2007.
\newblock

\bibitem{Hairer1993}
{\em Runge-Kutta and Extrapolation Methods},
  \href{http://dx.doi.org/10.1007/978-3-540-78862-1\_2}{pp.~129--353}.
\newblock Springer Berlin Heidelberg, Berlin, Heidelberg, 1993.
\newblock \url{https://doi.org/10.1007/978-3-540-78862-1\_2}.

\bibitem{LHCb-2007-140}
{LHCb} collaboration, E.~Bos and E.~Rodrigues, ``{The LHCb track extrapolator
  tools}'', 2007.
\newblock {CERN-LHCb-2007-140, GLAS-PPE-2007-24}.

\bibitem{Borsato:2021aum}
M.~Borsato {\em et~al.}, ``{Unleashing the full power of LHCb to probe stealth
  new physics}'', \href{http://dx.doi.org/10.1088/1361-6633/ac4649}{{\em Rept.
  Prog. Phys.} {\bfseries 85} no.~2, (2022) 024201},
  \href{http://arxiv.org/abs/2105.12668}{{\ttfamily arXiv:2105.12668
  [hep-ph]}}.

\bibitem{Aielli:2019ivi}
G.~Aielli {\em et~al.}, ``{Expression of interest for the CODEX-b detector}'',
  \href{http://dx.doi.org/10.1140/epjc/s10052-020-08711-3}{{\em Eur. Phys. J.
  C} {\bfseries 80} no.~12, (2020) 1177},
  \href{http://arxiv.org/abs/1911.00481}{{\ttfamily arXiv:1911.00481
  [hep-ex]}}.

\bibitem{Aielli:2022awh}
G.~Aielli {\em et~al.}, ``{The Road Ahead for CODEX-b}'',
  \href{http://arxiv.org/abs/2203.07316}{{\ttfamily arXiv:2203.07316
  [hep-ex]}}.

\bibitem{MATHUSLA:2020uve}
{MATHUSLA} collaboration, C.~Alpigiani {\em et~al.}, ``{An Update to the Letter
  of Intent for MATHUSLA: Search for Long-Lived Particles at the HL-LHC}'',
  \href{http://arxiv.org/abs/2009.01693}{{\ttfamily arXiv:2009.01693
  [physics.ins-det]}}.

\bibitem{Gligorov:2017nwh}
V.~V. Gligorov, S.~Knapen, M.~Papucci, and D.~J. Robinson, ``{Searching for
  Long-lived Particles: A Compact Detector for Exotics at LHCb}'',
  \href{http://dx.doi.org/10.1103/PhysRevD.97.015023}{{\em Phys. Rev. D}
  {\bfseries 97} no.~1, (2018) 015023},
  \href{http://arxiv.org/abs/1708.09395}{{\ttfamily arXiv:1708.09395
  [hep-ph]}}.

\bibitem{Alimena:2019zri}
J.~Alimena {\em et~al.}, ``{Searching for long-lived particles beyond the
  Standard Model at the Large Hadron Collider}'',
  \href{http://dx.doi.org/10.1088/1361-6471/ab4574}{{\em J. Phys. G} {\bfseries
  47} no.~9, (2020) 090501}, \href{http://arxiv.org/abs/1903.04497}{{\ttfamily
  arXiv:1903.04497 [hep-ex]}}.

\bibitem{Dreyer:2021aqd}
S.~Dreyer {\em et~al.}, ``{Physics reach of a long-lived particle detector at
  Belle II}'', \href{http://arxiv.org/abs/2105.12962}{{\ttfamily
  arXiv:2105.12962 [hep-ph]}}.

\bibitem{Schafer:2022shi}
R.~Sch\"afer, F.~Tillinger, and S.~Westhoff, ``{Near or Far Detectors?
  Optimizing Long-Lived Particle Searches at Electron-Positron Colliders}'',
  \href{http://arxiv.org/abs/2202.11714}{{\ttfamily arXiv:2202.11714
  [hep-ph]}}.

\bibitem{Alimena:2022hfr}
J.~Alimena {\em et~al.}, ``{Searches for Long-Lived Particles at the Future
  FCC-ee}'', \href{http://arxiv.org/abs/2203.05502}{{\ttfamily arXiv:2203.05502
  [hep-ex]}}.

\bibitem{LHCb:2014jgs}
{LHCb} collaboration, R.~Aaij {\em et~al.}, ``{Search for long-lived particles
  decaying to jet pairs}'',
  \href{http://dx.doi.org/10.1140/epjc/s10052-015-3344-6}{{\em Eur. Phys. J. C}
  {\bfseries 75} no.~4, (2015) 152},
  \href{http://arxiv.org/abs/1412.3021}{{\ttfamily arXiv:1412.3021 [hep-ex]}}.

\bibitem{LHCb:2015ujr}
{LHCb} collaboration, R.~Aaij {\em et~al.}, ``{Search for long-lived heavy
  charged particles using a ring imaging Cherenkov technique at LHCb}'',
  \href{http://dx.doi.org/10.1140/epjc/s10052-015-3809-7}{{\em Eur. Phys. J. C}
  {\bfseries 75} no.~12, (2015) 595},
  \href{http://arxiv.org/abs/1506.09173}{{\ttfamily arXiv:1506.09173
  [hep-ex]}}.

\bibitem{LHCb:2016inz}
{LHCb} collaboration, R.~Aaij {\em et~al.}, ``{Search for massive long-lived
  particles decaying semileptonically in the LHCb detector}'',
  \href{http://dx.doi.org/10.1140/epjc/s10052-017-4744-6}{{\em Eur. Phys. J. C}
  {\bfseries 77} no.~4, (2017) 224},
  \href{http://arxiv.org/abs/1612.00945}{{\ttfamily arXiv:1612.00945
  [hep-ex]}}.

\bibitem{LHCb:2016awg}
{LHCb} collaboration, R.~Aaij {\em et~al.}, ``{Search for long-lived scalar
  particles in $B^+ \to K^+ \chi (\mu^+\mu^-)$ decays}'',
  \href{http://dx.doi.org/10.1103/PhysRevD.95.071101}{{\em Phys. Rev. D}
  {\bfseries 95} no.~7, (2017) 071101},
  \href{http://arxiv.org/abs/1612.07818}{{\ttfamily arXiv:1612.07818
  [hep-ex]}}.

\bibitem{LHCb:2016buh}
{LHCb} collaboration, R.~Aaij {\em et~al.}, ``{Search for Higgs-like bosons
  decaying into long-lived exotic particles}'',
  \href{http://dx.doi.org/10.1140/epjc/s10052-016-4489-7}{{\em Eur. Phys. J. C}
  {\bfseries 76} no.~12, (2016) 664},
  \href{http://arxiv.org/abs/1609.03124}{{\ttfamily arXiv:1609.03124
  [hep-ex]}}.

\bibitem{LHCb:2017trq}
{LHCb} collaboration, R.~Aaij {\em et~al.}, ``{Search for Dark Photons Produced
  in 13 TeV $pp$ Collisions}'',
  \href{http://dx.doi.org/10.1103/PhysRevLett.120.061801}{{\em Phys. Rev.
  Lett.} {\bfseries 120} no.~6, (2018) 061801},
  \href{http://arxiv.org/abs/1710.02867}{{\ttfamily arXiv:1710.02867
  [hep-ex]}}.

\bibitem{LHCb:2017xxn}
{LHCb} collaboration, R.~Aaij {\em et~al.}, ``{Updated search for long-lived
  particles decaying to jet pairs}'',
  \href{http://dx.doi.org/10.1140/epjc/s10052-017-5178-x}{{\em Eur. Phys. J. C}
  {\bfseries 77} no.~12, (2017) 812},
  \href{http://arxiv.org/abs/1705.07332}{{\ttfamily arXiv:1705.07332
  [hep-ex]}}.

\bibitem{LHCb:2019vmc}
{LHCb} collaboration, R.~Aaij {\em et~al.}, ``{Search for $A'\to\mu^+\mu^-$
  Decays}'', \href{http://dx.doi.org/10.1103/PhysRevLett.124.041801}{{\em Phys.
  Rev. Lett.} {\bfseries 124} no.~4, (2020) 041801},
  \href{http://arxiv.org/abs/1910.06926}{{\ttfamily arXiv:1910.06926
  [hep-ex]}}.

\bibitem{LHCb:2020akw}
{LHCb} collaboration, R.~Aaij {\em et~al.}, ``{Search for long-lived particles
  decaying to $e^\pm \mu^\mp \nu$}'',
  \href{http://dx.doi.org/10.1140/epjc/s10052-021-08994-0}{{\em Eur. Phys. J.
  C} {\bfseries 81} no.~3, (2021) 261},
  \href{http://arxiv.org/abs/2012.02696}{{\ttfamily arXiv:2012.02696
  [hep-ex]}}.

\bibitem{LHCb:2021dyu}
{LHCb} collaboration, R.~Aaij {\em et~al.}, ``{Search for massive long-lived
  particles decaying semileptonically at $\sqrt {s}$ = 13 TeV}'',
  \href{http://dx.doi.org/10.1140/epjc/s10052-022-10186-3}{{\em Eur. Phys. J.
  C} {\bfseries 82} no.~4, (2022) 373},
  \href{http://arxiv.org/abs/2110.07293}{{\ttfamily arXiv:2110.07293
  [hep-ex]}}.

\bibitem{internalmeetingLLP}
{Tracking and Alignment Meeting, 29-March-2018 (LHCb internal)}.
  \url{https://indico.cern.ch/event/691552/}.

\bibitem{Chala:2019fdb}
M.~Chala, A.~Lenz, A.~V. Rusov, and J.~Scholtz, ``{$\Delta A_{CP}$ within the
  Standard Model and beyond}'',
  \href{http://dx.doi.org/10.1007/JHEP07(2019)161}{{\em JHEP} {\bfseries 07}
  (2019) 161}, \href{http://arxiv.org/abs/1903.10490}{{\ttfamily
  arXiv:1903.10490 [hep-ph]}}.

\bibitem{Dery:2019ysp}
A.~Dery and Y.~Nir, ``{Implications of the LHCb discovery of CP violation in
  charm decays}'', \href{http://dx.doi.org/10.1007/JHEP12(2019)104}{{\em JHEP}
  {\bfseries 12} (2019) 104}, \href{http://arxiv.org/abs/1909.11242}{{\ttfamily
  arXiv:1909.11242 [hep-ph]}}.

\bibitem{RICHMAN}
J.~D. Richman, ``{An experimenter's guide to the helicity formalism}'', Tech.
  Rep. CALT-68-1148, Calif. Inst. Technol., Pasadena, CA, Jun, 1984.
\newblock \url{http://cds.cern.ch/record/153636}.

\bibitem{LEADER}
E.~Leader, ``{Spin in particle physics}'',
{\em Camb. Monogr. Part. Phys. Nucl. Phys. Cosmol.} {\bfseries 15} (2011)
  pp.1--500.

\bibitem{Gronau:2015gha}
M.~Gronau and J.~L. Rosner, ``{Triple product asymmmetries in $\Lambda_b$ and
  $\Xi_b$ decays}'',
  \href{http://dx.doi.org/10.1016/j.physletb.2015.07.060}{{\em Phys. Lett. B}
  {\bfseries 749} (2015) 104--107},
  \href{http://arxiv.org/abs/1506.01346}{{\ttfamily arXiv:1506.01346
  [hep-ph]}}.

\bibitem{Wu:2009tu}
J.-J. Wu, S.~Dulat, and B.~S. Zou, ``{Evidence for a new Sigma* resonance with
  J**P = 1/2- in the old data of K- p ---\ensuremath{>} Lambda pi+ pi-
  reaction}'', \href{http://dx.doi.org/10.1103/PhysRevD.80.017503}{{\em Phys.
  Rev. D} {\bfseries 80} (2009) 017503},
  \href{http://arxiv.org/abs/0906.3950}{{\ttfamily arXiv:0906.3950 [hep-ph]}}.

\bibitem{Wu:2009nw}
J.-J. Wu, S.~Dulat, and B.~S. Zou, ``{Further evidence for the Sigma* resonance
  with J**P = 1/2- around 1380-MeV}'',
  \href{http://dx.doi.org/10.1103/PhysRevC.81.045210}{{\em Phys. Rev. C}
  {\bfseries 81} (2010) 045210},
  \href{http://arxiv.org/abs/0909.1380}{{\ttfamily arXiv:0909.1380 [hep-ph]}}.

\bibitem{Helminen:2000jb}
C.~Helminen and D.~O. Riska, ``{Low lying q q q q anti-q states in the baryon
  spectrum}'', \href{http://dx.doi.org/10.1016/S0375-9474(01)01294-5}{{\em
  Nucl. Phys. A} {\bfseries 699} (2002) 624--648},
  \href{http://arxiv.org/abs/nucl-th/0011071}{{\ttfamily
  arXiv:nucl-th/0011071}}.

\bibitem{Zhang:2004xt}
A.~Zhang, Y.~R. Liu, P.~Z. Huang, W.~Z. Deng, X.~L. Chen, and S.-L. Zhu,
  ``{J**P = 1/2- pentaquarks in Jaffe and Wilczek's diquark model}'', {\em
  HEPNP} {\bfseries 29} (2005) 250,
  \href{http://arxiv.org/abs/hep-ph/0403210}{{\ttfamily arXiv:hep-ph/0403210}}.

\bibitem{Gao:2010hy}
P.~Gao, J.-J. Wu, and B.~S. Zou, ``{Possible $\Sigma({1\over2}^-)$ under the
  $\Sigma^*(1385)$ peak in $K\Sigma^*$ photoproduction}'',
  \href{http://dx.doi.org/10.1103/PhysRevC.81.055203}{{\em Phys. Rev. C}
  {\bfseries 81} (2010) 055203},
  \href{http://arxiv.org/abs/1001.0805}{{\ttfamily arXiv:1001.0805 [nucl-th]}}.

\bibitem{redecay}
D.~M{\"u}ller, M.~Clemencic, G.~Corti, and M.~Gersabeck, ``{ReDecay: A novel
  approach to speed up the simulation at LHCb}'',
  \href{http://dx.doi.org/10.1140/epjc/s10052-018-6469-6}{{\em Eur. Phys. J.}
  {\bfseries C78} (2018) 1009} LHCb-DP-2018-004,
  \href{http://arxiv.org/abs/1810.10362}{{\ttfamily arXiv:1810.10362
  [hep-ex]}}.

\bibitem{Gassner:2004yda}
J.~Gassner, M.~Needham, and O.~Steinkamp, ``{Layout and Expected Performance of
  the LHCb TT Station}'',.

\bibitem{Skwarnicki:1986xj}
T.~Skwarnicki, {\em {A study of the radiative CASCADE transitions between the
  Upsilon-Prime and Upsilon resonances}}.
\newblock PhD thesis, Cracow, INP, 1986.

\bibitem{Hocker:2007ht}
A.~Hocker {\em et~al.}, ``{TMVA - Toolkit for Multivariate Data Analysis}'',
  \href{http://arxiv.org/abs/physics/0703039}{{\ttfamily
  arXiv:physics/0703039}}.

\bibitem{BESIII:2015bjk}
{BESIII} collaboration, M.~Ablikim {\em et~al.}, ``{Measurements of absolute
  hadronic branching fractions of $\Lambda_{c}^{+}$ baryon}'',
  \href{http://dx.doi.org/10.1103/PhysRevLett.116.052001}{{\em Phys. Rev.
  Lett.} {\bfseries 116} no.~5, (2016) 052001},
  \href{http://arxiv.org/abs/1511.08380}{{\ttfamily arXiv:1511.08380
  [hep-ex]}}.

\bibitem{FOCUS:2005sye}
{FOCUS} collaboration, J.~M. Link {\em et~al.}, ``{Study of Lambda+(c) Cabibbo
  favored decays containing a Lambda baryon in the final state}'',
  \href{http://dx.doi.org/10.1016/j.physletb.2005.08.014}{{\em Phys. Lett. B}
  {\bfseries 624} (2005) 22--30},
  \href{http://arxiv.org/abs/hep-ex/0505077}{{\ttfamily arXiv:hep-ex/0505077}}.

\bibitem{CLEO:1990unu}
{CLEO} collaboration, P.~Avery {\em et~al.}, ``{Inclusive production of the
  charmed baryon Lambda(c) from e+ e- annihilations at s**(1/2) = 10.55-GeV}'',
  \href{http://dx.doi.org/10.1103/PhysRevD.43.3599}{{\em Phys. Rev. D}
  {\bfseries 43} (1991) 3599--3610}.

\bibitem{Anjos:1989tc}
J.~C. Anjos {\em et~al.}, ``{A Study of Decays of the Lambda(c)+}'',
  \href{http://dx.doi.org/10.1103/PhysRevD.41.801}{{\em Phys. Rev. D}
  {\bfseries 41} (1990) 801--804}.

\bibitem{ACCMOR:1990gke}
{ACCMOR} collaboration, S.~Barlag {\em et~al.}, ``{Measurement of Frequencies
  of Various Decay Modes of Charmed Particles D0, $D^+$, $D(s$)+ and
  $\Lambda(c$)+ Including the Observation of New Channels}'',
  \href{http://dx.doi.org/10.1007/BF01565603}{{\em Z. Phys. C} {\bfseries 48}
  (1990) 29--46}.

\bibitem{ARGUS:1988hly}
{ARGUS} collaboration, H.~Albrecht {\em et~al.}, ``{Observation of the Charmed
  Baryon $\Lambda(c$) in $e^+ e^-$ Annihilation at 10-{GeV}}'',
  \href{http://dx.doi.org/10.1016/0370-2693(88)90896-9}{{\em Phys. Lett. B}
  {\bfseries 207} (1988) 109--114}.

\bibitem{LHCb:2017ygo}
{LHCb} collaboration, R.~Aaij {\em et~al.}, ``{Prompt and nonprompt J/$\psi$
  production and nuclear modification in $p$Pb collisions at
  $\sqrt{s_{\text{NN}}}= 8.16$ TeV}''
  \href{http://dx.doi.org/10.1016/j.physletb.2017.09.058}{{\em Phys. Lett. B}
  {\bfseries 774} (2017) 159--178},
  \href{http://arxiv.org/abs/1706.07122}{{\ttfamily arXiv:1706.07122
  [hep-ex]}}.

\bibitem{Becattini:2016gvu}
F.~Becattini, I.~Karpenko, M.~Lisa, I.~Upsal, and S.~Voloshin, ``{Global
  hyperon polarization at local thermodynamic equilibrium with vorticity,
  magnetic field and feed-down}'',
  \href{http://dx.doi.org/10.1103/PhysRevC.95.054902}{{\em Phys. Rev. C}
  {\bfseries 95} no.~5, (2017) 054902},
  \href{http://arxiv.org/abs/1610.02506}{{\ttfamily arXiv:1610.02506
  [nucl-th]}}.

\bibitem{Pich:1998xt}
A.~Pich, ``{Effective field theory: Course}'', in {\em {Probing the standard
  model of particle interactions. Proceedings, Summer School in Theoretical
  Physics, NATO Advanced Study Institute, 68th session, Les Houches, France,
  July 28-September 5, 1997. Pt. 1, 2}}, pp.~949--1049.
\newblock 1998.
\newblock
\href{http://arxiv.org/abs/hep-ph/9806303}{{\ttfamily arXiv:hep-ph/9806303
  [hep-ph]}}.
\newblock

\bibitem{Buras:1998raa}
A.~J. Buras, ``{Weak Hamiltonian, CP violation and rare decays}'', in {\em {Les
  Houches Summer School in Theoretical Physics, Session 68: Probing the
  Standard Model of Particle Interactions}}, pp.~281--539.
\newblock 6, 1998.
\newblock \href{http://arxiv.org/abs/hep-ph/9806471}{{\ttfamily
  arXiv:hep-ph/9806471}}.

\bibitem{Ilisie:2016jta}
V.~Ilisie, \href{http://dx.doi.org/10.1007/978-3-319-22966-9}{{\em {Concepts in
  Quantum Field Theory}}}.
\newblock UNITEXT for Physics. Springer, 2016.

\bibitem{Shtabovenko:2020gxv}
V.~Shtabovenko, R.~Mertig, and F.~Orellana, ``{FeynCalc 9.3: New features and
  improvements}'', \href{http://dx.doi.org/10.1016/j.cpc.2020.107478}{{\em
  Comput. Phys. Commun.} {\bfseries 256} (2020) 107478},
  \href{http://arxiv.org/abs/2001.04407}{{\ttfamily arXiv:2001.04407
  [hep-ph]}}.

\bibitem{Patel:2015tea}
H.~H. Patel, ``{Package-X: A Mathematica package for the analytic calculation
  of one-loop integrals}'',
  \href{http://dx.doi.org/10.1016/j.cpc.2015.08.017}{{\em Comput. Phys.
  Commun.} {\bfseries 197} (2015) 276--290},
  \href{http://arxiv.org/abs/1503.01469}{{\ttfamily arXiv:1503.01469
  [hep-ph]}}.

\bibitem{Hahn:2000kx}
T.~Hahn, ``{Generating Feynman diagrams and amplitudes with FeynArts 3}'',
  \href{http://dx.doi.org/10.1016/S0010-4655(01)00290-9}{{\em Comput. Phys.
  Commun.} {\bfseries 140} (2001) 418--431},
  \href{http://arxiv.org/abs/hep-ph/0012260}{{\ttfamily arXiv:hep-ph/0012260}}.

\bibitem{Alloul:2013bka}
A.~Alloul, N.~D. Christensen, C.~Degrande, C.~Duhr, and B.~Fuks, ``{FeynRules
  2.0 - A complete toolbox for tree-level phenomenology}'',
  \href{http://dx.doi.org/10.1016/j.cpc.2014.04.012}{{\em Comput. Phys.
  Commun.} {\bfseries 185} (2014) 2250--2300},
  \href{http://arxiv.org/abs/1310.1921}{{\ttfamily arXiv:1310.1921 [hep-ph]}}.

\bibitem{Tomonaga:1946zz}
S.~Tomonaga, ``{On a relativistically invariant formulation of the quantum
  theory of wave fields}'', \href{http://dx.doi.org/10.1143/PTP.1.27}{{\em
  Prog. Theor. Phys.} {\bfseries 1} (1946) 27--42}.

\bibitem{Peskin:1995ev}
M.~E. Peskin and D.~V. Schroeder, {\em {An Introduction to quantum field
  theory}}.
\newblock Addison-Wesley, Reading, USA, 1995.

\bibitem{Peccei:1977hh}
R.~D. Peccei and H.~R. Quinn, ``{CP Conservation in the Presence of
  Instantons}'', \href{http://dx.doi.org/10.1103/PhysRevLett.38.1440}{{\em
  Phys. Rev. Lett.} {\bfseries 38} (1977) 1440--1443}.

\bibitem{Weinberg:1977ma}
S.~Weinberg, ``{A New Light Boson?}'',
  \href{http://dx.doi.org/10.1103/PhysRevLett.40.223}{{\em Phys. Rev. Lett.}
  {\bfseries 40} (1978) 223--226}.

\bibitem{Wilczek:1977pj}
F.~Wilczek, ``{Problem of Strong $P$ and $T$ Invariance in the Presence of
  Instantons}'', \href{http://dx.doi.org/10.1103/PhysRevLett.40.279}{{\em Phys.
  Rev. Lett.} {\bfseries 40} (1978) 279--282}.

\bibitem{Pich:2012sx}
A.~Pich, ``{The Standard Model of Electroweak Interactions}'', in {\em {2010
  European School of High Energy Physics}}, pp.~1--50.
\newblock 1, 2012.
\newblock \href{http://arxiv.org/abs/1201.0537}{{\ttfamily arXiv:1201.0537
  [hep-ph]}}.

\bibitem{Higgs:1964pj}
P.~W. Higgs, ``{Broken Symmetries and the Masses of Gauge Bosons}'',
  \href{http://dx.doi.org/10.1103/PhysRevLett.13.508}{{\em Phys. Rev. Lett.}
  {\bfseries 13} (1964) 508--509}.

\bibitem{Englert:1964et}
F.~Englert and R.~Brout, ``{Broken Symmetry and the Mass of Gauge Vector
  Mesons}'', \href{http://dx.doi.org/10.1103/PhysRevLett.13.321}{{\em Phys.
  Rev. Lett.} {\bfseries 13} (1964) 321--323}.

\bibitem{Guralnik:1964eu}
G.~S. Guralnik, C.~R. Hagen, and T.~W.~B. Kibble, ``{Global Conservation Laws
  and Massless Particles}'',
  \href{http://dx.doi.org/10.1103/PhysRevLett.13.585}{{\em Phys. Rev. Lett.}
  {\bfseries 13} (1964) 585--587}.

\bibitem{Weinberg:1967tq}
S.~Weinberg, ``{A Model of Leptons}'',
  \href{http://dx.doi.org/10.1103/PhysRevLett.19.1264}{{\em Phys. Rev. Lett.}
  {\bfseries 19} (1967) 1264--1266}.

\bibitem{ATLAS:2012yve}
{ATLAS} collaboration, G.~Aad {\em et~al.}, ``{Observation of a new particle in
  the search for the Standard Model Higgs boson with the ATLAS detector at the
  LHC}'', \href{http://dx.doi.org/10.1016/j.physletb.2012.08.020}{{\em Phys.
  Lett. B} {\bfseries 716} (2012) 1--29},
  \href{http://arxiv.org/abs/1207.7214}{{\ttfamily arXiv:1207.7214 [hep-ex]}}.

\bibitem{CMS:2012qbp}
{CMS} collaboration, S.~Chatrchyan {\em et~al.}, ``{Observation of a New Boson
  at a Mass of 125 GeV with the CMS Experiment at the LHC}'',
  \href{http://dx.doi.org/10.1016/j.physletb.2012.08.021}{{\em Phys. Lett. B}
  {\bfseries 716} (2012) 30--61},
  \href{http://arxiv.org/abs/1207.7235}{{\ttfamily arXiv:1207.7235 [hep-ex]}}.

\bibitem{Cabibbo:1963yz}
N.~Cabibbo, ``{Unitary Symmetry and Leptonic Decays}'',
  \href{http://dx.doi.org/10.1103/PhysRevLett.10.531}{{\em Phys. Rev. Lett.}
  {\bfseries 10} (1963) 531--533}.

\bibitem{Kobayashi:1973fv}
M.~Kobayashi and T.~Maskawa, ``{CP Violation in the Renormalizable Theory of
  Weak Interaction}'', \href{http://dx.doi.org/10.1143/PTP.49.652}{{\em Prog.
  Theor. Phys.} {\bfseries 49} (1973) 652--657}.

\bibitem{Akhmedov:1999uz}
E.~K. Akhmedov, ``{Neutrino physics}'', in {\em {ICTP Summer School in Particle
  Physics}}, pp.~103--164.
\newblock 6, 1999.
\newblock \href{http://arxiv.org/abs/hep-ph/0001264}{{\ttfamily
  arXiv:hep-ph/0001264}}.

\bibitem{Manohar:2006ga}
A.~V. Manohar and M.~B. Wise, ``{Flavor changing neutral currents, an extended
  scalar sector, and the Higgs production rate at the CERN LHC}'',
  \href{http://dx.doi.org/10.1103/PhysRevD.74.035009}{{\em Phys. Rev. D}
  {\bfseries 74} (2006) 035009},
  \href{http://arxiv.org/abs/hep-ph/0606172}{{\ttfamily arXiv:hep-ph/0606172}}.

\bibitem{Chivukula:1987py}
R.~S. Chivukula and H.~Georgi, ``{Composite Technicolor Standard Model}'',
  \href{http://dx.doi.org/10.1016/0370-2693(87)90713-1}{{\em Phys. Lett. B}
  {\bfseries 188} (1987) 99--104}.

\bibitem{DAmbrosio:2002vsn}
G.~D'Ambrosio, G.~F. Giudice, G.~Isidori, and A.~Strumia, ``{Minimal flavor
  violation: An Effective field theory approach}'',
  \href{http://dx.doi.org/10.1016/S0550-3213(02)00836-2}{{\em Nucl. Phys. B}
  {\bfseries 645} (2002) 155--187},
  \href{http://arxiv.org/abs/hep-ph/0207036}{{\ttfamily arXiv:hep-ph/0207036}}.

\bibitem{Georgi:1974sy}
H.~Georgi and S.~L. Glashow, ``{Unity of All Elementary Particle Forces}'',
  \href{http://dx.doi.org/10.1103/PhysRevLett.32.438}{{\em Phys. Rev. Lett.}
  {\bfseries 32} (1974) 438--441}.

\bibitem{Georgi:1979df}
H.~Georgi and C.~Jarlskog, ``{A New Lepton - Quark Mass Relation in a Unified
  Theory}'',
\href{http://dx.doi.org/10.1016/0370-2693(79)90842-6}{{\em Phys. Lett.}
  {\bfseries 86B} (1979) 297--300}.

\bibitem{Dorsner:2006dj}
I.~Dorsner and P.~Fileviez~Perez, ``{Unification versus proton decay in
  SU(5)}'', \href{http://dx.doi.org/10.1016/j.physletb.2006.09.034}{{\em Phys.
  Lett.} {\bfseries B642} (2006) 248--252},
\href{http://arxiv.org/abs/hep-ph/0606062}{{\ttfamily arXiv:hep-ph/0606062
  [hep-ph]}}.

\bibitem{FileviezPerez:2013zmv}
P.~Fileviez~Perez and M.~B. Wise, ``{Low Scale Quark-Lepton Unification}'',
  \href{http://dx.doi.org/10.1103/PhysRevD.88.057703}{{\em Phys. Rev. D}
  {\bfseries 88} (2013) 057703},
  \href{http://arxiv.org/abs/1307.6213}{{\ttfamily arXiv:1307.6213 [hep-ph]}}.

\bibitem{Perez:2016qbo}
P.~Fileviez~Perez and C.~Murgui, ``{Renormalizable SU(5) Unification}'',
  \href{http://dx.doi.org/10.1103/PhysRevD.94.075014}{{\em Phys. Rev.}
  {\bfseries D94} no.~7, (2016) 075014},
\href{http://arxiv.org/abs/1604.03377}{{\ttfamily arXiv:1604.03377 [hep-ph]}}.

\bibitem{FileviezPerez:2019ssf}
P.~Fileviez~P\'erez, C.~Murgui, and A.~D. Plascencia, ``{Axion Dark Matter,
  Proton Decay and Unification}'',
  \href{http://dx.doi.org/10.1007/JHEP01(2020)091}{{\em JHEP} {\bfseries 01}
  (2020) 091}, \href{http://arxiv.org/abs/1911.05738}{{\ttfamily
  arXiv:1911.05738 [hep-ph]}}.

\bibitem{Bertolini:2013vta}
S.~Bertolini, L.~Di~Luzio, and M.~Malinsky, ``{Light color octet scalars in the
  minimal SO(10) grand unification}'',
  \href{http://dx.doi.org/10.1103/PhysRevD.87.085020}{{\em Phys. Rev.}
  {\bfseries D87} no.~8, (2013) 085020},
\href{http://arxiv.org/abs/1302.3401}{{\ttfamily arXiv:1302.3401 [hep-ph]}}.

\bibitem{Hisano:2012cc}
J.~Hisano, K.~Tsumura, and M.~J.~S. Yang, ``{QCD Corrections to Neutron
  Electric Dipole Moment from Dimension-six Four-Quark Operators}'',
  \href{http://dx.doi.org/10.1016/j.physletb.2012.06.038}{{\em Phys. Lett. B}
  {\bfseries 713} (2012) 473--480},
  \href{http://arxiv.org/abs/1205.2212}{{\ttfamily arXiv:1205.2212 [hep-ph]}}.

\bibitem{Buttazzo:2017ixm}
D.~Buttazzo, A.~Greljo, G.~Isidori, and D.~Marzocca, ``{B-physics anomalies: a
  guide to combined explanations}'',
  \href{http://dx.doi.org/10.1007/JHEP11(2017)044}{{\em JHEP} {\bfseries 11}
  (2017) 044}, \href{http://arxiv.org/abs/1706.07808}{{\ttfamily
  arXiv:1706.07808 [hep-ph]}}.

\bibitem{Sala:2013osa}
F.~Sala, ``{A bound on the charm chromo-EDM and its implications}'',
  \href{http://dx.doi.org/10.1007/JHEP03(2014)061}{{\em JHEP} {\bfseries 03}
  (2014) 061},
\href{http://arxiv.org/abs/1312.2589}{{\ttfamily arXiv:1312.2589 [hep-ph]}}.

\bibitem{Chang:1990jv}
D.~Chang, W.-Y. Keung, C.~S. Li, and T.~C. Yuan, ``{{QCD} Corrections to {\CP}
  Violation From Color Electric Dipole Moment of $b$ Quark}'',
\href{http://dx.doi.org/10.1016/0370-2693(90)91875-C}{{\em Phys. Lett.}
  {\bfseries B241} (1990) 589--592}.

\bibitem{Braaten:1990gq}
E.~Braaten, C.-S. Li, and T.-C. Yuan, ``{The Evolution of Weinberg's Gluonic
  {CP} Violation Operator}'',
\href{http://dx.doi.org/10.1103/PhysRevLett.64.1709}{{\em Phys. Rev. Lett.}
  {\bfseries 64} (1990) 1709}.

\bibitem{Boyd:1990bx}
G.~Boyd, A.~K. Gupta, S.~P. Trivedi, and M.~B. Wise, ``{Effective Hamiltonian
  for the Electric Dipole Moment of the Neutron}'',
  \href{http://dx.doi.org/10.1016/0370-2693(90)91874-B}{{\em Phys. Lett. B}
  {\bfseries 241} (1990) 584--588}.

\bibitem{Escribano:1993xr}
R.~Escribano and E.~Masso, ``{Constraints on fermion magnetic and electric
  moments from LEP-1}'',
  \href{http://dx.doi.org/10.1016/S0550-3213(94)80039-1}{{\em Nucl. Phys.}
  {\bfseries B429} (1994) 19--32},
\href{http://arxiv.org/abs/hep-ph/9403304}{{\ttfamily arXiv:hep-ph/9403304
  [hep-ph]}}.

\bibitem{ALEPH:2006bhb}
{ALEPH, DELPHI, L3, OPAL, LEP Electroweak Working Group} collaboration,
  J.~Alcaraz {\em et~al.}, ``{A Combination of preliminary electroweak
  measurements and constraints on the standard model}'',
  \href{http://arxiv.org/abs/hep-ex/0612034}{{\ttfamily arXiv:hep-ex/0612034}}.

\bibitem{CorderoCid:2007uc}
A.~Cordero-Cid, J.~M. Hernandez, G.~Tavares-Velasco, and J.~J. Toscano,
  ``{Bounding the top and bottom electric dipole moments from neutron
  experimental data}'',
  \href{http://dx.doi.org/10.1088/0954-3899/35/2/025004}{{\em J. Phys.}
  {\bfseries G35} (2008) 025004},
\href{http://arxiv.org/abs/0712.0154}{{\ttfamily arXiv:0712.0154 [hep-ph]}}.

\bibitem{Ema:2022pmo}
Y.~Ema, T.~Gao, and M.~Pospelov, ``{Improved indirect limits on charm and
  bottom quark EDMs}'', \href{http://arxiv.org/abs/2205.11532}{{\ttfamily
  arXiv:2205.11532 [hep-ph]}}.

\bibitem{Kuang:2012wp}
Y.-P. Kuang, J.-P. Ma, O.~Nachtmann, W.-P. Xie, and H.-H. Zheng, ``{Testing
  Anomalous Color-Electric Dipole Moment of the c-Quark from $\psi' \to \jpsi +
  \pip + \pim$ at Beijing Spectrometer}'',
  \href{http://dx.doi.org/10.1103/PhysRevD.85.114010}{{\em Phys. Rev.}
  {\bfseries D85} (2012) 114010},
\href{http://arxiv.org/abs/1202.3042}{{\ttfamily arXiv:1202.3042 [hep-ph]}}.

\bibitem{Konig:2014iqa}
M.~König, M.~Neubert, and D.~M. Straub, ``{Dipole operator constraints on
  composite Higgs models}'',
  \href{http://dx.doi.org/10.1140/epjc/s10052-014-2945-9}{{\em Eur. Phys. J.}
  {\bfseries C74} no.~7, (2014) 2945},
\href{http://arxiv.org/abs/1403.2756}{{\ttfamily arXiv:1403.2756 [hep-ph]}}.

\bibitem{Weinberg:1989dx}
S.~Weinberg, ``{Larger Higgs Exchange Terms in the Neutron Electric Dipole
  Moment}'', \href{http://dx.doi.org/10.1103/PhysRevLett.63.2333}{{\em Phys.
  Rev. Lett.} {\bfseries 63} (1989) 2333}.

\bibitem{Wilczek:1976ry}
F.~Wilczek and A.~Zee, ``{Delta I=1/2 Rule and Right-Handed Currents: Heavy
  Quark Expansion and Limitation on Zweig's Rule}'',
  \href{http://dx.doi.org/10.1103/PhysRevD.15.2660}{{\em Phys. Rev. D}
  {\bfseries 15} (1977) 2660}.

\bibitem{Degrassi:2005zd}
G.~Degrassi, E.~Franco, S.~Marchetti, and L.~Silvestrini, ``{QCD corrections to
  the electric dipole moment of the neutron in the MSSM}'',
  \href{http://dx.doi.org/10.1088/1126-6708/2005/11/044}{{\em JHEP} {\bfseries
  11} (2005) 044},
\href{http://arxiv.org/abs/hep-ph/0510137}{{\ttfamily arXiv:hep-ph/0510137
  [hep-ph]}}.

\bibitem{Jenkins:2017dyc}
E.~E. Jenkins, A.~V. Manohar, and P.~Stoffer, ``{Low-Energy Effective Field
  Theory below the Electroweak Scale: Anomalous Dimensions}'',
  \href{http://dx.doi.org/10.1007/JHEP01(2018)084}{{\em JHEP} {\bfseries 01}
  (2018) 084}, \href{http://arxiv.org/abs/1711.05270}{{\ttfamily
  arXiv:1711.05270 [hep-ph]}}.

\bibitem{Buchalla:1995vs}
G.~Buchalla, A.~J. Buras, and M.~E. Lautenbacher, ``{Weak decays beyond leading
  logarithms}'', \href{http://dx.doi.org/10.1103/RevModPhys.68.1125}{{\em Rev.
  Mod. Phys.} {\bfseries 68} (1996) 1125--1144},
  \href{http://arxiv.org/abs/hep-ph/9512380}{{\ttfamily arXiv:hep-ph/9512380}}.

\bibitem{Baker:2006ts}
C.~A. Baker {\em et~al.}, ``{An improved experimental limit on the electric
  dipole moment of the neutron}'',
  \href{http://dx.doi.org/10.1103/PhysRevLett.97.131801}{{\em Phys. Rev. Lett.}
  {\bfseries 97} (2006) 131801},
\href{http://arxiv.org/abs/hep-ex/0602020}{{\ttfamily arXiv:hep-ex/0602020
  [hep-ex]}}.

\bibitem{Cirigliano:2016njn}
V.~Cirigliano, W.~Dekens, J.~de~Vries, and E.~Mereghetti, ``{Is there room for
  CP violation in the top-Higgs sector?}'',
  \href{http://dx.doi.org/10.1103/PhysRevD.94.016002}{{\em Phys. Rev.}
  {\bfseries D94} no.~1, (2016) 016002},
\href{http://arxiv.org/abs/1603.03049}{{\ttfamily arXiv:1603.03049 [hep-ph]}}.

\bibitem{Fuyuto:2017xup}
K.~Fuyuto and M.~Ramsey-Musolf, ``{Top Down Electroweak Dipole Operators}'',
\href{http://arxiv.org/abs/1706.08548}{{\ttfamily arXiv:1706.08548 [hep-ph]}}.

\bibitem{Kamenik:2011dk}
J.~F. Kamenik, M.~Papucci, and A.~Weiler, ``{Constraining the dipole moments of
  the top quark}'', \href{http://dx.doi.org/10.1103/PhysRevD.88.039903,
  10.1103/PhysRevD.85.071501}{{\em Phys. Rev.} {\bfseries D85} (2012) 071501},
  \href{http://arxiv.org/abs/1107.3143}{{\ttfamily arXiv:1107.3143 [hep-ph]}}.
[Erratum: Phys. Rev.D88,no.3,039903(2013)].

\bibitem{Graner:2016ses}
B.~Graner, Y.~Chen, E.~G. Lindahl, and B.~R. Heckel, ``{Reduced Limit on the
  Permanent Electric Dipole Moment of Hg199}'',
  \href{http://dx.doi.org/10.1103/PhysRevLett.116.161601}{{\em Phys. Rev.
  Lett.} {\bfseries 116} no.~16, (2016) 161601},
  \href{http://arxiv.org/abs/1601.04339}{{\ttfamily arXiv:1601.04339
  [physics.atom-ph]}}. [Erratum: Phys.Rev.Lett. 119, 119901 (2017)].

\bibitem{Engel:2013lsa}
J.~Engel, M.~J. Ramsey-Musolf, and U.~van Kolck, ``{Electric Dipole Moments of
  Nucleons, Nuclei, and Atoms: The Standard Model and Beyond}'',
  \href{http://dx.doi.org/10.1016/j.ppnp.2013.03.003}{{\em Prog. Part. Nucl.
  Phys.} {\bfseries 71} (2013) 21--74},
  \href{http://arxiv.org/abs/1303.2371}{{\ttfamily arXiv:1303.2371 [nucl-th]}}.

\bibitem{Buras:2000dm}
A.~J. Buras, P.~Gambino, M.~Gorbahn, S.~Jager, and L.~Silvestrini, ``{Universal
  unitarity triangle and physics beyond the standard model}'',
  \href{http://dx.doi.org/10.1016/S0370-2693(01)00061-2}{{\em Phys. Lett. B}
  {\bfseries 500} (2001) 161--167},
  \href{http://arxiv.org/abs/hep-ph/0007085}{{\ttfamily arXiv:hep-ph/0007085}}.

\bibitem{Pich:2009sp}
A.~Pich and P.~Tuzon, ``{Yukawa Alignment in the Two-Higgs-Doublet Model}'',
  \href{http://dx.doi.org/10.1103/PhysRevD.80.091702}{{\em Phys. Rev. D}
  {\bfseries 80} (2009) 091702},
  \href{http://arxiv.org/abs/0908.1554}{{\ttfamily arXiv:0908.1554 [hep-ph]}}.

\bibitem{Penuelas:2017ikk}
A.~Pe\~nuelas and A.~Pich, ``{Flavour alignment in multi-Higgs-doublet
  models}'', \href{http://dx.doi.org/10.1007/JHEP12(2017)084}{{\em JHEP}
  {\bfseries 12} (2017) 084}, \href{http://arxiv.org/abs/1710.02040}{{\ttfamily
  arXiv:1710.02040 [hep-ph]}}.

\bibitem{Martinez:2016fyd}
R.~Martinez and G.~Valencia, ``{Top and bottom tensor couplings from a color
  octet scalar}'', \href{http://dx.doi.org/10.1103/PhysRevD.95.035041}{{\em
  Phys. Rev.} {\bfseries D95} no.~3, (2017) 035041},
\href{http://arxiv.org/abs/1612.00561}{{\ttfamily arXiv:1612.00561 [hep-ph]}}.

\bibitem{Cheng:2015lsa}
X.-D. Cheng, X.-Q. Li, Y.-D. Yang, and X.~Zhang,
  ``{${B}_{s,d}\;-\;{\bar{B}}_{s,d}$ mixings and ${B}_{s,d}\;\to \;{{\ell
  }}^{+}{{\ell }}^{-}$ decays within the Manohar-Wise model}'',
  \href{http://dx.doi.org/10.1088/0954-3899/42/12/125005}{{\em J. Phys. G}
  {\bfseries 42} no.~12, (2015) 125005},
  \href{http://arxiv.org/abs/1504.00839}{{\ttfamily arXiv:1504.00839
  [hep-ph]}}.

\bibitem{Miralles:2019uzg}
V.~Miralles and A.~Pich, ``{LHC bounds on coloured scalars}'',
  \href{http://dx.doi.org/10.1103/PhysRevD.100.115042}{{\em Phys. Rev.}
  {\bfseries D100} no.~11, (2019) 115042},
\href{http://arxiv.org/abs/1910.07947}{{\ttfamily arXiv:1910.07947 [hep-ph]}}.

\bibitem{Eberhardt:2021ebh}
O.~Eberhardt, V.~Miralles, and A.~Pich, ``{Constraints on coloured scalars from
  global fits}'', \href{http://dx.doi.org/10.1007/JHEP10(2021)123}{{\em JHEP}
  {\bfseries 10} (2021) 123}, \href{http://arxiv.org/abs/2106.12235}{{\ttfamily
  arXiv:2106.12235 [hep-ph]}}.

\bibitem{DeBlas:2019ehy}
J.~De~Blas {\em et~al.}, ``{$\texttt{HEPfit}$: a code for the combination of
  indirect and direct constraints on high energy physics models}'',
  \href{http://dx.doi.org/10.1140/epjc/s10052-020-7904-z}{{\em Eur. Phys. J. C}
  {\bfseries 80} no.~5, (2020) 456},
  \href{http://arxiv.org/abs/1910.14012}{{\ttfamily arXiv:1910.14012
  [hep-ph]}}.

\bibitem{Murgui:2019czp}
C.~Murgui, A.~Pe\~nuelas, M.~Jung, and A.~Pich, ``{Global fit to $b \to c \tau
  \nu$ transitions}'', \href{http://dx.doi.org/10.1007/JHEP09(2019)103}{{\em
  JHEP} {\bfseries 09} (2019) 103},
  \href{http://arxiv.org/abs/1904.09311}{{\ttfamily arXiv:1904.09311
  [hep-ph]}}.

\bibitem{Becirevic:2016yqi}
D.~Be\v{c}irevi\'c, S.~Fajfer, N.~Ko\v{s}nik, and O.~Sumensari, ``{Leptoquark
  model to explain the $B$-physics anomalies, $R_K$ and $R_D$}'',
  \href{http://dx.doi.org/10.1103/PhysRevD.94.115021}{{\em Phys. Rev. D}
  {\bfseries 94} no.~11, (2016) 115021},
  \href{http://arxiv.org/abs/1608.08501}{{\ttfamily arXiv:1608.08501
  [hep-ph]}}.

\bibitem{Cornella:2019hct}
C.~Cornella, J.~Fuentes-Martin, and G.~Isidori, ``{Revisiting the vector
  leptoquark explanation of the B-physics anomalies}'',
  \href{http://dx.doi.org/10.1007/JHEP07(2019)168}{{\em JHEP} {\bfseries 07}
  (2019) 168}, \href{http://arxiv.org/abs/1903.11517}{{\ttfamily
  arXiv:1903.11517 [hep-ph]}}.

\bibitem{Fajfer:2012jt}
S.~Fajfer, J.~F. Kamenik, I.~Nisandzic, and J.~Zupan, ``{Implications of Lepton
  Flavor Universality Violations in B Decays}'',
  \href{http://dx.doi.org/10.1103/PhysRevLett.109.161801}{{\em Phys. Rev.
  Lett.} {\bfseries 109} (2012) 161801},
  \href{http://arxiv.org/abs/1206.1872}{{\ttfamily arXiv:1206.1872 [hep-ph]}}.

\bibitem{Hiller:2016kry}
G.~Hiller, D.~Loose, and K.~Sch\"onwald, ``{Leptoquark Flavor Patterns \& B
  Decay Anomalies}'', \href{http://dx.doi.org/10.1007/JHEP12(2016)027}{{\em
  JHEP} {\bfseries 12} (2016) 027},
  \href{http://arxiv.org/abs/1609.08895}{{\ttfamily arXiv:1609.08895
  [hep-ph]}}.

\bibitem{Becirevic:2018afm}
D.~Be\v{c}irevi\'c, I.~Dor\v{s}ner, S.~Fajfer, N.~Ko\v{s}nik, D.~A. Faroughy,
  and O.~Sumensari, ``{Scalar leptoquarks from grand unified theories to
  accommodate the $B$-physics anomalies}'',
  \href{http://dx.doi.org/10.1103/PhysRevD.98.055003}{{\em Phys. Rev. D}
  {\bfseries 98} no.~5, (2018) 055003},
  \href{http://arxiv.org/abs/1806.05689}{{\ttfamily arXiv:1806.05689
  [hep-ph]}}.

\bibitem{Chupp:2017rkp}
T.~Chupp, P.~Fierlinger, M.~Ramsey-Musolf, and J.~Singh, ``{Electric dipole
  moments of atoms, molecules, nuclei, and particles}'',
  \href{http://dx.doi.org/10.1103/RevModPhys.91.015001}{{\em Rev. Mod. Phys.}
  {\bfseries 91} no.~1, (2019) 015001},
  \href{http://arxiv.org/abs/1710.02504}{{\ttfamily arXiv:1710.02504
  [physics.atom-ph]}}.

\bibitem{Martin:1997ns}
S.~P. Martin, ``{A Supersymmetry primer}'',
  \href{http://dx.doi.org/10.1142/9789812839657_0001}{{\em Adv. Ser. Direct.
  High Energy Phys.} {\bfseries 18} (1998) 1--98},
  \href{http://arxiv.org/abs/hep-ph/9709356}{{\ttfamily arXiv:hep-ph/9709356}}.

\bibitem{Aydin:2002ie}
Z.~Z. Aydin and U.~Erkarslan, ``{The Charm quark EDM and singlet P wave
  charmonium production in supersymmetry}'',
  \href{http://dx.doi.org/10.1103/PhysRevD.67.036006}{{\em Phys. Rev.}
  {\bfseries D67} (2003) 036006},
\href{http://arxiv.org/abs/hep-ph/0204238}{{\ttfamily arXiv:hep-ph/0204238
  [hep-ph]}}.

\bibitem{ATLAS:2019jvl}
{ATLAS} collaboration, G.~Aad {\em et~al.}, ``{Search for squarks and gluinos
  in final states with same-sign leptons and jets using 139 fb$^{-1}$ of data
  collected with the ATLAS detector}'',
  \href{http://dx.doi.org/10.1007/JHEP06(2020)046}{{\em JHEP} {\bfseries 06}
  (2020) 046}, \href{http://arxiv.org/abs/1909.08457}{{\ttfamily
  arXiv:1909.08457 [hep-ex]}}.

\bibitem{Zhao:2016jcx}
S.-M. Zhao, T.-F. Feng, Z.-J. Yang, H.-B. Zhang, X.-X. Dong, and T.~Guo, ``{The
  one loop contributions to $c(t)$ electric dipole moment in the \CP violating
  BLMSSM}'', \href{http://dx.doi.org/10.1140/epjc/s10052-017-4627-x}{{\em Eur.
  Phys. J.} {\bfseries C77} no.~2, (2017) 102},
\href{http://arxiv.org/abs/1610.07314}{{\ttfamily arXiv:1610.07314 [hep-ph]}}.

\bibitem{Yamanaka:2014nba}
N.~Yamanaka, T.~Sato, and T.~Kubota, ``{Linear programming analysis of the
  $R$-parity violation within EDM-constraints}'',
  \href{http://dx.doi.org/10.1007/JHEP12(2014)110}{{\em JHEP} {\bfseries 12}
  (2014) 110}, \href{http://arxiv.org/abs/1406.3713}{{\ttfamily arXiv:1406.3713
  [hep-ph]}}.

\bibitem{Panico:2016ull}
G.~Panico and A.~Pomarol, ``{Flavor hierarchies from dynamical scales}'',
  \href{http://dx.doi.org/10.1007/JHEP07(2016)097}{{\em JHEP} {\bfseries 07}
  (2016) 097},
\href{http://arxiv.org/abs/1603.06609}{{\ttfamily arXiv:1603.06609 [hep-ph]}}.

\bibitem{Iltan:2004xr}
E.~O. Iltan, ``{The Effects of nonuniversal extra dimensions on the fermion
  electric dipole moments in the two Higgs doublet model}'',
  \href{http://dx.doi.org/10.1088/1126-6708/2004/04/018}{{\em JHEP} {\bfseries
  04} (2004) 018}, \href{http://arxiv.org/abs/hep-ph/0403007}{{\ttfamily
  arXiv:hep-ph/0403007}}.

\bibitem{Yang:2019aao}
J.-L. Yang, T.-F. Feng, S.-K. Cui, C.-X. Liu, W.~Li, and H.-B. Zhang,
  ``{Electric dipole moments of neutron and heavy quarks in the B-LSSM}'',
  \href{http://dx.doi.org/10.1007/JHEP04(2020)013}{{\em JHEP} {\bfseries 04}
  (2020) 013}, \href{http://arxiv.org/abs/1910.05868}{{\ttfamily
  arXiv:1910.05868 [hep-ph]}}.

\bibitem{Yan:2020ocy}
B.~Yan, S.-M. Zhao, and T.-F. Feng, ``{Electric dipole moments of neutron and
  heavy quarks c, t in CP violating U(1)XSSM}'',
  \href{http://dx.doi.org/10.1016/j.nuclphysb.2022.115671}{{\em Nucl. Phys. B}
  {\bfseries 975} (2022) 115671},
  \href{http://arxiv.org/abs/2011.08533}{{\ttfamily arXiv:2011.08533
  [hep-ph]}}.

\bibitem{Cai:2022xha}
F.-M. Cai, S.~Funatsu, X.-Q. Li, and Y.-D. Yang, ``{Rare top-quark decays $t
  \to cg(g)$ in the aligned two-Higgs-doublet model}'',
  \href{http://arxiv.org/abs/2202.08091}{{\ttfamily arXiv:2202.08091
  [hep-ph]}}.

\bibitem{Heo:2008sr}
J.~H. Heo and W.-Y. Keung, ``{Electron Electric Dipole Moment induced by
  Octet-Colored Scalars}'',
  \href{http://dx.doi.org/10.1016/j.physletb.2008.02.021}{{\em Phys. Lett. B}
  {\bfseries 661} (2008) 259--262},
  \href{http://arxiv.org/abs/0801.0231}{{\ttfamily arXiv:0801.0231 [hep-ph]}}.

\bibitem{Fajfer:2014etr}
S.~Fajfer and J.~O. Eeg, ``{Colored scalars and the neutron electric dipole
  moment}'', \href{http://dx.doi.org/10.1103/PhysRevD.89.095030}{{\em Phys.
  Rev. D} {\bfseries 89} no.~9, (2014) 095030},
  \href{http://arxiv.org/abs/1401.2275}{{\ttfamily arXiv:1401.2275 [hep-ph]}}.

\bibitem{CDF:2022hxs}
{CDF} collaboration, T.~Aaltonen {\em et~al.}, ``{High-precision measurement of
  the W boson mass with the CDF II detector}'',
  \href{http://dx.doi.org/10.1126/science.abk1781}{{\em Science} {\bfseries
  376} no.~6589, (2022) 170--176}.

\bibitem{Pospelov:2000bw}
M.~Pospelov and A.~Ritz, ``{Neutron EDM from electric and chromoelectric dipole
  moments of quarks}'',
  \href{http://dx.doi.org/10.1103/PhysRevD.63.073015}{{\em Phys. Rev.}
  {\bfseries D63} (2001) 073015},
\href{http://arxiv.org/abs/hep-ph/0010037}{{\ttfamily arXiv:hep-ph/0010037
  [hep-ph]}}.

\bibitem{Lebedev:2004va}
O.~Lebedev, K.~A. Olive, M.~Pospelov, and A.~Ritz, ``{Probing CP violation with
  the deuteron electric dipole moment}'',
  \href{http://dx.doi.org/10.1103/PhysRevD.70.016003}{{\em Phys. Rev. D}
  {\bfseries 70} (2004) 016003},
  \href{http://arxiv.org/abs/hep-ph/0402023}{{\ttfamily arXiv:hep-ph/0402023}}.

\bibitem{Hisano:2012sc}
J.~Hisano, J.~Y. Lee, N.~Nagata, and Y.~Shimizu, ``{Reevaluation of Neutron
  Electric Dipole Moment with QCD Sum Rules}'',
  \href{http://dx.doi.org/10.1103/PhysRevD.85.114044}{{\em Phys. Rev. D}
  {\bfseries 85} (2012) 114044},
  \href{http://arxiv.org/abs/1204.2653}{{\ttfamily arXiv:1204.2653 [hep-ph]}}.

\bibitem{Haisch:2019bml}
U.~Haisch and A.~Hala, ``{Sum rules for CP-violating operators of Weinberg
  type}'', \href{http://dx.doi.org/10.1007/JHEP11(2019)154}{{\em JHEP}
  {\bfseries 11} (2019) 154}, \href{http://arxiv.org/abs/1909.08955}{{\ttfamily
  arXiv:1909.08955 [hep-ph]}}.

\bibitem{Yamanaka:2020kjo}
N.~Yamanaka and E.~Hiyama, ``{Weinberg operator contribution to the nucleon
  electric dipole moment in the quark model}'',
  \href{http://dx.doi.org/10.1103/PhysRevD.103.035023}{{\em Phys. Rev. D}
  {\bfseries 103} no.~3, (2021) 035023},
  \href{http://arxiv.org/abs/2011.02531}{{\ttfamily arXiv:2011.02531
  [hep-ph]}}.

\bibitem{Bhattacharya:2015esa}
T.~Bhattacharya, V.~Cirigliano, R.~Gupta, H.-W. Lin, and B.~Yoon, ``{Neutron
  Electric Dipole Moment and Tensor Charges from Lattice QCD}'',
  \href{http://dx.doi.org/10.1103/PhysRevLett.115.212002}{{\em Phys. Rev.
  Lett.} {\bfseries 115} no.~21, (2015) 212002},
\href{http://arxiv.org/abs/1506.04196}{{\ttfamily arXiv:1506.04196 [hep-lat]}}.

\bibitem{Bhattacharya:2015wna}
{PNDME} collaboration, T.~Bhattacharya, V.~Cirigliano, S.~Cohen, R.~Gupta,
  A.~Joseph, H.-W. Lin, and B.~Yoon, ``{Iso-vector and Iso-scalar Tensor
  Charges of the Nucleon from Lattice QCD}'',
  \href{http://dx.doi.org/10.1103/PhysRevD.92.094511}{{\em Phys. Rev. D}
  {\bfseries 92} no.~9, (2015) 094511},
  \href{http://arxiv.org/abs/1506.06411}{{\ttfamily arXiv:1506.06411
  [hep-lat]}}.

\bibitem{Bhattacharya:2016zcn}
T.~Bhattacharya, V.~Cirigliano, S.~Cohen, R.~Gupta, H.-W. Lin, and B.~Yoon,
  ``{Axial, Scalar and Tensor Charges of the Nucleon from 2+1+1-flavor Lattice
  QCD}'', \href{http://dx.doi.org/10.1103/PhysRevD.94.054508}{{\em Phys. Rev.
  D} {\bfseries 94} no.~5, (2016) 054508},
  \href{http://arxiv.org/abs/1606.07049}{{\ttfamily arXiv:1606.07049
  [hep-lat]}}.

\bibitem{Gupta:2018qil}
R.~Gupta, Y.-C. Jang, B.~Yoon, H.-W. Lin, V.~Cirigliano, and T.~Bhattacharya,
  ``{Isovector Charges of the Nucleon from 2+1+1-flavor Lattice QCD}'',
  \href{http://dx.doi.org/10.1103/PhysRevD.98.034503}{{\em Phys. Rev. D}
  {\bfseries 98} (2018) 034503},
  \href{http://arxiv.org/abs/1806.09006}{{\ttfamily arXiv:1806.09006
  [hep-lat]}}.

\bibitem{Gupta:2018lvp}
R.~Gupta, B.~Yoon, T.~Bhattacharya, V.~Cirigliano, Y.-C. Jang, and H.-W. Lin,
  ``{Flavor diagonal tensor charges of the nucleon from (2+1+1)-flavor lattice
  QCD}'', \href{http://dx.doi.org/10.1103/PhysRevD.98.091501}{{\em Phys. Rev.
  D} {\bfseries 98} no.~9, (2018) 091501},
  \href{http://arxiv.org/abs/1808.07597}{{\ttfamily arXiv:1808.07597
  [hep-lat]}}.

\bibitem{Dekens:2021bro}
W.~Dekens, L.~Andreoli, J.~de~Vries, E.~Mereghetti, and F.~Oosterhof, ``{A
  low-energy perspective on the minimal left-right symmetric model}'',
  \href{http://arxiv.org/abs/2107.10852}{{\ttfamily arXiv:2107.10852
  [hep-ph]}}.

\bibitem{Dai:1989yh}
J.~Dai and H.~Dykstra, ``{{QCD} Corrections to {CP} Violation in Higgs
  Exchange}'', \href{http://dx.doi.org/10.1016/0370-2693(90)91439-I}{{\em Phys.
  Lett. B} {\bfseries 237} (1990) 256--258}.

\bibitem{Shifman:1976de}
M.~A. Shifman, A.~I. Vainshtein, and V.~I. Zakharov, ``{On the Weak Radiative
  Decays (Effects of Strong Interactions at Short Distances)}'',
  \href{http://dx.doi.org/10.1103/PhysRevD.18.2583}{{\em Phys. Rev. D}
  {\bfseries 18} (1978) 2583--2599}. [Erratum: Phys.Rev.D 19, 2815 (1979)].

\bibitem{Brod:2018pli}
J.~Brod and E.~Stamou, ``{Electric dipole moment constraints on CP-violating
  heavy-quark Yukawas at next-to-leading order}'',
  \href{http://dx.doi.org/10.1007/JHEP07(2021)080}{{\em JHEP} {\bfseries 07}
  (2021) 080}, \href{http://arxiv.org/abs/1810.12303}{{\ttfamily
  arXiv:1810.12303 [hep-ph]}}.

\bibitem{Yamanaka:2017mef}
N.~Yamanaka, B.~K. Sahoo, N.~Yoshinaga, T.~Sato, K.~Asahi, and B.~P. Das,
  ``{Probing exotic phenomena at the interface of nuclear and particle physics
  with the electric dipole moments of diamagnetic atoms: A unique window to
  hadronic and semi-leptonic CP violation}'',
  \href{http://dx.doi.org/10.1140/epja/i2017-12237-2}{{\em Eur. Phys. J. A}
  {\bfseries 53} no.~3, (2017) 54},
  \href{http://arxiv.org/abs/1703.01570}{{\ttfamily arXiv:1703.01570
  [hep-ph]}}.

\bibitem{deVries:2019nsu}
J.~de~Vries, G.~Falcioni, F.~Herzog, and B.~Ruijl, ``{Two- and three-loop
  anomalous dimensions of Weinberg\textquoteright{}s dimension-six CP-odd
  gluonic operator}'',
  \href{http://dx.doi.org/10.1103/PhysRevD.102.016010}{{\em Phys. Rev. D}
  {\bfseries 102} no.~1, (2020) 016010},
  \href{http://arxiv.org/abs/1907.04923}{{\ttfamily arXiv:1907.04923
  [hep-ph]}}.

\bibitem{Iltan:2001vg}
E.~O. Iltan, ``{Top quark electric and chromo electric dipole moments in the
  general two Higgs doublet model}'',
  \href{http://dx.doi.org/10.1103/PhysRevD.65.073013}{{\em Phys. Rev. D}
  {\bfseries 65} (2002) 073013},
  \href{http://arxiv.org/abs/hep-ph/0111038}{{\ttfamily arXiv:hep-ph/0111038}}.

\bibitem{Jung:2013hka}
M.~Jung and A.~Pich, ``{Electric Dipole Moments in Two-Higgs-Doublet Models}'',
  \href{http://dx.doi.org/10.1007/JHEP04(2014)076}{{\em JHEP} {\bfseries 04}
  (2014) 076},
\href{http://arxiv.org/abs/1308.6283}{{\ttfamily arXiv:1308.6283 [hep-ph]}}.

\bibitem{Gisbert:2022lao}
H.~Gisbert, V.~Miralles, and J.~Ruiz-Vidal, ``{W-boson mass and electric dipole
  moments from colour-octet scalars}'', in {\em {30th International Symposium
  on Lepton Photon Interactions at High Energies}}.
\newblock 4, 2022.
\newblock \href{http://arxiv.org/abs/2204.12453}{{\ttfamily arXiv:2204.12453
  [hep-ph]}}.

\bibitem{deBlas:2022hdk}
J.~de~Blas, M.~Pierini, L.~Reina, and L.~Silvestrini, ``{Impact of the recent
  measurements of the top-quark and W-boson masses on electroweak precision
  fits}'', \href{http://arxiv.org/abs/2204.04204}{{\ttfamily arXiv:2204.04204
  [hep-ph]}}.

\bibitem{Lu:2022bgw}
C.-T. Lu, L.~Wu, Y.~Wu, and B.~Zhu, ``{Electroweak Precision Fit and New
  Physics in light of $W$ Boson Mass}'',
  \href{http://arxiv.org/abs/2204.03796}{{\ttfamily arXiv:2204.03796
  [hep-ph]}}.

\bibitem{Asadi:2022xiy}
P.~Asadi, C.~Cesarotti, K.~Fraser, S.~Homiller, and A.~Parikh, ``{Oblique
  Lessons from the $W$ Mass Measurement at CDF II}'',
  \href{http://arxiv.org/abs/2204.05283}{{\ttfamily arXiv:2204.05283
  [hep-ph]}}.

\bibitem{Burgess:2009wm}
C.~P. Burgess, M.~Trott, and S.~Zuberi, ``{Light Octet Scalars, a Heavy Higgs
  and Minimal Flavour Violation}'',
  \href{http://dx.doi.org/10.1088/1126-6708/2009/09/082}{{\em JHEP} {\bfseries
  09} (2009) 082},
\href{http://arxiv.org/abs/0907.2696}{{\ttfamily arXiv:0907.2696 [hep-ph]}}.

\bibitem{Brivio:2021yjb}
I.~Brivio, S.~Dawson, J.~de~Blas, G.~Durieux, P.~Savard, A.~Denner, A.~Freitas,
  C.~Hays, B.~Pecjak, and A.~Vicini, ``{Electroweak input parameters}'',
  \href{http://arxiv.org/abs/2111.12515}{{\ttfamily arXiv:2111.12515
  [hep-ph]}}.

\bibitem{He:2013tla}
X.-G. He, H.~Phoon, Y.~Tang, and G.~Valencia, ``{Unitarity and vacuum stability
  constraints on the couplings of color octet scalars}'',
  \href{http://dx.doi.org/10.1007/JHEP05(2013)026}{{\em JHEP} {\bfseries 05}
  (2013) 026},
\href{http://arxiv.org/abs/1303.4848}{{\ttfamily arXiv:1303.4848 [hep-ph]}}.

\bibitem{Cheng:2018mkc}
L.~Cheng, O.~Eberhardt, and C.~W. Murphy, ``{Novel theoretical constraints for
  color-octet scalar models}'',
  \href{http://dx.doi.org/10.1088/1674-1137/43/9/093101}{{\em Chin. Phys. C}
  {\bfseries 43} no.~9, (2019) 093101},
  \href{http://arxiv.org/abs/1808.05824}{{\ttfamily arXiv:1808.05824
  [hep-ph]}}.

\bibitem{Cao:2013wqa}
J.~Cao, P.~Wan, J.~M. Yang, and J.~Zhu, ``{The SM extension with color-octet
  scalars: diphoton enhancement and global fit of LHC Higgs data}'',
  \href{http://dx.doi.org/10.1007/JHEP08(2013)009}{{\em JHEP} {\bfseries 08}
  (2013) 009},
\href{http://arxiv.org/abs/1303.2426}{{\ttfamily arXiv:1303.2426 [hep-ph]}}.

\bibitem{Miralles:2022jnv}
V.~Miralles, O.~Eberhardt, H.~Gisbert, A.~Pich, and J.~Ruiz-Vidal, ``{Global
  fit on coloured scalars including the last W-boson mass measurement}'', in
  {\em Electroweak session of the 56th Rencontres de Moriond}.
\newblock 5, 2022.
\newblock \href{http://arxiv.org/abs/2205.05610}{{\ttfamily arXiv:2205.05610
  [hep-ph]}}.

\bibitem{Andreev1}
V.~A. Andreev {\em et~al.}, ``{SPATIAL FOCUSING OF 1-GeV PROTONS BY A BENT
  MONOCRYSTAL}'', {\em JETP Lett.} {\bfseries 41} (1985) 500.

\bibitem{Denisov1}
A.~S. Denisov, A.~I. Smirnov, V.~I. Baranov, Y.~A. Chesnokov, V.~I. Kotov, {\em
  et~al.}, ``{First results from a study of a 70 GeV proton beam being
  focused}'', {\em Nucl. Instrum. Meth. B} {\bfseries 69} (1992) 382.

\bibitem{Denison:1991vf}
A.~S. Denisov {\em et~al.}, ``{First results on studying 70-GeV proton beam
  focusing by bent crystal}'', {\em JETP Lett.} {\bfseries 54} (1991) 487--490.

\bibitem{Pivk:2004ty}
M.~Pivk and F.~R. Le~Diberder, ``{sPlot: A statistical tool to unfold data
  distributions}'', \href{http://dx.doi.org/10.1016/j.nima.2005.08.106}{{\em
  Nucl. Instrum. Meth.} {\bfseries A555} (2005) 356--369},
  \href{http://arxiv.org/abs/physics/0402083}{{\ttfamily arXiv:physics/0402083
  [physics.data-an]}}.

\bibitem{Sjostrand:2007gs}
T.~Sjostrand, S.~Mrenna, and P.~Z. Skands, ``{A Brief Introduction to PYTHIA
  8.1}'' \href{http://dx.doi.org/10.1016/j.cpc.2008.01.036}{{\em Comput. Phys.
  Commun.} {\bfseries 178} (2008) 852--867},
\href{http://arxiv.org/abs/0710.3820}{{\ttfamily arXiv:0710.3820 [hep-ph]}}.

\bibitem{Clemencic_2011}
M.~Clemencic, G.~Corti, S.~Easo, C.~R. Jones, S.~Miglioranzi, M.~Pappagallo,
  and P.~R. and, ``The {LHCb} simulation application, gauss: Design, evolution
  and experience'',
  \href{http://dx.doi.org/10.1088/1742-6596/331/3/032023}{{\em Journal of
  Physics: Conference Series} {\bfseries 331} no.~3, (Dec, 2011) 032023}.
  \url{https://doi.org/10.1088/1742-6596/331/3/032023}.

\bibitem{Dicus:1989va}
D.~A. Dicus, ``{Neutron Electric Dipole Moment From Charged Higgs Exchange}'',
  \href{http://dx.doi.org/10.1103/PhysRevD.41.999}{{\em Phys. Rev. D}
  {\bfseries 41} (1990) 999}.

\bibitem{Kublbeck:1990xc}
J.~Kublbeck, M.~Bohm, and A.~Denner, ``{Feyn Arts: Computer Algebraic
  Generation of Feynman Graphs and Amplitudes}'',
  \href{http://dx.doi.org/10.1016/0010-4655(90)90001-H}{{\em Comput. Phys.
  Commun.} {\bfseries 60} (1990) 165--180}.

\bibitem{Shtabovenko:2016sxi}
V.~Shtabovenko, R.~Mertig, and F.~Orellana, ``{New Developments in FeynCalc
  9.0}'' \href{http://dx.doi.org/10.1016/j.cpc.2016.06.008}{{\em Comput. Phys.
  Commun.} {\bfseries 207} (2016) 432--444},
  \href{http://arxiv.org/abs/1601.01167}{{\ttfamily arXiv:1601.01167
  [hep-ph]}}.

\bibitem{Ilisie:2015tra}
V.~Ilisie, ``{New Barr-Zee contributions to $\mathbf{(g-2)_\mu}$ in
  two-Higgs-doublet models}'',
  \href{http://dx.doi.org/10.1007/JHEP04(2015)077}{{\em JHEP} {\bfseries 04}
  (2015) 077}, \href{http://arxiv.org/abs/1502.04199}{{\ttfamily
  arXiv:1502.04199 [hep-ph]}}.

\bibitem{wikipediaSM}
V.~\href{https://ca.wikipedia.org/wiki/Model_est%C3%A0ndard_de_f%C3%ADsica_de_part%C3%ADcules}{https://ca.wikipedia.org/wiki/Model\_estàndard\_de\_física\_de\_partícules}.
\newblock
  \url{https://ca.wikipedia.org/wiki/Model_est%C3%A0ndard_de_f%C3%ADsica_de_part%C3%ADcules}.

\bibitem{Koppenburg:2015pca}
P.~Koppenburg and V.~Vagnoni, ``{Precision physics with heavy-flavoured
  hadrons}'', \href{http://dx.doi.org/10.1142/9789814644150_0002}{{\em Adv.
  Ser. Direct. High Energy Phys.} {\bfseries 23} (2015) 31--59},
  \href{http://arxiv.org/abs/1510.04466}{{\ttfamily arXiv:1510.04466
  [hep-ex]}}.

\end{thebibliography}\endgroup

\end{document}